\documentclass[a4paper,12pt]{book}
\usepackage{citeref}
\usepackage[nohug,heads=littlevee]{diagrams}
\diagramstyle[labelstyle=\scriptstyle]
\usepackage{pst-spectra}
\usepackage{pstricks-add}
\usepackage{pst-plot}
\usepackage{pst-all}
\psset{algebraic=true}

\advance\textheight2.7cm        
\advance\topmargin-2.0cm
\advance\textwidth2.6cm

\def\bsigma{{\bm\sigma}}
\def\bomega{{\bm\omega}}
\def\nn{\nonumber}

\def\at#1{[*** \att #1 ***]}
\def\at#1{}           
\def\MMt#1{~\\*MM$\to$~~~#1 ~\att\\}  
\def\MMt#1{}                      

\def\bebib{\begin{thebibliography}{999}\addcontentsline{toc}{chapter}{References}}
\def\ebib{\end{thebibliography}}

\begin{document}

    \frontmatter

    \thispagestyle{empty}
    \begin{titlepage}

\author{\bf \Large Arnold Neumaier\\~\\\bf \Large Dennis Westra\\~\\
University of Vienna, Austria} 
\title{{\bf \Huge Classical and quantum mechanics\\~\\via Lie algebras}}
\date{~\\~\\~\\
\vspace{2cm}
April 14, 2011\\
~\\
This is the draft of a book. The manuscript has not yet full book 
quality.\\ 
Please refer to the book once it is published.\\~\\
Until then, we'd appreciate suggestions for improvements; 
please send them to Arnold.Neumaier@univie.ac.at\\
\vspace{2cm}
copyright \copyright~ by Arnold Neumaier and Dennis Westra
}

\maketitle

\end{titlepage}
    \newpage~
    \thispagestyle{empty}
    \newpage
    \thispagestyle{empty}
    \pagenumbering{roman} 
    \tableofcontents 
    
{\Large \bf Preface}
\addcontentsline{toc}{chapter}{Preface}

This book presents classical mechanics, quantum mechanics, and 
statistical mechanics in an almost completely algebraic setting, 
thereby introducing mathematicians, physicists, and engineers to the 
ideas relating classical and quantum mechanics with Lie algebras and 
Lie groups.

The book should serve as an appetizer, inviting the reader to go more
deeply into these fascinating, interdisciplinary fields of science.

Much of the material covered here is not part of standard 
textbook treatments of classical or quantum mechanics (or is only 
superficially treated there). For physics students who want to 
get a broader view of the subject, this book may therefore serve 
as a useful complement to standard treatments of quantum mechanics.

We motivate everything as far as possible by classical mechanics.
This forced an approach to quantum mechanics close to Heisenberg's 
matrix mechanics, rather than the usual approach dominated by 
Schr\"odinger's wave mechanics. Indeed, although both approaches are 
formally equivalent, only the Heisenberg approach to quantum mechanics
has any similarity with classical mechanics; and as we shall see, 
the similarity is quite close. Indeed, the present book emphasizes the 
closeness of classical and quantum mechanics, and the material is
selected in a way to make this closeness as apparent as possible.

Almost without exception, this book is about precise concepts and 
exact results in classical mechanics, quantum mechanics, and 
statistical mechanics. The structural properties of 
mechanics are discussed independently of computational techniques for 
obtaining quantitatively correct numbers from the assumptions made. 
This allows us to focus attention on the simplicity and beauty of 
theoretical physics, which is often hidden in a jungle of techniques 
for estimating or calculating quantities of interests. 
The standard approximation machinery for calculating from first 
principles explicit thermodynamic properties of materials, or 
explicit cross sections for high energy experiments can be found in 
many textbooks and is not repeated here. 

Compared with the 2008 version, most of Chapters 2--3 and all of 
Chapters 14--18 are new; the remaining chapters were slightly 
improved. 

\bigskip
The book originated as course notes from a course given by the first
author in fall 2007, written up by the second author, and expanded
and polished by combined efforts, resulting in a uniform whole that
stands for itself. Parts II and IV are mainly based on earlier work 
by the first author (including \sca{Neumaier} \cite{Neu.ens,Neu.therm});
and large parts of Part I were added later.
The second author acknowledges support by the Austrian 
FWF-projects START-project Y-237 and IK 1008-N.
Thanks go to Roger Balian, Clemens Elster, Martin Fuchs, Johann Kim,
Mihaly Markot, Mike Mowbray, Hermann Schichl, Peter Schodl, and 
Tapio Schneider, 
who contributed through their comments on earlier versions of parts of 
the book.

The audience of the course consisted mainly of mathematics students
shortly before finishing their diploma or doctorate degree and a few
postgraduates, mostly with only a limited background knowledge in
physics.

Thus we assume some mathematical background knowledge, but only a
superficial acquaintance with physics, at the level of what is
available to readers of the Scientific American, say.
It is assumed that the reader has a good command of matrix algebra 
(including complex numbers and eigenvalues) and knows basic properties 
of vector spaces, linear algebra, groups, differential equations, 
topology, and Hilbert spaces. No background in Lie algebras, Lie groups,
or differential geometry is assumed. Rudiments of differential 
geometry would be helpful to expand on our somewhat terse treatment
of it in Part V; most material, however, is completely independent of 
differential geometry.

While we give precise definitions of all mathematical concepts
encountered (except in Chapter 4, which is taken verbatim from the 
theoretical physics FAQ), and an extensive index of concepts and 
notation, we avoid the deeper use of functional analysis and 
differential geometry without being mathematically inaccurate, 
by concentrating on situations that have 
no special topological difficulties and only need a single chart.
But we mention where one would have to be more careful
about existence or convergence issues when generalizing to infinite
dimensions.

On the physics side, we usually first present the mathematical models 
for a physical theory before relating these models to reality. 
This is adequate both for mathematically-minded readers without 
much physics knowledge and for physicists who know already on a more 
elementary level how to interpret the basic settings in terms of 
real life examples.

\bigskip
This is an open-ended book. It should whet the appetite for more, and 
lead the reader into going deeper into the subject.\footnote{
Some general references for further reading:
\sca{Barut \& Raczka} \cite{barutraczka}, 
\sca{Cornwell} \cite{cornwell},
\sca{Gilmore} \cite{gilmore}, and
\sca{Sternberg} \cite{sternberg},
for the general theory of Lie algebras, Lie groups, and 
their representations from a physics point of view,
\sca{Wybourne}  \cite{wybourne} and 
\sca{Fuchs \& Schweigert} \cite{fuchsandschweigert}
for a more application oriented view of Lie algebras,
\sca{Kac} \cite{kac} and 
\sca{Neeb} \cite{Nee} for infinite-dimensional Lie algebras,
\sca{ Papou{\v s}ek \& Aliev} \cite{PapA} for quantum mechanics and 
spectroscopy,
\sca{van der Waerden} \cite{vanderWaerden} for the history of quantum 
mechanics, 
and
\sca{Weinberg} \cite{weinberg} for a (somewhat) Lie algebra oriented 
treatment of quantum field theory.
}\,
Thus many topics
are discussed far too short for a comprehensive treatment, and often
only the surface is scratched. A term has only this many hours, and 
our time to extend and polish the lectures after they were given was 
limited, too. We added some material, and would have liked to be more 
complete in many respects. Nevertheless, we believe that the topics
treated are the fundamental ones, whose understanding gives a solid
foundation to assess the wealth of material on other topics.

We usually introduce physical concepts by means of informal historical
interludes, and only discuss simple physical situations in which the
relevant concepts can be illustrated.
We refer to the general situation only by means of remarks; however,
after reading the book, the reader should be able to go deeper into the
original literature that treats these topics in greater physical
depth.

Part I is an invitation to quantum mechanics, concentrating on giving
motivation and background from history, from classical mechanics, and 
from simple, highly symmetric quantum systems. The latter are used to 
introduce the most basic Lie algebras and Lie groups.
Part II gives a thorough treatment of the formal part of equilibrium 
statistical mechanics, viewed in the present context as the common 
core of classical and quantum mechanics, and discusses the 
interpretation of the theory in terms of models, statistics and 
measurements.
Part III introduces the basics about Lie algebras and Poisson algebras,
with an emphasis on the concepts most relevant to the conceptual side
of physics. 
Part IV discusses the dynamics of nonequilibrium phenomena, i.e., 
processes where the expectation changes with time, in as far as no 
fields are involved. This results in a dissipative dynamics.
Part V introduces the relevant background from 
differential geometry and applies it to classical Hamiltonian and 
Lagrangian mechanics, to a symplectic formulation of quantum mechanics,
and to Lie groups.
Part VI applies the concepts to the study of quantum oscillators 
(bosons) and spinning systems (fermions), and to the analysis of 
empirically observed spectra, concentrating on the mathematical
contents of these subjects.
The book concludes with numerous references and an index of all 
concepts and symbols introduced.
For a more detailed overview of the topics treated, see Section 
\ref{s.outline}.

\bigskip
We hope that you enjoy reading this book!

\bigskip
\hfill Wien, \today

\hfill Arnold Neumaier, Dennis Westra

\newpage
\at{{\bf Things still to be done:}\newline
The arXiv version must have ''today'' replaced by a fixed date!\newline
also remove showlabels.sty, disable the macros \at and \MMt,\newline
If Chapter or Part order changes, the preface, the last section of 
     Chapter 1, and the head of the first chapter in a part must be 
     updated.\newline
Add indexing for the new chapters.\newline
Address the at signs.\newline
Table of symbols -- as part of the index. \newline
Add explanations to some of the symbols (I did it for $B_{CK}$)\newline
Polish the references:\newline
some are duplicate\newline
many are incomplete\newline
many capitals in names or German titles etc. are missing\newline
some Umlauts are missing\newline
sometimes the last page is not given\newline
first author names shoud be initials only\newline
''Author1 [no1,no2], author [no3]'' statt 
''Author1, Author1, Author2 [no1,no2,no3]''\newline
''$\backslash sca\{A1 \& A2\}$ [no] in place of 
        $\backslash sca\{A1\mbox{ and }A2\}$ [no], etc.\newline
Sometimes parentheses around citations are needed.\newline
Check figures to print well in black and white\newline
modelled vs. modeled\newline
replace $\vec x$ by $\x$ (often)\newline
replace $\dagger$ by $*$ (often)\newline
avoid $\mapsto$ (often)\newline
opening apostrophs are often ' instead of `\newline
polish intros; work on the black marks\newline
section x.y should be capitalized, etc.\newline
Einstein convention -- note it in each context where it is used!\newline
Section \gzit{c.spin} should be Section \ref{c.spin}, etc.;\newline
also for Chapter references\newline
remove showlabels, confidential, variable dates in title and preface, 
showidx\newline
How to get the indices into the table of contents?\newline
double backslash does no longer act like newline\newline
~} 

    \mainmatter
    \part{An invitation to quantum mechanics}\label{p.invit}
\chapter{Motivation}\label{c.motivation}

Part \ref{p.invit} is an invitation to quantum mechanics, concentrating 
on giving
motivation and background from history, from classical mechanics, and 
from 2-state quantum mechanics.

\bigskip
The first chapter is an introduction and serves as a motivation for the
following chapters. We shall go over different areas of physics and
give a short glimpse on the mathematical point of view.

The final section of the chapter outlines the content of the whole
book.

For the mathematicians most of the folklore vocabulary of physicists 
may not be familiar, but later on in the book, precise definitions in 
mathematical language will be given. Therefore, there is no need to 
understand everything in the first chapter on
first reading; we merely introduce informal names for certain concepts
from physics and try to convey the impression that these have
important applications to reality and that there are many interesting 
solved and unsolved mathematical problems in many areas of theoretical 
physics.\footnote{
  We encourage readers to investigate for themselves some of the
  abundant literature to get a better feeling and more understanding
  than we can offer here.
  }

\section{Classical mechanics}

Classical mechanics is the part of physics whose development started 
around the time of Isaac Newton (1642-1727).

It was in the period of Newton, Leibniz, and Galileo that
classical mechanics was born, mainly studying planetary motion.
Newton wanted to understand why the earth seems to circle around the 
sun, why the moon seems to circle around the earth, and why apples 
(and other objects) fall down. By analyzing empirical
data, he discovered a formula explaining most of the observed
phenomena involving gravity.
Newton realized that the laws of physics here on earth are the
same as the laws of physics determining the motion of the planets.
This was a major philosophical breakthrough.

The motion of a planet is described by its position and velocity at
different times. With the laws of Newton it was possible to deduce a
set of differential equations involving the positions and velocities
of the different constituents of the solar system. Knowing exactly all
positions and velocities at a given time, one could in principle
deduce the positions and velocities at any other time.

\at{grep for every instance of the word
"could" in the entire book and check whether one can use
"can". Also "will" or "would" to present tense.}

Our solar
system is a well-posed initial value problem (IVP). However, an
initial error in position and velocities at time $t_0=0$ grows 
exponentially at time $t>0$ by a factor of $\sim e^{\lambda t}$. 
The value of $\lambda$ varies for different initial conditions; its
maximum is called the
\bfi{maximal Lyapunov exponent}. A system with maximal
Lyapunov exponent $\lambda=0$ is called \bfi{integrable}.
If $\lambda<0$ the solutions converge
to each other and if $\lambda>0$ the solutions move away from each
other. The solar system is apparently not quite integrable: according 
to numerical simulations, the maximal Lyapunov exponent for our solar 
system seems to be small but positive, with 
$\lambda^{-1}$ being about five million years
(\sca{Laskar} \cite{laskar1,laskar2}, \sca{Lissauer} \cite{lissauer}).

Frequently, instead of considering separately time-dependent positions 
and velocities of many objects (e.g., planets, atoms, or particles) 
in a system, it is more convenient to work with single 
\bfi{trajectories}, paths parameterized by time, in a high-dimensional 
space called the \bfi{phase space} of the system. 
In the case of the planetary system with $N$ planets, the points in 
phase space are described by vectors
which have $6N$ components grouped into $N$ pairs consisting of
three components for position and three components for \bfi{momentum},
velocity multiplied by mass, of each particle. One reason that one 
prefers momentum before velocity is that the total momentum of all 
particles is \bfi{conserved}, i.e., remains constant in time.
A deeper reason that will become apparent later is that on the level
of position and momentum, the similarity between classical and quantum 
mechanics is most apparent. For a single particle moving in space,
there are three spatial directions to specify its position and three
directions to specify the velocity. Hence the phase space of a (free)
particle is six-dimensional; for a system of $N$ astronomical bodies,
the dimension is $6N$.

Low-dimensional phase spaces are well-understood in general. Newton
showed that the configuration of a single planet moving around the sun
is stable (in fact, the system is integrable) and motion follows 
\bfi{Kepler's' laws}, which were already known before, thus giving 
these a theroetical basis. Higher-dimensional phase spaces tend to
cause problems. Indeed, for more planets (that is, more than 2 bodies
in the system), deviations from elliptic motions are predicted, and the
question of stability was open for a long time. The Swedish king Oskar
II was willing to reward with a big amount of money the scientist who
proved stability of our solar system.

However, Poincar\'e showed that already three objects (one sun, two
planets) cause big problems for a possible stability proof of our solar
system and received the prize in 1887. The numerical studies from the
end of the last century (quoted above) strongly indicate that the solar
system is unstable, though a mathematical proof is missing.

\bigskip
\at{Sub-section needed here. 
In fact, maybe a previous subsection heading should say
"finite-dimensional classical mechanics" explicitly. I.e., try to
signal the distinction with a more blatant signpost.}

We now turn from celestial mechanics, where phase space is
finite-dimensional, to \bfi{continuum mechanics}, which has to cope
with infinite-dimensional phase spaces. For example, to describe a
fluid, one needs to give the distribution of mass and energy and the
local velocity for all (infinitely many) points in the fluid. The
dynamics is now governed by partial differential equations. In
particular, fluid mechanics is dominated by the \bfi{Navier--Stokes
equations}, which still presents a lot of difficult mathematical 
problems.

Showing that solutions exist for all times (not only short-time
solutions) is one of the Clay Millennium problems (see, e.g.,
\sca{Ladyzhenskaya} \cite{lady}), and will be rewarded by one million
dollars.

The infinitely many dimensions of the phase space cause serious
additional problems. The Lyapunov exponents now depend on where the
fluid starts in phase space and for fast-flowing fluids, the
maximal Lyapunov exponent is much larger than zero in most regions of
phase space. This results in a phenomenon called
\bfi{turbulence}, well-known from the behavior of water.
The notion of turbulence is still not well understood mathematically.
Surprisingly enough, the problems encountered with turbulence are of
the same kind as the problems encountered in quantum field theories
(QFT's) -- one of the many instances where a problem in classical
mechanics has an analogue in quantum physics.

Another area of continuum mechanics is \bfi{elasticity theory}, where
solids are treated as continuous, nearly rigid objects, which deform
slightly under external forces. The rigidity assumption is easily
verified empirically; try to swim in metal at room temperature\dots.
Due to the rigidity assumption the behavior is much better understood
mathematically than in the fluid counterpart.

The configuration of a solid is close to equilibrium and the deviations
from the equilibrium position are strongly suppressed (this {\em is}
rigidity). Hence the rigidity assumption implies that linear Taylor
approximations work well since the remaining terms of the Taylor series
are small, and the Lyapunov exponent is zero.

Elasticity theory is widely applied in engineering practice.
Modern bridges and high rise buildings would be impossible without the
finite element analyses which determine their stability and their
vibration modes. In fact, the calculations are done in
finite-dimensional discretizations, where much of the physics is
reducible to linear algebra. Indeed, all continuum theories are
(and have to be) handled computationally in approximations with only
finitely many degrees of freedom; in most areas very successfully.
The mathematical difficulties are often related to establishing a
valid continuum limit.

\section{Relativity theory}

In the period between 1900 and 1920, classical mechanics was enriched 
with \bfi{special relativity theory} (SRT) and \bfi{general relativity
theory} (GRT). In SRT and GRT, space and time merge into a
four-dimensional manifold, called space-time.

In SRT, space-time is flat. 
Distances in space and time are measured with the Minkowski 
metric, an indefinite metric (discussed in more detail in 
Section \ref{s.lorentz}) which turns space-time a pseudo-Riemannian 
manifold. Different observers (in SRT) all see the
same speed of light. But they see the same distances only when
measured with the Minkowski metric -- not with the Euclidean spatial or
temporal metric (which also holds for the orthogonality mentioned
above). It follows that spatial separation and temporal separation
between localized systems (for example a chicken laying an egg and an 
atom splitting) are different for different observers! 
But the difference is observable only when the two systems move at 
widely different velocities,
hence Newton couldn't have noticed this deviation from his theory.

In classical mechanics,
time is absolute in the sense that there exists a global time up to
time shifts; the time difference between two events is the same in
every coordinate system. The
symmetries of classical space-time is thus the group generated by
time-translations, space translations and rotations. This group is the
Galilean group. Due to the experimental fact that the speed of light in
vacuum is the same for all observers led Einstein to the conclusion
that time is not absolute and that the Galilean group should be
enlarged with transformations that rotate space and time coordinates
into each other. The result was the theory of special relativity. 
Due to special relativistic effects in the quantum theory, the world
indeed looks different; for example, without special relativity,
gold would be white, and mercury would be solid at room temperature
\sca{Norrby} \cite{norrby}.

\bigskip
SRT is only valid if observers move at fixed velocities with
respect to each other. To handle observers whose relative velocities 
may vary requires the more general but also more complex GRT. 
The metric now depends on the space-time point; it becomes a
nondegenerate symmetric bilinear form on the space-time manifold. 
The transformations (diffeomorphisms)
relating the metric in one patch to the metric in another patch cannot
change the signature. Hence the signature is the same for all observers.

The changing metric has the effect that in GRT, space-time is no longer 
flat, but has curvature. That means that freely moving objects do
not follow real straight lines -- in fact the notion of what straight
means is blurred. The trajectory that an object in a free fall, where
no forces are exerted on the object, will follow is called a
\bfi{geodesic}. The geodesics are determined by the geometry by means 
of a second-order differential equation. The preferred
direction of time on a curved space-time is now no longer fixed, or as
mathematicians say 'canonical', but is determined by the
observer: The geodesic along the observers' 4-momentum vector
defines the world line of the observer (e.g., a measuring instrument) 
and with it its time; the space-like surfaces
orthogonal to the points on the world line define the observer's
3-dimensional space at each moment. When the observer also defines
a set of spatial coordinates around its position, and a measure of time
(along the observers' geodesic), one can say that a chart around
the observer has been chosen.

When time becomes an observer dependent quantity, so becomes energy.
Local energy conservation is still well defined, described by a 
conservation law for the resulting differential equations. The
differential equations are covariant,
meaning that they make sense in any coordinate system.  For a large
system in general relativity, the definition of a total energy which is
conserved, i.e., time-independent, is however problematic, and 
well-defined only if the system
satisfies appropriate boundary conditions such as asymptotic flatness,
believed to hold for the universe at large.
Finally, if the system is dissipative, there is energy
loss, and the local conservation law is no longer valid.
Not even the rate of energy loss is
well defined. Dissipative general relativity has not yet found
its final mathematical form. \at{add some references}

\section{Statistical mechanics and thermodynamics}\label{s.thstat}

\bfi{Thermodynamics} describes ordinary matter on the level of ordinary 
experience. Loosely speaking, it is the theory of the quantitative 
relations between volume and pressure, temperature and heat, and how 
this characterizes different substances in their different phases -- 
as solid, fluid, or gas.
\bfi{Statistical mechanics} is the part of physics that derives the 
macroscopic properties of matter -- which themselves are primarily 
described by thermodynamics -- from microscopic principles. 

An important ingredient in classical statistical mechanics is a
\bfi{phase space density} $\rho$ playing the role of a measure to
calculate probabilities; the expectation value of a function $f$
is given by
\lbeq{e.expect}
\< f \> := \sint \rho f,
\eeq
where the integral indicates integration with respect to the so-called
Liouville measure in phase space.

In the quantum version of statistical mechanics the density $\rho$
gets replaced by a linear operator $\rho$ on Hilbert space
called the \bfi{density matrix}, the functions become linear operators,
and we have again \gzit{e.expect}, except that the integral is now
interpreted as the \bfi{quantum integral},
\lbeq{e.quantint}
\sint f = \tr f,
\eeq
where $ \tr f$ denotes the trace of a trace class operator.

We shall see that the algebraic properties of the classical integral
and the quantum integral are so similar that using the same name and
symbol is justified.

A deeper justification for the quantum integral becomes visible
if we introduce the\index{$\lp$}\index{Lie product} 
\bfi{Lie product}\footnote{\label{f.lp}
The symbol $\lp$\index{$\lp$, Lie product}, frequently used in the 
following, is interpreted as a stylized capital letter L and should be 
read as ''Lie''.
}
\lbeq{e.lp}
f \lp g := \cases{
   \{g,f\} & in the classical case,\cr
   \frac{i}{\hbar}[f,g] & in the quantum case,
   }
\eeq
unifying the classical \bfi{Poisson bracket} 
\[
\{f,g\}:=\partial_q f \cdot \partial_p g - 
\partial_q g\cdot \partial_p f
\]
on the algebra $\Ez=C^\infty(\Omega)$ of smooth functions on phase 
space $\Omega=\Rz^{3N}\times\Rz^{3N}$, 
and the quantum \bfi{commutator} 
\[
[f,g]:=fg-gf
\]
on the algebra $\Ez=\Lin C^\infty(\Rz^{3N})$ of linear operators 
on the space of smooth functions on configuration space. 
(Here $i=\sqrt{-1}$ -- complex numbers will figure prominently in this
book! --, and $\hbar$ is {\bfi{Planck's constant}}\index{$\hbar$} 
in the form introduced by \sca{Dirac} \cite{diracbook}. 
Planck had used instead the constant $h=2\pi \hbar$ which caused many 
additional factors of $2\pi$ in formulas discovered later.)
The Lie product is in both cases an antisymmetric bilinear map from 
$\Ez\times \Ez$ to $\Ez$ and satisfies the \bfi{Jacobi identity}; 
see Chapter \ref{c.lie} for precise definitions.

In the classical case, the fact that integration and
differentiation are inverse operations implies that the integral of
a derivative of a function vanishing at infinity is zero.
The traditional definition of the Poisson bracket therefore implies
that, for functions $f,g$ vanishing at infinity,
\lbeq{e.intdiff}
 \sint f\lp g =0.
\eeq
Remarkably, in the quantum case, \gzit{e.intdiff} is valid for
Hilbert--Schmidt operators $f,g$ since then $\tr fg =\tr gf$, so that
$\sint f\lp g=\tr \frac{i}{\hbar}[f,g]
=\frac{i}{\hbar}(\tr fg - \tr gf)=0$. Thus the quantum integral
behaves just like the Liouville integral!

Thus we see that there is a very close parallel between the classical 
and the quantum case. Indeed, statistical mechanics is the area where 
classical mechanics and quantum mechanics meet most closely, and hence 
an area of central interest for our book. This field, growing out of 
the desire to seek a more fundamental understanding of thermodynamics,
was developed especially during the industrial
revolution in England. Maxwell wrote many papers on a mathematical
foundation of thermodynamics. With the establishment of a
molecular world view the thermodynamical machinery slowly got
replaced by statistical mechanics, where the macroscopic
properties like heat capacity, entropy, temperature were explained
through considerations of the statistical properties of a big
population of particles. The first definite treatise on statistical
thermodynamics is by \sca{Gibbs} \cite{Gib}\footnote{
Today, this book from 1902 is still easily readable.}
who also invented much of the modern mathematical notation in physics, 
especially the notation for vector analysis.

Quantum mechanics and classical mechanics look almost the same
when viewed in the context of statistical mechanics; indeed, Gibbs'
account of statistical mechanics had to be altered very little after
quantum mechanics revolutionized the whole of science.
In this course, we shall always emphasize the closeness of classical
and quantum reasoning, and put things in a way to make this closeness
as apparent as possible.

\section{Hamiltonian mechanics}\label{s.hammech}

Understanding statistical physics requires the setting of
\bfi{Hamiltonian mechanics}, which exists both in a classical and a
quantum version.

Much of classical mechanics can be understood in both a Hamiltonian 
formulation and a Lagrangian formulation; cf. Chapter \ref{c.pmani}.
In the Hamiltonian formulation, the basic object is the 
\bfi{Hamiltonian} function
$H$ on the phase space $\Omega$, which gives the value of the energy
at an arbitrary point of phase space. Specifying $H$ essentially
amounts to specifying the physical system under consideration.
Often the phase space $\Omega$ is the \bfi{cotangent bundle} $T^*M$ of 
a manifold $M$. 
In the Lagrangian formulation, the main object is a \bfi{Lagrangian
  function} $L$ on the \bfi{tangent bundle} $TM$ of a manifold $M$.
The Lagrangian is thus not a function on phase space, and has no
simple physical interpretation, but in some sense, it plays a more
fundamental role than the Hamiltonian since it survives the transition
to the currently most fundamental physical theory, quantum field theory.
The physics point of view on quantum field theory is expounded in many 
books, perhaps explained most thoroughly in the volumes by 
{\sc Weinberg} \cite{weinberg}. 
For an introduction to quantum field theory from a mathematician's 
point of view, see {\sc Zeidler} \cite{zeidler}; this book draws
connections to many topics of contemporary mathematics, and points out
basic unresolved marthematical issues in quantum field theory.
Both books together provide complementary perspectives on the subject.

The passage between the Hamiltonian and the Lagrangian formulation is
straightforward in simple cases but may cause problems, especially in
so-called \bfi{gauge theories} (pronounced in English transcription
as {\em gaidge}). Gauge theories are outside the scope of this book;
\at{except for a short glimpse in Chapter 12, cf. Example 
\gzit{ex.gauge}} the curious reader is referred to the vast literature.

In the Hamiltonian
formulation the time-dependence of a function $f$ on the
phase space is determined by the \bfi{classical Heisenberg equation}
\lbeq{e.cHeis} 
\frac{df}{dt} = H \lp f =  \left\{f,H\right\}\,, 
\eeq
where $H$ is the Hamiltonian function defined above. 
Important to note is that the Hamiltonian
function determines the time-evolution. If we can solve these
differential equations, this defines an operator $U(s,t)$ that
maps objects at a time $s$ to corresponding objects at a time $t$.

Clearly, the composition of the operators gives 
$U(s,s')U(s',t)=U(s,t)$. If the
Hamiltonian is independent of time (this amounts to assuming that
there are no external forces acting on the system) the maps $U$
form a so-called one-parameter group, since one can write
\[
U(s,t)=e^{(t-s)\ad_H}\,,
\]
with the associated {\bf Hamiltonian vector field} $\ad_H$, which is 
determined by $H$. The vector field $\ad_H$ generates shifts in time.
In terms of $\ad_H$, the multiplication is given by
\[
e^{t\ad_H}e^{s\ad_H}=e^{(t+s)\ad_H}\,,
\]
and the inverse is given by
\[
U(r,s)^{-1}=U(s,r)\,, \quad  (e^{t\ad_H})^{-1}=e^{-t\ad_H}\,.
\]
From a mathematical point of view, the physical Hamiltonian is  
just one of many  Hamiltonian functions that can be used in the
above discussion.
Given another Hamiltonian function $H'$ we get another Hamiltonian
vector field $\ad_{H'}$, another one-parameter group, and a
``time'' parameter $t$ with a different physical interpretation
(if one exists). For example, if $H'$ is a component of the
momentum vector (or the angular momentum vector) then $t$ corresponds to
a translation (or rotation) in the corresponding coordinate direction
by a distance (or angle) $t$.
Combining these groups for all $H'$, where the initial value problem
determined by $\ad_{H'}$ is well posed, we get an infinite-dimensional
Lie group. (See Sections \ref{s.LinLie} and \ref{s.Liegroups} for a
definition of Lie groups.) 
Thus, classical mechanics can be understood in terms of
infinite-dimensional groups!

\at{You say "...the physical Hamiltonian is
just one of many Hamiltonian functions..." which suggests that
the other "Hamiltonians" $H'$ are "unphysical", but then this
contradicts the subsequent words "...with a different physical
interpretation". My guess is that you're trying to introduce the
concept of generalized motions in phase space, their generators, and
associated parameters. If so, it's deeply confusing to say that
another "Hamiltonian" $H'$ might be a component of the (angular)
momentum vector. Better to mention displacements and rotations
more explicitly, and then talk about how there's a common idea
underlying them all, rather than blurring the meaning of
"Hamiltonian" as generator of time evolution.}

To avoid technical complications, 
we shall however mainly be concerned with the cases where we can 
simplify the system such that the groups are finite-dimensional. 
In the present case, to obtain a finite-dimensional group one either 
picks a nice subgroup (this involves understanding the
symmetries of the system) or one makes a partial discretization of
phase space.

\bigskip
Most of our discussions will be restricted to conservative systems,
which can be described by Hamiltonians. However, these only describe
systems which can be regarded as isolated from the environment,
apart from influences that can be specified as external forces.
Many real life systems (and strictly speaking all systems with the
exception of the universe as a whole) interact with the environment --
indeed, if it were not so, they would be unobservable!

Ignoring the environment is possible in so-called reduced descriptions,
where only the variables of the system under consideration are kept.
This usually results in differential equations which are dissipative.
In a {\bfi{dissipative system}}, the energy dissipates, which
means that some energy is lost into the unmodelled environment.
Due to the energy loss, going back in time
is not well defined in infinite-dimensional spaces; the initial value
problem is solvable only in the forward time direction. Hence we can
not find an inverse for the translation operators $U(s,t)$, and these 
are defined only for $s\le t$. Therefore, in the most general case of 
interest, the dissipative infinitesimal generators do not generate a 
group, but only a semigroup. A well-known example is heat propagation, 
described by the heat equation.
Its solution forward in time is a well-posed initial value
problem, whereas the solution backward in time suffers from
uncontrollable instability. Actually, many dissipative systems are
not even described by a Hamiltonian dynamics, but the semigroup
property of the flow they generate still remains valid.

\section{Quantum mechanics}\label{s.motQM}

The historical point of view on quantum mechanics is that it is a 
deformation of classical mechanics obtained by a process called 
quantization. The deformation parameter is Planck's constant
$\hbar$.

Since in daily life we do not really see so much of quantum
physics, one requires that the
so-called \bfi{correspondence principle} holds. The correspondence
principle states that the formulas of quantum physics should turn into
corresponding formulas of classical mechanics in the limit when 
$\hbar \rightarrow 0$. This limit is called the \bfi{classical limit}.
However, $\hbar$ is a constant of Nature and not a parameter that can 
be changed in experiments; thus this `limit' must be properly 
interpreted. The \bfi{action} of a physical system is a certain 
functional $S$ that gives rise to equations of motion and is measured in
units of $\hbar$. The right way to think of the classical limit is 
therefore the limit when the dimensionless quotient $S/\hbar$
becomes arbitrarily large. The classical limit therefore amounts to
considering the limiting case of very large $S$. 
Thus classical mechanics
appears as the limit of quantum mechanics when all values of the action 
involved are huge when measured in units of $\hbar$, which is the case 
when the systems are sufficiently large.

\at{To symmetries? 
Here you could perhaps point out a parallel with the Newton/SR
case: the deformation parameter there is $v/c$ which involves not only
the physical constant $c$ but also a variable characteristic velocity
$v$ of the system under study. In both cases the physical constant
introduced in the new theory establishes a "scale" associated with some
property of the system, where the differences between the old and new
theories become significant.}

Keeping only the linear orders of $\hbar$ one obtains so-called 
\bfi{semiclassical approximations}, which is intermediate between 
quantum mechanics and the classical limit, and often can be used for 
quite small systems.

The value of Planck's constant is approximately $6.6\cdot 10^{-34} Js$;
its smallness is the reason why we do not encounter quantum
phenomena on our length and time scales that often. 

A nice gedanken experiment, which goes under the name 
\bfi{Schr\"odinger's cat},
illustrates -- though in a philosophically questionable way -- quantum 
mechanics in terms of daily-life physics. In short the experiment goes 
as follows. Suppose we have put a cat in a box and in the same box we
have a single radioactive nucleus.
The state of the nucleus is determined by the laws of quantum
physics; the nucleus can disintegrate but we don't know when.
The process of disintegration is described by a probability distribution
that is dictated by the laws of quantum mechanics. Now suppose we link
to the nucleus a detector and a gun; if the nucleus disintegrates,
the gun, which is aimed at the cat, goes off and kills the cat.
The state of the cat, dead or alive, is now given by a quantum
mechanical probability distribution. But common sense expects the cat 
to be either dead or alive\dots. 

This sort of quantum weirdness is often propagated in the literature 
about quantum mechanics for the general public. But those who want to 
get a thorough understanding of quantum physics should try to forget 
all that weird stuff, which is only due to an inadequate visualization 
of what quantum mechanics is about. The science of quantum mechanics 
is very orderly and free from paradoxes, if everything is expressed in 
the proper -- mathematical -- language.

\bigskip
Many quantum observations, for example scattering processes, exhibit a
stochastic character, so probabilities are frequent in quantum
mechanics. Probabilities in quantum mechnaics are usually introduced in
terms of an abstract concept without an intuitive meaning, that of the 
wave function. In contrast, in this book, we shall hardly make use of 
wave functions, since our goal is to emphasize a view that shows how 
classical mechanics and quantum mechanics are very alike. Rather than 
postulating probabilities as fundamental, as in the usual approach, 
we derive the probabilistic view in Chapter \ref{c.models} in a manner 
exhibiting the conditions under which this view is appropriate.
Therefore, from the point of view of this book, the wave function is 
only a mathematical artifice to perform certain calculations which 
result in physics of interest. 
Nevertheless we give here a short introduction so that 
readers can easily make the connection to the standard presentation.

The \bfi{wave function} of a single particle is a complex-valued 
function of space and time coordinates that is square-integrable over 
the space coordinates for all time $t$. For example, the Uranium 
nucleus in the gedanken experiment described above is described by such 
a wave function. 

In quantum chemistry,
the square absolute value of the wave function of an electron is 
interpreted (apart from a constant factor) as its charge density.
One usually normalizes the wave function to have total weight $1$;
then the squared absolute value of the wave function integrates to one,
and can be viewed as a probability distribution in space. It is the 
distribution governing the random quantum response of an array of 
detectors placed at various position in space (when this is feasible 
to do). In the Copenhagen interpretation of quantum mechanics (a 
semiclassical view interpreting quantum aspects of a system with 
classical measurement equipment), one regards it as the distribution 
for ``finding'' the particle in the position where such a detctor 
responds. 

 In particular, in case of a single particle, the probability
density for observing a detector response at $x$ is $|\psi(x)|^2$.
Physicists and chemists occasionally view a scaled version of the
probability distribution $|\psi|$ as a charge
density. The justification is that in a population of a great number of
particles that are all subject to the same Schr\"odinger equation, the
particles will distribute themselves more or less according to the 
probability distribution of a single particle.

In a first course on quantum mechanics one postulates the
\bfi{time-dependent Schr\"odinger equation}
\lbeq{e.tschr}
i\hbar \frac{d}{dt}\psi = H\psi
\eeq
for a single particle described by the wave function $\psi$,
where $H$ is the Hamiltonian -- now an operator. The Schr\"odinger
equation describes the dynamics of the wave function and thus of the
particle. 
Given a solution $\psi$ to the Schr\"odinger equation, normalized to 
satisfy $\psi^*\psi=1$ (where $\psi^*$ is the adjoint linear 
functional, in finite dimensions the conjugate transpose), one obtains a
density operator $\rho = \psi \psi^*$, which is a Hermitian, positive
semidefinite rank-one operator of trace $\tr \rho = \psi^*\psi=1$. 
This type of density operator characterizes so-called \bfi{pure states};
the nomenclature coincides here with that of the mathematical theory 
of {\bf $\mathbf{C}^*$-algebras}.

\at{emphasize the connection between statistical mechanics and pure 
states; give
adequate signposts foreshadowing where you were going. You only a
mention the connection later in the para before eq\gzit{e.rhoexpand}.
There, you mention the canonical ensemble but if that had been
introduced back in the thermo section, you could now refer to it and
proceed.}

\bigskip
In quantum mechanics, the classical functions are replaced by
corresponding operators defined on a dense subspace of a suitable
separable Hilbert space. For example, the momentum in the
$x$-direction of a particle
described by a wave function $\psi(x)$ can be described by the operator
$-i\hbar \partial_x$. As we shall see, this process of
\bfi{quantization} has interesting connections to the representation
theory of Lie algebras. Using the correspondence between classical
functions and operators one deduces the Hamiltonian for an electron
of the hydrogen atom,
the basis for an explanation of atomic physics and the periodic system
of elements. 

The hydrogen atom is the quantum version of the 2-body
problem of Newton and is the simplest of a large class of problems
for which one can explicitly get the solutions of the Schr\"odinger
equation -- it is integrable in a sense paralleling the classical
notion. Unfortunately, integrable systems are not very frequent;
they seem to exist only for problems with finitely many degrees of
freedom, for quantum fields living on a 2-dimensional space-time,
and for noninteracting theories in higher dimensions.
(Whether there are exactly solvable interacting local 4-dimensional 
field theories is an unsolved problem.)
Nevertheless, the hydrogen atom and other integrable systems are 
very important since one can study in these simple models the features 
which in some approximate way still hold for more complicated physical 
 systems.

\section{Quantum field theory}

In $N$-particle quantum mechanics, the Hamiltonian is 
a second-order differential operator with respect to the spatial 
coordinates. According to the present state of knowledge, the 
fundamental description of nature is, however, given not in terms
of particles but in terms of fields.
The quantum mechanics of fields is called \bfi{quantum field theory};
the Hamiltonian is now an expression composed of field operators
rather than differential operators.  While nonrelativistic quantum 
mechanics may be treated equivalently as $N$-particle mechanics
or as nonrelativistic quantum field theory, the relativistic quantum
mechanics of more than one particle (see, e.g., the survey by
{Keister \& Polyzou} \cite{KeiP})
is somewhat clumsy, and one uses almost exclusively relativistic 
quantum field theory for the description of multiple relativistic 
particles.

Although very important in physics, the mathematical theory of
quantum fields is well developed only in the case of
space-time dimensions $<4$. While this covers important applications 
such as quantum wires or nanotubes (which have only one or two 
important spatial dimensions), it does not cover the most important 
case -- the 4-dimensional space-time we live in.
For example, quantum electrodynamics
(QED), the most accurate of all quantum theories, exists only in the
form of perturbative expansions (or other approximation schemes) and
provides mathematically well-defined coefficients of a series
in the fine structure constant $\alpha$ (whose experimental value
is about $1/137$), which is believed to be divergent
\sca{Dyson} \cite{dyson-divs}.
The first few terms provide approximations to $\alpha$-dependent
numbers like the \bfi{magnetic moment} $g(\alpha)$ of the electron,
which match experimental data to 12 digit accuracy, 
or the \bfi{Lamb shift} $\Delta E_\fns{Lamb}(\alpha)$ of hydrogen,
whose explanation in 1947 by Julian Schwinger ushered in the era of 
quantum field theory.
However, the value of a divergent asymptotic series at
a finite value of $\alpha$ is mathematically ill-defined;
no consistent definition of functions such as $g(\alpha)$
is known to which the series would be
asymptotic. \at{add references, cf. QML2}

Finding a mathematically consistent theory of a class of 4-dimensional
quantum field theories (quantum Yang--Mills gauge theory with a compact
simple, nonabelian gauge group, believed to be the most
accessible case) is another of the Clay Millennium problems whose
solution is worth a million dollars.  
\at{add references, take them from 
my FAQ -- is there a rigorous interacting QFT in 4 dimensions?}

In this book, we shall say very little about quantum field theory.
\at{still true?}

\section{The Schr\"odinger picture}

For conservative systems, the Hamiltonian is a self-adjoint linear
operator. We assume that the quantum system is confined to a large
box; a frequently employed artifice in quantum mechanics, which in
a rigorous treatment must be removed later by going to the so-called 
\bfi{thermodynamic limit} where the box contains an arbitrarily large 
ball. Under this circumstance the spectrum of the Hamiltonian is 
discrete. The eigenvalues of the Hamiltonian correspond to
energy levels, but only energy differences are observable (e.g., as
spectral lines), and one generally shifts the Hamiltonian operator
such that the lowest eigenvalue is zero. By the \bfi{spectral theorem},
the eigenvalues
\[
0=E_0\leq E_1\leq E_2\leq ...\,
\]
are real, and we can find a set of eigenvectors $\psi_k$, normalized to 
satisfy $\psi_k^*\psi_k=1$, such that
\[
H\psi_k = E_k\psi_k
\]
and 
\lbeq{e.specdec}
H=\sum E_k \psi_k\psi_{k}^{*}\,.
\eeq
$\psi_0$ is called a \bfi{ground state}, the $E_k$ are the 
\bfi{energy levels}, and $E_1=E_1-E_0\ge 0$ is called the 
\bfi{energy gap}. The energy gap is positive iff\footnote{
\bfi{iff} is the mathematician's abbreviation for ``if and only if''.
} 
 the smallest 
eigenvalue is simple, i.e., iff the ground state is unique up to a 
\bfi{phase}, a constant factor of absolute value 1.
The eigenvectors are the solutions of the 
\bfi{time-independent Schr\"odinger equation}
\lbeq{e.tis}
H \psi = E \psi,
\eeq
and the ground state is the solution of minimal energy $E_0$.
With our normalization of energies, $E_0=0$ and hence $H\psi_0=0$,
implying that the ground state is a time-independent solution of the
time-dependent Schr\"odinger equation \gzit{e.tschr}. The other 
eigenvectors $\psi_k$ lead to time-dependent solutions 
$\psi_k(t)=e^{itE_k/\hbar}\psi_k$, which oscillate with the angular
frequency $\omega_k=E_k/\hbar$. This gives \sca{Planck}'s 
\at{get history right} basic relation
\lbeq{e.Ehnu}
 E= \hbar \omega
\eeq
relating energy and angular frequency. (In terms of the ordinary 
frequency $\nu = \omega/2\pi$ and \sca{Planck}'s original constant
$h=2\pi \hbar$, this reads $E= h\nu$.) The completeness of the 
spectrum and the superposition principle now implies that
for a \bfi{nondegenerate spectrum} (where all energy levels are 
distinct),
\[
\psi(t)=\sum \alpha_k e^{i\omega_k t}\psi_k
\]
is the general solution of the time-dependent Schr\"odinger equation.
(In the degenerate case, a more complicated confluent formula is 
available.) Thus the time-dependent Schr\"odinger equation is solvable
in terms of the time-independent Schr\"odinger equation, or equivalently
with the spectral analysis of the Hamiltonian. This is the reason why
spectra are in the center of attention in quantum mechanics.
The relation to observed spectral lines, which gave rise to the name 
spectrum for the list of eigenvalues, is discussed in 
Chapter \ref{c.spec}.

The spectral decomposition \gzit{e.specdec} also
provides the connection to quantum statistical mechanics.
A thorough discussion of equilibrium statistical mechanics emphasizing
the quan\-tum-classical correspondence will be given in Part II. 
Here we only scratch the surface.
Under sufficiently idealized conditions, a thermal quantum system
is represented as a so-called \bfi{canonical ensemble}, characterized
by a
density operator of the form
\[
\rho = e^{-\beta H},  ~~~\beta =(\kbar T)^{-1}\,,
\]
with $T$ the temperature and $\kbar$\index{$\kbar$} the \bfi{Boltzmann 
constant}, a tiny constant with approximate value 
$1.38 \cdot 10^{-23} J/K$. Hence we get
\lbeq{e.rhoexpand}
\rho = \sum e^{-\beta E_k} \psi_k \psi_{k}^{*}
= \psi_0\psi_{0}^{*}+e^{-\beta E_1} \psi_1 \psi_{1}^{*}+\ldots
\eeq
At room temperature, $T\approx 300 K$, hence 
$\beta \sim 5 \cdot 10^{20}J^{-1}$. Therefore, if the energy gap 
$E_1-E_0$ is not too small,
it is enough to keep a single term, and we find that
$\rho \approx \psi_0\psi_{0}^{*}$. Thus the system is approximately in
the ground state. 

The fact that the ground state is the most relevant state is the basis
for fields like \bfi{quantum chemistry}:
For the calculation of electron configuration and the corresponding
energies of molecules at fixed
positions of the nuclei, it suffices to know the ground state. An
exception is to be made for
laser chemistry where a few excited states become relevant. To
compute the ground state, one must solve \gzit{e.tis} for the electron
wave function $\psi_0$, which, because of the minimality condition,
is a global optimization problem in an infinite-dimensional space.
The \bfi{Hartree--Fock method} and their generalizations are used to
solve these in some discretization, and the various solution techniques
are routinely available in modern quantum chemistry packages.

Applying the Schr\"odinger equation to the pure state
$\rho=\psi\psi^*$ and noting that $H^*=H$, one finds that
\[
i\hbar \frac{d\rho}{dt}
= i\hbar(\dot\psi \psi^* + \psi\dot\psi^*)
= (i\hbar\dot\psi) \psi^* - \psi(i\hbar\dot\psi)^*
=H \psi\psi^* - \psi(H\psi)^*=H\rho-\rho H,
\]
giving the \bfi{quantum Liouville equation}
\lbeq{e.qLiouv}
\frac{d\rho}{dt} = \frac{1}{i\hbar}(H\rho-\rho H) =
\frac{1}{i\hbar}[H,\rho]\,.
\eeq

\section{The Heisenberg picture}

\at{The common theme is that both classical 
mechanics and quantum mechanics have a Heisenberg picture, and in 
this picture they look similar. The Schr"odinger picture considered in
the previous section appears only since it is the one with which
most readers will be superficially familiar (at least through knowing 
the term wave function).\\
You didn't mention the phrase "Schr\"odinger Picture" in the
previous section, nor make a reference now to tell the reader that's
what you'd just been talking about. You just introduce the phrase below
without linking it to the foregoing. Instead, you could first tell the
reader that you've just been talking about the "Schr\"odinger Picture",
clarify what it is, and then introduce the Heisenberg picture as an
equivalent.}

In the beginning of quantum mechanics there were two independent
formulations; the \bfi{Heisenberg picture} (discovered in 1925 by
Heisenberg), and the \bfi{Schr\"odinger picture} (discovered in 1926 by
Schr\"odinger). Although the formulations seemed very different at
first, they were quickly shown to be completely equivalent.

In the Schr\"odinger picture, the physical configuration is described 
by a time-dependent state vector in a Hilbert space, and the 
observables are time-independent operators on this
Hilbert space. In the Heisenberg picture this is the other way around;
the vector is time-independent and the observables are
time-dependent operators.

The connection between the two pictures comes from
noting that everything in physics that is \bfi{objective} in the sense 
that it can be verified repeatedly is computable from expectation 
values (here representing averages over repeated observations), 
and that the time-dependent expectations
\lbeq{e.schr}
\<f\>_t=  \sint \rho(t) f =  \sint f \rho(t)
\eeq
(remember that the quantum integral \gzit{e.quantint} is a trace!)
in the Schr\"odinger picture can be alternatively written in the
Heisenberg picture as
\lbeq{e.heis}
\< f \>_t = \sint \rho f(t) = \sint f(t) \rho\,.
\eeq
The traditional view of classical mechanics corresponds to the
Heisenberg picture -- the observables depend on time and the density is
time-independent.  However, both pictures can  be used in
classical mechanics, too.

To transform the Heisenberg picture description to the Schr\"odinger
picture, we note that the Heisenberg expectations \gzit{e.heis}
satisfy
\[
\frac{d\<f\>_t}{dt}= \sint \rho \dot f(t)
=\sint  \rho\{f(t),H\} =\<\{f,H\}\>_t\,,
\]
giving the differential equation
\lbeq{e.dyn}
\frac{d\<f\>_t}{dt}=\<\{f,H\}\>_t
\eeq
for the expectations. An equivalent description in the Schr\"odinger
picture expresses the same dynamical law using the Schr\"odinger
expectations \gzit{e.schr}.
To deduce the dynamics of $\rho(t)$ we need the following formula
which can be justified for concrete Poisson brackets with
integration by parts,
\lbeq{e.kill}
\sint \{f,g\}h = \sint f \{g,h\}\,;
\eeq
cf. \gzit{e.intbyparts} below.
Using this, we find as consistency condition that
$ \frac{d}{dt}\<f\>_t=\sint f\dot \rho(t)$ and
$\<\{f,H\}\>_t= \sint \{f,H\}\rho(t)= \sint f \{H,\rho(t)\}$
must agree for all $f$.
This dictates the \bfi{classical Liouville equation}
\lbeq{e.cLiouv}
\dot \rho(t) = \{H,\rho(t)\}\,.
\eeq
In the quantum case, the Heisenberg and Schr\"odinger
formulations are equivalent if the dynamics of $f(t)$ is given by
the \bfi{quantum Heisenberg equation}
\[
i\hbar \frac{d}{dt}f(t) = [H,f(t)]\,.
\]
To check this, one may proceed in the same way as we
did for the classical case above.
Using the Lie product notation \gzit{e.lp},
the dynamics for expectations takes the form
\[
\frac{d}{dt} \<f\> = \<H\lp f\>,
\]
the Heisenberg equation becomes
\lbeq{e.Heis}
\frac{d}{dt} f = H\lp f,
\eeq
and the Liouville equation becomes 
\lbeq{e.Liou}
\dot \rho(t) = \rho(t)\lp H.
\eeq
(Note that here $H$ appears in the opposite order!)
These formulas hold whether we consider classical or quantum mechanics.

We find the remarkable result that these equations look identical
whether we consider classical or quantum mechanics; moreover, they are 
linear in $f$ although they encode all the intricacies of nonlinear
classical or quantum dynamics. Thus, \bfi{on the statistical level,
classical and quantum mechanics look formally simple and identical
in structure}.
The only difference lies in the definition of the Lie product 
and the integral.

The connection is in fact much deeper; we shall see that classical
mechanics and quantum mechanics are two applications of the same
mathematical formalism. At the present stage, we get additional hints
for this by noting that, as we shall see later, both the classical and
the quantum Lie product satisfies the \bfi{Jacobi identity}
\[
f\lp (g\lp h) + g \lp (h \lp f)+ h \lp (f \lp g)=0,
\]
hence define a Lie algebra structure; cf. Section \ref{s.lie}.
They also satisfy the \bfi{Leibniz identity}
\[
g\lp fh = (g\lp f)h+f(g\lp h)
\]
characteristic of a Poisson algebra; cf. Section \ref{s.poisson}.
Integrating the Leibniz identity and using \gzit{e.intdiff} gives
$0=\sint g\lp fh
= \sint((g\lp f)h+f(g\lp h))=-\sint(f\lp g)h + \sint f(g\lp h)$,
hence the \bfi{integration by parts} formula
\lbeq{e.intbyparts}
\sint f (g\lp h) = \sint (f\lp g) h.
\eeq
Readers having some background in Lie algebras will recognize 
\gzit{e.intbyparts}
as the property that $\sint fg$ defines a bilinear form with the
properties characteristic for the \idx{Killing form} 
of a Lie algebra -- again both in the classical and the quantum case.
Finally, the Poisson algebra of quantities carries both in the
classical case and in the quantum case an intrinsic real structure
given by an involutive mapping $*$ satisfying $f^{**}=f$ and
natural compatibility relations with the algebraic operations:
$f^*$ is the complex conjugate in the classical case, and the adjoint 
in the quantum case. 

Thus the common structure of classical mechanics
and quantum mechanics is encoded in the algebraic structure of a
Poisson $*$-algebra. This algebraic structure is built up in
the course of the book, and then exploited to analyze one of the
characteristic quantum features of nature -- the spectral lines visible
in light emanating from the sun, or from some chemical compound brought
into the flame of a Bunsen burner.

\section{Outline of the book}\label{s.outline}

The goal of this book is to introduce the ideas relating quantum
mechanics, Lie algebras and Lie groups, motivating everything as far 
as possible by classical mechanics. We shall mostly be concerned with 
systems described by a finite-dimensional
phase space; the infinite-dimensional case is too difficult for a
presentation at the level of this book. However, we present the
concepts in such a way that they are valid even in infinite dimensions,
and select the material so that it provides some insight into
the infinite-dimensional case.

Chapter \ref{c.matrixgl} discusses the mathematics and physics of the
2-level system, the simplest quantum system. It describes a number 
of important physical situations: Systems having only two 
energetically accessible energy eigenstates (e.g., 2-level atoms), the 
spin of a single particle such as an electron or a silver atom, the 
two polarization degrees of freedom of light, the isospin symmetry 
between proton and neutron, and the qubit, the 
smallest unit of quantum information theory. From a mathematical point 
of view, this is essentially the theory of the Lie group SU(2) and its 
Lie algebra su(2); therefore we introduce along the way basic concepts 
for matrix groups and their Lie algebras.

Chapter \ref{c.symm} 
discusses the mathematics of the most important symmetries found in our 
universe, and their associated Lie groups and Lie algebras:
The rotation group and the group of rigid motions in physical space 
$\Rz^3$, the Heisenberg groups describing systems of point particles, 
the Galilei group and the Poincare group describing the symmetry of 
Newtonian and Minkowski space-time, the Lorentz group describing basic 
features of the theory of relativity, and some more groups describing 
the hydrogen atom, the periodic system of elements, a model for nuclei,
and quarks.
Currently, parts of this chapter are only very sketchy.\at{}

Chapter \ref{c.FAQ} currently contain a number of sections quoted 
verbatim from the Theoretical Physics FAQ at 
\url{http://www.mat.univie.ac.at/~neum/physfaq/physics-faq.html};
the material there must be integrated into the other chapters of the 
book (mostly into Chapter \ref{c.symm}), a task still to be done.\at{}.

Chapter \ref{c.oscillating} 
discusses systems of classical oscillators, starting with
ordinary differential equations modeling nonlinearly coupled,
damped oscillators, and introducing some notions from classical
mechanics -- the Hamiltonian (energy), the notion of conservative
and dissipative dynamics, and the notion of phase space.
We then look in more detail into the single oscillator case, the
classical anharmonic oscillator, and show that the phase space dynamics
can be represented both in terms of Hamilton's equations, or in terms
of Poisson brackets and the classical Heisenberg equation of motion.
Since the Poisson bracket satisfies the Jacobi identity, this gives
the first link to Lie algebras.
Considering the special case of harmonic oscillators, we show that 
they naturally arise from eigenmodes of linear partial differential
equations describing important classical fields: The Maxwell equations 
for beams of light and gamma rays, the Schr\"odinger equation and the
Klein--Gordon equation for nonrelativistic and relativistic beams of 
alpha rays, respectively, and the Dirac equation for beams of beta rays.

Chapter \ref{c.blackbody} 
relates the dynamics of arbitrary systems to those of
oscillators by coupling the latter to the system, and exploring
the resulting frequency spectrum.
The observation that experimental spectra often have a pronounced
discrete structure (analyzed in more detail in Chapter 11)
is found to be explained by the fact that the discrete spectrum
of a quantum Hamiltonian is directly related to the observed spectrum
via the quantum Heisenberg equation of motion.
Indeed, the observed spectral lines have frequencies corresponding to
differences of eigenvalues of the Hamiltonian, multiplied by Planck's
constant.
This naturally explains the Rydberg--Ritz combination principle that
had been established about 30 years before the birth of modern quantum
theory. An excursion into the early history of quantum mechanics
paints a colorful picture of this exciting time when the modern world
view got its nearly definite shape.
We then discuss general properties of the spectrum of a system
consisting of several particles, and how it reflects the bound state
and scattering structure of the multiparticle dynamics.
Finally, we show how black body radiation, the phenomenon whose
explanation (by Planck) initiated the quantum era, is related to
the spectrum via elementary statistical mechanics.

\bigskip
Part II discusses statistical mechanics from an algebraic perspective,
concentrating on thermal equilibrium but discussing basic things in 
a more general framework.
A treatment of equilibrium statistical mechanics
and the kinematic part of nonequilibrium statistical mechanics
is given. From a single basic assumption (Definition 
\ref{3.1.}) the full structure of phenomenological thermodynamics and 
of statistical mechanics is derived, except for the third law 
which requires an additional quantization assumption.

Chapter \ref{c.ctherm} 
gives a description of standard phenomenological equilibrium 
thermodynamics for single-phase systems in the absence of chemical 
reactions and electromagnetic fields. 
Section \ref{s.phen} states the assumptions needed in an axiomatic
way that allows an easy comparison with the statistical mechanics
approach discussed in later chapters, and derives the basic formulas 
for the thermodynamic observables.
Section \ref{s.cons} then discusses the three fundamental laws of 
thermodynamics; their implications are discussed in the remainder of
the chapter. In particular, we derive the conventional formulas that
express the thermodynamic observables in terms of the Helmholtz and 
Gibbs potentials and the associated extremal principles.

Chapter \ref{c.quants}
introduces the technical machinery of statistical mechanics, 
Gibbs states and the partition function, in a uniform way common to 
classical mechanics and quantum mechanics.
Section \ref{s.quantities} introduces the algebra of quantities and 
their basic realizations in classical and quantum mechanics. 
Section \ref{s.gibbs} defines Gibbs states, their partition functions, 
and the related KMS condition, and illustrates the concepts by means 
of the canonical ensemble and harmonic oscillators. 
The abstract properties of Gibbs states are studied in Section 
\ref{s.gen}, using the Kubo product and the Gibbs-Bogoliubov inequality.
These are the basis of approximation methods, starting with mean field 
theory, and we indicate the connections.
However, since approximation methods are treated abundantly in common 
texts, we discuss elsewhere in the present book only exact results. 
The final Section \ref{s.limit} discusses limit resolutions for the 
values of quantities, and the associated uncertainty relations.

Chapter \ref{c.lawtherm}
rederives the laws of thermodynamics from statistical 
mechanics, thus putting the phenomenological discussion of Chapter 
\ref{c.ctherm} on more basic foundations.
Section \ref{zeroth} defines thermal states and discusses their 
relevance for global, local, and microlocal equilibrium. 
Section \ref{s.eos} deduces the existence of an equation of state and 
connects the results to the phenomenological exposition in 
Section \ref{s.phen}. Section  \ref{first} proves the first
law of thermodynamics. In Section \ref{second}, we compare thermal 
states with arbitrary Gibbs states and deduce the extremal principles 
of the second law. Section  \ref{third} shows that the third law is 
related to a simple quantization condition for the entropy and relates
it to the time-independent Schr\"odinger equation.

In Chapter \ref{c.models}
we discuss in more detail the relation between 
mathematical models of physical systems and reality. Through a 
discussion of the meaning of uncertainty, statistics, and probability, 
the abstract setting introduced in the previous chapters is given both 
a deterministic and a statistical interpretation.
Section \ref{s.detail} discusses questions relating to different 
thermal models constructed on the basis of the same Euclidean 
$*$-algebra by selecting different lists of extensive quantities. 
Section  \ref{s.model} discusses the hierarchy of equilibrium 
descriptions and how they relate to each other.
Section  \ref{s.stat} reviews the role of statistics in the algebraic 
approach to statistical mechanics.
Section  \ref{s.measurement}  gives an operational meaning to classical 
instruments for measuring the value of uncertain quantities, and to 
statistical instruments whose measurement results are only
statistically correlated to properties of the measured system. 
Section  \ref{s.qprob} extends the discussion to quantum systems, 
and discusses the deterministic and statistical features of quantum 
mechanics.
Section  \ref{s.complexity} relates the subject to information theory, 
and recovers the usual interpretation of the value of the entropy as a 
measure of unobservable internal complexity (lack of information).
The final Section  \ref{s.maxent} discusses the extent to which an 
interpretation in terms of subjective probability makes sense, and 
clarifies the relations with the maximum entropy principle.

\bigskip
Part III introduces the basics about Lie algebras and Lie groups,
with an emphasis on the concepts most relevant to the conceptual side
of physics. \at{replace Section numbers by references}

Chapter \ref{c.lie}
introduces Lie algebras. We introduce in Section \ref{s.lie} 
the basic definitions and tools for verifying the Jacobi identity, 
and establish the latter for the Poisson bracket of a single harmonic 
oscillator and in Section \ref{sec.calc.lie} for algebras of 
derivations in associative algebras.
Noncommutative associative algebras give rise to Lie algebras in a 
different way -- via commutators, discussed in Section \ref{s.LinLie}. 
The fact that linear operators on a vector space form a Lie algebra 
brings quantum mechanics into the picture. Differential equations in
associative algebras defining exponentials naturally produce Lie 
groups and the exponential map, which relates Lie groups and Lie 
algebras. In Section \ref{s.Liega}, we discuss classical groups and 
their Lie algebras. 
Taking as the vector space the space of n-dimensional 
column vectors gives as basic examples the Lie algebra of $n\times n$ 
matrices and its most important Lie subalgebras, the orthogonal, 
symplectic, and unitary Lie algebras. Many finite-dimensional Lie 
groups arise as groups of square invertible matrices, and we discuss 
the most important families, in particular the unitary, orthogonal,
and symplectic groups.
We then discuss Heisenberg algebras and Heisenberg groups and their 
relation to the Poisson bracket of harmonic oscillators via the 
canonical commutation relations. The product law in Heisenberg groups 
is given by the famous Weyl relations, which are an exactly 
representable case of the Baker--Campbell--Hausdorff formula valid for 
many other Lie groups, in particular for arbitrary finite-dimensional 
ones.
We end the Chapter with a treatment of the slightly richer structure 
of a Lie $*$-algebra usually encountered in the mechanical applications.
In traditional terms, Lie $*$-algebras are equivalent to 
complexifications of real Lie algebras, but the $*$-formulation is often
more suitable for discussing physics.

Chapter \ref{c.ppoisson} 
brings more physics into play by introducing Poisson 
algebras, the algebras in which it is possible to define Hamiltonian 
mechanics. Poisson algebras abstract the algebraic features of both 
Poisson brackets and commutators, and hence serve as a unifying
tool relating classical and quantum mechanics.
After defining the basic concepts in Section \ref{s.poisson}, we discuss
in Section \ref{s.rigid} rotating rigid bodies, in Section 
\ref{s.obs.rot} the concept of angular momentum, and the commutative 
Poisson algebra of smooth functions of angular momentum. 
It is directly related
to the group SO(3) of 3-dimensional rotations and the corresponding 
Lie algebra so(3) of infinitesimal rotations, which is generated by 
the components of the angular momentum. In particular, we obtain in 
Section \ref{sec-clas-rig} the Euler equations for a spinning rigid 
body from a Hamiltonian quadratic in the angular momentum. This example 
shows how the quantities of a classical Poisson algebra are naturally 
interpreted as physical observables.
The angular momentum Poisson algebra is a simple instance of 
Lie--Poisson algebras, a class of commutative Poisson algebras 
canonically associated with any Lie algebra and constructed in 
Section \ref{lie--poiss}. 
The Poisson bracket for the harmonic oscillator is another
instance, arising in this way from the Heisenberg algebra. 
Thus Hamiltonian mechanics on Lie--Poisson algebras generalizes the 
classical anharmonic oscillator, and gives for so(3) the dynamics of 
spinning rigid bodies.
Sections on classical symplectic mechanics and its application to the 
dynamics of molecules and an outlook to quantum field theory conclude 
the chapter.

Chapter \ref{c.repclass}
introduces representations of Lie algebras and Lie groups 
in associative algebras and in Poisson algebras.
A general physical system can be characterized in terms of a Poisson 
representation of the kinematical Lie algebra of distinguished 
quantities of interest, a Hamiltonian, a distinguished
Hermitian quantity in the Poisson algebra defining the dynamics, 
and a state defining a particular system at a particular time.
We also introduce Lie algebra and Lie group representations in 
associative algebras, which relate Lie algebras and Lie groups of 
matrices or linear operators to abstract Lie algebras and
Lie groups. These linear representations turn out to be most important 
for understanding the spectrum of quantum systems, as discussed later 
in Section \ref{s.chains}.
We then discuss unitary representations of the Poincar´e group, 
the basis for relativistic quantum field theory.
An overview over semisimple Lie algebras and their classification 
concludes the chapter.

\bigskip
Part IV discusses the dynamics of nonequilibrium phenomena, i.e., 
processes where the expectation changes with time, in as far as no 
fields are involved. This part is still in a preliminary,
somewhat sketchy form. It also lacks references (both to historical 
origins, the literature on the subject) and subject indexing, and must 
also better connected with the earlier parts.

Chapter \ref{c.markov} discusses general Markov processes, i.e., 
abstract (classical or quantum) stochastic processes without memory.
It will also contain the basic features of quantum dynamical semigroups 
and the associated Lindblad dynamics.

Chapter \ref{c.diffusion} discusses stochastic differential equations 
and associated diffusion processes, and their deterministic limits -- 
dissipative Hamiltonian systems.

Chapter \ref{c.coll} discusses collective processes described by a 
master equation, and their most prominent application -- stirred 
chemical reactions.

\bigskip
Part V gives an introduction to differential geometry 
from an algebraic perspective. 

Chapter \ref{c.manifolds}
starts with an introduction to basic concepts of differential geometry. 
We define (smooth, infinitely often differentiable) manifolds and the 
associatated algebra of scalar fields. Its derivations define vector 
fields, which form important examples of Lie algebras. The exterior 
calculus on alternating forms is developped. Finally, Lie groups are
interpreted as manifolds.

Chapter \ref{c.pmani}
discusses the construction of Poisson algebras related to 
manifolds, and associated Poisson manifolds, the arena for the most 
general classical dynamics. We show how classical symplectic mechanics 
(in flat phase space) and constrained Hamiltonian mechanics
fit into the general abstract picture. We end the chapter with a 
discussion of the Lagrangian approach to classical mechanics.

Chapter \ref{c.hquant}
is about Hamiltonian quantum mechanics. We discuss a 
classical symplectic framework for the Schr¨odinger equation. 
This is then generalized to a framework for quantum-classical dynamics, 
including important models such as the Born-Oppenheimer approximation 
for the quantum motion of molecules. A section on deformation 
quantization relates classical and quantum descriptions of a system, 
and the Wigner transform makes the connection quantitiatively useful 
in the special case of a canonical system with finitely many degrees 
of freedom.

\bigskip
Part VI applies these concepts to the study of the dominant 
kinds of elementary motion in a bound system, vibrations (described by 
oscillators, Poisson representations of the Heisenberg group), 
rotations (described by a spinning top, Poisson representations of 
the rotation group), and their interaction. On the quantum level, 
quantum oscillators are always bosonic systems, while spinning systems 
may be bosonic or fermionic depending on whether or not the spin is 
integral. The analysis of experimental spectra, concentrating on 
the mathematical contents of the subject, concludes our discussion.

Chapter \ref{c.harmonic} 
is a study of harmonic oscillators (bosons, 
elementary vibrations), both from the classical and the quantum point 
of view. We introduce raising and lowering operators in the symplectic 
Poisson algebra, and show that
the classical case is the limit $\hbar\to 0$ of the quantum harmonic 
oscillator. 
The representation theory of the single-mode Heisenberg 
algebra is particularly simple since by the Stone--von Neumann theorem, 
all unitary representations are equivalent. We find that the quantum 
spectrum of a harmonic oscillator is discrete and consists of the 
classical frequency (multiplied by $\hbar)$ and its nonnegative 
integral multiples (overtones, excited states).
For discussing the representation where the harmonic oscillator 
Hamiltonian is diagonal, we introduce Dirac's bra-ket notation, and
deduce the basic properties of the bosonic Fock spaces, first for
a single harmonic oscillator and then for a system of finitely many 
harmonic modes. 
We then introduce coherent states, an overcomplete basis representation
in which not only the Heisenberg algebra, but the action of the 
Heisenberg group is explicitly visible. The coherent state 
representation is particularly relevant for the study of quantum 
optics, but we only indicate its connection to the modes of the 
electromagnetic field.

Chapter \ref{c.spin}
discusses spinning systems, again 
from the classical and the quantum perspective. Starting with the 
Lie-Poisson algebra for the rotation group and a Hamiltonian quadratic 
in the angular momentum, we obtain the Euler equations for the 
classical spinning top.
The quantum version can be obtained by looking for canonical 
anticommutation relations, which naturally produce the Lie algebra
of a spinning top. As for oscillators, the canonical 
anticommutation relations have a unique irreducible unitary 
representation, which corresponds to a spin $1/2$ representation of 
the rotation group. The multimode version gives rise to fermionic
Fock spaces; in contrast to the bosonic case, these are 
finite-dimensional when the number of modes is finite. In particular,
the single mode fermionic Fock space is 2-dimensional.
Many constructions for bosons and fermions only differ in the signs 
of certain terms, such as commutators versus anticommutators.
For example, quadratic expressions in bosonic or fermionic Fock spaces 
form Lie algebras, which give natural representations of the universal 
covering groups of the Lie algebras $so(n)$ in the fermionic case and
$sp(2n,\Rz)$ in the bosonic case, the so-called spin groups and 
metaplectic groups, respectively.
In fact, the analogies apart from sign lead to a common generalization 
of bosonic and fermionic objects in form of super Lie algebras, which
are, however, outside the scope of the book. 
Apart from the Fock representation, the rotation group has a unique 
irreducible unitary representation of each finite dimension. We derive 
these spinor representations by restriction of corresponding nonunitary 
representations of the general linear group $GL(2,\Cz)$ on homogeneous 
polynomials in two variables, and find corresponding spin coherent 
states.

Chapter \ref{c.highest}
discusses highest weight representations, providing tools 
for classifying many irreducible representations of interest.
The basic ingredient is a triangular decomposition, which exists for all
finite-dimensional semisimple Lie algebras, but also in other cases
of interest such as the oscillator algebra, the Heisenberg algebra 
with the harmonic oscillator Hamiltonian adjoined.
We look at detail at 4-dimensional Lie algebras with a nontrivial
triangular decomposition (among them the oscillator algebra 
and $so(3)$), which  behave almost like the oscillator algebra. 
As a result, the analysis leading to Fock spaces generalizes without 
problems, and we are able to classify all irreducible unitary 
representations of the rotation group.

Chapter \ref{c.spec}
applies the Lie theoretic structure to the analysis
of quantum spectra. After a short history of some aspects of 
spectroscopy, we look at the spectrum of bound systems of particles.
We show how to obtain from a measured spectrum the spectrum of the
associated Hamiltonian, and discuss qualitative results on vibrations 
(giving discrete spectra) and chemical reactions (giving continuous 
spectra) that come from the consideration of simple systems and the 
consideration of approximate symmetries. The latter are shown to 
result in a clustering of spectral values.
The structure of the clusters is determined by how the irreducible 
representations of a dynamical Lie algebra split when the algebra is 
reduced to a subalgebra of generating symmetries. 
The clustering can also occur in a hierarchical fashion with fine 
splitting and hyperfine splitting, corresponding to a chain of 
subgroups.
As an example, we discuss the spectrum of the hydrogen atom.

\bigskip
The material presented should be complemented (in a later version of 
the book) by two further parts, one covering quantum field theory, and 
the other on nonequilibrium statistical mechanics, deriving space-time 
dependent thermodynamics from quantum field theory.

\chapter{The simplest quantum system}\label{c.matrixgl}

The simplest quantum system is a 2-level system. It describes a number 
of different situations important in practice: Systems having only two 
energetically accessible energy eigenstates (e.g., 2-level atoms), the 
spin of a single particle such as an electron or a silver atom, the 
two polarization degrees of freedom of light, the isospin symmetry 
between proton and neutron, and the qubit, the smallest unit of quantum 
information theory.  

The observable quantities of a 2-level system are $2\times 2$ matrices.
Matrices and their infinite-dimensional generalizations -- linear 
operators -- are the bread and butter of quantum mechanics. 

In mathematics and physics, \idx{symmetries} are described in terms of 
Lie groups and Lie algebras. An understanding of these concepts is
fundamental to appreciate the unity of modern physics.

This chapter introduces some basic concepts for matrix groups and their 
Lie algebras, concentrating on the case of $2\times 2$ matrices 
and their physical interpretation. 
In the next chapter we introduce in an elementary way a number 
of other Lie groups and Lie algebras that are important for physics, by 
means of concrete matrix representations, and relate them to concrete 
physics. 
A general, more abstract treatment of Lie groups and Lie algebras is 
given later in Chapter \ref{c.lie}. 

We assume that the reader already has a good command of matrix algebra 
(including complex numbers and eigenvalues) and knows basic properties 
of vector spaces, linear algebra, limits, and power series (quickly 
reviewed in Section \ref{s.matrix}).

The beginning is just matrix calculus with some new terminology, but 
the subject soon takes on a life of its own\dots.
Readers who see matrix groups for the first time may want to
skip forward to the sections with more physical content to get 
a better idea of how matrix group are used in physics, before reading 
the chapter in a linear order.

\at{informal introduction and preview of what is to come}

\section{Matrices, relativity and quantum theory}\label{s.matrix}

The early 20th century initiated two revolutions in physics that 
changed the nature of the mathematical tools used to describe physics.
Both revolutions gave matrices a prominent place in understanding
the new physics.

The transition from the old, Newtonian world view to the new, 
relativistic conception of the world culminated in the realization
that Nature is governed by symmetries that can be descibed in terms
of the Lorentz group, a group of $4\times 4$-matrices that 
mathematicians refer to by the symbolic name $SO(1,3)$ or $SL(2,\Cz)$. 
Since then, many other symmetry groups have found uses in physics. 
Indeed, symmetry considerations and the associated group theory have 
become a unifying theme and one of the most powerful tools in modern 
physics.

Independently of the theory of relativity, an increasing number of 
quantum phenomena defying an explanation in terms of classical physics
were noticed, beginning in 1900, when Max \sca{Planck} \cite{planck1900}
successfully used a quantization condition for his analysis of black 
body radiation.
After 25 years of groping in the dark to make classical sense of these 
quantum phenomena, Werner \sca{Heisenberg} \cite{Hei} laid
in 1925 the mathematical foundations of modern quantum mechanics.
The key was the insight that basic physical quantities such as the
components of position and momentum should be represented in terms of
matrices (in this case infinite arrays of numbers) rather than by 
single numbers as in classical mechanics.
Since then, matrices and linear operators, their infinite-dimensional 
generalizations, form the cornerstone of quantum mechanics.

Therefore, the language of matrices is an indispensible foundation 
for a deeper understanding of modern physics. To fix the 
notation and to remind the reader, we begin by repeating some 
definitions and properties of matrices and related concepts. 
Thorough treatments (from complementary points of views) are given 
in \sca{Lax} \cite{Lax} and \sca{Horn \& Johnson} \cite{HorJ}.

\bigskip
\idx{$\Kz$} denotes a field, usually the field $\Rz$\index{$\Rz$, 
field of real numbers} of real numbers or the field 
$\Cz$\index{$\Cz$, field of complex numbers} of complex numbers. 
Then \idx{$\Kz^n$} denotes the space of column vectors 
$x$ of length $n$ with entries $x_k\in\Kz$\index{$x_k$, vector entry} 
($k=1,\dots,n$), and
\idx{$\Kz^{m\times n}$} denotes the vector space of all $m\times n$ 
matrices $A$ with entries $A_{jk}\in\Kz$\index{$A_{jk}$, matrix entry} 
($j=1,\dots,m$; $k=1,\dots,n$).
We identify $1\times 1$ matrices with the entry they contain, and 
column vectors with matrices having a single column; thus 
$\Kz^{1\times 1}=\Kz$ and $\Kz^{m\times 1}=\Kz^m$. The \bfi{zero matrix}
of any size (with all entries zero) is denoted by $0$, or by 
$0_n$\index{$0_n$, zero matrix}
if it is square and its size $n\times n$ is emphasized.
The \bfi{identity matrix} of any size is denoted by $1$, or by 
$1_n$\index{$1_n$, identity matrix} if its size $n\times n$ is 
emphasized; multiples of the identity are identified with the 
corresponding elements of $\Kz$. 
The \bfi{transpose} of the matrix $A\in \Kz^{m\times n}$ is the matrix 
$A^T\in \Kz^{n\times m}$\index{$A^T$, transpose} with entries 
$(A^T)_{jk}:=A_{kj}$; the transpose of the column vector $x\in\Kz^n$ 
is the row vector $x^T\in\Kz^{1\times n}$. The matrix $A$ is called 
\bfi{symmetric} if $A^T=A$.
The \bfi{conjugate transpose}\index{transpose!conjugate} of the matrix 
$A\in \Cz^{m\times n}$ is the matrix 
$A^*\in\Cz^{n\times m}$\index{$A^*$, conjugate transpose} with entries 
$(A^*)_{jk}:=\ol A_{kj}$, where 
$\ol \lambda=\lambda^*$\index{$\ol\lambda$, complex conjugate}
denotes the \idx{complex conjugate} of $\lambda\in\Cz$. 
The matrix $A$ is called \bfi{Hermitian} if $A^*=A$.

The \bfi{product} of the $m\times n$ matrix $A$ and the $n\times r$ 
matrix $B$ is the $m\times r$ matrix $AB$ with entries 
$\D(AB)_{jk}=\sum_{l=1}^m A_{jl}B_{lk}$. The product is associative,
$(AB)C=A(BC)$, and distributive, $A(B+C)=AB+AC$, $(A+B)C=AB+AC$,
but in general not commutative. For a square matrix 
$A\in\Kz^{n\times n}$, the number
$\D\tr A:=\sum_{k=1}^n A_{kk}$\index{$\tr A$, trace} is the 
\bfi{trace} of $A$, and $\det A$\index{$\det A$, determinant} denotes
the \bfi{determinant} of $A$. The trace of a product is generally 
written without parentheses. The trace is a linear operation, and 
for $A,B\in \Kz^{n\times n}$, we have
\[
\tr A^T = \tr A, ~~\tr A^* = (\tr A)^*,~~\tr AB = \tr BA,
\]
\[
\det A^T = \det A, ~~\det A^* = (\det A)^*,~~\det(AB)=\det A \det B.
\]
The determinant $\det(\lambda-A)$ of a real or complex $n\times n$ 
matrix $A$ factors into a product of $n$ linear factors 
$\lambda-\lambda_k$ with complex numbers $\lambda_k$ called the 
\bfi{eigenvalues} of $A$; they are are unique up to ordering.
The trace is the sum of the eigenvalues, and the determinant is their 
product. All eigenvalues of a Hermitian matrix are real numbers.
The real or complex $n\times n$ matrix $A$ is called 
\bfi{positive semidefinite} (\bfi{positive semidefinite}) if 
$x^*Ax\ge 0$ for all $x\in\Cz^n$ (resp. $x^*Ax> 0$ for all nonzero 
$x\in\Cz^n$). The eigenvalues of a positive semidefinite (positive 
definite) matrix have nonnegative (positive) real part.

$|x|=\sqrt{x^Tx}$\index{$|x|$, length} is the \bfi{length} of the vector
$x\in\Rz^n$. $|\psi|=\sqrt{\psi^*\psi}$\index{$|\psi|$, norm} is the 
\bfi{norm} of the vector $\psi\in\Cz^n$.
For any real or complex matrix $A$, the matrix $A^*A$ is Hermitian and 
positive semidefinite; the square roots of the (real and nonnegative) 
eigenvalues of $A^*A$ are called the \bfi{singular values} of $A$.
The maximal singular value of $A$ is called the \bfi{spectral norm} 
$\|A\|$\index{\|A\|, spectral norm} of $A$. 
For $A,B\in \Cz^{n\times n}$ and $\alpha,\beta\in\Cz$, we have 
\[
\|\alpha A+\beta B\|\le|\alpha| \|A\|+|\beta| \|B\|, ~~~
\|AB\|\le \|A\|\,\|B\|,
\]
and
\[
|A\psi| \le \|A\|\,|\psi| \for \psi\in\Cz^n.
\]
Matrix functions of a real or complex square matrix $A$ are defined by 
power series with real or complex coefficients.
If a power series in $x$ has convergence radius $r$ then the series 
with $x$ replaced by a square matrix $A$ converges for $\|A\|<r$.
Any two matrix functions of the same matrix $A$ commute.
 If $A$ has the eigenvalues $\lambda_k$ then the matrix function 
$f(A)$ has the eigenvalues $f(\lambda_k)$.
Identities for power series involving only a single variable $x$ 
remain valid when $x$ is replaced by a square matrix; moreover,
$f(A^T)=f(A)^T$ and $f(A^*)=f(A)^*$.
In particular, the 
\bfi{matrix exponential}\index{$e^A$, matrix exponential}
\[
e^A=\sum_{k=0}^{\infty}\frac{1}{k!}A^k
\]
is defined for all real or complex square matrices $A$, and satisfies 
for $s,t\in\Cz$ the relations
\lbeq{e.1par}
e^{sA}e^{tA}=e^{(s+t)A},~~~ (e^{sA})^t=e^{stA}.
\eeq
On the other hand, $e^{A+B}$ is in general distinct from $e^Ae^B$; 
however, $e^{A+B}=e^ae^B$ if $A$ and $B$ \bfi{commute}, i.e., $AB=BA$.
We also note the formula
\lbeq{e.edet}
\det e^A = e^{\tr A},
\eeq
which follows from 
$\det e^A = \prod e^{\lambda_k} =e^{\sum \lambda_k} = e^{\tr A}$.

\section{Continuous motions and matrix groups}\label{s.mgroups}

A basic fact about square real and complex matrices is that they 
can often be interpreted in terms of motions in the underlying vector 
space on which they act. This gives them an intuitive meaning that 
makes it easy to interpret even very abstract applications.
Since motions can be combined and reversed they carry a natural group 
structure that gives rise to the concept of a matrix group.
Different matrix groups characterize different forms of permitted 
motions. 
Matrix groups are important examples of so-called \bfi{Lie groups}, 
defined in Section \ref{s.Liegroups}. 
Indeed, by the \bfi{theorem of Ado} (\sca{Ado} \cite{Ado}), every 
finite-dimensional Lie group is isomorphic to a matrix group.

\bigskip
A \bfi{matrix group} over $\Kz$ is a nonempty, closed set $\Gz$ of 
invertible matrices from $\Kz^{n\times n}$ with the property that the 
product of any two matrices from $\Gz$ and the inverse of a matrix 
from $\Gz$ are in $\Gz$. In particular, $1\in\Gz$, and the limit 
of any convergent sequence of elements $U_l\in\Gz$ ($l=1,2,3,\dots$)
is again in $\Gz$. If $\Kz=\Rz$ (or $\Kz=\Cz$) 
then $\Gz$ is called a \bfi{real}\index{matrix group!real} 
(or \bfi{complex}\index{matrix group!complex}) matrix group.
The matrix group $\Gz$ is called \bfi{abelian} if all its
elements commute, i.e., if $UV=VU$ for all $U,V\in\Gz$.

Let $\Gz$ be a real or complex matrix group. 
A \bfi{$\Gz$-motion} (by $V\in\Gz$) is an arbitrarily often 
differentiable map $U:[0,1]\to\Gz$ such that $U(0)=1$ (and $U(1)=V$).  
If the group consists of $n\times n$ matrices, the $\Gz$-motion moves 
a vector $x_0\in\Kz$ from $x(0)=x_0$ to $x(1)=Vx_0$, sweeping out a 
path $x(t)=U(t)x_0$\index{$U(t)$, motion} for $t\in[0,1]$. It is natural
to interpret $t$ as a time coordinate in suitable units of time.

\begin{expls}
We shall meet a large number of examples in this and the next chapter,
progressing from the matrix groups easiest to define to the ones most
useful in physics. We begin by naming the smallest and largest 
matrix groups of a given size.

(i) The set $\Id(n)$\index{$\Id(n)$, trivial group} consisting only of 
the $n\times n$ identity matrix is a matrix group, called a 
\bfi{trivial group}.

(ii) The set $L(n,\Kz)=GL(n,\Kz)$\index{$L(n,\Kz)$, general linear 
group}\index{$GL(n,\Kz)$, general linear group} of all invertible 
$n\times n$ matrices with entries in $\Kz$ is a matrix group, called a 
\bfi{general linear group} over $\Kz$. In particular, $L(1,\Kz)$ is 
the \bfi{multiplicative group} $\Kz^\times:=\Kz\setminus\{0\}$ of the 
field $\Kz$.
\end{expls} 

We now illustrate the geometric inplications of the definitions by means
of the complex plane and the motions corresponding to $\Cz^\times$ and 
some of its subgroups.

The first subgroup of $\Cz^\times$ of geometric interest is the group
$\Rz^\times_+$\index{$\Rz^\times_+$, set of positive real numbers} 
of positive real numbers. An $\Rz^\times_+$-motion 
stretches or compresses all vectors from the origin to a nonzero 
complex number by a time-varying factor. Such a stretching or 
compression is called a \bfi{dilatation} or \bfi{dilation}; 
thus $\Rz^\times_+$ is the 
group of dilatations (with respect to the origin) of the complex plane. 
It is well-known that the real multiplicative group and the real 
additive group $\Rz$ of translations along the real axis are isomorphic:
Vie the exponential function, one can associate to every translation by 
$f\in\Rz$ a dilatation $U=e^f\in\Rz^\times_+$, and conversely, find for 
every dilatation $U\in\Rz^\times_+$ a unique $f=\log U\in\Rz$ such that 
$U=e^f$. In particular, a uniform stretch (or compression) is 
obtained by the exponential motion $U(t)=e^{tf}$ when $f>0$ 
(resp. $f<0$). 

The subgroup $L(1,\Rz)=\Rz^\times$ of all nonzero real numbers 
contains dilations, the reflections at zero given by multiplication 
with $-1$, and their products, given by arbitrary negative real numbers.

Another important subgroup is the group consisting of all complex 
numbers with absolute value one, forming the unit circle in the 
complex plane. This group is the smallest of the unitary groups 
defined in Section \ref{s.nlevel}, and is therefore generally denoted by
$U(1)$\index{$U(1)$, complex unit circle}\footnote{The reader should not
confuse the occurrences of $U(1)$ as group with those of $U(1)$ as 
the final group element of a motion $U:[0,1]\to\Gz$.}. 
Using the Euler relation $e^{i\varphi}=\cos\varphi+i\sin\varphi$, one 
can again represent arbitrary group elements $U\in U(1)$ as exponential 
$U=e^{i\varphi}$ of a purely imaginary element $f=i\varphi$. 
As group elements acting by multiplication on the complex plane, 
the elements of $U(1)$ correspond to rotations around zero. 
Indeed, $U=e^{i\varphi}$ is a rotation by the angle $\varphi$.
In particular, a uniform rotational motion progresses by equal angles 
in equal time intervals, hence is given by the exponential motion
$U(t)=\cos(t\varphi)+i\sin(t\varphi)=e^{i\varphi t}$.

Since rotations by integral multiples of $2\pi$ have no net effect, 
the representation $U=e^{i\varphi}$ does not define $\varphi$ uniquely;
hence imaginary elements $f$ differing by a multiple of $2\pi i$ give 
the same group element $U=e^f$. Thus while the two groups behave the 
same locally, there is a global topological difference. 
This also shows in the fact that, as a manifold, $U(1)$ is compact, 
while $\Rz^\times_+$ is noncompact.

\section{Infinitesimal motions and matrix Lie algebras}

The matrix 
\[
U'(0):=\frac{d}{dt}U(t)\Big|_{t=0}
\]
is called the \bfi{infinitesimal motion} of the $\Gz$-motion 
$U:[0,1]\to\Gz$. Thus $f$ is the infinitesimal motion of $U:[0,1]\to\Gz$
iff, for small $t$,
\[
U(t)=1+t f + O(t^2).
\]
Here the \bfi{Landau symbol} $O(t^2)$\index{$O(t^2)$, Landau symbol} 
denotes an expression in $t$ whose norm is bounded for small $t$ by a 
constant multiple of $t^2$ 
(which may be different in each occurrence of the Landau symbol).
The \bfi{Lie algebra} of (or associated with) the real or complex 
matrix group $\Gz$ is the set $\log \Gz$\index{$\log \Gz$, Lie algebra 
of $\Gz$} of all infinitesimal motions of $\Gz$-motions.

The following fundamental theorem gives basic properties of the
Lie algebra $\log \Gz$ and describes the effect that a coordinate 
transformation in form of a $\Gz$-motion has on Lie algebra 
elements.

\begin{thm}~\\
(i) The Lie algebra $\Lz:=\log \Gz$ of any real or complex matrix group 
$\Gz$ is a vector space containing with any two matrices $f,g$ 
their \bfi{commutator}\index{$[f,g]$, commutator}
\[
[f,g]:=fg-gf.
\]
(ii) For every $U\in\Gz$,\index{$\Ad_U$, adjoint mapping}
\lbeq{e.Ad}
\Ad_U f:= UfU^{-1} \for f\in\Lz
\eeq
defines a linear mapping $\Ad_U:\Lz\to\Lz$, called the \bfi{adjoint
mapping} of $U$.

(iii) Every adjoint mapping is a Lie algebra 
{\bf automorphism}\index{automorphism!of a Lie algebra}\index{Lie 
algebra! automorphism}, i.e.,
\lbeq{e.Adaut}
 \Ad_U [f,g] = [\Ad_U f,\Ad_U g] \for f,g\in\Lz.
\eeq
\end{thm}

\bepf
Let $\Gz$ be a matrix group over $\Kz=\Rz$ or $\Kz=\Cz$, and 
$\alpha,\beta\in\Kz$. If $f,g\in\Lz$ then there are $\Gz$-motions
$V_1(t)=1+t f + O(t^2)$ and $V_2(t)=1+t g + O(t^2)$. The product
$V(t):=V_1(\alpha t)V_2(\beta t)$ is a $\Gz$-motion with 
$V(t)=(1+\alpha tf)(1+\beta tg) +O(t^2)=1+t(\alpha f+\beta g) +O(t^2)$,
so that $\alpha f+\beta g\in\log \Gz$. Therefore $\Lz$ is a vector 
space over $\Kz$.

Let $U\in\Gz$. Then  $W(t):=UV(t)U^{-1}$ is a $\Gz$-motion with
$W(t)=U(1+tf)U^{-1}+O(t^2)= 1 + tUfU^{-1}+O(t^2)$, whence 
$UfU^{-1} \in\Lz$. Therefore $\Ad_U$ maps $\Lz$ into itself.
The linearity of the adjoint mapping is straightforward.

In particular, since $V_2(t)^{-1}=(1+tg)^{-1}+O(t^2)=1-tg+O(t^2)$,
the Lie algebra $\Lz$ contains $\Ad_{V_2(t)}f-f = V_2(t)fV_2(t)^{-1}-f
= (1+tg)f(1-tg)-f +O(t^2) = t[g,f] +O(t^2)$. Dividing by $t$ and 
letting $t$ go to zero, we see that the commutator $[g,f]$ is also
in $\Lz$.

Finally, the property \gzit{e.Adaut} is again straightforward.
\epf

The above property (i) motivates to define in general a 
\bfi{matrix Lie algebra} over $\Kz$ to be a subspace $\Lz$ of 
$\Kz^{n\times n}$ closed under commutation. 
The matrix Lie algebra $\Lz$ is called \bfi{abelian} if all its
elements commute, i.e., if $[f,g]=0$ for all $f,g\in\Lz$.
A subset of a matrix Lie algebra $\Lz$ closed under commutation is 
again a matrix Lie algebra, and is called a \bfi{Lie subalgebra} of
$\Lz$. 

\begin{expls}
We shall meet a large number of examples in this and the next 
chapter, progressing from the matrix Lie algebras easiest to define 
to the ones most useful in physics. We begin by naming the smallest 
and largest matrix Lie algebras of a given size.

(i) The set $\id(n)$\index{$\id(n)$, trivial Lie algebra} consisting 
only of the $n\times n$ zero matrix 
is a matrix Lie algebra, called a \bfi{trivial Lie algebra}.
Clearly, if $\Kz$ is the real or complex field, $\id(n)$ is the Lie 
algebra of the trivial group $\Id(n)$.

(ii) $l(n,\Kz)=gl(n,\Kz)=\Kz^{n\times n}$\index{$l(n,\Kz)$, general 
linear Lie algebra}\index{$gl(n,\Kz)$, general linear Lie algebra}
is a matrix Lie algebra, called a \bfi{general linear Lie algebra} 
over $\Kz$. If $\Kz$ is the real or 
complex field, $l(n,\Kz)$ is the Lie algebra of the general linear 
group $L(n,\Kz)$ since for every $f\in l(n,\Kz)$, the mapping $U$ 
defined by $U(t)=e^{tf}$ is an $L(n,\Kz)$-motion with infinitesimal 
motion $f$.
\end{expls} 

We are mainly interested in matrix groups whose associated
Lie algebra has interesting properties. However, there are important 
matrix groups with a trivial infinitesimal structure. A matrix group 
is called \bfi{discrete} if its Lie algebra is trivial.
Discrete matrix groups that play an important role for the 
description of symmetris of molecules and crystals. For example,
a \bfi{permutation group} is a group $\Gz$ of bijective mappings of a 
finite set $X$. If the members of $X$ are the atoms of a molecule with
given chemical structure, its \bfi{symmetry group} $\Gz$ consits of 
the permutations that preserve the chemical nature of the atoms and 
the chemical bonds between them; for example, the 
benzene ring has a dihedral symmetry group with 12 elements.
Assocoated with each permutation group is a finite group of $n\times n$
\bfi{permutation matrices} $U\in\Rz^{n\times n}$, where $n$ is the size 
of $X$, and $U_{jk}=1$ iff, in a fixed ordering of the elements of $X$, 
the $j$th element is permuted to the $k$th element, $U_{jk}=0$ 
otherwise. The representation theory
of these discrete matrix groups gives important information about the 
chemical properties of symmetric molecules. In this book, we shall
meet discrete groups only in passing; for a deeper treatment we refer to
{\sc Cornwell} \cite{cornwell}, {\sc Cotton} 
\cite{Cot}, {\sc Kim} \cite{Kim}, or {\sc Weyl} \cite{Wey}.
\at{check Weyl's book}

\section{Uniform motions and the matrix exponential}

We now generalize the construction of uniform rotations in $U(1)$
to arbitrary real or complex matrix groups. 

Let $\Kz=\Rz$ or $\Kz=\Cz$, and $f\in \Kz^{n\times n}$. 
Because of \gzit{e.1par},
the set of $\exp(tf)=e^{tf}$ with $t\in\Rz$ is a matrix group, called
the \bfi{one-parameter group} with \bfi{infinitesimal generator} $f$.
The infinitesimal generator is determined only up to a nonzero scalar 
multiple. Because of the property \gzit{e.1par} and the analogy to 
uniform rotations in the complex plane, $\Gz$-motions of the form 
$U(t)=e^{tf}$ are called \bfi{uniform motions}. Since 
$U(t)=1+tf+O(t^2)$, the infinitesimal generator of a uniform motion
belongs to the Lie algebra $\log \Gz$. In view of 
\[
\frac{d}{dt} e^{tf} = f e^{tf}=e^{tf}f
\]
and the unique solvability of the intitial-value problems for ordinary 
differential equations in finite-dimensional spaces, uniform 
$\Gz$-motions
are characterized by the property
\[
U(1)=0, ~~~ \frac{d}{dt} U(t) = f U(t)
\]
for some $f\in\log \Gz$, which is the infinitesimal generator.

The set of limits of sequences of products 
$U_1^{k_1}\cdot\dots\cdot U_s^{k_s}$ 
of an arbitrary number $s\ge 0$ of arbitrary powers $U_j^{k_j}$ of 
arbitrary elements $U_j$ ($j=1,\dots,s$) from a set $S$ is a matrix
group, called the group \bfi{generated} by $S$. If $\Lz$ is a matrix 
Lie algebra, \bfi{$\exp\Lz$} denotes the group generated by the 
exponentials $e^f$ with $f\in\Lz$. Whether the exponentials themselves 
form a group depends on the Lie algebra $\Lz$.; see \at{refs below
for $sl(2,\Rz)$ and $so(3)$.}

By joining and inverting paths, it is easy to see \at{match 
derivatives when joining by infinite slowdown}
that the set $\Gz_0$\index{$\Gz_0$, connected subgroup} consisting of 
all $V\in\Gz$ for which a motion 
by $V$ exists is a matrix group, called the \bfi{connected subgroup} 
of $\Gz$. The matrix group $\Gz$ is called \bfi{connected} if 
$\Gz_0=\Gz$.  Clearly, the group generated by a 
set of elements form $\Gz_0$ is contained $\G_0$. 
A matrix group $\Gz$ and its connected subgroup $\Gz_0$ have the same 
associated Lie algebra. Moreover, due to the existence of uniform 
motions, $\exp \Lz$ is connected. Thus one often concentrates in 
physics on connected groups.

The \bfi{dimension} of a real or complex matrix group $\Gz$ is the 
dimension of the associated Lie algebra $\log \Gz$, considered as a 
real vector spece. Finite groups are 0-dimensional.
The only connected 0-dimensional matrix groups are the groups $\Id(n)$.
The connected 1-dimensional group are just the one-parameter groups.

\bigskip

\begin{thm}~\\
(i) Let $\Gz$ be a matrix group. Then the exponentials $e^{f}$ with
 $f\in\log\Gz$ belong to $\Gz$ and generate its connected subgroup 
$\Gz_0$; thus, $\exp\log\Gz=\Gz_0$.

(ii)  Let $\Lz$ be a matrix Lie algebra. Then $\log\exp\Lz=\Lz$.
\end{thm}

\bepf
(i) This is one of the few places where we need explicit analysis to 
establish a limit. To show that $e^f\in\Gz$, we need to represent $e^f$
as a limit of group elements. We construct these by noting that a
uniform motion can be considered as a composition of many identical 
tiny, almost infinitesimal motions. The eigenvalues of the matrices 
$e^{f/k}$ have the form $e^{\lambda/k}$ with eigenvalues 
$\lambda$ of $f$. For any fixed $c>\|f\|$, the exponentials 
$e^{\lambda/k}$ can be bounded in absolute value by $e^{c/k}$ since 
$|\lambda|\le \|f\|<c$. Therefore the $E_k:=e^{f/k}$ satisfy 
\[
\|E_k\|\le e^{c/k}.
\]
We know already that there is some $\Gz$-motion $U:[0,1]\to\Gz$ with 
infinitesimal motion $f\in\Lz:=\log\Gz$. We now put $U_k:= U(1/k)$
and show that
\lbeq{e.explim}
e^f = \lim_{k\to\infty} U_k^k.
\eeq
Since $\|U(t)\|=\|1+tf+O(t^2)\|=1+t\|f\|+O(t^2)\le 1+ct \le e^{ct}$
for sufficiently small $t>0$, we have 
\[
\|U_k\|\le e^{c/k}
\]
for sufficiently large $k$. Since the Taylor expansions of $U(t)$ and 
$e^{tf}$ agree up to first order, we have $U(t)-e^{tf}=O(t^2)$; hence 
there is a constant $C>0$ such that
\[
\|U_k-E_k\|=\|U(1/k)-e^{f/k}\| \le C/k^2
\]
for sufficiently large $k$. Now
\[
\bary{lll}
\|U_k^k-e^f\| &=&\|U_k^k-E_k^k\|
=\D\Big\|\sum_{j=1}^k E_k^{j-1}(U_k-E_k)E_k^{k-j}\Big\|\\
&\le& \D\sum_{j=1}^k \| E_k\|^{j-1}\|U_k-E_k\|\|E_k\|^{k-j}\\
&\le& \D\sum_{j=1}^k e^{(j-1)c/k} C/k^2 e^{(k-j)c/k}
=\D\sum_{j=1}^k Ce^{(k-1)c/k}/k^2 \le Ce^c/k,
\eary
\]
which tends to zero as $k\to \infty$. This establishes the limit 
\gzit{e.explim} and proves that $e^f\in\Gz$.

Since every $e^{f}$ with $f\in\Lz$ is part of a uniform motion, 
it is in $\Gz_0$. hence the group generated by these exponentials is 
contained in $\Gz_0$. 
\at{missing: every element of $\Gz_0$ is a limit of exponential 
products. It suffices to do the small $t$ case.}

(ii) \at{to be done}
\epf

The theorem implies that connected matrix groups are characterized 
completely by their Lie algebras. Since Lie algebras are vector 
spaces, their structure can be studied with the help of linear algebra,
while most matrix groups are intrinsically nonlinear. This explains the
importance of Lie algebras in the study of connected groups.

\section{Volume preservation and special linear groups}

\at{determinants and \bfi{oriented volume}, 
 may be a negative and even a complex number!}
The oriented volume is preserved iff the determinant is one.
The unoriented volume is preserved iff the determinant has absolute 
value one.

If $\Gz$ is a matrix group then the set $S\Gz$\index{$S\Gz$, 
determinant 1 subgroup} consisting
of all elements in $\Gz$ with determinant one is a matrix group. 
Indeed, if $U,V\in S\Gz$ then $\det (UV)=\det U\det V = 1$ and
$\det U^{-1} = (\det U)^{-1} =1$, so that $UV,U^{-1}\in S\Gz$.
In particular, the \bfi{special linear group} 
$SL(n,\Kz)$\index{$SL(n,\Kz)$, special linear group} consisting
of all $n\times n$ matrices with entries in $\Kz$ and determinant one
is a matrix group.

\at{define center, normal subgroup, quotient, (re)move?} The center of
$SL(n,\Cz)$ is the group $\Zz_n =\{\lambda\in\Cz\mid\lambda^n=1\}$ 
of $n$th roots of unity, and the quotients
$PSL(n,\Cz)=SL(n,\Cz)/\Zz_n$\index{$PSL(n,\Cz)$} form a family of simple
Lie groups. \at{how does this square with Ado's theorem?}
The group $PSL(2,\Cz)=SL(2,\Cz)/\Zz_2$ is isomorphic to the restricted 
Lorentz group defined in Section \ref{s.lorentz}.

If $\Lz$ is a matrix Lie algebra then the set $s\Lz$\index{$s\Lz$,
traceless Lie subalgebra} consisting
of all elements in $\Lz$ with zero trace is a matrix Lie algebra.
Indeed, if $f,g\in s\Lz$ then $\tr [f,g] = \tr fg - \tr gf =0$, 
so that $[f,g]\in s\Lz$.
In particular, the \bfi{special linear Lie algebra} 
$sl(n,\Kz)$\index{$sl(n,\Kz)$, special linear Lie algebra}
consisting of all $n\times n$ matrices with entries in $\Kz$ and zero 
trace is a matrix Lie algebra. Since
\[
\det (1+tf+O(t^2))=1+\tr tf+O(t)^2\,,
\]
the trace of infinitesimal generators of $S\Gz$ vanishes; conversely,
the property \gzit{e.edet} implies that the exponentials of elements of 
$s\Lz$ have determinant one, hence belong to $S\Gz$. Therefore
$s\Lz$ is the Lie algebra corresponding to the matrix group $S\Gz$.

We consider the algebraic properties of the special linear group 
\idx{$SL(2,\Cz)$} and its Lie algebra \idx{$sl(2,\Cz)$} in some detail, 
since the group $SL(2,\Cz)$, its subgroups, and the Lie algebra 
\idx{$sl(2,\Cz)$} and its Lie subalgebras play a very important role 
in physics.
$SL(2,\Cz)$ and/or $sl(2,\Cz)$ are implicitly present even in 
applications not mentioning Lie groups or Lie algebras explicitly:
In special relativity, $SL(2,\Cz)$ appears because of its relation
to the Lorentz group. The Dirac equation for electrons and 
positrons (see Section \ref{s.beta}) uses properties of Pauli matrices 
(or their cousins, the Dirac matrices), whose relation to $SL(2,\Rz)$ 
is now established.

\bigskip
\bfi{3-vectors and 4-vectors.}
As traditional in physics, we usually use fat letters to write column 
vectors $\a$\index{$\a$, 3-dimensional vector} with three 
components $a_1,a_2,a_3$. Depending on the context, these three 
components may be real or complex numbers, matrices, linear 
operators, or elements from an arbitrary associative algebra $\Az$.
We write $\Az^{3}$ for the set of all vectors with three components 
from $\Az$. The \bfi{inner product} of $\a,\b\in\Az^3$ is the 
element\index{$\a \cdot\b$, scalar product}
\[
\a\cdot \b:=\a^T\b = a_1b_1+a_2b_2+a_3b_3 \in \Az;
\]
clearly $\a \cdot \b = \b \cdot \a$.
We write\index{$\a^2$, square of 3-dimensional vector}
\[
\a^2:=\a\cdot\a = a_1^2+a_2^2+a_3^2.
\]
The \bfi{length} of a vector $\a\in\Cz^3$ is
\[
|\a|:=\sqrt{\a^*\a} =\sqrt{|a_1|^2+|a_2|^2+|a_3|^2},
\]
so $\a^2=|\a|^2$ if $\a\in\Rz^3$. 

We write $\Az^{1,3}$ for the set of vectors $p$ with four 
components $p_0,p_1,p_2,p_3\in\Az$, arranged as
\[
\p = {p_0 \choose \p},~~~p_0\in\Az,~\p\in\Az^3.
\]
Using the traditional terminology from relativity theory, we call such 
vectors $p$ \bfi{4-vectors}, and call $p_0$\index{$p_0$, time part of 
4-vector $p$} the \bfi{time part} and $\p$\index{$\p$, 
space part of 4-vector $p$} the \bfi{space part} of $p$.
The \bfi{Minkowski inner product} of $p,q\in\Az^{1,3}$ is the 
element\index{$p\cdot q$, Minkowski inner product} 
\lbeq{e.mink}
p\cdot q:=p_0q_0-\p\cdot\q = p_0q_0-p_1q_1-p_2q_2-p_3q_3, 
\eeq
\at{use uniform signs $(\pm\mp\mp\mp---)$ everywhere!}
and 
\[
p^2 := p\cdot p = p_0^2-\p^2.
\]
Note that $p^2$\index{$p^2$, Minkowski square} may be negative!

\bigskip
\bfi{Pauli matrices.}
With these preparations, we define the 
\bfi{Pauli matrices}\index{$\sigma_k$, Pauli matrices} 
\lbeq{e.pauli}
\sigma_0 = \pmatrix{1 & 0\cr 0&1}\,, ~~~
\sigma_1 = \pmatrix{0 & 1\cr 1&0}\,, ~~~
\sigma_2 = \pmatrix{0&-i\cr i&0}\,,~~~
\sigma_3 = \pmatrix{1&0\cr 0 &-1}\,.
\eeq
Assembling the last three in the \bfi{Pauli vector}\index{$\bsigma$, 
Pauli vector} 
\[
{\bsigma} = (\sigma_1,\sigma_2,\sigma_3)^T \in l(2,\Cz)^3,
\] 
we write for any $\p\in\Rz^3$
\lbeq{paulimatrices}
\p\cdot \bsigma:=p_1\sigma_1+p_2\sigma_2+p_3\sigma_3 =
\pmatrix{p_3 & p_1+ip_2\cr p_1-ip_2 & -p_3}\,.
\eeq
This matrix has zero trace, hence belongs to  $sl(2,\Cz)$, and it
is easily seen that every element of $sl(2,\Cz)$ can be written 
uniquely in this form. Similarly, each complex $2\times 2$-matrix can 
be written as a complex linear combination of all four Pauli matrices.
Defining the \bfi{Pauli 4-vectors}\index{$\sigma_\pm$, Pauli 4-vectors}
\[
\sigma_\pm:={\sigma_0 \choose \pm \bsigma} \in l(2,\Cz)^{1,3},
\]
we may write the general element of $l(2,\Cz)$ as
\lbeq{pauli-vec}
p\cdot\sigma_\pm = p_0\sigma_0\mp\p\cdot\bsigma
= \pmatrix{p_0\mp p_3      & \mp p_1 \mp ip_2\cr 
           \mp p_1\pm ip_2 & p_0\pm p_3}\,,\mbox{~~~for some }
p={p_0 \choose \p}\in \Cz^{1,3}.
\eeq
We note that, for $p,q\in\Cz^{1,3}$,
\[
(p\cdot \sigma_-)(q\cdot \sigma_+)=p\cdot q \sigma_0\,,
\]
\[
\det (p\cdot \sigma_\pm) = p^2.
\]

\bigskip
\bfi{Clifford algebras.}
\at{Clifford algebras and Dirac $\gamma$ matrices; here or perhaps 
later} \gzit{e.pauliprod} also implies the anti-commutation rule
\[
[\p\cdot \bsigma,\q\cdot\bsigma]_+ = 2 \p\cdot \q \sigma_0\,.
\]
Here 
\[
[f,g]_+:=fg+gf
\]
denotes\index{$[f,g]_+$, anticommutator} the \bfi{anticommutator} of 
$f$ and $g$.
\[
[\sigma_k,\sigma_l]_+ = 2\delta_{kl}\sigma_0
\for k,l=1,2,3\,.
\]

\section{The vector product, quaternions, and $SL(2,\Cz)$}
\label{s.vecq}

\bigskip
\bfi{The vector product.}
The structure of the Lie algebra $sl(2,\Cz)$ is intimately 
tied up with the vector product in $\Rz^3$.

The \bfi{vector product} of $\a,\b \in \Az ^{3}$ is the 
vector\index{$\a \times \b$, vector product}
\[
  \a \times \b:= 
 \left( \bary{c}
    a_{2}b_{3} - a_{3}b_{2}\\
    a_{3}b_{1} - a_{1}b_{3}\\
    a_{1}b_{2} - a_{2}b_{1}
 \eary \right)
  \in\Az^3,
\]
One easily checks that
\[
  \a \times \b = -\b \times \a,
\]
and the determinant formula for the \
bfi{triple product}\index{$\det (\a,\b,\cc)$, triple product}
\[
 (\a \times \b) \cdot \cc
  =\a \cdot (\b \times \cc) 
  = \det (\a,\b,\cc)
  := \det \left( \bary{ccc}
    a_{1} & b_{1} & c_{1}\\
    a_{2} & b_{2} & c_{2}\\
    a_{3} & b_{3} & c_{3}
 \eary \right)
 ,
\]
The most common case is that all three components are real or complex 
vectors. In this case, the following rules, which will be used in the 
following without comment, hold. 
\[
  \a \cdot (\b \times \cc)=\b \cdot (\cc \times \a)
=\cc \cdot (\a \times \b),
\]
\[
(\a\times \b)\cdot(\cc\times \d)
=(\a\cdot \cc)(\b \cdot \d)-(\b \cdot \cc)(\a\cdot \d),
\]
\[
  \a \times (\b \times \cc)=\b(\a \cdot \cc)-\cc(\a \cdot \b),
\]
\[
  (\a \times \b )\times \cc=\b(\a \cdot \cc)-\a(\b \cdot \cc),
\]
\[
  \a \times (\b \times \cc)
=(\a \times \b) \times \cc+\b \times (\a \times \cc),
\]
\[
  (\a \times \cc)\times (\b \times \cc)=\det(\a,\b,\cc)\cc.
\]
Indeed, each property follows by a simple 
computation either directly or from the previous property. 

A simple calculation with \gzit{paulimatrices} 
verifies the product formula
\lbeq{e.pauliprod}
(\p\cdot\bsigma)(\p\cdot\bsigma)
=(\p\cdot\q)\sigma_0 + i (\p\times \q)\cdot\bsigma,
\eeq
from which we obtain the commutation rule
\[
[\p\cdot \bsigma,\q\cdot \bsigma] = 2i\p\times \q \cdot \bsigma
\]
and the trace formula
\[
\tr (\p\cdot \bsigma)( \q\cdot\bsigma) = 2 \p\cdot\q.
\]

Although measurements usually involve real numbers only, 
the need for complex matrices in physics is undisputable.

Less familiar than complex numbers are the quaternions, obtained
by extending the complex number system by adjoining a further square 
root. This is possible without introducing zero divisors by renouncing
the commutative law of multiplication (which is lost for matrices 
anyway). After their discovery by William Hamilton in 1843 (cf. 
\sca{Hamilton} \cite{Ham.vec}), quaternions had an important role to 
play in physics because of their usefulness in describing rotations. 
But with the introduction of the vector product by 
\sca{Gibbs} \cite{Gib.vec} in 1881, quaternions 
declined in popularity and later almost disappeared from physics. 
However, they were kept alive in mathematics, and found new and 
important applications in modern computational geometry, where they 
are the method of choice for working with rotational motion and 
describe time-dependent rotations, and in astrodynamics, where they are
used for spacecraft attitude control systems.

In the context of Lie groups, quaternions and matrices whose 
coefficients are quaternions still play a significant role in the 
classification of real simple Lie groups and associated symmetric 
spaces; see., e.g., \sca{Chevalley} \cite{Che}, 
\sca{Gilmore} \cite{gilmore}, or \sca{Helgason} \cite{helgason}.

The Pauli matrices satisfy the product rule \at{check repeat 
\gzit{e.pauliprod}, make sigma's etc bold} 
\lbeq{rep6}
  (a \cdot \sigma )(b \cdot \sigma )= a \cdot b +i(a \times b) \cdot 
  \sigma \for a,b \in \Cz^3.
\eeq
\at{match the following with the theorem below!}
\gzit{rep6} shows that the vector space $\Qz'$ of complex $2\times 2$ 
matrices of the form
$a_0+ia\cdot\sigma$ ($a_0\in\Rz$, $a\in\Rz^3$) is an algebra. Indeed,
if we embed $\Cz$ into $\Rz^{2\times 2}$ using the imaginary unit 
$i=\Big(\bary{rr}0&1\\-1&0\eary\Big)$ (which satisfies $i^2=-1$),
\at{a construction that embeds $L(n,\Cz)$ into $L(2n,\Rz)$}
we can write the quaternions \gzit{quaijk} as \at{index i,j,k}
\[
\ii=\Big(\bary{rr}i&0\\0&-i\eary\Big)=i\sigma_3,~~~
\j=\Big(\bary{rr}0&1\\-1&0\eary\Big)=i\sigma_2,~~~
\k=\Big(\bary{rr}0&i\\i&0\eary\Big)=i\sigma_1,
\]
\at{check, factors $i$?!} 
which exhibits the isomorphism. $\Qz'$ can also be desribed as the 
set of complex $2\times 2$ matrices $U$ satisfying 
$U_{22}=U_{11}^*$ and $U_{21}=-U_{12}^*$, and the matrices in $\Qz'$
of determinant 1 are just the unitary $2\times 2$ matrices, hence
form the Lie group $U(2)$. This proves that $U(2)$ is isomorphic to
a double covering of the group $SO(3)$. An explicit isomorphism is 
given by 
\lbeq{rot37}
  (r_0+ir\cdot\sigma)(x \cdot \sigma )(r_0+ir\cdot\sigma)^*
=(Q[r]x) \cdot \sigma,~~~r_0=\sqrt{1-r^2}.
\eeq
\at{S/U p.171/2 have the following formulas 
(or $U^T$ in place of $U^*$?) -- can I derive them?}
\[
a^TQb=\half \tr (a\cdot\sigma)U(b\cdot\sigma)U^*,
\]
\[ 
U=\frac{1+\sigma\cdot Q\sigma}{2\sqrt{1+\tr Q}}
\]

\begin{thm}           
The set $\Qz$ of quaternions is a \bfi{skew field}, i.e., an associative
algebra in which every nonzero element has an inverse. We have 
\lbeq{qua2as}
  U(r_0,\rr)+U(s_0,\s)=U(r_0+s_0,\rr+\s),
\eeq
\lbeq{qua2bs}
  \lambda U(r_0,\rr)=U(\lambda r_0,\lambda \rr),
\eeq
\lbeq{qua2s}
  U(r_0,\rr)^T=U(r_0,-\rr),
\eeq
\lbeq{qua3s}
  U(r_0,\rr)U(s_0,\s)=U(r_0s_0-r\cdot \s,~s_0\rr+r_0\s+\rr \times \s ),
\eeq
\lbeq{qua4s}
  U(r_0,\rr)^{-1}= \frac{1}{r_0^2+\rr^2} U(r_0,-\rr) ~~~ 
\mbox{if~} r_0^2+\rr^2 \ne 0.
\eeq
\end{thm}   

\bepf                 
\gzit{qua2as}--\gzit{qua2s} are trivial, 
and \gzit{qua3s} follows by direct computation,
using \at{what}
Specializing \gzit{qua3s} to $s_0=r_0,\s=-\rr$ gives
\lbeq{qua5s}
  U(r_0,\rr)U(r_0,-\rr)=U(r_0^2+\rr^2,0)=(r_0^2+\rr^2) 1,
\eeq
which implies \gzit{qua4s}. Therefore $\Qz$ is a vector space closed 
under multiplication, and every nonzero element in $\Qz$ has an inverse.
Since matrix multiplication is associative, $\Qz$ is a skew field.
\epf

In the standard treatment, quaternions are treated like complex 
numbers, as objects of the form
\[
  \q (r_0, \rr)=r_0 1+r_1\ii+r_2\j+r_3\k
\]
with special unit quaternions $1,\ii,\j,\k$.
The correspondence is given by the identification 
\lbeq{quaijks}
\ii=\sigma_1, ~~\j=\sigma_2, ~~\k=-\sigma_3
\eeq
in terms of which $\q (r_0,\rr)=U(r_0,\rr)$. 
\at{check, factors $i$?!}

\at{tensor product = Kronecker product -- combining two systems -- 
matrices with matrix entries}

\section{The Hamiltonian form of a Lie algebra}

\at{Motivate and rephrase using the Schr\"odinger equation, 
where the infintesimal generator is scaled to make it Hermitian.}

In the Hamiltonian form, one takes Hermitian matrices and uses the 
Lie product i/hbar[f,g], to match things with quantum mechanical usage.
Expressed in terms of commutators, as usual, the structure constants 
(e.g., for su(2)-so(3)) become purely imaginary,
although the Lie algebra is real. 
\at{Thus there exist two incompatible 
traditions side by side, and this must be disentangled.}

In the applications, distinguished generators typically are Hermitian 
and represent 
important real-valued observables. \at{but these need not be in 
the Lie algebra}
Therefore they tend
to replace the matrix $A$ by $iA$. This is
one of the reasons why the structure constants for real algebras
appear in the physics literature with an $i$ when written in terms of 
commutators.

\at{
Lie algebra isomorphism,\\
Abstract Lie algebras,\\
$\lp$, and factor $\iota=i/\hbar$, $\iota$-representations,\\
needed for the Schr\"odinger equation, script p.13,\\
group isomorphism,\\
linear group representations}

Every one-parameter group is isomorphic to either $L(1,\Rz)$ or $U(1)$.
Two connected matrix groups are called \bfi{locally isomorphic} if 
their associated Lie algebras are isomorphic. For example, $L(1,\Rz)$ 
and $U(1)$ are locally isomorphic but not isomorphic.

Generators, commutation relations, and structure constants

\at{classical limit $\hbar\to 0$, script p.20f}

Introduce the vector $\LL$ of generators for $SL(2,\Cz)$ and its 
commutation relations.

The components of the vector product satisfy
\[
(\a\times \b)_j=a_{j+1}b_{j-1}-a_{j-1}b_{j+1}~~~
\mbox{(indices $j=1,2,3\mod 3$)}.
\]
We may express this as \at{what}
in terms of the {\bfi{Levi--Civit\`a symbol}},\index{$\epsilon_{jkl}$,
Levi--Civit\`a symbol}
defined by
\lbeq{e.leviciv}
\epsilon_{jkl} = \left\{ \begin{array}{l@{~~}l} 1 & 
\textrm{if $jkl$ is an even permutation of }123 \,,\\
-1 &
\textrm{if $jkl$ is an odd permutation of }123\,,\\
0 & \textrm{ otherwise.} \end{array}\right.
\eeq
Thus $\epsilon_{jkl}$ is completely antisymmetric in the
indices $j,k,l$ and $1\leq j,k,l\leq 3$. 

\at{give the contraction properties of $\eps_{ijk}$
(i.e., the rules for converting them to sums of deltas}

The Pauli matrices \gzit{e.pauli} satisfy (componentwise version of 
\gzit{e.pauliprod})
\[
\sigma_i\sigma_j = \delta_{ij} + i\epsilon_{ijk}\sigma_k\,,
\]
\[
\tr \sigma_i=0\,,~~ \tr \sigma_i\sigma_j = 2\delta_{ij}\,,~~
\tr[\sigma_i,\sigma_j]\sigma_k = 4i\epsilon_{ijk}\,.
\]
The Pauli matrices satisfy 
\[
[\sigma_k,\sigma_l]=2i\sum_{m=1}^3 \epsilon_{klm}\sigma_m
\for k,l=1,2,3\,.
\]
The summation
above contains only one nonzero term. For example, 
$[\sigma_1,\sigma_2]=2i\sigma_3$,
and all other Lie products can be found using a cyclic permutation.

\section{Atomic energy levels and unitary groups}\label{s.nlevel}

\at{describe semiclassical view of an atom}

In the terminology to be systematically developed in Chapter 
\ref{c.quants}, the \bfi{quantities} are the elements of the algebra
$\Ez=\Cz^{N\times N}$ of square complex $N\times N$ matrices, the 
\bfi{constants} are the multiples of the identity matrix.
The \bfi{Hamiltonian} $H$ represents the \bfi{energy} and is a diagonal 
matrix $H=\Diag(E_1,\dots,E_N)$
whose diagonal entries $E_k$ are the \bfi{energy levels} of the system.
In the \bfi{nondegenerate} case (the only case considered in this 
section), all $E_k$ are distinct, and the diagonal 
matrices comprise all functions of $H$. Quantities represented by 
arbitrary nondiagonal matrices are less easy to interpret. However,
an important class of quantities are the matrices of the form
$P=\psi\psi^*$, where $\psi$ is a vector of norm 1; they satisfy 
$P^2=P=P^*$ and are the quantities observed in binary measurements
such as detector clicks; see Section \ref{s.qprob}.
The \bfi{states} of the $N$-level system are the linear mappings
that map a quantity $f\in\Ez$ to its \bfi{value} 
$\<f\>=\tr \rho f$,\index{$\<f\>$, value of a quantity}
where $\rho\in\Ez$ is a positive semidefinite Hermitian 
matrix with trace one, called the \bfi{density matrix} of the state.
(Frequently, one talks somewhat sloppily of the ''state'' $\rho$.) 
The diagonal entries $p_k:=\rho_{kk}$ 
represent the probability for obtaining a response in a binary test for 
the $k$th quantum level; the off-diagonal entries $\rho_{jk}$
represent deviations from a classical mixture of quantum levels.

\at{eigenstates; ground state and excited states; preparation;
spectrum; forward ref to systems with continuous and mixed spectrum, 
script p.52-54}
The standard basis consisting of the $N$ \bfi{unit vectors} 
\idx{$\mbox{$\mid$}k\>$}
with a one in component $k$ and zeros in all other component 
corresponds to the $N$ levels of the quantum systems. 

\at{properties of the value map, script p.17}

\at{normalization and trace 1; beams have unnormalized states, and 
$\tr \rho$ is the \bfi{beam intensity}, the average number of particles 
that an ideal particle detector would register per unit time}

Extra structure of a \bfi{Euclidean $*$-algebra}, important for 
the applications; see Section \at{and script p.20}: The 
\bfi{conjugate} $f^*$\index{$f^*$, conjugate} of $f$ is given by 
conjugate transposition, and the \bfi{integral} 
$\sint g = \tr g$\index{$\sint g$, quantum integral} is the 
\bfi{trace}, the sum of the diagonal entries or, equivalently, 
the sum of the eigenvalues. 

\bigskip
Closely related to $n$-level quantum systems are the unitary groups
and their Lie algebras.

A matrix $U\in \Cz^{n\times n}$ is called \bfi{unitary}
if $U^*U=1$, equivalently, if $U$ is invertible and $U^{-1}=U^*$. 
The set $U(n)$\index{$U(n)$, unitary group}
of all unitary $n\times n$ matrices is a matrix group, 
called a (full) \bfi{unitary group}. Indeed, if 
$U,V\in U(n$ then $(UV)^*UV=V^*U^*UV=V^*V=1$ and 
$(U^{-1})^*U^{-1}=(U^*)^*U^{_1}=UU^{-1}=1$, so that 
$UV, U^{-1} \in U(n)$. The unitary ${n\times n}$ matrices with 
determinant one form a matrix group $SU(n)$\index{$SU(n)$, special 
unitary group}, called a \bfi{special unitary group}.

\at{move up: $SL(n,\Cz)$ acts but does not preserve trace}
$U(n)$ acts on the matrices of fixed trace and determinant 
by mapping $\rho$ to 
\[
U\{\rho\}:=U\rho U^*.
\]
Show preservation of semidefiniteness, trace and determinant.
$UV\{\rho\} =U\{V\{\rho\}\}$.

A matrix $A\in \Cz^{n\times n}$ is called \bfi{antihermitian}
if $A^*=-A$. The set $u(n)$\index{$u(n)$, unitary Lie algebra} of all 
antihermitian $n\times n$ matrices is a matrix Lie algebra, called a
\bfi{unitary Lie algebra}. Indeed, if $f,g\in u(n)$ then 
$[f,g]^*=(fg-gf)^*=(g^*f^*-f^*g^*=(-g)(-f)-(-f)(-g)=-(fg-gf)=-[f,g]$, 
hence $[f,g]\in u(n)$. The Lie algebra of infinitesimal generators of 
$U(n)$ is $u(n)$. Indeed, if $U(t)=1+tf+O(t^2)$ is unitary, we have
$1=U(t)U(t)^* = (1+tf+O(t^2))(1+tf+O(t)^2)^*=1+t(f+f^*)+O(t)^2$,
implying that $f$ is antihermitian. \at{Conversely,...}
The antihermitian ${n\times n}$ matrices with 
trace zero form a Lie algebra $su(n)$\index{$su(n)$, special unitary 
Lie algebra}, called a \bfi{special unitary Lie algebra}.
$su(n)$ is the Lie algebra of the matrix group $SU(n)$.

\section{Qubits and Bloch sphere}\label{s.qubits}

The smallest quantum systems have two levels only and are called 
\bfi{qubits}; they play an fundamental role in quantum information 
theory and quantum computing; cf.  {\sc Nielsen \& Chuang} \cite{NieC}.

\at{Bloch sphere for $\rho$ = Poincar\'e sphere, script p.23; 
adapt the following}
We have
\[
(p\cdot \sigma_\pm)^*=\ol p\cdot \sigma_\pm \,,
\]
so that $p\cdot \sigma_\pm$ is Hermitian if and only if the components 
of $p$ are real, and antihermitian if and only if the components 
of $p$ are purely imaginary. Therefore 
\[
u(2)=\{ i p\cdot \sigma_\pm \mid p\in\Rz^{1,3} \}\,,
\]
and letting $p$ take complex values we get the whole of $l(2,\Cz)$.

Similarly, the matrices $i\sigma_0,i\sigma_1,i\sigma_2,i\sigma_3$ form 
a basis of the Lie algebra $u(2)$, considered as a real vector space; 
indeed, any Hermitian $2\times 2$ matrix can be written in a unique 
way as $p\cdot \sigma_+$ for some $p\in\Rz^{1,3}$.

We obtain $su(2)$ for $p$ real and $p_0=0$.

\bigskip
\bfi{The Lie algebras \idx{$u(2)$} and \idx{$su(2)$}.}
The matrices $i\sigma_1,i\sigma_2,i\sigma_3$ form a basis for the Lie 
algebra $su(2)$, considered as a real vector space; indeed, any 
traceless and Hermitian $2\times 2$ matrix can be written in a unique 
way as $\p\cdot \bsigma$ for some $\p\in\Rz^3$. 
\at{relate to quaternion units}
Clearly, $i\sigma_0$ spans the center of the Lie algebra $u(2)$. 
As a consequence, we can write $u(2)\cong\Rz\oplus su(2)$. 
\at{needs center and quotient}

\bigskip
\bfi{The Lie group \idx{$U(2)$}.}
In the case $n=2$ it is a nice exercise to show that each special
unitary matrix $U$ can be written as
\[
U=\pmatrix{ x& y\cr -\ol y & \ol x }\,,~~~x,y\in \Cz\,,~
|x|^2+|y|^2=1\,.
\]
Writing $x=a+ib$ and $y=c+id$ for $a,b,c,d\in\Rz$ we see that 
$a^2+b^2+c^2+d^2=1$. This implies that there is
a one-to-one correspondence between $SU(2)$ and the set of points 
on the unit sphere $S^3$ in $\Rz^4$.
Thus $SU(2)$ is as a manifold homeomorphic to $S^3$. 
(The manifold point of view of matrix groups may be used to give a 
definition of abstract Lie groups; see \at{}.) 

\at{more detailed version. Combine!}
We now show that \idx{$SU(2)$} is a real manifold that is isomorphic
to the three sphere $S^3$. We do this by finding an explicit
parametrization of $SU(3)$ in terms of two complex numbers $x$ and $y$
satisfying $|x|^2+|y|^2 = 1$, which defines the three-sphere.

We write an element $g\in SU(2)$ as
\[
g = \pmatrix{ a & b \cr c& d }\,.
\]
Writing out the equation $g^* g=1$ and $\det g=1$ one finds the
following equations:
\begin{eqnarray}
&&|a|^2 + |c|^2 =1 \,, ~~~~  |b|^2 + |d|^2=1\,,\nn\\
&& \ol a b + \ol c d =0\,,~~~~~~~ ad-bc=1\,.\nn
\end{eqnarray}

 We first assume $b=0$ and find then that $ad=1$ and $\ol cd=0$,
implying that $c=0$ and $U$ is diagonal with $a=\ol d$.
 Next we suppose $b\neq 0$ and use $a =-c\ol d/\ol b$ to
 deduce that $|b|=|c|$ and $|a|=|d|$; we thus have $b\neq 0
 \Leftrightarrow c\neq 0$. We also see that we can use the ansatz
 \begin{eqnarray}
 a &=& e^{i\alpha} \cos \theta\,,~~~~~ b =e^{i\beta}\sin\theta \,,\nn\\
 c&=& - e^{i\gamma}\sin\theta\,,~~~ d= e^{i\delta}\cos\theta\,.\nn
 \end{eqnarray}
 Using again $a = -c\ol d/\ol b$ we see $\alpha +\delta=
 \beta+\gamma $ and writing out $ad-bc = 1$ we find $\alpha = -\delta$
 and $\beta = -\gamma$. We thus see $a=\ol d$ and $b= -\ol c$. Hence
 the most general element of $SU(2)$ can be written as
 \[
 g(x,y)= \pmatrix{ x & y \cr -\ol y & \ol x }\,, ~~~
\mbox{with}~~~ |x|^2+ |y|^2 = 1\,.
 \]
The map $S^3\to SU(2)$ mapping $(x,y)$ to $g(x,y)$ is clearly
injective, and from the above analysis bijective. Furthermore the map
is smooth. Hence we conclude that $SU(2)\cong S^3$ as a real
manifold.

\section{Polarized light and beam transformations}\label{s.polarized}

\at{channels = activities}

Qubits are closely related to the polarization of light.
Since polarization phenomena show the basic principles 
of quantum mechanics in a clean and transparent way,
we use polarization to derive the basic equations of
quantum mechanics, the Liouville equation and the Schr\"odinger 
equation, thus giving them an easily understandable meaning.
 
Polarized light was discovered by Christiaan \sca{Huygens} \cite{Huy}
in 1690. The transformation behavior of beams of completely polarized 
light was first described by Etienne-Louis \sca{Malus}\cite{Mal} in 
1809 (who coined the name ''polarization''), and that of partially 
polarized light  by George \sca{Stokes} \cite{Sto} in 1852.
The transverse nature of polarization was discovered 
by Augustin \sca{Fresnel} \cite{Fre2} in 1866, and the description in 
terms of (what is now called) the Bloch sphere by Henri \sca{Poincare} 
\cite{Poi} in 1892. 

It is instructive to read Stokes' 1852 paper \cite{Sto} in the light
of modern quantum mechanics. One finds there all quantum phenomena
for modern qubits, explained in classical terms!

Splitting polarized monochromatic beams into two beams with different, 
but orthogonal polarization corresponds to writing 
a wave functions as superposition of preferred basis vectors.
Mixtures are defined (in Stokes' paragraph 9) as arising 
from ''groups of independent polarized streams'' 
and give rise to partially polarized beams.
What is now called the polarization matrix is represented by Stokes
with four real parameters comprising, in today's terms, 
the Stokes vector, or, equivlently, the polarization matrix.
Stokes asserts (in his paragraph 16) the impossibility 
of recovering from a mixture of several distinct pure states 
any information about these states beyond what is encoded 
in the Stokes vector (i.e., the polarization matrix).
The latter can be linearly decomposed in many
essentially distinct ways into a sum of pure states, 
but all these decompositions are optically indistinguishable.

If one interprets the normalized polarization matrix as density matrix 
of a qubit, a polarized monochromatic beam of classical light 
behaves exactly like a modern qubit, which shares all the features
mentioned.
Polarized light is therefore the simplest quantum phenomenon,
and the only one that was understood quantitatively 
already before the birth of quantum mechanics in 1900.

Experiments with polarization filters are easy to perform; 
probably they are already known from school. 
Since polarization is a macroscopic phenomenon,
the counterintuitive features of quantum mechanics
irritating the untrained intuition are still absent.
But polarization was recognized as a quantum phenomenon only 
when quantum mechanics was already fully developed.
Norbert \sca{Wiener} \cite{Wie} 1930 exhibited a description in terms 
of the Pauli matrices and wrote: ``It is the conviction of the author 
 that this analogy'' between classical optics and quantum mechanics 
``is not merely an accident, but is due to a deep-lying connection 
between the two theories''. This is indeed the case; see, e.g.,
\sca{Neumaier} \cite{Neu.optslides}.

\bigskip
\bfi{The mathematics of polarization.} 
A beam of polarized light of fixed frequency is characterized 
by a state, described equivalently by the \bfi{Stokes vector},
a real 4-dimensional vector 
\[
S=(S_0,S_1,S_2,S_3)^T={S_0\choose\SS} \in \Rz^{1,3}
\]
with 
\lbeq{e.Sforw}
S_0\ge |\SS|, 
\eeq
or by a \bfi{polarization matrix} (also called  \bfi{coherence matrix}) 
a complex positive semidefinite Hermitian $2\times 2$ matrix $C$. 
These are related by 
\[
C  = \half \pmatrix{S_0+S_3 & S1+iS_2\cr S_1-iS_2 & S_0-S_3}
=\half S \cdot \sigma_- = \half(S_0\sigma_0 + \SS \cdot\bsigma)
\]
in terms of the Pauli matrices \gzit{e.pauli}.
(In the literature, the signs and order of the components may differ.)

\at{\gzit{e.Sforw} as forward cone in Minkowski space; 
the cone property; perhaps elsewhere}

The trace $\tr C=S_0$ of the polarization matrix is the \bfi{intensity} 
of the beam. If $S_0=0$, the beam is \bfi{dark} and contains no light.
Otherwise, one may normalize the intensity by dividing the polarization 
matrix by $S_0$, resulting in a \bfi{density matrix} of trace one,
\[
\rho = C/\tr C =\half r \cdot \sigma_-, ~~~r = S/S_0 = {1 \choose \rr};
\]
it contains the intensity-independent information about the beam. 
The intensity-independent quotient
\[
d:=|\rr|=|\SS|/S_0 \in [0,1]
\]
is called the \bfi{degree of polarization}, and allows the
determinant of the polarization matrix to be written as 
$\det C = \frac{1}{4}(S_0^2-\SS^2)=\frac{1}{4}S_0^2(1-d^2)$.

The extremal case $d=0$ characterizes \bfi{unpolarized light}, 
which therefore has a polarization matrix $C=\half S_0\sigma_0$.
At the other extreme, 
a fully polarized beam (a pure polarization state) has $d=1$;
it corresponds to a so-called \bfi{pure} polarization state.
Since $d=1$ characterizes singular polarization matrices, a pure 
polarization state can be written in the form $C=\psi\psi^*$ 
with a \bfi{state vector} $\psi$ determined up to a phase. 
In this case, the intensity of the beam is $S_0=|\psi|^2=\psi^*\psi$.
In particular, a normalized state vector has
norm $|\psi|=\sqrt{\psi^*\psi}=1$.

\bigskip
\bfi{Beam transformations.}
Optical instruments may transform beams by letting them pass through
a filter. A linear, non-mixing (not depolarizing) \bfi{filter} is 
characterized by a complex $2\times 2$ \bfi{Jones matrix} $U$.
(In the literature, many authors call $U^*$ the Jones matrix.)
The instrument transforms an in-going beam in the state $C$ 
into an out-going beam in the state $C'=UC U^*$.
If the instrument is lossless, the intensities of 
the in-going and the out-going beam are identical. 
This is the case if and only if the Jones matrix $U$ is unitary.

A linear, mixing (depolarizing) filter transforms $C$ 
instead into a sum of several terms of the form $UC U^*$.
It is therefore described by a \bfi{completely positive} linear map 
on the space of $2\times 2$ matrices, or a corresponding real 
$4\times 4$ matrix acting linearly on the Stokes vector, called the 
\bfi{M\"uller matrix}. For definitions and details, see, e.g., 
{\sc Aiello} et al. \cite{AiePW} and {\sc Benatti \& Floreanini} 
\cite{BenF}.

\at{relate to $L(2,\Cz)$. Filters preserve pureness; a polarizer
 has $U=tt^*$ and creates fully polarized light. script p.19;
semigroup property p. 15. Note that $U=\lambda\sigma_0$ acts trivial 
when $|\lambda|=1$; hence only the action of $U(2)/U(1)=SU(2)$ is 
relevant.}

\bigskip
\bfi{The Liouville equation.}
Passage through inhomogeneous media can be modelled 
by means of slices consisting of many very thin filters 
with Jones matrices close to the identity. 

\at{see script p.16, combine with next par., and relate to QML 
treatment. Liouville preserves pureness, giving Schr\"odinger, 
script p.19}

\bigskip
\bfi{The Schr\"odinger equation.}
If $\Delta t$ is the time needed to pass through one slice 
and $\psi(t)$ denotes the pure state at time $t$ then 
$\psi(t+\Delta t) = U(t,\Delta t) \psi(t)$, where $U(t,\pdot)$ 
is a $L(2,\Cz)$-motion parameterized by the transition time $\Delta t$.
We therefore introduce its infinitesimal generator
\[
H(t):=
i\hbar\partial U(t,\Delta t)/\partial \Delta t)\Big|_{\Delta t = 0},
\]
called the \bfi{Hamiltonian}\index{Hamiltonian!of a filter} of the 
filter.  Thus we have 
\lbeq{e.infU}
U(t)=1-\frac{i\Delta t}{\hbar}H(t) +O(\Delta t^2)
\eeq

In the lossless case, $U(t)=U(t,\Delta t)$ is unitary, which implies 
that $H(t)$ is Hermitian. \at{relate to general Hermitian Lie algebra}

A linear, non-mixing (not depolarizing) instrument with Jones matrix 
$U$ transforms an in-going beam in the pure state with state vector 
$\psi$ into an out-going beam in a pure state with state vector 
$\psi'=U\psi$. 
\gzit{e.infU} implies \at{check sign!}
\[
i\hbar\frac{d}{dt}\psi(t) 
\approx \frac{i\hbar}{\Delta t} (\psi(t+\Delta t)-\psi(t))
= \frac{i\hbar}{\Delta t} (U(t)-1) \psi(t).
\]
In a continuum limit we thus recover the time-dependent 
\bfi{Schr\"odinger equation}
\[
   i\hbar\frac{d}{dt}\psi(t) = H(t) \psi(t).
\]

\section{Spin and spin coherent states}
\label{s.spinor}

In this section,
we discuss the {\bfi{spinor representations}} of $L(2,\Cz)$, 
see also \sca{Sternberg} \cite{sternberg}. By restricting to the unitary
matrices we get unitary representations of the group $SU(2)$.
As we shall see later in Section \ref{s.su2rep}, these representations
comprise all irreducible unitary representations of $SU(2)$. 

For $0\le s\in\shalf \Zz$ (the factor $\half$ appears here for 
historical reasons only) we denote with \idx{$\Pz _s$} the complex 
vector space of all
homogeneous polynomials of degree $2s$ in $z =(z_1,z_2) \in \Cz ^2$.
The space $\Pz_s$ has dimension $2s+1$ since the monomials $z_1^k
z_2^{2s-k}$ ($k=0,1,\ldots,2s$) form a basis of $\Pz _s$.
The group $L(2,\Cz)$ of invertible complex $2\times 2$ matrices acts
on $\Cz^2$ in the natural way. On $\Pz _s$ we get a
representation of $L(2,\Cz)$ by means of the formula\index{$U(g)$,
representation of group element}
\lbeq{spr10}
  (U(g) \psi )(z):= \psi (g^{-1}z) \mbox{~~~for~}g \in L(2,\Cz)\,.
\eeq
Then indeed
$U(g)U(h)\psi(z) = U(h)h\psi(g^{-1}z) = \psi (h^{-1}g^{-1}z)
= \psi( (gh)^{-1}z)=U(gh)\psi(z)$.
Taking infinitesimal group elements, we find that the Lie algebra
$l(2,\Cz)$ acts on $\Pz _s$ by means of the representation $J$
\at{concept undefined}
defined by\index{$J(f)$, representation of Lie algebra element}
\lbeq{spr18}
  (J(f) \psi )(z) = -(fz) \cdot \nabla \psi (z)\,,\mbox{~~~for~}f\in
  l(2,\Cz)\,.
\eeq
Note that this is again a homogeneous polynomial of degree $2s+1$.

In nonrelativistic quantum mechanics, an elementary particle with 
\bfi{spin} $s$ is described 
by an element of the space $L^2(\Rz^3,\Pz_s)$ of square integrable 
mappings from $\Rz^3$ to $\Pz_s$. The Hamiltonian for a particle with 
spin $s$ in a magnetic field $\B\in\Rz^3$ is given in terms of the 
Pauli matrices by
\lbeq{e.sham}
H= - \B\cdot \bsigma 
= - \pmatrix{ B_3 & B_1 + iB_2 \cr B_1 - iB_2 & -B_3}\,,  
\eeq
where the action of $H$ is given by \gzit{spr18}. The dynamics is
described by the Schr\"odinger equation 
\lbeq{e.ssch}
  i \hbar \dot{\psi } =H \psi\,.
\eeq

\bigskip
\bfi{The unitary case.}
By restricting in \gzit{pauli-vec} to real-valued $p$,
we represent $u(2)$. \at{improve formulation} The resulting
representation turns out to be unitary. To give $\Pz_s$ the appropriate
Hilbert space structure, we define on the unit disk
\[
D=\{z\in\Cz^2 \mid z^*z\le 1\}
\]
of $\Cz ^2$ the measure $Dz$\index{$Dz$, invariant measure}
 by
\lbeq{spr1}
  \int Dz f(z^*,z)= \int _D dz ^2 f(z^*,z)\,.
\eeq
Explicitly we thus have $Dz =dz_1d\bar z_1 dz_2 d\bar z_2$, so that
for example
\beqar
\int Dz \bar z_{1}^{k}z_{1}^{l} \bar z_{2}^{m}z_{2}^{n}
&=& 4\pi^2\delta_{kl}\delta_{mn}\int_{0\leq x^2+y^2\leq 1}
x^{2k+1}y^{2m+1} dxdy\nn\\
&=& \pi^2 \delta_{kl}\delta_{mn}\int_{0\leq  s+t\leq 1}
s^{k}t^{m} \,ds\,dt \nn \\
&=& \frac{\pi^2 k!m!}{(k+m+2)!} \delta_{kl}\delta_{mn}\nn\,,
\eeqar
where in the last step we used
\[
\int_{0}^{1} x^a(1-x)^b \,dx = \frac{a!b!}{(a+b+1)!}\,.
\]

\begin{prop}
$Dz$ is an $SU (2)$ invariant measure satisfying
\lbeq{spr2}
  \int Dz (z^*x)^{2s}(z^Ty)^{2s}= \gamma _s (x^Ty)^{2s},
   ~~~ 0 \le s \in \half \Zz ,
\eeq
where
\lbeq{spr3}
  \gamma _s = \pi ^2 /(2s+1)(2s+2).
\eeq
\end{prop}

\bepf
Under a change of integration $z'=gz$ we have
$dz_{1}'d\bar z_{1}' =\det g dz_{1}d\bar z_{1} = dz_{1}d\bar z_{1}$.
Hence if we use for $g \in SU(2)$ the substitution $z=gz'$, then the
integral in \gzit{spr2} transforms into the same integral with
$(x',y')=(\bar{U}x, Uy)$ in place of $(x,y)$. Thus it is invariant
under $SU (2)$ and depends therefore only on $x^Ty$. Indeed, we can
always rotate $x$ such that $x=(x_1,0)$ and then clearly the
right-hand side is a polynomial with terms
$x_{1}^{2s}y_{1}^{m}y_{2}^{2s-m}$, which is only invariant under the 
diagonal $U(1)$-subgroup if $m=2s$. Hence the right-hand side of 
\gzit{spr2} is fixed up to the constant $\gamma_s$, which is found by 
looking at the special case $x=y= {1 \choose 0}$:
\lbeq{rcl}
  \gamma _s=\D\int Dz \,(z^*_1)^{2s} z_1 ^{2s} = \int _D dz ^2
  |z_1|^{4s} = \frac{\pi^2 (2s)!}{(2s+2)!}=\frac{\pi^2}{(2s+1)(2s+2)}\,.
\eeq
\epf

We make $\Pz_s$ into a Hilbert space by giving it the inner product
\lbeq{spr4}
  \phi ^* \psi = \< \phi | \psi \> := \gamma_{s}^{-1} \int Dz
  \,\ol{\phi (z)} \psi (z)\,.
\eeq
We introduce the basis vectors $\pi_{k}^{(s)} = z_{1}^{k}z_{2}^{2s-k}$
for $\Pz_s$, in terms of which the inner product reads
\at{check and relate to Pauli set stuff below!}
\[
\< \pi_{k}^{(s)}| \pi_{l}^{(s)}\> 
= {2s \choose k}^{-1}\delta_{kl}\,. 
\]
For $x \in \Cz ^2$, we define the {\bfi{coherent state}}
$|x,s\>\in\Pz_s$ to be the functions
\lbeq{spr5}
  |x,s \> (z) := (x^*z)^{2s}=(\bar x_1 z_1+\bar x_2 z_2)^{2s}\,.
\eeq
Then we can restate \gzit{spr2} as
\lbeq{spr6}
  \< x,s|y,s \> = (y^*x)^{2s} \mbox{~~~for~~~}x,y \in \Cz ^2.
\eeq
In particular, the coherent state $|x,s \>$ is normalized to norm $1$
if and only if $x$ has norm $1$. Directly from \gzit{spr5}, we see that
\lbeq{spr12}
  | 0,s \> =0,~~~| \lambda x,s \> = \lambda ^{2s} |x,s \> ,
\eeq
so that it suffices in principle to look at coherent states with $x$ of
norm $1$. In particular, choosing the parametrization
$x={1 \choose w}$ gives the traditional
{\bfi{spin coherent states}} of \sca{Radcliffe} \cite{radcliffe}.
For coherent states, \gzit{spr10} implies
\lbeq{spr11}
  U(g) |x,s \>= | g^{-1*}x,s \> \mbox{~~~for~}g \in SL(2,\Cz)\,,
\eeq
Thus coherent states define a representation of $L(2,\Cz)$, the
{\bfi{spinor representation}} of $L(2,\Cz)$. We verify that we
correctly have $U(g)U(h)|x,s\>= | g^{-1*} h^{-1*}
x,s\>=|(gh)^{-1*}x,s\>=U(gh)|x,s\>$. One sees easily that only the
subgroup $SU(2)$ is represented unitarily and we have
\[
 U(g) |x,s \>= | g x,s \> \mbox{~~~for~}g \in SU(2)\,.
\]
Note that the Schr\"odinger equation \gzit{e.ssch} implies that 
$\psi(t)=U(t)\psi(0)$, where $U(t)=e^{-itH/\hbar}$.
Since the Hamiltonian \gzit{e.sham} is an element of $su(2)$, 
we have $U(t)\in SU(2)$, and 
equation \gzit{spr11} implies the \bfi{temporal stability} of coherent 
states. This means that if the initial state vector is a coherent state,
then under the time evolution determined by $H$ the state vector 
remains for all times a coherent state. Since the norm of the wave 
function is invariant under the dynamics, too, one can work
with normalized coherent states throughout.

In general, let $\Hz$ be a Hilbert space of functions on some space
$\Omega$. If we can write function evaluation as inner product,
i.e., if for every $x\in \Omega$ there is an element $g_x\in \Hz$
such that $f(x)=\< g_x | f\>$ for some , then we say that $\Hz$ has 
the  {\bfi{reproducing kernel property}}. 

We show that the space $\Pz_s$ has the reproducing kernel property.
Expanding $(x^*z)^{2s}$ using the binomial series we obtain
\[
|x,s\> = \sum_{m=0}^{2s} {2s \choose m}
\ol{\pi_{m}^{(s)}(x)}\,\pi_{m}^{(s)}\,,
\]
from which it follows 
\[
\pi_{m}^{(s)} = \gamma_{s}^{-1}\int_D Dx\,\pi_{m}^{(2s)}(x)|x,s\>\,,
\]
so that the coherent states span $\Pz_s$. From \gzit{spr6} we find
\lbeq{spr7}
  \< x,s | y,s \> = |y,s\> (x)
\eeq
for all coherent states $| y,s \>$ and since these span $\Pz_s$,
we have for all $\psi \in \Pz _s$
\[
\<x,s|\psi\> =\psi(x)\,,\mbox{~~~for all~}\psi\in\Pz_s\,,
\]
which is the reproducing kernel property.
This implies that we can reproduce elements as follows. 
For all $\phi \in \Pz _s$ we have 
\lbeq{rclk}
  \< \phi | \psi \> = \gamma_{s}^{-1} \int Dz \,\ol{\phi (z)}
  \psi (z)= \gamma_{s}^{-1}  \int Dz \, \< \phi|z,s\> \<z,s|\psi\>\,,
\eeq
from which it follows that we can reproduce $\psi$
\lbeq{spr8}
  \psi = \gamma ^{-1} _s \int Dx \psi (x) | x,s \> \mbox{~~~for all~}
  \psi \in \Pz _s\,.
\eeq
Equation \gzit{spr8} implies the {\bfi{completeness
relation}}
\lbeq{spr9}
  \gamma _s^{-1} \int Dz |z,s \> \< z,s | =1\,.
\eeq
These properties characterize coherent states in general.
For an extension of the coherent state concept to semisimple 
Lie groups see \sca{Perelomov} \cite{Per} and \sca{Zhang} et al. 
\cite{ZhaFG}. Coherent states for Heisenberg groups are called 
\bfi{Glauber coherent states}, and are basic for modern quantum optics.
See Section \at{} and the book by \sca{Mandel \& Wolf} \cite{ManW}.

\at{The distinction between a non-relativistic (Pauli)
electron and a relativistic (Dirac) electron is muddied. By focusing
on $SU(2)$ it seems like the former, but later you mention the
restricted Lorentz group (which suggests the latter). Maybe the whole
section needs a careful re-think and general overhaul. The whole spin
coherent state thing seems a bit premature when you haven't yet
discussed ordinary coherent states, nor even the unirreps of so(3).
Maybe the so(3) quantum numbers should be derived first -- and maybe
that's part of what you had in mind for the earlier "atomic energy
levels" section? On the other hand, the Stern-Gerlach experiment
is the usual physical motivation for spins being quantized, but
currently that seems to be in a subsequent section.}

\section{Particles and detection probabilities}\label{s.sterng}

\at{The Stern-Gerlach experiment, Silver atoms, spin, 
coherent states, superpositions}

Entanglement

{The Stern--Gerlach Experiment.}
The Stern--Gerlach experiment is one of the most prominent and best 
known experiments in the history of quantum mechanics. The experiment 
provided a first experimental verification of the discrete nature of 
quantum mechanics. At the time of the experiment, which took place in 
1922, the phenomenon of spin was not well-understood and, from the 
point of view of our present knowledge,  a wrong model was used. 
Fortunately, the outcome of the experiment was in concordance with this 
model and the discrete nature of quantum mechanics was accepted as a 
fact. 

When later a better model was invented, the theory and the 
Stern--Gerlach experiment showed discrepancies. It was perhaps 
partially because of these discrepancies that Goudsmit and Uhlenbeck 
postulated that the electron had half-integer spin: with the 
half-integer spin of the electron the experiment of Stern and Gerlach 
was again in agreement with the theory \footnote{
As more often in the history 
of physics, it was a coincidence that determined the acceptance of a 
theory. Another such example was the measurement of the deflection of 
rays of the stars that can be seen close to the sun during a solar 
eclipse done by Eddington in 1919, thereby verifying the general theory 
of relativity of Einstein. The actual deflections are too small to be 
measured and hence the deflections found by Eddington have to be 
ascribed to noise; luckily the noise gave a pattern in agreement with 
the theoretical results.}. 

The setup of the Stern--Gerlach experiment is quite easy. To understand 
the physics behind the experiment, one only has to know that the 
energy of a small object with magnetic moment $\mu$ in an magnetic 
field $B$ is given by the equation
\[
U = -\mu\cdot B\,.
\]
The energy is measured relative to the energy far away towards 
infinity where there is no magnetic field. Note that the magnetic 
moment is a vector. Hence, classically it lives in a representation 
of $SO(3)$, the representations of which are labelled by integers 
$l=0,1,2,..$. The dimension of the $l$th representation is $2l+1$. 
In the Stern--Gerlach experiment a beam of particles with some fixed 
absolute value of $\mu$ is sent through an inhomogeneous magnetic 
field pointing in, say, the $z$-direction. Behind the magnetic field 
a screen is placed that will light up whenever a particle hits the 
screen. The force $F$ exerted on a particle with magnetic moment 
$\mu=(\mu_x,\mu_y,\mu_z)$ is given by
\[
F = -\nabla U = \mu_z \frac{\partial B_z}{\partial z}\,.
\] 
Thus, classically, the beam will be smeared out; the particles with 
$\mu$ pointing in the $+z$-direction will be deflected upwards, those 
with $\mu$ pointing in the $-z$-direction will be deflected downwards. 
Classically all positions of $\mu$ are possible and distributed in a 
Gaussian way, so that the screen will show a bounded strip, most 
intense in the center and fading out towards the ends. However, the 
result of the Stern--Gerlach experiment showed very clearly two blobs, 
centered at the positions corresponding to $\mu$ pointing up and down. 
Both blobs had the same intensity.

Assume that we have a bunch of particles (for example electrons), then 
they all have the same value of $l$, but the $z$-component of the 
magnetic moment might be different. Since the $z$-value can take 
$2l+1$ values, the beam will split in $2l+1$ different parts.

In their experiment, Stern and Gerlach used silver atoms, of which we 
now know that there is one electron in the outmost orbit and it is 
this electron which gives rise to the magnetic moment. The spin of an 
electron is however not in an $SO(3)$-representation, but in an 
$SU(2)$-representation and this correspondsn to $l=1/2$. 
This representation is two-dimensional and thus the general state 
$\psi$ of an electron can be described as
\[
\psi = a \vert + \rangle + b \vert - \rangle\,, ~~~ |a|^2 + |b|^2 =1\,,
\]
where $\vert+\rangle $ is the state with $\mu$ pointing in the 
$+z$-direction and $\vert - \rangle $ is the state with $\mu$ pointing 
in the $-z$-direction. When one measures the $z$-component of the 
magnetic moment, one finds with probability $|a|^2$ the value $+1/2$ 
and with probability $|b|^2$ the value $-1/2$. In a sample of heated 
silver atoms, there is no preferred direction for $\mu$ and thus in 
the end, the possibility that the value of the magnetic moment of a 
single silver atom is $+1/2$ is more or less $1/2$. This explains why 
the two blobs in the Stern--Gerlach experiment are equally bright.

\section{Photons on demand}\label{s.pod}

\at{maybe}

In this section we consider a quantum model for photons on demand,
and its realization through laser-induced emission by a single calcium 
ion in a cavity. The exposition is based on {\sc Keller} et al. 
\cite{KelLH}.

\begin{figure}[htb]
\begin{center}
\resizebox{!}{4cm}{\includegraphics{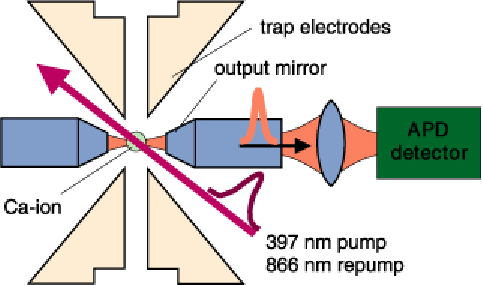}}
\caption{Experimental set-up for the generation of single-photon 
pulses with an ion-cavity system. The drawing shows a cross-section 
through the trap, perpendicular to the trap axis. 
(Figure 16 from \cite{KelLH})}
\label{f.setup}
\end{center}
\end{figure}

\begin{figure}[htb]
\begin{center}
\resizebox{!}{4cm}{\includegraphics{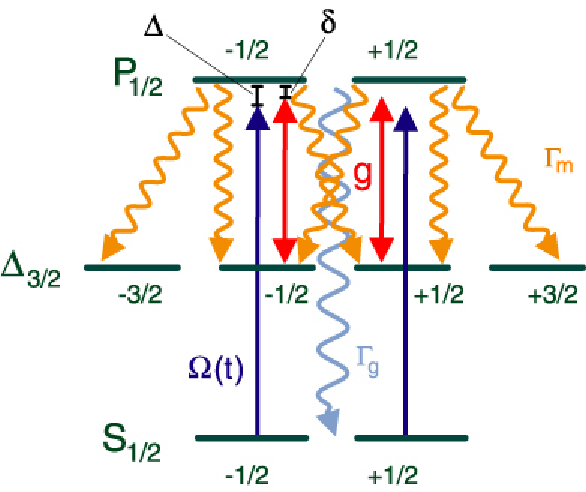}}
\caption{Scheme of the eight-level model on which we base our 
numerical calculations. Pump and cavity field are assumed to be 
linearly polarized in the direction of the quantization axis. 
For clarity, the four possible spontaneous decay transitions to the 
ground state are represented by a single arrow. 
(Figure 5 from \cite{KelLH})}
\label{f.8level}
\end{center}
\end{figure}

\begin{figure}[htb]
\begin{center}
\resizebox{!}{4cm}{\includegraphics{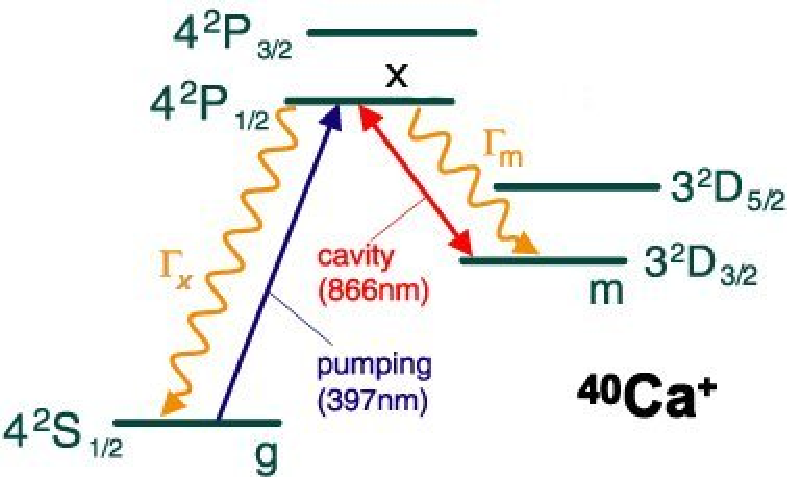}}
\label{f.3level}
\end{center}
\end{figure}

In their paper, {\sc Keller} et al. discuss in detail a model based on 
the simplified level scheme given in Figure \ref{f.3level}
which ignores the fine structure of the $Ca^+$ states.

\begin{itemize}
\item
A {\bf single} ion is localized in the cavity 
for many hours
\item
pulsating external fields (lasers) with
a total cycle time 100kHz give a 
predictable rate of single photons
\item 
pump laser at 397nm close to the
excitation frequency $S\to P$
\item
Repeated excitation to $P$ and decay to $S$ until 
decay into the metastable $D$ state; then inactive
\item
$\implies$ produces {\bf exactly} one photon (not counting losses)
\item
reexcite ion into excited state with a reset laser at 866nm,
until it falls back into the ground state
\item
ground state $g$, metastable state $m$,
excited state $x$ of $Ca^+$ 
\item
photons 
$\gamma_\fns{cavity}, \gamma_\fns{pump}, \gamma_\fns{reset}$ 
\item
electron $e$ bound in detector
\end{itemize}

\begin{figure}[htb]
\begin{center}
\resizebox{!}{4cm}{\includegraphics{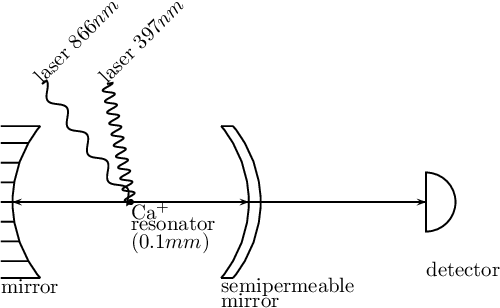}}
\end{center}
\end{figure}

\bfi{Active processes}
\begin{itemize}
\item
$a$:~~~ $\gamma_\fns{cavity}\rightleftharpoons\gamma_\fns{cavity}$
~(cavity detuning)
\item
$b$:~~~ $g + \gamma_\fns{pump}\rightleftharpoons x$ 
~~~(excitation)
\item 
$c$:~~~ $x \rightleftharpoons m + \gamma_\fns{cavity}$ 
~~(decay to metastable state)
\item
$d$:~~~ $\gamma_\fns{cavity} + e \rightleftharpoons \emptyset$
~~~(photodetection)
\item
$e$:~~~ $m + \gamma_\fns{reset}\rightleftharpoons x$
~~~(ion reset)
\end{itemize}
Only $a,b,c$ are modelled explicitly by {\sc Keller} et al..

But $d,e$ can be modelled similarly.

\bigskip
\bfi{Interaction picture model}
by {\sc Keller} et al. 
(without reset and photodetection)
\begin{itemize}
\item
$a$ = annihilator of cavity mode of photon
\item
$b=|g\>\<x|$
\item
$c=|m\>\<x|$
\item
Hamiltonian
$H=\hbar(\delta a^*a+\Delta |g\>\<g| + 2 \re(\Omega(t)b^*+\mu ac^*))$
\item
$\delta=\omega_\fns{cavity}-\omega_{xm}$ cavity detuning
\item
$\Delta=\omega_\fns{pump}-\omega_{gx}$ pump detuning
\item
$\Omega(t)$ classical pulse shape of pump laser
\item 
$\mu$ ({\sc Keller}'s $g$) ion--cavity coupling strength
\end{itemize}

\begin{itemize}
\item
$\kappa=0.02\,\Gamma_x$ cavity loss rate 
\item
$\Gamma_x\approx 138$ MHz ({\sc Keller}'s $\Gamma_g$) 
spontaneous decay rate into ground state 
\item
$\Gamma_m\approx 11$ MHz spontaneous decay rate into metastable state 
\end{itemize}

To account for losses, the dynamics of the density matrix 
is set up in the form of a

\bigskip
\bfi{Lindblad master equation} 
\[
\bary{lll}
\dot\rho &=-\frac{i}{\hbar}[H,\rho]
 & + \kappa(2a\rho a^* - a^*a\rho -\rho a^*a)\\
&& + \frac{\Gamma_x}{2}(2b\rho b^* - b^*b\rho -\rho b^*b)\\
&& + \frac{\Gamma_m}{2}(2c\rho c^* - c^*c\rho -\rho c^*c)
\eary
\]
Note that the master equation is 
an equation for transition rates; 
probabilities are obtained by 
integration over time.
\begin{itemize}
\item time-dependent expectations
$\<f\>_t = \tr f \rho(t)$
\item time dependent emission rate
$p(t)=2 \kappa_\fns{tr} \<a^*a\>_t$
($2 \kappa_\fns{tr}$ intensity transmission rate)
\item probability of photon emerging from the cavity
$\eta_\fns{photon} = \int_0^\infty p(t)dt$
\item single-photon efficiency
$\eta_\fns{abs} = (\kappa/\kappa_\fns{tr} -1)\eta_\fns{photon}$
\end{itemize}

The Hilbert space on which the master equation is based
is the tensor product of a single mode Fock space 
for the cavity photon and a 3-mode space for the $Ca^+$ ion. 

An orthonormal basis of the space is given by the kets $|n,k\>$,
where $n=0,1,\dots$ is the photon occupation number and
 $k\in\{g,x,m\}$ labels the ion level.

The structure of the Hamiltonian and the dissipation terms 
in the master equation is such that if the system is started 
in the ground state $|0,g\>$, it evolves to a mixed state 
in which the photon number is never larger than 1. 

Thus multiphoton states do not contribute at all,
and one can truncate the cavity photon Fock space 
to the two modes with occupation number $n=0,1$, 
without changing the essence of the model.
 
Of interest for the photon production is the projection 
of the density matrix to the photon space, obtained 
by tracing over the ion degrees of freedom. This results 
in an effective time-dependent photon density matrix
\[
\rho_\fns{photon}(t) = 
\pmatrix{\rho_{00}(t) & \rho_{01}(t) \cr \rho_{10}(t) & \rho_{11}(t)},
\]
where 
$\rho_{11}(t)=p(t)$ is the probability density of finding 
a photon,
$\rho_{00}(t)$ is the probability density of finding no photon, and
$\rho_{01}(t)=\rho_{10}(t)^*$ measures the amount of entanglement
between the 1-photon state and the vacuum state. 

Semidefiniteness of the state requires $|\rho_{01}|\le \sqrt{p(1-p)}$.

Assuming for simplicity that we have approximate equality,
$\rho_\fns{photon}$ is essentially rank one, 
\hspace*{0.5cm}
$\rho_\fns{photon}(t) \approx \psi(t)\psi(t)^*,~~~
\psi(t)=s(t)|0\>+c(t)|1\>$,
where $s(t)$ and $c(t)$ are functions with $|s(t)|^2+|c(t)|^2=1$,
determined only up to a time-dependent phase factor. 
In particular, we may take $c(t)$ to be real and nonnegative.

Thus, in the approximation considered, 
the quantum electromagnetic field is in a superposition 
of the vacuum mode and the single-photon field mode, 
with a 1-photon amplitude $c(t)=\sqrt{p(t)}$ 
that varies with time and encodes the 
probability density $p(t)$ of detecting a photon particle.

In the actual experiments, $p(t)$ has a bell-shaped form,
and the total photon detection probability, 
referred to as the \bfi{efficiency}, is significant, 
but smaller than 1.

Discarding the vacuum contribution corresponding 
to the dark, unexcited cavity, and giving up 
the interaction picture by inserting the field description 
$|1\>_t =e^{-i\omega t}\psi_0(\x)$ of the photon mode, 
the (now time-dependent) 1-photon state takes the form 
\hspace*{0.5cm}
$\A_\fns{1photon}(\x,t)=\sqrt{p(t)}e^{-i\omega t}\psi_0(\x)$. 

(At this stage one notices a minor discrepancy 
with the field description, since the 1-photon state is
no longer an exact solution of the Maxwell equations. 
To correct this deviation from Maxwell's equations, 
one has to work with quasi-monochromatic modes 
and the paraxial approximation.)

We now add the reset mechanism to get a 
continuous pulsed photon stream. 
Thus we consider a periodic sequence
of excitation-reset cycles of the ion in the cavity.
As before, we find that the electromagnetic field corresponding 
to the sequence of pulses is a single, periodically excited 
1-photon mode of the electromagnetic field.
Thus what appears at the photodetector as 
a {\bf sequence} of photon particles arriving 
is from the perspective of quantum electrodynamics
the manifestation of a {\bf single} nonstationary, 
pulsed 1-photon state of the electromagnetic field!

\at{Dynamical semigroups}

\section{Unitary representations of $SU(2)$}\label{s.uniSU2}

\at{This is an alternative version of the spin construction.
To be combined with the above. Make 3-vectors fat!}

We call a set of linear operators 
$ L_0$, $ L_{1}$, $ L_{2}$, $ L_{3}$ 
on a Euclidean space $\Hz$ a {\bf Pauli set} if $ L_0$ is
Hermitian positive definite and 
\lbeq{rep1}
  [  L_{1},  L_{2}]=2i  L_{3},~~~~
  [  L_{2},  L_{3}]=2i  L_{1},~~~~
  [  L_{3},  L_{1}]=2i  L_{2},
\eeq
\lbeq{rep3}
  ( L_0  L_{\mu})^{*}=  L_0  L_{\mu} \for
  \mu = 1:3.
\eeq
(In the infinite-dimensional case, we also require that $ L_0$
and \gzit{rep3} are self-adjoint.) 
We say the Pauli set has {\bf spin} $j$ if 
\lbeq{rep0}
 L_{1}^2+ L_{2}^2+ L_{3}^2=4j(j+1).
\eeq
Definiteness of $ L_0$ implies that the Hermitian inner 
product on $\Cz^s$ defined by
\lbeq{rep4}
   \overline{\phi}\psi := \phi^* L_0\psi \for \phi,\psi\in\Cz^s
\eeq
is positive definite, and
\lbeq{rep5}
   \overline{ L_{\mu }\phi } \psi = 
   \overline{\phi }  L_{\mu } \psi \for \mu =1,2,3.
\eeq
We write 
\[
a\cdot  L:=a_1 L_1+a_2 L_2+a_3 L_3 \for a\in\Cz^3.
\]

\begin{prop}~\\
(i) Any Pauli set satisfies
\lbeq{rep2}
  [a \cdot  L , b \cdot  L ] =2i(a \times b) \cdot  L \for
  a,b \in \Cz ^{3},
\eeq
and hence defines a representation $\hat X$ of $so(3)$ by
\lbeq{rep2i}
\hat X(a):=a\cdot  L/2i,
\eeq
which is unitary in the inner product \gzit{rep4}.

(ii) If $s \ne 1$ then $\Lz=\{a\cdot  L \mid a\in\Cz^3\}$
is a Lie algebra isomorphic to $so(3)$.

(iii) $C:= L_1^2+ L_2^2+ L_3^2$ is a Casimir operator of 
$\Lz$, i.e., 
\lbeq{rep2b}
  [C,a \cdot  L] = 0 \Forall a \in \Rz ^{3}.
\eeq

(iv) For any Pauli set and an arbitrary rotation 
$Q=(e_1,e_2,e_3) \in SO(3)$, the $ L_k'=e_k\cdot  L$ form 
together with $ L_0$ another Pauli set. 
\end{prop}

\bepf
(i) \gzit{rep2} follows from
\[
  [a \cdot  L , b \cdot  L ]= \D \sum_{\mu ,\nu =1:3} a_{\mu}
  b_{\nu } [ L_{\mu },  L_{\nu }]= 
  \D \sum_{\mu < \nu } (a_{\mu } b_{\nu }-b_{\mu } a_{\nu })
  [ L_{\mu },  L_{\nu }]= 2i (a \times b) \cdot  L,
\]
and (iv) follows directly from \gzit{rep2}.

(ii) \gzit{rep2} implies that $\Lz$ is closed under formation of 
commutators. Hence $\Lz$ is a Lie algebra. The isomorphism follows from 
Example (i) below.

(iii-iv) \at{}
\epf

In contrast to (iv), the spin equation 
\gzit{rep0} is {\it not} preserved under general rotations. 
\at{Or do we get it by a similarity transform?}

\begin{expls}\label{ex.pauli}
The following examples all have $ L_0=1$.

(i) On $\Cz^3$, a Pauli set of spin $1$ is given by
$a\cdot  L=2iX(a)$.

(ii) {\bf (Angular momentum)} Let $J=q\times p$, where 
the components of position $q$ and momentum $p$ are the linear
operators on $\Hz=C^\infty(\Rz^3)$ defined by
\[
(p_\mu\psi)(x)=i\hbar\frac{\partial\psi(x)}{\partial x_\mu},~~~
(q_\mu\psi)(x)=x_\mu\psi(x) \for \mu=1,2,3.
\] 
Then $ L_\mu:=2J_\mu/\hbar$ for $\mu=1,2,3$
defines a Pauli set on $\Hz$.

(iii) On any Euclidean space, a Pauli set of spin 0 is given by 
$ L_{1}= L_{2}= L_{3}=0$.

(iv) A Pauli set of spin $1/2$ on $\Cz^2$, exhibiting the 
isomorphism between the Lie algebras $u(3)$ and $so(3)$, is given by
the {\bf Pauli matrices}
\lbeq{rep2a}
  s=2,~~~
   L_{1}= 
  \Big( \bary{cc}
   0 &1 \\
   1&0
  \eary \Big),~~~
   L_{2}=
  \Big( \bary{cc}
   0 &-i \\
   i&0
  \eary \Big),~~~
   L_{3}=
  \Big( \bary{cc}
   1&0 \\
   0&-1
  \eary \Big).
\eeq
\end{expls}

\bigskip
(iii) and (iv) from Example \ref{ex.pauli} are the first two cases 
of an infinite family of Pauli sets with arbitrary nonnegative 
half-integral spin:

\begin{thm}\label{t.pauli}
The matrices $ L_0,  L_{1},  L_{2},  L_{3}
\in \Cz ^{s \times s}$ defined by
\[
  ( L_{1}\psi)_{k}=(k-1)\psi_{k-1}+(s-k)\psi_{k+1},
\]
\[
  ( L_{2}\psi)_{k}=i(k-1)\psi_{k-1}-i(s-k)\psi_{k+1},
\]
\[
  ( L_{3}\psi)_{k}=(s+1-2k)\psi_{k,}
\]
\[
  ( L_0 \psi)_{k} = {s-1 \choose k-1}\psi_{k}
\]
for $\psi \in \Cz ^{s}$ form a Pauli set of spin $j=\frac{s-1}{2}$,
called the {\bf canonical Pauli set} over $\Cz^s$.
Written as matrices, the $ L_k$ and hence the $a\cdot L$
are tridiagonal, $ L_0$ is diagonal, and the inner 
product \gzit{rep4} takes the form
\[
\overline{\phi}\psi = \sum_{k}{s-1\choose k-1}\phi_{k}^*\psi_{k}.
\]

\end{thm}

\bepf
For $\mu = 1,2$ we have
\lbeq{rep3a}
  ( L_{\mu }\psi )_{k}= \alpha_{\mu } (s-k)\psi_{k +1}+ \beta
 _{\mu }(k-1) \psi_{k -1},
\eeq
where
\lbeq{rep3b}
  \alpha_{1} = \beta_{1}=1,~ \alpha_{2} = -i,~ \beta_{2} = i.
\eeq
Therefore, for $\mu,\nu\in\{1,2\}$,
\lbeq{rep3c}
 \bary{rcl}
  (  L_\mu  L_\nu \psi )_{k}&=& \alpha_\mu (s-k)
  (\alpha_\nu(s-k-1) \psi_{k+2} + \beta_\nu k \psi_{k}) \\
  ~&~&+\beta_\mu (k-1)(\alpha_\nu(s-k+1) \psi_{k} + \beta_\nu (k-2)
  \psi_{k-1}) \\
  ~&=& \alpha_\mu \alpha_\nu (s-k)(s-k-1) \psi_{k+2}+ \beta_\mu
  \beta_\nu(k-1)(k-2) \psi_{k-2} \\
  ~&~&+( \alpha_\mu \beta_\nu k (s-k)+ \alpha_\nu\beta_\mu (k-1)
  (s-k+1))\psi_{k}.
 \eary
\eeq
Similarly,
\lbeq{rep3d}
  ( L_\mu  L_{3} \psi )_{k}= \alpha_\mu (s-k)(s-1-2k)\psi
 _{k+1} + \beta_\mu (k-1)(s+3-2k)\psi_{k-1},
\eeq
\lbeq{rep3e}
  ( L_{3}  L_\mu \psi )_{k}= (s+1-2k) \alpha_\mu (s-k)
  \psi_{k+1} + (s+1-2k) \beta_\mu (k-1) y_{k-1},
\eeq
Since $\alpha_1^2+\alpha_2^2=\beta_1^2+\beta_2^2=0$,
$\alpha_1\beta_1+\alpha_2\beta_2=2$, \gzit{rep3c} implies
\[
(( L_{1}^2+ L_{2}^2)\psi )_{k} 
= (2k(s-k)+2(k-1)(s-k+1))\psi_k = (4k(s+1-k)-2s-2)\psi_k,
\]
hence
\[
(( L_{1}^2+ L_{2}^2+ L_{3}^2)\psi )_{k} 
=(4k(s+1-k)-2s-2+(s+1-2k)^2)\psi_k =(s^2-1)\psi_k,
\]
giving \gzit{rep0}.
Taking differences in \gzit{rep3c}--\gzit{rep3e} gives
\[
\bary{rcl}
  ([  L_{1} ,  L_{2} ] \psi )_{k}
&=& ( L_{1} L_{2}\psi)_{k}-( L_{1} L_{2}\psi)_{k}\\
&=& (\alpha_{1} \beta_{2}-\alpha_{2} \beta_{1})
(k(s-k)-(k-1)(s-k+1))\psi_{k}\\
&=& 2i(s+1-2k) \psi_{k}= (2i  L_{3} \psi )_{k},
\eary
\]
\[
  ([  L_{1},  L_{3}] \psi )_{k}= -2 \alpha_{1} (s-k) \psi 
 _{k+1} + 2 \beta_{1} (k-1)\psi_{k-1}=-(2i  L_{2}\psi )_{k},
\]
\[
  ([ L_{2} , L_{3}] \psi )_{k} = -2 \alpha_{2}(s-k) \psi
 _{k+1}+ 2 \beta_{2}(k-1) \psi_{k-1}=(2i  L_{1} \psi )_{k}.
\]
This gives the commutation relations \gzit{rep1}. 
By writing the tridiagonal matrices out in full, using
\[
  (s-k){s-1 \choose k-1}= \frac{(s-1)!}{(k-1)!(s+1-k)!}=k 
  {s-1 \choose k},
\]
\gzit{rep3} is easily verified. Since $ L_0$ is positive 
definite, we have a Pauli set of spin $j$.
\epf

\chapter{The symmetries of the universe}\label{c.symm}

An understanding of the symmetries of the universe is necessary
to be able to appreciate the modern concept of elementary particles. 

~\at{informal introduction and preview of what is to come}

The special orthogonal group $SO(3)$ of 3-dimensional rotations and 
the related special Euclidean group $ISO(3)$ of distance and 
orientation preserving affine mappings of 3-dimensional space are of 
exceptional importance in physics and mechanics. Indeed, 
the corresponding symmetries are inherent in many systems of interest
and in the building blocks of most larger systems. 
The associated Lie algebra $so(3)$ of real, antisymmetric 
$3\times 3$ matrices describes angular velocity and angular momentum, 
both in classical and in quantum mechanics; see Section \ref{s.quarks}.
From a mathematical point 
of view, 3-dimensional rotations are also interesting due to the
sporadic isomorphism between the Lie algebras $so(3)$ and $u(2)$ and 
the resulting isomorphism between $SO(3)$ and a quotient of $SU(2)$,
see Section\ref{s.rotSU2}.

\at{annotate for index (done for rotation part only)}

\at{this chapter is still very raw and needs much additional input}

\section{Rotations and $SO(n)$}\label{s.rotn}

A matrix $U\in \Kz^{n\times n}$ is called \bfi{orthogonal}
if $U^TU=1$, equivalently, if $U$ is invertible and $U^{-1}=U^T$. 
The set $O(n,\Kz)$\index{$O(n,\Kz)$, orthogonal group} of all 
orthogonal $n\times n$ matrices with entries 
in $\Kz$ is a matrix group, called a (full) \bfi{orthogonal group} 
over $\Kz$.
Indeed, if $U,V\in O(n,\Kz)$ then $(UV)^TUV=V^TU^TUV=V^TV=1$ and 
$(U^{-1})^TU^{-1}=(U^T)^TU^{-1}=UU^{-1}=1$, so that 
$UV, U^{-1} \in O(n,\Kz)$. It is customary to write 
$O(n):=O(n,\Rz)$.\index{$O(n)$, real orthogonal group}
Note that $O(n)$ is a subgroup of $U(n)$. The orthogonal ${n\times n}$ 
matrices with determinant one form a matrix group 
$SO(n)$\index{$SO(n)$, special orthogonal group}, called a 
\bfi{special orthogonal group}. 

Every element $U\in O(n,\Kz)$ satisfies 
$(\det U)^2 = det U^T \det U = \det (U^TU)=1$, hence either 
$\det U=1$ or $\det U=-1$. Therefore for odd $n$, we have 
$O(n,\Kz)=\{U,-U\mid U\in SO(n,\Kz)$, and for any $n$,
$O(n,\Kz)=\{U,MU\mid U\in SO(n,\Kz)$, where $M$ is the diagonal matrix 
with a $M_{nn}=-1$ and other diagonal entries one.

The elements of the special orthogonal group $SO(n)$ are called 
rotations. Thus, a \bfi{rotation} is a real orthogonal matrix 
$Q \in \Rz ^{n \times n}$ with determinant one,
\lbeq{rot1}
   Q^TQ=QQ^T=1, ~~~ \det Q=1.
\eeq
Since $|Qx|^2=(Qx)^T(Qx)=x^TQ^TQx=x^Tx$ for $Q\in SO(n)$, rotations 
preserve the length 
$|x|=\sqrt{x^Tx}$ of a vector, 
\[
|Qx|=|x|\Forall x\in\Rz^n.
\]
Since $\det (QA)=\det Q\det A = \det A$ for $Q\in SO(n)$, rotations
also preserve the orientation of volumes. Conversely, these condition 
together imply that $Q^TQ=1$ and $\det Q=1$, hence $Q\in SO(n)$.
It can be shown that $SO(n)$ is a connected matrix group; \at{why? 
We prove this below for $SO(2)$ and $SO(3)$ only; spectral 
factorization gives the general result.} hence every rotation is 
obtainable by a rotational motion. 
Since $\det$ is a continuous function of its entries and $\det 1= 1$,
$SO(n)=O(n)_0$ is the connected part of $O(n)$.

The natural metric for rotations is the 
\bfi{Frobenius distance}\index{ $d(Q_1,Q_2)$, Frobenius distance},
\lbeq{qua15}
d(Q_1,Q_2)=\frac{1}{\sqrt{2n+2}}\|Q_1-Q_2\|_F.
\eeq
Here $\|A\|_F:=\sqrt{A:A}$\index{ $\|A\|_F$, Frobenius norm}, 
where $A:B=\tr A^TB$\index{$A:B=\tr A^TB$} denotes the standard 
inner product between matrices, is the \bfi{Frobenius norm} of a matrix 
$A\in\Rz^{n\times n}$. (The reader is invited to check the triangle 
inequality $d(Q_1,Q_3)\le d(Q_1,Q_2)+d(Q_2,Q_3)$.)

\begin{prop}
The Frobenius distance has the invariance property
\lbeq{qua16a}
   d(Q_1,Q_2)=d(QQ_1,QQ_2)=d(Q_1Q,Q_2Q) ~~~\mbox{if }Q\in SO(3),
\eeq
and satisfies
\lbeq{qua16}
   d(Q_1,Q_2)=\half\sqrt{n-Q_1:Q_2} \in [0,1].
\eeq
\end{prop}

\bepf
The inner product $A:B$ is orthogonal invariant,
\[
QA:QB=\tr(QA)^TQB=\tr A^TQ^TQB=\tr A^TB =A:B,
\]
\[
AQ:BQ=\tr(AQ)^TBQ=\tr Q^TA^TBQ=\tr QQ^TA^TB
=\tr A^TB =A:B.
\]
Therefore $\|QA-QB\|_F^2=Q(A-B):Q(A-B)=(A-B):(A-B) = \|A-B\|_F^2$
and  $\|AQ-BQ\|_F^2=(A-B)Q:(A-B)Q=(A-B):(A-B) = \|A-B\|_F^2$,
so that
\[
\|QA-QB\|_F=\|A-B\|_F=\|AQ-BQ\|_F.
\]
This implies \gzit{qua16a}. \gzit{qua16} follows from
\[
\bary{lll}
(2n+2)d(Q_1,Q_2)^2&=&\tr(Q_1-Q_2)^T(Q_1-Q_2)\\
&= &\tr Q_1^TQ_1-\tr Q_1^TQ_2-\tr Q_2^TQ_1+\tr Q_2^TQ_2\\
&=&2\tr  1 -2\tr Q_1^TQ_2=2n-2Q_1:Q_2\\
&\le& 2n+2
\eary
\]
by \gzit{rot17a}. \at{$\tr Q \ge -1$ only holds for $n=3$! 
check the final inequality and adapt factors}
\epf

\bigskip
A matrix $f\in \Kz^{n\times n}$ is called \bfi{antisymmetric}
if $A^T=-A$. The set $o(n,\Kz)${$o(n,\Kz)$, orthogonal Lie algebra} 
of all antisymmetric $n\times n$ matrices with entries in $\Kz$ is a 
matrix Lie algebra, called an \bfi{orthogonal Lie algebra} over $\Kz$. 
Indeed, if $f,g\in o(n,\Kz)$ then 
$[f,g]^T=(fg-gf)^T=(g^Tf^T-f^Tg^T=(-g)(-f)-(-f)(-g)=-(fg-gf)=-[f,g]$, 
hence $[f,g]\in o(n,\Kz)$.  It is customary to write 
$o(n):=o(n,\Rz)$.\index{$o(n)$, real orthogonal Lie algebra}
Note that $o(n)$ is a Lie subalgebra of $u(n)$.
The antisymmetric ${n\times n}$ matrices with 
trace zero form a Lie algebra $so(n)$\index{$so(n)$, special orthogonal 
Lie algebra}, called a \bfi{special orthogonal Lie algebra}.
$so(n)$ is the Lie algebra of the matrix group $SO(n)$.
Note that $so(n,\Kz)=o(n,\Kz)$ since $f\in o(n,\Kz)$ inplies 
$ \tr f = \tr f^T =\tr (-f) = -\tr f$, hence the trace is automatically 
zero.

\bigskip
We briefly look at the smallest orthogonal groups and their Lie algebra.
For $n=1$, we have $O(1)=\{1,-1\}$, $SO(1)=\Id(1)=\{1\}$, and 
$o(1)=so(1)=\id(1)=\|0\}$. 

For $n=2$, the Lie algebra $o(2)=so(2)$ is 1-dimensional and consists 
of the antisymmetric $2\times 2$ matrices
\[
\pmatrix{0 & \alpha\cr-\alpha& 0 } =\alpha\ii,
\]
where 
\lbeq{e.imag}
\ii:=\pmatrix{0 & \alpha\cr-\alpha& 0 }
\eeq
is the canonical generator. $SO(2)$ consists of the rotations 
\lbeq{e.2rot1}
Q[\alpha]=\pmatrix{\cos\alpha & \sin\alpha\cr-\sin\alpha& \cos\alpha}
=e^{\alpha\ii}
\eeq
the result of a uniform rotation $U(t):=e^{t\alpha\ii}$
around zero by some angle $\alpha$ in counter-clockwise direction.
The product of rotations is a rotation by the sum of the angles, 
\lbeq{e.2rot2}
Q[\alpha]Q[\beta]=Q[\alpha+\beta],
\eeq
and the Frobenius distance of two rotations is a function of the 
difference of the angles,
\lbeq{e.2rot2a}
d(Q[\alpha],Q[\beta])=\sin (|\alpha-\beta|/2),
\eeq
correctly taking account of the fact that angles differing by an 
integral multiple of $2\pi$ determine the same rotation.
Note that $\ii^2=-1$, hence we may identify $\ii$ with the imaginary 
unit $i$. This identification provides the isomorphisms 
$U(1) \cong SO(2)$ and $u(1)\cong so(2)$, reflecting the 
fact that the complex number plane is isomorphic to the
2-dimensional real plane. 

The full orthogonal group $O(2)$ consists of the rotations and the 
matrices
\lbeq{e.2refl}
R[\alpha]=\pmatrix{\cos\alpha & \sin\alpha\cr\sin\alpha& -\cos\alpha}
=R[0]e^{\alpha\ii}
\eeq
describing 2-dimensional \bfi{reflections} at the axis \at{}.

\section{3-dimensional rotations and $SO(3)$}\label{s.rot3}

The general form of an antisymmetric $3\times 3$ matrix\index{$X(\a)$, 
antisymmetric $3\times 3$ matrix} is
\[
  X(\a):= \left( \bary{ccc}
                 0 & - a_3 & a_2 \\
                 a_3 & 0 & -a_1 \\
                 -a_2 & a_1 & 0
         \eary \right),~~~a \in \Kz ^3;
\]
therefore
\[
so(3,\Kz)=\{X(\a)\mid \a \in\Kz^3\}.
\]
We note the rules
\lbeq{rot2}
  X(\a)^T=-X(\a),~~~X(\a)\b=\a \times \b=-X(\b)\a, ~~~ X(\a)\a=0,
\eeq
\lbeq{rot2b}
  X(\a \times \b) = \b\a^T-\a\b^T,
\eeq
\lbeq{rot2a}
  X(\a)X(\b)=
  \left( \bary{ccc}
   -a_3b_3-a_2b_2 & a_2b_1 & a_3b_1 \\
   a_1 b_2 & -a_3b_3-a_1b_1 & a_3 b_2 \\
   a_1 b_3 & a_2b_3 &-a_2b_2-a_1b_1
  \eary \right)
   =\b\a^T-(\a\cdot \b)1.
\eeq
From \gzit{rot2a} for $\a=\b$, we find by repeated 
multiplication with $X(\a)$, using \gzit{rot2},
\lbeq{rot4}
  X(\a)^2=\a\a^T-\a^21,~~~X(\a)^3=-\a^2X(\a),~~~X(\a)^4=-\a^2X(\a)^2;
\eeq
 and since $(1-X(\a))((1+\a^2)1+X(\a)+X(\a)^2)=(1+\a^2) 1$, we have
\lbeq{rot4a}
  (1-X(\a))^{-1}=1+(1+\a^2)^{-1}(X(\a)+X(\a)^2).
\eeq

We use these relations for $\Kz=\Rz$ to prove the following
explicit characterization of 3-dimensional rotations.

\begin{thm} \label{t.qrot}~\\
(i) For all $\rr\in\Rz^3$ with $|\rr|\le 1$, the matrix\index{$Q[\rr]$, 
3-dimensional rotation}
\lbeq{rot5}
  Q[\rr]:=1 +2\rr_0X(\rr)+2X(\rr)^2,~~~
\mbox{where } \rr_0= \sqrt{1-\rr^2},
\eeq
is a rotation.

(ii) If $\rr=0$ then $Q[\rr]$ is the identity; otherwise, $Q[\rr]$ 
describes a rotation around the axis through the vector $\rr$ by the 
angle 
\lbeq{rot9}
  \alpha = 2 \arcsin |\rr|,
\eeq
and we have
\lbeq{rot10}
Q{\rr}\rr=\rr,~~~
  |\rr| = \sin \frac{\alpha }{2},~~~ r_0 = \cos \frac{\alpha }{2} \ge 0.
\eeq

(iii) Conversely, every rotation $Q$ has the form $Q=Q[r]$ for some 
$\rr\in\Rz^3$ with $\rr^2\le 1$.
\end{thm}

\bepf
(i) Writing $X=X(\rr)$, $Q=Q[\rr]$, we find from \gzit{rot4} that 
$X^4=-\rr^2X^2=(1-r_0^2)X^2$, hence
\[
  \bary{rcl}
   Q^TQ&=&(1-2r_0X+2X^2)(1+2r_0X+2X^2) \\
   ~&=&1+(4-4r_0^2)X^2+4X^4=1.
  \eary
\]
Thus $Q^{-1}=Q^T,~QQ^T=QQ^{-1}=1$. Since 
$(\det Q)^2=\det Q\det Q^T = \det (QQ^T)=1$,
we have $\det Q[\rr]= \pm 1$. Since the sign is positive for $\rr=0$, 
continuity of \gzit{rot5} implies that $\det Q[\rr]=1$ for all $\rr$. 
Thus $Q [\rr]$ is a rotation. Moreover, 
\lbeq{rot5x}
X(Q+1)=r_0(Q-1),
\eeq
since $(r_01-X)Q=(r_01-X)(1+2r_0X+2X^2)=r_01+(2r_0^2-1)X^2-2X^3=r_01+X$.

(ii) The case $\rr=0$ is obvious; hence assume that $\rr\ne 0$.
Specializing the relation
\lbeq{rot7a}
  \a^TQ[\rr]\b=\a\cdot \b-2(r_0(\a\times \b)\cdot \rr
+(\a\times \rr)\cdot (\b\times \rr)),
\eeq
which follows using \gzit{rot2}, to a unit vector $\a$ and $\b=\a$, 
we see that the angle $\alpha$ between a vector $\a$ and its rotated 
image $Q[\rr]\a$ is $\cos \alpha = \a^TQ[\rr]\a=1-2|\a\times \rr|^2$,
hence 
\lbeq{rot7aa}
|\a\times \rr|=\sqrt{(1-\cos\alpha)/2}=\sin(\alpha/2).
\eeq
In particular, a unit vector $\a$ orthogonal to $\rr$ is rotated by the 
angle \gzit{rot9} since then $|\a\times \rr|=|\a|\cdot|\rr|=|\rr|$. 
Since \gzit{rot2} implies $Q[\rr]\rr=\rr$, the vector $\rr$ is fixed 
by the rotation, and $Q[\rr]$ describes a rotation around the vector 
$\rr$ by the angle $\alpha$ given by \gzit{rot9}.
  
(iii) Let $Q$ be an arbitrary rotation.

{\sc Case 1.}
If $Q+1$ is nonsingular, we define, motivated by \gzit{rot5x},
\[
  \bary{rcl}
   \tilde X&:=&(Q-1)(Q+ 1)^{-1}=(Q-Q^TQ)(Q+Q^TQ)^{-1} \\
   ~&=& (1 -Q^T)(1 +Q^T)^{-1}=-(1-Q^T)^{-1}(Q^T-1)=-\tilde X^T.
  \eary
\]
Hence $\tilde X$ is antisymmetric, $\tilde X=X(\a)$ for some $\a$.
Now $\tilde X(Q+ 1)=Q- 1$, hence we have $(1 -\tilde X)Q=1+\tilde X$.
Writing
\[
  r_0:=1/ \sqrt{1+\a^2},~~\rr:=r_0\a,
\]
we find from \gzit{rot4a} and \gzit{rot4} that
\[
  \bary{rcl}
   Q&=& (1+r^2_0\tilde X+r^2_0\tilde X)(1 +\tilde X)
     =1+(r^2_0+1-r_0^2\a^2)\tilde X+2r^2_0\tilde X^2 \\
   ~&=& 1+2r^2_0X(\a)+2r_0^2X(\a)^2=1+2r_0X(\rr)+2X(\rr)^2.
  \eary
\]
Since $1-\rr^2=1-r^2_0\a^2=1-\a^2/(1+\a^2)=1/(1+\a^2)=r^2_0\ge 0$, we 
conclude that $Q=Q[\rr]$.

{\sc Case 2.}
If $Q+1$ is singular then $-1$ is an eigenvalue of $Q$.
The other two eigenvalues must have product
$-1$ since the determinant is the product of all eigenvalues,
counted with their algebraic multiplicity. Since two complex conjugate 
eigenvalues have a positive product, this implies that the eigenvalues 
are all real. Any real eigenvalue $\lambda$ has an associated real 
eigenvector $\x \ne 0$ to with $Q\x= \lambda \x$, and
$\x^2=(Q\x)^2= \lambda ^2 \x^2$ implies $\lambda ^2 = 1$, hence 
$\lambda= \pm 1$. Thus $-1$ is a double eigenvalue 
and $Q+1$ has rank $1$, $Q+1=\rr\s^T$ with a unit vector $\rr$. Now 
orthogonality implies $\s=2\rr$, giving $Q=-1+2\rr\rr^T=Q[\rr]$, using 
\gzit{rot4}. 
\epf

For angles $|\alpha| \le \frac{\pi}{2}$ (corresponding to 
$\rr^2 \le \half \le r_0^2$), one may also use 
\lbeq{rot20}
   \q=2r_0\rr,~~~~q_0= \cos \alpha =2r^2_0-1= \sqrt{1-\q^2},
\eeq
to rewrite $Q=Q[\rr]$ as
\lbeq{rot21}
  Q=1+X(\q)+ \frac{1}{1+\q_0}X(q)^2
   =(1-\shalf X(\q))^{-1}(1+\shalf X(\q)),
\eeq
which has nonlinearities only in the higher order term.
Since $1$ and $X(\q)$ are symmetric, we see that
\lbeq{rot17}
  \q=\half \left( \bary{c}
    Q_{32} - Q_{23} \\
    Q_{13} - Q_{31} \\
    Q_{21} - Q_{12}
   \eary \right)
\eeq
is linear in the coefficients of $Q$. Therefore, \gzit{rot21}
is referred to as the \bfi{linear parameterization}.
From \gzit{rot8}, one easily checks that $Q=Q[\rr]$ satisfies 
\lbeq{rot17a}
\tr Q =4 r_0^2-1\ge-1.
\eeq
Using also \gzit{rot17}, we see that if $\tr Q>-1$ then $\rr,r_0$ 
(and hence the rotation axis and angle)
can be uniquely recovered from $Q$ by
\lbeq{rot16}
  \rr= \frac{\q}{\sqrt{1+ \tr Q}},~~~r_0= \half \sqrt{1+ \tr Q}.
\eeq
And if $\tr Q=-1$ then $r_0=0$ (corresponding to a rotation by 
$180^o$) and $Q+1$ has rank $1$, hence $\rr$ 
is one of the two unit vectors parallel to the columns of $Q$.

\bigskip
\bfi{The exponential map.} 
The Lie algebra $so(3)$ of all real, antisymmetric matrices 
consists of infinitesimal rotations, and the following result shows 
explicitly that every rotation can be written 
as a matrix exponential of an antisymmetric matrix.
In particular, this implies that $SO(3)$ is connected.

\begin{prop}
If $\e$ is the unit vector in the direction of $\rr$ (and an arbitrary
unit vector if $\rr=0$) then, with $\alpha$ defined by \gzit{rot9},
\lbeq{rot11}
  Q[\rr]= e^{\alpha X(\e)}.
\eeq
\end{prop}

\bepf 
Using the relations
\[
  \sin \alpha = 2 \sin \frac{ \alpha }{2} \cos \frac{ \alpha }{2},~~~
  \cos \alpha = 1-2 \sin ^2 \frac{\alpha }{2},~~~
  \rr= \Big( \sin \frac{\alpha }{2} \Big) \e, \]
we find
\lbeq{rot12}
  Q[\rr]=1+(\sin \alpha )X(\e)+(1- \cos \alpha )X(\e)^2=:Q(\alpha ).
\eeq
Now $Q (0)=1$ and
\[
  \frac{d}{d\alpha} Q(\alpha )=(\cos\alpha)X(\e)+(\sin\alpha)X(\e)^2
=X(\e)Q(\alpha),
\]
using \gzit{rot4}. Solving this differential equation gives 
\gzit{rot11}.
\epf

If we write \gzit{rot12} in terms of $\a=\alpha \e$, we conclude from 
\gzit{rot11} the \bfi{Rodrigues formula}
\lbeq{rot13x}
  e^{X(\a)}=1+\frac{\sin |\a|}{|\a|}X(\a)
                +\frac{1-\cos |\a|}{|\a|^2} X(\a)^2
  ~~~\mbox{if~} \a \not= 0
\eeq
for the exponential of a real, antisymmetric matrix. It describes a 
rotation along an axis parallel to $\a$ by an angle $\alpha = t|\a|$.
The Rodrigues formula can also be obtained by writing the exponential 
as a power series  and simplifying using \gzit{rot4}.
In particular, for small $\a$ we find $e^{X(\a)}=1+X(\a)+O(|\a|^2)$ 
for small $\a$, showing explicitly that the $X(\a)\in so(3)$ are 
infinitesimal rotations.

As a useful application of the exponential form, we prove:

\begin{prop}          
For any rotation $Q$,
\lbeq{rot23}
   X(Q\a)=Q^TX(\a)Q,
\eeq
\lbeq{rot22}
   Q\a \times Q\b=Q(\a \times \b),
\eeq
\end{prop}

\bepf                 
By differentiation, using \gzit{rot11}. 
\at{details, or is it $X(Q\a)=QX(\a)Q^T$?}.
\epf

\section{Rotations and quaternions}\label{s.rquat}

In view of \gzit{rot5} and \gzit{rot4}, we have
\lbeq{rot8}
  Q[\rr]=1+2
   \left( \bary{ccc}
     -r_2^2-r_3^2 & r_1r_2-r_0r_3 & r_1r_3+r_0r_2 \\
     r_1r_2+r_0r_3 & -r^2_1-r^2_3 & r_2r_3-r_0r_1 \\
     r_1r_3-r_0r_2 & r_2r_3+r_0r_1 & -r^2_1 - r^2_2 
   \eary \right) ,~~~r_0=\sqrt{1-|\rr|^2},
\eeq
where $1$ denotes the identity matrix and $\rr\in\Rz^3$ satisfies 
$|\rr| \le 1$.
Alternatively, we may write \gzit{rot8} in the homogeneous 
\bfi{quaternion parameterization}\index{$Q[r_0,\rr]$, quaternion 
parameterization of a rotation} 
\lbeq{ecs14a}
  Q[r_0,\rr]=1+\frac{2}{r_0^2+r_1^2+r_2^2+r_3^2}\left( \bary{ccc}
  -r_2^2-r_3^2  & r_1r_2-r_0r_3 & r_1r_3+r_0r_2 \\
  r_1r_2+r_0r_3 & -r_1^2-r_3^2  & r_2r_3-r_0r_1 \\
  r_1r_3-r_0r_2 & r_2r_3+r_0r_1 & -r_1^2-r_2^2
  \eary \right)
\eeq
of rotations,
with independent $r_0\in \Rz$, $\rr\in \Rz^3$, not both zero. 
$Q[r_0,\rr]$ satisfies
\lbeq{e.hom}
Q[r_0,\rr]=Q[\lambda r_0,\lambda \rr] \Forall \lambda\ne 0
\eeq
and reduces to $Q[\rr]$ if the arbitrary scale is chosen such that 
$r_0^2+r_1^2+r_2^2+r_3^2=1$ and $r_0\ge0$.
Because of \gzit{e.hom}, parallel vectors $(r_0,\rr)$ in the quaternion 
parameterization give the same rotation. This shows that the 
3-dimensional rotation group has the topology of a 3-dimensional 
projective space. 
(Note also that the linear parameterization \gzit{rot21}
 can be obtained from the homogeneous form \gzit{ecs14a}
by choosing the arbitrary scale such that $r_0^2+\rr^2=2r_0$.)

In computational geometry, the quaternion parameterization of rotations 
is preferable to the frequently discussed (and more elementary) 
parameterization by Euler angles, since it does not need expensive 
trigonometric functions, its parameters have a geometric meaning 
independent of the coordinate system used, and it has significantly 
better interpolation properties (\sca{Shoemake} \cite{Sho}, 
\sca{Ramamoorthi \& Barr} \cite{RamB}).
Note that the projective identification mentioned above has to be 
taken into account when constructing smooth motions joining two close 
rotations $Q[\rr]$ with nearly opposite $\rr$ of length close to 1.

\bigskip
\bfi{Quaternions.}
\at{what has the above to do with quaternions? match with Chapter 2}

A \bfi{quaternion} is a $4 \times 4$ matrix of the 
form\index{$U(r_0,\rr)$, quaternion matrix, $4 \times 4$ form}
\lbeq{qua1}
  U(r_0,\rr):= 
  \left( \bary{cc}
   r_0 & \rr^T \\
   -\rr & r_0  1 +X(\rr)
  \eary \right) ,~~~r_0 \in \Rz,~\rr \in \Rz ^3.
\eeq

\begin{thm}           
The set $\Qz$ of quaternions is a \bfi{skew field}, i.e., an associative
algebra in which every nonzero element has an inverse. We have 
\lbeq{qua2a}
  U(r_0,\rr)+U(s_0,\s)=U(r_0+s_0,\rr+\s),
\eeq
\lbeq{qua2b}
  \lambda U(r_0,\rr)=U(\lambda r_0,\lambda \rr),
\eeq
\lbeq{qua2}
  U(r_0,\rr)^T=U(r_0,-\rr),
\eeq
\lbeq{qua3}
  U(r_0,\rr)U(s_0,\s)=U(r_0s_0-r\cdot \s,~s_0\rr+r_0\s+\rr \times \s ),
\eeq
\lbeq{qua4}
  U(r_0,\rr)^{-1}= \frac{1}{r_0^2+\rr^2} U(r_0,-\rr) ~~~ 
\mbox{if~} r_0^2+\rr^2 \ne 0.
\eeq
\end{thm}   

\bepf                 
\gzit{qua2a}--\gzit{qua2} are trivial, 
and \gzit{qua3} follows by direct computation,
using \gzit{rot2}, \gzit{rot2a} and \gzit{rot2b}.
Specializing \gzit{qua3} to $s_0=r_0,\s=-\rr$ gives
\lbeq{qua5}
  U(r_0,\rr)U(r_0,-\rr)=U(r_0^2+\rr^2,0)=(r_0^2+\rr^2) 1,
\eeq
which implies \gzit{qua4}. Therefore $\Qz$ is a vector space closed 
under multiplication, and every nonzero element in $\Qz$ has an inverse.
Since matrix multiplication is associative, $\Qz$ is a skew field.
\epf

\at{adapt, in view of the spinor version}      
In the standard treatment, quaternions are treated like complex 
numbers, as objects of the form
\[
  \q (r_0, \rr)=r_0 1+r_1\ii+r_2\j+r_3\k
\]
with special unit quaternions $1,\ii,\j,\k$.
The correspondence is given by the identification 
\lbeq{quaijk}
  \ii= 
   \left( \bary{cccc}
    0 & 1 & 0 & 0 \\
   -1&0&0&0 \\
    0&0&0&-1 \\
    0&0&1&0
   \eary \right) ,~~~
  \j=
   \left( \bary{cccc}
    0&0&1&0 \\
    0&0&0&1\\
   -1&0&0&0\\
    0&-1&0&0
   \eary \right) ,~~~
  \k=
   \left( \bary{cccc}
    0&0&0&1\\
    0&0&-1&0\\
    0&1&0&0 \\
   -1&0&0&0
   \eary \right) ,
\eeq
in terms of which $\q (r_0,\rr)=U(r_0,\rr)$. 

\at{relation to spinors; check signs!}
Identifying the imaginary unit $i$ with the real $2\times 2$ matrix 
\gzit{e.imag}, we expand  \at{what} to get the above real form:
\[
\bary{lll}
r_0 + (r_1,r_2,-r_3)^T*\bsigma &=& 
    \pmatrix{r_0+ir_1 & r_2+ir_3 \cr -r_2+ir_3 & r_0-ir_1} \\
&=& \pmatrix{r_0 & r_1 & r_2 & r_3 \cr -r_1 & r_0 & -r_3 & r_2 \cr 
          -r_2 & r_3 & r_0 & -r_1 \cr -r_3 & -r_2 & r_1 & r_0}\\
&=& r_0 + r_1\ii + r_2\j + r_3\k.
\eary
\]

\section{Rotations and $SU(2)$}\label{s.rotSU2}

\bigskip
\bfi{The isomorphism $su(2) \cong so(3)$.}
From \gzit{rot2a} and \gzit{rot2b}, one finds immediately that
\lbeq{rot2c}
  [X(\a),X(\b)]=X(\a \times \b).
\eeq
This implies that the spaces of real and complex antisymmetric
matrices,
\[
so(3)=\{X(\a)\mid \a\in \Rz^3\},~~~
so(3,\Cz)=\{X(\a)\mid \a\in \Cz^3\},
\]
are closed under forming commutators, and hence form a Lie algebra.
We shall see soon that the elements of $so(3)$ are infinitesimal 
rotations. \at{} Introducing
\[
L_1=\pmatrix{0&0&0\cr 0&0&1\cr 0&-1&0}\,,~~~L_2
=\pmatrix{0&0&-1\cr0&0&0\cr 1&0&0}\,,~~~L_3 = \pmatrix{0&-1&0\cr
1&0&0\cr 0&0&0}\,,
\]
which form a basis of $so(3)$,
we see that the correspondence $\frac{1}{2i}\sigma_k \mapsto L_k$ for 
$k=1,2, 3$ defines an isomorphism of Lie algebras. 
\at{avoid $\mapsto$}

\at{match at the right place}
Taking a closer look at $X(\a)$ we see that there is a basis of
$so(3)$ consisting of three elements $J_i$ with $X(\a) =
\sum_{i=1}^{3}\a_i J_i$, corresponding to infinitesimal rotations 
around the coordinate axes. \at{use fat $\J$!}
We assemble the three $J$'s in a
column vector and (ab-)use the notation $X(\a) = \a\cdot J$.
Writing out $X(\a)\lp X(\a')=X(\a \times \a')$ we get
\lbeq{so.3.first}
J_k\lp J_l = \sum_m\epsilon_{klm}J_m\,,
\eeq
with the Levi--Civit\'a symbol $\epsilon_{klm}$ defined in
\gzit{e.leviciv}.

\bigskip
\bfi{The group version.}
If $g$ is an element of $SU(2)$ and $p\in su(2)\cong \Rz^3$ we see that
\[
\tr (gpg^{-1}) = 0\,, ~~~(gpg^{-1})^*
= (g^{-1})^* p g^* = gpg^{-1}\,,
\]
and we conclude that $g$ induces a map $\Rz^3\to \Rz^3$. Since
\[
\tr(gpg^{-1}gqg^{-1}) =\tr(pq)\,,
\]
the map induced by $g\in SU(2)$ defines an element of $O(3)$. 
But also $\p\times \q \cdot \rr$ is invariant under the
action of $g$ and thus we found a map $R:SU(2)\to SO(3)$, 
whereby $g\in SU(2)$ gets mapped to the element
$R(g)$ in $SO(3)$ corresponding to $p\mapsto gpg^{-1}$. 
\at{avoid $\mapsto$} 
The map $g\to R(g)$ is a group homomorphism; that is, 
$R(g_1g_2) = R(g_1)R(g_2)$. 

\at{connect}
and hence the map $R:SU(2)\to SO(3)$ is surjective. Suppose now that
$g(x,y)$ is mapped to the identity element in $SO(3)$. 
The kernel of $R:SU(2)\to SO(3)$ is easily checked to consist of 
$\{ \pm 1\}$, which is the central
$\Zz_2$ subgroup of $SU(2)$. (Easy exercise: Prove that
$\Zz_2$ is the center of $SU(2)$.)
As any kernel of group homomorphisms, the
kernel is a normal subgroup. All in all we have shown
\[
SO(3) \cong SU(2)/\Zz_2\,.
\]

\at{to do:\\
script p.23f:\\
forward ref. to universal covering, simply connected, etc. \\
$C\to TC T^*$ from polarization exhibits the quotient structure.
}

\section{Angular velocity}\label{s.angvel}

Quaternions are the most elegant way to derive a 3-dimensional analogue 
of the formulas \gzit{e.2rot1} and \gzit{e.2rot2} for 2-dimensional 
rotations in terms of rotation angles. The resulting product 
formula for 3-dimensional rotations, Theorem \ref{t.prod} 
in Section \ref{s.angvel}, allows us to derive the properties
of angular velocity. 

\begin{thm}\label{t.prod} \bfi{(Product formula)}\\
Let $|\rr|, |\s| \le 1$. Then\index{$\rr \oplus \s$, addition of 
infinitesimal rotations}
\lbeq{qua8}
  Q[\rr]Q[\s]=Q[\rr \oplus \s],
\eeq
where with $r_0= \sqrt{1-\rr^2}$, $s_0= \sqrt{1-\s^2}$, and the sign 
chosen such that $\pm (r_0s_0-\rr\cdot \s) \ge 0$,
\lbeq{qua6}
  \rr \oplus \s := \pm(s_0\rr+r_0\s +\rr \times \s).
\eeq
Moreover,
\lbeq{qua8a}
  Q[\rr]^{-1}=Q[\rr]^T=Q[-\rr],~~~ \rr \oplus (-\rr)=0.
\eeq

\end{thm}

Note that $\oplus$ is not commutative!

\bepf       
\at{make part of the proof independent of normalization, and present
the part leading to \gzit{qua10} as separate proposition in the 
previous section!}          
Since $r_0^2+\rr^2=s_0^2+\s^2=1$,
\gzit{qua4} implies that $U(r_0,\rr)$ and $U(s_0,\s)$ are orthogonal 
$4\times 4$ matrices. Writing
\lbeq{qua9a}
\q=\rr \oplus \s,~ q_0=|r_0s_0-\rr\cdot \s|,
\eeq
we may write \gzit{qua3} as
\lbeq{qua9}
  U(r_0,\rr)U(s_0,\s)=U(q_0,\q),
\eeq
hence $U(q_0,\q)$ is also orthogonal, and \gzit{qua4} implies 
$q_0^2+\q^2=1$, giving
\lbeq{qua7}
  (\rr \oplus \s)^2 \le 1,~~~ 
\sqrt{1-(\rr \oplus \s)^2}=|r_0s_0-\rr\cdot \s|.
\eeq
Now
\[
  \bary{rcl}
   U(s_0,\s)U(0,\x)U(s_0,\s)^T
   &=&U(s_0,\s)U(0,\x)U_(s_0,-\s)=U(s_0,\s)U(\s^T\x,s_0\x+X(\s)\x) \\
   ~&=&U(0,s_0^2\x+2s_0X(\s)\x+\s\s^T\x+X(\s)^2\x),
  \eary
\]
hence
\lbeq{qua10}
  U(s_0,\s)U(0,\x)U(s_0,\s)^T=U(0,Q[\s]\x).
\eeq
Multiplication by $U(r_0,\rr)$ on the left and by $U(r_0,\rr)^T$ on the 
right gives, using \gzit{qua9}, 
\lbeq{qua11}
  U(q_0,\q)U(0,\x)U(q_0,\q)^T=U(0,Q[\rr]Q[\s]\x).
\eeq
On the other hand 
\[
  U(q_0,\q)U(0,\x)U(q_0,\q)^T=U(q_0,\q)U(0,\x) U(q_0,\q)^T=U(0,Q[\q]\x),
\]
and comparing this with \gzit{qua11} implies \gzit{qua8}.
Finally, \gzit{qua8a} is immediate.
\epf

Computationally, \gzit{qua9a} is numerically stable in finite
precision arithmetic, while the
direct formula $q_0=\sqrt{1-\q^2}$ suffers from loss of accuracy 
if $q_0$ is tiny, due to cancellation of leading digits.

\bigskip
Differentiation of the product formula gives a useful formula for
the derivative of a rotation.

\begin{thm} \bfi{(Differentiation formula)}\\   
If $\rr$ is a function of $t$ then
\lbeq{qua12}
  \frac{d}{dt} Q[\rr]=X(\bomega )Q[\rr],~~~ 
  \bomega =2(\rr \times \dot{\rr}+ r_0 \dot{\rr}-\dot{\rr}_0\rr),
\eeq
and we have
\lbeq{qua13}
  \dot{\rr}= \half (r_0 \bomega -\rr \times \bomega ), ~~~
  \dot{r}_0= \half \rr\cdot \bomega .
\eeq
\end{thm}

\bepf                 
Writing
\[
  \rr=\rr(t),~~~\tilde{\rr}=\rr(t+h)=\rr+h \dot{\rr} +O(h^2),
\]
we have
\[
  \bary{rcl}
  Q[\tilde{\rr} ] Q [-\rr]
  &=&Q[\tilde{\rr} \oplus (-\rr)]=Q[-\tilde{r}_0\rr+r_0 \tilde{\rr}
  - \tilde{\rr} \times \rr ] \\
  ~&=&
Q[-(r_0+h \dot{r}_0)\rr+r_0(\rr+h\dot{\rr})-(\rr+h \dot{\rr})\times \rr 
  +O(h^2)] \\
  ~&=& 
Q[h(\rr \times \dot{\rr} + r_0 \dot{\rr} - \dot{r} _0 \rr)]+O(h^2) \\
  ~&=&  1 +hX(\bomega )+O(h^2).
  \eary
\]
Multiplication by $Q[\rr]$ gives
\[
  \bary{rcl}
   Q[\rr(t+h)]&=&Q[\tilde{\rr}]=( 1+hX(\bomega )+O(h^2))Q[\rr] \\
   ~&=&Q[\rr]+hX(\bomega )Q[\rr]+O(h^2),
  \eary
\]
and \gzit{qua12} follows. Now
\lbeq{qua14}
   2r_0\dot{r}_0=(r^2_0)^\pdot =(1-\rr^2)^\pdot =-2\rr^T \dot{\rr},
\eeq
hence
\[
  \bary{rcl}
    \rr \times \bomega 
&=& 2(\rr \times (\rr \times \dot{\rr})+r_0 \rr \times \dot{\rr}) \\
    &=& 2(\rr\rr^T\dot{\rr}-\rr^2 \dot{\rr}+r_0\rr \times \dot{\rr})
=r_0 \bomega -2  \dot{\rr},
  \eary
\]
giving $\dot{\rr}= \half (r_0 \bomega -\rr \times \bomega )$. 
Multiplication by $\rr^T$ gives $\rr^T \dot{\rr} = \half r_0 \bomega $, 
and the formula for $\dot{r}_0$ follows from \gzit{qua14} if 
$r_0 \not= 0$. For $r_0=0$, the formula follows by continuity.
\epf       

\begin{prop}
In the quaternion parameterization, we have
\lbeq{qua17}
   d(Q[\rr],Q[\s])=|(-\rr)\oplus \s|=\sqrt{1-(r_0s_0+\rr\cdot \s)^2}.
\eeq
\end{prop}

\bepf
Since 
\[
\tr Q[\rr]^TQ[\s]=\tr Q[-\rr]Q[\s]=\tr Q[(-\rr)\oplus \s]
=4((-\rr)\oplus \s)_0^2-1=3-4((-\rr)\oplus \s)^2
\]
by \gzit{rot17a}, \gzit{qua17} follows from \gzit{qua16} and 
\gzit{qua9a}.
\epf

\section{Rigid motions and Euclidean groups}\label{s.rigidM}

Translations
$\Tr(n)=T(\Id(1),\id(n))$ and $\exp \pmatrix{0 & 0\cr; x& 1_n}$

This motivates a more general triangular construction for
Lie groups $T(\Gz_1,\dots,\Gz_m)$ \at{script p.27} and Lie algebras 
$t(\Lz_1,\dots,\Lz_m)$ \at{script p.34}, which will later also produce 
the Galilean group and the Poincare group. $T(n)=T(\Rz,\dots,\Rz)$

$D(\Gz_1,\dots,\Gz_m)$ diagonal, direct produt, $D(n)$.
and corresponding Lie algebras.

$ISO(n)$ as subgroup of $L(n+1,\Rz)$, and their Lie algebra
\at{define more general $I\Gz$ and $i\Lz$.}

The \bfi{Euclidean group} or \bfi{inhomogeneous special orthogonal 
group} $ISO(n)$ consists of all distance
preserving affine mappings
\[
  x \in \Rz ^n \to x'=Qx+c \in \Rz ^n,
\]
with $Q \in SO(n)$ and $c \in \Rz^n$. In homogeneous coordinates,
\[
  { x' \choose 1} = 
  \left( \bary{cc}
    Q & c \\
    0 & 1 
  \eary \right)
   {x \choose 1},
\]
hence we can write $ISO(n)$ as a matrix group,
\[
  ISO(n)= 
   \Bigg\{ \left( \bary{cc}
    Q & c \\
    0 & 1 
   \eary \right) \Bigg| ~
    Q \in SO(n),~~~c \in \Rz ^n 
   \Bigg\} .
\]
\at{script p.8: product in pair version}

In the special case special case $n=3$, the corresponding Lie algebra 
of infinitesimal generators is the Lie algebra
\[
  iso(3)= 
    \Bigg\{ \left( \bary{cc}
     X(\omega ) & v \\
     0 & 0
    \eary \right) \Bigg|~ 
     \omega ,v \in \Rz ^3 
    \Bigg\} ,
\]
parameterizing one-parameter families of Euclidean motions defined by 
the differential equation
\[
  \dot{x} (t)=X(\omega (t)) x(t)+v(t),
\]
or short
\[
  \dot{x} = \omega \times x+v.
\]
If we write
\lbeq{hel1}
  \hat{X}(a)= \left( \bary{cc}
    X(a) & 0 \\
    0 & 0
    \eary \right) , ~~~
    b \cdot p = \left( \bary{cc}
    0 & b \\
    0 & 0 
    \eary \right) .
\eeq
the commutator relations of $iso(3)$ can be compactly written as 
\lbeq{hel2}
  [\hat{X}(a), \hat{X}(b)]=\hat{X}(a \times b)~~~ \mbox{for~} a,b \in 
  \Rz ^3,
\eeq
\lbeq{hel3}
  [\hat{X}(a),b \cdot p]=(a \times b) \cdot p ~~~\mbox{for~} a,b \in 
   \Rz ^3 ,
\eeq
\lbeq{hel4}
  [a \cdot p,b \cdot p]=0 ~~~\mbox{for~} a,b \in \Rz ^3 .
\eeq

Spinors, vectors, tensors in Lie algebras containing a distinguished 
$so(3)$.

\at{this may go elsewhere or nowhere}
Levi-Malcev theorem (Kirillov p.91, Barut \& Raczka p.19)

A maximal chain of ideals 
\[
\Lz=\Lz_0 \supseteq \Lz_1 \supseteq\dots \supseteq\Lz_s
\]
with $\Lz_{k-1}\lp \Lz_k\subseteq\Lz_k$ should give the triangular 
structure; cf. Lie's theorem. Is this related to Ado's theorem?

\section{Connected subgroups of $SL(2,\Rz)$}
\label{s.conn2}

~\at{this needs the triangular construction, hence only here}

$SL(2,\Rz)$ is a matrix group of dimension 3.
Up to conjugacy \at{define}, proper connected subgroups of are:\\ 
For $d=1$: $ISO(1)$, $O(2)$, $O(1,1)$, 
$SD(2)=\Big\{\pmatrix{t & 0 \cr 0 \& t^{-1}}\Big\}$,\\
for $d=2$ apparently only 
$ST(2)=\Big\{\pmatrix{t & 0 \cr s \& t^{-1}}\Big\}$,\\
Is this list complete, irredundant?

$SL(2,\Rz)$ is an example of a connected matrix group in which not 
every element can be written as an exponential of an element of its 
Lie algebra.
Indeed, any $f\in sl(2,\Rz)$ has trace zero, hence its eigenvalues 
are $\pm\lambda$ for some $\lambda\in\Cz$. The eigenvalues of $e^f$
are therefore $e^{\pm\lambda}$, and 
$\tr e^f = e^{\lambda} + e^{-\lambda}$.
Since the product of the
eigenvalues is the real determinant, $\lambda$ is either real or 
purely imaginary. In the first case, $\tr e^f\ge 0$, while in the 
second case, $\lambda=i\omega$ and $\tr e^f=e^{i\omega} + e^{-i\omega}
=2 \cos \omega \ge -2$. Since the element 
$\pmatrix{-2 & 0 \cr 0 & -1/2}\in sl(2,\Rz)$ has trace $<-2$, it cannot
be written as $e^f$ with $f\in sl(2,\Rz)$. \at{why is it connected? 
$QR$ with positive diagonal elements in $R$; needs connectedness of 
$ST(2)$}

\section{Connected subgroups of $SL(3,\Rz)$}
\label{s.conn3}

$SL(3,\Rz)$ is a matrix group of dimension 6.
Up to conjugacy, proper connected subgroups of dimension $0<d<6$ are
$ISO(2)$, $O(3)$, $O(1,2)$, $H(1)$. Is this list complete?

\section{Classical mechanics and Heisenberg groups}\label{s.heis}

$H(n)$ and its Lie algebra; CCR  script p.37f

Poisson representations script p.4-7

units p.29-30

restricted direct sum

The oscillator group $Os(n)$

The Schr\"odinger group?

\section{Angular momentum, isospin, quarks}\label{s.quarks}

Angular momentum, commutation relations

$SU(n)$ for $n$ flavors, and their Lie algebra

Isospin ($n=2$), quarks ($uds\dots$)

Structure hints from the mass spectrum of elementary particles

Standard model gauge group: $S(U(2)\times U(3))$

The local symmetry group of the universe: 
$ISO(1,3) \times S(U(2)\times U(3))$

\section{Connected subgroups of $SL(4,\Rz)$}
\label{s.conn4}

$SL(4,\Rz)$ is a matrix group of dimension 15.
Up to conjugacy, the connected subgroups of dimension $d=10$ are:
 
$ISO(3)$ (homogeneous Galilei transformations) 

$O(4)$; $SO(4)\cong SO(3)\times SO(3)$ (hydrogen)\\
\at{first one proves $so(4)\cong so(3)\oplus so(3)$ and then notes that
$so(4)$ is connected.} 

$O(1,3)$ (Lorentz group)

$O(2,2)$ (no physical relevance)

Find the symmetry group of a Lorentz cone (also needed for completely 
positive maps)

Other dimensions?

\section{The Galilean group}\label{s.galilei}

 script p.24-26,28-30,34f

\bfi{Galilean spacetime.}
Until the beginning of the twentieth century, one thought that time for
all observers was the same in the following sense: if two events take
place at two different places in space, then the question whether the
events took place at the same time has an observer independent answer. 
Space was thought of as a grid on which the motions of all objects took 
place and time was thought to be completely independent from space. The
`distance' between two events therefore consisted of two numbers: a
difference in time and a spatial distance. For example, the distance
between when I woke up and when I took the subway to work is
characterized by saying that from the moment I woke up it took me half
an hour to reach the subway station, which is 500 meter from my bed.
We call the spacetime described in this manner the \bfi{Galilean
spacetime}.

There are three important kinds of symmetries in the Galilean
spacetime and the group that these symmetries generate is called the
\bfi{Galilean symmetry group}\footnote{
The group is also called the \bfi{Galilei group} or the 
\bfi{Galileo group}. We follow the tradition that proceeds in analogy
with the use of Euclidean space or Hermitian matrix.
}. 
 If we shift the clock an hour globally,
which is possible in Galilean spacetime, the laws of nature cannot
alter. Hence one symmetry generator is the time-shift: $t\mapsto t+a$
\at{avoid $\mapsto$ 2x}
for some fixed number $a$. Likewise, the laws of nature should not
change if we shift the origin of our coordinate system; hence a second
symmetry is the shift symmetry $(x^1,x^2,x^3)\mapsto
(x^1+b^1,x^2+b^2,x^3+b^3)$ for some fixed vector $(b^1,b^2,b^3)$. The
third kind of symmetries are rotations, that is, the group $SO(3)$,
which we have seen before. There are some additional discrete
symmetries, like space reflection, where a vector $(x^1,x^2,x^3)$ is
mapped to $(-x^1,-x^2,-x^3)$. We focus, however, on the connected part
of the Galilean symmetry group. The subgroup of the Galilean
symmetry group obtained by discarding the time translations is the
group $ISO(3)$. 
Below, \at{this now goes to Section \ref{s.rigidM}} 
when we discuss the Poincar\'e group, we give
more details on the group $ISO(3)$ as it is a subgroup of the
Poincar\'e group.

\at{explain name, \\
give definitions, \\
generators, \\
structure constants, \\
physical interpretation, \\
primal and dual Poisson representation, \\
the central extension, \\
and CPT. \\
Show that it leaves the N-particle 
Hamiltonian equations of motions invariant.}

\section{The Lorentz groups $O(1,3)$, $SO(1,3)$, $SO(1,3)_0$}
\label{s.lorentz}

 script p.30-33

~\at{
$SO(m,n)$ in general, \\
$SL(2,\Cz)$ and $SO(1,3)$\\
Take up the Stokes vector,\\
add interpretation of boosts, \\
length contracton and time dilation,\\
and the classical limit (contraction)}

When $\Kz=\Rz$, one has for symmetric bilinear forms another
 subdivision, since $B$
can have a definite \idx{signature} $(p,q)$ where $p+q$ is the dimension
of $V$. If $B$ is of signature $(p,q)$, this means that there exists
a basis of $V$ in which $B$ can be represented as
\[
B(v,w) = v^T A w\,,~~~\textrm{where}~~~A =
\textrm{diag}(\underbrace{-1,\dots, -1}_{p \mbox{\scriptsize ~times}},
\underbrace{1,\dots, 1}_{q\mbox{\scriptsize~times}})\,.
\]
The group of all linear
transformations that leaves $B$ invariant is denoted by \idx{$O(p,q)$}. 
The subgroup of $O(p,q)$ of transformations with determinant one is 
the so-called \bfi{special orthogonal group}\index{orthogonal 
group!special} and is denoted by \idx{$SO(p,q)$}.
The associated real Lie algebra is denoted \idx{$so(p,q)$} and its
elements are linear transformations $A:V\to V$ such that
for all $v,w\in V$ we have $B(Av,w)+B(v,Aw)=0$. The Lie product
is given by the commutator of matrices.

More general, the standard representation of
$so(p,q)$ is the one that defines $so(p,q)$ and is thus given by
$(p+q)\times (p+q)$-matrices that leave a metric of signature $(p,q)$
invariant; in Lie algebra theory the standard representation is called
the 
{\bf\idx{fundamental representation}}\index{representation!fundamental}.
In the fundamental representation of $so(3,1)$ (which is not unitary),
the Minkowski inner product is invariant. 

The group $SO(3)$ is a subgroup of $SO(3,1)$ and consists of all those
$SO(3,1)$-rotations that act trivially on the time-component of
four-vectors. The Galilean symmetry group is the subgroup of
$ISO(3,1)$ consisting of the $SO(3)$-rotations together with the time
translations.

An element of $SO(3,1)$ is called a Lorentz boost if the element acts
nontrivially on the zeroth component of four-vectors. By multiplying
with an appropriate element of the $SO(3)$ subgroup we may assume that
a Lorentz boost only mixes the zeroth and first component of
four-vectors. Then a Lorentz boost $L$ takes the following form
(recall $c=1$):
\lbeq{lor.trs}
L(v)^0 = \frac{x^0-vx^1}{\sqrt{1-v^2}}\,,~~~
L(v)^1= \frac{x^1-vx^0}{\sqrt{1-v^2}}\,,
\eeq
and $L(v)^2=v^2$, $L(v)^3=v^3$. Physically the Lorentz boost
\gzit{lor.trs} describes
how coordinates transform when one goes from one coordinate system to
another coordinate system that moves with respect to the first system
in the positive $x^1$-direction with velocity $v$. Since $v$ has to be
smaller than one, as is apparent from \gzit{lor.trs}, one concludes
that special relativity excludes superluminal velocities. The number
\lbeq{e.gammav}
\gamma = \frac{1}{\sqrt{1-v^2}}\,,
\eeq
is called the \bfi{$\gamma$-factor}. The $\gamma$-factor gives an 
indication
whether we should treat a physical situation with special relativity
or whether a nonrelativistic treatment would suffice. The {\bfi{Lorentz
  contraction}} factor is the inverse of $\gamma$ and measures how
distances shrink when measured in another coordinate system, moving at
a velocity $v$ with respect to the original coordinate system. For
$\alpha$-particles, moving with a typical speed of 15,000 kilometers per
second, we have $v=0.05$ and so $\gamma \sim 1.03$ and
$\gamma^{-1}\sim 0.97$, which implies that if we take a rod of 100
meter and let an $\alpha$-particle fly along the rod, it measures only
$97m$ (assuming that $\alpha$-particles can measure). The
$\gamma$-factor thus tells us that if we want accuracy of more than
$3\%$,  we need to treat the $\alpha$-particle
relativistically.

\bigskip
\bfi{The \idx{nonrelativistic limit}.} \at{script p.42}
In order to discuss the
nonrelativistic limit, we restore the presence of the velocity of
light $c$ in the formulas. For a particle at rest, the space
momentum $\p$ vanishes. The formula $p^2=-(mc)^2$ therefore implies
that, at rest, $p_0=mc$ and the {\bfi{rest energy}} is seen to be
$E=mc^2$. This suggests to define the {\bfi{kinetic energy}}
(which vanishes at rest) by the formula
\[
H:= p_0c-mc^2.
\]
Introducing \bfi{velocity} $\vv$ and \bfi{speed} $v$ by
\[
\vv=\p/m,~~~v=|\vv|=\sqrt{\vv^2},
\]
we find from $\p^2-p_0^2= - (mc)^2$ that $p_0=\sqrt{(mc)^2+\p^2}
=mc\sqrt{1+(v/c)^2}$,
so that
\[
H = mc^2(\sqrt{1+(v/c)^2}-1)
=mc^2\frac{1+(v/c)^2 -1}{\sqrt{1+(v/c)^2}+1}
=\frac{mv^2}{\sqrt{1+(v/c)^2}+1}\,.
\]
Similarly, the energy becomes
\[
E=p_0c = \frac{mc^2}{\sqrt{1-(v/c)^2}}\,.
\]
Taking the limit $c\to \infty$ we find that $H$ becomes the
kinetic energy $\half mv^2$ of a nonrelativistic particle of mass $m$,
The {\bfi{nonrelativistic approximation}} $H \approx \half mv^2$
for the kinetic energy is valid for small velocities $v=|\p/m|\ll c$,
where we may neglect the term $(v/c)^2$ in the square root of the
denominator.

\bigskip
\bfi{Lorentz group as $SL(2,\Cz)$.}
We mention \at{merge and refer back} some further properties of spin
coherent states. Because of the identity 
\[
|-x,s\>=(-1)^{2s}|x,s\>
\]
fermionic representations ($s\not\in\Zz$) are called {\bfi{chiral}}. 
\at{relate to PSL above}
Since fermions are chiral, they are not invariant under the
$\Zz_2$-subgroup of $SL(2,\Cz)$ and thus \at{???} fermions do not 
constitute a representation of the restricted Lorentz group.

\bigskip
We use the notation introduced in Section
\ref{s.spinor} and identify four-vectors $p\in \Rz^{1,3}$ with the
$2\times 2$-matrices $p\cdot \sigma_+$. For any four-vector $p\in
\Rz^{1,3}$ the Minkowski norm is given by
\[
\det (p\cdot \sigma_+) = p\cdot p\,.
\]
The group $SL(2,\Cz)$ acts on $\Rz^{1,3}$ through  
\[
A (p\cdot \sigma_+) A^*\,,\mbox{~~~for ~} A\in SL(2,\Cz)\,.
\]
Clearly this defines for each $A\in SL(2,\Cz)$ an element of
$SO(3,1)$, and hence we have a map $SL(2,\Cz)\to SO(3,1)$. The group
$SL(2,\Cz)$ is a real connected manifold of dimension $6$. Indeed, any
complex $2\times 2$ matrix has $4$ complex entries making $8$ real
numbers. The constraint $\det A=1$ gives two equations, for the real
and imaginary part, and hence removing two dimensions. 

Let us show that $SL(2,\Cz)$ is connected. For $A\in SL(2,\Cz)$ we can
apply the Gram--Schmidt proces to the column vectors of $A$. Looking
at how the Gram--Schmidt procedure works, we see that any element of
$A\in SL(2,\Cz)$ 
can be written as a product of an upper triangular matrix $N$ with
positive entries on the diagonal and a unitary matrix $U\in U(2)$. We
can write $U= e^{i\phi}U'$ with $U'\in SU(2)$ making clear that
$U(2)\cong S^2\times S^1$ so that $U(2)$ is connected and the matrix
$U$ can be smoothly connected to the identity. For $N$ we may write
\[
N=\pmatrix{a & b \cr 0 &c}
\]
with $ac=1$ and $a>0$ and $c>0$. \at{avoid $\mapsto$ 2x}
Then $t\mapsto tN+(1-t)1_{2\times 2}$
is a smooth path in $L(2,\Cz)$ for $t\in [0,1]$ that connects the 
unit matrix to $N$. Dividing by the square root of the determinant 
gives the required path in $SL(2,\Cz)$. Hence $SL(2,\Cz)$ is connected. 

The map $SL(2,\Cz)\to SO(3,1)$ is a smooth group homomorphism and 
thus any two points in
the image can be joined by a smooth path. Hence the image is a
connected subgroup of $SO(3,1)$. Since the dimensions of $SO(3,1)$ and
$SL(2,\Cz)$ are the same, the image contains an open connected 
neighborhood $O$ of the identity 
(this is nothing more than the statement that the induced
map $sl(2,\Cz)\to so(3,1)$ is an isomorphism). But the subgroup of
$SO(3,1)$ generated by a small open 
neighborhood of the identity is the connected component containing the
identity. Indeed, call $G'$ the group generated by the open
neighborhood $O$. We may assume $O^{-1}:=\{g^{-1},g\in O\}=O$, 
since if not we just replace $O$ by $O\cap O^{-1}$. 
If $x\in G'$ then $x O\subset G'$ is an open
neighborhood containing $x$ so that $G'$ is
open. If $x\notin G'$, then $xO\cap G' = \emptyset$, since if $y\in
xO\cap G'$, then  there is $z\in O$ such that $xz=y$, but then
$x=yz^{-1}$ lies in $G'$. Hence $G'$ is an open and closed subgroup of
the component that contains the identity, but then $G'$ is the
component that contains the identity. 

Hence the map $SL(2,\C)\to SO(3,1)_0$ is surjective and we only have
to check the kernel. An element $A$ is in the kernel if and only if
$Ap\cdot\sigma_+A^*=p\cdot\sigma_+$ for all $p\in \Rz^{1,3}$. 
This is a linear equation in $p$
so we may as well take $p\in \Cz^{1,3}$. Choosing $p\cdot \sigma_+=
\sigma_1+i\sigma_2$ and writing
\[
A=\pmatrix{ a & b \cr c& d}
\]
we find $a\bar b=0$, $a\bar d=1$ and $|c|=0$. Hence $b=c=0$ and $a\bar
d=1$. But since $\det A=1$ we have $ad=1$ so that $a^2=1$ giving
$a=\pm 1$. The kernel is therefore the normal subgroup $\Zz_2=\{1,-1\}$.

\section{The Poincare group $ISO(1,3)$}\label{s.poincare}

\at{script p.36}

The group of all translations in $V$ generates together with $SO(p,q)$
the group of {\bfi{inhomogeneous special orthogonal
transformations}}, which is denoted \idx{$ISO(p,q)$}. One can obtain
$ISO(p,q)$ from $SO(p,q+1)$ by performing a contraction; that is,
by rescaling some generators with some parameter $\epsilon$ and then
choosing a singular limit $\epsilon\to 0$ or $\epsilon \to \infty$.
The group $ISO(p,q)$ can also be seen as the group of
$(p+q+1)\times (p+q+1)$-matrices of the form
\[
\pmatrix{ Q & b \cr 0 & 1 }  ~~~ \textrm{with }~ Q\in SO(p,q)\,,~ b\in
V\,.
\]
The Lie algebra of $ISO(p,q)$ is denoted \idx{$iso(p,q)$} and can be
described as the
Lie algebra of $(p+q+1)\times (p+q+1)$-matrices of the form
\[
\pmatrix{ A & b \cr 0 & 0 }  ~~~ \textrm{with }~ A\in so(p,q)\,,~ b\in
V\,.
\]
Again, the Lie product in $iso(p,q)$ is the commutator of matrices.

\bigskip
\bfi{Minkowski spacetime.}
With the advent of special relativity, the classical spacetime view was
altered in the sense that time and space made up one spacetime, called
\bfi{Minkowski spacetime}. As a topological vector space Minkowski
spacetime is nothing more than $\Rz^4$, but it is equipped with the
\at{change signs in this section, or use $\pm$.
Note the two traditions. Change $3,1$ to $1m3$. 
See also \gzit{e.mink}}
{\bfi{Minkowski metric}}\footnote{We choose units such  
that $c=1$, and work with the signature $(-,+,+,+)$.} (also
see Section \ref{s.gamma} and Example \ref{ex.SOpq}):

\[
(x-y)^2 = -(x^0-y^0)^2 + (x^1-y^1)^2+(x^2-y^2)^2+(x^3-y^3)^2\,
=(\x-\y)^2-(x^0-y^0)^2.
\]
The time component of the \bfi{four-vectors} is the zeroth component.
We write a general four-vector as
\[
v = \pmatrix{ v^0 \cr \mathbf{v} \cr }\,,~~~
\mathbf{v}=\pmatrix{v^1\cr v^2\cr v^3\cr}\,,
\]
where $\mathbf{v}$ is the space-like part of $v$ and $v^0$ is the
time-like component of $v$. With the notation introduced we see that
the Minkowski metric can be written as $v^2= -(v^{0})^{2}+\mathbf{v}^2$,
where $\mathbf{v}^2$ is the usual Euclidean norm for
three-vectors. The Minkowski inner product is derived from the
Minkowski metric and given by
\[
\pmatrix{ v^0 \cr \mathbf{v} \cr }
\pmatrix{ w^0 \cr \mathbf{w} \cr }=-v^0w^0+\mathbf{v}\mathbf{w}\,.
\]
Note that in a strict sense \at{clarify}
the Minkowski inner product, the Minkowski
norm and the Minkowski metric are not an inner product, norm and
metric respectively as the positivity condition is clearly not 
satisfied.

The \bfi{Poincar\'e group} is a subgroup of the group of all
symmetries that leave the Minkowski metric invariant. The Poincar\'e
group is often denoted as $ISO(3,1)$. On a four-vector
$v$ the Poincar\'e group acts as $v\mapsto Av+b$, where $A$ is an
element of $SO(3,1)$ and $b$ is some four-vector. Hence the Poincar\'e
group consists of rotations and translations. 
An explicit representation of $ISO(3,1)$ can be given in terms of the
matrices
\[
\pmatrix{ A & b \cr 0 & 1 \cr }\,,
\]
where $A$ is a $4\times 4$-matrix in $SO(3,1)$ and $b$ is a
four-vector. Recall that $A$ is in $SO(3,1)$ if $A$ satisfies
\[
A^T\eta A=\eta\,,~~~\eta=\pmatrix{ -1 & 0&0&0\cr 0 &1&0&0 \cr
0&0&1&0\cr 0&0&0&1}\,.
\]
The affine linear transformations contain the translations and
$SO(3,1)$-rotations. The generators of the translations we call the
momenta, and since they have four components, we sometimes refer to
them as four-momenta.

The (real) Lie algebra of $ISO(3,1)$ is described by the matrices of 
the form
\[
(A,b):=\pmatrix{ A & b \cr 0 & 0\cr},
\]
with $A\in so(3,1)$ and $b\in \Rz^3$. The Lie product is given by the
commutator of matrices, and takes the form
\[
(A,b)\lp (A',b') = ([A,A'],Ab'-A'b)\,,
\]
where $Ab'$ is the usual matrix action of $A$ on $b'$. In particular,
we have
\[
(0,b)\lp (0,b')=0\,,~~~ (A,0)\lp (0,b')=(0,Ab')\,,
\]
from which we read off that the translations form a commutative
subalgebra. The translations form an ideal such that the
momenta form the standard representation of $so(3,1)$, that is, the
defining representation. 

\bigskip
\bfi{General spacetime.}
The generalization of Minkowski spacetime is a manifold with a 
\bfi{pseudo-Riemannian metric} $g$; the latter turns the tangent space 
at each point of the manifold into a Minkowski space. Thus around 
every point there is a
chart and a coordinate system such that $g$ takes the form of a
Minkowski metric. It is clear that a proper description of general
relativity requires differential geometry and the development of
tensor calculus. 

In general relativity but also already in special
relativity physicists use some conventions that are worth
explaining. Spacetime indices indicating components of four-vectors
are indicated by Greek letters $\mu,\nu,\ldots$. To denote a
four-vector $x=(x^\mu)$ one writes simply $x^\mu$. If an index appears
ones upstairs and once downstairs, it is to be summed over; this is
called the \bfi{Einstein convention}. Derivatives are objects with
indices downstairs; $ \partial_\mu = \partial / \partial x^\mu$. The
\bfi{Kronecker delta} $\delta^{\mu}_{\nu}$ is an invariant tensor and
we have $\partial_\mu x^\nu=\delta^{\mu}_{\nu}$ and 
$\partial_\mu x^\mu=4$.
The Minkowski metric is usually denoted by the Greek letter $\eta$ and
again one usually just writes $\eta_{\mu\nu}$ to denote the metric and
not just the $\mu\nu$-component; as a matrix the Minkowski metric is
given by:
\[
\eta = \pmatrix{- 1&0&0&0\cr
                0& 1&0&0\cr 
                0&0& 1&0\cr 
                0&0&0& 1}\,.
\]
The Minkowski inner product is now
written as $v\cdot w = v^\mu w^\nu \eta_{\mu\nu}$.
If $v$ and $w$ are two elements of the tangent space at a point $x$
their inner product in general relativity is given by $v^\mu w^\nu
g_{\mu\nu}(x)$, from which it is clear that general relativity is the
curved generalization of special relativity. To denote the metric $g$
physicists often describe a line element, which is to mean the
distance of an infinitesimal displacement $ds=(dx^\mu)$;
\[
ds^2=g_{\mu\nu}(x)dx^\mu dx^\nu\,.
\]
The metric $g_{\mu\nu}$ and its pointwise inverse $g^{\mu\nu}$ are 
used to lower and to raise indices; indeed, the
metric gives an isomorphism between the tangent space and the
cotangent space. Hence $\partial^\mu$ is defined as
$g^{\mu\nu}\partial_\nu$, and a check of consisteny gives
$g^{\mu\nu}=g^{\mu\lambda}g^{\nu\rho}g_{\rho\lambda}$. As a further
exercise in the conventions the reader might verify
$g^{\mu\lambda}g_{\lambda\nu}=\delta^{\mu}_{\nu}$, 
$g^{\mu\nu}g_{\mu\nu}=4$.
The described conventions are used a lot in physics literature and 
more on its nature and why it works can be found in many text books 
on relativity, e.g., in the nice introductory textbook by 
\sca{d'Inverno} \cite{dinverno}).

The symmetry group of a manifold $M$ with a pseudo-Riemannian metric
$g$ is
huge; it consists of all diffeomorphisms of the manifold, as any
diffeomorphism preserves a metric. The vector
fields on $M$ describe the infinitesimal generators of the group of
diffeomorphisms.

\at{to do:\\
$ISL(2,\Cz)/\Zz_2\cong ISO(1,3)$ extending the hom. iso\\
Causality, script p.50f\\
classical limit; contraction}

\section{A Lorentz invariant measure}

The reason that one uses the integration measure
\lbeq{e.invmeas}
\frac{d^3\k}{(2\pi)^32\omega(k)}, ~~~
\omega(k)=\sqrt{c^2\k^2 + \frac{m^2c^4}{\hbar^2}}. 
\eeq
is due to Lorentz covariance. The
integration measure is clearly rotation invariant. Hence to study the
behavior under a general Lorentz transformation $L$ we may assume 
that $L$
only mixes the $x$-direction and the time-direction. In that case we
have, using the $\gamma$-factor \gzit{e.gammav} 
\at{not really needed for proving invariance, since the subsequent 
delta-argun=ment suffices}
\beqar
L(k_x)&=& \frac{k_x - v k_t}{\sqrt{1-\frac{v^2}{c^2}}}
=\gamma(k_x - v k_t)\,,\nn\\
L(k_y)&=&k_y\,,~~L(k_z)=k_z\,,\nn\\
L(k_t)&=&  \frac{k_t - \frac{v}{c^2} k_x}{\sqrt{1-\frac{v^2}{c^2}}}
=\gamma(k_t - \frac{v}{c^2} k_x)\,.\nn
\eeqar
One easily checks that $L(k)^2=k^2$. Since $\omega(k)$ is the zeroth
component of the wave vector $k$, we see that the factors
$\sqrt{1-\frac{v^2}{c^2}}$ cancel out. Another way to see the
 covariance is to note the equality
\lbeq{cov.Lor.mu}
\frac{d^3\k}{(2\pi)^32\omega(k)}
= \frac{d^4 k}{(2\pi)^3}\delta\Big({k^2+\frac{m^2c^2}{\hbar^2}}\Big)
\theta(k_0)\,,
\eeq
where $\theta$ is the \bfi{Heaviside function} defined by 
 $\theta(x)=1$ if $x\geq 0$ and $\theta(x)=0$ if $x<0$. 
The Heaviside function selects the positive sign of the square root in 
\gzit{e.invmeas}. The equality of the
two integration measures in \gzit{cov.Lor.mu} is proven by integration
of both sides. It is clear that the expression \gzit{cov.Lor.mu} is
invariant under Lorentz transformations.

\section{Kepler's laws, the hydrogen atom, and $SO(4)$}\label{s.so4}

\at{The 2-body problem, Lenz-Runge vector, SO(4), hydrogen}

\section{The periodic systemand the conformal group $SO(2,4)$}
\label{s.conformal}

$SO(2,4)\cong SU(4)?$ containing $SO(1,3)$ 

conformal transformations and Poisson representation 
\at{Thirring II, p.52}

$H(3)$ as subgroup and the hydrogen atom

The periodic system \at{maybe}

\section{The interacting boson model and $U(6)$}\label{s.IBM}

\at{maybe}

\section{Casimirs}\label{s.casimir}

faithful representations \at{script p.22}

sum and product of reps

universal envelope (classical and quantum), 

Casimirs, \at{script p.39-42}

splitting representations through common eigenspaces of Casimirs

$so(3)$, $sl(2)$, spin

\section{Unitary representations of the Poincar\'e group}
\label{s.poincareR}

This section is neither in a good form nor complete in contents.
\at{this section needs at least the Foldy construction added}

Knowing the simple and semisimple Lie algebras is of course
interesting, but in physics there are also important non-semisimple
Lie algebras that play an important role. In this section we
treat one of the most important Lie algebras in physics, the
Poincar\'e Lie algebra. The Poincar\'e Lie algebra is not semisimple
since it contains an abelian (and thus solvable) ideal.

In physics, the irreducible unitary representations of the
Poincar\'e algebra correspond to {\bfi{elementary particles}},
more precisely to particles considered at distances so large that
their internal structure can be safely ignored. Using the Casimir 
operators in the universal enveloping algebra, one can label the 
different representations, since a Casimir must take a constant value in
any irreducible representation. \at{where are Casimirs defined?}

There are two independent Casimirs. One is the number $p^2$, which is
an invariant since the Minkowski inner product is invariant.
With our choice of signature  $(-+++)$, one has
$p^2=\p^2-p_0^2= - (mc)^2$ for a constant $m$ called the \bfi{mass}
of the representation (or the associated particles); for the signature
$(+---)$, we have instead $p^2=p_0^2-\p^2=(mc)^2$. In physically
relevant representations, $m\geq 0$ and $p_0>0$. 

A second Casimir accounts for the {\bfi{spin}} $s$ of the 
representation; for unitary representations, it is quantized, and takes 
nonnegative, half integral values. \at{relate to $so(3)$ spin!}
The particles are called \bfi{bosons}\index{boson} if the spin of this 
representation is integral, and \bfi{fermions}\index{fermion} otherwise,
i.e., if the spin is half an odd integer. For example, electrons have 
spin $s=1/2$ and are fermions, while photons have spin $s=1$ and are 
bosons. The name ''spin'' derives from relations to the representation 
theory of the rotation group; see Section \ref{s.qanti}, where also
the dichotomic nature of integral and nonintegral spin is explained, 
which justifies using different names for bosons and fermions.

Clearly, representations which differ in mass or spin are nonequivalent.
Less trivial is the fact that, among the \bfi{physical 
representations}\index{Poincar\'e group! physical representations}
 (i.e., those with $m\ge 0$ and $p_0>0$), there is an up to
equivalence unique irreducible representation for each combination
$m>0$ and $s \in\half\Zz_+$. In the massless case $m=0$, there are
precisely two for each $s \in\half\Zz_+$, a right-handed and a
left-handed one.

Given an irreducible unitary representation, we can choose a basis
such that the components of $p$ act diagonally, since they are
Hermitian and commute. Thus we can assign to a vector
in the representation the four momentum components.
The momentum components will also be
denoted $p_\nu$. The number $E=p_0c$ is called the \bfi{energy}, and
depends on the basis chosen, since the $so(3,1)$
rotations mix the momenta.  Having fixed a basis of the
translations, there is only a $SO(3)$ subgroup that leaves the energy
invariant. Intuitively this is clear, rotating a reference frame does
not change the energies. In general, for a given basis, the subgroup
of $so(3,1)$ that leaves the vector $(1,0,0,0)$ invariant is $SO(3)$
and the elements of the $SO(3)$ subgroup are rotations. There are
three independent $SO(3,1)$ elements that do not leave $(1,0,0,0)$
invariant, these transformations and their linear combinations
are called {\bfi{Lorentz boosts}} in the
physics literature. The Lorentz boosts mix time and space
coordinates. A basis of Poincar\'e Lie algebra thus consists of the
generators of three rotations, three Lorentz boosts and four
translations.

\section{Some representations of the Poincare group}\label{s.foldy}

mass and spin

The Foldy construction

relation to the invariant measure form; Foldy-Wouthuisen transform?

Galilei version and classical limit

\section{Elementary particles}\label{s.particles}

Elementary particle = 
irreducible unitary representation of the Poincare group with 
quantized spin, $p^2\ge 0$, and $p_0>0$.

\at{script p.43-46}

massless particles and gauge freedom

\at{the following must be replaced by a simpler argument based on the 
information from the proof of the unitary representation theorem}

\begin{prop}
For real $a \in \Rz ^{3}$ and any Pauli set of spin $j$,
\lbeq{rep18}
  | \overline\psi(a\cdot\sigma)\psi| \le 2j|a|\overline\psi\psi 
  \Forall \psi \in \Cz ^{s}.
\eeq
Equality holds iff $\psi$ satisfies the {\bf Weyl equation}
\[
(a\cdot\sigma)\psi = \pm 2j|a|\psi.
\]
\end{prop}

\bepf
We need to show that the symmetric matrix 
$\sigma_0(2j|a|-a\cdot\sigma)$ is positive semidefinite.
The matrix 
$A=\sigma_0((s-1)|a|-a \cdot \sigma )$ is tridiagonal with
nontrivial entries
\[
  A_{kk}= \D {s-1 \choose k-1}((s-1)|a|-(s+1-2k) a_{3}),
\]
\[
  A_{kk-1}=-\D {s-1 \choose k-1}(k-1)(a_{1}+ia_{2}),
\]
\[
  A_{kk+1}=-\D {s-1 \choose k-1}(s-k)(a_{1}-ia_{2})=A^{*}_{k+1k}.
\]
If $a_{3} \not= |a|$ then $a_{3} < |a|$, and we may define the 
lower triangular tridiagonal matrix $L$ and the diagonal matrix $D$ 
with nontrivial entries
\[
  L_{kk}=1,~~~L_{k+1k}=-\frac{a_{1}+ia_{2}}{|a|-a_{3}},~~~
  D_{kk}=(|a|-a_{3})k \D {s-1 \choose k} \ge 0.
\]
Now \at{no!} $A=LDL^{*}$; 
therefore, $A$ is Hermitian positive semidefinite.

If $a_{3}=|a|$ then $a_{1}=a_{2}=0$ and $A$ is diagonal with nonnegative
diagonal entries $A_{kk}= {s-1 \choose k-1}2(k-1)|a|$, and again 
positive semidefinite.
Now 
\[
 0\le\psi^{*} A\psi= 
(s-1)|a|\overline\psi\psi - \overline\psi(a\cdot\sigma)\psi, 
\]
and replacing $a$ by $-a$ gives the desired inequality. Equality holds
iff $A\psi=0$, which is the Weyl equation.
\epf

Note that the Weyl equation is solved for $a_{3} < |a|$ by $\psi$
iff $L^*\psi$ is zero except in the last component (since $A=LDL^{*}$
and the other diagonal entries of $D$ are positive).

More precisely, $a\cdot\sigma$ has the simple eigenvalues 
$(s+1-2l)|a|$ ($l=1:s$). \at{rotate $a$ to $a_3'=|a|$!}

\section{The position operator}\label{s.position}

\at{maybe; what else from the FAQ should go into this chapter?}

\chapter{From the theoretical physics FAQ}\label{c.FAQ}

\section{To be done}

The present chapter will be merged into the preceding chapters.
Some of the sections in later chapters, whose content is already in 
Part I will be eliminated.

\bigskip
The section on "Heisenberg groups and Poisson representations" is 
at the start of Chapter \ref{c.symm} since to define the Lie algebra 
of angular momentum already requires the CCR and Poisson brackets.

Perhaps the two chapters could be
integrated better via a sequence like:

- Reflections, Rotations and classical angular
  momentum (which could contain a lot of math
  on SO(3) including some Lie stuff).

- Galilei group, which builds on rotations.

- Symplectic/Hamiltonian stuff (classical
  non-relativistic dynamical groups)

- Classical relativistic stuff (Poincare)
  Maybe also classical electromagnetism in
  here somewhere, since it stands astride
  both the classical and quantum worlds.
  It's also the natural place to introduce the
  concept of (classical) gauge invariance.

- Non-relativistic QM (Heisenberg, Oscillator,
  Schrodinger, etc).

- Re-visit SO(3) in the quantum context and
  show how the requirement of being a
  symmetry of a positive-definite inner
  product is enough to imply stunningly
  unexpected facts about the spectrum
  of angular momentum experiments.
  This then becomes the archetype for
  how representations, Casimirs, etc,
  are at the heart of modern physics.
  This is also a good place to emphasize
  how Schr\"odinger wave functions, etc,
  are not the last word and about how a
  more general algebraic framework is
  cleaner and powerful (having shown
  that this is sufficient to make
  impressive predictions).

- Continue on to Isospin and gauge symmetries.

- Quantum dynamical groups (conformal,
  H-atom, etc).

Thus, each step in the mathematical sequence
of ideas is presented and developed in a
suitable physics context.

\section{Postulates for the formal core of quantum mechanics}

\begin{verbatim}

Quantum mechanics consists of a formal core that is universally agreed 
upon (basically being a piece of mathematics with a few meager pointers 
on how to match it with experimental reality) and an interpretational 
halo that remains highly disputed even after 85 years of modern quantum 
mechanics. The latter is the subject of the foundations of quantum 
mechanics; it is addressed elsewhere in this FAQ. 

Here I focus on the formal side. The relativistic case is outside the 
scope of the present axioms, though presumably very little needs to be 
changed. 

As in any axiomatic setting (necessary for a formal discipline),
there are a number of different but equivalent sets of axioms
or postulates that can be used to define formal quantum mechanics.
Since they are equivalent, their choice is a matter of convenience.

My choice presented here is the formulation which gives most direct 
access to statistical mechanics but is free from allusions to 
measurement. The reason for the first is that statistical mechanics is 
the main tool for applications of quantum mechanics to the macroscopic 
systems we are familiar with. The reason for the second is that real 
measurements consitute a complex process involving macroscopic 
detectors, hence should be explained by quantum statistical mechanics 
rather than be part of the axiomatic foundations themselves. (This is 
in marked contrast to other foundations, and distinguishes the present
axiom system.) 

Thus the following describes nonrelativistic quantum statistical 
mechanics in the Schroedinger picture. (As explained later, the 
traditional starting point is instead the special case of this setting 
where all states are assumed to be pure.) 

For brevity, I assume the knowledge of some basic terms from functional 
analysis, which are precisely defined in many mathematics books. 
[For a discussion of the difference between a Hermitian and a 
self-adjoint operator, see e.g., Definition 3 in 
http://arxiv.org/pdf/quant-ph/9907069 . The importance of this 
difference is that Hermitian operators have a real spectrum if and 
only if they are self-adjoint. Moreover, the Hille-Yosida theorem says 
that e^{iX) exists (and is unitary) for a Hermitian operator X if and 
only iff X is self-adjoint. A detailed discussion and the HY theorem 
itself are discussed in Vol.3 of the math physics treatise by Thirring.] 
The statements of my axioms contain in parentheses some additional 
explanations that, strictly speaking, are not part of the axioms but 
make them more easily intelligible; the list of examples given only 
has illustrative character and is far from being exhaustive. 


Quantum mechanics is governed by the following six axioms:

A1. A generic system (e.g., a 'hydrogen molecule') is defined by 
specifying a Hilbert space K and a (densely defined, self-adjoint) 
Hermitian linear operator H called the _Hamiltonian_ or the _energy_. 

A2. A particular system (e.g., 'the ion in the ion trap on this 
particular desk') is characterized by its _state_ rho(t)
at every time t in R (the set of real numbers). 
Here rho(t) is a Hermitian, positive semidefinite, linear trace class 
operator on K satisfying at all times the conditions
   trace rho(t) = 1.   (normalization)

A3. A system is called _closed_ in a time interval [t1,t2]
if it satisfies the evolution equation
   d/dt rho(t) = i/hbar [rho(t),H] for t in [t1,t2],
and _open_ otherwise. (hbar is Planck's constant, and is often set 
to 1.) 
If nothing else is apparent from the context, a system is assumed to 
be closed.

A4. Besides the energy H, certain other (densely defined, self-adjoint) 
Hermitian operators (or vectors of such operators) are distinguished 
as _observables_. 
(E.g., the observables for a system of N distinguishable particles 
conventionally include for each particle several 3-dimensional vectors: 
the _position_ x^a, _momentum_ p^a, _orbital_angular_momentum_ L^a 
and the _spin_vector_ (or Bloch vector) sigma^a of the particle with 
label a. If u is a 3-vector of unit length then u dot p^a, u dot L^a 
and u dot sigma^a define the momentum, orbital angular momentum,
and spin of particle a in direction u.)

A5. For any particular system, and for every vector X of observables 
with commuting components, one associates a time-dependent monotone 
linear functional <.>_t defining the _expectation_
      <f(X)>_t:=trace rho(t) f(X)
of bounded continuous functions f(X) at time t. 
(This is equivalent to a multivariate probability measure dmu_t(X) 
on a suitable sigma algebra over the spectrum spec(X) of X) defined by
   integral dmu_t(X) f(X) := trace rho(t) f(X) =<f(X)>_t.
The signma algebra is uniquely determined.)

A6. Quantum mechanical predictions consist of predicting properties 
(typically expectations or conditional probabilities) of the measures 
defined in Axiom A5, given reasonable assumptions about the states 
(e.g., ground state, equilibrium state, etc.) 

Axiom A6 specifies that the formal content of quantum mechanics is 
covered exactly by what can be deduced from Axioms A1-A5 without 
anything else added (except for restrictions defining the specific 
nature of the states and observables), and hence says that 
Axioms A1-A5 are complete.

The description of a particular closed system is therefore given by 
the specification of a particular Hilbert space in A1, the 
specification of the observable quantities in A4, and the
specification of conditions singling out a particular class of 
states (in A6). Given this, everything else is determined by the theory,
and hence is (in principle) predicted by the theory.

The description of an open system involves, in addition, the
specification of the details of the dynamical law. (For the basics,
see the entry 'Open quantum systems' in this FAQ.)


In addition to these formal axioms one needs a rudimentary 
interpretation relating the formal part to experiments.
The following _minimal_interpretation_ seems to be universally
accepted.

MI. Upon measuring at times t_l (l=1,...,n) a vector X of observables 
with commuting components, for a large collection of independent 
identical (particular) systems closed for times t<t_l, all in the same 
state 
    rho_0 = lim_{t to t_l from below} rho(t)
(one calls such systems _identically_prepared_), the measurement
results are statistically consistent with independent realizations
of a random vector X with measure as defined in axiom A5.


Note that MI is no longer a formal statement since it neither defines
what 'measuring' is, nor what 'measurement results' are and what 
'statistically consistent' or 'independent identical system' means. 
Thus MI has no mathematical meaning - it is not an axiom, but already
part of the interpretation of formal quantum mechanics.

MI relates the axioms not to a hypothetical classical realm but to a 
nonphysical entity: the social conventions of the community of 
physicists. The terms 'measuring', 'measurement results', 
'statistically consistent', and 'independent' already have informal 
meaning in the reality as perceived by a physicist. Everything stated 
in Axiom MI is understandable by every trained physicist. 
Thus statement MI is not an axiom for formal logical reasoning but 
a bridge to informal reasoning in the traditional cultural setting 
that defines what a trained physicist understands by reality.


The lack of precision in statement MI is on purpose, since it allows
the statement to be agreeable to everyone in its vagueness; different
philosophical schools can easily fill it with their own understanding
of the terms in a way consistent with the remainder.

Interpretational axioms necessarily have this form, since they must
assume some unexplained common cultural background for perceiving
reality. (This is even true in pure mathematics, since the language
stating the axioms must be assumed to be common cultural background.)


MI is what _every_ interpretation I know of assumes (and has to assume) 
at least implicitly in order to make contact with experiments. 
Indeed, all interpretations I know of assume much more, but they 
differ a lot in what they assume beyond MI.

Everything beyond MI seems to be controversial. In particular,
already what constitutes a measurement of X is controversial.
(E.g., reading a pointer, different readers may get marginally
different results. What is the true pointer reading?)

On the other hand there is an informal consensus on how to
perform measurements in practice. Good foundations including a 
good measurement theory should be able to properly justify this
informal consensus by defining additional formal concepts that
behave within the theory just as their informal relatives with 
the same name behave in reality.

In complete foundations, there would be formal objects in the
mathematical theory corresponding to all informal objects discussed 
by physicists, such that talking about the formal objects
and talking about the real objects is essentially isomorphic.
We are currently far from such complete foundations.


A state rho is called _pure_ at time t if rho(t) maps the Hilbert 
space K to a 1-dimensional subspace, and _mixed_ otherwise.

Although much of traditional quantum mechanics is phrased in terms of
pure states, this is a very special case; in most actual experiments 
the systems are open and the states are mixed states. Pure states
are relevant only if they come from the ground state of a 
Hamiltonian in which the first excited state has a large energy gap.
Indeed, assume for simplicity that H has discrete spectrum. In an 
orthonormal basis of eigenstates psi_k,
   f(H) = sum_k f(E_k) psi_k psi_k^*
for every function f defined on the spectrum. Setting the Boltzmann 
constant to 1 to simplify the formulas, the equilibrium density is
the canonical ensemble,
   rho(T) = 1/Z(T) exp(-H/T) = sum_k exp(-E_k/T)/Z(T) psi_k psi_k^*.
(Of course, equating this ensemble with equilibrium in a closed system
is an additional step beyond our axiom system, which would require
justification.) Taking the trace (which must be 1) gives 
   Z(T) = sum_k exp(-E_k/T),
and in the limit T -> 0, all terms exp(-E_k/T)/Z(T) become 0 or 1,
with 1 only for the k corresponding to the states with least energy 
Thus, if the ground state psi_1 is unique,
   lim_{T->0} rho(T) = psi_1 psi_1^*.
This implies that for low enough temperatures, the equilibrium state 
is approximately pure. The larger the gap to the second smallest
energy level, the better is the approximation at a given nonzero
temperature. In particular (reinstalling the Boltzmann constant kbar),
the approximation is good if the energy gap exceeds a small multiple 
of E^* := kbar T. 

States of simple enough systems with a few levels only 
can often be prepared in nearly pure states, by realizing a source 
governed by a Hamiltonian in which the first excited state has a much 
larger energy than the ground state. Dissipation then brings the
system into equilibrium, and as seen above, the resulting equilibrium 
state is nearly pure.


To see how the more traditional setting in terms of the 
Schroedinger equation arises, we consider the case of a closed 
system in a pure state rho(t) at some time t. 

If psi(t) is a unit vector in the range of the pure state rho(t) 
then psi(t), called the _state_vector_ of the system at time t, 
is determined up to a phase, and one easily verifies that 
   rho(t) = psi(t)psi(t)^*.
Remarkably, under the dynamics for a closed system specified in the 
above axioms, this property persists with time (only) if the system 
is closed, and the state vector satisfies the Schroedinger equation
  i hbar psi(t) = H psi(t) 
Thus the state remains pure at all times. Conversely, for every pure 
state, the phases of psi(t) at all times t can be chosen such that the
Schroedinger equation holds.

Moreover, if X is a vector of observables with commuting components
and the spectrum of X is discrete, then the measure from Axiom A5
is discrete,
   integral dmu(X) f(X) = sum_k p_k f(X_k)
with nonnegative numbers p_k summing to 1, commonly called 
_probabilities_. Associated with the p_k are eigenspaces K_k such that
   X psi = X_k psi      for  psi in K_k,
and K is the direct sum of the K_k. Therefore, every state vector psi
can be uniquely decomposed into a sum
   psi = sum_k psi_k    with  psi_k in K_k.
psi_k is called the _projection_ of psi to the eigenspace K_k.

A short calculation using Axiom A5 now reveals that for a pure state 
rho(t)=psi(t)psi(t)^*, the probabilities p_k are given by the
so-called _Born_rule_
   p_k = |psi_k(t)|^2,           (B)
where psi_k(t) is the projection of psi(t) to the eigenspace K_k.

Deriving the Born rule (B) from Axioms A1-A5 makes it completely
natural, while the traditional approach starting with (B) 
makes it an irreducible rule full of mystery and only justifiable 
by its miraculous agreement with experiment.

Note that Born's 1926 paper (reprinted in English translation in 
pp.52-55 of the reprint volume ''Quantum Theory and Measurement'' by 
Wheeler and Zurek) - which introduced the probabilistic interpretation 
that earned him a Nobel prize - didn't relate his interpretation to
measurement. Born's formulation doesn't depend on anything being 
measured (let alone to be assigned a precise numerical measurement 
value): ''gives the probability for the electron, arriving from the 
z-direction, to be thrown out into the direction designated by the 
angles alpha, beta, gamma, with the phase change delta''.

Nevertheless, it is often (see, e.g., 
    http://en.wikipedia.org/wiki/Born_rule )
claimed as part of Born's rule that the results of the measurement 
should equal exactly the eigenvalues. But unless the lambda_i are 
(as for polarization, spin or angular momentum in a particular 
direction - the common subjects of experiments involving Alice and Bob) 
system-independent, discrete, and known a priori - in which case one 
can label each measurement record with these numbers -, this form of the
rule is highly unrealistic.



I didn't mention indistinguishable particles in my examples 
illustrating the axioms, for two reasons: 
 1. One cannot easily specify the set of relevant observables without 
introducing lots of additional notation or terminology - whereas the 
explanations of the axioms should be very short.
 2. I think that the concept of indistinguishable particles is 
completely superseded by the concept of a quantum field. 
The latter gives much better intuition about the meaning of the 
formalism, and the former (which is difficult to justify and even more 
difficult to interpret intuitively) is then completely dispensable.
 
\end{verbatim}

\section{Lie groups and Lie algebras}

\begin{verbatim}
Lie groups can be illustrated by continuous rigid motion of a ball
with painted patterns on it in 3-dimensional space. The Lie group ISO(3)
consists of all rigid transformations.

A rigid transfromation is essentially the act of picking the ball and
placing it somewhere else, ignoring the detailed motion in between and
the location one started.
Special transformations are for example a translation in northern 
direction by 1 meter, or a rotation by one quarter around the vertical 
axis at some particular point (think of a ball with a string attached).
'Rigid' means that the distances between marked points on the ball
remains the same; the mathematician talks about 'preserving distances',
and the distances are therefore labeled 'invariants'.

One can repeat the same transformation several times, or two different
transformations and get another one - This is called the product of
these transformations. For example, the product of a translations
by 1 meter and another one by 2 meters in the same direction gives one
of 1+2=3 meters in the same direction. In this case, the distances add,
but if one combines rotations about different axes the result is no
longer intuitive. To make this more tractable for calculations,
one needs to take some kind of logarithms of transformations - these
behave again additively and make up the corresponding Lie algebra
iso(3) [same letters but in lower case]. The elements of the Lie algebra
can be visualized as very small, or 'infinitesimal', motions.


General Lie groups and Lie algebras extend these notions to to more
general manifolds. A manifold is just a higher-dimensional version
of space, and transformations are generalized motions preserving
invariants that are important in the manifold. The transformations
preserving these invariants are also called 'symmetries', and the
Lie group consisting of all symmetries is called a 'symmetry group'.
The elements of the corresponding Lie algebra are 'infinitesimal
symmetries'.

For example, physical laws are invariant under rotations and 
translations, and hence unter all rigid motions. But not only these: 
If one includes time explicitly, the resulting 4-dimensional space 
has more invariant motions or ''symmetries''.
The Lie group of all these symmetry transformations is called the
Poincar'e group, and plays a basic role in the theory of relativity.
The transformations are now about space-time frames in uniform motion.
Apart from translations and rotations there are symmetries called
'boosts' that accelerate a frame in a certain direction, and 
combinations obtained by taking products. All infinitesimal symmetries 
together make up a Lie algebra, called the Poincar'e algebra.

Much more on Lie groups and Lie algebras from the perspective of 
classical and quantum physics can be found in:
    Arnold Neumaier and Dennis Westra,
    Classical and Quantum Mechanics via Lie algebras,
    Cambridge University Press, to appear (2009?).
    http://www.mat.univie.ac.at/~neum/papers/physpapers.html#QML
    arXiv:0810.1019 
\end{verbatim}

\section{The Galilei group as contraction of the Poincare group}

\begin{verbatim}
The group of symmetries of special relativity is the Poincare group.

However, before Einstein invented the theory of relativity,
physics was believed to follow Newton's laws, and these have a
different group of symmetries - the Galilei group, and its
infinitesimal symmetries form the Galilei algebra.

Now Newton's physics is just a special case of the theory of relativity
in which all motions are very slow compared to the speed of light.
Physicists speak of the 'nonrelativisitic limit'.
Thus one would expect that the Galilei group is a kind of
nonrelativistic limit of the Poincar'e group.

This notion has been made precise by Inonu. He looked at the
Poincar'e algebra and 'contracted' it in an ingenious way
to the Galilei algebra. The construction could then be lifted to
the corresponding groups. Not only that, it turned out to be a
general machinery applicable to all Lie algebras and Lie groups,
and therefore has found many applications far beyond that for which
it was originally developed.
\end{verbatim}

\section{Representations of the Poincare group}

\begin{verbatim}
Whatever deserves the name ''particle'' must move like a single, 
indivisible object. The Poincare group must act on the description of 
this single object; so the state space of the object carries a 
unitary representation of the Poincare group. This splits into a direct 
sum or direct integral of irreducible reps. But splitting means 
divisibility; so in the indivisible case, we have an irreducible 
representation.

On the other hand, not all irreducible unitary reps of the Poincare 
group qualify. Associated with the rep must be a consistent and causal 
free field theory. As explained in Volume 1 of Weinberg's book on 
quantum field theory, this restricts the rep further to those with 
positive mass, or massless reps with quantized helicity.


Weinberg's book on QFT argues for gauge invariance from
causality + masslessness. He discusses massless fields in 
Chapter 5, and observes (probably there, or in the beginning 
of Chapter 8 on quantum electrodynamics) roughly the following:

Since massless spin 1 fields have only two degrees of freedom,
the 4-vector one can make from them does not transform correctly
but only up to a gauge transformation making up for the missing
longitudinal degree of freedom. Since sufficiently long range
elementary fields (less than exponential decay) are necessarily
massless, they must either have spin <=1/2 or have gauge behavior.

To couple such gauge fields to matter currents, the latter
must be conserved, which means (given the known conservation laws)
that the gauge fields either have spin 1 (coupling to a conserved
vector current), or spin 2 (coupling to the energy-momentum tensor).
[Actually, he does not discuss this for Fermion fields,
so spin 3/2 (gravitinos) is perhaps another special case.]

Spin 1 leads to standard gauge theories, while spin 2 leads
to general covariance (and gravitons) which, in this context,
is best viewed also as a kind of gauge invariance.

There are some assumptions in the derivation, which one can find
out by reading Weinberg's papers 
   Phys.Rev. 133 (1964), B1318-B1322  any spin (massive)    
   Phys.Rev. 134 (1964), B882-B896    any spin II (massless)
   Phys.Rev. 135 (1964), B1049-B1056  grav. mass = inertial mass
   Phys.Rev. 138 (1965), B988-B1002   derivation of Einstein 
   Phys.Rev. 140 (1965), B516-B524    infrared gravitons
   Phys.Rev. 181 (1969), 1893-1899    any spin III (general reps.)
on 'Feynman rules for any spin' and some related questions, which 
contain a lot of important information about applying the irreducible 
representations of the Poincare group for higher spin to field
theories, and their relation to gauge theories and general relativity.
A perhaps more understandable version of part of the material is in
   D.N. Williams,
   The Dirac Algebra for Any Spin,
   Unpublished Manuscript (2003)
   http://www-personal.umich.edu/~williams/papers/diracalgebra.pdf

Note that there are plenty of interactions that can be constructed
using the representation theory of the Lorentz group (and Weinberg's
constructions), and there are plenty of (compound) particles with 
spin >2. See the tables of the particle data group, e.g., Delta(2950)
(randomly chosen from http://pdg.lbl.gov/2003/bxxxpdf.html ).
   R.L. Ingraham,
   Prog. Theor. Phys. 51 91974), 249-261,
   http://ptp.ipap.jp/link?PTP/51/249/
constructs covariant propagators and complete vertices for spin J 
bosons with conserved currents for all J. See also
   H Shi-Zhong et al.,
   Eur. Phys. J. C 42 (2005), 375-389
   http://www.springerlink.com/content/ww61351722118853/
\end{verbatim}

\section{Forms of relativistic dynamics}

\begin{verbatim}
Relativistic multiparticle mechanics is an intricate subject, 
and there are no-go theorems that imply that the most plausible
possibilities cannot be realized. However, these no-go theorems
depend on assumptions that, when questioned, allow meaningful 
solutions. The no-go theorems thus show that one needs to be careful 
not to introduce plausible but inappropriate intuition into the 
formal framework.


To pose the problem, one needs to distinguish between kinematical
and dynamical quantities in the theory. Kinematics answers the 
question "What are the general form and properties of objects that 
are subject to the dynamics?" Thus it tells one about conceivable 
solutions, mapping out the properties of the considered representation 
of the phase space (or what remains of it in the quantum case).
Thus kinematics is geometric in nature. But kinematics does not know 
of equations of motions, and hence can only tell general (kinematical) 
features of solutions. 

In contrast, dynamics is based on an equation of motion (or an 
associated variational principle) and answers the question 'What 
characterizes the actual solution?', given appropriate initial or 
boundary conditions. Although the actual solution may not be available 
in closed form, one can discuss their detailed properties and devise 
numerical approximation schemes.

The difference between kinematical and dynamical is one of convention,
and has nothing to do with the physics. By choosing the representation,
i.e., the geometric setting, one chooses what is kinematical; 
everything else is dynamical.

Since something which is up to the choice of the person describing 
an experiment can never be distinguished experimentally, the physics
is unaffected. However, the formulas look very different in different
descriptions, and - just as in choosing coordinate systems - choosing 
a form adapted to a problem may make a huge difference for actual
computations.


Dirac distinguishes in his seminal paper 
    Rev. Mod. Phys. 21 (1949), 392-399
three natural forms of relativistic dynamics, the instant form, 
the point form, and the fromt form. They are distinguished by 
what they consider to be kinematical quantities and what are the
dynamical quantities. 


The familiar form of dynamics is the instant form, 
which treats space (hence spatial translations and rotations)
as kinematical and time (and hence time translation and Lorentz boosts)
as dynamical. This is the dynamics from the point of view of a
hypothetical observer (let us call it an 'instant observer')
who has knowledge about all information at some time t (the present), 
and asks how this information changes as time proceeds.

Because of causality (the finite bound of c on the speed of material 
motion and communication), the resulting differential equations 
should be symmetric hyperbolic differential equations for which the 
initial-value problem is well-posed. 
Because of Lorentz invariance, the time axis can be 
any axis along a timelike 4-vector, and (in special relativity) 
space is the 3-space orthogonal to it. For a real observer, 
the natural timelike vector is the momentum 4-vector of the material 
system defining its reference frame (e.g., the solar system).

While very close to the Newtonian view of reality, it involves 
an element of fiction in that no real observer can get all the 
information needed as intial data. Indeed, causality implies that
it is impossible for a physical observer to know the present anywhere
except at its own position.


A second, natural form of relativistic dynamics is, according to Dirac, 
the point form. This is the form of dynamics in which a particular
space-time point x=0 (the here and now) in Minkowski space is 
distinguished, and the kinematical object replacing space is, 
for fixed L, a hyperboloid x^2=L^2 (and x_0<0) in the past 
of the here and now.
The Lorentz transformations, as symmetries of the hyperboloid,
are now kinematical and take the role that space translations and 
rotations had in the instant form. On the other hand, _all_ space and
time translations are now dynamical, since they affect the position 
of the here-and-now.

This is the form of dynamics which is manifestly
Lorentz invariant, and in which space and time appear on equal footing.
An observer in the here and now  (let us call it a 'point observer')
can - in principle, classically - have arbitrarily accurate 
information about the particles and/or fields on the past 
hyperboloid; thus causality is naturally accounted for. 
Information given on the past hyperboloid of a point can be propagated 
to information on any other past hyperboloid using the dynamical 
equations that are defined via the momentum 4-vector P, which is a 
4-dimensional analogue of the nonrelativistic Hamiltonian. 
The Hamiltonian corresponding to motion in a fixed timelike 
direction u is given by H=u dot P. The commutativity of the components
of P is the condition for the uniqueness of the resulting state
at a different point x independent of the path x is reached from 0.


In principle, there are many other forms of relativistic dynamics: 
As Dirac mentions on p. 396 of his paper, any 3-dimensional surface 
in Minkowski space works as kinematical space if it meets 
every world line with time like tangents exactly once. 
In general, those transformations are kinematical which 
are also symmetries of the surface one treats as kinematical reference 
surface. By choosing a surface without symmetries _all_ 
transformations become dynamical. For reasons of economy, one wants 
however, a large kinematical symmetry group. The full Poincare group
is possible only for free dynamics. 

This leaves as interesting large subgroups two with 6 linearly 
independent generators, the Euclidean group ISO(3), leading to the 
instant form, and the Lorentz group SO(1,3), leading to the point form, 
and one with 7 linearly independent generators, the stabilizer of 
a front (or infinite momentum plane), a 3-space with lightlike normal, 
leading to the front form. This third natural form of relativistic 
dynamics according to Dirac, has many uses in quantum field theory,
but here I won't discuss it further.


All forms are equivalent, related classically by canonical 
transformations preserving algebraic operations and the Poisson bracket,
and quantum mechanically by unitary transformations preserving 
algebraic operations and hence the commutator. This means that any 
statement about a system in one of the forms can be translated into 
an equivalent statement of an equivalent system in any of the other 
forms.

Preferences are therefore given to one form over the other depending 
solely on the relative simplicity of the computations one wants to do. 
This is completely analogous to the choice of coordinate systems 
(cartesian, polar, cylindric, etc.) in classical mechanics.


For a multiparticle theory, however, the different forms and the 
need to pick a particular one seem to give different pictures of 
reality. This invites paradoxes if one is not careful.

This can be seen by considering trajectories of classical relativistic 
many-particle systems. There is a famous theorem by 
    Currie, Jordan and Sudarshan 
    Rev. Mod. Phys. 35 (1963), 350-375
which asserts that interacting two-particle systems cannot have 
Lorentz invariant trajectories in Minkowski space. Traditionally, 
this was taken by mainstream physics as an indication that the
multiparticle view of relativistic mechanics is inadequate, 
and a field theoretical formulation is essential. 
However, as time proceeded, several approaches to valid relativistic 
multi-particle (quantum) dynamics were found (see the FAQ entry on 
'Is there a multiparticle relativistic quantum mechanics?'), 
and the theorem had the same fate as von Neumann's proof that 
hidden-variable theories are impossible. Both results are now simply 
taken as an indication that the assumptions under which they were 
made are too strong. 

In particular, once the assumption by Currie, Jordan and Sudarshan
that all observers see the same trajectories of a system of interacting 
particles is rejected, their no-go theorem no longer applies.
The question then is how to find a consistent and covariant description
without this at first sight very intuitive property. But once it is
admitted that different observers see the same world but represented
in different personal spaces, the formerly intuitive property becomes
meaningless. For objectivity, it is enough that one can consistently
translate the views of any observer into that of any other observer.
Precisely this is the role of the dynamical Poincare transformations.

Thus nothing forbids an instant observer to observe 
particle trajectories in its present space, or a
point observer to observe particle trajectories in its past hyperboloid.
However, the present space (or the past hyperboloid) of two different 
observers is related not by kinematical transforms but dynamically,
with the result that trajectories seen by different observers on 
their different kinematical 3-surface look different. 
Classically, this looks strange on first sight, although
the Poincare group provides well-defined recipes for translating 
the trajectories seen by one observer into those seen by another
observer.

Quantum mechanically, trajectories are fuzzy anyway, due to the 
uncertainty principle, and as various successful multiparticle 
theories show, there is no mathematical obstacle for such a description.

The mathematical reason of this superficially paradoxical situation 
lies in the fact that there is no observer-independent definition 
of the center of mass of relativistic particles, and the related fact 
that there is no observer-independent definition of space-time 
coordinates for a multiparticle system. 
The best one can do is to define either a covariant position operator 
whose components do not commute (thus definig a noncommutative 
space-time), or a spatial position operator, the so-called 
Newton-Wigner position operator, which has three commuting coordinates 
but is observer-dependent.
(See the FAQ entry on 'Localization and position operators'.)
\end{verbatim}

\section{Is there a multiparticle relativistic quantum mechanics?}

\begin{verbatim}
In his QFT book, Weinberg says no, arguing that there is no way to 
implement the cluster separation property. But in fact there is:

There is a big survey by Keister and Polyzou on the subject
    B.D. Keister and W.N. Polyzou,
    Relativistic Hamiltonian Dynamics in Nuclear and Particle Physics,
    in: Advances in Nuclear Physics, Volume 20,
    (J. W. Negele and E.W. Vogt, eds.)
    Plenum Press 1991.
    www.physics.uiowa.edu/~wpolyzou/papers/rev.pdf 
that covered everything known at that time. This survey was quoted 
at least 116 times, see
    http://www.slac.stanford.edu/spires/find/hep?c=ANUPB,20,225
looking these up will bring you close to the state of the art 
on this.

They survey the construction of effective few-particle models. 
There are no singular interactions, hence there is no need for 
renormalization.

The models are _not_ field theories, only Poincare-invariant few-body
dynamics with cluster decomposition and phenomenological terms
which can be matched to approximate form factors from experiment or 
some field theory. (Actually many-body dynamics also works, but the
many particle case is extremely messy.) 
They are useful phenomenological models, but somewhat limited; 
for example, it is not clear how to incorporate external fields.

The papers by Klink at
   http://www.physics.uiowa.edu/~wklink/
and work by Polyzou at 
   http://www.physics.uiowa.edu/~wpolyzou/
contain lots of multiparticle relativistic quantum mechanics,
applied to real particles. See also the Ph.D. thesis by Krassnigg at
   http://physik.uni-graz.at/~ank/dissertation-f.html

Other work in this direction includes Dirac's many-time quantum 
theory, with a separate time coordinate for each particle; see, e.g., 
   Marian Guenther, Phys Rev 94, 1347-1357 (1954) 
and references there. Related multi-time work was done under the 
name of 'proper time quantum mechanics' or 'manifestly covariant 
quantum mechanics', see, e.g., 
   L.P. Horwitz and C. Piron, Helv. Phys. Acta 48 (1973) 316,
but it apparently never reached a stage useful to phenomenology.
\end{verbatim}

\section{What is a photon?}

\begin{verbatim}
According to quantum electrodynamics, the most accurately verified 
theory in physics, a photon is a single-particle excitation of the 
free quantum electromagnetic field. More formally, it is a state of 
the free electromagnetic field which is an eigenstate of the photon 
number operator with eigenvalue 1.

The pure states of the free quantum electromagnetic field
are elements of a Fock space constructed from 1-photon states.
A general n-photon state vector is an arbitrary linear combinations 
of tensor products of n 1-photon state vectors; and a general pure 
state of the free quantum electromagnetic field is a sum of n-photon 
state vectors, one for each n. If only the 0-photon term contributes, 
we have the dark state, usually called the vacuum; if only the 
1-photon term contributes, we have a single photon.

A single photon has the same degrees of freedom as a classical vacuum
radiation field. Its shape is characterized by an arbitrary nonzero
real 4-potential A(x) satisfying the free Maxwell equations, which in 
the Lorentz gauge take the form
   nabla dot nabla A(x) = 0,
   nabla dot A(x) = 0,
expressing the zero mass and the transversality of photons. Thus for
every such A there is a corresponding pure photon state |A>.
Here A(x) is _not_ a field operator but a photon amplitude;
photons whose amplitude differ by an x-independent phase factor are 
the same. For a photon in the normalized state |A>, the observable 
electromagnetic field expectations are given by the usual formulas 
relating the 4-potential and the fields,
   <\E(x)> = <A|\E(x)|A> 
           = - partial \A(x)/partial x_0 - c nabla_\x A_0(x),
and 
   <\B(x)> = <A|\B(x)|A> = nabla_\x x \A(x)
[hmmm. check if this really is the case...]
Here \x (fat x) and x_0 are the space part and the time part of a
relativistic 4-vector, \E(x), \B(x) are the electromagnetic
field operators (related to the operator 4-potential by analogous
formulas), and c is the speed of light. Amplitudes A(x) producing 
the same \E(x) and \B(x) are equivalent and related by a gauge 
transformation, and describe the same photon. 


In momentum space (frequently but not always the appropriate choice), 
single photon states have the form 
   |A> = integral d\p^3/p_0 A(\p)|\p>,
where |\p> is a single particle state with definite 3-momentum 
\p (fat p), p_0=|\p| is the corresponding photon energy divided by c,
and the photon amplitide A(\p) is a polarization 4-vector.
Thus a general photon is a superposition of monochromatic waves with 
arbitrary polarizations, frequencies and directions.
(The Fourier transform of A(\p) is the so-called analytic signal
A^(+)(x), and by adding its complex conjugate one gets the real 
4-potential A(x) in the Lorentz gauge.)

The photon amplitude A(\p) can be regarded as the photon's 
wave function in momentum space. Since photons are not localizable
(though they are localizable approximately), there is no 
meaningful photon wave function in coordinate space; see the
next entry in this FAQ. One could regard the 4-potential A(x) as 
coordinate space wave function, but because of its gauge dependence, 
this is not really useful.

[
This is second quantized notation, as appropriate for quantum fields.
This is how things always look in second quantization, even for a 
harmonic oscillator. The wave function psi(x) or psi(p) in standard 
(first quantized) quantum mechanics becomes the state vector
   psi = integral dx psi(x) |x>   or   integral dp psi(p) |p> 
in Fock space; the wave function at x or p turns into the coefficient 
of |x> or |p>. In quantum field theory, x, A (the photon amplitude), and
E(x) (the electric field operator) correspond to  k (a component of the
momentum), x, and p_k. Thus the coordinate index k is inflated to the 
spacetime position x, the argument of the wave function is inflated to 
a solution of the free Maxwell equations, the momentum operator is 
inflated to a field operator, and the integral over x becomes a 
functional integral over photon amplitudes,
   psi = integral dA psi(A) |A>.
Here psi(A) is the most general state vector in Fock space; for a 
single photon, psi depends linearly on A,
   psi(A) = integral d\p^3/p_0 A(\p)|\p> = |A>.
Observable electromagnetic fields are obtained as expectation values
of the field operators \E(x) and \B(x) constructed by differentiation of
the textbook field operator A(x). As the observed components of 
the mean momentum, say, in ordinary quantum mechanics are
   <p_k> = integral dx psi(x)^* p_k psi(x),
so the observed values of the electromagnetic field are
   <\E(x)> = <psi|\E(x)|psi> = integral dA psi(A)^* \E(x) psi(A).
   <\B(x)> = <psi|\B(x)|psi> = integral dA psi(A)^* \B(x) psi(A).
]
In a frequently used interpretation (valid only approximately),
the term A(\p)|\p> represents the one-photon part of a monochromatic 
beam with frequency nu=cp_0/h, direction \n(\p)=\p/p_0, and 
polarization determined by A(\p). Here h = 2 pi hbar, where hbar is 
Planck's number; omega=cp_0/hbar is the angular frequency.

The polarization 4-vector A(\p) is orthogonal to the 4-momentum p 
composed of p_0 and \p, obtained by a Fourier transform of the 
4-potential A(x) in the Lorentz gauge. (The wave equation translates 
into the condition p_0^2=\p^2, causality requires p_0>0, hence 
p_0=|\p|, and orthogonality p dot A(\p) = 0 expresses the Lorentz 
gauge condition. For massless particles, there remains the additional 
gauge freedom to shift A(\p) by a multiple of the 4-momentum p, which 
can be used to fix A_0=0.) 

A(\p) is usually written (in the gauge with vanishing time component) as
a linear combination of two specific polarization vectors eps^+(p) and 
eps^-(p) for circularly polarized light (corresponding to helicities +1 
and -1), forming together with the direction vector \n(\p) an 
orthonormal basis of complex 3-space. In particular,
   eps^+(p) eps^+(p)^* + eps^-(p)eps^-(p)^* + \n(\p)\n(\p)^* = 1
is the 3x3 identity matrix. (This is used in sums over helicities for 
Feynman rules.) Specifically, eps^+(p) and eps^-(p) can be obtained by 
finding normalized eigenvectors for the eigenvalue problem 
[check. The original eigenvalue problem is p dot J eps = lambda eps.]
   p x eps = lambda eps 
with lambda = +-i|p|. For example, if p is in z-direction then 
   eps^+(p) = (1, -i, 0)/sqrt(2),
   eps^-(p) = (i, -1, 0)/sqrt(2),
and the general case can be obtained by a suitable rotation.
An explicit calculation gives almost everywhere
    eps^+(p) = u(p)/p_0
where p_0=|p| and 
    u_1(p) = p_3 - i p_2 p'/p'',
    u_2(p) = -i p_3 - i p_1 p'/p''
    u_3(p) = p'
with
   p' = p_1+ip_2, 
   p''= p_3+p_0.
[what is eps^-(p)?]
These formulas become singular along the negative p_3-axis,
so several charts are needed to cover 

For experiments one usually uses nearly monochromatic light bundled 
into narrow beams. If one also ignores the directions (which are 
usually fixed by the experimental setting, hence carry no extra
information), then only the helicity degrees of freedom remain,
and the 1-photon part of the beam behaves like a 2-level quantum 
system ('a single spin'). 

A general monochromatic beam with fixed direction in a pure state is 
given by a second-quantized state vector, which is a superposition of 
arbitrary multiphoton states in the Bosonic Fock space generated by 
the two helicity degrees of freedom. This is the basis for most 
quantum optics experiments probing the foundations of quantum 
mechanics. 


The simplest state of light (generated for example by 
lasers) is a coherent state, with state vector proportional to
    e(A) = |vac> + |A> + 1/sqrt(2!) |A> tensor |A> 
                 + 1/sqrt(3!) |A> tensor |A> tensor |A> + ...
where |A> is a one-photon state. Thus coherent states also have the 
same degrees of freedom as classical electromagnetic radiation. 
Indeed, light in coherent states behaves classically in most respects.

At low intensity, the higher order terms in the expansion are
negligible, and since the vacuum part is not directly observable,
a low intensity coherent states resembles a single photon state.

On the other hand, true single photon states are very hard to produce 
to good accuracy, and were created experimentally only recently:
   B.T.H. Varcoe, S. Brattke, M. Weidinger and H. Walther,
   Preparing pure photon number states of the radiation field, 
   Nature 403, 743--746 (2000).
see also 
   http://www.qis.ucalgary.ca/quantech/fock.html 
Ordinary light is essentially never, and high-tech light almost never,
describable by single photons.


A good informal discussion of what a photon is from a more practical 
perspective was given by Paul Kinsler in
   http://www.lns.cornell.edu/spr/2000-02/msg0022377.html
But this does not tell the whole story. An interesting collection of 
articles explaining different current views is in 
    The Nature of Light: What Is a Photon?
    Optics and Photonics News, October 2003
    http://www.osa-opn.org/Content/ViewFile.aspx?Id=3185


The standard reference for quantum optics is
   L. Mandel and E. Wolf,
   Optical Coherence and Quantum Optics, 
   Cambridge University Press, 1995.
Mandel and Wolf write (in the context of localizing photons),
about the temptation to associate with the clicks of a photodetector
a concept of photon particles. [If there is interest, I can try to 
recover the details.] The wording suggests that one should resist the
temptation, although this advice is usually not heeded. However,
the advice is sound since a photodetector clicks even when it
detects only classical light! This follows from the standard analysis
of a photodetector, which treats the light classically and only 
quantizes the detector. Thus the clicks are an artifact of 
photodetection caused by the quantum nature of matter, rather than
a proof of photons arriving!!!


A coherent light source (laser) produces a coherent state of light, 
which is a superposition of the vacuum state, a 1-photon state,
a 2-photon state, etc, with squared amplitudes given by a Poisson 
distribution. At low intensity, this is misinterpreted in practice 
as random single photons arriving at the end of the beam in a 
random Poisson process, because the photodetector produces clicks 
according to this distribution. 

Incoherent light sources usually consist of thermal mixtures and 
produce other distributions, but otherwise the description (and 
misinterpretation) is the same. 
Nevertheless, one must understand this misinterpretation in order
to follow much of the literature on quantum optics.

Thus the talk about photons is usually done inconsistently;
almost everything said in the literature about photons should be taken 
with a grain of salt. 
There are even people like the Nobel prize winner Willis E. Lamb 
(the discoverer of the Lamb shift) who maintain that photons don't 
exist. See towards the end of
   http://web.archive.org/web/20040203032630/www.aro.army.mil/phys/proceed.htm
The reference mentioned there at the end appeared as 
   W.E Lamb, Jr., 
   Anti-Photon, 
   Applied Physics B 60 (1995), 77--84
This, together with the other reference mentioned by Lamb, is reprinted 
in 
   W.E Lamb, Jr., 
   The interpretation of quantum mechanics, 
   Rinton Press, Princeton 2001.

I think the most apt interpretation of an 'observed' photon as used
in practice (in contrast to the photon formally defined as above) is 
as a low intensity coherent state, cut arbitrarily into time slices 
carrying an energy of h*nu = hbar*omega, the energy of a photon at 
frequency nu and angular frequency omega. 
Such a state consists mostly of the vacuum (which is not directly 
observable hence can usually be neglected), and the contributions of 
the multiphoton states are negligible compared to the single photon 
contribution. 
With such a notion of photon, most of the actual experiments done make 
sense, though it does not explain the quantum randomness of the 
detection process (which comes from the quantized electrons in the 
detector).


A nonclassical description of the electromagnetic field where states of 
light other than coherent states are required is necessary mainly for
special experiments involving recombining split beams, squeezed
state amplification, parametric down-conversion, and similar 
arrangements where entangled photons make their appearance.
There is a nice booklet on this kind of optics:
   U. Leonhardt,
   Measuring the Quantum State of Light,
   Cambridge, 1997.

Nonclassical electromagnetic fields are also relevant in the
scattering of light, where there are quantum corrections 
due to multiphoton scattering. These give rise to important effects 
such as the Lamb shift, which very accurately confirm the quantum
nature of the electromagnetic field. They involve no observable 
photon states, but only virtual photon states, hence they are unrelated 
to experiments involving photons. Indeed, there is no way to observe 
virtual particles, and their name was chosen to reflect this. 
(Observed particles are always onshell, hence massless for photons, 
whereas it is an easy exercise that the virtual photon mediating 
electromagnetic interaction of two electrons in the tree approximation 
is never onshell.)
\end{verbatim}

\section{Particle positions and the position operator}

\begin{verbatim}
The standard probability interpretation for quantum particles 
is based on the Schr"odinger wave function psi(x), a square integrable 
single- or multicomponent function of position x in R^3. 
Indeed, with ^* denoting the conjugate transpose,
    rho(x) := psi(x)^*psi(x)
is generally interpreted as the probability density to find (upon 
measurement) the particle at position x. Consequently,
    Pr(Z) := integral_Z dx |psi(x)|^2
is interpreted as the probability of the particle being in the open 
subset Z of position space. Particles in highly localized states
are then given by wave packets which have no appreciable size
|psi(x)| outside some tiny region Z. 

If the position representation in the Schr"odinger picture exists,
there is also a vector-valued position operator x, whose components
act on psi(x) by multiplication with x_j (j=1,2,3). In particular,
the components of x commute, satisfy canonical commutation relations 
with the conjugate momentum 
   p = -i hbar partial_x,
and transform under rotations like a 3-vector, so that the commutation
relations with the angular momentum J take the form
   [J_j,x_k] = i eps_{jkl} x_l.
Moreover, in terms of the (unnormalizable) eigenstates |x,m> of the 
position operator correponding to the spectral value x (and a label m 
to distinguish multiple eigenstates) we can recover the position 
representation from an arbitrary representation by defining psi(x) 
to be the vector with components
   psi_m(x) := <x,m|psi>.
Therefore, if we have a quantum system defined in an arbitrary 
Hilbert space in which a momentum operator is defined, the necessary 
and sufficient condition for the existence of a spatial probability 
interpretation of the system is the existence of a position operator 
with commuting components which satisfy standard commutation 
relations with the components of the momentum operator and the
angular momentum operator.  

Thus we have reduced the existence of a probability interpretation
for particles in a bounded region of space to the question of the
existence of a position operator with the right properties.
We now investigate this existence problem for elementary particles,
i.e., objects represented by an irreducible representation of the 
full Poincare group. We consider first the case of particles of 
mass m>0, since the massless case needs additional considerations.


A. Massive case, m>0:

Let M := R^3 be the manifold of 3-momenta p. On the Hilbert space 
H_m^d obtained by completion of the space of all C^infty functions 
with compact support from M to the space C^d of d-component vectors 
with complex entries, with inner product defined by
   <phi|psi> := integral d\p/sqrt(p^2+m^2) phi(p)^*psi(p),
we define the position operator 
   q := i hbar partial_p,
which satisfies the standard commutation relations, the momentum in
time direction,
   p_0 := sqrt(m^2+|p|^2),
where m>0 is a fixed mass, and the operators
   J := q x p  + S,
   K := (p_0 q + q p_0)/2 + p x S/(m+p_0),
where S is the spin vector in a unitary representation of so(3) on 
the vector space C^d of complex vectors of length d, with the same 
commutation relations as J. 

This is a unitary representation of the Poincare algebra;
verification of the standard commutation relations (given, 
e.g., in Weinberg's Volume 1, p.61) is straightforward. 
It is not difficult to show that this representation is irreducible 
and extends to a representation of the full Poincare group.
Obviously, this representation carries a position operator.

Since the physical irreducible representations of the Poincare group
are uniquely determined by mass and spin, we see that in the massive 
case, a position operator must always exist. An explicit formula in 
terms of the Poincare generators is obtained through division by m 
in the formula 
    mq = K - ((K dot p) p/p_0 + J x p)/(m+p_0),
which is straightforward, though a bit tedious to verify from the above.
That there is no other possibility follows from
    T.F. Jordan
    Simple derivation of the Newton-Wigner position operator
    J. Math. Phys. 21 (1980), 2028-2032.

Note that the position operator is always observer-dependent, in the 
sense that one must choose a timelike unit vector to distinguish
space and time coordinates in the momentum operator. This is due to 
the fact that the above construction is not invariant under Lorentz 
boosts (which give rise to equivalent but different representations).

Note also that in case of the Dirac equation, the position operator is 
_not_ the operator multiplying a solution psi(x) of the Dirac equation 
by the spacelike part of x (which would mix electron and positron 
states), but a related operator obtained by first applying a so-called
Foldy-Wouthuysen transformation. 
   L.L. Foldy and S.A. Wouthuysen,
   On the Dirac Theory of Spin 1/2 Particles and Its Non-Relativistic 
   Limit,
   Phys. Rev. 78 (1950), 29-36.


B. Massless case, m=0:

Let M_0 := R^3\{0} be the manifold of nonzero 3-momenta p, and let 
   p_0 := |p|, n := p/p_0.
The Hilbert space H_0^d (defined as before but now with m=0 and with 
M_0 in place of M)
obtained by completion of the space of all C^infty functions 
with compact support from M to the space C^d of d-component vectors 
with complex entries, with inner product defined by
   <phi|psi> := integral d\p/sqrt(p^2+m^2) phi(p)^*psi(p),
carries a natural massless representation of the Poincare algebra, 
defined by
   J := q x p  + S,
   K := (p_0 q + q p_0)/2 + n x S,
where q = i hbar partial_p is the position operator, and S is the 
spin vector in a unitary representation of so(3) on C^d, with the 
same commutation relations as J. 
Again, verification of the standard commutation relations is 
straightforward. (Indeed, this representation is the limit of the 
above massive representation for m --> 0.)

It is easily seen that the helicity 
   lambda := n dot S
is central in the (suitably completed) universal envelope of the 
Lie algebra, and that the possible eigenvalues
of the helicity are s,s-1,...,-s, where s=(d-1)/2. Therefore, the 
eigenspaces of the helicity operator carry by restriction unitary 
representations of the Poincare algebra, which are easily seen to be
irreducible. They extend to a representation of the connected 
Poincare group. Moreover, the invariant subspace H_s formed by the 
direct sum of the eigenspaces for helicity s and -s form a massless 
irreducible spin s representation of the full Poincare group.

(It is easy to see that changing K to K-t(p_0)p for an arbitrary 
differentiable function t of p_0 preserves all commutation relations,
hence gives another representation of the Poincare algebra.
Since the massless irreducible representations of the Poincare group
are uniquely determined by their spin, the resulting representations 
are equivalent. This corresponds to the freedom below in choosing a
position operator.)

Now suppose that a Poincare invariant subspace H of L^2(M_0)^d has a 
position operator x satisfying the canonical commutation relations 
with p and the above commutator relations with J. Then F=q-x commutes 
with p, hence its components must be a (possibly matrix-valued) 
function F(p) of p. Commutation with p implies that partial_p x F = 0,
and, since M_0 is simply connected, that F is the gradient of a scalar 
function f. Rotation invariance then implies that this function 
depends only on p_0=|p|. Thus 
    F = partial_p f(p_0) = f'(p_0) n.
Thus the position operator takes the form 
   x = q - f'(p_0) n.
In particular,
   x x p = q x p. 
Now the algebra of linear operators on the dense subspace of C^infty 
functions in H contains the components of p, J, K and x, hence those of
   J - x x p = J - q x p = S. 
Thus the (p-independent) operators from the spin so(3) act on H. 
But this implies that either H=0 (no helicity) or H = L^2(M_0)^d
(all helicities between s and -s). 

Since the physical irreducible representations of the Poincare group
are uniquely determined by mass and spin, and for s>1/2, the spin s 
Hilbert space H_s is a proper, nontrivial subspace of L^2(M_0)^d, 
we proved the following theorem:


Theorem.
An irreducible representations of the full Poincare group with 
mass m>=0 and finite spin has a position operator transforming 
like a 3-vector and satisfying the canonical commutation relations 
if and only if either m>0 or m=0 and s<=1/2 (but s=0 if only
the connected poincare group is considered).

This theorem was announced without giving details in 
    T.D. Newton and E.P. Wigner,
    Localized states for elementary systems,
    Rev. Mod. Phys. 21 (1949), 400-406.
A mathematically rigorous proof was given in
   A. S. Wightman,
   On the Localizability of Quantum Mechanical Systems, 
   Rev. Mod. Phys. 34 (1962), 845-872.
See also
    T.F. Jordan
    Simple proof of no position operator for quanta with zero mass 
    and nonzero helicity
    J. Math. Phys. 19 (1980), 1382-1385.
who also considers the massless representations of continuous spin,
and
    D Rosewarne and S Sarkar,
    Rigorous theory of photon localizability,
    Quantum Opt. 4 (1992), 405-413.


For spin 1, the case relevant for photons, we have d=3, and the 
subspace of interest is the space H obtained by completion of the 
space of all vector-valued C^infty functions A(p) of a nonzero 
3-momentum p with compact support satisfying the transversality 
condition p dot A(p)=0,
with inner product defined by
   <A|A'> := integral dp/|p| A(p)^* A'(p).
It is not difficult to see that one can identify the wave functions
A(p) with the Fourier transform of the vector potential in the 
radiation gauge where its 0-component vanishes. This relates the
present discussion to that given in the FAQ entry ''What is a photon?''.


As a consequence of our discussion, photons (m=0, s=1) and gravitons 
(m=0, s=2) cannot be given natural probabilities for being in any given 
bounded region of space. Chiral spin 1/2 particles also do not have 
a position operator and hence have no such probabilities, by the same 
argument, applied to the connected Poincare group.

(Note that measured are only frequencies, intensities and 
S-matrix elements; these don't need a well-defined position concept 
but only a well-defined momentum concept, from which frequencies
can be found via omega=p_0/hbar - since c=1 in the present setting,
and directions via n = p/p_0.)


However, assuming there are scalar massless Higgs particles (s=0),
one could combine such a higgs, a photon, and a graviton into
a single reducible representation on L^2(M_0)^5, using the above
construction. By our derivation, one can find position eigenstates
which are superpositions of Higgs, photon, and graviton. Thus to
be able to regard photons and gravitons as particles with a proper
probability interpretation, one must consider Higgs, photons, and 
gravitons as aspects of the same localizable particle, which we 
might call a graphoton. (Without gravity, a phiggs particle would 
also do.)


Related papers:
    M.H.L. Pryce,
    Commuting Co-ordinates in the new field theory,
    Proc. Roy. Soc. London Ser. A 150 (1935), 166-172.
    (first construction of position operators in the massive case)

    B. Bakamjian and L.H. Thomas,
    Relativistic Particle Dynamics. II,
    Phys. Rev. 92 (1953), 1300-1310.
    (first construction of massive representations along the above 
    lines)

    L.L. Foldy,
    Synthesis of Covariant Particle Equations,
    Physical Review 102 (1956), 568-581.
    (nice and readable version of the Bakamjian-Thomas construction 
    for massive representations of the Poincare group)

    R. Acharya and E. C. G. Sudarshan,
    ''Front'' Description in Relativistic Quantum Mechanics,
    J. Math. Phys. 1 (1960), 532-536.
    (a ''most local'' description of the photon by wave fronts)

    I. Bialynicki-Birula,
    Photon wave function,
    http://arxiv.org/abs/quant-ph/0508202
    (A 53 page recent review article, covering various possibilities 
    to define photon wave functions without a position operator 
    acting on them. The best is (3.5), with a nonstandard inner 
    product (5.8). What is left of the probability interpretation is 
    (5.28) and its subsequent discussion.)

See also the entry ''Localization and position operators'' in this FAQ.


There are a few papers by M. Hawton, e.g.
    http://arxiv.org/abs/quant-ph/0101011 
    http://arxiv.org/abs/0711.0112v1
on a nonstandard position operator which does not transform like a 
3-vector. This is unphysical since it does not give orientation 
independent probabilities for observing a photon in a given region of 
space. Claims to the contrary in 
    http://lanl.arxiv.org/pdf/0804.3773v2,
supposedly constructing a Lorentz invariant photon number density,
are erroneous; see 
    http://groups.google.at/group/sci.physics.research/browse_thread/thread/815435df4bf2ea93?hl=en#

Other nonstandard position operators violating the conditions
necessary for a probability interpretation were discussed earlier,
starting with 
    M.H.L. Pryce,
    The Mass-Centre in the Restricted Theory of Relativity and Its 
    Connexion with the Quantum Theory of Elementary Particles,
    Proc. Roy. Soc. London, Ser. A, 195 (1948), 62-81.
\end{verbatim}

\section{Localization and position operators}

\begin{verbatim}
Position operators are part of the toolkit of relativistic quantum 
mechanics. 

In a relativistic setting, one always has a representation of the 
Poincare algebra. From the generators of the Poincare algebra 
(namely the 4-momentum p, the angular momentum \J, and the 
boost generators \K) one can make up (in massive representations) 
a nonlinear expression for a 3-dimensional \x (the position operator) 
that together with the space part \p of the 4-momentum has canonical 
commutation rules and hence gives a Heisenberg algebra. 
(The backslash is a convenient ascii notation to indicate bold face 
letters, corresponding to 3-vectors.)

The position operator so constructed is unique, once the time coordinate
is fixed, and is usually called the Newton-Wigner position operator,
although it appears already in earlier work of Pryce. Relevant
applications are related to the names Foldy and Wuythousen
(for their transform of the Dirac equation, widely used in relativistic
quantum chemistry) and Bakamjian and Thomas (for their relativistic
multi-particle theories); both groups rediscovered the Newton-Wigner 
results independently, not being aware of their work.

That the time coordinate has to be fixed means that the position 
operator is observer-dependent. Each observer splits space-time 
into its personal time (in direction of its total 4-momentum) and 
personal 3-space (orthogonal to it), and the position operator 
relates to this 3-space. By a Lorentz transformation, one can 
transform the 4-momentum to the vector (E_obs 0 0 0), which makes time
the 0-component. Most papers on the subject work in the latter setting.

For massless representations of spin >1/2, the construction breaks down.
This is related to the fact that massless particles with spin >1/2 
don't have modes of all helicities allowed by the spin 
(e.g., photons have spin 1 but no longitudinal modes), 
which makes them being always spread out, and hence not completely 
localizable. For details, see the FAQ entry 
    ''Particle positions and the position operator''


Here are a few references:

J.P. Costella and B.H.J. McKellar,
The Foldy-Wouthuysen transformation,
arXiv:hep-ph/9503416
* This paper discusses the physical relevance of the Newton-Wigner
representation, and its relation to the Foldy-Wouthuysen transformation

T. D. Newton, E. P. Wigner,
Localized States for Elementary Systems,
Rev. Mod. Phys. 21 (1949) 400-406
* The original paper on localization

L. L. Foldy and S. A. Wouthuysen,
On the Dirac Theory of Spin 1/2 Particles and Its Non-Relativistic 
Limit, 
Phys. Rev. 78 (1950), 29-36.
* On the transform of the Dirac equation now carrying the author's name

B. Bakamjian and L. H. Thomas
Relativistic Particle Dynamics. II
Phys. Rev.  92 (1953), 1300-1310.
and related papers in
Phys. Rev. 85 (1952), 868-872.
Phys. Rev. 121 (1961), 1849-1851. 
* First constructive papers on relativistic multiparticle dynamics,
based on a 3D position operator

L. L. Foldy,
Synthesis of Covariant Particle Equations,
Phys. Rev. 102 (1956), 568-581
* A lucid exposition of Poincare representations which start with
a 3D position operator, and a discussion of electron localization 
Before eq. (189), he notes that an observer-independent localization 
of a Dirac electron (which generally is considered to be a pointlike
particle since it can be exactly localized in a given frame) 
necessarily leaves a fuzziness of the order of the Compton wavelength 
of the particle. (This is also related to the so-called Zitterbewegung,
see, e.g., the discussion in Chapter 7 of Paul Strange's 
"Relativistic Quantum Mechanics".)

A. S. Wightman,
On the Localizability of Quantum Mechanical Systems,
Rev. Mod. Phys. 34 (1962) 845-872
* A group theoretic view in terms of systems of imprimitiviy

T. O. Philips,
Lorentz invariant localized states,
Phys. Rev. 136 (1964), B893-B896.
* A covariant coherent state alternative which does not require
to single out a time coordinate

V. S. Varadarajan,
Geometry of Quantum Theory
(second edition), Springer, 1985
* A book discussing some of this stuff

L. Mandel and E. Wolf,
Optical Coherence and Quantum Optics,
Cambridge University Press, 1995.
* The bible on quantum optics, a thick but very useful book. 
Relevant here since it contains a good discussion of the 
localizability of photons (which can be done only approximately, 
in view of the above) from a reasonably practical point of view. 

G.N. Fleming,
Reeh-Schlieder meets Newton-Wigner
http://philsci-archive.pitt.edu/archive/00000649/
* This paper gives some relations to quantum field theory


\end{verbatim}

\section{$SO(3)=SU(2)/\Zz_2$}\label{su2.app}

In this appendix we wish to show that $SO(3)SU(2)/\Zz_2$. First we collect some basics on $SO(3)$ and $SU(2)$.

A real $3\times 3$ matrix $R$ is called special orthogonal if
\[
R^TR=1\,,~~~\det R=1\,.
\]
Note that $1$ here denotes the $3\times 3$ identity matrix in the first equation.
It is easy to check that the special orthogonal matrices form a group; we denote this group by
$SO(3)$ and call it the special orthogonal group, or the rotation
group. An element of $SO(3)$ is also called a rotation.

If $\lambda$ is an eigenvalue of $R\in SO(3)$ we see that
$\lambda=\pm 1$. We want to show that there is always an eigenvector
with eigenvalue $1$. The characteristic polynomial of $R$ has three
roots $\lambda_1$, $\lambda_2$ and $\lambda_3$. The modulus of the
roots has to be $1$ and if there is a imaginary eigenvalue $\mu$, then
so is its conjugate $\bar\mu$ an eigenvalue. If all the three
eigenvalues are real, then the only possibilities are that all three
are $1$ or that two are $-1$ and the third is $1$. Let now $\lambda_1$
be imaginary and take $\lambda_2=\bar\lambda_1$. Then $\lambda_3 =
\frac{1}{\lambda_1\bar\lambda_1}$ is real and positive and since it
has to be of unit modulus $\lambda_3=1$. We see that there is always
an eigenvalue $1$. If $R$ is a rotation and not the identity there is
just one eigenvector with eigenvalue one; we denote this eigenvector
by $e_R$. We thus have $Re_R=e_R$ and if $R\neq 1$ then $e_R $ is
unique. The vector $e_R$ determines a one-dimensional subspace of
$\Rz$ that is left invariant under the action of $R$. We call this
one-dimensional invariant subspace the axis of rotation. 

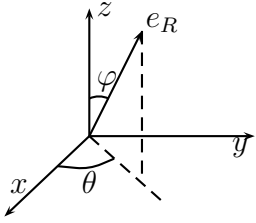
\begin{figure}
\begin{centering}
\caption{Rotation around $e_R$. Any axis is characterized by two angles $\varphi$ and $\theta$.}\label{rotation}
\begin{pspicture}(0,1)(4,4)
\psline{->}(1.5,2)(3.7,2)
\psline{->}(1.5,2)(0.37,0.93)
\psline{->}(1.5,2)(1.5,3.7)
\put(3.4,1.8){$y$}\put(0.45,1.25){$x$}\put(1.6,3.6){$z$}%
\psline{->}(1.5,2)(2.2,3.4)
\put(2.25,3.45){$e_R$}
\psline[linestyle=dashed](2.2,3.4)(2.2,1.5)\psline[linestyle=dashed](1.5,2)(2.45,1.15)
\put(1.4,1.25){$\theta$}
\psline[linearc=0.7](1.0822,1.6)(1.5,1.56)(1.83,1.7)
\psline[linearc=0.3](1.5,2.5)(1.63,2.56)(1.755,2.5)
\put(1.6,2.68){$\varphi$}
\end{pspicture}
\end{centering}
\end{figure}

Consider an arbitrary $SO(3)$ element $R$
with axis of rotation determined by $e_R$ over an angle $\psi$ and
denote the rotation by $R(e_R,\psi)$. Call the angle between the plane
in which $e_R$ and the $z$-axis lie and the plane in $xz$-plane
$\theta$. Call the angle between $e_R$ and the $z$-axis $\varphi$. See
figure \ref{rotation}.
The rotation can now be broken down into three rotations. First we use
two rotations two go to a coordinate system with coordinates $x'$,
$y'$ and $z'$ in which the $e_R$ points in the $z'$-direction, and
then we rotate around the $z'$-axis over an angle $\psi$. The two
rotation to go to the new coordinate system are: (a) a rotation around
the $z$-axis around an angle $\theta$ to align the $x$-axis with the
projection of $e_R$ onto the $xy$-plane, (b) a rotation over an angle
$\varphi$ around the image of the $y$-axis under the first
rotation. Hence we can write $R(e_R,\psi)$ as a product 
\[
R(e_R,\psi) = \pmatrix{ \cos \theta & -\sin\theta & 0 \cr
  \sin\theta &\cos\theta &0 \cr
  0&0&1 }\pmatrix{\cos\varphi & 0 & \sin\varphi\cr 0 &  1& 0 \cr
  -\sin\varphi & 0 &\cos\varphi }\pmatrix{
  \cos\psi & -\sin\psi & 0 \cr \sin\psi &\cos\psi & 0 \cr
  0&0&1}\,. 
\]
In this way we obtain a system of coordinates on the manifold
$SO(3)$. The three angles are then called the Euler angles. 
 
We note in particular the following: The group $SO(3)$ is generated by
all elementary rotations $R_x(\alpha)$, $R_y(\beta)$ and $R_z(\gamma)$
given by 
\begin{eqnarray}
R_x(\alpha) &=& \pmatrix{ 1 & 0 & 0 \cr 0&\cos\alpha &-\sin\alpha
  \cr 0& \sin\alpha&\cos\alpha }\nn\,,~~~ 
R_y(\beta) =  \pmatrix{ \cos\beta & 0 & \sin\beta\cr 0 & 1& 0 \cr
  -\sin\beta & 0 &\cos\beta }\nn\,,\\ 
R_z(\gamma) &=& \pmatrix{ \cos \gamma & -\sin\gamma & 0 \cr
  \sin\gamma &\cos\gamma &0 \cr 0&0&1 }\nn\,. 
\end{eqnarray} 

A $2\times 2$ complex matrix $U$ is called special unitary if it
satisfies
\[
U^\dagger U=1\,,~~~\det U=1\,.
\] 
It is easily checked that the special unitary matrices form a group,
which is called the special unitary group and is denoted $SU(2)$.

We now wish to show that $SU(2)$ is a real manifold that is isomorphic
to the three sphere $S^3$. We do this by finding an explicit
parametrization of $SU(3)$ in terms of two complex numbers $x$ and $y$
satisfying $|x|^2+|y|^2 = 1$. If one splits up $x$ and $y$ in a real
and imaginary parts, one sees that $x$ and $y$ define a point on
$S^3$. 

We write an element $U\in SU(2)$ as
\[ 
U = \pmatrix{ a & b \cr c& d }\,.
\]
Writing out the equation $U^\dagger U=1$ and $\det U=1$ one finds the following equations:
\begin{eqnarray}
&&|a|^2 + |c|^2 =1 \,, ~~~~  |b|^2 + |d|^2=1\,,\nn\\
&& \bar a b + \bar c d =0\,,~~~~~~~ ad-bc=1\,.\nn
\end{eqnarray}
 
 We first assume $b=0$ and find then that $ad=1$ and $\bar cd=0$, implying that $c=0$ and $U$ is diagonal with $a=\bar d$.
 Next we suppose $b\neq 0$ and use $a = -\frac{c\bar d}{\bar b}$ to
 deduce that $|b|=|c|$ and $|a|=|d|$; we thus have $b\neq 0
 \Leftrightarrow c\neq 0$. We also see that we can use the ansatz 
 \begin{eqnarray}
 a &=& e^{i\alpha} \cos \theta\,,~~~~~ b =e^{i\beta}\sin\theta \,,\nn\\
 c&=& - e^{i\gamma}\sin\theta\,,~~~ d= e^{i\delta}\cos\theta\,.\nn
 \end{eqnarray}
 Using again $a = -\frac{c\bar d}{\bar b}$ we see $\alpha +\delta=
 \beta+\gamma $ and writing out $ad-bc = 1$ we find $\alpha = -\delta$
 and $\beta = -\gamma$. We thus see $a=\bar d$ and $b= -\bar c$. Hence
 the most general element of $SU(2)$ can be written as  
 \[
 U(x,y)= \pmatrix{ x & y \cr -\bar y & \bar x }\,, ~~~{\rm with }~~ |x|^2+ |y|^2 = 1\,.
 \]
The map $S^3\to SU(2)$ given by $(x,y)\mapsto U(x,y)$ is clearly
injective, and from the above analysis bijective. Furthermore the map
is smooth. Hence we conclude that $SU(2)\cong S^3$ as a real
manifold. 

We introduce the Pauli matrices\footnote{Actually they are rescaled
  versions of the Pauli-matrices}
\[
\sigma^1 = \pmatrix{ 0& 1\cr1 & 0 }\,,~~~\sigma^2 = \pmatrix{ 0& -i\cr
  i& 0 }\,,~~~\sigma^3 = \pmatrix{ 1 & 0\cr 0& -1 }\,.
\]
Note that the Pauli-matrices are precisely all the traceless Hermitian
$2\times 2$ complex matrices and make up a three-dimensional vector
space. Therefore they provide a realization of the Lie algebra
$su(2)$. It is easy to check that the Pauli-matrices satisfy the
relations 
\[
\sigma^i\sigma^j = \delta^{ij} + i\epsilon^{ijk}\sigma^k\,,
\]
where we also used the Levi--Civit\'a symbol $\epsilon^{ijk}$; if
$(ijk)$ is not a permutation of $(123)$  $\epsilon^{ijk}$ is zero and
if $(ijk)$ is a permutation of $(123)$ then $\epsilon^{ijk}$ is the
sign of the permutation. In particular we note the commutator
relations
\[
[\sigma^i,\sigma^j] = \sigma^i\sigma^j - \sigma^j\sigma^i = 2i\epsilon^{ijk}\sigma^k\,,
\]
which resembles the vector product in $\Rz^3$. We also note the identities
\[
\tr (\sigma^i)=0\,,~~ \tr(\sigma^i\sigma^j) = 2\delta^{ij}\,,~~ \tr([\sigma^i,\sigma^j]\sigma^k) = 4i\epsilon^{ijk}\,.
\]

For every vector $\vec x \in\Rz$ we identify an element $x$ of $su(2)$ as follows
\[
\vec x = \pmatrix{x_1 \cr x_2 \cr x_3 } \leftrightarrow x = \sum_{i=1}^{3}x_i\sigma^i\,.
\]
From now on we simply identify the elements with each other and thus write equality signs instead of arrows.
We see that $\vec x\times \vec y$ corresponds to the element $\frac{1}{2i}[x,y]$;
\[
\vec x \times \vec y = \frac{1}{2i}[x,y]\,.
\]
And similarly we find
\[
\vec x \cdot \vec y = \frac{1}{2}\tr (xy)\,,~~~ \vec x\times \vec y
\cdot \vec z = \frac{1}{4i}\tr ([x,y]z)\,.
\]

If $U$ is an element of $SU(2)$ and $x\in su(2)\cong \Rz^3$ we see that 
\[
\tr (UxU^{-1}) = 0\,, ~~~(UxU^{-1})^\dagger = (U^{-1})^\dagger x U^\dagger = UxU^{-1}\,,
\] 
and we conclude that $U$ induces a map $\Rz^3\to \Rz^3$. In fact, $U$ defines a special orthogonal transformation since
\[
\vec x\cdot \vec y \mapsto \tr(UxU^{-1}UyU^{-1}) = \vec x\cdot \vec y
\]
is invariant, so that $U$ preserves the inner product on $\Rz^3$ and
similarly $\vec x\times \vec y \cdot \vec z$ is invariant so that the
action of $U$ preserves the orientation. We thus found a map $R:
SU(2)\to SO(3)$, whereby $U\in SU(2)$ gets mapped to the element
$R(U)$ in $SO(3)$ corresponding to $x\mapsto UxU^{-1}$. Since
$U_1(U_2x U_{2}^{-1})U_{1}^{-1} = (U_1U_2)x(U_1U_2)^{-1}$ the map
$SU(2)\to SO(3)$ is a group homomorphism; $R(U_1U_2) = R(U_1)R(U_2)$. 

Explicitly we find
\begin{eqnarray}
U(x,y)\sigma^1 U(x,y)^{-1} &=& {\rm Re}(x^2 - y^2)\sigma^1 - {\rm Im}(x^2-y^2)\sigma^2 + 2{\rm Re}(x\bar y)\sigma^3 \,,\nn\\
U(x,y)\sigma^2 U(x,y)^{-1} &=& {\rm Im}(x^2+y^2)\sigma^1+{\rm Re}(x^2+y^2)\sigma^2+2{\rm Im}(x\bar y)\sigma^3\,,\nn\\
U(x,y)\sigma^3 U(x,y)^{-1} &=&  -2{\rm Re}(xy)\sigma^1+2{\rm Im}(xy)\sigma^2+ (|x|^2 - |y|^2)\sigma^3\,.\nn
\end{eqnarray}
Therefore
\[
R(U(x,y)) = \pmatrix{ {\rm Re}(x^2 -y^2) & {\rm Im}(x^2+y^2) & -2{\rm Re}(xy) \cr -{\rm Im}(x^2-y^2) & {\rm Re}(x^2+y^2) & 2{\rm Im}(xy) \cr 2{\rm Re}(x\bar y) & 2{\rm Im}(x\bar y) & |x|^2-|y|^2 }\,.
\]
We find
\begin{eqnarray}
R(U(\cos \alpha/2,-i\sin\alpha/2)) &=& \pmatrix{ 1&0&0\cr 0 &
  \cos\alpha & -\sin\alpha \cr 0 & \sin\alpha &\cos\alpha
} = R_x(\alpha) \,,\nn\\ 
R(U(\cos\beta/2,-\sin\beta/2))&=& \pmatrix{ \cos\beta & 0 & \sin\beta
  \cr 0 &1&0 \cr -\sin\beta&0& \cos\beta } = R_y(\beta) \,,\nn\\ 
R(U(e^{-i\gamma/2},0)) &=& \pmatrix{ \cos\gamma & -\sin\gamma & 0 \cr
  \sin\gamma & \cos\gamma & 0 \cr 0 &0&1 } = R_z(\gamma)\,,\nn
\end{eqnarray}
and hence the map $R:SU(2)\to SO(3)$ is surjective. Suppose now that
$U(x,y)$ is mapped to the identity element in $SO(3)$. We see then
that $x\bar y=0$, so that either $x=0$ or $y=0$. Since also $|x|^2 -
|y|^2 = 1$, we cannot have $x=0$ and hence $y=0$. Furthermore from $
{\rm Re}x^2 = |x|^2 = 1$ we see $x=\pm 1$: indeed we see that
$R(U(x,y)) = R(U(-x,-y))$. The kernel of $R:SU(2)\to SO(3)$ is thus
given by $\pm 1$ times the identity matrix, which is the
$\Zz_2$-subgroup of $SU(2)$\footnote{Easy exercize: Prove that there
  is only one $\Zz_2$-subgroup in $SU(2)$.}. As any kernel of group homomorphisms, the
kernel is a normal subgroup. All in all we have shown 
\[
SU(2) \cong SO(3)/\Zz_2\,.
\]

\chapter{Classical oscillating systems}\label{c.oscillating}

In this chapter, we discuss in detail an important family of 
classical physical systems: harmonic or 
anharmonic oscillators, and their multivariate generalization, 
which describe systems of coupled oscillators such as macromolecules or
planetary systems.

Understanding classical oscillators is of great importance in
understanding many other physical systems. The reason is that
an arbitrary classical system behaves close to equilibrium like a
system of coupled linear oscillators. The equations we deduce are
therefore approximately valid in many other systems. For example,
a nearly rigid mechanical structure such as a high-rise building 
always remains close enough to equilibrium so that it can be
approximately treated as a linear oscillating system for the
elements into which it is decomposed for computational purposes via
the finite element method.

\bigskip
We shall see that the equations of motion of coupled oscillators can
 be cast in a form that suggest a Lie algebra
structure behind the formalism. This will provide the connection to
Part III of the book, where Lie algebras are in the center of our 
attention.

Besides the (an-)harmonic oscillators we discuss some basic linear
partial differential equations of physics: the Maxwell equations
describing (among others) light and gamma rays,
the Schr\"odinger equation and the Klein--Gordon equation (describing 
alpha rays), and the Dirac equation (describing beta rays). 
The solutions of these equations can be represented in terms of 
infinitely many harmonic oscillators, whose quantization (not treated
in this book) leads to quantum field theory.

\section{Systems of damped oscillators}

For any quantity $x$ depending on time $t$, differentiation with 
respect to time is denoted by \idx{$\dot x$}:
\[
\dot x = \frac{d}{dt}x\,.
\]
Analogously, $n$ dots over a quantity represents differentiating this
quantity $n$ times with respect to time.

The {\bfi{configuration space}} is the space of possible positions
that a physical system may attain, including external constraints.
For the moment, we think of it as a subset in $\Rz^n$. A
point in configuration space is generally denoted $q$. For
example, for a system of $N$ point masses, $q$ is an $N$-tuple of
vectors $\q_k\in \Rz^3$ arranged below each other; each $\q_k$
denotes the spatial coordinates of the $k$th moving point (planet,
atom, node in a triangulation of the body of a car or building, etc.), 
so that
$n=3N$.

A {\bfi{system of damped oscillators}} is defined by the 
differential equation
\lbeq{damposc}
M\ddot q + C\dot q + \nabla V(q)=0\,.
\eeq
The reader wishing to see simple examples should turn to Section 
\ref{s.cao}; here we explain the contents of equation \gzit{damposc}
in general terms. As before, $q$
is the configuration space point $q\in \Rz^n$. The $M$ and $C$ are real
$n\times n$-matrices, called the {\bfi{mass matrix}} and the
{\bfi{friction matrix}}, respectively. The mass matrix is always 
symmetric and positive definite (and often diagonal, the diagonal 
entries being the masses of the components). The friction matrix 
need not be symmetric but is always positive semidefinite. 
The \bfi{potential} $V$ is a smooth function from $\Rz^n$ to
$\Rz$, i.e., $V\in C^{\infty}(\Rz^n,\Rz)$, and $\nabla V$ is the
gradient of $V$,
\[
\nabla V = \left(\frac{\partial V}{\partial q_1}, \ldots ,
\frac{\partial V}{\partial q_n}\right)^T \,.
\]
Here the \bfi{gradient operator} $\nabla$ is considered as a vector 
whose components are the differential operators $\partial/\partial q_k$.
In finite-element applications in structural mechanics, the mass matrix 
is created by the discretization procedure for a corresponding partial 
differential equation. In general, the mass may here be distributed 
by the discretization over all adjacent degrees of freedom. However,  
in many applications the mass matrix is diagonal;
\[
M_{ij} = m_i \delta_{ij}\,,
\]
where $m_i$ is the mass corresponding to the coordinate $q_i$, and 
\idx{$\delta_{ij}$} is the {\bfi{Kronecker symbol}} 
(or Kronecker delta), 
which is $1$ if $i=j$ and zero otherwise. In the
example where $\q_k$ is a three-vector denoting the position of an
object, then $i$ is a multi-index
$i=(k,j)$ where
$k$ denotes an object index and $j=1,2,3$ is the index of the coordinate
of the $k$th object which sits in position $i$ of the vector $q$.
Then $m_i$ is the mass of the $k$th object. 

The quantity $F$ defined by
\[
F(q):=-\nabla V(q)\,,
\]
is the {\bfi{force}} on the system at the point $q$ due to the {
\idx{potential}} $V(q)$. We define the {\bfi{velocity}} $v$ of the
oscillating
system by
\[
v:=\dot q\,.
\]
The {\bfi{Hamiltonian energy}} $H$ is then defined by
\lbeq{ham}
H := \half v^T M v + V(q)\,.
\eeq
The first term on the right-hand side is called the {\bfi{kinetic
energy}} since it depends solely on the velocity of the system. The
second term on the right-hand side is called the {\bfi{potential
energy}} and it depends on the position of the system. For more complex
systems the potential energy can also depend on the velocities.
Calculating the time-derivative of the Hamiltonian energy $H$ we
get
\lbeq{dissip} \dot H = v^T M\dot v + \nabla V (q) \cdot \dot q
=\dot q^T (M\ddot q +\nabla V(q)) =-\dot q^T C\dot q \leq 0\,,
\eeq
where the last equality follows from the differential equation
\gzit{damposc} and
the final inequality follows since $C$ is assumed to be positive
semidefinite. If $C=0$ (the idealized case of no friction) then
the Hamiltonian energy is constant, $\dot H=0$, and in this case
we speak of {\bfi{conservative dynamics}} (the Hamiltonian $H$ is
conserved). If $C$ is positive definite we have $\dot H<0$ unless
$\dot q=0$ and
there is energy loss. This is called {\bfi{dissipative dynamics}}.
In the dissipative case, the sum
of the kinetic and potential energy has to decrease.

If the potential
$V$ is unbounded from below, it might happen that the system starts
falling in a
direction in which the potential is unbounded from below and the system
becomes unphysical; the velocity could increase without limits. Thus,
in a realistic and manageable physical system, the potential is always
bounded from below, and we shall make this assumption throughout. It
follows that the Hamiltonian is bounded from below.

Since in the dissipative case the Hamiltonian energy is decreasing
and is bounded below,
it will approach a limit as $t\to \infty$. Therefore, $\dot H\to
0$, and by \gzit{dissip}, $\dot q^T C\dot q \to 0$. Since $C$ is
positive definite for a dissipative system, this forces $\dot q
\to 0$. Thus, the velocities will get smaller and smaller, and
asymptotically the system will approach the configuration of being
in a state with $\dot q =0$, at the level of the accuracy of the
model. Typically this
implies that $q$ tends to some constant value $q_0$. Note that it does
not follow rigorously that $q$ tends to a constant value; it is
possible that $q\to \infty$. Nevertheless we assume that $q$ does not
walk away to infinity and then it follows from $\dot q =0$ that
$\ddot q=0$, so that \gzit{damposc} implies $\nabla V(q_0)=0$, and
we conclude that $q$ tends to a stationary point $q_0$ of the
potential. If this is a saddle point, small perturbations can (and will)
cause the system to move towards another stationary point. Because
of such stability reasons, the system ultimately moves towards a
local minimum.

In practice, the perturbations come from imperfections in the model.
Remember that the deterministic equation \gzit{damposc} is a 
mathematical idealization of the real world situation. A more 
appropriate model (but still an approximation) is the equation 
\[
M\ddot q + C\dot q + \nabla V(q) = \eps\,,
\]
where $\eps$ is a stochastic force, describing the imperfections
of the model. Typically, these are already sufficient to guarantee with
probability 1 that the system will not end up in a saddle point.
Usually, imperfections are small, irregular jumps due to friction,
see, e.g., \sca{Bowden \& Leben} \cite{BowL}, or Brownian motion due
to kicks by molecules of a solvent. See, e.g., 
\sca{Brown} \cite{brown}, \sca{Einstein \& Brown} \cite{einsteinbrown}, 
\sca{Garcia \& Palacios} \cite{garciapalacios}, 
\sca{Hanggi \& Marchesoni} \cite{hanggi05}, for an
overview on Brownian motion with lots of historical references and
citations \sca{Duplantier} \cite{duplantier05}, and for a discussion 
in the context of
protein folding \sca{Neumaier} \cite[Section 4]{Neu.protein}.

In many cases, the potential $V(q)$ has several local
minima. Our argument so far says that the state of the system will
usually move towards one of these local minima. Around the local
minimum it can oscillate for a while, and in the absence of
stochastic forces it will ultimately get into one of the local
minima. If we assume that there are stochastic
imperfections, we can say even more!

Suppose that the local minimum towards which the system tends is not 
a global minimum. Then occasional stochastic perturbations may
suffice to push (or kick) the system over a barrier separating the
local minimum from a valley leading to a different minimum. Such a
barrier is characterized by a \bfi{saddle point}, a stationary point 
where the Hessian of the potential has exactly one negative eigenvalue.
The energy needed to pass the barrier, called the {\bfi{activation 
energy}}, is simply the difference between the potential energy of the 
separating saddle point and the potential energy of the minimum. 
In a simple, frequently
used approximation, the negative logarithm of the probability of
exceeding the activation energy in a given time span is
proportional to the activation energy. This implies that small
barriers are easy to cross, while high barriers are difficult to
cross. In particular, if a system can cross a barrier
between a high-lying minimum to a much lower lying minimum, it is
much more likely to cross it in the direction of the lower minimum
than in the other direction. This means that (averaged over a
population of many similar systems) most systems will spend most of
their time near low minima, and if the energy barriers between the
different minima are not too high, most systems will be most of
the time close to the global minimum. Thus a global minimum
characterizes an {\bfi{absolutely stable}} equilibrium, while other
local minima are only {\bfi{metastable}} equilibrium positions,
which can become unstable under sufficiently large stochastic
perturbations.

There are famous relations called {\bfi{fluctuation-dissipation
theorems}} that assert (in a quantitative way) that friction is
related to stochastic (i.e., not directly modeled high frequency)
interactions with the environment. In particular, if a system is
sufficiently well isolated, both friction and stochastic
forces become negligible, and the system can be described as
a conservative system. Of course, from a fundamental point of
view, the only truly isolated system is the universe as a whole,
since at least electromagnetic radiation escapes from all systems
not enclosed in an opaque container, and systems confined to a
container interact quite strongly with the walls of the container
(or else the wall would not be able to confine the system).

Thus on a fundamental level, a conservative system describes the
whole universe from the tiniest microscopic details to the largest
cosmological facts. Such a system would have to be described by a
quantum field theory that combines the successful standard model
of particle physics with general relativity. At present, no such
theory is available.

On the other hand, conservative systems form a good first
approximation to many small and practically relevant systems, which
justifies that most of the book looks at the conservative case only.
However, in Part \ref{p.noneq}, the dissipative case is in the center 
of the discussion.

\bigskip
\bfi{The phase space formulation.}
So far, our discussion was framed in terms of position and velocity.
As we shall see, the Hamiltonian description is most powerful in
phase space coordinates. Here everything is expressed in terms of
the phase space observables $q$ and $p$, where
\[
p:=M v\,,
\]
is called the {\bfi{momentum}} $p$ of the oscillating system.
The {\bfi{phase space}} for a system of oscillators is the space of
 points
$(q,p)\in \Rz^n\times \Rz^n$. A {\bfi{state}} (in the classical 
sense) is a point $(p,q)$ in phase space.  The {\bfi{Hamiltonian
function}}\index{Hamiltonian}
(or simply the \bfi{Hamiltonian}) is the function defining the
Hamiltonian energy in terms of
the phase space observables $p$ and $q$. In our case, since a positive
definite matrix is always invertible, we can express $v$ in terms of
$p$ as $v=M^{-1}p$, and find that
\lbeq{e.hamil}
H(p,q)= \half p^T M^{-1} p + V(q)\,.
\eeq
Note that $H$ does not depend explictly on time. (In this book, we
only treat such cases; but in problems with time-dependent external 
fields, an explicit time dependence would be unavoidable.)

\section{The classical anharmonic oscillator}\label{s.cao}

To keep things simple, we concentrate on the case of a single degree 
of freedom. Everything said has a corresponding generalization to 
systems of coupled oscillators, but the essentials are easier to see
in the simplest case. 

The {\bfi{simple anharmonic oscillator}}\index{oscillator!simple
anharmonic} is obtained by taking $n=1$. The
differential equation \gzit{damposc} reduces to a scalar equation
\lbeq{e.anharm_damped}
m\ddot q + c \dot q + V'(q) = 0\,,
\eeq
where the prime denotes differentiation with respect to $q$. This
describes for example the behavior of an object attached to a
spring; then $q$ is the length of the spring, $m=M$ is the mass of
the object, $c=C$ is the friction constant (collective of the air,
some friction in the spring itself, or of a surface if the object
is lying on a surface) and $V(q)$ describes the potential energy
(see below) the spring has when extended or contracted to length
$q$. Note that a constant shift in the potential does not alter the 
equations of motion of an anharmonic oscillator; hence the potential is 
determined only up to a constant shift.

 The {\bfi{harmonic oscillator}}\index{oscillator!harmonic} 
is the special case
of the anharmonic oscillator defined by a potential of the form
\[
V(q) = \frac{k}{2}(q-q_0)^2\,, \quad k>0\,,
\]
where $q_0$ is the equilibrium position of the spring. 
(Strictly speaking, only oscillators that are not harmonic should be
called anharmonic, but we follow the mathematical practice where
limiting cases are taken to be special cases of the generic concept:
A linear function is also nonlinear, and a real number is also 
complex.) In this case, the force becomes
\lbeq{hooke}
F(q)=-\nabla V(q) = -k(q-q_0)
\eeq
The equation \gzit{hooke} is sometimes called {\bfi{Hooke's law}},
which asserts that the force needed to pull a spring from
equilibrium is linear in the deviation $q-q_0$ from equilibrium, a
valid approximation when $q-q_0$ is small. Since the force is
minus the gradient of the potential, the potential has to be
quadratic to reproduce Hookes' law. It is customary to shift the
potential such that it vanishes in global equilibrium; then one gets
the above form, and stability of the equilibrium position dictates the
sign of $k$. Note that the shift does not change the force, hence has no
physical effect.

The {\bfi{mathematical pendulum}} is described by the equation
\lbeq{1dpend}
V(q)=- k(1-\cos q)\,, \quad k>0\,,
\eeq
where $q$ is now the angle of deviation from the equilibrium, 
measured in radians.
Looking at small $q$ we can approximate as follows:
\[
V(q)= \frac{k}{2}q^2+O(q^4)\,,
\]
and after dropping the error term, we end up with a harmonic oscillator.
The same argument allows one to approximate an arbitrary anharmonic 
oscillator by a harmonic oscillator as long as the oscillations around 
a stable equilibrium position are small enough.

For $q$ not small the mathematical pendulum is far from being harmonic.
Physically this is
clear; stretching a (good) spring further and further is harder and
harder, but pushing the one-dimensional pendulum `far' from its
equilibrium position is really different. After rotating it over
$\pi$ radians the pendulum is upside down and pushing it further no
longer costs energy. 

\bigskip
\bfi{Dynamics in phase space.} We now restrict to conservative
systems and analyze the conservative anharmonic oscillator ($\dot
H=0$) a bit more. Since $c=0$, the differential equation
\gzit{e.anharm_damped} simplifies to

\lbeq{e.anharm} m\ddot q + V'(q) = 0\,. \eeq

The Hamiltonian energy is given by
\[
H= \half mv^2 + V(q).
\]
Note the form of the kinetic energy familiar from school.
Expressed in terms of the phase space observables $p$ and $q$
(which are now scalar variables, not vectors), we have
\[
H = \frac{p^2}{2m}+V(q),
\]
\lbeq{eom}
\dot q = v = \frac{p}{m} \,,~~~
\dot p = m\dot v = m\ddot q = -V'(q)\,.
\eeq
An observable is something you can calculate
from the state; simple examples are the velocity and the kinetic,
potential, or Hamiltonian energy. Thus arbitrary observables can be
written as smooth functions $f(p,q)$ of the phase space observables.
In precise terms, an {\bfi{observable}} is (for an anharmonic 
oscillator)
a function $f\in C^\infty(\Rz\times \Rz)$.  The required amount of
smoothness can be reduced in practical applications; on the
fundamental theoretical level, it pays to require infinite
differentiability to get rid of troubling exceptions.

Introducing the shorthand notation
\[
f_p := \frac{\partial f}{\partial p}\,, 
\quad f_q:= \frac{\partial f}{\partial q}\,,
\]
for partial derivatives, we can write the equations \gzit{eom} in
the form
\lbeq{coordform}
\dot q = H_p\, , \quad \dot p= -H_q \,.
\eeq
The equations \gzit{coordform} are called the {\bfi{Hamilton
equations}} in \bfi{state form}. Although derived here only
for the anharmonic oscillator, the Hamilton equations are of great
generality; the equations of motions of many (unconstrained)
conservative physical systems can be cast in this form, with more
complicated objects in place of $p$ and $q$, and more complex
Hamiltonians $H(p,q)$. A dynamical system governed by the Hamilton
equations is called an {\bfi{isolated Hamiltonian
system}}\index{Hamiltonian system!isolated}. If there
are external forces, the system is not truly isolated, but the
Hamilton equations are still valid in many cases, provided one
allows the Hamiltonian to depend explicitly on time, $H=H(p,q,t)$;
in this case, there would appear additional partial derivatives with 
respect to time in various of our formulas.

Calculating the time-dependence of an arbitrary observable $f$ we get
\[
\dot f
=\frac{\partial f}{\partial p}\dot p+\frac{\partial f}{\partial q}\dot q
= f_p \dot p + f_q \dot q\,,
\]
hence
\lbeq{poisson}
\dot f = H_p f_q - H_q f_p \,.
\eeq
In particular for $f=q$ or $f=p$ we recover \gzit{coordform}.
Thus this formulation is equivalent to the Hamilton equations.
We call \gzit{poisson} the Hamilton equations in \bfi{general
form}\index{Hamilton equations!general form}.
Let us apply \gzit{poisson} to $f=H$ and calculate the change of the
Hamiltonian:
\[
\dot H = H_p H_q - H_q H_p =0\,,
\]
which is consistent since we knew from the start that energy is 
conserved, $\dot H = 0$.
But now this relation has been derived for arbitrary isolated
Hamiltonian systems!

When external forces
are present, we have to consider time-dependent observables $f(p,q,t)$.
In this case, we have in place of \gzit{poisson}
\lbeq{poisson.t}
\dot f
= f_p \dot p + f_q \dot q +f_t \dot t
= H_p f_q - H_q f_p +f_t\,,
\eeq
and in particular for the Hamiltonian,
\[
\dot H = H_t = \partial H/\partial t.
\]
However, we restrict the subsequent discussion to the isolated case.

The Hamiltonian equations can be cast in a form that turns out to
be even more general and very useful. It brings us directly to the
heart of the subject of the present book. We define a binary
operation $ \lp$ on $C^\infty(\Rz\times\Rz)$ as follows:
\[
f\lp g := f_p g_q - g_p f_q\,.
\]
Physicists write $\{g,f\}$ for $f\lp g$ and call it the
{\bfi{Poisson bracket}}. Our alternative notation will turn out 
to be very useful, and generalizes in many unexpected ways.
The equation \gzit{poisson} can then be written in form of a 
\bfi{classical \idx{Heisenberg equation}}
\lbeq{e.Heisc}
\dot f = H\lp f.
\eeq
It turns out that this equation, appropriately interpreted, is extremely
general. It covers virtually {\em all}
conservative systems of classical and quantum mechanics.

A basic and most remarkable fact, which we shall make precise in the
following chapter, is that the vector space $C^\infty(\Rz\times\Rz)$
equipped with the binary operation $\lp$ is a Lie algebra.
We shall take this up systematically in Section \ref{s.poisson}.

\section{Harmonic oscillators and linear field equations}
\label{sec-monochrom}

Historically, radiating substances which produce rays of $\alpha$-, 
$\beta$- or $\gamma$-particles were fundamental for gaining an 
understanding of the structure of matter.
Even today, many experiments in physics are performed by rays 
(or beams, which is essentially the same) generated by
some source and then manipulated in the experiments.

The oldest, most familiar rays are light rays, $\alpha$-rays, 
$\beta$-rays, and $\gamma$-rays. (Nowadays, we also have neutron rays,
etc., and cosmic rays contain all sorts of particles.)

All kinds of rays are described by certain quantum fields,
obtained by quantizing corresponding classical field equations, 
linear partial differential equations whose time-periodic solutions 
provide the possible \bfi{single-particle modes} of the quantum fields.
In the following sections we look at these field equations in some 
detail; here we just make some introductory comments.

$\alpha$-rays are modes (realizations) of the field of doubly ionized 
helium, $He^{++}$, which is modeled on the classical
level by a Schr\"odinger wave equation or a Klein--Gordon wave equation.
$\beta$-rays are modes of a charged field of electrons or positrons,
modeled on the classical level by a Dirac wave equation. For
radiation of only positrons one uses the notation $\beta^+$, and for
rays with only electrons one uses $\beta^-$. Both
light rays and $\gamma$-rays are modes of the 
electromagnetic field
which are modeled on the classical level by the Maxwell wave equations.
Their quantization (which we do not treat in this book) produces the 
corresponding quantum fields.
 
In the present context, the Schr\"odinger, Klein--Gordon, Dirac, and 
Maxwell equations are all regarded as classical field equations for
waves in $3+1$ 
dimensions, though they can also be regarded as the equations for a 
single quantum particle (a nonrelativistic or relativistic scalar
particle, an electron, or a photon, respectively). This dual use is 
responsible for calling \bfi{second quantization} the quantum field 
theory, the quantum version of the classical theory of these equations.
It also accounts for the \bfi{particle-wave duality}, the puzzling 
property that rays sometimes (e.g., in photodetection or a Geiger 
counter) behave like a beam of particles, sometimes (in diffraction
experiments, of which the double slit experiments are the most famous 
ones) like a wave -- in the case of light a century-old conflict dating 
back to the times of Newton and Huygens. 

In the quantum field setting, quantum particles arise as eigenstates 
of an operator $N$ called the \bfi{number operator}. This operator has 
a discrete spectrum with nonnegative integer eigenvalues, counting the
number of particles in an eigenstate. The ground state, with zero 
eigenvalue, is essentially unique, and defines the \bfi{vacuum};
a quantum particle has an eigenstate corresponding to the eigenvalue $1$
of $N$, and eigenstates with eigenvalue $n$ correspond to systems of $n$
particles. If a quantum system contains particles of different types, 
each particle type has its own number operator. 

The states that are easy to prepare correspond to beams. The fact that
beams have a fairly well-defined direction translates into the formal
fact that beams are approximate eigenstates of the momentum operator.
Indeed, often beams are well approximated by exact eigenstates of the 
momentum operator, which describe so-called \bfi{monochromatic beams}.
(Real beams are at best quasi-monochromatic, a term we shall not 
explain.) Since the states of beams are not eigenstates of the number 
operator $N$, they contain an indefinite number of particles.

\bigskip
All equations mentioned are linear partial differential 
equations, and behave just like a set of infinitely many coupled 
harmonic oscillators, one at each space position. They describe 
non-interacting fields in a homogeneous medium. 
The definition of interacting fields leaves the linear regime
and leads into the heart of nonlinear field theory, both in a classical 
and a quantum version. This is outside the scope of the present book.
However, when position space (or momentum space) is discretized so that 
only a finite number of degrees of freedon remain to describe a field,
one is back to nonlinear oscillators, which can be understood completely
on the basis of the treatment given here, and indeed, number operator
will play a prominent role in Part V of this book.

Fortunately, for understanding beam experiments, it usually suffices 
to quantize a 
few modes of the classical field, and these are harmonic oscillators. 
Indeed, by separation of variables, the linear field equations can be 
decoupled in time, leading to a system 
of uncoupled harmonic oscillators forming the Fourier modes. 
Beams correspond to solutions which have a significant intensity only 
in a small neighborhood of a line in 3-space. Frequently, beams 
correspond to solutions that have an (almost) constant frequency.
Interactions with such (quasi-)monochromatic beams can be modelled 
in many situations simply as interactions with a harmonic oscillator. 

On the other hand, when a beam containing all frequencies interacts 
with a system which oscillates only with certain frequencies, the
beam will resonate with these frequencies. This allows the detection of 
a system's eigenfrequencies by observing its interaction with 
light or other radiation. This is the basis of spectroscopy, and 
will be discussed in more detail in Chapter \ref{s.specanal} and
Chapter \ref{c.spec}.

\bigskip
Note that, just as the free Maxwell equations describe both 
classical electromagnetic waves (in particular light and $\gamma$-rays)
or single photons (particles of the corresponding quantum field), so
the Schr\"odinger equation, the Klein--Gordon equation, and the Dirac 
equation describe both classical fields for $\alpha$- and $\beta$-rays,
or single $\alpha$-particles or electrons and positrons (particles of 
the corresponding quantum field), respectively. 

In the following, we consider four kinds of classical fields and their
associated quantum particles, differing in spin and hence in the way 
rotations and Lorentz transformations affect the fields.
\begin{itemize}
\item Slow bosons of spin zero, such as slow $\alpha$-particles.
The equation describing them is the Schr\"odinger equation.
\item Fast bosons of spin zero, such as fast $\alpha$-particles.
The equation describing them is the Klein--Gordon equation.
\item Fermionic particles of spin $1/2$, like electrons,
positrons and neutrinos. The dynamical equation in this case is the
Dirac equation.
\item Light and $\gamma$-rays; electromagnetic radiation. The
corresponding particles are photons, which have spin 1. 
The field describing these particles is the electromagnetic field. 
The equations governing their dynamics are the Maxwell equations.
\end{itemize}

Because of the differing spin, there is a significant difference 
between $\beta$-rays and the others: $\alpha$-particles and photons 
have integral spin and are therefore so-called {\bfi{bosons}}, 
while electrons and positrons have non-integral spin and are therefore 
so-called {\bfi{fermions}}. Only fermions are subject to the 
so-called \bfi{Pauli exclusion principle} which is responsible for the
extensivity of matter. This difference is reflected by the fundamental
requirement that the fields of bosons, in particular of 
$\alpha$-particles and photons, are quantized by imposing \bfi{canonical
commutation relations} (discussed in Section \ref{s.qharm}), while 
fermions, and hence positrons and electrons, are quantized by imposing 
\bfi{canonical anticommutation relations} (discussed in Section
\ref{s.qanti}). \at{rays or particles?}

\section{Alpha rays}\label{s.alpha}

We first consider rays consisting of \bfi{$\alpha$-particles}, helium 
atoms stripped of their two electrons, and consist of two protons and 
two neutrons. 
$\alpha$-particles are released by other heavier nuclei during
certain processes in the nucleus. For example, some elements are
$\alpha$-radioactive, which means that a nucleus of type $A$ will want
to go to a lower energy level, which can then be done by emitting two
of its protons and two of its neutrons. The result is thus two nuclei,
a Helium nucleus and a nucleus of type $A'\neq A$; schematically $A\to
A'+\alpha$. But also during nuclear
splitting $\alpha$-particles are released. Yet even more, the sun is
emitting $\alpha$-particles all the time; the sun produces heat by
means of a chain of nuclear fusion reactions, during which some
$\alpha$-particles are produced. If the atmosphere would not be there,
life on earth would be impossible due to the bombardment of
$\alpha$-particles. That $\alpha$-particles are not healthy has been
in the news lately (in 2007), since the former Russian spy Litvinenko is
said to have been killed by a small amount of polonium, which is an
$\alpha$-emitter.

An $\alpha$-particle emitted from a radioactive nucleus typically has
a speed of 15,000 kilometers per second. Although this might look very
fast, it is only 5\% of the speed of light, which means that for 
a lot of calculations $\alpha$-particles can be considered {\bf
 nonrelativistically}, that is, without using special
relativity. For some more accurate calculations though, special
relativity is required. 

For the nonrelativistic $\alpha$-particle we have to use the
Schr\"odinger equation. For a particle of mass $m$ moving in a
potential $V(x)$ the Schr\"odinger equation is given by
\[
i\hbar\frac{\partial}{\partial t}\psi(x,t)=
-\frac{\hbar^2}{2m}\Nabla^2 \psi(x,t) + V(x)\psi(x,t)
\]
where $\psi$ is the wave function of the particle, and 
$\nabla^2=\nabla \cdot \nabla$ is the \bfi{Laplace operator}.
The wave function
contains the information about the particle.
The quantity $|\psi(x,t)|^2$
is the probability density for finding the particle at time $t$ 
in a given position $(x)$. For beam
considerations, we take $V(x)=0$. Since we shoot the
$\alpha$-particles just in one direction, we assume
$\psi(x,t)=\phi(t)\chi(x)$. We obtain
\lbeq{schrod.ch7}
\frac{i\dot\phi(t)}{\phi(t)} =
-\frac{\hbar}{2m}\frac{\chi''(x)}{\chi(x)}\,,
\eeq
where the dot denotes the derivative with respect to time $t$ and the 
prime ' the derivative with respect to the coordinate $x$. 
The left-hand side of
\gzit{schrod.ch7} only depends on time $t$ and the right-hand side
only on $x$, which implies that both sides are a constant (with the
dimensions of time$^{-1}$) independent of $t$ and $x$. 
We denote this constant by $\omega$ and obtain two linear 
ordinary differential equations for $\phi$ and $\chi$ with the solutions
\[
\phi(t)= e^{-i\omega t}\,,~~~ \chi(x) =
ae^{ikx}+b^*e^{-ikx}\,,~~~k=\sqrt{\frac{2m\omega }{\hbar}}\,,
\]
where $a$ and $b$ are some constants; we have normalized the constant in
front of $\phi$ to $1$ since we are only interested in the product of
$\phi$ and $\chi$. Note that we wrote the solution suggestively as
if $\omega\ge 0$ and in
fact, on physical grounds it is; the solutions with $\omega<0$ are not
integrable and hence cannot determine a probability distribution.

We can express $E$ in terms of $k$, which plays the role of the inverse 
wavelength, getting $\omega(k) = \frac{\hbar k^2}{2m}$. Reintroducing 
an arbitrary direction unit vector $\n$ and the wave vector $\k=k\n$,
we obtain the \bfi{dispersion relation} of the 
Schr\"odinger equation,
\[
\omega_\k =  \frac{\hbar \k^2}{2m}\,.
\]
Therefore the general solution can be expanded as
\[
\psi(x,t)= \int d\k\,\Big(a(\k)e^{-i\omega_\k t-i\k\dot\x}
+b(\k)^*e^{-i\omega_k t+i\k\dot\x}\Big)\,.
\]
If we have an experiment with a great number of non-interacting 
particles, all of which have the same wave function $\psi$, the 
quantity $|\psi(x,t)|^2$ is proportional to the particle density. 
However, $\alpha$-particles interact and thus
the Hamiltonian is different. If we assume the particle density is not
too high we can still assume that the $\alpha$-particles move as if
there were no other $\alpha$-particles. Under this assumption we may
again take $|\psi(x,t)|^2$ as the particle density. The energy
density is then proportional to $|\psi(x,t)|^2$. Putting as before
the whole experiment in a box of finite volume $V$ one can again
arrive at a Hamiltonian corresponding to a collection of independent 
harmonic oscillators.

Now we look at relativistic $\alpha$-particles, and remind the reader of
the notation introduced in Section \ref{s.lorentz}.
The dynamics of relativistic $\alpha$-particles of mass $m$
is described by a real-valued function $\psi(x^\mu)$ whose
evolution is
governed by the {\bfi{Klein--Gordon equation}}, which is given by
\lbeq{e.kleing}
\left(\Box - \frac{m^2c^2}{\hbar^2}\right)\psi = 0 
\eeq
with  the second order differential operator
\[
\Box:= -\frac{\partial^2}{c^2\partial t^2}
+\frac{\partial^2}{\partial x_1^2}+\frac{\partial^2}{\partial x_2^2}
+\frac{\partial^2}{\partial x_3^2}\,,
\]
called the {\bfi{d'Alembertian}}. Here $c$ is the speed of light. 
We look for wave like solutions $\psi\sim e^{ik\cdot x}$ for some
vector $\k^\mu$. Note that $\partial_\mu e^{ik\cdot x} = ik_{\mu}
e^{ik\cdot x}$ and hence $\Box  e^{ik\cdot x}= \mp k^2 e^{ik\cdot x}$,
where $k^2=k\cdot k$. Hence we obtain the condition on $k$
\lbeq{e.om}
\pm k^2 +\frac{m^2c^2}{\hbar^2} =0\,.
\eeq
Writing $k^0=\omega$ and denoting the spatial parts of $k$ with bold
$\mathbf{k}$ we thus get the \idx{dispersion relation}
\lbeq{dis.per.KG}
\omega = \pm \sqrt{c^2\k^2 + \frac{m^2c^4}{\hbar^2}}\,.
\eeq
We see that $\hbar|\omega| = E =  mc^2$, combining Einstein's famous 
formula $E=mc^2$ and Planck's law $E=\hbar \omega$. The solution for a 
given choice of sign of $\omega$ is expanded
in Fourier terms and most often written as
\[
\psi(x,t) = \int \frac{d^3\k}{(2\pi)^3 2\omega(k)}\left(
  a(\k)e^{i\omega(k)t-i\k\cdot \x}+
  a(\k)^*e^{-i\omega(k)t+i\k\cdot \x}\right)\,,
\]
where we used the Lorentz-invariant measure 
\gzit{e.invmeas} involving 
$\omega(k)=\sqrt{c^2\k^2 + \frac{m^2c^4}{\hbar^2}}$. 

\section{Beta rays}\label{s.beta}

We now discuss beams composed of spin $\half$ particles, the
$\beta$-rays. $\beta$-radiation is emitted by radioactive
material. Unstable nuclei can lose some of their energy and go to a
more stable nucleus under the emission of $\beta$-rays. There are
two kinds of $\beta$-rays -- those with positive charge and those with 
negative charge. 
The negatively charged version consists of nothing more
than \bfi{electrons}. The positively charged counterpart consists of the
antiparticles of the electrons, the so-called \bfi{positrons}.

Other examples of fermions are neutrinos. The sun emits a stream of 
neutrinos; in each second, there are
approximately $10^{13}$ neutrinos flying through your body (it
depends on which latitude you are, how big you are and whether you are
standing or lying down, making a difference of a factor of $100$
perhaps). Neutrinos fly very fast; the solar neutrinos travel at
the speed of light (or very close).  The reason they travel that fast is
that neutrinos have a zero or very tiny mass, and massless particles
(such as photons) always travel at the speed of light. For a long time,
neutrinos were believed to be massless; only recently it became an
established fact that at least one of the three generations of 
neutrinos must have a tiny positive mass. We do not feel anything 
of the many neutrinos coming from the sun and steadily passing through 
our body, because -- unlike protons and electrons -- they hardly
interact with matter; for example, to absorb half of the solar 
neutrinos, one would need a solid lead wall of around $10^{16}$ 
meters thick! The reason is that they do not have charge: 
they are electrically neutral.  

To discuss the case that the particles shot by the beam are fermions,
we have to use the Dirac equation. It is convenient to use the same
conventions for dealing with relativistic particles. In addition to
the previously introduced symbols, we now introduce the so-called
\idx{$\gamma$-matrices}. In four dimensions there are four of them,
called $\gamma^0,\ldots,\gamma^3$, and they satisfy
\lbeq{clifford}
\gamma^\mu\gamma^\nu+\gamma^\nu\gamma^\mu = 2\eta^{\mu\nu}\,.
\eeq
The associative algebra generated by the $\gamma$-matrices subject to
the above relation is a {\bfi{Clifford algebra}}. There are several
possibilities to find a representation for the $\gamma$-matrices in
terms of $4\times 4$-matrices; a frequently made choice is
\[
\gamma^0 = i\sigma_1\otimes 1\,,~~~\gamma^1=\sigma_2\otimes 1
\]
\[
\gamma^2 = \sigma^3\otimes \sigma^1\,,~~~\gamma^3= \sigma^3\otimes
\sigma^2\,,
\]
where the $\sigma_k$ are the Pauli matrices \gzit{e.pauli}.
however, we only need the defining relation \gzit{clifford}. 
We assemble the
$\gamma$-matrices in a vector $(\gamma^0,\gamma^1,\gamma^2,\gamma^3)$
and inner products with vectors $p^\mu$ are given by $\gamma\cdot p =
\gamma^\mu p_\mu = \gamma^\mu p^\nu \eta_{\mu\nu}$.

A fermion is described by a vector-like object $\psi$, which takes
values in the spinor representation of the Lie algebra
$so(3,1)$. Hence we can think of $\psi$
as a vector with
four-components. In this case, the $\gamma$-matrices are $4\times
4$-matrices; that such a representation exists is shown by the
explicit construction above. We need a property of the
$\gamma$-matrices, namely that they are traceless (in any
representation). To prove this, take any $\gamma^\mu$ and choose
another $\gamma$-matrix $\gamma^\nu$, $\mu\neq \nu$. Then we have
\[
\tr \gamma^\mu = \tr \Bigl( \gamma^\mu \gamma^\nu
(\gamma^\nu)^{-1}\Bigr)= -\tr \Bigl( \gamma^\nu \gamma^\mu
(\gamma^{\nu})^{-1}\Bigr)=-\tr \gamma^\mu \,.
\]
Hence $\tr\gamma^\mu=0$.

With these
preliminaries the Dirac equation is given by
\[
\left(\gamma\cdot \partial + \frac{mc}{\hbar}\right)\psi =0\,.
\]
Acting on the Dirac equation with $(\gamma\cdot \partial -
\frac{mc}{\hbar})$ and using $\gamma\cdot p \gamma\cdot p = p^2$ for
any four-vector $p^\mu$, we see that each component of the spinor 
obeys the Klein--Gordon equation \gzit{e.kleing}.

We look for solutions of the form $\psi = u e^{ik\cdot x}$. Putting
this ansatz in the Dirac equation we obtain
\lbeq{dirac.2}
\left(i\gamma\cdot k - \frac{mc}{\hbar}\right) u =0\,,
\eeq
and the additional constraint $k^2+\frac{m^2c^2}{\hbar^2}=0$ follows 
from the Klein--Gordon equation. Equation \gzit{dirac.2} can be written 
as
\[
\left(1 - i\frac{\hbar}{mc}\gamma\cdot k\right)u=0\,,
\]
and it is easy to see that
\[
P=\frac{1}{2}\left(1 - i\frac{\hbar}{mc}\gamma\cdot k\right)
\]
satisfies $P^2=P$. Hence $P$ is a projection operator and $\Cz^4$
splits as $V\oplus V'$ with $V=P\Cz^4$ and $V'=(1-P)\Cz^4$. The Dirac
equation thus tells us that $u$ has to be in $V'$. Denote $p^\mu=
\frac{mc}{\hbar}k^\mu$, then $p^2=-1$. We now choose a frame
moving along with the particle, so that in that frame the particle
is not moving, hence we may choose $p^\mu=(1,0,0,0)$ in the chosen
frame. It follows that $2P=1+i\gamma^0$. The eigenvalues of
$i\gamma^0$ are $\pm 1$ since $(i\gamma^0)^2=1$. But
the $\gamma$-matrices are traceless, and hence the
eigenvalues add up to $0$. Therefore the eigenvalues of
$i\gamma^0$ are $-1,-1,+1,+1$. Thus $P$ can be cast in the form
\[
P = \pmatrix{ 1 & 0 & 0 & 0 \cr 0 & 1 & 0 & 0 \cr 0&0&0&0\cr
0&0&0&0 }\,.
\]
We conclude that there are two independent degrees of freedom for
a fermion; similar to the case of light one speaks of two
polarizations. For a particular choice of the sign of $\omega_k$ we can
thus specialise the expansion of the fermion
to
\[
\psi = \int ~ \frac{d^3k}{(2\pi)^3} ~\left(v_+(k) e^{i\omega_k
t-\mathbf{k}\cdot x}+v_-(k) e^{-i\omega_k t+\mathbf{k}\cdot
x}\right)\,,
\]
where $\omega_k= \sqrt{\bfi{k}\cdot\bfi{k}+m^2c^2/\hbar^2}$
and where the $v_\pm(k)$ are linear combinations of the two basis
polarization vectors $u_1$ and $u_2$:
\[
v_\pm(k) = \alpha_\pm(k) u_1 + \beta_{\pm}(k)u_2\,.
\]

\section{Light rays and gamma rays}\label{s.gamma}

Lasers produce light of a high intensity and with almost only one
frequency. That is, the light of a laser is almost {\bf
\idx{monochromatic}}. We assume that the laser is perfect and thus emits
only radiation of one particular wavelength. Also we consider `general
lasers', which can radiate electrons, $\alpha$-particles,
$\beta^\pm$-radiation and so on. We shortly comment on the nature of
the different kinds of radiation and see how the modes come into
play. To make life easy for us, we imagine the laser is placed such
that the medium through which the beam is shot, is the vacuum.

\begin{figure}
\begin{centering}
\includegraphics[scale=0.7]{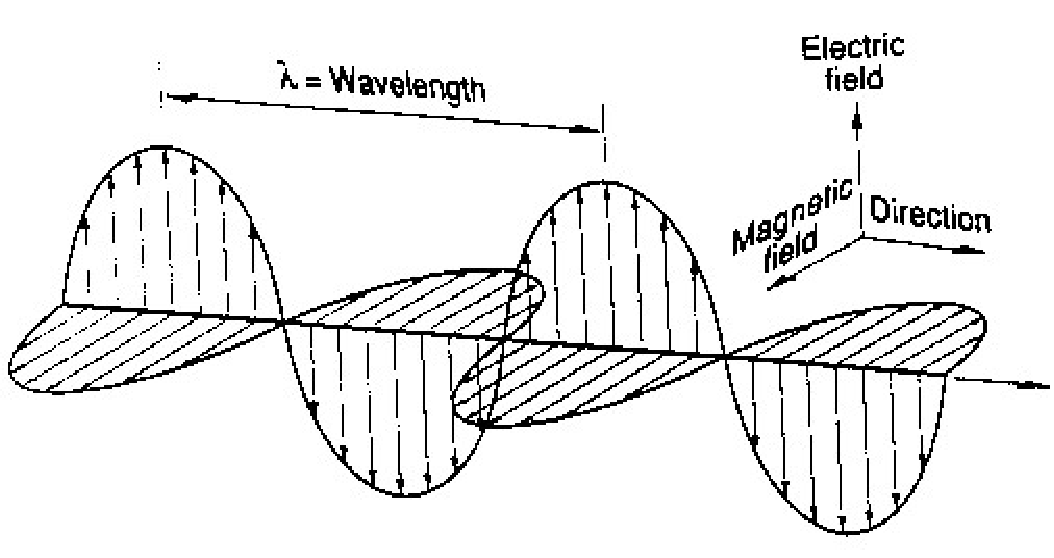}
\caption{An image of a solution of the electromagnetic wave
  equations. The Poynting vector gives the direction of the wave and
  is perpendicular to the electric and the magnetic field.}
\label{e_mag.pic}
\end{centering}
\end{figure}

First we consider the common situation where light is radiated.
Light waves
are particular solutions to the {\bfi{Maxwell equations}} in 
vacuum, or any other homogeneous medium. The Maxwell equations in 
vacuum are given by
\beqar
&&\Nabla\cdot \E = 0\,, ~~~ \Nabla\times\E = -\frac{\partial
  \B}{\partial t}\,,\nonumber\\
&& \Nabla\cdot \B =0\,, ~~~\Nabla\times\B =
\frac{1}{c^2}\frac{\partial \E}{\partial t}\,,\nonumber
\eeqar
where $\E$ is the electric field strength, $\B$ is the magnetic field
strength and $t$ is the time, and $c$ is again the speed of light.
As usual in physics, boldface symbols 
denote 3-dimensional vectors, while their components are not written 
in bold; 
\[
\nabla \cdot \A = \div \A := \frac{\partial A_1}{\partial x_1} + 
\frac{\partial A_2}{\partial x_2} + \frac{\partial A_3}{\partial x_3}  
\]
and
\[
\nabla \times \A = \curl \A: =\pmatrix{
\D\frac{\partial A_3}{\partial x_2}-\frac{\partial A_2}{\partial x_3}\cr
\D\frac{\partial A_1}{\partial x_3}-\frac{\partial A_3}{\partial x_1}\cr
\D\frac{\partial A_2}{\partial x_1}-\frac{\partial A_1}{\partial x_2}
}
\] 
denote the \bfi{divergence} and \bfi{curl} of a vector field $A$,
respectively. Using the generally valid relation 
\[
\Nabla\times(\Nabla\times \X)= \Nabla(\Nabla\cdot \X) - \Nabla^2 \X
\]
and the fact that the divergence of $\B$ and $\E$ vanishes we obtain 
from the Maxwell equations the wave equations
\[
\Nabla^2\E = \frac{1}{c^2}\frac{\partial^2}{\partial
  t^2}\E,,~~~\Nabla^2\B = \frac{1}{c^2}\frac{\partial^2}{\partial
  t^2}\B\,.
\]
To solve we use the ansatz
\[
\E(x,t)=\E e^{i\omega t-i\k\cdot \x}\,,~~~ \B(x,t)=\B
e^{i\omega t-i\k\cdot \x}\,,
\]
where $\E$ and $\B$ are now fixed vectors. The ansatz represents waves
propagating in the $\k$-direction and at any fixed point in space the
measured frequency is $\omega$. From the wave equations
we immediately find the {\bfi{dispersion relations}} for the 
Maxwell equations that relate $\omega$ and $\k=(k_x,k_y,k_z)^T$;
\[
\omega = c|\k|,
\]
where 
\[
|\k|:= \k\cdot \k = k_{x}^{2}+k_{y}^{2}+k_{z}^{2}.
\]
We compute
\[
\Nabla\cdot \E(x,t) = 0 ~\Rightarrow ~ \k\cdot \E = 0\,,
\]
and similarly $\k\cdot \B=0$. Thus $\k$ is perpendicular to both
$\E$ and $\B$. We find for the outer products
\[
\Nabla\times\E(x,t) = -i \k\times
\E(x,t)\,,~~~\Nabla\times\B(x,t) = -i\k\times
\B(x,t)\,,
\]
and thus it follows
\[
\k\times\E = \omega \B\,,~~~ \k\times \B = -\frac{\omega}{c^2} \E\,.
\]
We see that $\E$ and $\B$ are perpendicular to each other, and $\k$
is perpendicular to $\E$ and $\B$, hence $\k$ is parallel
to the so-called \bfi{Poynting vector} $\P = \E\times\B$.
Figure \ref{e_mag.pic} displays an image of a solution.

Without loss of generality we may change the coordinates so that 
$\k$ points into the $z$-direction; then $k_x=k_y=0$ and only $k_z$ is
nonzero. Then, since $\omega\B = \k\times\E$ and $\E$ is 
orthogonal to $\k$, light is completely determined by giving the 
$x$- and $y$-components of $\E$. Thus light has two degrees of freedom; 
put in other words, light has two {\bfi{polarizations}}. Linearly
polarized light is light where $\E$ oscillates in a constant direction
orthogonal to the light ray. Circularly polarized light is light
where $\E$ rotates along the path of light; this can be achieved
by superimposing two linearly polarized light beams. Since the
Maxwell equations are linear, any sum of solutions is again a
solution. Note that to actually get the solution for
$\E(x,t)$, one has to take the real part.

So we have seen that a light beam is determined by giving two
polarizations. These polarizations can be interpreted as modes of an
oscillator. One can write the general solution in terms of coefficient 
functions $a(\k)$ as\footnote{We are not
  taking all details into account here, since we only want
  to convey the general picture of what is happening and don't use the
  material outside this section.}
\lbeq{max.exp}
\E (x,t) =  \int d\k ~\Big(a(\k)\mathbf{u}_\k(\x)e^{-i\omega_\k t}
+a^*(\k)\mathbf{u}_{\k}(\x)^*e^{i\omega_\k t }\Big),
\eeq
where the frequency is given by the dispersion relation 
$\omega_\k=c|\k|$, and
\[
\mathbf{u}_\k(\x)=\eps_1(\k)e^{i\k\cdot \x}+\eps_2(\k)e^{i\k\cdot \x}\,,
\]
where $\eps_1(\k)$ and $\eps_2(\k)$ are {\bfi{polarization vectors}}
chosen to satisfy
\[
\eps_1(\k)\cdot \eps_2(\k)=\eps_1(\k)\cdot \k=\eps_2(\k)\cdot\k=0\,,~
\eps_1(\k)^2=\eps_2(\k)^2=1\,.
\]
\at{add explicit realizations of $ \eps_{1,2}(\k)$.}
Note the similarity with the Maxwell equation. The main difference
is in the dispersion relation. In addition, since now the fields 
functions are real, the coefficients of the positive 
frequency part and the negative frequency part are related.
The positive frequency part 
\[
\widehat\E:=\int d\k a^*(\k)\mathbf{u}_{k}(\x)^*\e^{i\omega_\k  t }
\] 
of the solution \gzit{max.exp} is called the {\bfi{analytic signal}}
of $\E$; clearly $\E=\ol{\widehat\E}+\widehat\E=2\re \widehat\E$.

In the quantum theory one promotes the modes $a(\k)$ and $a^*(\k)$ to
operators. We treat the transition from the classical theory to the 
quantum theory in detail only for the harmonic oscillator, 
corresponding to a single monochromatic mode; see 
Chapter \ref{c.harmonic}.

To motivate the connection, we rewrite the Hamiltonian into a specific
form that we will later recognize as the Hamiltonian of a harmonic
oscillator, thereby showing that the Maxwell equations give rise to (an
infinite set of) harmonic oscillators.

First we consider the system in a finite volume $V$ to avoid some 
questions of
finiteness. In that case (since one has to impose appropriate
boundary conditions), the integral over wave vectors $\k$ for the
electric field becomes a sum over a discrete (but infinite) set of
wave vectors. To get a sum over finitely many terms, one also has to 
remove wave vectors with very large momentum; this corresponds to
discretizing space\footnote{
Getting a proper limit is the subject of
renormalization theory, which is beyond the scope of our presentation. 
The mathematical details for interactive
fields are still obscure; indeed, whether \bfi{quantum electrodynamics} 
(QED) exists as a mathematically well-defined theory is one of the big 
open questions in mathematical physics.
}. 

The functions $\mathbf{u}_\k(\x)$ can then be normalized as
\[
\int_V d\x ~\mathbf{u}_\k(\x)^*\mathbf{u}_{\k'}(\x)=\delta_{\k\k'}\,.
\]
The energy density of the electromagnetic field is
proportional to $\E^2+\B^2$. Hence classically the Hamiltonian is
given by
\[
H=\half\int_V d\x~ \left(\E^2+\B^2\right)\,.
\]
Inserting the expansions \gzit{max.exp} of $\E$ and $\B$ into the 
expression for the
Hamiltonian and taking into account the normalization of the
$\mathbf{u}_\k$ one obtains after shifting the ground state energy 
to zero and performing the so-called thermodynamic limit $V\to \infty$  
a Hamiltonian of the form
\[
H=\shalf\sum_\k 
\hbar\omega_\k\left(a^{*}_{\k}a_\k+a_\k a^{*}_{\k}\right)\,.
\]
In Chapter \ref{c.harmonic} we will 
show that the quantum mechanical Hamiltonian of the harmonic 
oscillator is given by $H= \hbar \omega a^*a$
for some constant $\omega$ and operators $a$ and $a^*$. For light we
thus obtain for each possible $\k$-vector a quantum oscillator. In
practice, a laser admits only a selection of possible $\k$-vectors. In
the ideal case that there is only one possible $\k$-vector, that is,
the Poynting vector can only point in one direction and only one
wavelength is allowed, the Hamiltonian reduces to the Hamiltonian of
one harmonic oscillator.

\chapter{Spectral analysis}\label{s.specanal}

In this chapter we show that the spectrum of a quantum
Hamiltonian (defining the admissible energy levels)
contains very useful information about a conservative quantum system.
It not only allows one to solve the Heisenberg equations of motion
but also has a direct link to experiment, in that the differences
of the energy levels are directly observable, since they can be
probed by coupling the system to a harmonic oscillator with
adjustable frequency.

\section{The quantum spectrum}\label{s.qspec}

In quantum mechanics the classical Hamiltonian becomes an operator on
some Hilbert space. Formally, instead of a function of $q$ and $p$
defined by a formal expression $H(q.p)$
defining a classical $N$-particle Hamiltonian as a function of 
position $q$ and momentum $p$, one has a similar expression where now 
$q$ and $p$ are vectors whose components are linear operators on the 
Hilbert space $\Rz^{3N}$. 
The main difference is the lack of commutativity; so the order of 
operators in the expressions matters. 

Which operators are used to encode the components of $q$ and $p$ depends
on the representation used. In the 
\bfi{position representation}, the components of $q$ act as 
multiplication by position coordinates, while the components of $p$
are multiples of the differentiation operators with respect to the 
position coordinates; in the \bfi{momentum representation}, this also
holds but with position and momentum interchanged. Both representations 
are equivalent.

The collection of eigenvalues of the Hamiltonian
$H$ of a quantum system is referred to as the spectrum of $H$ (or of 
the system). Formally, the {\bfi{spectrum}} of a linear operator 
$H$ is the set of all
$E\in\Cz$ such that $H-E$ is {\em not} invertible.
In finite dimensions, this implies the nontrivial solvability of the
equation $(H-E)\psi=0$, and hence of the existence of an eigenvector
$\psi\ne 0$ satisfying the {\bfi{time-independent Schr\"odinger
equation}}\index{Schr\"odinger
equation}
\lbeq{e.Schroe}
H\psi = E\psi.
\eeq
In infinite dimensions, things are a bit more complicated and require
the spectral theorem from functional analysis. If the spectrum of $H$ 
is, however, discrete then \gzit{e.Schroe} remains valid.

As we shall show in Section \ref{section-reps-heis},
the Hamiltonian of a quantum harmonic oscillator
in normal mode form is given by $H= E_0 + \hbar \omega n$,
where $n$ is the so-called \bfi{number operator} whose spectrum 
consists of the nonnegative integers. Hence the eigenvectors
of $H$ (also called \bfi{eigenfunctions} if, as here, the Hilbert 
space consists of functions) are eigenvectors of $n$,
and the eigenvalues $E_k$ of $H$ are related to the eigenvalues 
$k\in\Nz_0$ of $n$ by the formula
\[
E_k=E_0+k\hbar \omega.
\]
This shows that the eigenvalues of the quantum harmonic
oscillator are quantized, and the eigenvalue differences are
integral multiples of the energy quantum $\hbar\omega$. That the
spectrum of the Hamiltonian is discrete is sometimes
rephrased as `$H$ is quantized'.

In this and the next section we investigate the experimental meaning
of the spectrum of the Hamiltonian of an arbitrary quantum system.
Since the Hamiltonian describes the evolution of the system via the
quantum Heisenberg equation \gzit{e.Heis}, i.e.,
\[
\dot f = H\lp f = \frac{i}{\hbar}[H,f]\,,
\]
one expects that the spectrum will be related to the time dependence of
$f(t)$. To solve the Heisenberg equation, we need to find a
representation where the Hamiltonian acts diagonally.

In the case where the Hilbert space $\Hz$ is finite-dimensional, we can
always diagonalize $H$, since $H$ is Hermitian. There is an orthonormal
basis of eigenvectors of $H$, and fixing such a basis we may
represent all $\psi\in\Hz$ by their components $\psi_k$ in this basis,
thus identifying $\Hz$ with $\Cz^n$ with the standard inner product.
In this representation, $H$ acts as a diagonal matrix whose diagonal
entries are the eigenvalues corresponding to the basis of eigenvectors;
\[
(H\psi)_k=E_k\psi_k\,.
\]

In the case where the Hilbert space is infinite-dimensional
and $H$ is self-adjoint, an analogous representation is possible, using
the {\bfi{Gel'fand--Maurin theorem}}, also known
under the name {\bfi{nuclear spectral theorem}}. The theorem asserts
that if $H$ is self-adjoint, then $H$ can be extended into the
dual space of the domain of definition of $H$; there it has a complete
family of eigenfunctions, which can be used to coordinatize the
Hilbert space. The situation is slightly complicated by the fact that
the spectrum may be partially or fully continuous, in which case
the concept of a basis of eigenvectors no longer makes sense since the
eigenvectors corresponding to points in the continuous spectrum
are no longer square integrable and hence lie outside the Hilbert space.

In the physics literature, the rigorous mathematical exposition is 
usually abandoned at this stage, and one simply
proceeds by analogy, choosing a set $\Omega$ of labels of the
eigenstates and treating them ''formally'' as if they form a discrete
set. Often, the discreteness of the spectrum is enforced verbally by
artificially ''putting the particles in a finite box'' and going to an
infinite volume limit at the very end of the computations. The
justification for the approach is that most experiments are indeed
very well localized; in letting two protons collide in CERN we do not
take interaction with particles on Jupiter into
account. Mathematically we thus put our system in a box. Since we do
not want our system to interact too much with the walls of the box we
take the box large enough. Having met the final requirement one
observes that the physical quantities do not depend on the precise
form and size of the box. To simplify the equations one then takes the
size of the box to infinity. Making this mathematically precise is
quite difficult, though well-understood for nonrelativiastic systems.
In particular, for the part of the spectrum that becomes continuous 
in this limit, the limits of the eigenvectors become \bfi{generalized 
eigenvectors} lying no longer in the Hilbert space itself but in a 
distributional extension of the Hilbert space which must be discussed 
in the setting of a so-called \bfi{Gelfand triple} or 
\bfi{rigged Hilbert space}; cf. Section \ref{s.braket}.
 
In many cases of physical interest, these generalized eigenvectors
come in two flavors, depending on the boundary conditions imposed,
resulting in two families of {\bfi{in-eigenstates}} $|k\>_-$ and
{\bfi{out-eigenstates}} $|k\>_+$ labelled by a set $\Omega$ which
in the case of the harmonic oscillator is $\Omega=\Nz_0$.
(The bra-ket notation used here informally is made precise in Section 
\ref{s.braket}.)
The in- and out-states are called so because they have a natural
geometric interpretation in scattering experiments (see Section
\ref{s.manypart}).
In addition to these eigenstates, there is a measure
$d\mu(k)$ on $\Omega$, and a {\bfi{spectral density}} $\rho(k)$ with
real positive values such that every vector in the Hilbert space has
a unique representation in the form
\[
\psi = \int_\Omega d\mu(k) \psi_+(k)|k\>_+=\int_\Omega d\mu(k)
\psi_-(k)|k\>_-\,.
\]
For any fixed choice of the sign in $\psi(k):=\psi_\pm(k)$,
the inner product is given by
\lbeq{e.diagH}
\phi^*\psi = \int_\Omega d\mu(k) \rho(k) \ol\phi(k)\psi(k)\,.
\eeq
The spectral measure $d\mu(k)$ may also have a discrete part
corresponding to square integrable eigenstates, in which case
$|k\>_+=|k\>_-$. If all eigenvectors are square integrable, the
spectrum is completely discrete. In particular, this is the case
for the harmonic oscillator, for which we construct the diagonal
representation explicitly in Section \ref{c.harmonic}.

Since the $|k\>_\pm$ are eigenvectors with corresponding eigenvalue
$E(k)=E_k$, that is, the Hamiltonian satisfies
\lbeq{e.diagH2}
(H\psi)(k)=E(k)\psi(k)\,,
\eeq
we say that $H$ \bfi{acts diagonally} in the representation defined by
the $\psi(k)$.
Thus one can identify the Hilbert space with the space $L^2(\Omega)$
of coefficient functions $\psi_\pm$ with finite
$\int_\Omega d\mu(k) \rho(k) |\psi_\pm(k)|^2$; the Hamiltonian is then
determined by \gzit{e.diagH2}.
The in- and out-states are related by the so-called 
{\bfi{S-matrix}}, a unitary matrix $S\in \Lin L^2(\Omega)$ such that
\[
\psi_+(k)=(S  \psi_-)(k)\,.
\]
As a consequence of the time-symmetric nature of
conservative quantum dynamics and the time-asymmetry of scattering
eigenstates, the in-representation and
the out-representation are both equivalent to the original
representation on which the Hamiltonian is defined. In many cases of
interest, one can then rigorously prove existence and uniqueness of
the S-matrix.

The transformation from an arbitrary
Hilbert space representation to the equivalent representation in
terms of which $H$ is diagonal, is an analogue of a Fourier
transformation; the latter corresponds to the special case where
$\Hz=L^2(\Rz)$ and $H$ is a differential operator with constant
coefficients.

In general, the Gel'fand--Maurin theorem guarantees the existence of
a topological space $\Omega$ and a Borel measurable spectral density
function $\rho:\Omega\to \Rz_+$ such that the original Hilbert
space is $L^2(\Omega,\rho)$ with inner product \gzit{e.diagH}
and such that \gzit{e.diagH2} holds. Indeed, $\Omega$ can be 
constructed as the set of {\bfi{characters}},
\at{this is treated repatedly! Collect to one place}
that is, $*$-homomorphisms into the complex numbers, of a maximal
commutative C${}^*$-algebra of bounded linear operators containing the
bounded operators $e^{itH}$ ($t\in\Rz$). (Since we don't use this 
construction further, the concepts involved will not be explained in 
detail.)

The above reasoning is completely parallel to the finite-dimensional 
case, where $\Hz=\Cz^n$. There one would write $\psi = \sum_k
\psi_k|k\>$ and have $(H\psi)_k = E_k\psi_k$. An arbitrary quantity
$f\in \Lin\Hz=\Cz^{n\times n}$ would then be represented by a matrix, 
acting as $(f\psi)_k = \sum_l f_{kl}\psi_l$.
In the infinite-dimensional setting, $k$ takes values in the
label space $\Omega$. The quantities of primary interest
are represented by integral operators defined by a kernel function
\[
(f\psi)(k) := \int_\Omega d\mu(l)f(k,l)\psi(l)\,;
\]
the $f(k,l)$ are the analogues of the matrix entries $f_{kl}$.

\bigskip
Finding a diagonal representation for a given quantum system (i.e.,
given a Hilbert space and a Hamiltonian) is in general quite
difficult; succeeding is virtually equivalent with ``solving'' the
theory of the system. Indeed, in terms of the diagonal representation,
we can obtain a full solution of the Heisenberg dynamics. We have
\beqar
\int_\Omega d\mu(l) \dot f(k,l,t)\psi(l) &=&
\frac{i}{\hbar} \Bigl( (Hf\psi)(k) - (fH\psi)(k)\Bigr) \nn\\
&=& \frac{i}{\hbar} \Bigl( E(k) \int_\Omega d\mu(l) (f\psi)(l) -
\int_\Omega d\mu f(k,l,t)E(l)\psi(l) \Bigr)\nn\\\
&=& \frac{i}{\hbar} \int_\Omega d\mu(l) f(k,l,t)\Bigl( E(k) -
E(l)\Bigr)\psi(l)\nn\,,
\eeqar
from which it follows that
\lbeq{f-operator}
\dot f(k,l,t) = \frac{i}{\hbar} \Bigl(E(k) -E(l)\Bigr) f(k,l,t)\,.
\eeq
In
\gzit{f-operator} we recognize a linear differential equation with
constant coefficients, whose general solution is
\[
f(k,l,t) = e^{\frac{i}{\hbar}(E(k)-E(l))t}f(k,l,0)\,.
\]
Thus the kernel function of the operator $f$ has oscillatory behavior
with frequencies
\lbeq{e.RR}
\omega_{kl} = \frac{E(k)-E(l)}{\hbar}\,.
\eeq
This relation, the modern form of the
{\bfi{Rydberg--Ritz combination principle}} found in 1908 by Walter
Ritz \cite{ritz},
may be expressed in the form
\lbeq{planck-E}
\Delta E = \hbar\omega,
\eeq
The formula \gzit{planck-E} appears first in Planck's famous paper
\cite{planck1900} from 1900
where he explained the radiation spectrum of a black body. Planck
wrote it in the form $\Delta E = h\nu$, where $h=2\pi\hbar $ and
$\nu=\omega/2\pi$ is the linear frequency.
The symbol for the quotient $\hbar=h/2\pi$, which
translates this into our formula was invented much later, in 1930,
by Dirac in his famous book\footnote{The
book contains the Dirac equation but also Dirac's famous mistake
(cf. Section \ref{s.history}) -- he had wrongly interpreted the 
antiparticle of the electron predicted by his equation (later named 
the positron) to be the proton.
} 
on quantum mechanics \cite{diracbook}.

\section{Probing the spectrum of a system}\label{s.probing}

All physical systems exhibit small (and sometimes large) oscillations
of various frequencies, collectively referred to as the
{\bfi{spectrum}}
of the system. By observing the size of these oscillations and their
dependence on the frequency, valuable information can be obtained
about intrinsic properties of the system. Indeed, the resulting science
of {\bfi{spectroscopy}} is today one of the indispensable means for
obtaining experimental information on the structure of chemical
materials and the presence of traces of chemical compounds.

To probe the spectrum of a  quantum system, we bring it
into contact with a macroscopically observable (hence classical)
weakly damped harmonic oscillator.
That we treat just a single harmonic oscillator
is for convenience only. In practice, one often observes many
oscillators simultaneously, e.g., by observing the oscillations of the
electromagnetic field in the form of electromagnetic radiation --
light, X-rays,  or microwaves. However, the oscillators do not
interact that strongly in most cases and in the case of
electromagnetic radiation not at all. In that case the result of
probing a system with multiple oscillators results in a linear
superposition of the results of probing with a single oscillator.
This is a special case of the general fact that solutions of linear
differential equations depend linearly on the right hand side.

From the point of view of the macroscopically observable classical 
oscillator, the probed quantum system appears simply as a time-dependent
external force $F(t)$ that modifies the dynamics of the free
harmonic oscillator. Instead of the equation 
$m \ddot q + c \dot q + k q =0$ we get the differential equation 
describing the 
{\bfi{forced harmonic oscillator}}\index{oscillator!forced}, given 
by
\[
m\ddot q + c\dot q + kq = F(t)\,.
\]
The external force $F$ is usually the value
\[
F(t)=\<f(t)\>
\]
of a quantity $f$ from the algebra of quantities of the probed system,
as discussed in more detail in Part II.
This follows from the general principles of Section \ref{s.qc}
for modeling interactions of a quantum system with 
a classical, macroscopic system (only the latter are directly 
measurable). How classical measurements are to be interpreted in a 
pure quantum context will be discussed in Section \ref{s.measurement}.

If the measurement is done far from the probed system, such as a 
measurement of light (electromagnetic radiation) emitted by a 
far away source (e.g., a star, but also a Bunsen flame observed by 
the eye), the back reaction of the classical oscillator on the probed 
system can be neglected. Then the probed system can be considered as a 
Hamiltonian system and evolves according to the Heisenberg 
equation \gzit{e.heis}. In particular, the analysis of 
Section \ref{s.qspec}
applies, and since expectations are linear, the external
force $F$ evolves as a superposition of exponentials $e^{i\omega t}$,
where the $\omega$ are differences of eigenvalues of $H$. In the
quantum case the spectrum may have a discrete part, leading to a sum
of different exponentials $e^{i\omega t}$ that, as we shall see,
leads to conspicuous spikes in the Fourier transform of the response 
and a continuous part that 
leads to an integral over such terms which typically provide a 
smooth background response. In the
following, we shall assume for simplicity a purely discrete spectrum,
and hence an expansion of $F$ of the form
\[
F(t) = \sum_l F_l e^{i\omega_l t}\,,
\]
with distinct, real and nonzero frequencies. However, the analysis
holds with obvious changes also for a (partly or fully) continuous
spectrum if the sums are replaced by appropriate integrals.

The solution to the differential
equation consists of a particular solution and a solution to the
homogeneous equation. Due to damping, the latter is transient
and decays to zero. To get a particular solution, we note that common
experience shows that forced oscillations typically have the same
frequency as the force. We therefore make the ansatz
\[
q(t) = \sum_l q_l e^{i\omega_l t}\,.
\]
Inserting both sums into the differential equation, we obtain the
relation
\[
\sum_l \Bigl( -m\omega_{l}^{2} + ic\omega_l + k\Bigr)
q_le^{i\omega_l t} = \sum_l F_l e^{i\omega_lt}\,,
\]
from which we conclude that we have a solution precisely when
\[
q_l = \frac{F_l}{k-m\omega_{l}^{2} + ic\omega_l}\,,~~~ \Forall l.
\]
Since the frequencies are real and distinct, the denominator cannot 
vanish. The energy in the $l${th} mode is therefore proportional to
\lbeq{e.lorentzshape}
|q_l|^2 =\frac{|F_l|^2}{(k-m\omega_{l}^{2})^2 + (c\omega_l)^2}\,.
\eeq

Now first imagine that the system under study has only one
frequency, that is, $F_l\neq 0$ for only one $l$. For example, the
system under study is also an oscillator that is swinging with a
certain frequency. In this case the oscillator with which we probe the
system will also swing with that same frequency as the probed system,
but with an amplitude given by \gzit{e.lorentzshape}. We see that for
$\omega=\sqrt{\frac{k}{m}}$ close to $\omega_l$ the oscillator
responds most to the force it feels from the probed system. The
frequency $\omega=\sqrt{\frac{k}{m}}$ is called the {\bfi{resonance
frequency}} of the oscillator. The above we know from phenomena of
daily (or not so daily) life, as pushing a swing (or riding a car with 
a defect shock absorber); if you push with the
`right' frequency the result will be that the swing goes higher
and higher, pushing with another frequency results in a seemingly
chaotic incoherent swinging.

Returning to the case that there are more $F_l$ nonzero, we see that
the oscillator will swing with the same frequencies as the probed
system. But the intensity with which the oscillator swings depends on
the positions of the $\omega_l$ relative to the resonance
frequencies. Suppose that $c$ is relatively small, so that we can
ignore the term $c\omega$ in the denominator of
\gzit{e.lorentzshape}. Then the $q_l$ for which $\omega_l$ is close to
$\omega$ show a higher intensity.

Looking for resonances with an oscillator that has an adjustable
frequency $\omega$ therefore gives a way to experimentally find the
frequencies in the force incident to the oscillator. If the frequency
$\omega$ passes over one of the frequencies of the probed system, the
oscillator will swing more intensively.

\begin{figure}[h]
\begin{center}
\begin{pspicture}(-3,-1)(5,5)
\psset{xunit=1.5cm,yunit=1.5cm}
\psaxes[labels=none,ticks=none](1,0)(-2,0)(4,3.3)
\psplot{-2}{4}{1.2/((x-1)^2+0.4)}
\rput(1,-0.5){$\omega_0$}
\rput(3,-0.5){$\omega\rightarrow$}
\end{pspicture}
\caption{Lorentz-shape. The absorbed energy of the oscillator with
  varying frequency $\omega$.
  }\label{fig1}
\end{center}
\end{figure}
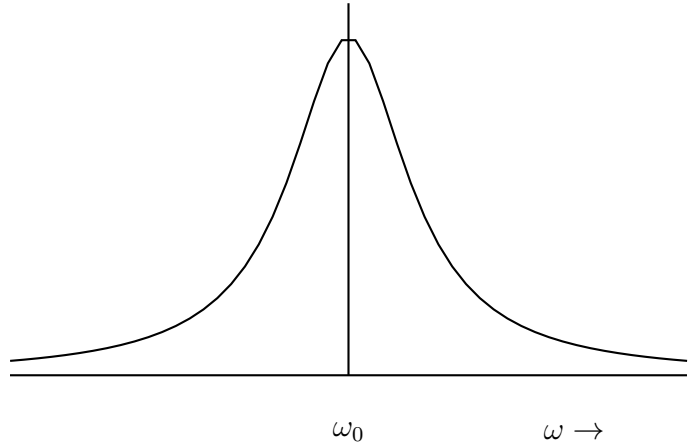

The resonances occur around a natural frequency but also the width
of the interval in which the system shows a resonance has
information. If the interval is small, one speaks of a {\bfi{sharp
resonance}}\index{resonance!sharp} and this corresponds to a discrete 
or nearly discrete spectrum of the frequencies. 
If the resonance is not sharp, the
response corresponds to a continuous spectrum. The graph that shows
the absorbed energy (which is proportional to $|q_l|^2$) as a function
of the frequency ($\sim \omega_l$) for a system with
one resonance frequency typically has a {\bfi{Lorentz shape}},
according to the formula \gzit{e.lorentzshape}: There
is a peak around $\omega_0 = \sqrt{k/m}$ with a certain width,
and on both sides of the peak the
function tends to zero at plus and minus infinity. In Figure \ref{fig1}
\at{add description to both axes; misplaced  $\omega$ in figure}
we displayed a graph of a Lorentz shape for a harmonic oscillator with
varying frequency $\omega$ in contact with a probed system that has
one $F_l$ nonzero for the frequency $\omega_0$. 

\at{It might be better to place the latex source of the figure
just after this point.}

For general systems
with more resonance frequencies, the graph is a superposition of such
curves and the peaks around the resonance frequencies can have
different widths and different heights. This graph is recorded by
typical spectrometers, and the shapes and positions of characteristic
pieces of the graph contain important information about the system.
We shall assume that the peaks have already been translated
into resonance frequencies (a nontrivial task in case of overlapping 
resonances), and concentrate on relating these frequencies to the
Hamiltonian of the system. This is done in Section \ref{s.manypart}.

\section{The early history of quantum mechanics}\label{s.history}

In this section we remark on some important aspects of the history of
quantum mechanics. We focus on the physics of the atom, which was one
of the main reasons to develop quantum mechanics. 
In Section \ref{c.blackbody} we
discuss the physics of the black body and the history of the formula
of Planck, which describes black body radiation. For an interesting
historical account we refer to for example \sca{van der
  Waerden}\cite{vanderWaerden} or \sca{Zeidler} \cite{zeidler}.

The importance of the spectrum in quantum physics is not only due to
the preceding analysis, which allows a complete solution of the
dynamics, but also to the fact that the spectrum can easily be probed
experimentally. Indeed, spectral data (from black body radiation and
the spectral absorption and emission lines of hydrogen) were
historically the trigger for the development of modern quantum theory.
Even the name spectrum for the set of eigenvalues was derived from this 
connection to experiment.

Probing the spectrum through contact with a damped harmonic oscillator
has been discussed in Section \ref{s.probing}.
Note that the observed frequencies give the spectrum of the force,
not the spectrum of the Hamiltonian. As derived above, the spectrum
of the force consists of the spectral differences of the Hamiltonian
spectrum. This is in accordance with the fact that (in nonrelativistic
mechanics) absolute energy is meaningless and only energy differences
are observable.

In case of the harmonic oscillator, the spectrum of the Hamiltonian $H$
is discrete (see Chapter \ref{c.harmonic} for the details and 
derivation), consisting of the nonnegative integral multiples $k\omega$
of the base frequency $\omega$. Thus the set of labels for the
eigenvectors $|k\>$ is discrete, $\Omega = \Nz_0$.
The number of allowed frequencies is thus
countable and the external force may be expanded into a sum of the form
\[
F(t) = \sum e^{i\omega_{kl}t}F_{kl}\,.
\]
Explicitly, the frequencies are given by $\omega_{kl} = \omega
(k-l)\in \Zz\omega$. Thus quantum mechanics produces overtones. This
is not an authentic quantum mechanical feature; in classical mechanics
one finds overtones in a similar setting -- for example, in the
pitching of a guitar string.

A historically more interesting system is the hydrogen atom, where
the energies are given by an equation of the form
\[
E_k = E_0 - \frac{C}{k^2}\,,
\]
for some constant $C$. Then the frequencies are given by
the {\bfi{Rydberg formula}}
\lbeq{balmer}
\omega_{kl} = R_H \Bigl( \frac{1}{k^2} -
\frac{1}{l^2} \Bigr)\,,
\eeq
where $R_H\sim 1.1\cdot 10^7 m^{-1}$ is the Rydberg constant.
The Rydberg formula correctly gives the observed spectral lines of
the hydrogen atom. The formula was discovered by Rydberg in 1889 
(\sca{Martinson and Curtis} \cite{MarC}) after
preliminary work of Balmer, who found the formula for the
{\bfi{Balmer series}} of spectral lines (given by $k=2$). 
Schr\"odinger derived this formula using the theoretical framework of 
quantum mechanics.

Let us review the situation of the time where quantum mechanics
was conceived.
Around 1900 physicists were experimentally exploring the atom, which 
until then was (since antiquity) only a philosophically disputable part 
of Nature. The experiments clearly indicated
that atoms existed and that matter was built up from
atoms. The physicist Boltzmann had argued that atoms existed, but his
point of view had not been accepted; only after his death in 1906,
the existence of atoms was unarguably proved by experiments by Perrin 
around 1909. This lead to the problem of finding the
constituents of the atom and its structure. In 1897 Thompson
had discovered the electron as a subatomic particle. Since the atom
is electrically neutral, the atom has to contain positively charged
particles. Thompson thought of a model in which the atom
was a positively charged sphere with the electrons being in this ``plum
pudding'' of positive charge. But then in 1911 Rutherford put
Thompson's model to the test; Marsden and Geiger, who were working under
the supervision of Rutherford, shot $\alpha$-particles at a thin foil
 of gold and looked at the scattering pattern \cite{geigermarsden}.
The experiment is therefore called the {\bfi{Geiger--Marsden 
experiment}}.
At that time, $\alpha$-particles were considered a special
radiation emitted by some 'radio-active' elements; now we know
that these are the nuclei of Helium with the electrons being stripped
off.

Since the $\alpha$-particles are positive, they have a particular
kind of interaction with the positively charged sphere of Thompson's
model. But since the electrons swim around in the positive charge,
the net charge is zero and most interaction is screened off.
Therefore it was expected that the $\alpha$-particles would be only
slightly deflected. However, the pattern was not at all like that!
It rather looked as if almost all $\alpha$-particles went straight
through and a small percentage was deflected by a concentrated
positive charge. Most $\alpha$-particles that were
deflected were scattered backwards, implying that they had an almost
head-on collision with a positive charge. 

The very small percentage of
scattered $\alpha$-particles indicated that the chance that an
$\alpha$-particle meets a positively charged nucleus on its way is very
small, which implies that the nucleus is very small compared to the
atom. Therefore Rutherford (who wrote a paper to explain the
results of the Geiger--Marsden experiment)
concluded that the nucleus of an atom is positively charged and the
electrons circle around the nucleus, and furthermore, the size of the
nucleus is very small compared to the radii at which the electrons
circle around the atom \cite{rutherford1911}.
If one imagines the atomic nucleus to have the
size of a pea and one would place it at the top of the Eiffel tower,
the closest electrons would circle around in an orbit that touches the
ground; the atom is mostly empty.

In 1918 it was again Rutherford who performed an important
experiment from which he concluded that the electric charge of the
atomic nucleus was carried by little particles, called protons. The
hydrogen atom was found to be the simplest atom; it consists of a
proton and one electron circling around the proton. Because of this
experiment the discovery of the proton is attributed to Rutherford.

Classically, if an electron circles around in an electric field it
radiates and thus loses energy. The question thus arises why the
hydrogen atom is stable.
Again classically, an electron can circle around a positive charge with
arbitrary energy. If the electron changes its orbit, this happens 
gradually, hence the energy changes continuously and the absorption or 
emissions patterns of the hydrogen atom should be continuous.
But experiments done by Rydberg in 1888 and Balmer in 1885 showed
that hydrogen absorbed or emitted light at well-defined frequencies,
visible as lines in the spectrum obtained by refraction.
For the atomic model this implies that the electron can only have
well-defined energies separated by gaps (forbidden energies). In 1913
Bohr wrote a series of papers \cite{bohr1,bohr3,bohr2,bohr4} in which he
postulated a model to account for this. Bohr postulated that angular
momentum is quantized
(if $p$ is the momentum of the electron and $r$ the radius, then the
angular momentum is $L=r\times p$, where the cross denotes the vector
product) and that the electron does not lose energy
continuously. With these assumptions he could explain the spectrum
observed by Rydberg.

The model of Bohr did not explain the behavior of atoms, it only gave
rules the atom had to obey. In 1925 Werner Heisenberg wrote a paper
\cite{Hei}
where he tried to give a fundamental basis for the rules of quantum
mechanics. Heisenberg described the dynamics of the transitions of an
electron in an atom by using the `states' of the electron as labels.
For example, he wrote the frequency emitted by an electron jumping from
a state $n$ to a state $n-\alpha$ as $\nu(n,n-\alpha)$. Just
two months later Max Born and Pascal Jordan wrote a paper
\cite{bornjordan1925} about the paper of Heisenberg, in which they
made clear that what Heisenberg actually did was promoting observables
to matrices. The three of them, Born, Jordan and Heisenberg, wrote in
the same year a paper \cite{bornjordanheisenberg} where they
elaborated on the formalism they developed. Also in the same year 1925
Paul Dirac wrote a paper in response to the paper of Heisenberg, in
which the remarkable relation $q_rp_s-p_sq_r = \delta_{rs}i\hbar$
appeared. Dirac tried to find the relation between a classical theory
and the corresponding quantum theory. In fact, Dirac postulated this
equation: ``we make the fundamental assumption that the difference
between the Heisenberg product of two quantum quantities is equal to
$ih/2\pi$ times their
Poisson bracket expression''.

So, in the beginning years of quantum mechanics, the dynamics of the
observables was described by a kind of matrix mechanics. (A modern
version of this is the view presented in the present book.)
Based on work of de Broglie, Schr\"odinger came up with a differential
equation for the nonrelativistic electron \cite{schroedinger1926}.
A probability interpretation for Schr\"odinger's wave function
was found by Born. In 1927, Pauli
reformulated his exclusion principle in terms of spin and antisymmetry.
In 1928, Dirac discovered the Dirac equation for the relativistic
electron. In 1932, the early years concluded with the discovery of the
positron by Anderson and the neutron by Chadwick, which were enough
to explain the behavior of ordinary matter
and radioactivity. But the forces that hold the nucleus together were
still unknown, and already in 1934, Yukawa predicted the existence of
new particles, the mesons. Since then the particle zoo has increased
further and further.

A number of Nobel prizes (most of them in physics, but one in chemistry
-- early research on atoms was interdisciplinary)
for the pioneers accompanied the early development of quantum 
mechanics\footnote{The remarks to each Nobel laureate are the official
wordings in the announcements of the Nobel prizes.
For press announcements, Nobel lectures of the laureates,
and their biographies, see the web site
\url{http://nobelprize.org/physics/laureates}.}:
\begin{itemize}
\item
1908 Ernest Rutherford, (Nobel prize in chemistry)
     for his investigations into the disintegration of the elements,
     and the chemistry of radioactive substances
\item
1918 Max Planck,
     in recognition of the services he rendered to the advancement
     of physics by his discovery of energy quanta
\item
1921 Albert Einstein,
     for his services to theoretical physics, and especially for his
     discovery of the law of the photoelectric effect
\item
1922 Niels Bohr,
     for his services in the investigation of the structure of atoms
     and of the radiation emanating from them
\item
1929 Louis de Broglie,
     for his discovery of the wave nature of electrons
\item
1932 Werner Heisenberg
     for the creation of quantum mechanics, the application of which
     has led among others to the discovery of the allotropic forms of
     hydrogen
\item
1933 Erwin Schr\"odinger and Paul A.M. Dirac,
     for the discovery of new productive forms of atomic theory
\item
1935 James Chadwick,
     for the discovery of the neutron
\item
1936 Carl D. Anderson,
     for his discovery of the positron

\end{itemize}

and belatedly, but still for work done before 1935,

\begin{itemize}
\item
1945 Wolfgang Pauli,
     for the discovery of the exclusion principle, also called the
     Pauli principle
\item
1949 Hideki Yukawa,
     for his prediction of the existence of mesons on the basis of
     theoretical work on nuclear forces
\item
1954 Max Born,
     for his fundamental research in quantum mechanics, especially
     for his statistical interpretation of the wave function

\end{itemize}

\bigskip
The story of the discovery of antimatter is interesting.
Though Dirac called it a prediction in his Nobel lecture,
{\em
``There is one other feature of these equations which I should now like
to discuss, a feature which led to the prediction of the positron'',}
it was only a postdiction. Yes, he had a theory in which there were
antiparticles. But before the positron was discovered,
Dirac thought the antiparticles had to be protons (though there was a
problem with the mass) since new particles were inconceivable at
that time. Official history seems to have followed
Dirac's lead in his Nobel lecture, and tells the story as it should
have happened from the point of the theorist, namely that he
(i.e., theory) actually predicted the positron.
The truth is a little different.

Anderson discovered and named the positron in 1932. He wrote the
announcement of his discovery in Science \cite{anderson32},
``with due reserve in interpretation''. The proper publication
\cite{anderson33-1}, where he also predicted ``negative protons''
(now called antiprotons), was still without any awareness of Dirac's
theory. It is in the subsequent paper \cite{anderson33-2} that
Anderson relates the positron to Dirac's theory.

Heisenberg, Dirac, and Anderson were all 31 years old when they got
the Nobel prize. The fact that Anderson's paper \cite{anderson33-1}
is very rarely cited\footnote{
\url{http//www.prola.aps.org/} lists only 37 citations, and
only 5 before 1954. The paper \cite{anderson33-2} is cited 35 times.
} 
should cast some doubt on the relevance of citation counts for
actual impact in science.

\section{The spectrum of many-particle systems}\label{s.manypart}

To give a better intuition for what kind of spectra quantum systems
can be expected to have, we discuss here the spectrum of many-particle
systems from an informal point of view.

There are {\bfi{bound states}}, where all particles of the system
stay together, and there are {\bfi{scattering states}}, where the 
system
is broken up into several fragments moving independently but possibly
influencing each other. The nomenclature comes from the scattering
experiments in physics; shooting particles at each other can result
in the formation of a system where the particles are bound together or
where the particles scatter off from each other. In the case of a
scattering process, different 
\bfi{arrangements\index{arrangement}}, (i.e., partitions of the set 
of individual particles into fragments which form a subsystem moving 
together) describe the combination of particles before a collision and 
their recombination in the debris after a collision.

The discrete spectrum of a Hamiltonian $H$ corresponds to the
bound states; each discrete eigenvalue to a different mode of
the bound system. The study of the discrete spectrum of compound
systems is the domain of {\bfi{spectroscopy}}. We shall return to
this topic in Chapter \ref{c.spec}, when 
the machinery to understand a spectrum is fully developed.

The continuous part of the spectrum corresponds to
the scattering states. In general, the spectrum is discrete till a
certain energy level, called the {\bfi{dissociation threshold}}, and
after the dissociation threshold the spectrum is continuous. For
the hydrogen atom, the dissociation threshold is $13.6eV$. For the
harmonic oscillator, the dissociation threshold is infinite. In
such a case, where the dissociation threshold is infinite, there
is no continuous spectrum and the system is always bound; we call
this {\bfi{confinement}}. For example, three quarks always form a
bound state, that is, they are confined. A single quark can not
get loose from its partners. It may also be the case that there is
no bound state; for example, the atoms in inert gases don't form
bound states, hence a system consisting of more than one of such atoms
has only a continuous spectrum.

In scattering experiments the ingoing particles and the outgoing
particles can be different. Hence one needs to keep track of what
precisely went where. After the scattering the particles separate from
each other in different clusters. The constituents in cluster $i$ form a
bound state, which can be in an excited state, which we denote
$E_i$. If the cluster $i$ is moving with a momentum, the total kinetic
energy of cluster $i$ is $p_{i}^{2}/2m_i$, where $m_i$ is the mass of
cluster $i$. If there are $N$ clusters after a collision (scattering),
the resulting total energy is 
\[
E=\sum_{i=1}^{N}\Bigl(\frac{p_{i}^{2}}{2m_i} + E_i\Bigr)\,.
\]
In scattering experiments a possible outcome of clusters and their
constituents is called a \bfi{channel}. It is very common in particle
physics that a single reaction -- like shooting two protons at each
other -- has more than one channel. We see that in each channel, there
is a continuous spectrum above a certain energy level $\Delta$, which
is the sum of the ground state energies of the different clusters. To
theoretically disentangle the spectrum, one uses an analytic
continuation of the scattering amplitudes. We thus view the spectrum
as a subset of the complex plane. When multiplying the
momenta with a complex phase that has a nonzero imaginary part, the
continuous part of the spectrum becomes imaginary and is tilted away
from the real axis. The bound states still appear on the real line as
isolated points, that is, discrete. But now at each bound state with
energy above $\Delta$ there is a line connected representing the
continuously varying momentum of the corresponding cluster. The
technique of disentangling the spectrum using analytic continuation is
called {\bfi{complex scaling}}. For more
background and rigorous mathematical arguments, see, e.g., 
\sca{Simon} \cite{simon-78},
\sca{Moiseyev} \cite{moiseyev-98}, or 
\sca{Bohm} \cite{bohm.03}.

{\bfi{Dissipation}.}
If we admit dissipation, the Hamiltonian is no longer Hermitian,
since there is typically an antihermitian contribution to the potential,
generally called an {\bfi{optical potential}} since it was first 
used in optics. Also, the dynamics need no longer be governed by the
Heisenberg equation, but can be both in the classical and in the 
quantum case of the more general form
\lbeq{e.lindblad}
\dot f = H\lp f + \sum_{j,k} L_j^* \lp G_{jk} (L_k \lp f)
\eeq
with {\bfi{Lindblad operators}} $L_j$ encoding interactions
with the unmodelled environment into which the lost energy dissipates,
and complex coefficients $G_{jk}$ forming a symmetric, positive definite
matrix. Remembering that $\lp$ acts as a derivation, the additional
terms can be viewed as generalized diffusion terms; indeed, the
dynamics \gzit{e.lindblad} describes classically for example 
reaction-diffusion equations, and its quantum version is the quantum 
equivalent of stochastic differential equations, which model systems 
like Brownian motion and give microscopic models of diffusion processes.
For details, see, e.g., \sca{Gardiner} \cite{Gar},
\sca{Breuer \&  Petruccione} \cite{BreP}.

Assuming that the terms in the sum of \gzit{e.lindblad}
are negligible, the dynamics satisfies
the Heisenberg equation, and the above analysis applies with small
changes. However, since $H$ is no longer Hermitian, the energy levels
typically acquire a possibly nonzero (and then positive) imaginary part.
Isolated eigenvalues with positive imaginary parts are called
{\bfi{resonances}}. The oscillation frequencies are still of of the 
form $\hbar \omega =\Delta E$, but since the energies have a  positive 
imaginary part, the oscillations will be damped, as can be seen by
looking at the form of $e^{i\omega t}$. That this does not
lead to a decay of the response of the oscillator is due to stochastic
contributions modelled by the Lindblad terms and neglected in our
simplified analysis.

Resonances with tiny imaginary parts behave almost like bound
states, and represent {\bfi{unstable particles}}, which decay in a
stochastic manner; the value $\Gamma = 2 \im \omega$ gives their
{\bfi{lifetime}}, defined as the time where (in a large sample of
unstable particles) the number of undecayed particles left is reduced
by a factor of $e$, the basis of the exponential function.

Thus the spectrum of a Hamiltonian contains valuable experimentally
observable information about a quantum system.

\section{Black body radiation}\label{c.blackbody}

In the remainder of this chapter, we discuss the spectrum of a 
\idx{black body} and some of its consequences.

In the history the `black body' plays an important role. Applying some
basic concepts of quantum mechanics and statistical mechanics one
arrives at the distribution formula first derived by Max Planck in
December 1900 \cite{planck1901}. According to Van der Waerden in his
(partially autobiographical) book \cite{vanderWaerden}
the presentation of Planck in December 1900 was the birth of quantum
mechanics.

What is a black body? A body that looks black does not reflect any
light, it absorbs all incoming light. Hence if some radiation comes
from a perfectly black body, it needs to be due to the interaction of
the internal degrees of freedom with light. It is hard to
experimentally construct a black body. The theoretical idea is to have
a hollow box with a single little hole, through which the box can emit
radiation outwards. Since the hole is assumed to be very small, no
light will fall inwards and then be reflected through the hole
again. Thus no light will be reflected (or at least almost no
light). In practice many objects behave like black bodies above a
certain temperature. The sun does not reflect a substantial amount of
light (where should it come from?) compared to the amount it
radiates. Therefore one of the best black bodies is the
sun.

Given a black body, there is a positive integrable function
$f(\omega)$ of the frequency $\omega$, such that the amount of energy
radiated in the frequency interval $[\omega_1,\omega_2]$ is
\[
E([\omega_1,\omega_2]) = \int_{\omega_1}^{\omega_2}d\omega\ f(\omega)\,.
\]
The function $f(\omega)$ is the \idx{radiation-energy density}. The main
object of this section is the function $f(\omega)$. The importance of
the black body lies in the fact that the radiation emitted is only due 
to its internal energy and its interaction with light. 
In practice a system
has always interaction with the environment and light falling onto it
(since we want to `see' where the black body is, the latter is often
inevitable). What would we expect from the radiation-energy density?
First, since $\omega=0$ means that the energy of the photons emitted
is $\hbar \omega =0$ and $\omega<0$ is not a possibility, we have
$f(0)=0$. Second, the function $f$ has to be integrable, hence
$\lim_{\omega\to\infty} f(\omega)=0$. The total integral 
$\int_0^\infty d\omega f(\omega)$
represents the total energy of the body. Therefore we certainly want
$f$ to be integrable, i.e., $f\in L^1(\Rz^+)$.

We know from experience that
black bodies (like dark metals) do not radiate any thermal energy when
they are at room temperature, but heating them up makes them glow
red. When we rise the temperature, the color shifts more and more in
the blue direction. This phenomenon can also be seen in flames; the
outer, cooler side is red while more inwards, the flame gets
lighter, reaches white and goes over to blue, and then becomes
invisible. Empirically one concludes that the function $f(\omega)$ has
a maximum at a frequency $\omega_{max}$, where $\omega_{max}$ is
temperature dependent; the larger the temperature $T$, the larger
$\omega_{max}$. Before 1900 it was already found that the fraction
$\omega_{max}/T$ was almost independent of the body that was heated
up. In 1893 the physicist Wilhelm Wien\footnote{His real name is
  rather long: Wilhelm Carl Werner Otto Fritz Franz Wien.} used the
statistical mechanics developed by Maxwell and Boltzmann to the laws
of thermodynamics to derive {\bfi{Wien's displacement law}} 
\cite{wien}
\[
\frac{\omega_{max}}{T}=\frac{2\pi c}{b}\,,
\]
where $c$ is the speed of light and $b$ is a
constant whose numerical value is approximately $2.9 \cdot 10^{-3}
m\cdot K$. In 1896 Wien derived a formula, called {\bfi{Wien's
approximation}} for the radiation density
\lbeq{wien}
f(\omega) = a \omega^3 e^{-b \frac{\omega}{T}}\,,
\eeq
for some parameters $a,b>0$. It is clear that the proposed
$f$ is integrable and satisfies $f(0)=0$. For large $\omega$ the
radiation-energy density matches the observed densities, however, for
small $\omega$ the radiation density of Wien does not match the
experiments.

On the other hand, there were other radiation laws. First, there was
{\bfi{Stefan's law}} (or Stefan--Boltzmann law) derived on basis
of empirical results in 1879 \cite{stefan}. 
The statement of Stefan's law is that
the total energy radiated per second of a hot radiating body is
proportional to the fourth power of the temperature:
\[
\frac{dE}{dt}=\sigma A T^4\,,
\]
where $A$ is the area of the body and $\sigma $ is a constant. In 1884
Boltzmann gave a theoretical derivation of Stefan's law using the
theoretical tools of statistical mechanics \cite{boltzmannstefan}. 
The second radiation law known in 1900 was {\bfi{Rayleigh's law}}. 
Lord Rayleigh used classical
mechanics to derive a better description of the radiation density for
low values of $\omega$ \cite{rayleigh}. He proposed
\[
f(\omega) = \gamma \omega^2\,,
\]
which is clearly wrong for large $\omega$ and is not even
integrable. Later in 1905, Lord Rayleigh improved the derivation of his
proposal in a collaboration with Sir James Jeans, again based on
purely classical arguments. Although their discovery was interesting,
it did not match the experiments for high $\omega$. 
In December 1900, Max Planck had given a seminar and
gave a derivation of $f(\omega)$ that resulted in a radiation-energy
density that matched the experiments both for low and for high
$\omega$. Even more was true, the formula of Planck reproduced Wien's
displacement law, Wien's approximation, Rayleigh's proposal and
Stefan's law. The formula Planck\index{Planck's formula} derived was 
giving the energy density
of a black body in thermal equilibrium, from which one obtains the
radiation-energy density
\[
f(\omega) = \frac{\hbar V}{\pi^2 c^3}\frac{\omega^3}{e^{\hbar \omega
    \beta}-1}\,,
\]
where $V$ is the volume of the black body (the cavity actually) and
$\beta=1/kT$, $k $ is \idx{Boltzmann's constant}. 
Indeed for low $\omega$ we get an expression that is
quadratic in $\omega$, for high $\omega$ we get Wien's law and
integrating the expression over $\omega$ one sees that the integral is
proportional to $T^4$. The accordance with Wien's displacement law
will be shown later -- we will also remark on the agreement of Planck's
law later.

So what precisely did Planck do that the others did wrong? The key
ingredient in Planck's derivation is to consider the constituents of
the black body as follows: the black body is just a cavity where the
inner walls can have an interaction with light. The walls of the
cavity are made of molecules that behave like compounds of harmonic
oscillators. Planck assumed that the energies of the molecules take
values in some discrete set: the states of the molecules do not vary
continuously but are discrete. Hence we can put the states in
bijection with the natural numbers.
Furthermore he assumed that the light
inside the cavity induces transitions in the molecules by absorbing or
emitting radiation. A transition from a state labelled with $n$ and
with energy $E_n$ and a state labelled with $m$ and with energy $E_m$
is only possible if the energy differences and the frequencies are
related by $|E_n - E_m |=\hbar \omega$. Thus by discretizing the
states of the interior of the black body the interaction with light
varies over a discrete set of frequencies. Planck at the moment saw
the discretization as a purely theoretical and mathematical tool that
would bear no relation with reality. It just reproduced the correct
results, which was most important: it gave a formula that fitted all
experiments. Very puzzling at the time was the necessary assumption 
that the energy was quantized -- an assumption that marked the start
of the quantum era. It took some time until the derivation of Planck's
law was given a clear meaning. 

\section{Derivation of Planck's law}

In 1905 Einstein gave a comprehensible derivation, which we
shall present below. In modern textbooks one can find a
one-page-derivation and we will present such a proof below as well.
For both derivations we need a basic fact from statistical mechanics,
called the Boltzmann distribution. 
\at{derive this later using part II?}

Suppose that we have a physical system consisting of many identical 
molecules (or atoms, or any other smaller subsystems). Each molecule 
can attain different states that are labelled 
with integers $n=0,1,2,\ldots$. In a modern treatment, these states
are identified with the eigenstates of the quantum Hamiltonian, and 
we shall use this terminology, though it was not available when Einstein
wrote his paper. We thus 
assume that the spectrum of the molecules is discrete and there is a
bijection between the eigenstates of the molecule and the natural 
numbers. Each eigenstate $n$ of the molecule corresponds to an
eigenvalue $E_n$  of the Hamiltonian, giving the energy the molecule 
has in eigenstate $n$. The Boltzmann distribution gives the relative 
frequency of eigenstates of the molecules. Writing $N(n)$ for the 
number of molecules in state $n$, the {\bfi{Boltzmann distribution}}
dictates that
\lbeq{boltzmann}
\frac{N(n)}{N(m)} = e^{- \frac{(E_n - E_m)}{kT}}\,.
\eeq
when the system is in thermal equilibrium with itself and with the 
surrounding system. Thus, the temperature $T$ has to be constant.
Such a `mixed' state, where the volume, the temperature, and the 
number
of particles are kept constant and in thermal equilibrium with its
environment is called a {\bfi{canonical ensemble}}.
A derivation of the
Boltzmann distribution can be found in many elementary textbooks on
statistical physics, e.g., \sca{Reichl} \cite{Rei}, 
\sca{Mandl} \cite{mandl}, \sca{Huang} \cite{kersonhuang}, 
or \sca{Kittel} \cite{kittel}.

The probability $p_n$ of measuring an arbitrary molecule to be in
state $n$ is
\[
p_n = \frac{e^{- \frac{E_n}{kT}}}{Z}\,,
\]
where $Z$ is the {\bfi{partition function}}
\[
Z= \sum_n e^{- \frac{E_n}{kT}}\,.
\]
One can thus rewrite
\[
p_n =  e^{- \frac{(E_n - F)}{kT}}\,,
\]
where we defined the {\bfi{Helmholtz free energy}} $F$ as
\[
F= - kT \ln Z\,.
\]
One often regroups the states into states that have equal energy. Then
to each natural number corresponds an energy $E_n$, and a natural
number $g_n$ counting the number of states with energy $E_n$. The
number $g_n$ is called the {\bfi{degeneracy}} of the energy $E_n$.
Thus we have
\[
Z= \sum_{n} g_ne^{- \frac{E_n}{kT}}\,,
\]
and the probability $p_n$ of measuring a molecule with energy $E_n$ is
\[
p_n = g_n e^{- \frac{(E_n - F)}{kT}}\,.
\]
The constant $\beta = 1/kT$ is called the {\bf\idx{inverse temperature}}
and plays a fundamental role; in statistical physics, it is customary 
to express all quantities in terms of $\beta$.
The average of the energy, denoted $\bar E$, is found by
\[
\bar E = \sum_{E_n} E_n p_n = -\frac{\partial}{\partial \beta }\ln
Z(\beta)\,.
\]

\bigskip
\bfi{Einstein's derivation.}
We now focus on two energies in the molecule, $n$ and $m$ with
$E_m>E_n$ and degeneracies $g_n$ and $g_m$, and assume
the molecules have interaction with light. There are three types of
processes that might happen: 
(i) A molecule in state $m$ might decay to state $n$ while omitting 
light with the frequency 
\lbeq{e.rfreq}
\hbar \omega = E_m -E_n;
\eeq
this process is called {\bfi{spontaneous decay}}. 
(ii) A molecule might jump from $n$ to
$m$ by absorbing light with the right frequency \gzit{e.rfreq}. 
(iii) A molecule decays from $m$ to $n$  by being kicked by light 
having the right frequency  \gzit{e.rfreq}; this process is called
{\bfi{induced emission}}. 
Thus, there is one transition which happens even in the absence of 
light: in a spontaneous emission the molecule may jump from $m$ to 
$n$, thereby emitting light. 
The other two transitions take place under the influence of light;
they are therefore dependent of how much light is present and thus
depends on the radiation-energy density $f(\omega)$.

The probabilities of transitions are given as transition rates
$dW/dt$; $dW$ is the infinitesimal difference in molecules in a
certain state and $dt$ is an infinitesimal time interval. 
Now spontaneous emission is independent of the presence of light and 
only depends on the characteristics of the molecule and the number of 
molecules in state $m$. Therefore,
\[
dW_1 = N(m)A_{mn}dt\,,
\]
where $dW_1$ is the number of molecules undergoing spontaneous
emission from $m$ to $n$ during a time interval $dt$, and where
$A_{mn}$ is some number depending on the states $n$ and $m$ (not on
temperature in particular). We denote
$dW_2$ the amount of molecules absorbing light and jumping from $n$ to
$m$ during a time interval $dt$ and $dW_3$ the number of molecules
jumping from $m$ to $n$ under influence of light (getting the right
kick). The probabilities are determined by some constants $B_{mn}$ and
$C_{mn}$, which are characteristic for the states $m$ and $n$ and the
amount of light that has the right frequency. Thus $dW_2$ and $dW_3$
are proportional to $f(\omega)$;
\[
dW_2 = N(n)B_{mn}fdt\,, ~~~ dW_3 = N(m)C_{mn}fdt\,.
\]

Now we consider we have an enclosed system of molecules that are in
equilibrium with the light in the system. Being in equilibrium means
\[
dW_1 + dW_3 = dW_2\,.
\]
Using the Boltzmann distribution we get
\lbeq{boltz.ein}
g_me^{-\frac{E_m}{kT}}(A_{mn}+fC_{mn})= g_n
e^{-\frac{E_n}{kT}}fB_{mn}\,.
\eeq
Now comes a basic assumption that Einstein does;
if $T$ becomes larger the system gets very
hot and transitions will be more and more frequent. Therefore one
assumes that as $T\to \infty$ that also $f\to \infty$. In this case
the exponentials in \gzit{boltz.ein} become $1$ and the term with $A$
can be neglected and we obtain
\lbeq{einst.assump}
g_m C_{mn}=g_nB_{mn}\,.
\eeq
From another point of view the assumption Einstein makes is
natural. The relation $g_m C_{mn}=g_nB_{mn}$ is
representing that the processes $m\to n$ under induced emission, or
$n\to m $ under absorption are symmetric; the numbers $C_{mn}$ and
$B_{mn}$ only differ by the ratio of number of states with energy
$E_n$ to the number of states with energy $E_m$. Indeed, taking
$g_m=g_n=1$, the process of induced emission is the time-reversed
process of absorption. Since the equations in nature show a
{\bfi{time-reversal symmetry}} (in this case) we find in this case
$C_{mn}=B_{mn}$. If now
$g_n$ and $g_m$ are not equal one has to correct for this and
multiply the probabilities with the corresponding multiplicities to
get \gzit{einst.assump}.
With the assumption \gzit{einst.assump} we find
\[
f = \frac{A_{mn}/C_{mn}}{e^{(E_m-E_n)/kT}-1}\,.
\]
Inserting now $E_m-E_n=\hbar \omega$ and requiring that Wien's law
\gzit{wien} holds in the limit where $\omega$ is large we obtain
\[
f(\omega) = \frac{a \omega^3}{e^{\hbar\omega/kT}-1}\,.
\]
In particular we find that $A_{mn}/C_{mn} = a\omega^3$, which relates
the constants $A_{mn}$ and $C_{mn}$ to the energy difference
$E_m-E_n$. The constant $a$ does not depend on the frequency and the
temperature.

\bigskip
\bfi{Modern derivation.}
We now discuss a relatively fast derivation that in addition gives a
value for the constant $a$ in Wien's law \gzit{wien}.
We consider a box with the
shape of a cube with sides $L$. Later we then require that the precise
shape of the box is not relevant in the limit where the typical sizes
are much larger then the wavelength. Then the only relevant parameter
is the volume $V=L^3$. We assume the walls of the box can absorb and
emit light; we furthermore assume that the walls are made of a 
conducting
material. Away from the walls light satisfies Maxwell equations, but
at the walls the perpendicular components of the electric field have
to vanish; if the electric field would not vanish, the electrons in
the material of the wall would be accelerated, but then the system is
not in equilibrium. A plane wave solution to the Maxwell equations is of
the form
\[
e^{i\omega t - ik_xx-ik_yy-ik_zz}\,, ~~~\omega^2 =
c^2(k_{x}^{2}+k_{y}^{2}+k_{z}^{2})\,.
\]
We can always chose a coordinate system that is aligned with the
box. Then the boundary conditions imply $e^{ik_xL}=0$ and thus $k_x =
\frac{\pi n_x}{L}$ for some integer $n_x$. The wave functions with
negative $n_x$ are identical to the corresponding wave functions with
positive $n_x$; they just differ by a phase. Therefore we may assume
$n_x\geq 0$. For the other coordinate directions the discussion is
similar.

Thus we find that for each triple of integer numbers
$\n= (n_x,n_y,n_z)$ we have a harmonic oscillator with frequency
\[
\omega_{n} = \frac{c\pi}{L}n,~~~n=\sqrt{\n \cdot \n}\,.
\]
We now use the fact (proved below in Section \ref{section-reps-heis}) 
that for each harmonic oscillator the energies are 
$E_{n}(r) = \hbar \omega_{n } (r+\shalf)$.
Since energy is 
defined only up to a constant shift, we subtract the zero-point energy 
$E_0=\shalf \hbar \omega_{n}$ and take 
$E_{n}(r)= \omega_{n}r$. The partition function is then
\[
Z(\n,\beta)= \sum_{r=0}^{\infty} e^{r\hbar\omega_n\beta}
= \frac{1}{e^{\hbar\omega_n\beta}-1}\,.
\]
Therefore the average energy in the mode corresponding to $\n$ is
\beqar
\bar E_{n} &=& -\frac{\partial}{\partial \beta }\ln (1-e^{\hbar
  \omega_{n} \beta})\nn\\
&=& \frac{\hbar\omega_{n}}{e^{\hbar\omega_{n}\beta} - 1}\nn\,.
\eeqar
We now have to sum up all the energies for all modes. Since we are
interested in the behavior of $f(\omega)$ in the regime where the number
$L$ is much larger than the wavelength we replace the sum over $\vec
n$ by an integral. We have to integrate over the positive octant where
$n_x\geq 0$, $n_y\geq 0$ and $n_z\geq 0$. Since all expressions are
rotationally symmetric in $\n$, we can also integrate over all of
$\Rz^3$ and divide by $8$. We have not yet taken into account that
light has two polarizations. Therefore, for each $\n$ there are
two harmonic oscillators. The total energy enclosed in the box is thus
\[
E=\frac{2}{8}\int_{\Rz^3} \,dn_xdn_ydn_z\, \bar E_{n}
= \pi \int_{0}^{\infty}\,n^2 dn
\frac{\hbar \omega_n}{e^{\hbar \omega_n\beta }-1}\,.
\]
Here we transformed to polar coordinates and wrote $\omega_n =
\frac{c\pi}{L}\sqrt{n_{x}^{2}+n_{y}^{2}+n_{z}^{2}}$.

We now exchange the integral over $n$ to an integral over $\omega$. We
have
\[
\omega = \frac{c\pi n}{L}\, \Rightarrow \, n^2 dn = \left(\frac{L}{\pi
    c}\right)^3\omega^2d\omega\,,
\]
from which we find
\[
E=
\frac{L^3\hbar}{\pi^2c^3}\int_{0}^{\infty}\,\frac{\omega^3}{e^{\hbar
    \omega\beta}-1}d\omega\,.
\]
With $L^3=V$ being the volume we thus find
\lbeq{mandl.f}
f(\omega)= \frac{V\hbar
}{\pi^2c^3}\frac{\omega^3}{e^{\hbar\omega\beta}-1}\,.
\eeq
Of course, this $f$ only represents the radiation-energy density
inside the black body. However, up to some overall constants the above
$f$ is the radiation-energy density of a black body since the emitted
radiation is proportional to the energy density. 

In the following
section we shall derive Stefan's law.

\section{Stefan's law and Wien's displacement law}
\index{Stefan's law}\index{Wien's displacement law}

From the calculated density \gzit{mandl.f} we can draw some
conclusions, which we now shortly treat.

To calculate the total radiation that is
emitted, we first calculate the total energy by integrating
\gzit{mandl.f} over all $\omega$. We get for the total
energy of the light inside the black body
\[
E =
\frac{Vk^4T^4}{\pi^2c^3\hbar^3}\int_{0}^{\infty}\,\frac{x^3dx}{e^x-1}\,.
\]
But we have
\[
\int_{0}^{\infty}\,\frac{x^3dx}{e^x-1} = \frac{\pi^4}{15}\,,
\]
and thus the energy density $u(T)$ is given by
\[
u(T)=\frac{E}{V} = \frac{\pi^2 k^4 T^4}{15\hbar^3c^3}\,.
\]
We see that the energy density is expressible by fundamental constants
and the fourth power of the temperature. Since the energy density
determines the total radiation emitted per time interval we see that
the total energy a black body radiates per time interval is
proportional to $T^4$. This already explains Stefan's law, but in
order to derive Stefan's law we have to be a bit more careful.

In order to see how much a black body will radiate, we pinch a small
hole in the black body. Let us say that the area of the hole is
$dA$. Now the question is how many photons will hit the hole from
inside out? 
We fix a time $t$ and a small time interval $dt$. Only the photons that 
are within a distance between $ct$ and $ct+cdt$ away from the hole are
eligible to pass through the hole in a time interval
$dt$ after time $t$. We thus consider a thin shell of a half sphere
inside the black body a distance $ct$ away from the hole and of
thickness $cdt$. Light however spreads in all directions and so not
all the photons inside the shell are going in the direction of the
hole. Our task is to find the ratio of the total that does go through
the hole. This is a purely geometric question.

We introduce spherical coordinates around the hole; an angle
$\varphi$ ranging from $0$ to $2\pi$ that goes around the hole,
and a polar angle $\theta$ ranging from $0$ to $\pi/2$ (values below zero
correspond to points outside the black body). We cut the half sphere
of radius $ct$
in little stripes by cutting for fixed $\theta$ along the angle
$\varphi$; each stripe is a thin band of thickness $ctd\vartheta$ and 
of length $2\pi\sin\theta$. Consider a little `cube' of size 
$dV=(ct)^2 \sin\theta d\theta d\varphi cdt$ in the shell. The fraction of
radiation going in the right direction is given by the solid angle
$d\Omega$ that $dA$ describes seen from the little cube. But $d\Omega$
is given by the projection of the surface $dA$ onto the surface of the
sphere of radius $ct$ around the little cube:
\[
d\Omega = \frac{dA \cos\theta}{4\pi c^2t^2}\,.
\]
The cube of volume emits all the radiation present in the cube (since
the light waves just pass through), and that amounts to an energy 
$dE=u(T)dV$.
From the little cube under consideration the amount of radiation going
in the right direction is thus
\[
u d\Omega dV = u(T) \frac{c dA \cos\theta \sin\theta }{4\pi}d\theta
d\varphi dt\,.
\]
Note that the amount of radiation is independent of the radius of the
half sphere. Since the question is of a purely geometric nature, that is
to be expected.
We now get the total amount of radiation from summing up all the
$d\Omega$ contributions: Denoting by $dU$ the energy that leaves the hole
during the time interval $dt$, we have
\beqar
\frac{dU}{dt} &=& 
\int_{0}^{\pi/2}d\theta\int_{0}^{2\pi}d\varphi\,  
u(T) \frac{c dA  \cos\theta \sin\theta }{4\pi} \nn\\
&=& \frac{c}{4}u(T)dA\nn\,.
\eeqar
For a black body that radiates over all its surface, and not only
through one little hole, we sum up the contributions over all little
surfaces $dA$. In order that the above analysis still holds the shape
of the black body needs to be such that radiation that exits the
black body does not enter again. If the black body is convex this
requirement is met, e.g., we could take a sphere. We then find Stefan's
law in the form given by
\[
\frac{dU}{dt} =  \frac{\pi^2 k^4 T^4 A}{60\hbar^3c^2} =  \sigma A T^4\,,
\]
with \bfi{Stefan's constant} 
\[
\sigma = \frac{\pi^2k^4}{60\hbar^3c^2}\sim 5.7\cdot
10^{-8}J.s^{-1}.m^{-2}K^{-4}\,.
\]
We now turn to Wien's displacement law. We write the radiation-energy
density as
\[
f(\omega)=A\frac{\omega^3}{e^{\hbar\omega\beta}-1}\,.
\]
Differentiation with respect to $\omega$ and putting the result to zero
to obtain the position of the maximum  
gives the equations
\[
3-x = 3e^{-x}\,,~~~x=\hbar \omega_{max}\beta\,.
\]
We discard the trivial solution $x=0$ since this corresponds to the
behavior at $\omega=0$. One finds the other solution by solving the
equation $3-x = 3e^{-x}$ with numerical methods and finds $x \sim
2.82$. Hence we have
\[
\omega_{max} \sim \frac{2.82\,T}{\hbar k}\,.
\]

    \part{Statistical mechanics}\label{p.statmech}

\chapter{Phenomenological thermodynamics}\label{c.ctherm}

Part \ref{p.statmech} discusses statistical mechanics from an algebraic 
perspective, concentrating on thermal equilibrium but discussing basic 
things in a more general framework.
A treatment of equilibrium statistical mechanics
and the kinematic part of nonequilibrium statistical mechanics
is given which derives from a single basic assumption (Definition 
\ref{3.1.}) the full structure of phenomenological thermodynamics and 
of statistical mechanics, except for the third law 
which requires an additional quantization assumption.

\bigskip
This chapter gives a concise description of standard phenomenological 
equilibrium thermodynamics for single-phase systems in the absence 
of chemical reactions and electromagnetic fields. 
From the formulas provided, it is an easy step to go to various 
examples and applications discussed in standard textbooks such as
\sca{Callen} \cite{Cal} or \sca{Reichl} \cite{Rei}.
A full discussion of global equilibrium would also involve the
equilibrium treatment of multiple phases and chemical reactions. 
Since their discussion offers no new aspects compared with 
traditional textbook treatments, they are not treated here.

Our phenomenological approach is similar to that of \sca{Callen} 
\cite{Cal}, who introduces the basic concepts by means of a few 
postulates from which everything else follows. 
The present setting is a modified version designed to
match the more fundamental approach based on statistical 
mechanics. By specifying the kinematical properties of states 
outside equilibrium, his informal thermodynamic 
stability arguments (which depend on a dynamical assumption close to 
equilibrium) can be replaced by rigorous mathematical arguments.

\section{Standard thermodynamical systems}\label{s.phen}

We discuss here the special but very important case of thermodynamic 
systems describing the single-phase global equilibrium of matter 
composed of one or several kinds of substances in the absence of 
chemical reactions and electromagnetic fields. We call such systems 
\bfi{standard thermodynamic systems}; they are ubiquitous in 
applications. 
In particular, a standard system is considered to be uncharged, 
homogeneous, and isotropic, so that each finite 
region looks like any other and is very large in microscopic units.

The substances of fixed chemical composition are labeled by an index 
$j\in J$. A standard thermodynamic system is completely characterized 
by\footnote{
In the terminology, we mainly follow the IUPAC convention 
(\sca{Alberty} \cite[Section 7]{Alb}), except that we use the letter
$H$ to denote the Hamilton energy, as customary in quantum mechanics. 
In equilibrium, $H$ equals the internal energy $U$. The Hamilton energy 
should not be confused with the enthalpy which is usually denoted by 
$H$ but here is given in equilibrium by $H+PV$.
For a history of thermodynamics notation, see 
\sca{Battino} et al. \cite{BatSW}.
} 
the \bfi{mole number} $N_j$ of each substance $j$, the corresponding 
\bfi{chemical potential} $\mu_j$ of substance $j$, the \bfi{volume} $V$,
the \bfi{pressure} $P$, the \bfi{temperature} $T$, the \bfi{entropy} 
$S$, and the \bfi{Hamilton energy} $H$. These variables, the 
\bfi{extensive} variables $N_j,V,S,H$ and the \bfi{intensive} variables 
$\mu_j,P,T$, are jointly called the \bfi{basic thermodynamic variables}.
We group the $N_j$ and the $\mu_j$ into vectors $N$ and $\mu$ indexed 
by $J$ and write $\mu\cdot N = \sum_{j\in J} \mu_jN_j$. 
In the special case of a \bfi{pure substance}, there is just a single 
kind of substance; then we drop the indices and have
$\mu\cdot N = \mu N$. In this section, all numbers are real.

The mathematics of thermodynamics makes essential use of the concept 
of convexity. A set $X\subseteq \Rz^n$ is called \bfi{convex} if 
$tx+(1-t)y\in X$  for all $x,y\in X$ and all $t\in[0,1]$.
A real-valued function $\phi$ is called \bfi{convex} on the convex set
$X\subseteq \Rz^n$ if $\phi$ is defined on $X$ and, for all $x,y\in X$,
\[
\phi(tx+(1-t)y) \le t\phi(x)+(1-t)\phi(y) \for 0\le t\le 1.
\]
Clearly, $\phi$ is convex iff for all $x,y\in X$, the function 
$\mu:[0,1]\to \Rz$ defined by
\[
\mu(t):=\phi(x+t(y-x))
\]
is convex. It is well-known that, for twice continuously 
differentiable $\phi$, this is the case iff the second derivative 
$\mu''(t)$ is nonnegative for $0\le t\le 1$. 
Note that by a theorem of Aleksandrov (see \sca{Aleksandrov} \cite{Ale},
\sca{Alberti \& Ambrosio} \cite{AlbA}, \sca{Rockafellar} \cite{Roc.ca}),
convex functions are almost everywhere twice continuously 
differentiable: For almost every $x\in X$, there exist a 
unique vector $\partial\phi(x)\in \Rz^n$, the \bfi{gradient} of $\phi$ 
at $x$, and a unique symmetric, positive semidefinite matrix 
$\partial^2\phi(x)\in \Rz^{n\times n}$, the \bfi{Hessian} of $\phi$ 
at $x$, such that 
\[
\phi(x+h)=\phi(x)+h^T\partial\phi(x)
+\half h^T\partial^2\phi(x)h + o(\|h\|^2)
\]
for sufficiently small $h \in \Rz^n$.
A function $\phi$ is called \bfi{concave} if $-\phi$ is convex. 
Thus, for a twice continuously differentiable function $\phi$ of a 
single variable $\tau$, $\phi$ is concave iff $\mu''(\tau)\le 0$ for 
$0\le \tau\le 1$. 

\begin{prop}\label{e.convex}
If $\phi$ is convex on the convex set $X$ then the function $\psi$ 
defined by
\[
\psi(s,x):=s\phi(x/s)
\]
is convex in the set $\{(s,x)\in \Rz \times X \mid s> 0\}$ 
and concave in the set $\{(s,x)\in \Rz \times X \mid s< 0\}$. 
\end{prop}

\bepf
It suffices to show that $\mu(t):=\psi(s+tk,x+th)$ is convex (concave)
for all $s,x,h,k$ such that $s+tk>0$ (resp. $<0$).
Let $z(t):=(x+th)/(s+tk)$ and $c:=sh-kx$. Then
\[
z'(t)=\frac{c}{(s+tk)^2},~~~\mu(t)=(s+tk)\phi(z(t)),
\]
hence
\[
\mu'(t)=k\phi(z(t))+\phi'(z(t))\frac{c}{s+tk},
\]
\[
\mu''(t)=k\phi'(z(t))\frac{c}{(s+tk)^2}
+\frac{c^T}{(s+tk)^2}\phi''(z(t))\frac{c}{s+tk}
+\phi'(z(t))\frac{-ck}{(s+tk)^2}
=\frac{c^T\phi''(z(t))c}{(s+tk)^3},
\]
which has the required sign.
\epf

Equilibrium thermodynamics is about characterizing so-called
equilibrium states in terms of intensive and extensive variables 
and their relations, and comparing them with similar nonequilibrium 
states. In a nonequilibrium state, only extensive variables have a 
well-defined meaning; but these are not sufficient to characterize 
system behavior completely.

All valid statements in the equilibrium thermodynamics of standard
systems can be deduced from the following definition.

\begin{dfn}\bfi{(Phenomenological thermodynamics)}
\label{d.phen}\\
(i) Temperature $T$, pressure $P$, and volume $V$ are positive,
mole numbers $N_j$ are nonnegative. 
The \bfi{extensive variables} $H,S,V,N_j$ are additive under the 
composition of disjoint subsystems. We combine the $N_j$ into a column 
vector with these components.

(ii) There is a convex \bfi{system function} $\Delta$ of the 
\bfi{intensive variables} $T,P,\mu$ which is monotone increasing in $T$ 
and monotone decreasing in $P$. The intensive variables are related by 
the \bfi{equation of state}
\lbeq{e.def}
\Delta(T,P,\mu)=0.
\eeq
The set of $(T,P,\mu)$ satisfying $T>0$, $P>0$ and the equation of 
state is called the \bfi{state space}. 

(iii) The Hamilton energy $H$ satisfies 
the \bfi{Euler inequality}
\lbeq{e.Hi}
H\ge T S - P V + \mu \cdot N
\eeq
for all $(T,P,\mu)$ in the state space.

(iv) \bfi{Equilibrium states} have well-defined intensive and extensive 
variables satisfying equality in \gzit{e.Hi}.
A system is in \bfi{equilibrium} if it is completely characterized by 
an equilibrium state.
\end{dfn}

This is the complete list of assumptions defining phenomenological 
equilibrium thermodynamics for standard systems; the system function 
$\Delta$ can be determined either by fitting to experimental data, 
or by calculation from a more fundamental description, cf. Theorem 
\ref{t.eos}. All other properties follow from the system function.
Thus, all equilibrium properties of a material are characterized by 
the system function $\Delta$.

Surfaces where the system function is not differentiable 
correspond to so-called \bfi{phase transitions}. 
The equation of state shows that, apart from possible phase 
transitions, the state space has the 
structure of an $(s-1)$-dimensional manifold in $\Rz ^{s}$, where 
$s$ is the number of intensive variables; in case of a standard 
system, the manifold dimension is therefore one higher than the number 
of kinds of substances.

Standard systems describe only a single phase of a substance 
(typically the solid, liquid, or gas phase), and changes between these 
as some thermodynamic variable(s) change.
Thermodynamic systems with multiple phases (e.g., boiling water, or 
water containing ice cubes) are only piecewise homogeneous. Each phase 
may be described separately as a standard thermodynamic system. 
But discussing the equilibrium at the interfaces between different 
phases needs some additional effort. (This is described in all 
common textbooks on thermodynamics.) Therefore, we consider only 
regions of the state space where the system function $\Delta$ is twice 
continuously differentiable.

\bigskip
Each equilibrium instance of the material is characterized by a
particular state $(T,P,\mu)$,
from which all equilibrium properties can be computed:

\begin{thm}~

(i) In any equilibrium state, the extensive variables are given by 
\lbeq{e.Sp}
S=\Omega\frac{\partial \Delta}{\partial T}(T,P,\mu),~~~
V=-\Omega\frac{\partial \Delta}{\partial P}(T,P,\mu),~~~
N=\Omega\frac{\partial \Delta}{\partial \mu}(T,P,\mu),
\eeq
and the \bfi{Euler equation}
\lbeq{e.Hp}
H=T S - P V + \mu \cdot N.
\eeq
Here $\Omega$ is a positive number called the \bfi{system size}. 

(ii) In equilibrium, we have the \bfi{Maxwell reciprocity relations}
\lbeq{7-10p}
-\frac {\partial V} {\partial T}
=\frac {\partial S} {\partial P},~~~
\frac {\partial N_j} {\partial T}
=\frac {\partial S} {\partial\mu_j},~~~
\frac {\partial N_j} {\partial P}
=-\frac {\partial V} {\partial\mu_j},~~~
\frac {\partial N_j} {\partial\mu_k}
=\frac {\partial N_k} {\partial\mu_j},
\eeq
and the \bfi{stability conditions}
\lbeq{7-13p}
\frac {\partial S} {\partial T}\ge 0,~~~ 
\frac {\partial V} {\partial P}\le 0,~~~ 
\frac {\partial N_j} {\partial\mu_j}\ge 0.
\eeq
\end{thm}

\bepf
At fixed $S,V,N$, inequality \gzit{e.Hi} holds in equilibrium with 
equality, by definition. Therefore the triple $(T,P,\mu)$ is a maximizer
of $TS-PV+\mu\cdot N$ under the constraints $\Delta(T,P,\mu)=0$, $T>0$, 
$P>0$. A necessary condition for a maximizer is the stationarity of 
the Lagrangian
\[
L(T,P,\mu)=TS-PV+\mu\cdot N -\Omega\Delta(T,P,\mu)
\]
for some Lagrange multiplier $\Omega$. Setting the partial derivatives 
to zero gives \gzit{e.Sp}, and since the maximum is attained in 
equilibrium, the Euler equation \gzit{e.Hp} follows. The system size
$\Omega$ is positive since $V>0$ and $\Delta$ is decreasing in $P$.
Since the Hessian matrix of $\Delta$, 
\[
\Sigma= 
\left(\begin{array}{rrr}
\vspace{0.3cm}
\D\frac{\partial^2\Delta}{\partial T^2}& 
\D\frac{\partial^2\Delta}{\partial P\partial T}& 
\D\frac{\partial^2\Delta}{\partial \mu\partial T} \\
\vspace{0.3cm}
\D\frac{\partial^2\Delta}{\partial T\partial P}& 
\D\frac{\partial^2\Delta}{\partial P^2}& 
\D\frac{\partial^2\Delta}{\partial \mu\partial P} \\
\D\frac{\partial^2\Delta}{\partial T \partial \mu}& 
\D\frac{\partial^2\Delta}{\partial P \partial \mu}& 
\D\frac{\partial^2\Delta}{\partial\mu^2}
\end{array}\right)
= \Omega^{-1}
\left(\begin{array}{rrr}
\vspace{0.3cm}
\D\frac{\partial S}{\partial T}& 
\D\frac{\partial S}{\partial P}& 
\D\frac{\partial S}{\partial \mu} \\
\vspace{0.3cm}
\D-\frac{\partial V}{\partial T}& 
\D-\frac{\partial V}{\partial P}& 
\D-\frac{\partial V}{\partial \mu} \\
\D\frac{\partial N}{\partial T}& 
\D\frac{\partial N}{\partial P}& 
\D\frac{\partial N}{\partial\mu}
\end{array}\right),
\]
is symmetric, the Maxwell reciprocity relations follow. Since 
$\Delta$ is convex, $\Sigma$ is positive semidefinite; hence the 
diagonal elements of $\Sigma$ are nonnegative, giving the stability 
conditions.
\epf

Note that there are further stability conditions since the determinants
of all principal submatrices of $\Sigma$ must be nonnegative. 
In addition, since $N_j\ge 0$, \gzit{e.Sp} implies that $\Delta$ is 
monotone increasing in each $\mu_j$.

\begin{expl}\label{ex.ideal}
The equilibrium behavior of electrically neutral gases at sufficiently 
low pressure can be modelled as ideal gases.
An \bfi{ideal gas} is defined by a system function of the form
\lbeq{e.ideal}
\Delta(T,P,\mu)=\sum_{j\in J}\pi_j(T) e^{\mu_j/ R T}-P,
\eeq
where the $\pi_j(T)$ are positive functions of the temperature,
\lbeq{e.R}
R\approx 8.31447\, JK^{-1}\mbox{mol}^{-1}
\eeq
is the \bfi{universal gas constant}\footnote{
For the internationally recommended values of this and other constants, 
their accuracy, determination, and history, 
see \sca{CODATA} \cite{CODATA}.
}, 
and we use the bracketing convention $\mu_j/ R T =\mu_j/( R T)$.
Differentiation with respect to $P$ shows that $\Omega=V$ is the 
system size, and from \gzit{e.def}, \gzit{e.Sp}, and \gzit{e.Hp}, we 
find that, in equilibrium,
\[
P=\sum_j \pi_j(T) e^{\mu_j/ R T},~~~
S=V\sum_j \Big(\frac{\partial}{\partial T}\pi_j(T) 
-\frac{\mu_j\pi_j(T)}{ R T^2}\Big) e^{\mu_j/ R T},
\]
\[
N_j =\frac{V\pi_j(T)}{ R T} e^{\mu_j/ R T},~~~
H=V\sum_j \Big(T\frac{\partial}{\partial T}\pi_j(T) 
-\pi_j(T)\Big) e^{\mu_j/ R T}.
\]
Expressed in terms of $T,V,N$, we have
\[
PV= R T \sum_j N_j,~~~\mu_j= R T\log\frac{ R T N_j}{V\pi_j(T)},
\]
\[
H=\sum_j h_j(T)N_j,~~~
h_j(T)= R T\Big(T\frac{\partial}{\partial T}\log\pi_j(T) -1\Big),
\]
from which $S$ can be computed by means of the Euler equation 
\gzit{e.Hp}. 
In particular, for one \bfi{mole} of a single substance, defined by 
$N=1$, we get the \bfi{ideal gas law} 
\lbeq{e.ilaw}
PV=RT
\eeq
discovered by \sca{Clapeyron} \cite{Clap}; cf. \sca{Jensen} \cite{Jen}.

In general, the difference $h_j(T)-h_j(T')$ can be found 
experimentally by measuring the energy needed for raising or lowering 
the temperature of pure substance $j$ from $T'$ to $T$ while keeping 
the $N_j$ constant. In terms of infinitesimal increments, the 
\bfi{heat capacities} 
\[
C_j(T)=dh_j(T)/dT,
\]
we have
\[
h_j(T)=h_j(T')+\int_{T'}^T dT\, C_j(T).
\]
From the definition of $h_j(T)$, we find that
\[
\pi_j(T)=\pi_j(T')\exp \int_{T'}^T \frac{dT}{T}
\Big(1+\frac{h_j(T)}{ R T}\Big).
\]
Thus there are two undetermined integration constants for each kind of
substance. These cannot be determined experimentally as long
as we are in the range of validity of the ideal gas approximation.
Indeed, if we pick arbitrary constants $\alpha_j$ and $\gamma_j$ and 
replace $\pi_j(T),\mu_j,H$, and $S$ by 
\[
\pi_j'(T):=e^{\alpha_j-\gamma_j/ R T}\pi_j(T),~~~
\mu_j'=\mu_j+\gamma_j- R T \alpha_j,
\]
\[
H'=H+\sum_j \alpha_jN_j,~~~S'=S+ R \sum_j \gamma_jN_j,
\]
all relations remain unchanged. Thus, the Hamilton energy and the 
entropy of an ideal gas are only determined up to an arbitrary linear 
combination of the mole numbers. This is an instance of the deeper 
problem to determine under which conditions thermodynamic variables 
are controllable; cf. the discussion in the context of Example 
\ref{ex.gibbs} below. 

This gauge freedom 
(present only in the ideal gas) can be fixed by choosing a particular 
\bfi{standard temperature} $T_0$ and setting arbitrarily $h_j(T_0)=0$, 
$\mu_j(T_0)=0$. 
Alternatively, at sufficiently large temperature $T$, heat capacities 
are usually nearly constant, and making use of the gauge freedom, 
we may simply assume that 
\[
h_j(T)=h_{j0} T,~~~\pi_j(T)=\pi_{j0} T \mbox{~~~for large~} T.
\]
\end{expl}

\section{The laws of thermodynamics}\label{s.cons}

In global equilibrium, all thermal variables are constant 
throughout the system, except at phase boundaries, where the extensive 
variables may exhibit jumps and only the intensive variables remain 
constant. This is sometimes referred to as the \bfi{zeroth law of 
thermodynamics}  (\sca{Fowler \& Guggenheim}\cite{FowG}) and 
characterizes global equilibrium; it allows one to measure intensive 
variables (like temperature) by bringing a calibrated instrument that 
is sensitive to this variable (for temperature a thermometer) into 
equilibrium with the system to be measured. 

For example, the ideal gas law \gzit{e.ilaw} can be used as a basis for 
the construction of a \bfi{gas thermometer}: The amount of expansion of 
volume in a long, thin tube can easily be read off from a scale along 
the tube. 
We have $V=aL$, where $a$ is the cross section area and $L$ is the 
length of the filled part of the tube, hence $T=(aP/R)L$. Thus, at 
constant pressure, the temperature of the gas is proportional to $L$. 
For the history of temperature, see 
\sca{Roller} \cite{Rol} and \sca{Truesdell} \cite{Tru}.

We say that two thermodynamic systems are brought in good \bfi{thermal 
contact} if the joint system tends after a short time to an equilibrium 
state. To measure the temperature of a system,
one brings it in thermal contact with a thermometer and waits until 
equilibrium is established.
The system and the thermometer will then have the same temperature, 
which can be read off from the thermometer. If the system is much larger
than the thermometer, this temperature will be essentially the same 
as the temperature of the system before the measurement.
For a survey of the problems involved in defining and
measuring temperature outside equilibrium, see 
\sca{Casas-V\'asquez \& Jou} \cite{CasJ}.

\bigskip
To be able to formulate the first law of thermodynamics we need the
concept of a reversible change of states, i.e., changes
preserving the equilibrium condition. For use in later sections,
we define the concept in a slightly more general form,
writing $\alpha$ for $P$ and $\mu$ jointly. We need to assume that the 
system under study is embedded into its environment in such a way that, 
at the boundary, certain thermodynamic variables are kept constant 
(and independent of position). This determines the \bfi{boundary 
conditions} of the thermodynamic system; see the discussion in 
Section \ref{s.c1}.

\begin{dfn}
A \bfi{state variable} is an almost everywhere continuously 
differentiable function $\phi(T,\alpha)$ defined on the
state space (or on a subset of it). 
Temporal changes in a state variable that 
occur when the boundary conditions are kept fixed are called 
\bfi{spontaneous changes}. 
A \bfi{reversible transformation} is a continuously differentiable 
mapping 
\[
\lambda \to (T(\lambda ),\alpha(\lambda ))
\]
from a real interval into the state space; thus
$\Delta(T(\lambda ),\alpha(\lambda ))=0$. The \bfi{differential}
\lbeq{3-12}
d\phi=\frac {\partial \phi} {\partial T}dT
+\frac {\partial \phi} {\partial \alpha} \cdot d\alpha, 
\eeq
obtained by multiplying the chain rule by $d\lambda$,
describes the change of a state variable $\phi$ under 
arbitrary (infinitesimal) reversible transformations. 
In formal mathematical terms, differentials are exact linear forms on 
the state space manifold; cf. Chapter \ref{c.manifolds}.
\end{dfn}

Reversible changes per se have nothing to do with changes in time. 
However, by sufficiently slow, quasistatic changes of the boundary 
conditions, reversible changes can often be realized approximately as 
temporal changes. The degree to which this is possible determines the 
efficiency of thermodynamic machines. The analysis of the efficiency
by means of the so-called \bfi{Carnot cycle} was the historical origin 
of thermodynamics.

The state space is often parameterized by 
different sets of state variables, as required by the application. 
If $T=T(\kappa,\lambda)$, $\alpha=\alpha(\kappa,\lambda)$ is such a
parameterization then the state variable $g(T,\alpha)$ can be written
as a function of $(\kappa,\lambda)$,
\lbeq{e.partial0}
g(\kappa,\lambda) = g(T(\kappa,\lambda),\alpha(\kappa,\lambda)).
\eeq
This notation, while mathematically ambiguous, is common in the
literature; the names of the argument decide which function is intended.
When writing partial derivatives without arguments, this leads to
serious ambiguities. These can be resolved by writing 
$\D\Big(\frac{\partial g}{\partial \lambda}\Big)_\kappa$ for the 
partial derivative of \gzit{e.partial0} with respect to $\lambda$;
it can be evaluated using \gzit{3-12}, giving the \bfi{chain rule}
\lbeq{e.partial}
\Big(\frac{\partial g}{\partial \lambda}\Big)_\kappa
=\frac{\partial g} {\partial T}
\Big(\frac{\partial T}{\partial \lambda}\Big)_\kappa
+\frac {\partial g} {\partial \alpha} \cdot
\Big(\frac{\partial \alpha}{\partial \lambda}\Big)_\kappa.
\eeq
Here the partial derivatives in the original 
parameterization by the intensive variables are written without 
parentheses.

Differentiating the equation of state \gzit{e.def}, using the chain 
rule \gzit{3-12}, and simplifying using \gzit{e.Sp} gives the 
\bfi{Gibbs-Duhem equation}
\lbeq{e.GDp}
0=SdT- VdP+N\cdot d\mu
\eeq
for reversible changes. If we differentiate the 
Euler equation \gzit{e.Hp}, we obtain
\[
dH=TdS+SdT-PdV-VdP+\mu\cdot dN+N\cdot d\mu,
\]
and using \gzit{e.GDp}, this simplifies to the 
\bfi{first law of thermodynamics} 
\lbeq{3.1stp}
dH=TdS-Pd V +\mu \cdot dN.
\eeq
Historically, the first law of thermodynamics took on this form only 
gradually, through work by \sca{Mayer} \cite{May},
\sca{Joule} \cite{Jou}, \sca{Helmholtz} \cite{Hel}, and
\sca{Clausius} \cite{Cla}.

Considering global equilibrium from a fundamental point of view, the 
extensive variables are the variables that are conserved or at least 
change so slowly that they may be regarded as time independent on the 
time scale of interest. In the absence of chemical reactions, the 
mole numbers, the entropy, and the Hamilton energy are conserved; 
the volume is a system size variable which, in the fundamental view, 
must be taken as infinite (thermodynamic limit) to exclude the 
unavoidable interaction with the environment. However, real systems 
are always in contact with their environment,
and the conservation laws are approximate only. In thermodynamics, 
the description of the system boundary is generally reduced to the 
degrees of freedom observable at a given resolution.

The result of this reduced description (for derivations, see, e.g., 
\sca{Balian} \cite{Bal}, \sca{Grabert} \cite{Gra}, 
\sca{Rau \& M\"uller} \cite{RauM}) is a dynamical effect called 
\bfi{dissipation} (\sca{Thomson} \cite{Tho}). It is described by the 
\bfi{second law of thermodynamics}, which was discovered by
(\sca{Clausius} \cite{Cla2}.
The Euler inequality \gzit{e.Hi} together with the Euler equation 
\gzit{e.Hp} only express the nondynamical part of the second law since, 
in equilibrium thermodynamics, dynamical questions are ignored: 
Parts (iii)-(iv) of Definition \ref{d.phen} say
that if $S,V,N$ are conserved (thermal, mechanical and 
chemical isolation) then the \bfi{internal energy},
\lbeq{e.int}
U:=TS-PV+\mu\cdot N
\eeq
is minimal in equilibrium; 
\at{formulate more clearly; $T,P,\mu$ have no meaning outside
 equilibrium!}
if $T,V,N$ are conserved (mechanical and 
chemical isolation of a system at constant 
temperature $T$) then the \bfi{Helmholtz (free) energy},
\[
F:=U-TS=-PV+\mu\cdot N 
\]
is minimal in equilibrium; and if $T,P,N$ are conserved (chemical 
isolation of a system at constant temperature $T$ and pressure $P$) 
then the \bfi{Gibbs (free) energy},
\[
G:=F+PV=\mu\cdot N
\] 
is minimal in equilibrium.

The \bfi{third law of thermodynamics}, due to \sca{Nernst} \cite{Ner},
says that entropy is nonnegative. In view of \gzit{e.Sp}, 
this is equivalent to the monotonicity of $\Delta(T,P,\mu)$.

\section{Consequences of the first law}\label{s.c1}

The first law of thermodynamics describes the observable
energy balance in a reversible process. 
The total energy flux $dH$ into the system is composed of the 
\bfi{thermal energy flux} or \bfi{heat flux} $TdS$,
the \bfi{mechanical energy flux} $-PdV$, and the 
\bfi{chemical energy flux} $\mu \cdot dN$. 

The Gibbs-Duhem equation \gzit{e.GDp} describes the energy balance 
necessary to compensate the changes $d(TS)=TdS+SdT$ of
thermal energy, $d(PV)=Pd V + V dP$ of
mechanical energy, and $d(\mu \cdot N)=\mu \cdot dN+N\cdot d\mu$ 
of chemical energy in the energy contributions to the Euler equation 
to ensure that the Euler equation
remains valid during a reversible transformation. Indeed, both
equations together imply that $d(TS-PV+\mu\cdot N -H)$
vanishes, which expresses the preservation of the Euler equation.

Related to the various energy fluxes are the \bfi{thermal work}
\[
Q = \int T(\lambda)dS(\lambda),
\]
the \bfi{mechanical work}
\[
W_\fns{mech} = -\int P(\lambda)dV(\lambda),
\]
and the \bfi{chemical work}
\[
W_\fns{chem} = \int \mu(\lambda)\cdot dN(\lambda)
\]
performed in a reversible transformation. The various kinds of work 
generally depend on the path through the state space; however, the
mechanical work depends only on the end points if the associated
process is conservative. 

As is apparent from the formulas given, thermal work is done by
changing the entropy of the system, mechanical work by changing the 
volume, and chemical work by changing the mole numbers. 
In particular, in case of thermal, mechanical, or chemical
\bfi{isolation}, the corresponding fluxes vanish identically. 
Thus, constant $S$ characterizes thermally isolated, 
\bfi{adiabatic} systems, constant $V$ characterizes mechanically 
isolated, systems, and constant $N$ characterizes chemically isolated,  
\bfi{closed}\footnote{
Note that the terms 'closed system' has also a 
much more general interpretation -- which we do {\em not} use in this 
chapter --, namely as a conservative dynamical system.
} 
or \bfi{impermeable} systems. 
Note that this constancy only holds when all assumptions for a 
standard system are valid: global equilibrium, a single phase, and 
the absence of chemical reactions.
Of course, these boundary conditions are somewhat idealized situations, 
which, however, can be approximately realized in practice and are of 
immense scientific and technological importance.

The first law shows that, in appropriate units, the temperature $T$ is 
the amount of energy needed to increase in a mechanically and 
chemically isolated system the entropy $S$ by one unit. 
The pressure $P$ is, in appropriate units, the amount of energy needed 
to decrease in a thermally and chemically isolated system the volume 
$V$ by one unit. In particular, increasing pressure decreases the 
volume; this explains the minus sign in the definition of $P$.
The chemical potential $\mu_j$ is, in appropriate units, the amount of 
energy needed to increase in a thermally and mechanically isolated 
system the mole number $N_j$ by one. With the traditional units, 
temperature, pressure, and chemical potentials are no longer energies.

We see that the entropy and the volume behave just like the mole
number. This analogy can be deepened by observing that mole numbers 
are the natural measure of the amounts of ``matter'' of each kind in a 
system, and chemical energy flux is accompanied by adding or removing 
matter.
Similarly, volume is the natural measure of the amount of ``space'' a
system occupies, and mechanical energy flux in a standard system is
accompanied by adding or removing space.
Thus we may regard entropy as the natural measure of the amount of 
``heat'' contained in a system\footnote{
Thus, entropy is the modern replacement for the historical concepts of 
\bfi{phlogiston} and \bfi{caloric}, which failed to give a correct 
account of heat phenomena. 
Phlogiston turned out to be ``missing oxygen'', an early 
analogue of the picture of positrons as holes, ``missing electrons'', 
in the Dirac sea. Caloric was a massless substance of heat which had 
almost the right properties, explained many effects correctly, and 
fell out of favor only after it became known that caloric could be
generated in arbitrarily large amounts from mechanical energy, thus
discrediting the idea of heat being a substance. (For the precise 
relation of entropy and caloric, see \sca{Kuhn} \cite{Kuh1,Kuh2},
\sca{Walter} \cite{Walt}, and the references quoted there.) 
In the modern picture,
the extensivity of entropy models the substance-like properties of
the colloquial term ``heat''. But as there are no particles 
of space whose mole number is proportional to the volume, so there are 
no particles of heat whose mole number is proportional to the entropy.
Nevertheless, the introduction of heat particles on a formal level 
has some uses; see, e.g., \sca{Streater} \cite{Str}.
},  
since thermal energy flux is accompanied by adding or removing heat.
Looking at other extensive quantities, we also recognize energy as the 
natural measure of the amount of ``power'' (colloquial), 
momentum as the natural measure of the amount of ``force'' (colloquial),
and mass as the natural measure of the amount of ``inertia'' 
(colloquial) of a system. 
In each case, the notions in quotation marks are the colloquial terms 
which are associated in ordinary life with the more precise, formally 
defined physical quantities. For historical reasons, the words heat, 
power, and force are used in physics with a meaning different from the 
colloquial terms ``heat'', ``power'', and ``force''.

\section{Consequences of the second law}\label{s.c2}

The second law is centered around the impossibility of perpetual 
motion machines due to the inevitable loss of energy by  
dissipation such as friction (see, e.g., \sca{Bowden \& Leben} 
\cite{BowL}), uncontrolled radiation, etc..
This means that -- unless continually provided from the outside --
energy is lost with time until a metastable state is attained,
which usually is an equilibrium state. Therefore, the energy at
equilibrium is minimal under the circumstances dictated by the 
boundary conditions. In a purely kinematic setting as in our treatment, 
the approach to equilibrium cannot be studied, and only the  
minimal energy principles -- one for each set of boundary conditions -- 
remain. 

Traditionally, the second law is often expressed in the form of
an extremal principle for some thermodynamic potential.
We derive here the extremal principles for the Hamilton energy,
the Helmholtz energy, and the Gibbs energy\footnote{
The different potentials are related by so-called Legendre transforms;
cf. \sca{Rockafellar} \cite{Roc} for the mathematical properties of 
Legendre transforms, \sca{Arnol'd} \cite{Arn} for their application 
in mechanics, and \sca{Alberty} \cite{Alb} for their application in 
chemistry.
}, 
which give rise to the \bfi{Hamilton potential}
\[
U(S,V,N) :=\max_{T,P,\mu}\,
\{TS-PV+\mu\cdot N\mid \Delta(T,P,\mu)=0;T>0;P>0\},
\]
the \bfi{Helmholtz potential}
\[
F(T,V,N):=\max_{P,\mu}\,
\{-PV+\mu\cdot N\mid \Delta(T,P,\mu)=0;T>0;P>0\},
\]
and the \bfi{Gibbs potential}
\[ 
G(T,P,N):=\max_\mu\,
\{\mu\cdot N\mid \Delta(T,P,\mu)=0;T>0;P>0\}.
\]
The Gibbs potential is of particular importance for everyday 
processes since the latter frequently happen at approximately constant 
temperature, pressure, and mole number. 
(For other thermodynamic potentials used in practice, see 
\sca{Alberty} \cite{Alb}; for the maximum entropy 
principle, see Section \ref{s.maxent}.)

\begin{thm}\label{t.extstd}
\bfi{(Extremal principles)}\\
(i) In an arbitrary state, 
\lbeq{e.2ndU}
H \ge U(S,V,N),
\eeq
with equality iff the state is an equilibrium state. 
The remaining thermodynamic variables are then given by
\[
T = \frac{\partial}{\partial S}U(S,V,N),~~~
P = -\frac{\partial}{\partial V}U(S,V,N),~~~
\mu = \frac{\partial}{\partial N}U(S,V,N),~~~
H = U(S,V,N).
\]
In particular, an equilibrium state is uniquely determined by 
the values of $S$, $V$, and $N$.

(ii) In an arbitrary state, 
\lbeq{e.2ndF}
H-TS \ge F(T,V,N),
\eeq
with equality iff the state is an equilibrium state.
The remaining thermodynamic variables are then given by
\[
S=-\frac{\partial F}{\partial T}(T,V,N),~~~
P=-\frac{\partial F}{\partial V}(T,V,N),~~~
\mu=\frac{\partial F}{\partial N}(T,V,N),
\]
\[
H=F(T,V,N)+TS.
\]
In particular, an equilibrium state is uniquely determined by 
the values of $T$, $V$, and $N$.

(iii) In an arbitrary state,
\lbeq{e.2ndG}
H-TS+PV \ge G(T,P,N),
\eeq
with equality iff the state is an equilibrium state.
The remaining thermodynamic variables are then given by
\[
S=-\frac{\partial G}{\partial T}(T,P,N),~~~
V=\frac{\partial G}{\partial P}(T,P,N), ~~~
\mu=\frac{\partial G}{\partial N}(T,P,N),
\]
\[
H=G(T,P,N)+TS-PV.
\]
In particular, an equilibrium state is uniquely determined by 
the values of $T$, $P$, and $N$.
\end{thm}

\bepf
We prove (ii); the other two cases are entirely similar.
\gzit{e.2ndF} and the statement about equality is a direct consequence 
of Axiom \ref{d.phen}(iii)--(iv). Thus, the difference $H-TS-F(T,V,N)$ 
takes its minimum value zero at the equilibrium value of $T$. 
Therefore, the derivative with respect to $T$ vanishes, which gives the
formula for $S$. To get the formulas for $P$ and $\mu$, we note that 
for constant $T$, the first law \gzit{3.1stp} implies
\[
dF=d(H-TS)=dH-TdS=-PdV+\mu\cdot dN.
\]
For the reversible transformation which only changes $P$ or $\mu_j$,
we conclude that $dF=-PdV$ and $dF=\mu\cdot dN$, respectively.
Solving for $P$ and $\mu_j$, respectively, implies the formulas for
$P$ and $\mu_j$.
\epf

The above results imply that one can regard each thermodynamic 
potential as a complete alternative way to describe the manifold of 
thermal states and hence all equilibrium properties.
This is very important in practice, where one usually describes
thermodynamic material properties in terms of the Helmholtz or Gibbs 
potential, using models like NRTL (\sca{Renon \& Prausnitz} \cite{RenP},
\sca{Prausnitz} et al. \cite{PraLA})
or SAFT (\sca{Chapman} et al. \cite{ChaGJR,ChaGJR2}).

The additivity of extensive quantities is reflected in the 
corresponding properties of the thermodynamic potentials:

\begin{thm}\label{t.ext}
The potentials $U(S,V,N)$, $F(T,V,N)$, and $G(T,P,N)$
satisfy, for real $\lambda,\lambda^1,\lambda^2\ge 0$,
\lbeq{e.homUx}
U(\lambda S,\lambda V,\lambda N)=\lambda U(S,V,N),
\eeq
\lbeq{e.homFx}
F(T,\lambda V,\lambda N)=\lambda F(T,V,N),
\eeq
\lbeq{e.homGx}
G(T,P,\lambda N)=\lambda G(T,P,N),
\eeq
\lbeq{e.convUx}
U(\lambda^1 S^1+\lambda^2S^2,\lambda^1 V^1+\lambda^2V^2,
\lambda^1 N^1+\lambda^2N^2)
\le \lambda^1 U(S^1,V^1,N^1)+\lambda^2 U(S^2,V^2,N^2),
\eeq
\lbeq{e.convFx}
F(T,\lambda^1 V^1+\lambda^2V^2,\lambda^1 N^1+\lambda^2N^2)
\le \lambda^1 F(T,V^1,N^1)+\lambda^2 F(T,V^2,N^2),
\eeq
\lbeq{e.convGx}
G(T,P,\lambda^1 N^1+\lambda^2N^2)
\le \lambda^1 G(T,P,N^1)+\lambda^2 G(T,P,N^2).
\eeq
In particular, these potentials are convex in $S$, $V$, and $N$.
\end{thm}

\bepf
The first three equations express homogeneity and are a direct 
consequence of the definitions. Inequality \gzit{e.convFx} holds since, 
for suitable $P$ and $\mu$,
\[
\bary{lll}
F(T,\lambda^1 V^1+\lambda^2V^2,\lambda^1 N^1+\lambda^2N^2)
&=&-P(\lambda^1 V^1+\lambda^2V^2)+\mu\cdot(\lambda^1 N^1+\lambda^2N^2)\\
&=&\lambda^1(-PV^1+\mu\cdot N^1)+\lambda^2(-PV^2+\mu\cdot N^2)\\
&\le& \lambda^1 F(T,V^1,N^1)+\lambda^2 F(T,V^2,N^2);
\eary
\]
and the others follow in the same way.
Specialized to $\lambda^1+\lambda^2=1$, the inequalities express the
claimed convexity.
\epf

For a system at constant temperature $T$, pressure $P$, and mole
number $N$, consisting of
a number of parts labeled by a superscript $k$ which are separately
in equilibrium, the Gibbs energy is extensive, since
\[
\bary{lll}
G&=&H-TS+PV= \D\sum H^k-T\sum S^k+P\sum V^k \\
&=& \D\sum (H^k-TS^k+PV^k)=\sum G^k.
\eary
\]
Equilibrium requires that $\sum G^k$ is minimal among all choices 
with $\sum N^k=N$, and by introducing a Lagrange multiplier vector 
$\mu^*$ for the constraints, we see that in equilibrium, the derivative 
of $\sum (G(T,P,N^k)-\mu^*\cdot N^k)$ with respect to each $N^k$ 
must vanish. This implies that 
\[
\mu^k= \frac{\partial G}{\partial N^k}(T,P,N^k)=\mu^*.
\]
Thus, in equilibrium, all $\mu^k$ must be the same.  
At constant $T$, $V$, and $N$, one can apply the same argument to the 
Helmholtz potential, and at constant $S$, $V$, and $N$ to the 
Hamilton potential. In each case, the equilibrium is characterized
by the constancy of the intensive parameters.

The degree to which 
macroscopic space and time correlations are absent characterizes
the amount of \bfi{macroscopic disorder} of a system.
Global equilibrium states are therefore macroscopically highly uniform; 
they are the most ordered macroscopic states in the universe rather 
than the most disordered ones. 
A system not in global equilibrium is characterized by macroscopic 
local inhomogeneities, indicating that the space-independent global 
equilibrium variables alone are not sufficient to describe the system.
Its intrinsic complexity is apparent only in a microscopic treatment;
cf. Section \ref{s.complexity} below.
The only macroscopic shadow of this complexity is the critical
opalescence of fluids near a critical point (\sca{Andrews} \cite{And}, 
\sca{Forster} \cite{For}). The contents of the second law of
thermodynamics for global equilibrium states may therefore be phrased 
informally as follows:
{\em In global equilibrium, macroscopic order (homogeneity) is perfect 
and microscopic complexity is maximal}.
In particular, the traditional interpretation of entropy
as a measure of disorder is often misleading.
Much more carefully argued support for this statement, with numerous
examples from teaching practice, is in \sca{Lambert} \cite{Lam}.

\begin{thm} \label{4.1.} (\bfi{Entropy form of the second law})\\
In an arbitrary state of a standard thermodynamic system 
\[
S \le S(H,V,N)
:=\min\,\{T^{-1}(H+PV-\mu\cdot N)\mid \Delta(T,P,\mu)=0\},
\]
with equality iff the state is an equilibrium state.
The remaining thermal variables are then given by
\lbeq{e.ent1x}
T^{-1}=\frac{\partial S}{\partial H}(H,V,N),~~~
T^{-1}P=\frac{\partial S}{\partial V}(H,V,N),~~~
T^{-1}\mu=-\frac{\partial S}{\partial N}(H,V,N),
\eeq
\lbeq{e.ent2x}
U=H=TS(T,V,N)-PV+\mu\cdot N.
\eeq
\end{thm}

\bepf
This is proved in the same way as Theorem \ref{t.extstd}.
\epf

This result implies that when a system in which $H$, $V$ and $N$ are 
kept constant reaches equilibrium, the entropy must have increased.
Unfortunately, the assumption of constant $H$, $V$ and $N$
is unrealistic; such constraints are not easily realized in nature.
Under different constraints\footnote{
For example, if one pours milk into a cup of coffee, stirring mixes
coffee and milk, thus increasing complexity. Macroscopic order is
restored after some time when this increased complexity has become
macroscopically inaccessible. Since $T,P$ and $N$ are constant, 
the cup of coffee ends up in a 
state of minimal Gibbs energy, and not in a state of maximal entropy!
More formally, the first law shows that, for standard systems at fixed 
value of the mole number, the value of the entropy decreases 
when $H$ or $V$ (or both) decrease reversibly; this shows that the 
value of the entropy 
may well decrease if accompanied by a corresponding decrease of 
$H$ or $V$. The same holds out of equilibrium (though our 
equilibrium argument no longer applies); for example, the reaction 
2 H${}_2$ $+$ O${}_2$ $\to$ 2 H${}_2$O (if catalyzed) 
may happen spontaneously at constant $T=25\,^\circ$C and $P=1$~atm, 
though it decreases the entropy. 
}, 
the entropy is no longer maximal. 

In systems with several phases, a naive interpretation of the second 
law as moving systems towards increasing disorder is even more 
inappropriate: 
A mixture of water and oil spontaneously separates, thus ''ordering'' 
the water molecules and the oil molecules into separate phases! 

Thus, while the second law in the 
form of a maximum principle for the entropy has some theoretical and 
historical relevance, it is not the extremal principle ruling nature.
The irreversible nature of physical processes is instead manifest as 
\bfi{energy dissipation} which, in a microscopic interpretation, 
indicates 
the loss of energy to the unmodelled microscopic degrees of freedom.
Macroscopically, the global equilibrium states are therefore states 
of least free energy, the correct choice of which depends on the 
boundary condition, with the least possible freedom for change. 
This macroscopic immutability is another intuitive explanation for the 
maximal macroscopic order in global equilibrium states.

\section{The approach to equilibrium}\label{s.appeq}

Using only the present axioms, one can say a little bit about the
behavior of a system close to equilibrium in the following,
idealized situation.
Suppose that a system at constant $S$, $V$, and $N$ which is close to 
equilibrium at some time $t$ reaches equilibrium at some later time 
$t^*$. Then the second law implies 
\[
0\le H(t)-H(t^*) \approx (t-t^*)\frac{dH}{dt},
\]
so that $dH/dt\le 0$. We assume that the system is composed of two 
parts, which are both in equilibrium at times $t$ and $t^*$. Then
the time shift induces on both parts a reversible transformation, 
and the first law can be applied to them. Thus
\[
dH=\sum_{k=1,2} dH^k =\sum_{k=1,2} (T^kdS^k-P^kdV^k+\mu^k\cdot dN^k).
\]
Since $S$, $V$, and $N$ remain constant, we have $dS^1+dS^2=0$,
$dV^1+dV^2=0$, $dN^1+dN^2=0$, and since for the time shift $dH\le 0$,
we find the inequality
\[
0\ge (T^1-T^2)dS^1 - (P^1-P^2)dV^1 +(\mu^1-\mu^2)\cdot dN^1.
\]
This inequality gives information about the direction of the flow 
in case that all but one of the extensive variables are known to be 
fixed.

In particular, at constant $V^1$ and $N^1$, we have $dS^1\le 0$ if 
$T^1>T^2$; i.e., ''heat'' (entropy) flows from the hotter part towards 
the colder part. At constant $S^1$ and $N^1$, we have $dV^1\le 0$ if 
$P^1<P^2$; i.e., ''space'' (volume) flows from lower pressure to 
higher pressure: the volume of the lower pressure part decreases and 
is compensated by a corresponding increase of the volume in the higher 
pressure part. And for a pure substance at constant $S^1$ and 
$V^1$, we have $dN^1\le 0$ if $\mu^1>\mu^2$; i.e., ''matter'' (mole 
number) flows from higher chemical potential towards lower chemical 
potential. These qualitative results give temperature, pressure,
and chemical potential the familiar intuitive interpretation. 

This glimpse on nonequilibrium properties is a shadow of the far 
reaching fact that, in nonequilibrium 
thermodynamics, the intensive variables behave like potentials whose
gradients induce forces that tend to diminish these gradients,
thus enforcing (after the time needed to reach equilibrium) agreement 
of the intensive variables of different parts of a system. 
In particular, temperature acts as a thermal potential, whose 
differences create thermal forces which induce thermal currents, 
a flow of ''heat'' (entropy), in a similar way as differences in 
electrical potentials create electrical currents, a flow of 
''electricity'' (electrons)\footnote{
See Table \ref{3.t.} for more parallels in other thermodynamic systems,
and \sca{Fuchs} \cite{Fuc} for a thermodynamics course (and for a 
German course \sca{Job} \cite{Job}) 
thoroughly exploiting these parallels. 
}. 
While these dynamical issues are outside the scope 
of the present work, they motivate the fact that one can control some
intensive parameters of the system by controlling the corresponding 
intensive parameters of the environment and making the walls permeable 
to the corresponding extensive quantities. This
corresponds to standard procedures familiar to everyone from ordinary 
life, such as: heating to change the temperature; applying pressure 
to change the volume; immersion into a substance to change the chemical 
composition; or, in the more general thermal models discussed in 
Section \ref{s.detail}, applying forces to displace an object.

The stronger nonequilibrium version of the second law says that 
(for suitable boundary conditions) equilibrium is actually attained 
after some time (strictly speaking, only in the limit of infinite time).
This implies that the energy difference 
\[
\delta E:=H-U(S,V,N)=H-TS-F(S,V,N)=H-TS+PV=G(S,V,N)
\]
is the amount of energy that is dissipated in order to reach 
equilibrium. In an equilibrium setting, we 
can only compare what happens to a system prepared in a nonequilibrium 
state assuming that, subsequently, the full energy difference 
$\delta E$ is dissipated so that the system ends up in an equilibrium 
state. Since few variables describe everything of interest, this 
constitutes the power of equilibrium thermodynamics. But this power is 
limited, since equilibrium thermodynamics is silent about when -- or 
whether at all -- equilibrium is reached. Indeed, in many cases, only 
metastable states are reached, which change too slowly to ever reach 
equilibrium on a human time scale. Typical examples of this are crystal 
defects, which constitute nonglobal minima of the free energy -- the 
globasl minimum would be a perfect crystal.

\section{Description levels}

As we have seen, extensive and intensive variables play completely 
different roles in equilibrium thermodynamics. Extensive variables 
such as mass, charge, or volume depend additively on the size of the 
system. The conjugate intensive variables act as parameters defining 
the state. 

A system composed of many small subsystems, each in equilibrium,
needs for its complete characterization the values of the extensive and
intensive variables in each subsystem. Such a system
is in global equilibrium only if its intensive variables are independent
of the subsystem. On the other hand, the values of the extensive
variables may jump at phase space boundaries, if (as is the case for 
multi-phase systems) the equations of state allow multiple values for 
the extensive variables to correspond to the same values of the 
intensive variables.
If the intensive variables are not independent of the subsystem then,
by the second law, the differences in the intensive variables of 
adjacent subsystems give rise to thermodynamic forces trying to move 
the system towards equilibrium. 

A real nonequilibrium system does not actually consist of subsystems in
equilibrium; however, typically, smaller and smaller pieces behave 
more and more like equilibrium systems. Thus we may view a real system
as the continuum limit of a larger and larger number of smaller and 
smaller subsystems, each in approximate equilibrium. As a result, the 
extensive and intensive variables become fields depending on the 
continuum variables used to label the subsystems. 
For extensive 
variables, the integral of their fields over the label space gives 
the bulk value of the extensive quantity; thus the fields themselves 
have a natural interpretation as a density. For intensive variables,
an interpretation as a density is physically meaningless; instead,
they have a natural interpretation as field strengths. 
The gradients of their fields have physical significance as the 
sources for thermodynamic forces. 

From this field theory perspective, the extensive variables in the 
single-phase global equilibrium case have constant densities, and 
their bulk values are the densities multiplied by the 
system size (which might be mass, or volume, or another additive 
parameter), hence scale linearly with the size of the system, while 
intensive variables are invariant under a change of system size. 
We do {\em not} use the alternative convention to call extensive any
variable that scales linearly with the system size, and intensive
any variable that is invariant under a change of system size.

We distinguish four nested levels of 
thermal descriptions, depending on whether the system is considered 
to be in global, local, microlocal, or quantum 
equilibrium. The highest and computationally simplest level, 
\bfi{global equilibrium}, is concerned with macroscopic situations 
characterized by finitely many space- and time-independent variables. 
The next level, \bfi{local equilibrium}, treats macroscopic situations 
in a continuum mechanical description, where the equilibrium 
subsystems are labeled by the space coordinates. Therefore the relevant 
variables are finitely many space- and time-dependent fields.
The next deeper level, \bfi{microlocal}\footnote{
The term microlocal for a phase space dependent analysis is taken
from the literature on partial differential equations; see, e.g., 
\sca{Martinez} \cite{Mar}.
}~ 
\bfi{equilibrium}, treats mesoscopic situations 
in a kinetic description, where the equilibrium subsystems are labeled 
by phase space coordinates. The relevant variables are now finitely 
many fields depending on time, position, and momentum; 
cf. \sca{Balian} \cite{Bal2}. The bottom level is the 
microscopic regime, where we must consider \bfi{quantum equilibrium}.
This no longer fits a thermodynamic framework but must be described 
in terms of quantum dynamical semigroups; see Section \ref{s.model}.

The relations between the different description levels are
discussed in Section \ref{s.model}. Apart from descriptions on these 
clear-cut levels, there are also various hybrid descriptions, where 
some part of a system is described on a more detailed 
level than the remaining parts, or where, as for stirred chemical 
reactions, the fields are considered to be spatially homogeneous and 
only the time-dependence matters. 

What was said at the beginning of Section \ref{s.cons} about measuring 
intensive variables like temperature applies in principle also in local 
or microlocal equilibrium, but with fields in place of variables. 
The extensive variables are now densities represented by distributions 
that can be meaningfully integrated over bounded regions (the domains 
of contact with a measuring instrument), whereas intensive variables 
are nonsingular fields (e.g., pressure) whose integrals denote -- 
after divistion by the size of the domain of contact with an instrument
-- a local mean value of the fields.

\chapter{Quantities, states, and statistics}\label{c.quants}

When considered in sufficient 
detail, no physical system is truly in global equilibrium; one can 
always find smaller or larger deviations. To describe these deviations, 
extra variables are needed, resulting in a more complete but also more 
complex model. At even higher resolution, this model is again 
imperfect and an approximation to an even more complex, better model. 
This refinement process may be repeated in several stages. 
At the most detailed stages, we transcend the frontier of 
current knowledge in physics, but even as this frontier recedes, 
deeper and deeper stages with unknown details are imaginable.

Therefore, it is desirable to have a meta-description of thermodynamics
that, starting with a detailed model, allows to deduce the properties 
of each coarser model, in a way that all description levels are 
consistent with the current state of the art in physics. 
Moreover, the results should be as independent as possible of unknown 
details at the lower levels.
This meta-description is the subject of \bfi{statistical mechanics}. 

This chapter introduces the technical machinery of statistical 
mechanics, Gibbs states and the partition function, in a uniform 
way common to classical mechanics and quantum mechanics. 
As in the phenomenological case, the intensive variables determine 
the state (which now is a more abstract object), whereas the extensive 
variables now appear as values of other abstract objects called 
quantities. This change of setting allows the natural incorporation 
of quantum mechanics, where quantities need not commute, while 
values are numbers observable in principle, hence must 
satisfy the commutative law.

The operational meaning of the abstract concepts of quantities, 
states and values introduced in the following becomes apparent once we 
have recovered the phenomenological results of Chapter \ref{c.ctherm} 
from the abstract theory developped in this and the next chapter. 
Chapter \ref{c.models} discusses in more detail how the theory relates 
to experiment.

\at{adapt Section 1.5 to match the contents}

\section{Quantities}\label{s.quantities}

Any fundamental description of physical systems must give account of 
the numerical values of quantities observable in experiments when the 
system under consideration is in a specified state. Moreover, the form 
and meaning of states, and of what is observable in principle, must be 
clearly defined. We consider an axiomatic conceptual 
foundation on the basis of quantities\footnote{
Quantities are formal, numerical properties associated to a given 
system in a given state.
We deliberately avoid the notion of observables, since it is not clear 
on a fundamental level what it means to `observe' something, and since 
many things (such as the fine structure constant, neutrino masses, 
decay rates, scattering cross sections) which can be observed in nature 
are only indirectly related to what is traditionally called an 
`observable' in quantum mechanics. The related problem of how to 
interpret measurements is discussed in Section \ref{s.measurement}.
} 
and their values, consistent with the conventions adopted by the 
International System of Units (SI) \cite{SI}, who declare:
''{\em A quantity in the general sense 
is a property ascribed to phenomena, bodies, or substances that can 
be quantified for, or assigned to, a particular phenomenon, 
body, or substance. [...] 
The value of a physical quantity is the quantitative expression
of a particular physical quantity as the product of a number and a
unit, the number being its numerical value.}'' 

In different states, the quantities of a given system may have 
different values; the state (equivalently, the values determined by it) 
characterizes an individual system at a particular time.
Theory must therefore define what to consider as quantities, 
what as states, and how a state assigns values to a quantity.
Since quantities can be added, multiplied, compared, and integrated, 
the set of all quantities has an elaborate structure whose properties 
we formulate after the discussion of the following motivating example.

\begin{example}\label{ex.Nlevel}
As a simple example satisfying the axioms to be 
introduced, the reader may think of an $N$-level quantum system.
\at{The reaminder of this example is now in the new Chapter 
on matrix groups.}
The \bfi{quantities} are the elements of the algebra
$\Ez=\Cz^{N\times N}$ of square complex $N\times N$ matrices, the 
\bfi{constants} are the multiples of the identity matrix, the 
\bfi{conjugate} $f^*$ of $f$ is given by conjugate transposition, and 
the \bfi{integral} $\sint g = \tr g$ is the \bfi{trace}, the sum of the 
diagonal entries or, equivalently, the sum of the eigenvalues. 
The standard basis consisting of the $N$ \bfi{unit vectors} \idx{$e^k$} 
with a one in component $k$ and zeros in all other component 
corresponds to the $N$ levels of the quantum systems. The Hamiltonian
$H$ is represented by a diagonal matrix $H=\Diag(E_1,\dots,E_N)$
whose diagonal entries $E_k$ are the \bfi{energy levels} of the system.
In the nondegenerate case, all $E_k$ are distinct, and the diagonal 
matrices comprise all functions of $H$. Quantities given by
arbitrary nondiagonal matrices are less easy to interpret. However,
an important class of quantities are the matrices of the form
$P=\psi\psi^*$, where $\psi$ is a vector of norm 1; they satisfy 
$P^2=P=P^*$ and are the quantities observed in binary measurements
such as detector clicks; see Section \ref{s.qprob}.
The \bfi{states} of the $N$-level system are mappings defined by a 
\bfi{density matrix} $\rho\in\Ez$, a positive semidefinite Hermitian 
matrix with trace one, assigning to each quantity $f\in\Ez$ the 
\bfi{value} $\<f\>=\tr \rho f$ of $f$ in this state. The diagonal 
entries $p_k:=\rho_{kk}$ represent the probability for obtaining a 
response in a binary test for the $k$th quantum level; the off-diagonal 
entries $\rho_{jk}$ represent deviations from a classical mixture of 
quantum levels. 
\end{example}

\begin{dfn} ~\\
(i) A \bfi{$*$-algebra} is a set $\Ez$ whose elements are called 
\bfi{quantities}, together 
with operations on $\Ez$ defining for any two quantities $f,g\in\Ez$ 
the \bfi{sum} $f+g\in\Ez$, the 
\bfi{product} $fg\in\Ez$, and the \bfi{conjugate} $f^*\in\Ez$,
such that the following axioms (Q1)--(Q4) hold for all $\alpha\in\Cz$ 
and all $f,g,h\in\Ez$:

(Q1) 
~$\Cz \subseteq \Ez$, i.e., complex numbers are special elements
called \bfi{constants}, for which addition, multiplication and 
conjugation have their traditional meaning. 

(Q2)
~{$(fg)h=f(gh)$,~~ $\alpha f=f\alpha $,~~ $0f=0$,~~ $1f=f$.}

(Q3)
~{$(f+g)+h=f+(g+h)$,~~ $f(g+h)=fg+fh$,~~ $f+0=f$.}

(Q4)
~{$f^{**}=f$,~~ $(fg)^* =g^* f^* $,~~ $(f+g)^* =f^* +g^*$.}

(ii) A $*$-algebra $\Ez$ is called \bfi{commutative} if $fg=gf$ for 
all $f,g\in\Ez$, and  \bfi{noncommutative} otherwise.
The $*$-algebra $\Ez$ is called \bfi{nondegenerate} if

(Q5)
~{$f^* f =0 \implies f =0$.}

(iii) We introduce the notation
\[
-f:=(-1)f,~~ f-g:=f+(-g), ~~~[f,g]:=fg-gf,
\]
\[
f^0:=1,~~ f^l:=f^{l-1}f~~~ (l=1,2,\dots ),
\]
\[
\re f := \half(f+f^*),~~~\im f := \frac{1}{2i}(f-f^*),
\]
for $f,g\in\Ez$.
$[f,g]$ is called the \bfi{commutator} of $f$ and $g$, and $\re f$, 
$\im f$ are referred to as the \bfi{real part} (or \bfi{Hermitian part})
and \bfi{imaginary part} of $f$, respectively. 
$f\in\Ez$ is called \bfi{Hermitian} if $f^*=f$.

(iv) A \idx{$*$-homomorphism} is a mapping $\phi$ from a $*$-algebra 
$\Ez$ 
with unity to another (or the same) $*$-algebra $\Ez'$ with unity
such that
\[
\phi(f+g)=\phi(f)+\phi(g),~~~\phi(fg)=\phi(f)\phi(g),~~~
\phi(\alpha f)=\alpha\phi(f),
\]
\[
\phi(f^*)=\phi(f)^*,~~~\phi(1)=1.
\]
for all $f,g$ in $\Ez$ and $\alpha\in\Cz$.
\end{dfn}

Note that we assume commutativity only for the product of complex 
numbers and elements of $\Ez$. In general, the product of two elements 
of $\Ez$ is indeed noncommutative.
However, general commutativity of the addition is a consequence of our 
other assumptions. We prove this together with some other useful 
relations. 

\begin{prop}\label{p5.1.2}~\\
(i) For all  $f$, $g$, $h\in \Ez$,
\lbeq{e.p1}
(f+g)h=fh+gh,~~f-f=0,~~ f+g=g+f
\eeq
\lbeq{e.p2}
[f,f^*]=-2i[\re f,\im f].
\eeq
(ii) For all $f\in\Ez$, $\re f$ and $\im f$ are Hermitian. $f$ is 
Hermitian iff $f=\re f$ iff $\im f=0$. If $f,g$ are commuting 
Hermitian quantities then $fg$ is Hermitian, too.
\end{prop}

\bepf
(i) The right distributive law follows from
\[
\begin{array}{lll}
(f+g)h&=&((f+g)h)^{* *}=(h^* (f+g)^* )^* =(h^* (f^* +g^* ))^* \\
&=&(h^* f^* +h^* g^* )^* =(h^* f^* )^* +(h^* g^* )^* \\
&=&f^{* * }h^{* * }+g^{* * }h^{* * }=fh+gh.
\end{array}
\]
It implies $f-f=1f-1f=(1-1)f=0f=0$. From this, we may deduce that 
addition is commutative, as follows. The quantity $h:=-f+g$
satisfies
\[
-h=(-1)((-1)f+g)=(-1)(-1)f+(-1)g=f-g, 
\]
and we have
\[
f+g=f+(h-h)+g=(f+h)+(-h+g)=(f-f+g)+(f-g+g)=g+f. 
\]
This proves \gzit{e.p1}. If $u=\re f$, $v=\im f$ then $u^*=u,v^*=v$
and $f=u+iv, f^*=u-iv$. Hence 
\[
[f,f^*]=(u+iv)(u-iv)-(u-iv)(u+iv)=2i(vu-uv)=-2i[\re f,\im f],
\]
giving \gzit{e.p2}. 

(ii) The first two assertions are trivial, and the third holds since
$(fg)^*=g^*f^*=gf=fg$ if $f,g$ are Hermitian and commute.
\epf

\at{motivate order relation: observed values are ordered, and this 
must be reflected on the level of the observables themselves.}

\begin{dfn}~\\
(i) The $*$-algebra $\Ez$ is called \bfi{partially ordered} if there is
a partial order $\geq$ satisfying the following axioms (Q6)--(Q9)
for all $f,g,h\in\Ez$:

(Q6) 
~$\geq$ is reflexive ($f\geq f$),
antisymmetric ($f\geq g \geq f \Rightarrow f=g$),
and transitive ($f\geq g \geq h \Rightarrow f \geq h$).

(Q7)
~{$f\geq g \Leftrightarrow f+h\geq g+h$.}

(Q8)
~{$f\geq 0 \implies f=f^*$ and $g^*fg\geq 0$.}

(Q9)
~ $1 \geq 0$.

We introduce the notation
\[ 
f \leq g :\Leftrightarrow g\geq f,
\]
\[
\|f\|:=\inf\{\alpha\in\Rz \mid f^*f \leq \alpha^2, \alpha\geq0 \},
\]
where the infimum of the empty set is taken to be $\infty$. The number
$\|f\|$ is referred to as the \bfi{(spectral) norm} of $f$. 
An element $f\in\Ez$ is called \bfi{bounded} if $\|f\|<\infty$.
The \bfi{uniform topology} is the topology induced on $\Ez$ by
declaring as open sets arbitrary unions of finite intersections 
of the \bfi{open balls} $\{f\in\Ez \mid \|f-f_0\|<\eps\}$ for some 
$\eps>0$ and some $f_0 \in\Ez$.
\end{dfn}

\begin{prop}\label{p1.3}~\\
(i) For all quantities  $f$, $g$, $h\in \Ez$ and $\lambda \in\Cz$,
\lbeq{e.p3}
f^*f\geq 0,~~ ff^*\geq 0,
\eeq
\lbeq{e.p4}
f^*f\leq 0 \implies \|f\|=0 \implies f=0,
\eeq
\lbeq{e.p5}
f\leq g \implies h^*fh\leq h^*gh,~|\lambda|f\leq|\lambda|g,
\eeq
\lbeq{e.p6}
f^*g+g^*f\leq 2\|f\|~\|g\|,
\eeq
\lbeq{e.p7}
\|\lambda f\|=|\lambda| \|f\|,~~~ \|f\pm g\|\leq \|f\|\pm \|g\|,
\eeq
\lbeq{e.p8}
\|f g\|\leq \|f\|~ \|g\|.
\eeq
(ii) Among the complex numbers, precisely the nonnegative real numbers
$\lambda$ satisfy $\lambda\geq 0$.

\end{prop}

\bepf
(i) \gzit{e.p3} follows from the case $f=1$ of (Q8) by substituting
then $f$ or $f^*$ for $g$. 
\gzit{e.p4} follows from \gzit{e.p3}, the definition of the norm, and
(Q5). 
To prove \gzit{e.p5}, we deduce from $f\le g$ and (Q7) that $g-f\ge 0$,
then use (Q8) to find $h^*gh-h^*fh=h^*(g-f)h\ge 0$, hence 
$h^*fh\leq h^*gh$. Specializing to $g=sqrt{|\lambda|}$ then gives 
$|\lambda|f\leq|\lambda|g$.

To prove \gzit{e.p5}, let $\alpha=\|f\|$, $\beta=\|g\|$. Then 
$f^*f\leq \alpha^2$ and $g^*g\leq \beta^2$. Since
\[
\begin{array}{lll}
0\leq (\beta f - \alpha g)^*(\beta f - \alpha g)&=&
\beta^2f^*f-\alpha\beta(f^*g+g^*f)+\alpha^2 g^*g\\
&\leq& \beta^2\alpha^2 -\alpha\beta(f^*g+g^*f) +\alpha^2\beta^2,
\end{array}
\] 
$f^*g+g^*f\leq 2\alpha\beta$ if $\alpha\beta\neq 0$, and for
$\alpha\beta=0$, the same follows from \gzit{e.p4}. Therefore
\gzit{e.p6} holds. The first half of \gzit{e.p7} is trivial, and
the second half follows for the plus sign from 
\[
(f+g)^*(f+g)=f^*f+f^*g+g^*f+g^*g
\leq \alpha^2+ 2\alpha\beta+\beta^2=(\alpha+\beta)^2,
\]
and then for the minus sign from the first half.
Finally, by \gzit{e.p5},
\[
(fg)^*(fg)=g^*f^*fg\leq g^*\alpha^2g=\alpha^2g^*g\leq\alpha^2\beta^2.
\]
This implies \gzit{e.p8}.

(ii) If $\lambda$ is a nonnegative real number then $\lambda=f^*f\geq0$ 
with $f=\sqrt{\lambda}$. If $\lambda$ is a negative real number then 
$\lambda=-f^*f\leq0$ with $f=\sqrt{-\lambda}$, and by antisymmetry,
$\lambda\geq0$ is impossible. If $\lambda$ is a nonreal number then 
$\lambda\neq\lambda^*$ and $\lambda\geq0$ is impossible by (Q8).
\epf

\begin{dfn}
A \bfi{Euclidean $*$-algebra} is a nondegenerate, partially ordered 
$*$-algebra $\Ez$, whose elements are called 
\bfi{quantities}, together with a complex-valued 
\bfi{integral} $\sint$ defined on a subspace $\Sz$ of $\Ez$,
whose elements are called \bfi{strongly integrable}, satisfying
the following axioms (EA1)--(EA6) for all bounded $g\in\Ez$, all 
strongly integrable $h,h',h_l\in \Ez$, and all $\alpha,\beta\in\Cz$:

(EA1) ~ 
$h^*$, $\alpha h$, $gh$, and $hg$ are strongly integrable,

(EA2) ~
$\sint (\alpha h + \beta h') = \alpha\sint h+\beta\sint h', ~~~
\sint gh = \sint hg,~~~(\sint h) ^* = \sint h^*$,

(EA3) ~
$ \sint h^* h > 0$ ~if $h \not= 0$,

(EA4) ~
$\sint h^* g h= 0$ for all strongly integrable $h~~\Rightarrow~~ g=0$~~~
\bfi{(nondegeneracy)},

(EA5) ~
$ \sint h_l^* h_l \to 0 ~~\Rightarrow~~ \sint g h_l \to 0$,~
$\sint h_l^* g h_l \to 0$,

(EA6) ~
$h_l\downto 0~~\Rightarrow~~ \inf\sint h_l=0$~~~
\bfi{(Dini property)}.

Here, integrals extend over the longest following product or quotient
(in contrast to differential operators, which act on the shortest 
syntactically meaningful term), the \bfi{monotonic limit} is defined by
$g_l \downarrow 0$ iff, for every strongly integrable $h$, the sequence 
(or net) $\sint h^*g_lh$ consists of real numbers converging 
monotonically decreasing to zero. 
\end{dfn}

Note that the integral can often be naturally extended from strongly 
integrable quantities to a significantly larger space of integrable 
quantities.

\begin{prop}
\lbeq{e.ean4}
g\in\Ez,~~\sint gf = 0 \Forall f \in \Ez \implies g=0.
\eeq
For strongly integrable $f,g$,
\lbeq{e.intcs}
\sint (gh)^*(gh)\le \sint g^*g~\sint h^*h.~~~
\mbox{\bf (\bfi{Cauchy-Schwarz inequality})}
\eeq
In particular, every strongly integrable quantity is bounded.
\end{prop}

\bepf

If $\sint gf = 0$ for all $f \in \Ez$ then this holds in particular 
for $f=hh^*$. Thus $0=\sint ghh^*=\sint h^*gh$ by (EA2), and
(EA4) gives the desired conclusion \gzit{e.ean4}.
\gzit{e.intcs} holds since by (EA3), $\sint g^*h$ defines a positive 
definite inner product on $\Sz$, and directly implies the final 
statement.
\epf

We now describe the basic Euclidean $*$-algebras relevant in 
nonrelativistic physics. However, the remainder is completely
independent of details how the axioms are realized; a specific 
realization is needed only when doing specific quantitative 
calculations.

\begin{expls}\label{e3.1}~\\
(i) \bfi{($N$-level quantum systems)}
The simplest family of Euclidean 
$*$-algebras is the algebra $\Ez=\Cz^{N\times N}$ of 
square complex $N\times N$ matrices; cf. Example \ref{ex.Nlevel}.
Here the quantites are square matrices, the constants are the multiples 
of the identity matrix, the conjugate is conjugate transposition, and 
the integral is the trace, the sum of the diagonal entries or, 
equivalently, the sum of the eigenvalues. In particular, all quantities 
are strongly integrable.

(ii) \bfi{(Nonrelativistic classical mechanics)}
An atomic $N$-particle system is described in classical mechanics by
the phase space $\Rz^{6N}$ with six coordinates -- position 
$x^a\in\Rz^3$ and momentum $p^a\in\Rz^3$ -- for each particle.
The algebra 
\[
\Ez_N:= C^\infty(\Rz^{6N})
\]
of smooth complex-valued, arbitrarily often differentiable functions 
$g(x^{1:N},p^{1:N})$  of 3-dimensional positions and momenta is a 
commutative Euclidean $*$-algebra with complex conjugation as conjugate
and the \bfi{Liouville integral}
\[
\sint g=C^{-1} \int dp^{1:N}dx^{1:N} g_N(x^{1:N},p^{1:N}),
\]
where $C$ is a positive constant.
Strongly integrable quantities are the Schwartz functions in 
$\Ez$.\footnote{A \bfi{Schwartz function} of $n$ variables $z\in\Rz^n$
is a function $f\in C^\infty(\Rz^n)$ such that the expressions
$z_1^{k_1}\dots z_n^{k_n}\frac{\partial^{l_1+\dots+l^n}f(z)}
{\partial z_1^{l_1}\dots\partial z_n^{l_n}}$ 
are bounded in $z$ for any choice of nonnegative integers 
$k_1,\dots,k_n,l_1,\dots,l_n$.
} 
The axioms are easily verified.

(iii) \bfi{(Classical fluids)}
A fluid is classically described by an atomic system with an
indefinite number of particles. The appropriate Euclidean $*$-algebra 
for a single species of monatomic particles is the 
direct sum $\Ez=\D\oplus_{N\ge 0} \Ez_N$ whose quantities are 
infinite sequences $g=(g_0,g_1,...)$ of $g_N\in\Ez_N^{\sym}$, with 
$\Ez_N^{\sym}$ consisting of all permutation-invariant functions 
functions form $\Ez_N$ as in (ii), and weighted Liouville integral
\[
\sint g=\sum_{N\ge 0} 
C_N^{-1}\int dp^{1:N}dx^{1:N} g_N(x^{1:N},p^{1:N}).
\]
Here $C_N$ is a symmetry factor for the symmetry group of the
$N$-particle systen, which equals $h^{3N}N!$ for indistinguishable
particles; $h= 2\pi \hbar$ is Planck's constant.
This accounts for the Maxwell statistics and gives the correct entropy 
of mixing. Classical fluids with monatomic particles of several 
different kinds require a tensor product of several such algebras, and 
classical fluids composed of molecules require additional degrees
of freedom to account for the rotation and vibration of the molecules.

(iv) \bfi{(Nonrelativistic quantum mechanics)} 
Let $\Hz$ be a Euclidean space, a dense subspace of a Hilbert space.
Then the algebra $\Ez:= \Lin \Hz$ of continuous linear operators 
on $\Hz$ is a Euclidean $*$-algebra with the adjoint as conjugate and
the \bfi{quantum integral}
\[
  \sint g= \tr g,
\]
given by the trace of the quantity in the integrand.
Strongly integrable quantities are the operators $g\in\Ez$ which 
are trace class; this includes all linear operators of finite rank. 
Again, the axioms are easily verified. In the quantum context, 
Hermitian quantities $f$ are often referred to as \bfi{observables};
but we do not use this term here.

\end{expls}

We end this section by stating some results needed later.
The exposition in this and the next chapter is fully rigorous if the 
statements of Proposition \ref{app2a.} and Proposition \ref{app1.}
are assumed in addition to (EA1)--(EA6).
We prove these propositions only in case that $\Ez$ 
is finite-dimensional\footnote{
We'd appreciate to be informed about possible proofs in general that 
only use the properties of Euclidean $*$-algebras (and perhaps further, 
elementary assumptions).
}. 
But they can also be proved if the quantities 
involved are smooth functions, or if they have a spectral 
resolution; cf., e.g., \sca{Thirring} \cite{Thi} (who works in the
framework of $C^*$-algebras and von Neumann algebras). 

\bigskip
\begin{prop} \label{app2a.} 
For arbitrary quantities $f$, $g$,
\[
e^{\alpha f}e^{\beta f}=e^{(\alpha+\beta)f}~~(\alpha,\beta\in\Rz),
\]
\[
(e^f)^*=e^{f^*},
\]
\[
e^f g = g e^f ~~~\mbox{if $f$ and $g$ commute},
\]
\[
f^*=f \implies \log e^f=f,
\]
\[
f\ge 0 \implies \sqrt{f}\ge 0,~~(\sqrt{f})^2 =f,
\]
For any quantity $f=f(s)$ depending continuously on $s\in[a,b]$,
\[
\int_a^b ds \sint f(s) = \sint \Big(\int_a^b ds f(s)\Big),
\]
and for any quantity $f=f(\lambda)$ depending continuously 
differentiably on a parameter vector $\lambda$, 
\[
\frac{d}{d\lambda} \sint f = \sint df/d\lambda.
\]
\end{prop}

\bepf 
In finite dimensions, the first five assertions are standard 
matrix calculus, and the remaining two statements hold since $\sint f$ 
must be a finite linear combination of the components of $f$.
\epf

\begin{prop} \label{app1.}
Let $f,g$ be quantities depending continuously differentiably on a
parameter or parameter vector $\lambda $, and suppose that
\lbeq{e.commute}
[f(\lambda ),g(\lambda )]=0\mbox { for all }\lambda.
\eeq
Thus, for any continuously differentiable function $F$ of two
variables,
\lbeq{app1}
\frac {d} {d\lambda }\sint F(f,g)
=\sint\partial _1F(f,g)\frac {df} {d\lambda }
+\sint\partial _2F(f,g)\frac {dg } {d\lambda }\ .
\eeq
Here $\partial _1F$ and $\partial _2 f$ denote differentiation by the 
first and second argument of $F$, respectively.
\end{prop}

\bepf
We prove the special case $F(x,y)=x^my^n$, where (\ref{app1}) reduces
to
\lbeq{app2}
\frac {d} {d\lambda }\sint f^mg^n
=\sint mf^{m-1}g^n\frac {df} {d\lambda }
+\sint nf^mg^{n-1}\frac {dg} {d\lambda}.
\eeq
The general case then follows for polynomials $F(x,y)$ by taking
suitable linear combinations, and for arbitrary $F$ by a limiting
procedure. To prove (\ref{app2}), we note that, more generally,
\[ 
\begin{array}{lll}
\D\frac {d} {d\lambda }\sint f_1\dots f_{m+n}
&=\sint\frac {d} {d\lambda }(f_1\dots f_{m+n})\\
&\D=\sint\sum _{j=1} ^{m+n}f_1\dots f_{j-1}\frac {df_j} {d\lambda }
f_{j+1}\dots f_{m+n} \\
&\D=\sum _{j=1} ^{m+n}\sint f_1\dots f_{j-1}\frac {df_j} {d\lambda }
f_{j+1}\dots f_{m+n} \\
&\D=\sum _{j=1} ^{m+n}\sint f_{j+1}\dots f_{m+n}f_1\dots f_{j-1}
\frac {df_j} {d\lambda }\ , 
\end{array}
\]
using the cyclic commutativity (EA2) of the integral.
If we specialize to $f_j=f$ if $j\le m$, $f_j=g$ if $j>m$, and use
\gzit{e.commute}, we arrive at (\ref{app2}).
\epf

Of course, the proposition generalizes to families of more than two
commuting quantities; but more important is the special case $g=f$:

\begin{cor} \label{app2.}
For any quantity $f$ depending continuously differentiably on a
parameter vector $\lambda $, and any continuously differentiable 
function $F$ of a single variable,
\lbeq{app3}
\frac {d} {d\lambda }\sint F(f)=\sint F'(f)\frac {df} {d\lambda }.
\eeq
\end{cor}

\section{Gibbs states}\label{s.gibbs}

Our next task is to specify the formal properties of the value of a 
quantity. 

\begin{dfn}\label{d.state}
A \bfi{state} is a mapping $^-$ that assigns to all quantities $f$
from a subspace of $\Ez$ containing all bounded quantities
its \bfi{value} $\< f\> \in \Cz$ 
such that for all $f,g \in \Ez$, $\alpha \in \Cz$,

(E1)~ $\<1\> =1, ~~\<f^*\>=\<f\>^*,~~ \< f+g\> =\<f\> +\<g\> $, 

(E2)~ $\<\alpha f\> =\alpha\<f\>$, 

(E3)~ If $f \ge 0$ then $\<f\> \ge 0$,

(E4)~ If $f_l\in\Ez,~ f_l \downarrow 0$ then $\<f_l\> \downarrow  0$.
\end{dfn}

Note that this formal definition of a state -- always used in the 
remainder of the book -- differs from the phenomenological 
thermodynamic states defined in Section \ref{s.phen}. 
The connection between the two notions will be made in
Section \ref{s.eos}. 

Statistical mechanics essentially originated with Josiah Willard Gibbs,
whose 1902 book \sca{Gibbs} \cite{Gib} on (at that time of course 
only classical) statistical mechanics is still readable. See
\sca{Uffink} \cite{Uff} for a history of the subject. 

All states arising in thermodynamics have the following
particular form.

\begin{dfn} \label{2.7.}
A \bfi{Gibbs state} is defined by assigning to any $g\in\Ez$ the value
\lbeq{2-10a}
\<g\>:=\sint e^{-S/\kbar} g, 
\eeq
where $S$, called the \bfi{entropy} of the state, is a Hermitian 
quantity with strongly integrable $e^{-S/\kbar}$, satisfying the 
normalization condition
\lbeq{2-10}
\sint e^{-S/\kbar}=1,
\eeq
and $\kbar$ is the Boltzmann constant
\lbeq{e.kbar}
\kbar \approx 1.38065 \cdot 10^{-23} J/K.
\eeq
Theorem \ref{2.6.} below implies that a Gibbs state is indeed a state.
\end{dfn}

The Boltzmann constant defines the units in which the entropy is 
measured. In analogy\footnote{
As we shall see in \gzit{e.qmunc} and \gzit{e.thunc}, $\hbar$ and 
$\kbar$ play indeed analogous roles in quantum mechanical and 
thermodynamic uncertainty relations.
} 
with Planck's constant $\hbar$,
we write $\kbar$ in place of the customary $k$ or $k_B$, in order to 
be free to use the letter $k$ for other purposes. 
By a change of units one can enforce any value of $\kbar$.
Chemists use instead of particle number $N$ the corresponding \bfi{mole 
number}, which differs by a fixed numerical factor, the \bfi{Avogadro
constant} 
\[
N_A=R/\kbar \approx 6.02214 \cdot 10^{23} {\fct{mol}}^{-1}, 
\]
where $R$ is the  universal gas constant \gzit{e.R}.
As a result, all results from statistical mechanics may be translated 
to phenomenological thermodynamics by setting $\kbar = R$, 
corresponding to setting $1 \fct{mol} =  6.02214 \cdot 10^{23}$,
the number of particles in one mole of a pure substance.

What is here called entropy has a variety of alternative names in the 
literature on statistical mechanics. For example,
\sca{Gibbs} \cite{Gib}, who first noticed the rich thermodynamic 
implications of states defined by \gzit{2-10a}, called $-S$ the 
{\em index of probability};  
\sca{Alhassid \& Levine} \cite{AlhL} and \sca{Balian} \cite{Bal2}
use the name {\em surprisal} for $S$.  Our terminology is close to 
that of \sca{Mrugala} et al. \cite{MruNSS}, who call 
$S$ the {\em microscopic entropy}, and \sca{Hassan} et al. \cite{HasVL},
who call $S$ the {\em information(al) entropy operator}. 
What is traditionally (and in Section \ref{s.phen}) called entropy 
and denoted by $S$
is in the present setting the value $\ol S=\<S\>=\sint e^{-S/\kbar} S$.

\begin{thm} \label{2.6.}~\\
(i) A Gibbs state determines its entropy uniquely.

(ii) For any Hermitian quantity $f$ with strongly integrable $e^{-f}$, 
the mapping $\<\cdot\>_f$ defined by 
\lbeq{2-6a}
\< g \>_f:=Z_f^{-1}\sint e^{-f} g,~~~\mbox{where } Z_f:=\sint e^{-f},
\eeq
is a state. It is a Gibbs state with entropy 
\lbeq{2-8}
S_f:=\kbar (f+\log Z_f).
\eeq
(iii) The \bfi{KMS condition} (cf. \sca{Kubo} \cite{Kub0},
\sca{Martin \& Schwinger} \cite{MarS})
\lbeq{e.KMS}
\<gh\>_f = \<hQ_f g\> ~~~\mbox{for bounded } g,h
\eeq
holds. Here $Q_f$ is the linear mapping defined by
\[
Q_f g :=e^{-f}ge^{f}.
\]
\end{thm}

\bepf 
(i) If the entropies $S$ and $S'$ define the same Gibbs state then 
\[
\sint (e^{-S/\kbar}-e^{-S'/\kbar}) g = \<g\>-\<g\>=0 
\]
for all $g$, hence  \gzit{e.ean4} gives $e^{-S/\kbar}-e^{-S'/\kbar}=0$. 
This implies that $e^{-S/\kbar}=e^{-S'/\kbar}$, hence $S=S'$ by 
Proposition \ref{app2a.}.

(ii) The quantity $d:=e^{-f/2}$ is nonzero and satisfies $d^*=d$, 
$e^{-f}=d^*d\geq 0$. Hence $Z_f>0$ by (EA3), and $\rho:=Z_f^{-1}e^{-f}$ 
is Hermitian and nonnegative. For $h\ge 0$, the quantity $g=\sqrt{h}$
is Hermitian (by Proposition \ref{app2a.}) and satisfies 
$g\rho g^*=Z_f^{-1}(gd)(gd)^* \ge 0$, hence 
by (EA2),
\[
\<h\>_f=\<g^*g\>_f= \sint \rho g^*g =\sint g\rho g^* \ge 0. 
\] 
Moreover, $\<1\>_f =Z_f^{-1}\sint e^{-f}=1$. Similarly, if $g\ge 0$ 
then $g=h^*h$ with $h=\sqrt{g}=h^*$ and with $k:=e^{-f/2}h$, we get 
\[
Z_f\<g\>_f = \sint e^{-f}hh^*=\sint h^*e^{-f}h = \sint k^*k \ge 0.
\]
This implies (E3). the other axioms (E1)--(E4) follow easily from the 
corresponding properties of the integral. Thus $\<\cdot\>_f$ is a state.
Finally, with the definition \gzit{2-8}, we have
\[
Z_f^{-1}e^{-f}=e^{-f-\log Z_f}=e^{-S_f/\kbar}, 
\]
whence $\<\cdot\>_f$ is a Gibbs state.

(iii) By (EA2),
$\<hQ_fg\>_f=\sint e^{-f}hQ_fg=\sint Q_fge^{-f}h =\sint e^{-f}gh 
=\<gh\>_f$.
\epf

Note that the state \gzit{2-6a} is unaltered when $f$ is 
shifted by a constant. $Q_f$ is called the \bfi{modular automorphism}
of the state $\<\cdot\>_f$ since $Q_f(gh)=Q_f(g)Q_f(h)$; for a classical
system, $Q_f$ is the identity. In the following, we shall not make use
of the KMS condition; however, it plays an important role in the 
mathematics of the thermodynamic limit (cf. \sca{Thirring} \cite{Thi}).

$Z_f$ is called the \bfi{partition function} of $f$; it is a function of
whatever parameters appear in a particular form given to $f$ in the
applications. A large part of traditional statistical mechanics is 
concerned with the calculation, for given $f$, of the partition 
function $Z_f$ and of the values $\<g\>_f$ for selected quantities $g$. 
As we shall see, the basic results of 
statistical mechanics are completely independent of the details 
involved, and it is this basic part that we concentrate upon in this 
book.

\begin{expl}\label{ex.canonical}
A \bfi{canonical ensemble}\footnote{\label{f.ensemble} 
Except in the traditional notions of a microcanonical, canonical, or 
grand canonical ensemble, we avoid the term \bfi{ensemble} which in 
statistical mechanics is de facto used as a synonym for state but
often has the connotation of a large real or imagined 
collection of identical copies of a systems. The latter interpretation 
has well-known difficulties to explain why each single 
macroscopic system is described correctly by thermodynamics;
see, e.g., \sca{Sklar} \cite{Skl}.
}, 
is defined as a Gibbs state whose entropy is an affine function of a 
Hermitian quantity $H$, called the \bfi{Hamiltonian}:
\[
S=\beta H + \const,
\]
with a constant depending on $\beta$, computable from \gzit{2-8} and
the partition function
\[
Z=\sint e^{-\beta H}
\]
of $f=\beta H$.
In particular, in the quantum case, where $\sint$ is the trace, the
finiteness of $Z$ implies that 
$S$ and hence $H$ must have a discrete spectrum that is bounded below. 
Hence the partition function takes the familiar form
\lbeq{e3.3}
Z=\tr e^{-\beta H} = \sum_{n \in \cal N} e^{-\beta E_n},
\eeq
where the $E_n$ ($n\in\cal N$) are the \bfi{energy levels}, the 
eigenvalues of $H$.
If the spectrum of $H$ is known, this leads to explicit formulas for 
$Z$. For example, a \bfi{two level system} is defined by the energy 
levels $0,E$ (or $E_0$ and $E_0+E$, which gives the same results), 
and has
\lbeq{e.2level}
Z=1+e^{-\beta E}.
\eeq
It describes a single \bfi{Fermion mode}, but also many other systems
at low temperature; cf. \gzit{e.2levelapprox}. In particular, it is the 
basis of laser-induced chemical reactions in photochemistry (see, e.g., 
\sca{Karlov} \cite{Kar}, \sca{Murov} et al. \cite{MurCH}), where 
only two electronic energy levels (the ground state and the first 
excited state) are relevant; cf. the discussion of 
\gzit{e.2levelapprox} below.

For a \bfi{harmonic oscillator}, defined by the energy levels $nE$, 
$n=0,1,2,\dots$ and describing a single \bfi{Boson mode}, we have
\[
Z=\sum_{n=0}^\infty e^{-n\beta E} = (1-e^{-\beta E})^{-1}.
\]
Independent modes are modelled by taking tensor products of single 
mode algebras and adding their Hamiltonians, leading to spectra which 
are obtained by summing the eigenvalues of the modes in all possible 
ways. The resulting partition function is the product of the 
single-mode partition functions. 
\at{expand? treat Maxwell case? $\sint f = \sum_n f(n)/n!$}] 
From here, a thermodynamic limit 
leads to the properties of ideal gases. Then nonideal gases due to 
interactions can be handled using the cumulant expansion, as 
indicated at the end of Section \ref{s.gen}. The details are outside 
the scope of this book.
\end{expl}

Since the Hamiltonian can be any Hermitian quantity, the quantum 
partition function formula \gzit{e3.3} can in principle be used to
compute the partition function of arbitrary quantized Hermitian 
quantities.

\section{Kubo product and generating functional} \label{s.gen}

The negative logarithm of the partition function, the so-called
generating functional, plays a fundamental role in statistical 
mechanics.

We first discuss a number of general properties, discovered by 
\sca{Gibbs} \cite{Gib}, \sca{Peierls} \cite{Pei}, 
\sca{Bogoliubov} \cite{Bog}, \sca{Kubo} \cite{Kub}, 
\sca{Mori} \cite{Mor}, and \sca{Griffiths} \cite{Gri}. 
The somewhat technical setting involving the Kubo inner product is
necessary to handle noncommuting quantities correctly; 
everything would be much easier in the classical case.
On a first reading, the proofs in this section may be skipped.

\begin{prop} Let $f$ be Hermitian such that $e^{sf}$ is strongly
integrable for all $s\in[-1,1]$. Then 
\lbeq{e.kubo}
\<g;h\>_f:=\<g E_f h\>_f,
\eeq
where $E_f$ is the linear mapping defined for Hermitian $f$ by
\[
E_f h:=\int_0^1 ds\, e^{-sf}he^{sf},
\]
defines a bilinear, positive definite inner product 
$\<\cdot\,;\cdot\>_f$ on the algebra of quantities, 
called the \bfi{Kubo} (or \bfi{Mori} or \bfi{Bogoliubov}) 
\bfi{inner product}.
For all $f,g$, the following relations hold:
\lbeq{e.kubo2}
\<g;h\>_f^* =\<h^*;g^*\>_f,
\eeq
\lbeq{e.definit}
\<g^*;g\>_f > 0 ~~~\mbox{if } g \ne 0,
\eeq
\lbeq{e.kubo1}
\<g;h\>_f =g\<h\>_f ~~~\mbox{if $g\in \Cz$},
\eeq
\lbeq{e.kubo0}
\<g;h\>_f =\<gh\>_f ~~~\mbox{if $g$ or $h$ commutes with $f$},
\eeq
\lbeq{e.E0}
E_f g = g ~~~\mbox{if $g$ commutes with $f$}.
\eeq
If $f=f(\lambda)$ depends continuously differentiably on the
real parameter vector $\lambda$ then 
\lbeq{e.deriv0}
\frac{d}{d\lambda} e^{-f} = - \Big(E_f \frac{df}{d\lambda}\Big)e^{-f}.
\eeq
\end{prop}

\bepf
(i) We have
\[
\<g;h\>_f^* =\<(gE_fh)^*\>_f = \<(E_fh)^*g^*\>_f
=\Big\<\int_0^1 ds\,e^{sf}h^*e^{-sf}g^*\Big\>_f
=\int_0^1 ds\<e^{sf}h^*e^{-sf}g^*\>_f.
\]
The integrand equals
\[
\sint e^{-f}e^{sf}h^*e^{-sf}g^* = \sint e^{sf}e^{-f}h^*e^{-sf}g^* 
=\sint e^{-f}h^*e^{-sf}g^*e^{sf} = \<h^*e^{-sf}g^*e^{sf}\>_f
\]
by (EA2), hence
\[
\<g;h\>_f^* = \int_0^1 ds\<h^*e^{-sf}g^*e^{sf}\>_f
= \Big\<h^* \int_0^1 ds\,e^{-sf}g^*e^{sf}\Big\>_f
= \<h^*E_fg^*\>_f=\<h^*;g^*\>_f.
\]
Thus \gzit{e.kubo2} holds.

(ii) Suppose that $g\ne 0$. For $s\in[0,1]$, we define $u=s/2,v=(1-s)/2$
and $g(s):= e^{-uf}ge^{vf}$. Since $f$ is Hermitian, 
$g(s)^*= e^{vf}g^*e^{-uf}$, hence by (EA2) and (EA3),
\[
\sint  e^{-f}g^*e^{-sf}ge^{sf}=\sint e^{vf}ge^{-2uf}g^*e^{vf}
=\sint g(s)^*g(s)>0, 
\]
so that
\[
\<g^*;g\>_f=\<g^*E_fg\>_f
=\int_0^1 ds\,\sint e^{-f}g^*e^{-sf}ge^{sf} > 0.
\]
This proves \gzit{e.definit}, and shows that the Kubo inner product is 
positive definite.

(iii) If $f$ and $g$ commute then $ge^{sf}=e^{sf}g$, hence 
\[
E_fg=\int_0^1 ds e^{-sf}e^{sf} g = \int_0^1 ds g = g,
\]
giving \gzit{e.E0}. The definition of the Kubo inner product then
implies \gzit{e.kubo0}, and taking $g\in\Cz$ gives \gzit{e.kubo1}.

(iv) The function $q$ on $[0,1]$ defined by
\[
q(t):= \int_0^t ds\, e^{-sf}\frac{df}{d\lambda}e^{sf}
+\Big(\frac{d}{d\lambda}e^{-tf}\Big) e^{tf}
\]
satisfies $q(0)=0$ and 
\[
\frac{d}{dt}q(t) = e^{-tf}\frac{df}{d\lambda}e^{tf}
+\Big(\frac{d}{d\lambda}e^{-tf}\Big)f e^{tf}
+\frac{d}{d\lambda}(-e^{-tf}f) e^{tf} = 0.
\]
Hence $q$ vanishes identically. In particular, $q(1)=0$, giving 
\gzit{e.deriv0}.  
\epf

As customary in thermodynamics, we use differentials to express
relations involving the differentiation by arbitrary parameters.
To write \gzit{e.deriv0} in differential form, we formally multiply by 
$d\lambda$, and obtain the \bfi{quantum chain rule} for exponentials,
\lbeq{e.chain}
d e^{-f} = (- E_fd f) e^{-f}.
\eeq
If the $f(\lambda)$ commute for all values of $\lambda$
then the quantum chain rule reduces to the classical chain rule.
Indeed, then $f$ commutes also with $\frac{df}{d\lambda}$; hence 
$E_f\frac{df}{d\lambda} = \frac{df}{d\lambda}$, and $E_fd f = df$.

\bigskip
{\em The following theorem is central to the mathematics of 
statistical mechanics.}
As will be apparent from the discussion in the next chapter, 
part (i) is the 
abstract mathematical form of the second law of thermodynamics, 
part (ii) allows the actual computation of thermal properties from
microscopic assumptions, and part (iii) is the abstract form of the 
first law.

\begin{thm} \label{t3.3}
Let $f$ be Hermitian such that $e^{sf}$ is strongly
integrable for all $s\in[-1,1]$. 

(i) The \bfi{generating functional}
\lbeq{e.gen}
W(f):=- \log \sint e^{-f}
\eeq
is a concave function of the Hermitian quantity $f$.
In particular,
\lbeq{e.GB}
W(g) \le W(f)+\<g-f\>_f.~~~
\mbox{\bf (\bfi{Gibbs-Bogoliubov inequality})} 
\eeq
Equality holds in \gzit{e.GB} iff $f$ and $g$ differ by an additive 
constant. 

(ii) For Hermitian $g$, we have
\lbeq{e.starh}
W(f+\tau g)=W(f)-\log\<e^{-f-\tau g}e^f\>_f.
\eeq
Moreover, the \bfi{cumulant expansion}
\lbeq{e.cumulant}
W(f+\tau g)
= W(f)+\tau\<g\>_f + \frac{\tau^2}{2}(\<g\>_f^2-\<g;g\>_f) + O(\tau^3)
\eeq
holds if the coefficients are finite.

(iii) If $f=f(\lambda)$ and $g=g(\lambda)$ depend continuously 
differentiably on $\lambda$ then the following \bfi{differentiation 
formulas} hold:
\lbeq{e.diff}
d\<g\>_f = \<dg\>_f-\<g;df\>_f+\<g\>_f\<df\>_f, 
\eeq
\lbeq{e.diffW}
dW(f)=\<df\>_f.
\eeq
(iv) The entropy of the state $\<\cdot\>_f$ is
\lbeq{e.ent}
S=\kbar(f-W(f)).
\eeq
\end{thm}

\bepf
We prove the assertions in reverse order.

(iv) Equation \gzit{e.gen} says that $W(f)=-\log Z_f$, which together 
with \gzit{2-8} gives \gzit{e.ent}.

(iii) We have 
\[
\bary{lll}
d\sint ge^{-f} &=& \sint dg e^{-f} + \sint gde^{-f}
=\sint dge^{-f}-\sint gE_fd fe^{-f}\\
&=&\sint(dg-gE_fd f)e^{-f} = Z_f\<dg-gE_fd f\>_f.
\eary
\]
On the other hand, 
$d\sint ge^{-f} = d(Z_f\<g\>_f)=dZ_f\<g\>_f+Z_fd\<g\>_f$, so that
\lbeq{e.s1}
dZ_f\<g\>_f+Z_fd\<g\>_f = Z_f\<dg-gE_fd f\>_f 
= Z_f\<dg\>_f-Z_f\<g;df\>_f.
\eeq
In particular, for $g=1$ we find by \gzit{e.kubo1} that 
$dZ_f=-Z_f\<1;df\>_f=-Z_f\<df\>_f$. Now \gzit{e.diffW} follows from
$dW(f)=-d\log Z_f =-dZ_f/Z_f = \<df\>_f$, and solving \gzit{e.s1} for 
$d\<g\>_f$ gives \gzit{e.diff}.

(ii) Equation \gzit{e.starh} follows from
\[
e^{-W(h)} = \sint e^{-h} = \sint e^{-h} e^f e^{-f}
= \sint e^{-f} e^{-h} e^f = (\sint e^{-f}) \<e^{-h} e^f\>_f
= e^{-W(f)}  \<e^{-h} e^f\>_f
\]
by taking logarithms and setting $h=f+\tau g$. To prove the cumulant 
expansion, we introduce the function $\phi$ defined by
\[
\phi(\tau):=W(f+\tau g),
\]
From \gzit{e.diffW}, we find $\phi'(\tau) = \<g\>_{f+\tau g}$
for $f,g$ independent of $\tau$, and by differentiating this again,
\[
\phi''(\tau)=\D\frac{d}{d\tau}  \<g\>_{f+\tau g}
=\D-\Big\<g\frac{E_fd (f+\tau g)}{d\tau}\Big\>_{f+\tau g}
+\<g\>_{f+\tau g}^2.
\]
In particular, 
\lbeq{e.x5}
\phi'(0) = \<g\>_f,~~~
\phi''(0) = \<g\>_f^2-\<gE_f g\>_f= \<g\>_f^2-\<g;g\>_f.
\eeq
A Taylor expansion now implies \gzit{e.cumulant}. 

(i) Since the Cauchy-Schwarz equation 
for the Kubo inner product implies 
\[
\<g\>_f^2=\<g;1\>_f^2\le \<g;g\>_f\<1;1\>_f= \<g;g\>_f, 
\]
\gzit{e.x5} implies that 
\[
\frac{d^2}{d\tau^2} W(f+\tau g)\Big|_{\tau=0}\le 0
\]
for all $f,g$. This implies that $W(f)$ is concave.
Moreover, replacing $f$ by $f+sg$, we find that $\phi''(s)\le 0$ for
all $s$. The remainder form of Taylor's theorem therefore gives
\[
\phi(\tau)=\phi(0)+\tau\phi'(0)+\int_0^\tau ds (\tau-s)\phi''(s)
\le \phi(0)+\tau\phi'(0),
\]
and for $\tau=1$ we get
\lbeq{e.x6}
W(f+g)\le W(f)+\<g\>_f.
\eeq
\gzit{e.GB} follows for $\tau=1$ upon replacing $g$ by $g-f$.

By the derivation, equality holds in \gzit{e.x6} only if $\phi''(s)=0$ 
for all $0<s<1$. By \gzit{e.x5}, applied with $f+sg$ in place of $f$, 
this
implies $\<g\>_{f+sg}^2 = \<g;g\>_{f+sg}$. Thus we have equality in 
the Cauchy-Schwarz argument, forcing $g$ to be a multiple of $1$.
Therefore equality in the Gibbs-Bogoliubov inequality \gzit{e.GB} 
is possible only if $g-f$ is constant.
\epf

As a consequence of the  Gibbs-Bogoliubov inequality, we derive an
important inequality for the entropy in terms of that of a given 
reference state.

\begin{thm} \label{t4.5} 
Let $S_c$ be the entropy of a reference state. Then, for an arbitrary 
Gibbs state $\<\cdot\>$ with entropy $S$,
\lbeq{e4.5}
\< S \> \le \< S_c\>,
\eeq
with equality only if $S_c =S$.
\end{thm}

\bepf
Let $f=S/\kbar$ and $g=S_c/\kbar$. Since $S$ and $S_c$ are 
entropies, $W(f)=W(g)=0$, and the Gibbs-Bogoliubov inequality 
\gzit{e.GB} gives $0\le \<g-f\>_f = \<S_c-S\>/\kbar$.
This implies \gzit{e4.5}. If equality holds then equality holds in 
\gzit{e.GB}, so that $S_c$ and $S$ differ only by a constant.
But this constant vanishes since the values agree.
\epf

The difference
\lbeq{4-5}
\< S_c-S\>  =\< S_c\> -\< S\>  \ge 0
\eeq
is known as \bfi{relative entropy}.
In an information theoretical context (cf. Section \ref{s.complexity}),
the relative entropy may be interpreted as the amount of information
in a state $\< \cdot \>$ which cannot be explained by
the reference state. This interpretation makes sense since 
the relative entropy vanishes precisely for the reference state. 
A large relative entropy therefore indicates that the state contains 
some important information not present in the reference state.

\bfi{Approximations.} 
The cumulant expansion is the basis of a well-known
approximation method in statistical mechanics. Starting from special
reference states $\<\cdot\>_f$ with explicitly known $W(f)$ and $E_f$ 
(corresponding to so-called \at{give some!} explicitly solvable models),
one obtains 
inductively expressions for values in these states by 
applying the differentiation rules. (In the most important cases,
the resulting formulas for the values are commonly
referred to as a \bfi{Wick theorem}, cf. \sca{Wick} \cite{Wic},
although in the classical case, the formulas are much older and were 
derived in 1918 by \sca{Isserlis} \cite{Iss}.
For details, see textbooks on statistical mechanics, 
e.g., \sca{Huang} \cite{Hua}, \sca{Reichl} \cite{Rei}.)

From these, one can calculate the coefficients in the cumulant 
expansion; note that higher order terms can be found 
by proceeding as in the proof, using further differentiation. 
\at{Alternatively, one may proceed on the basis of BCH-formulas for 
the Lie groups defining the exactly solvable model.}
This gives approximate generating functions (and by 
differentiation associated values) for Gibbs states 
with an entropy close to the explicitly solvable reference state.
From the resulting generating function and the differentiation 
formulas \gzit{e.diff}--\gzit{e.diffW},
one gets as before the values for the given state.

The best tractable reference state  $\<\cdot\>_f$ to be used for a 
given Gibbs state $\<\cdot\>_g$ can be obtained by minimizing the 
upper bound in the Gibbs-Bogoliubov inequality \gzit{e.GB} over 
all $f$ for which an explicit generating function is known.
Frequently, one simply approximates $W(g)$ by the minimum of this
upper bound,
\lbeq{e.meanfield}
W(g) \approx W_m(g):=\inf_f \Big(W(f)+\<g-f\>_f\Big).
\eeq
Using $W_m(g)$ in place of $W(g)$ defines a so-called 
\bfi{mean field theory}; cf. \sca{Callen} \cite{Cal}.
For computations from first principles (quantum field theory), see, 
e.g., the survey by \sca{Berges} et al. \cite{BerTW}.

\section{Limit resolution and uncertainty} \label{s.limit}

Definition \ref{d.state} generalizes the expectation axioms of 
\sca{Whittle} \cite[Section 2.2]{Whi} for classical probability theory.
Indeed, the values of our quantities are traditionally called 
expectation values, and refer to the mean over an ensemble of (real or 
imagined) identically prepared systems. 

In our treatment, we keep the notation with pointed brackets familiar 
from statistical mechanics, but use the more neutral term \bfi{value} 
for $\<f\>$ to avoid any reference to probability or statistics.
This keeps the formal machinery completely independent of controversial
issues about the interpretation of probabilities. Statistics and 
measurements, where the probabilistic aspect enters directly, are 
discussed separately in Chapter \ref{s.model}.

Our analysis of the uncertainty inherent in the description of a system 
by a state is based on the following result.

\begin{prop} 
For Hermitian $g$, 
\lbeq{e.res0}
\<g\>^2 \le \<g^2\>.
\eeq
Equality holds if $g=\<g\>$. 
\end{prop}

\bepf
Put $\ol g = \<g\>$. Then $0\le\<(g-\ol g)^2\>=\<g^2-2\ol g g+\ol g^2\>
=\<g^2\>-2\ol g \<g\>+\ol g^2=\<g^2\>-\<g\>^2$.
This gives \gzit{e.res0}. If $g=\ol g$ then equality holds in this
argument.
\epf

\begin{dfn}
The number
\[
\cov(f,g):=\re \<(f-\overline{f})^*(g-\overline{g}) \>
\]
is called the \bfi{covariance} of $f,g\in\Ez$. Two quantities $f,g$ are 
called \bfi{uncorrelated} if $\cov(f,g)=0$, and \bfi{correlated} 
otherwise. The number 
\[
\sigma(f):=\sqrt{\cov(f,f)}
\]
is called the \bfi{uncertainty} of $f\in\Ez$ in the state $\<\cdot\>$. 
The number
\lbeq{e.res}
\res(g):=\sqrt{\<g^2\>/\<g\>^2-1},
\eeq
is called the \bfi{limit resolution}
of a Hermitian quantity $g$ with nonzero value $\<g\>$.
\end{dfn}

Note that (E3) and \gzit{e.res0} ensure that $\sigma(f)$ and $\res(g)$ 
are nonnegative real numbers that vanish if $f,g$ are constant, 
i.e., complex numbers, and $g\ne 0$.
This definition is analogous to the definitions of elementary classical 
statistics, where $\Ez$ is a commutative algebra of random variables, 
to the present, more general situation; in a statistical context, 
the uncertainty 
$\sigma(f)$ is referred to as \bfi{standard deviation}.

There is no need to associate an intrinsic statistical 
meaning to the above concepts. We treat the uncertainty 
$\sigma(f)$ and the limit resolution $\res(g)$ simply as an absolute 
and relative uncertainty measure, respectively, specifying 
how accurately one can treat $g$ as a sharp number, given by this 
value. 

In experimental practice, the limit resolution is a lower bound 
on the relative accuracy with which one can expect $\<g\>$ to be 
determinable reliably\footnote{
The situation is analogous to the limit resolution with which one can
determine the longitude and latitude of a city such as Vienna.
Clearly these are well-defined only up to some limit resolution
related to the diameter of the city. No amount of measurements can 
reduce the uncertainty below about 10km. For an extended object,
the uncertainty in its position is conceptual, 
not just a lack of knowledge or precision. Indeed, a point may be 
{\em defined} in these terms: It is an object in a state where the 
position has zero limit resolution.
}\ 
from measurements of a single system at a single time. 
In particular, a quantity $g$ is considered to be 
\bfi{significant} if $\res(g)\ll 1$, while it is \bfi{noise} if 
$\res(g)\gg 1$. If $g$ is a quantity and $\widetilde g$ is a good 
approximation of its value then $\Delta g:=g-\widetilde g$ is 
noise. Sufficiently significant quantities can be treated as 
\bfi{deterministic}; the analysis of noise is the subject of 
\bfi{statistics}.

\begin{prop} \label{p5.2}

For any state,

(i) $f\leq g \implies \<f\> \leq \<g\>$.

(ii) For $f,g\in\Ez$,
\[
\cov(f,g)=\re(\<f^*g\>-\<f\>^*\<g\>),
\]
\[
\<f^*f\>=\<f\>^*\<f\>+\sigma(f)^2,
\]
\[
|\<f\>|\leq\sqrt{\<f^*f\>}.
\]

(iii) If $f$ is Hermitian then $\bar f = \<f\>$ is real and
\[
\sigma(f)=\sqrt{\<(f-\overline{f})^2 \>}
=\sqrt{\<f^2\>-\<f\>^2}.
\]

(iv) Two commuting Hermitian quantities $f,g$ are uncorrelated iff
\[
\<fg\>=\<f\>\<g\>.
\]

\end{prop}

\bepf
(i) follows from (E1) and (E3).

(ii) The first formula holds since
\[
\<(f-\bar f)^*(g-\bar g)\>
=\<f^*g\>-\bar f^*\<g\>-\<f\>^*\bar g +\bar f^*\bar g 
= \<f^*g\>-\<f\>^*\<g\>.
\]
The second formula follows for $g=f$, using (E1), and the third 
formula is an immediate consequence.

(iii) follows from (E1) and (ii).

(iv) If $f,g$ are Hermitian and commute then $fg$ is Hermitian by 
Proposition \ref{p5.1.2}(ii), hence $\<fg\>$ is real. By (ii),
$\cov(f,g)=\<fg\>-\<f\>\<g\>$, and the assertion follows.
\epf

Formally, the essential difference between classical mechanics 
and quantum mechanics in the latter's lack of commutativity.
While in classical mechanics there is in principle no lower
limit to the uncertainties with which we can prepare the quantities
in a system of interest,
the quantum mechanical uncertainty relation for noncommuting 
quantities puts strict limits on the uncertainties in the preparation
of microscopic states. Here, {\em preparation} is defined informally 
as bringing the system into a state such that measuring certain 
quantities $f$ gives numbers that agree with the values $\<f\>$ to an 
accuracy specified by given uncertainties.

We now discuss the limits of the accuracy to which this 
can be done.

\begin{prop} \label{p5.1}~\\
(i) The \bfi{Cauchy--Schwarz inequality}  
\[
|\< f^*g \>|^2 \le \< f^*f \>\< g^*g \>
\]
holds for all $f,g\in\Ez$.

(ii) The \bfi{uncertainty relation}
\[
\sigma(f)^2\sigma(g)^2 
\geq |\cov(f,g)|^2+\left|\shalf\<f^*g-g^*f\>\right|^2
\]
holds for all $f,g\in\Ez$.

(iii) For $f,g\in\Ez$, 
\lbeq{ecov1}
\cov(f,g)=\cov(g,f)=\shalf(\sigma(f+g)^2-\sigma(f)^2-\sigma(g)^2),
\eeq
\lbeq{ecov}
|\cov(f,g)| \leq \sigma(f)\sigma(g), 
\eeq
\lbeq{esig}
\sigma(f+g) \leq \sigma(f)+\sigma(g).
\eeq
In particular,
\lbeq{e.prodbound}
|\<fg\>-\<f\>\<g\>|\leq\sigma(f)\sigma(g) 
~~~\mbox{for commuting Hermitian } f,g. 
\eeq

\end{prop}

\bepf
(i) For arbitrary $\alpha ,\beta\in \Cz$ we have
\[
\begin{array}{ll}
0&\le \<(\alpha f-\beta g)^*(\alpha f-\beta g )\> \\
&=\alpha ^* \alpha \< f^*f \>-\alpha ^* \beta \< f^*g \>
-\beta ^*\alpha \< g^*f \>+\beta\beta^* \< g^*g \>\\
&=|\alpha |^2\< f^*f \>-2\re(\alpha ^* \beta \< f^*g \>)
+|\beta|^2\< g^*g \>
\end{array}
\]
We now choose $\beta=\< f^*g \>$, and obtain for arbitrary
real $\alpha $ the inequality
\lbeq{f.8}
0\le \alpha ^2\< f^*f \>
-2\alpha |\< f^*g \>|^2+|\< f^*g \>|^2\< g^*g \>.
\eeq
The further choice $\alpha=\< g^*g \>$ gives
\[
0\le \< g^*g \>^2\< f^*f \>-\< g^*g \>|\< f^*g \>|^2.
\]
If $\< g^*g \>>0$, we find after division by $\< g^*g \>$ that (i) 
holds. And if $\< g^*g \>\le 0$ then $\< g^*g \>=0$ and we have 
$\< f^*g \>=0$ since otherwise a tiny $\alpha $ produces a negative
right hand side in \gzit{f.8}. Thus (i) also holds in this case.

(ii) Since $(f-\bar f)^*(g-\bar g)-(g-\bar g)^*(f-\bar f)=f^*g-g^*f$,
it is sufficient to prove the uncertainty relation for the case of
quantities $f,g$ whose value vanishes. In this case, (i) implies
\[
(\re \<f^*g\>)^2 +(\im \<f^*g\>)^2 =|\<f^*g\>|^2 \leq 
\< f^*f \>\< g^*g \> = \sigma(f)^2\sigma(g)^2.
\]
The assertion follows since $\re \<f^*g\>=\cov(f,g)$ and
\[
i\im \<f^*g\>=\shalf(\<f^*g\>-\<f^*g\>^*)=\shalf\<f^*g-g^*f\>.
\]

(iii) Again, it is sufficient to consider the case of
quantities $f,g$ whose value vanishes. Then
\lbeq{esig1}
\begin{array}{lll}
\sigma(f+g)^2 &=& \<(f+g)^*(f+g)\>
=\<f^*f\>+\<f^*g+g^*f\>+\<g^*g\>\\
&=& \sigma(f)^2+2\cov(f,g)+\sigma(g)^2,
\end{array}
\eeq
and \gzit{ecov1} follows. \gzit{ecov} is an immediate consequence of
(ii), and \gzit{esig} follows easily from \gzit{esig1} and 
\gzit{ecov}. Finally, \gzit{e.prodbound} is a consequence of 
\gzit{ecov} and Proposition \ref{p5.2}(iii).
\epf

If we apply Proposition \ref{p5.1}(ii) to scalar position $q$ and 
momentum $p$ variables satisfying the 
\bfi{canonical commutation relation}
\lbeq{ccr}
[q,p]=i\hbar,
\eeq
we obtain 
\lbeq{e6.unc0}
\sigma(q)\sigma(p)\geq \shalf\hbar,
\eeq
the \bfi{uncertainty relation} of \sca{Heisenberg} \cite{Hei,Rob}.
It implies that no state exists where both position $q$ and momentum 
$p$ have arbitrarily small uncertainty.

\chapter{The laws of thermodynamics}\label{c.lawtherm}

This chapter rederives the laws of thermodynamics from statistical 
mechanics, thus putting the phenomenological discussion of 
Chapter \ref{c.ctherm} on more basic foundations.

We confine our attention to a restricted but very important class 
of Gibbs states, those describing thermal states.
We introduce thermal states by selecting the quantities 
whose values shall act as extensive variables in a 
thermal model. On this level, we shall be able to reproduce the
phenomenological setting of the present section from first principles;
see the discussion after Theorem \ref{t7.3}.
If the underlying detailed model is assumed to be known then the
system function, and with it all thermal properties,
are computable in principle, although we only hint at the ways to do
this numerically. We also look at a hierarchy of
thermal models based on the same bottom level description
and discuss how to decide which description levels are appropriate.

Although dynamics is important for systems not in global equilibrium, 
we completely ignore dynamical issues in this chapter. 
We take a strictly kinematic point of view, and look as before only at 
a single phase without chemical reactions.
In principle, it is possible to extend the present setting to cover the 
dynamics of the nonequilibrium case and deduce quantitatively the 
dynamical laws of  nonequilibrium thermodynamics 
(\sca{Beris \& Edwards} \cite{BerE}, \sca{Oettinger} \cite{Oet}) 
from microscopic properties, including phase formation, chemical 
reactions, and the approach to equilibrium; see, e.g., 
\sca{Balian} \cite{Bal2}, \sca{Grabert} \cite{Gra}, 
\sca{Rau \& M\"uller} \cite{RauM}, \sca{Spohn} \cite{Spo}.

\section{The zeroth law: Thermal states}\label{zeroth}

Thermal states are special Gibbs states, used in statistical mechanics 
to model macroscopic physical systems that are homogeneous on the 
(global, local, microlocal, or quantum) level 
used for modeling. They have all the properties traditionally 
postulated in thermodynamics. While we discuss the lower levels
on an informal basis, we consider in the formulas for notational 
simplicity mainly the case of global equilibrium, where there are 
only finitely many extensive variables. 
Everything extends, however, with (formally trivial but from a 
rigorous mathematical view nontrivial) changes to local and 
microlocal equilibrium, where extensive variables are fields, 
provided the sums are replaced by appropriate integrals;
cf. \sca{Oettinger} \cite{Oet}. 

In the setting of statistical mechanics, the intensive variables are, 
as in Section \ref{s.phen}, numbers parameterizing the entropy and
characterizing a particular system at a particular time.
To each admissible combination of intensive variables there is a 
unique thermal state providing values for all quantities. 
The extensive variables then appear as the values of 
corresponding extensive quantities. 

A basic extensive quantity present in each thermal system is
the \bfi{Hamilton energy} $H$; it is identical to the \bfi{Hamiltonian}
function (or operator) in the underlying dynamical description of the 
classical (or quantum) system. In addition, there are further basic
extensive quantities which we call $X_j$ ($j\in J$) and collect in a
vector $X$, indexed by $J$. All other extensive quantities are 
expressible as linear combinations of these basic extensive quantities.
The number and meaning of the extensive
variables depends on the type of the system; typical examples are
given in Table \ref{3.t.} in Section \ref{s.model}.

In the context of statistical mechanics (cf. Examples \ref{e3.1}), 
the Euclidean $*$-algebra $\Ez$ is typically an algebra of functions 
(for classical physics) or linear operators 
(for quantum physics), and $H$ is a particular function or linear 
operator characterizing the class of systems considered.
The form of the operators $X_j$ depends on the level of thermal 
modeling; for further discussion, see Section \ref{s.model}.

For qualitative theory and for deriving semi-empirical 
recipes, there is no need to know details about $H$ or $X_j$; 
it suffices to treat them as primitive objects. The advantage we gain 
from such a less detailed setting is that to reconstruct all of 
phenomenological thermodynamics, a much simpler machinery suffices than
what would be needed for a detailed model

It is intuitively clear from the informal definition of extensive 
variables in Section \ref{s.appeq} that the only functions of
independent extensive variables that are again extensive can be
linear combinations, and it is a little surprising that {\em the whole
machinery of equilibrium thermodynamics follows from a formal version 
of the simple assumption that in thermal states the entropy is
extensive}. We take this to be the mathematical expression of the 
zeroth law and formalize this assumption in a precise mathematical
definition. 

\begin{dfn} \label{3.1.}
A \bfi{thermal system} is defined by a family of Hermitian 
\bfi{extensive variables} $H$ and $X_j$ ($j\in J$) from a Euclidean 
$*$-algebra. A \bfi{thermal state} of a thermal system is 
a Gibbs state whose entropy $S$ is a linear combination of the 
basic extensive quantities of the form
\lbeq{3-2}
S=T^{-1}\Big(H-\sum _{j\in J}\alpha_jX_j\Big) 
=T^{-1}(H-\alpha\cdot X) ~~~
\mbox{\bf (\bfi{zeroth law of thermodynamics})} 
\eeq
with suitable real numbers $T\not=0$ and $\alpha_j$ ($j\in J$).
Here $\alpha$ and $X$ are the vectors with components $\alpha_j$ 
($j\in J$) and $X_j$ ($j\in J$), respectively.
\end{dfn}

Thus the value of an arbitrary quantity $g$ is
\lbeq{e.thermal}
\ol g:=\<g\>=\sint e^{-\beta(H-\alpha\cdot X)}g,
\eeq
where 
\lbeq{3-8}
\beta =\frac {1} {\kbar T}\ .
\eeq
The numbers $\alpha_j$ are called the \bfi{intensive
variables conjugate to} $X_j$, the number $T$ is called the
\bfi{temperature}, and $\beta$ the \bfi{coldness}.
$\ol S,\ol H,\ol X,T$, and $\alpha$ are called the
\bfi{thermal variables} of the system.
Note that the extensive variables of traditional thermodynamics are 
in the present setting not represented by the extensive quantities 
$S,H,X_j$ themselves but by their values $\ol S,\ol H,\ol X$.

Since we can write the zeroth law (\ref{3-2}) in the form
\lbeq{e.euler}
H=TS+\alpha \cdot X,
\eeq
called the \bfi{Euler equation}, the temperature $T$ is considered to 
be the intensive variable conjugate to the entropy $S$. 

\begin{rems}\label{r7.2} 
(i) As already discussed in Section \ref{s.cons} for the case of 
temperature, measuring intensive variables is based upon the empirical 
fact that two systems in contact where the free exchange of some 
extensive quantity is allowed tend to relax to a joint equilibrium 
state, in which the corresponding intensive variable is the same in 
both systems. If a small measuring 
device is brought into close contact with a large system, the joint 
equilibrium state will be only minimally different from the original 
state of the large system; hence the intensive variables of the 
measuring device will move to the values of the intensive variables 
of the large system in the location of the measuring device. This 
allows to read off their value from a calibrated scale.

(ii) Many treatises of equilibrium thermodynamics take the possibility 
of measuring temperature to be the contents of the zeroth law of 
thermodynamics. The present, different choice for the zeroth law has 
far reaching consequences.
Indeed, as we shall see, the definition implies the first and 
second law, and (together with a quantization condition) the third law 
of thermodynamics. Thus these become theorems rather than separately 
postulated laws.

(iii) We emphasize that the extensive quantities $H$ and $X_j$ are 
independent of the intensive quantities $T$ and $\alpha$, while $S$, 
defined by \gzit{3-2}, is an extensive quantity defined only when 
values for the intensive quantities are prescribed.
From \gzit{3-2} it is clear that values also depend 
on the particular state a system is in. It is crucial to distinguish 
between the quantities $H$ or $X_j$, which are part of the definition 
of the system but independent of the state (since they are 
independent of $T$ and $\alpha$), and their values 
$\ol{H}=\< H \>$ or $\ol{X}_j=\< X_j \>$, which change with the state. 

(iv) In thermodynamics, the interest is restricted to the values of 
the thermal variables. In statistical mechanics, the values of 
the thermal variables determine a state of the microscopic system.
In particular, the knowledge of the intensive variables allows one
to compute the values \gzit{e.thermal} of {\em arbitrary} microscopic 
quantities, not only the extensive ones. Of course, these values 
don't give information about the position and momentum of individual 
particles but only about their means. For example, the mean velocity 
of an ideal monatomic gas at temperature $T$ turns out to be 
$\<\vv\> = 0$, and the mean velocity-squared is $\<\vv^2\> =3\kbar T$.
(We don't derive these relations here; usually they are obtained from 
a starting point involving the Boltzmann equation.)

(v) A general Gibbs state has an incredibly high complexity.
Indeed, in the classical case, the specification of an arbitrary
Gibbs state for 1 mole of a pure, monatomic substance such as Argon 
requires specifying the entropy $S$, a function of 
$6N_A \approx 36 \cdot 10^{23}$ degrees of freedom. In comparison, 
a global equilibrium state of Argon is specified by three numbers 
$T,p$ and $\mu$, a local equilibrium state by three fields depending
on four parameters (time and position) only, and a microlocal 
equilibrium state by three fields depending on seven parameters
(time, position, and momentum). Thus global, local, and microlocal
equilibrium states form a small minority in the class of all Gibbs 
states. It is remarkable that this small class of states suffices
for the engineering accuracy description of all macroscopic phenomena.

(vi) Of course, the number of thermal variables or fields needed to 
describe a system depends on the true physical situation. For example, 
a system that is in local equilibrium only cannot be adequately 
described by the few variables characterizing global equilibrium. 
The problem of selecting the right set of extensive quantities for 
an adequate description is discussed in Section \ref{s.model}.

(vii) The formulation \gzit{3-2} is almost universally used in practice.
However, an arbitrary linear combination
\lbeq{3-6}
S=\gamma H+h_0X_0+\dots +h_sX_s
\eeq
can be written in the form \gzit{3-2} with $T=1/\gamma $ and 
$\alpha_j=-h_j/\gamma $, provided that $\gamma \not=0$; indeed, 
\gzit{3-6} is mathematically the more natural form, which also allows 
states of infinite temperature that are excluded in \gzit{3-2}. 
This shows that the coldness $\beta$ is a more natural variable than 
the temperature $T$; it figures prominently in statistical mechanics.
Indeed, the formulas of statistical mechanics are continuous in
$\beta$ even for systems such as those considered in 
Example \ref{ex.schottky}, where $\beta$ may become zero or negative.
The temperature $T$ reaches in this case infinity, then jumps to 
minus infinity, and then continues to increase. According to 
\sca{Landau \& Lifschitz} \cite[Section 73]{LanL}, states of negative 
temperature, i.e., negative coldness, must therefore be considered 
to be hotter, i.e., less cold, than states of any positive temperature. 
On the other hand, in the limit $T \to 0$, a system becomes infinitely 
cold, giving intuition for the unattainability of zero absolute 
temperature. 

(viii) In mathematical statistics, there is a large body of work on
{\em exponential families}, 
which is essentially the mathematical equivalent 
of the concept of a thermal state over a commutative algebra; 
see, e.g., \sca{Barndorff-Nielsen} \cite{BarN}.
In this context, the values of the extensive quantities 
define a {\em sufficient statistic}, from which the whole distribution
can be reconstructed (cf. Theorem \ref{3.3.} below 
and the remarks on objective probability in Section \ref{s.limit}). 
This is one of the reasons why
exponential families provide a powerful machinery for statistical 
inference; see, e.g., \sca{Bernardo \& Smith} \cite{BerS}. 
For recent extensions to quantum statistical inference, see, e.g.,  
\sca{Barndorff-Nielsen} et al. \cite{BarN2} and the references there.

(ix) For other axiomatic settings for deriving thermodynamics, which 
provide different perspectives, see 
\sca{Carath\'eodory} \cite{Car},
\sca{Haken} \cite{Haken}, \sca{Jaynes} \cite{Jaynes}, 
\sca{Katz} \cite{Kat}, \sca{Emch \& Liu} \cite{EmcL},
and \sca{Lieb \& Yngvason} \cite{LieY}.

\end{rems}

\section{The equation of state}\label{s.eos}

Not every combination $(T,\alpha)$ of intensive variables defines
a thermal state; the requirement that $\<1\>=1$ enforces a restriction
of $(T,\alpha)$ to a manifold of admissible thermal states.

\begin{thm} \label{t.eos}
Suppose that $T>0$.

(i) For any $\kappa>0$, the \bfi{system function} $\Delta$ defined by
\lbeq{e.eosa}
\Delta(T,\alpha):= \kappa T \log \sint e^{-\beta (H-\alpha\cdot X)}
\eeq
is a convex function of $T$ and $\alpha$. It vanishes only if $T$
and $\alpha$ are the intensive variables of a thermal state.

(ii) In a thermal state, the intensive variables are related by the 
\bfi{equation of state}
\lbeq{e.eos}
\Delta(T,\alpha)=0.
\eeq
The \bfi{state space} is the set of $(T,\alpha)$ satisfying 
\gzit{e.eos}.

(iii) The values of the extensive variables are given by
\lbeq{e.Sx}
\ol S=\Omega\frac{\partial \Delta}{\partial T}(T,\alpha),~~~
\ol X=\Omega\frac{\partial \Delta}{\partial \alpha}(T,\alpha)~~~
\mbox{for some } \Omega>0,
\eeq
and the \bfi{phenomenological Euler equation}
\lbeq{e.H}
\ol H=T \ol S + \alpha \cdot \ol X.
\eeq

(iv) Regarding $\ol S$ and $\ol X$ as functions of $T$ and $\alpha$, 
the matrix
\lbeq{7-12}
\Sigma:= 
\left(\begin{array}{cc}
\D\frac{\partial\ol{S}}{\partial T}& 
\D\frac{\partial\ol{S}}{\partial\alpha } \\
 ~\\
\D\frac{\partial\ol{X}}{\partial T}& 
\D\frac{\partial\ol{X}}{\partial\alpha }
\end{array}\right)
\eeq
is symmetric and positive semidefinite; in particular, we have the 
\bfi{Maxwell reciprocity relations}
\lbeq{7-10}
\frac {\partial\ol{X}_i} {\partial\alpha_j}
=\frac {\partial\ol{X}_j} {\partial\alpha_i},\quad
\frac {\partial\ol{X}_i} {\partial T}
=\frac {\partial\ol{S}} {\partial\alpha_i},
\eeq
and the \bfi{stability conditions}
\lbeq{7-13}
\frac {\partial\ol{S}} {\partial T}\ge 0,~~ 
\frac {\partial\ol{X}_j} {\partial\alpha_j}\ge 0~~~(j\in J).
\eeq
\end{thm}

\bepf
By Theorem \ref{t3.3}(i), the function $\phi$ defined by
\[
\phi(\alpha_0,\alpha):=\log\sint e^{-(\alpha_0H-\alpha\cdot X)} 
= -W(\alpha_0H-\alpha\cdot X)
\]
is a convex function of $\alpha_0$ and $\alpha$. Put 
$\Omega=\kbar/\kappa$. Then, by Proposition \ref{e.convex}, 
\lbeq{e.co}
\Delta(T,\alpha)=-\kappa T W(\beta(H-\alpha\cdot X))
= \kappa T \phi\Big(\frac{1}{\kbar T},\frac{\alpha}{\kbar T}\Big)
\eeq
is also convex. The condition $\Delta(T,\alpha)=0$ is equivalent to
\[
\sint e^{-S/\kbar} = \sint e^{-\beta(H-\alpha\cdot X)} 
=e^{\Delta/\kappa T} =1,
\]
the condition for a thermal state. This proves (i) and (ii).

(iii) The formulas for $\ol S$ and $\ol X$ follow by differentiation of 
\gzit{e.co} with respect to $T$ and $\alpha$, using \gzit{e.diffW}.
Equation \gzit{e.H} follows by taking values in 
\gzit{e.euler}, noting that $T$ and $\alpha$ are real numbers. 

(iv) By (iii), the matrix 
\[
\Sigma= 
\left(\begin{array}{cc}
\D\frac{\partial^2\Delta}{\partial T^2}& 
\D\frac{\partial^2\Delta}{\partial T\partial\alpha} \\
 ~\\
\D\frac{\partial^2\Delta}{\partial\alpha \partial T}& 
\D\frac{\partial^2\Delta}{\partial\alpha^2}
\end{array}\right)
\]
is the Hessian matrix of the convex function $\Delta$. Hence $\Sigma$ 
is symmetric and positive semidefinite. \gzit{7-10} expresses the 
symmetry of $\Sigma$, and \gzit{7-13} holds since the diagonal entries
of a  positive semidefinite matrix are nonnegative.
\epf

\begin{rems}\label{r5.2}
(i) For $T<0$, the same results hold, with the change that $\Delta$ is 
concave instead of convex, $\Sigma$ is negative semidefinite, and
the inequality signs in \gzit{7-13} are reversed.
This is a rare situation; it can occur only in (nearly) massless 
systems embedded out of equilibrium within (much heavier) matter, 
such as spin systems (cf. \sca{Purcell \& Pound} \cite{PurP}), 
radiation fields in a cavity (cf. \sca{Hsu \& Barakat} \cite{HsuB}),
or vortices in 2-dimensional 
fluids (cf. \sca{Montgomery \& Joyce} \cite{MonJ}, 
\sca{Eyinck \& Spohn} \cite{EyiS}). A massive thermal system  
couples significantly to kinetic energy. In this case, the total 
momentum $p$ is an extensive quantity, related to the velocity $v$, 
the corresponding intensive variable, by $p = M v$, where $M$ is the 
extensive total mass of the system. From \gzit{e.Sx}, we find that 
$\ol p = \Omega\partial \Delta/\partial v$, which implies that
$\Delta = \Delta|_{v=0} + \frac{\ol M}{2\Omega} v^2$. Since the mass is
positive, this expression is convex in $v$, not concave; hence $T>0$.
Thus, in a massive thermal system, the temperature must be 
positive. 

(ii) In applications, the free scaling constant $\kappa$ is usually
chosen as 
\lbeq{e.Deltascale}
\kappa=\kbar/\Omega, 
\eeq 
where $\Omega$ is a measure of 
\bfi{system size}, e.g., the total volume or total mass of the system.
In actual calculations from statistical mechanics, the integral is 
usually a function of the shape and size of the system. To make the 
result independent of it, one performs the so-called thermodynamic 
limit $\Omega\to\infty$; thus $\Omega$ must be chosen in such a way 
that this limit is nontrivial. Extensivity in single phase global 
equilibrium then justifies treating $\Omega$ as an arbitrary positive 
factor. 
\end{rems}

In phenomenological thermodynamics (cf. Section \ref{s.phen}), 
one makes suitable, more or 
less heuristic assumptions on the form of the system function, while in
\bfi{statistical mechanics}, one derives its form from \gzit{e.eos}
and specific choices for the quantities $H$ and $X$ within one of the
settings described in Example \ref{e3.1}. Given these choices, the 
main task is then the evaluation of the system function \gzit{e.eosa},
from which the values of all quantities can be computed.
\gzit{e.eosa} can often be approximately evaluated from the 
cumulant expansion \gzit{e.cumulant} and/or a mean field approximation 
\gzit{e.meanfield}.

An arbitrary Gibbs state is generally not a thermal state. However, 
we can try to approximate it by an equilibrium state in which the
extensive variables have the same values. The next result shows that
the slack (the difference between the left hand side and the right 
hand side) in \gzit{e.excess}, which will turn out to be the 
microscopic form of the Euler inequality \gzit{e.Hi}, is always 
nonnegative and
vanishes precisely in equilibrium. Thus it can be used as a measure of 
how close the Gibbs state is to an equilibrium state.

\begin{thm} \label{t7.3}
Let $\<\cdot\>$ be a Gibbs state with entropy $S$.
Then, for arbitrary $(T,\alpha)$ satisfying $T>0$ and the equation of 
state \gzit{e.eos}, the values 
$\ol H=\<H\>$, $\ol S=\<S\>$, and $\ol X=\<X\>$ satisfy
\lbeq{e.excess}
\ol H \ge T\ol S-\alpha\cdot \ol X.
\eeq
Equality only holds if $S$ is the entropy of a thermal state with 
intensive variables $(T,\alpha)$.
\end{thm}

\bepf
The equation of state implies that $S_c:=T^{-1}(H-\alpha\cdot X)$
is the entropy of a thermal state. Now the assertion follows from 
Theorem \ref{t4.5}, since
$\<S\>\le \<S_c\>=T^{-1}(\<H\>-\alpha\cdot \<X\>)$,
with equality only if $S=S_c$.
\epf

As the theorem shows, everything of macroscopic interest is 
deducible from an explicit formula for the system function.
Hence one can use thermodynamics in many situations very successfully 
as a phenomenological theory 
without having to bother about microscopic details. It suffices that
a phenomenological expression for $\Delta(T,\alpha)$ is available.
In particular, the phenomenological axioms from Section \ref{s.phen}
now follow by specializing the above to a \bfi{standard system}, 
characterized by the extensive quantities
\lbeq{3-1}
H, X_0=V,\quad X_j=N_j\ (j\ne 0),
\eeq
where, as before, $V$ denotes the (positive) \bfi{volume} of the 
system, and each $N_j$ denotes the (nonnegative) number of molecules 
of a fixed chemical composition (we shall call these \bfi{particles of 
kind} $j$). However, $H$ and the $N_j$ are now quantities from $\Ez$,
rather than thermal variables. We call
\lbeq{3-4}
P:=-\alpha_0 
\eeq
the \bfi{pressure} and 
\lbeq{3-4z}
\mu _j:=\alpha_j\quad (j\not=0)
\eeq
the \bfi{chemical potential} of kind $j$; hence
\[
\alpha \cdot X = -PV+\mu\cdot N.
\]
Specializing the theorem, we find the phenomenological Euler equation 
\lbeq{e.Hx}
\ol{H}=T\ol{S}-P V +\mu \cdot \ol{N}.
\eeq
Note that $\ol V=V$ since we took $V$ as system size.
For reversible changes, we have the first law of thermodynamics 
\lbeq{3.1stx}
d\ol{H}=Td\ol{S}-Pd V +\mu \cdot d\ol{N}
\eeq
and the Gibbs-Duhem equation
\lbeq{e.GDx}
0=\ol{S}dT- V dP+\ol{N}\cdot d\mu.
\eeq
A comparison with Section \ref{s.phen}
shows that dropping the bars from the values 
reproduces for $T>0$, $P>0$  and $\ol S\ge 0$ the axioms of 
phenomenological thermodynamics, except for the extensivity outside 
equilibrium (which has local equilibrium as its justification).
The assumption $T>0$ 
was justified in Remark \ref{r5.2}(i), and $\ol S\ge 0$ will be 
justified in Section \ref{third}. But there seem to be no theoretical 
arguments which shows that the pressure of a standard system 
in the above sense must always be positive. (At $T<0$, negative 
pressure is possible; see Example \ref{ex.schottky}.) We'd appreciate 
getting information about this from readers of this book.

Apart from boundary effects, whose role diminishes  
as the system gets larger, the extensive quantities scale linearly with 
the volume. In the thermodynamic limit, corresponding 
to an idealized system infinitely extended in all directions, the
boundary effects disappear and the linear scaling 
becomes exact, although this can be proved rigorously only in simple 
situations, e.g., for hard sphere model systems
(\sca{Yang \& Lee} \cite{YanL}) or spin systems 
(\sca{Griffiths} \cite{Gri}).
A thorough treatment of the thermodynamic limit (e.g., \sca{Ruelle}
\cite{Rue1,Rue2}, \sca{Thirring} \cite{Thi}, or, in the framework of 
large deviation theory, \sca{Ellis} \cite{Ell}) in general needs 
considerably more algebraic 
and analytic machinery, e.g., the need to work in place of thermal 
states with more abstract KMS-states (which are limits of sequences of 
thermal states still satisfying a KMS condition \gzit{e.KMS}).
Moreover, proving the existence of the limit requires detailed 
properties of the concrete microscopic description of the system.

For very small systems, typically atomic clusters or molecules, 
$N$ is fixed and a \bfi{canonical ensemble} without the $\mu \cdot N$ 
term is more appropriate. For the thermodynamics of small systems 
(see, e.g., (\sca{Bustamente} et al. \cite{BusLR}, 
\sca{Gross} \cite{Gro}, \sca{Kratky} \cite{Kra})
such as a single cluster of atoms, $V$ is still taken as a fixed 
reference volume, but now changes in the physical volume (adsorption 
or dissociation at the surface) are not represented in the system, 
hence need not respect the thermodynamic laws. For large 
surfaces (e.g., adsorption studies in chromatography; see 
{\sc Karger} et al. \cite{KarSH}, {\sc Masel} \cite{Mas}), 
a thermal description is achievable by including additional variables 
(surface area and surface tension) to account for the boundary effects;
but clearly, surface terms scale differently with system size 
than bulk terms.

Thus, whenever the thermal description is valid, computations can be 
done in a fixed reference volume which we take as system size
$\Omega$. (Formulas for an arbitrary volume $V$ are then derived by 
extensivity, scaling every extensive quantity with $V/\Omega$.)
The reference volume may be represented in the Euclidean $*$-algebra 
as a real number, so that in particular $\ol V=V$. 
Then \gzit{e.eosa} together with {e.Deltascale} implies that
\[
\Delta(T,P,\mu) = V^{-1}\kbar T
\log (e^{-\beta PV})\sint e^{-\beta (H-\mu\cdot N)},
\]
hence 
\lbeq{e.eosb}
\Delta(T,P,\mu)= V^{-1}\kbar T(\log Z(T,V,\mu) - PV)
=P(T,\mu)-P,
\eeq
where
\lbeq{3-10}
Z(T,V,\mu):=\sint e^{-\beta (H-\mu \cdot N)}
\eeq
is the so-called \bfi{grand canonical partition function} of the system
and 
\lbeq{3-9a}
P(T,\mu):=V^{-1}\kbar T\log Z(T,V,\mu),
\eeq
while $P$ without argument is the parameter in the left hand side of 
\gzit{e.eosb}.
With our convention of considering a fixed reference volume and treating
the true volume by scaling extensive variables, this expression is 
independent of $V$, since it relates intensive variables unaffected by 
scaling. (A more detailed argument would have to show that the  
\bfi{thermodynamic limit}
$P(T,\mu):=\lim_{V\to\infty}V^{-1}\kbar T\log Z(T,V,\mu)$
exists, and argue that thermodynamics is applied in practice only to 
systems where $V$ is so large that the difference to the limit is 
negligible.

The equation of state \gzit{e.eos} therefore takes the form
\lbeq{e.eos2}
P= P(T,\mu).
\eeq
Quantitative expressions for the equation of state can often be 
computed from \gzit{3-10}--\gzit{3-9a} using the cumulant expansion 
\gzit{e.cumulant} and/or a mean field approximation \gzit{e.meanfield}.
Note that these relations imply that
\[
e^{-\beta P(T,\mu)V} = \sint e^{-\beta(H-\mu\cdot N)}.
\]
Traditionally (see, e.g., \sca{Gibbs} \cite{Gib}, 
\sca{Huang} \cite{Hua}, \sca{Reichl} \cite{Rei}), 
the thermal state corresponding to \gzit{e.eosb}--\gzit{3-9a}
is called a \bfi{grand canonical ensemble}, and the following results 
are taken as the basis for microscopic calculations from 
statistical mechanics.

\begin{thm} \label{3.3.}
For a standard system in global equilibrium, values of an
arbitrary quantity $g$ can be calculated from \gzit{3-10} and
\lbeq{3-11}
\< g\> =Z(T,\mu)^{-1}\sint e^{-\beta (H-\mu \cdot N)}g.
\eeq
The values of the extensive quantities are given in terms of 
the equation of state \gzit{3-9a} by
\lbeq{3-19}
\ol{S}= V \frac{\partial P}{\partial T}(T,\mu ),~~~
\ol{N}_j= V \frac{\partial P}{\partial \mu _j}(T,\mu )
\eeq
and the phenomenological Euler equation \gzit{e.Hx}.
\end{thm}

\bepf 
Equation \gzit{3-9a} implies that $Z(T,V,\mu)=e^{\beta PV}$, hence
\[ 
\begin{array}{lll}
\<g\> &=\sint e^{-S/\kbar }g=\sint e^{-\beta (H+PV-\mu \cdot N)}g \\ 
&=e^{-\beta PV}\sint e^{-\beta (H-\mu \cdot N)}g=
Z(T,V,\mu )^{-1} \sint e^{-\beta (H-\mu \cdot N)}g,
\end{array} 
\]
giving \gzit{3-11}. The formulas in \gzit{3-19} follow from 
\gzit{e.Sx} and \gzit{e.eosb}. 
\epf

No thermodynamic limit was needed to derive the above results. 
Thus, everything holds -- though with large limit resolutions in 
measurements -- even for single small systems 
(\sca{Bustamente} et al. \cite{BusLR}, \sca{Gross} \cite{Gro}, 
\sca{Kratky} \cite{Kra}).

\begin{expl}\label{ex.schottky}
We consider the two level system from Example \ref{ex.canonical},
using $\Omega=1$ as system size. From \gzit{3-10} and  \gzit{3-9a},
we find $Z(T,\mu)=1+e^{-E/\kbar T}$, hence
\[
P(T,\mu)=\kbar T\log(1+e^{-E/\kbar T})
=\kbar T\log(e^{E/\kbar T}+1)-E.
\] 
From \gzit{3-11}, we find 
\[
\ol H = \frac{Ee^{-E/\kbar T}}{1+e^{-E/\kbar T}}
=\frac{E}{e^{E/\kbar T}+1},~~~\kbar T = \frac{E}{\log(E/\ol H -1)}.
\]
(This implies that a two-level system has negative temperature and 
negative pressure if $\ol H>E/2$.)
The \bfi{heat capacity} $C:=d\ol H/dT$ takes the form
\[
C=\frac{E^2}{\kbar T^2}\frac{e^{E/\kbar T}}{(e^{E/\kbar T}+1)^2}.
\]
It exhibits a pronounced maximum, the so-called \bfi{Schottky bump}
(cf. \sca{Callen} \cite{Cal}), from which $E$ can be determined.
In view of \gzit{e.2levelapprox} below, this allows the experimental
estimation of the spectral gap of a quantum system.
The phenomenon persists to some extent for
multilevel systems; see \sca{Civitarese} et al. \cite{CivHH}.
\end{expl}

\section{The first law: Energy balance} \label{first}

We now discuss relations between 
changes of the values of extensive or intensive variables,
as expressed by the first law of thermodynamics. To derive the first 
law in full generality, we use the concept of reversible 
transformations introduced in Section \ref{s.phen}. Corresponding
to such a transformation, there is a family of thermal states 
$\<\cdot\>_\lambda$ defined by
\[
\<f\>_\lambda = \sint e^{-\beta(\lambda)(H-\alpha(\lambda)\cdot X)}f,~~~
\beta(\lambda)=\frac{1}{\kbar T(\lambda)}.
\]
\bfi{Important:} In case of local or microlocal equilibrium, 
where the thermal system carries a dynamics, it is important
to note that reversible transformations are ficticious transformations
which have nothing to do with how the system changes with time,
or whether a process is reversible in the dynamical sense that both
the process and the reverse process can be realized dynamically.
The time shift is generally {\em not} a reversible transformation.

We use differentials corresponding to reversible transformations;
writing $f=S/\kbar$, we can delete the index $f$ from the formulas
in Section \ref{s.gibbs}. In particular, we write the Kubo 
inner product \gzit{e.kubo} as
\lbeq{e.kuboS}
\<g;h\> := \<g;h\>_{S/\kbar}.
\eeq

\begin{prop}
The value $\ol g(T,\alpha):=\<g(T,\alpha)\>$ of every 
(possibly $T$- and $\alpha$-dependent) quantity 
$g(T,\alpha)$ is a state variable satisfying the  \bfi{differentiation 
formula}
\lbeq{e.diffS}
d\<g\>=\<dg\>-\<g-\ol g;dS\>/\kbar.
\eeq
\end{prop}
\bepf
That $\ol g$ is a state variable is an immediate consequence of the 
zeroth law \gzit{3-2} since the entropy depends on $T$ and $\alpha$ 
only. The differentiation formula follows from \gzit{e.diff} and
\gzit{e.kuboS}.
\epf

\begin{thm} \label{3.6.}
For reversible changes, we have the \bfi{first law of thermodynamics}
\lbeq{3.1st}
d\ol{H}=Td\ol{S}+\alpha\cdot d\ol{X}
\eeq
and the \bfi{Gibbs-Duhem equation}
\lbeq{e.GD}
0=\ol{S}dT+\ol{X}\cdot d\alpha.
\eeq

\end{thm}

\bepf 
Differentiating the equation of state \gzit{e.eos}, using the chain 
rule \gzit{3-12}, and simplifying using \gzit{e.Sx} gives the 
Gibbs-Duhem equation (\ref{e.GD}). If we differentiate the 
phenomenological Euler equation \gzit{e.H}, we obtain
\[
d\ol{H}=Td\ol{S}+\ol{S}dT+\alpha\cdot d\ol{X}+
\ol{X}\cdot d\alpha,
\]
and using (\ref{e.GD}), this simplifies to the first law of
thermodynamics.
\epf

Because of the form of the energy terms in the first law (\ref{3.1st}), 
one often uses the analogy to mechanics and calls the intensive 
variables \bfi{generalized forces}, and differentials of extensive 
variables \bfi{generalized displacements}.

For the Gibbs-Duhem equation, we give a second proof which provides 
additional insight. Since $H$ and $X$ are fixed
quantities for a given system, they do not change under reversible
transformations; therefore 
\[
dH=0,~~~dX = 0.
\]
Differentiating the Euler equation \gzit{e.euler}, therefore gives 
the relation
\lbeq{3-16}
0=TdS+SdT+X\cdot d\alpha.
\eeq
On the other hand, $S$ depends explicitly on $T$ and $\alpha$, 
and by Corollary \ref{app2.},
\lbeq{3-17}
\< dS\> =\int e^{-S/\kbar }dS=
\kbar d \left(\int e^{-S/\kbar }\right)=
\kbar d1=0,
\eeq
taking values in (\ref{3-16}) implies again the Gibbs-Duhem
equation. By combining equation \gzit{3-16} with the Kubo product we get
information about limit resolutions:

\begin{thm} \label{7.1.}~\\
(i) Let $g$ be a quantity depending continuously differentiable on the 
intensive variables $T$ and $\alpha$. Then
\lbeq{7-4}
\< g-\ol{g};S-\ol{S}\> =
\kbar T\Big(\frac {\partial \ol{g}} {\partial T}-
\Big\<\frac{\partial g}{\partial T}\Big\>\Big), 
\eeq
\lbeq{7-5}
\<g-\ol{g};X_j-\ol{X}_j\> =
\kbar T\Big(\frac {\partial \ol{g}} {\partial \alpha_j}-
\Big\<\frac{\partial g}{\partial\alpha_j}\Big\> \Big),
\eeq
(ii) If the extensive variables $H$ and $X_j$ ($j\in J$) are pairwise 
commuting then
\lbeq{7-7}
\< (S-\ol{S})^2\> =
\kbar T\ \frac {\partial\ol{S}} {\partial T},
\eeq
\lbeq{7-8}
\< (X_j-\ol{X}_j)
(S-\ol{S})\> =\kbar T
\ \frac {\partial\ol{X}_j} {\partial T}\quad \quad (j\in J),
\eeq
\lbeq{7-9}
\< (X_j-\ol{X}_j)
(X_k-\ol{X}_k)\> =\kbar T
\ \frac {\partial\ol{X}_j} {\partial\alpha_k}\quad \quad (j,k\in J),
\eeq
\lbeq{e.limitSX}
\res(S)
=\sqrt{\frac{\kbar T}{\ol S^2}\frac{\partial \ol S}{\partial T}},~~~
\res(X_j)
=\sqrt{\frac{\kbar T}{\ol X_j^2}
\frac{\partial \ol X_j}{\partial \alpha_j}},
\eeq
\lbeq{e.limitH}
\res(H)
=\sqrt{\frac{\kbar T}{\ol H^2}
\Big(T\frac{\partial \ol H}{\partial T}+
\alpha\cdot\frac{\partial \ol H}{\partial \alpha}\Big)}.
\eeq
\end{thm}

\bepf 
Multiplying the differentiation formula \gzit{e.diffS} by $\kbar T$ 
and using \gzit{3-16}, we find, for arbitrary reversible 
transformations,
\[
\kbar T(d\<g\>-\<dg\>)=\<g-\ol g;S\>dT + \<g-\ol g;X\>\cdot d\alpha.
\]
Dividing by $d\lambda$ and choosing $\lambda =T$ and 
$\lambda =\alpha_j$, respectively, gives
\[
\< g-\ol{g};S\> =
\kbar T\Big(\frac {\partial \ol{g}} {\partial T}-
\Big\<\frac{\partial g}{\partial T}\Big\>\Big),~~~
\<g-\ol{g};X_j\> =
\kbar T\Big(\frac {\partial \ol{g}} {\partial \alpha_j}-
\Big\<\frac{\partial g}{\partial\alpha_j}\Big\> \Big).
\]
(i) follows upon noting that $\<g-\ol g;h-\ol h\>= \<g-\ol g;h\>$ since
by \gzit{e.kubo1},
\[
\<g-\ol g;\ol h\>=\<g-\ol g\>\ol h=(\<g\>-\ol g)=0.
\]
If the extensive variables $H$ and $X_j$ ($j\in J$) are pairwise 
commuting then we can use \gzit{e.kubo0} to eliminate the Kubo inner
product, and by choosing $g$ as $S$ and $X_j$, respectively, we find 
\gzit{7-7}--\gzit{7-9}. The limit resolutions \gzit{e.limitSX} now
follow from \gzit{e.res} and the observation that $\<(g-\ol g)^2\>
=\<(g-\ol g)g\>-\<g-\ol g\>\ol g = \<(g-\ol g)g\> =\<g^2\>-\ol g^2$.
The limit resolution \gzit{e.limitH} follows similarly from
\[
\bary{lll}
\ol H^2\res(H)^2&=&\<H-\ol H;H-\ol H\> 
= T\<H-\ol H;S-\ol S\>+\alpha\cdot\< H-\ol H;X-\ol X\>\\
&=&\D\kbar T \Big(T\frac{\partial \ol H}{\partial T}+
\alpha\cdot\frac{\partial \ol H}{\partial \alpha}\Big).
\eary
\] 
\epf

Note that higher order central moments can be obtained in the same 
way, substituting more complicated expressions for $f$ and using the
formulas for the lower order moments to evaluate the right hand side
of (\ref{7-4}) and (\ref{7-5}). 

The extensive variables scale linearly
with the system size $\Omega$ of the system. Hence, the limit 
resolution of the extensive quantities is $O(\sqrt{\kbar/\Omega})$ in 
regions of the state space where the
extensive variables depend smoothly on the intensive variables. 
Since $\kbar$ is very small, they are negligible unless the system 
considered is very tiny. Thus, macroscopic thermal variables 
can generally be obtained with fairly high precision.
The only exceptions are states close to \bfi{critical points} 
where the extensive variables need not be differentiable, 
and their derivatives may therefore become huge. 
In particular, in the thermodynamic limit $\Omega\to\infty$, 
uncertainties are absent except close to a critical point, where
they lead to critical opacity.

\begin{cor} \label{7.4.}
For a standard thermal system, 
\lbeq{e.limitSN}
\res(S)
=\sqrt{\frac{\kbar T}{\ol S^2}\frac{\partial \ol S}{\partial T}},~~~
\res(N_j)
=\sqrt{\frac{\kbar T}{\ol N_j^2}
\frac{\partial \ol N_j}{\partial \mu_j}},
\eeq
\lbeq{e.limitH2}
\res(H)
=\sqrt{\frac{\kbar T}{\ol H^2}
\Big(T\frac{\partial \ol H}{\partial T}
+P\frac{\partial \ol H}{\partial P}
+\mu\cdot\frac{\partial \ol H}{\partial \mu}\Big)}.
\eeq
\end{cor}

\bepf
Apply \gzit{7-7}, \gzit{7-9} and \gzit{e.limitH2} to a standard system.
\epf

Note that $\res(V)=0$ since we regarded $V$ as the system size,
so that it is just a number.

The above results imply an approximate \bfi{thermodynamic 
uncertainty relation} 
\lbeq{e.thunc}
\Delta S \Delta T \ge \kbar T
\eeq
for entropy $S$ and the logarithm $\log T$ of temperature, 
analogous to the Heisenberg 
uncertainty relation \gzit{ccr} for position and momentum, in which the
Boltzmann constant $\kbar$ plays a role analogous to Planck's constant 
$\hbar$. Indeed (\sca{Gilmore} \cite{Gil.th}),
\gzit{e.thunc} can be derived by observing that 
\gzit{e.limitSN} may be interpreted approximately as
$(\Delta S)^2\ge  \kbar T \frac{\partial S}{\partial T}$;
together with the first order Taylor approximation 
$\Delta S =  \frac{\partial S}{\partial T}\Delta T$, we find that
$\Delta S \Delta T
 = (\Delta S)^2 \Big(\frac{\partial S}{\partial T}\Big)^{-1}
\ge  \kbar T$. A similar argument gives the approximate uncertainty 
relation
\lbeq{e.thunc2}
\Delta N_j \Delta \mu_j \ge \kbar T.
\eeq

\section{The second law: Extremal principles} \label{second}

The extremal principles of the second law of thermodynamics
assert that in a nonthermal state, some energy expression depending 
on one of a number of standard  boundary conditions
is strictly larger than that of related thermal states.
The associated thermodynamic potentials can
be used in place of the system function to calculate all 
thermal variables given half of them.
Thus, like the system function, thermodynamic potentials give a 
complete summary of the equilibrium properties of homogeneous materials.
We only discuss the \bfi{Hamilton potential}
\[
U(\ol S,\ol X)
:=\max_{T,\alpha}\,
\{T\ol S+\alpha\cdot \ol X\mid \Delta(T,\alpha)=0, T>0\}
\]
and the \bfi{Helmholtz potential}
\[
 A(T,\ol X)
:=\max_\alpha\,\{\alpha\cdot \ol X\mid \Delta(T,\alpha)=0\};
\]
other potentials can be handled in a similar way.

\begin{thm} \label{t.var}
\bfi{(Second law of thermodynamics)}\\
(i) In an arbitrary state, 
\[
\ol H \ge U(\ol S,\ol X),
\]
with equality iff the state is a thermal state of positive 
temperature. The remaining thermal variables are then given by
\lbeq{e.entint1}
T = \frac{\partial U}{\partial \ol S}(\ol S,\ol X),~~~
\alpha = \frac{\partial U}{\partial \ol X}(\ol S,\ol X),
\eeq
\lbeq{e.entint2}
U=\ol H = U(\ol S,\ol X).
\eeq
In particular, a thermal state of positive temperature is uniquely 
determined by the values of $\ol S$ and $\ol X$.

(ii) Let $T>0$. Then, in an arbitrary state, 
\[
\ol H-T\ol S \ge A(T,\ol X),
\]
with equality iff the state is a thermal state of temperature $T$.
The remaining thermal variables are then given by
\lbeq{e.enthelm1}
\ol S=-\frac{\partial A}{\partial T}(T,\ol X),~~~
\alpha=\frac{\partial A}{\partial \ol X}(T,\ol X),
\eeq
\lbeq{e.enthelm2}
\ol H=T \ol S + \alpha \cdot \ol X = A(T,\ol X)+T\ol S.
\eeq
In particular, a thermal state of positive temperature is uniquely 
determined by the values of $T$ and $\ol X$.
\end{thm}

\bepf 
This is proved in the same way as Theorem \ref{t.extstd};
thus we give no details.
\epf

The additivity of extensive quantities is again reflected in 
corresponding properties of the thermodynamic potentials:

\begin{thm}~\\
(i) The function $U(\ol S,\ol X)$ is a convex function 
of its arguments which is positive homogeneous of degree 1, i.e., 
for real $\lambda,\lambda^1,\lambda^2\ge 0$,
\lbeq{e.homU}
U(\lambda \ol S,\lambda \ol X)=\lambda U( \ol S,\ol  X),
\eeq
\lbeq{e.convU}
U(\lambda^1 \ol S^1+\lambda^2\ol S^2,\lambda^1 \ol X^1+\lambda^2\ol X^2)
\le \lambda^1 U(S^1,X^1)+\lambda^2 U(S^2,X^2).
\eeq

(ii) The function $A(T, \ol X)$ is a convex function 
of $X$ which is positive homogeneous of degree 1, i.e., 
for real $\lambda,\lambda^1,\lambda^2\ge 0$,
\lbeq{e.homA}
A(T,\lambda \ol X)=\lambda A(T,\ol X),
\eeq
\lbeq{e.convA}
A(T,\lambda^1 \ol X^1+\lambda^2\ol X^2)
\le \lambda^1 A(T,X^1)+\lambda^2 A(T,X^2).
\eeq
\end{thm}

\bepf  
This is proved in the same way as Theorem \ref{t.ext};
thus we give no details.
\epf

The extremal principles imply energy dissipation properties for 
time-dependent states. Since the present kinematical setting does not 
have a proper dynamical framework, it is only possible to outline the 
implications without going much into details.

\begin{thm} \label{4.4.}~\\
(i) For any time-dependent system for which $S$ and $X$ remain
constant and which converges to a thermal state with positive 
temperature, the Hamilton energy $\<H\>$ attains its global minimum in 
the limit $t\to\infty$.

(ii) For any time-dependent system maintained at fixed temperature 
$T>0$, for which $X$ remains constant and which converges to a thermal 
state, the Helmholtz energy $\<H-TS\>$ attains its global minimum 
in the limit $t\to\infty$.
\end{thm}

\bepf
This follows directly from Theorem \ref{t.var}.
\epf

This result is the shadow of a more general, dynamical observation
(that, of course, cannot be proved from kinematic assumptions alone 
but would require a dynamical theory).
Indeed, it is a universally valid empirical fact that in all natural
time-dependent processes, energy is lost or dissipated, i.e., 
becomes macroscopically unavailable, unless compensated by energy
provided by the environment. Details go beyond the present framework,
which adopts a strictly kinematic setting.

\section{The third law: Quantization} \label{third}

The third law of thermodynamics asserts that the value of 
the entropy is always nonnegative. 
But it cannot be deduced from our axioms without making
a further assumption, as a simple example demonstrates.

\begin{expl} \label{5.1.} 

The algebra $\Ez=\Cz^m$ with pointwise operations is a
Euclidean $*$-algebra for any integral of the form
\[
\sint f=\frac {1} {N}\sum _{n=1} ^{N} w_n f_n~~~ (w_n>0);
\]
the axioms are trivial to verify. For this integral the state defined by
\[
\< f\> =\frac {1} {N}\ \sum _{n=1} ^{N} f_n,
\]
is a state with entropy $S$ given by $S_n=\kbar \log w_n$.
The value of the entropy
\[
\ol{S}=\frac {1} {N} \sum_{n=1}^{m} S_n =
\frac {\kbar } {N}\log \prod _{n=1} ^{N}w_n,
\]
is negative if we choose the $w_n$ such that $\prod w_n < 1 $.
\end{expl}

Thus, we need an additional condition which guarantees the validity
of the third law. 
Since the third law is also violated in classical statistical mechanics,
which is a particular case of the present setting, we need a condition
which forbids the classical interpretation of our axioms.
 
We take our inspiration from a simple information theoretic model of 
states discussed in Section \ref{s.complexity} below,
which has this property. (Indeed, the third law is a necessary 
requirement for the interpretation of the value of the entropy 
as a measure of internal complexity, as discussed there.)
There, the integral is a sum over the components, and, since functions 
were defined componentwise,
\lbeq{5-1}
\sint F(f)=\sum _{n\in {\cal N}} F(f_n).
\eeq
We say that a quantity $f$ is \bfi{quantized} iff \gzit{5-1}
holds with a suitable \bfi{spectrum} $\{f_n\mid n\in{\cal N}\}$ 
for all functions $F$ for which $F(f)$ is strongly integrable; 
in this case, the $f_n$ are called the \bfi{levels} of $f$. 
For example, in the quantum setting 
all trace class linear operators are quantized quantities, since these 
always has a discrete spectrum. 

Quantization is the additional ingredient needed to derive the 
third law:

\begin{thm} \label{5.3.}
\bfi{(Third law of thermodynamics)} \\
If the entropy $S$ is quantized then $\ol{S}\ge 0$. 
Equality holds iff the entropy has a single level only ($|{\cal N}|=1$).
\end{thm}

\bepf
We have
\lbeq{5-2}
1=\sint e^{-S/\kbar }=\sum _{n\in {\cal N}}
e^{-S_n / \kbar },
\eeq
hence
\lbeq{5-3}
\ol{S}=\sint Se^{-S/\kbar }
=\sum _{n\in {\cal N}}S_ne^{-S_n/\kbar }.
\eeq
If ${\cal N}=\{n\}$ then (\ref{5-2}) implies $e^{-S_n/\kbar }=0$, hence 
$S_n=0$, and \gzit{5-3} gives $\ol{S}=0$. And if $|{\cal N}|>1$ then
\gzit{5-2} gives $e^{-S_n/\kbar }<1$, hence $S_n>0$ for all 
$n\in {\cal N}$, and \gzit{5-3} implies $\ol{S}>0$.
\epf

In quantum chemistry, energy $H$, volume $V$, and particle
numbers $N_1,\dots ,N_s$ form a quantized family of pairwise commuting
Hermitian variables. Indeed, the Hamiltonian $H$ has discrete energy 
levels if the system is confined to a finite volume, $V$ is a number,
hence has a single level only, and $N_j$ counts particles hence has 
as levels the nonnegative integers.
As a consequence, the entropy $S=T^{-1}(H+PV-
\mu \cdot N)$ is quantized, too, so that the third law of
thermodynamics is valid. The number of levels is infinite, so that
the value of the entropy is positive. 

A zero value of the entropy (\bfi{absolute zero}) is therefore
an idealization which cannot be realized in practice.
But Theorem \ref{5.3.} implies in this idealized situation that
entropy and hence the joint spectrum of $(H,\ V,\ N_1,\dots ,\ N_S)$
can have a single level only. 

This is the situation discussed in ordinary
quantum mechanics (pure energy states at fixed particle numbers).
It is usually associated with the limit $T\to 0$, though at absolute 
temperature $T=0$, i.e., infinite coldness $\beta$, the thermal 
formalism fails (but for low $T$ asymptotic expansions are possible).

To see the behavior close to this limit, we consider for simplicity a 
canonical ensemble with Hamiltonian $H$ (Example \ref{ex.canonical});
thus the particle number is fixed. Since $S$ is quantized, the spectrum 
of $H$ is discrete, so that there is
a finite or infinite sequence $E_0<E_1<E_2<\dots$ of distinct energy 
levels. Denoting by $P_n$ the (rank $d_n$) orthogonal projector to the 
$d_n$-dimensional eigenspace with energy $E_n$, we have the spectral 
decomposition
\[
\phi(H)= \sum_{n\ge 0} \phi(E_n) P_n
\]
for arbitrary functions $\phi$ defined on the spectrum. In particular,
\[
e^{-\beta H}= \sum e^{-\beta E_n} P_n.
\]
The partition function is 
\[
Z= \tr e^{-\beta H} = \sum e^{-\beta E_n} \tr P_n.
= \sum e^{-\beta E_n} d_n.
\]
As a consequence, 
\[
e^{-S/\kbar}= Z^{-1} e^{-\beta H} 
= \frac{\D\sum e^{-\beta E_n} P_n}{\D\sum e^{-\beta E_n} d_n}
= \frac{\D\sum e^{-\beta(E_n-E_0)}P_n}{\D\sum e^{-\beta(E_n-E_0)}d_n},
\]
hence values take the form
\lbeq{e.low}
\<f\> = \sint e^{-S/\kbar}f = \sint\Big(
 \frac{\sum e^{-\beta(E_n-E_0)}P_n}{\sum e^{-\beta(E_n-E_0)}d_n}
\Big).
\eeq
From this representation, we see that only the energy levels $E_n$
with 
\[
E_n \le E_0 +O(\kbar T)
\]
contribute to a canonical ensemble of temperature $T$.
If the temperature $T$ is small enough, so that $\kbar T \ll E_2-E_0$,
the exponentials $e^{-\beta (E_n-E_0)}$ with $n\ge 2$ can be neglected,
and we find
\lbeq{e.2levelapprox}
e^{-S/\kbar} \approx  
\frac{P_0+ e^{-\beta (E_1-E_0)} P_1}{d_0+e^{-\beta (E_1-E_0)} d_1}
= \frac{P_0}{d_0} 
+ \frac{d_0 P_1-d_1P_0}{d_0(e^{\beta (E_1-E_0)}d_0+d_1)}.
\eeq
Thus, the system behaves essentially as the two level system discussed 
in Examples \ref{ex.canonical} and \ref{ex.schottky}; the 
\bfi{spectral gap} $E_1-E_0$ takes the role of $E$.
In particular, if already  $\kbar T \ll E_1-E_0$, we find that 
\[
e^{-S/\kbar} = d_0^{-1}P_0 + O(e^{-\beta (E_1-E_0)})
\approx d_0^{-1}P_0
\]
is essentially the projector to the
subspace of minimal energy, scaled to ensure trace one.

In the \bfi{nondegenerate} case, where the lowest energy eigenvalue is 
simple, there is a corresponding normalized eigenvector $\psi$, 
unique up to a phase, satisfying the \bfi{Schr\"odinger equation}
\lbeq{e.schroe}
H \psi = E_0 \psi,~~|\psi|=1~~~(E_0 \mbox{ minimal}).
\eeq
In this case, the projector is $P_0=\psi\psi^*$ and has rank $d_0=1$.
Thus 
\[
e^{-S/\kbar} = \psi\psi^* + O(e^{-\beta (E_1-E_0)}).
\]
has almost rank one, and the value takes the form
\lbeq{e.quant0}
\<g\>=\tr e^{-S/\kbar} g \approx \tr \psi\psi^*g = \psi^*g\psi.
\eeq
In the terminology of quantum mechanics, $E_0$ is the \bfi{ground state 
energy}, the solution $\psi$ of \gzit{e.schroe} is called the 
\bfi{ground state}, and 
\lbeq{e.pureex}
\<g\>= \psi^*g\psi
\eeq 
is the expectation of the observable $g$ in the ground state.

For a general \bfi{state vector} $\psi$ normalized to satify
$\psi^*\psi=1$, the formula \gzit{e.pureex} defines the values in the 
\bfi{pure state} $\psi$. It is easily checked that
\gzit{e.pureex} indeed defines a state in the sense of Definition 
\ref{d.state}. These are not Gibbs states, but their idealized limiting 
cases.

Our derivation therefore shows that -- unless the ground state is 
degenerate -- {\em a canonical ensemble at sufficiently low temperature 
is in an almost pure state described by the quantum mechanical 
ground state}. 

Thus, the third law directly leads to the conventional form of quantum 
mechanics, which can therefore be understood as the low temperature 
limit of thermodynamics. It also indicates when a quantum mechanical 
description by a pure state is appropriate, namely always when the
gap between the ground state energy and the next energy level is
significantly larger than the temperature (measured in units where the 
Boltzmann constant is set to 1). This is the typical situation in most 
of quantum chemistry and justifies the use of the Born-Oppenheimer 
approximation in the absence of level crossing; cf. \sca{Smith} 
\cite{Smi}, \sca{Yarkony} \cite{Yar}. 
Moreover, it gives the correct (mixed) form of the state
in case of ground state degeneracy, and the form of the correction
terms when the energy gap is not large enough for the ground state 
approximation to be valid.

\chapter{Models, statistics, and measurements}\label{c.models}

In this chapter, we discuss the relation between models and reality.
This topic is difficult and to some extent controversial since 
it touches on unresolved foundational issues about the meaning
of probability and the interpretation of quantum mechanics. 
By necessity, the ratio between the number of words and the number of 
formulas is higher than in other chapters.

We discuss in more detail the relation between 
different thermal models constructed on the basis of the same 
Euclidean $*$-algebra by selecting different lists of extensive 
quantities.

Moreover, a  discussion of the meaning of uncertainty and probability 
gives the abstract setting introduced in the previous chapters 
both a deterministic and a statistical interpretation.

\at{improve overview; adapt Section 1.5 to match the contents}

The interpretation of probability, statistical mechanics, and -- 
today intrinsically interwoven -- of quantum mechanics has a long 
history, resulting in a huge number of publications. 
Informative sources for the foundations of probability in general
include \sca{Fine} \cite{Fin} and \sca{Hacking} \cite{Hac}.
For statistical mechanics, see
\sca{Ehrenfest} \cite{Ehr}, 
\sca{ter Haar} \cite{tHaa}, \sca{Penrose} \cite{Pen}, 
\sca{Sklar} \cite{Skl}, \sca{Grandy} \cite{Gran}, and
\sca{Wallace} \cite{Wal}.
For the foundations of quantum mechanics, see 
\sca{Stapp} \cite{Sta}, \sca{Ballentine} \cite{Ball},
\sca{Home \& Whitaker} \cite{HomW}, \sca{Peres \& Terno} \cite{PerT},
\sca{Schlosshauer} \cite{Schl} and the reprint collection by 
\sca{Wheeler \& Zurek} \cite{WheZ}.

\section{Description levels}\label{s.detail}

There is no fully objective way of defining how quantities and states 
are related to reality, since the observer modeling a particular 
situation may describe the same object from different perspectives 
and at different levels of faithfulness. Different observers may 
choose to study different 
materials or different experiments, or they may study the same 
material or the same experiment in different levels of detail, 
or draw the system boundary differently. For example, one observer 
may regard a measuring instrument as part of the system of interest, 
while for another observer it only serves as a recording device. 

All this affects the choice of the system of interest and its 
mathematical model in a subjective manner. In particular, silently 
changing the definition of what constitutes the system of interest 
is a major reason for apparent paradoxes discussed in the literature, 
and it requires care to disentangle the problems involved and to arrive 
at a clear view. 

On the other hand, once the basic choices are made that unambiguously  
specify the system of interest, everything else can be described 
objectively. 

\bigskip
In practice, relevant quantities and corresponding states are assigned 
to real life situations by well-informed judgment concerning the 
behavior of the equipment used. The validity of the assignment is 
experimentally tested by comparing experimental results with the 
chosen mathematical model. The model defines the meaning of the 
concepts: the theory defines what an object is. 

For example, a substance is regarded as an ideal gas if it behaves 
to a satisfactory degree like the mathematical model of an ideal gas
with certain values of temperature, pressure and volume. 
Similarly, a solid is regarded as a crystal if it behaves to a 
satisfactory degree like the mathematical model of a crystal for 
suitable numerical values of the model parameters.

In general, as put by the author of one of the most influential 
textbooks of thermodynamics: ``Operationally, a system is in an 
equilibrium state if its properties are consistently described by 
thermodynamic theory.'' (\label{p.callen}\sca{Callen} \cite[p.15]{Cal})
At first sight, this sounds like a circular definition.
But this is not the case since the formal meaning of ''consistently 
described by thermodynamic theory'' is already known. The operational
definition simply moves it from the domain of theory to the 
domain of reality by defining when a system deserves the designation
''is in an equilibrium state''. In particular, this definition allows
one to determine experimentally whether or not a system is in 
equilibrium.

In general, we know or assume on the basis of past experience, 
claims of manufacturers, etc., that certain materials or machines 
reliably produce states that, to a satisfactory degree for the 
purpose of the experiment or application,
depend only on variables that are accounted for in our theory and 
that are, to a satisfactory degree, either fixed or controllable.
The nominal state of a system can be checked and, if necessary, 
corrected by \bfi{calibration}, using appropriate measurements that
reveal the parameters characterizing the state.

\bigskip
\at{clarify and streamline the following}
We first emphasize the flexibility of the thermal setting.
While the zeroth law may look very restrictive at first sight, by 
choosing a large enough family of extensive quantities the entropy of 
an {\em arbitrary}\/ Gibbs state can be approximated arbitrarily well
by a linear combination of these quantities. This does not solve the
selection problem but gives a useful perspective:

The zeroth law appears simply as an embodiment of \bfi{Ockham's 
razor} ``frustra fit per plura quod potest fieri per pauciora''
(\sca{Ockham} \cite{Ock}), freely paraphrased in modern form: that we 
should opt for the most economic model explaining a phenomenon -- 
by restricting attention to the relevant extensive quantities only.
At each time $t$, there is -- except in degenerate cases --
a {\em single} Gibbs state, with entropy $S(t)$, say, which best 
describes the system under consideration at the chosen level of 
modeling. Taking the description
by the Gibbs state as fundamental, its value is the objective,
true value of the entropy, relative only to the algebra of quantities 
chosen to model the system. A description of the state 
in terms of a thermal system is therefore adequate if (and, under
an observability qualification to be discussed below, only if), 
for all relevant times $t$, the entropy $S(t)$ can be adequately 
approximated by a linear combination of the extensive quantities 
available at the chosen level of description.

In the preceding chapter, we assumed a fixed selection of extensive 
quantities defining the thermal model. 

As indicated at the end
of Section \ref{s.phen}, observable differences from the conclusions
derived from a thermal model known to be valid on some level imply 
that one or more conjugate 
pairs of thermal variables are missing in the model. So, how should
the extensive quantities be selected?

The set of extensive variables depends on the application 
and on the desired accuracy of the model; it must be chosen in such 
a way that knowing the measured values of the extensive variables 
determines (to the accuracy specified) the complete behavior of the
thermal system. The choice of extensive 
variables is (to the accuracy specified)
completely determined by the level of accuracy with which 
the thermal description should fit the system's behavior. 
This forces everything \at{} else: 
The theory must describe the freedom available to
characterize a particular thermal system with this set of
extensive variables, and it must describe how the numerical values of
interest can be computed for each state of each thermal system.

Clearly, physics cannot be done without approximation,
and the choice of a resolution is unavoidable.
(To remove even this trace of subjectivity, inherent in
any approximation of anything, the entropy would have to be 
represented without any approximation, which would require to use 
the algebra of quantities of the still unknown theory of everything,
and to demand that the extensive quantities exhaust this algebra.)
Once the (subjective) choice of the 
resolution of modeling is fixed, this fixes the amount of approximation 
tolerable in the ansatz, and hence the necessary list of extensive 
quantities. 
This is the only subjective aspect of our setting. 
In contrast to the information theoretic approach where the choice of 
extensive quantities is considered to be the subjective matter of which 
observables an observer happens to have knowledge of.

\begin{table}[htb]
\caption{Typical conjugate pairs of thermal variables
and their contribution to the Euler equation. 
The signs are fixed by tradition. (In the gravitational term,
$m$ is the vector with components $m_j$, the mass of a particle of 
kind $j$, $g$ the acceleration of gravity, and $h$ the height.)
}
\label{3.t.}

\begin{center}
{\small
\begin{tabular}{|l|l|l|}
\hline
extensive $X_j$ & intensive $\alpha_j$ 
& contribution $\alpha_jX_j$ \\
\hline
\hline
entropy $S$ & temperature $T$ 
& thermal, $TS$ \\
\hline
particle number $N_j$ & chemical potential $\mu _j$ 
& chemical, $\mu _jN_j$\\
conformation tensor $C$ & relaxation force $R$
& conformational $\sum R_{jk} C^{jk}$\\
\hline
strain $\eps^{jk}$ & stress $\sigma_{jk}$
& elastic, $\sum \sigma_{jk}\eps^{jk}$\\
volume $V$ & pressure $-P$ 
& mechanical, $-PV$ \\
surface $A_S$ & surface tension $\gamma $ 
& mechanical, $\gamma A_S$ \\
length $L$ & tension $J$ 
& mechanical, $JL$ \\
displacement $q$ & force $-F$
& mechanical, $-F\cdot q$\\
momentum $p$ & velocity $v$
& kinetic, $v\cdot p$ \\
angular momentum $J$ & angular velocity $\Omega$
& rotational, $\Omega \cdot J$\\
\hline
charge $Q$ & electric potential $\Phi$
& electrical, $\Phi Q$\\
polarization $P$ & electric field strength $E$ 
& electrical, $E\cdot P$\\
magnetization $M$ & magnetic field strength $B$ 
& magnetical, $B\cdot M$ \\
electromagnetic field $F$ & electromagnetic field strength $-F^s$ 
& electromagnetic, $-\sum F^s_{\mu\nu}F^{\mu\nu}$\\
\hline
mass $M=m\cdot N$ & gravitational potential $gh$
& gravitational, $ghM$ \\
energy-momentum $U$ & metric $g$ 
& gravitational, $\sum g_{\mu\nu}U^{\mu\nu}$ \\
\hline
\end{tabular}
} 
\end{center}
\end{table}

In general, which quantities need 
to be considered depends on the resolution with which the system is 
to be modeled \at{only repeated?} -- 
the higher the resolution, the larger the family of 
extensive quantities. 
Thus -- whether we describe bulk matter, surface effects, 
impurities, fatigue, decay, chemical reactions, or transition states, 
-- the thermal setting remains the same since it is a universal 
approximation scheme, while the number of degrees of freedom 
increases with increasingly detailed models. 

In phenomenological thermodynamics, the relevant extensive 
quantities are precisely those variables that are
observed to make a difference in modeling the phenomenon of interest.
Table \ref{3.t.} gives typical extensive variables ($S$ and $X_j$), 
their intensive conjugate variables ($T$ and $\alpha_j$), and their 
contribution ($TS$ and $\alpha_j X_j$) to the Euler equation 
\gzit{e.euler}\footnote{\label{f.euler} 
The Euler equation looks like an  
energy balance. But since $S$ is undefined, this formal balance has no 
contents apart from defining the entropy $S$ in terms of the energy 
and other contributions. 
The energy balance is rather given by the first law discussed later, 
and is about {\em changes} in energy. Conservative work contributions 
are exact differentials. For example, the mechanical force 
$F=-dV(q)/dq$ translates into the term 
$-F\cdot dq=dV(q)$ of the first law, corresponding to the
term $-F\cdot q$ in the Euler equation.
The change of the kinetic energy $E_\kin=mv^2/2$ contribution of 
linear motion with velocity $v$ and momentum $p=mv$ is 
$dE_\kin=d(mv^2/2)= mv \cdot dv = v \cdot dp$, which is exactly what 
one gets from the $v \cdot p$ contribution in the Euler 
equation. Since $v \cdot p = mv^2$ is larger than the kinetic
energy, this shows that motion implies a contribution to the
entropy of $(E_\kin-v \cdot p)/T=-mv^2/2T$.
A similar argument applies to the angular motion of a rigid 
body in its rest frame, providing the term involving angular velocity 
and angular momentum.
}. 
Some of the extensive variables and their intensive conjugates are 
vectors or (in elasticity theory, the theory of complex fluids, and in 
the relativistic case) tensors; cf. \sca{Balian} \cite{Bal} for the 
electromagnetic field and \sca{Beris \& Edwards} \cite{BerE}, 
\sca{\"Ottinger} \cite{Oet} for complex fluids.
 
\bigskip
To analyze the relation between two different thermal description 
levels, we compare a coarse system and a more detailed system 
quantitatively, taking for simplicity the temperature constant, 
so that the $T$-dependence can be suppressed in the formulas.
When the Hamiltonian $H$ and the $\X_j$ are fixed, the states are 
completely determined by $\alpha$.

The variables and quantities of the fine system are written as before, 
but the variables and quantities associated with the coarser system
get an additional index $c$. That the fine system is a refinement of 
the coarse system means that the extensive quantities 
of the coarse system are $X_c=CX$, with a fixed matrix $C$ with 
linearly independent rows, whose components tell how the components of 
$X_c$ are built from those of $X$. The entropy of the coarse system
is then given by
\[
S_c=T^{-1}(H-\alpha_c\cdot X_c)
=T^{-1}(H-\alpha_c\cdot CX)
=T^{-1}(H-\alpha\cdot X),
\]
where 
\lbeq{e.star}
\alpha=C^T\alpha_c.
\eeq
We see that the thermal states of the coarse model are just the states 
of the detailed model for which the intensive parameter vector $\alpha$ 
is of the form $\alpha=C^T\alpha_c$ for some $\alpha_c$. Thus the
coarse state space can simply be viewed as a lower-dimensional subspace 
of the detailed state space. Therefore, one expects the coarse 
description to be adequate precisely when the detailed state is close 
to the coarse state space, with an accuracy determined by the 
desired fidelity of the coarse model. 
Since the relative entropy \gzit{4-5},
\lbeq{e.relent}
\< S_c-S\>= \<T^{-1}(H-\alpha_c\cdot CX)-T^{-1}(H-\alpha\cdot X)\>
=\<T^{-1}(\alpha-C^T\alpha_c)\cdot X\>,
\eeq
measures the amount of information in the detailed state which cannot 
be explained by the coarse state, we associate to an arbitrary detailed
state $\alpha$ the coarse state $\alpha_c$ determined as a function 
of $\alpha$ by minimizing \gzit{e.relent}. \at{check. This is a 
single linear expression???}
If $\alpha^*=C^T\alpha_c\approx \alpha$ then 
\[
S_c=T^{-1}(H-\alpha^*\cdot X)\approx 
T^{-1}(H-\alpha\cdot X)=S,
\]
and the coarse description is adequate. 
If $\alpha^*\not\approx \alpha$, there is no a priori reason to trust
the coarse model, and we have to investigate to which extent its 
predictions will significantly differ from those of the detailed model.
One expects the differences to be significant; however, 
in practice, there are difficulties if there are limits on our
ability to prepare particular detailed states. The reason is that
the entropy and chemical potentials can be prepared and measured
only by comparison with sufficiently known states. For ideal gases,
they are inherently ambiguous because of the gauge freedom discussed
in Example \ref{ex.ideal}, which implies that different models of the 
same situation may have nontrivial differences in Hamilton energy, 
entropy, and chemical potential. A similar ambiguity persists in more 
perplexing situations:

\begin{expl} \label{ex.gibbs}
\bfi{(The Gibbs paradox)}\newline
Suppose that we have an ideal gas of two 
kinds $j=1,2$ of particles which are experimentally indistinguishable.
Suppose that in the samples available for experiments, the two kinds 
are mixed in significantly varying proportions $N_1:N_2=q_1:q_2$ which,
by assumption, have no effect on the observable properties; in 
particular, their values are unknown but varying.
The detailed model treats them as distinct, the coarse model as 
identical. 
Reverting to the barless notation of Section \ref{s.phen}, we have
\[
X=\pmatrix{V\cr N_1 \cr N_2},~~~\alpha = \pmatrix{-P\cr \mu_1\cr \mu_2},
\]
and, assuming $C=\pmatrix{1 & 0 & 0\cr 0 & c_1 & c_2}$ for suitable
$c_1,c_2>0$,
\[
X_c=\pmatrix{V\cr N_c}=\pmatrix{V\cr c_1N_1+c_2N_2},~~~
\alpha_c= \pmatrix{-P \cr \mu_c}.
\]
From the known proportions, we find
\[
N_j=x_jN_c,~~~ x_j=\frac{q_j}{c_1q_1+c_2q_2}.
\]
The mixture behaves like an ideal gas of a single kind, hence 
\[
PV=\kbar T N_c,~~~H=h_c(T)N_c,~~~
\mu_c=\kbar T \log\frac{\kbar TN_c}{V\pi_c}.
\]
Now $N_c=(\kbar T)^{-1}PV=\sum N_j=\sum x_j N_c$ implies that
$x_1+x_2=1$. Because of indistinguishability, this must hold for
any choice of $q_1,q_2\ge 0$; for the two choices $q_1=0$ and $q_2=0$,
we get $c_1=c_2=1$, hence $N_c=\sum N_j$, and the $x_j$ are mole 
fractions. Similarly, if we use for all kinds $j$ of substances
the same normalization for fixing the gauge freedom discussed in 
Example \ref{ex.ideal}, the relation 
$h_c(T)N_c=H=\sum h_j(T)N_j=\sum h_j(T)x_jN_c$ implies for varying 
mole fractions that $h_j(T)=h_c(T)$ for $j=1,2$. From this, we get 
$\pi_j(T)=\pi_c(T)$ for $j=1,2$. Thus
\[
H-H_c = \sum h_j(T)N_j - h_c(T)N_c = 0,
\]
\[
\mu_j-\mu_c =\kbar T \log\frac{\kbar TN_j}{V\pi_j}-
\kbar T \log\frac{\kbar TN_c}{V\pi_c}
=\kbar T \log x_j,
\]
the Gibbs energy satisfies
\[
G-G_c=\sum \mu_jN_j-\mu_cN_c = \sum (\mu_j-\mu_c)N_j
=\kbar T N_c \sum x_j\log x_j,
\]
and the entropy satisfies 
\[
\bary{lll}
S-S_c&=&T^{-1}(H-PV+G) - T^{-1}(H_c-PV+G_c) \\
&=& T^{-1}(G-G_c) =\kbar N_c \sum x_j\log x_j.
\eary
\]
The latter term is called the \bfi{entropy of mixing}.
Its occurence is referred to as the \bfi{Gibbs paradox}
(cf. \sca{Jaynes} \cite{Jay.G}, \sca{Tseng \& Caticha} \cite{TseC}, 
\sca{Allahverdyan \& Nieuwenhuizen} \cite{AllN},
\sca{Uffink} \cite[Section 5.2]{Uff}).
It seems to say that there are two different entropies, depending on 
how we choose to model the situation. For fixed mole fractions, there 
is no real paradox since the fine and the coarse description differ 
only by a choice of the unobservable gauge parameters, and only gauge 
invariant quantities (such as entropy differences) have a physical 
meaning.

If the mole fractions may vary, the fine and the coarse 
description differ significantly. But the difference in the
descriptions is observable only if we know processes which affect the 
different kinds differently.

Fixed chemical potentials can be prepared only through chemical 
contact with substances with known chemical potentials, and the latter 
must be computed from observed mole fractions. Therefore, the chemical 
potentials can be calibrated only if we can prepare equilibrium states 
at fixed mole fraction. This requires that we are able to separate to 
some extent particles of different kinds. \at{place this correctly!}

Examples are a difference in mass, which allows a mechanical separation,
a difference in molecular size or shape, which allows their separation 
by a semipermeable membrane, a difference in spin, which allows a 
magnetic separation, or a difference in scattering properties of the 
particles, which allows a chemical or radiation-based differentiation.
In each of these cases, the particles become distinguishable; 
the coarse description is therefore inadequate and gives a wrong 
description for the entropy and the chemical potentials.
\end{expl}

Generalizing from the example, we conclude that even when both a 
coarse model and a 
more detailed model are faithful to all experimental information 
possible at a given description level, there is no guarantee that 
they agree in the values of all thermal variables of the
coarse model. In the language of control theory (see, e.g., 
\sca{Ljung} \cite{Lju}), agreement is guaranteed only when all  
parameters of the more detailed models are observable. 

On the other hand, all observable state functions of the detailed 
system that depend only on the coarse state have the 
same value within the experimental accuracy, if both models are adequate
descriptions of the situation. Thus, while the values of some variables 
need not be experimentally determinable, 
{\em the validity of a model is an objective property}. 
\at{justify,move?} 

Therefore, preferences for one or the other of two 
valid models can only be based on other criteria. The criterion usually
employed in this case is Ockham's razor, although
there may be differences of opinion on what counts as the most
economic model. In particular, a fundamental description of macroscopic 
matter by means of quantum mechanics is hopelessly overspecified in 
terms of the number of degrees of freedom needed for comparison with 
experiment, most of which are in principle unobservable by equipment 
made of ordinary matter. But it is often the most economical model in 
terms of description length (though extracting the relevant 
information from it may be difficult).
Thus, different people may well make different rational choices,
or employ several models simultaneously.

As soon as a discrepancy of model predictions with experiment is 
reliably found, the model is inadequate and must be
replaced by a more detailed or altogether different model. 
This is indeed what happened with the textbook example of the
Gibbs paradox situation, ortho and para hydrogen, cf. 
\sca{Bonhoeffer \& Harteck} \cite{BonH}, \sca{Farkas} \cite{Far}.
Hydrogen seemed at first to be a single substance, but then 
thermodynamic data forced a refined description.
Similarly, in spin echo experiments (see, e.g., 
\sca{Hahn} \cite{Hah,Hah2}, \sca{Rothstein} \cite{Rot},
\sca{Ridderbos \& Redhead} \cite{RidR}), the specially 
prepared system appears to be in equilibrium but, according to 
Callen's empirical definition quoted on \pageref{p.callen}
it is not -- 
the surprising future behavior (for someone not knowing the special 
preparation) shows that some correlation variables were 
neglected that are needed for a correct description. 

\sca{Grad} \cite{Grad} speaks of ''the adoption of a new entropy is 
forced by the discovery of new information''. More precisely, the 
adoption of a new {\em model} (in which the entropy has different 
values) is forced, since the old model is simply wrong under the new 
conditions and remains valid only under some restrictions. 

Observability issues aside, the coarser description usually has a more 
limited range of applicability; with the qualification discussed in 
the example, it is generally restricted to those systems whose 
detailed intensive variable vector $\alpha$ is close to 
the subspace of vectors of the form $C^T\alpha$ reproducible in the 
coarse model.

\at{move?}
Finding the right family of thermal variables is therefore a 
matter of discovery, not of subjective choice. This is further 
discussed in Section \ref{s.model}.

\section{Local, microlocal, and quantum equilibrium} \label{s.model}

\at{reorder the first 5 pars.}
As we have seen in Section \ref{s.detail}, when descriptions on 
several levels are justified empirically, they differ significantly 
only in quantities that are negligible in the more detailed models
and vanish in the coarser models, or by terms that are not observable 
in principle. \at{phrase this more precisely!}
We now apply the above considerations to various levels of equilibrium
descriptions.

A global equilibrium description is adequate at some resolution 
if and only if only the nonequilibrium forces present in the finer 
description are small, and a more detailed local equilibrium 
description will (apart from variations of the Gibbs paradox, which 
should be cured on the more detailed level) agree with the global 
equilibrium description to the accuracy within which the differences 
in the corresponding approximations to the entropy, as measured by the 
relative entropy \gzit{4-5}, are negligible. Of course, if the
relative entropy of a thermal state relative to the true Gibbs state
is large then the thermal state cannot be regarded as a faithful 
description of the true state of the system, and the thermal  
model is inadequate.

In statistical mechanics, where the microscopic dynamics is
given, the relevant extensive quantities are those whose 
values vary slowly enough to be macroscopically observable at a given 
spatial or temporal resolution (cf. \sca{Balian} \cite{Bal2}). 
Which ones must
be included is a difficult mathematical problem that has been solved 
only in simple situations (such as monatomic gases) where a weak 
coupling limit applies. In more general situations, the selection is 
currently based on phenomenological consideration, without any formal 
mathematical support.

In equilibrium statistical mechanics, which describes 
time-independent, {\em global} equilibrium situations, 
the relevant extensive quantities are the additive conserved quantities 
of a microscopic system and additional parameters describing order 
parameters that emerge from broken symmetries or various defects
not present in the ideal model.
\bfi{Phase equilibrium} needs, in addition, copies of the extensive 
variables (e.g., partial volumes) for each phase, since the phases are 
spatially distributed, while the intensive variables are shared by all 
phases.
\bfi{Chemical equilibrium} also accounts for exchange of atoms through 
a list of permitted chemical reactions whose length is again 
determined by the desired resolution.

In states not corresponding to global equilibrium -- usually called 
\bfi{non-equilibrium states}, a thermal description is still 
possible assuming so-called \bfi{local equilibrium}. There, 
the natural extensive quantities are those whose 
values are locally additive and slowly varying in space 
and time and hence, reliably observable at the scales of interest. 
In the statistical mechanics of local equilibrium, the thermal 
variables therefore become space- and time-dependent fields 
(\sca{Robertson} \cite{Rob}).
On even shorter time scales, phase space behavior becomes relevant, 
and the appropriate description is in terms of \bfi{microlocal 
equilibrium} and position- and momentum-dependent phase space densities.
Finally, on the microscopic level, a linear operator description in 
terms of \bfi{quantum equilibrium} is needed.

\bigskip
The present formalism is still applicable to local, microlocal, and
quantum equilibrium (though most products now become inner products in 
suitable function spaces), but the relevant quantities are now 
time-dependent and additional dynamical issues (relating states at 
different times) arise; these are outside the scope of the present 
book.

In local equilibrium, one needs a 
hydrodynamic description by Navier-Stokes equations and their 
generalizations; see, e.g., \sca{Beris \& Eswards} \cite{BerE}, 
\sca{Oettinger} \cite{Oet}, \sca{Edwards} et al. \cite {EdwOJ}. 
In the local view, one gets the interpretation of 
extensive variables as locally conserved (or at least slowly varying) 
quantities (whence additivity) and of intensive variables as 
parameter fields, which cause non-equilibrium currents when they are 
not constant, driving the system towards global equilibrium.
In microlocal equilibrium, one needs a kinetic  description by
the Boltzmann equation and its generalizations; see, e.g., 
\sca{Bornath} et al. \cite{BorKKS},
\sca{Calzetta \& Hu} \cite{CalH}, 
\sca{M\"uller \& Ruggeri} \cite{MueR}.

\bigskip
\bfi{Quantum equilibrium.} 
Fully realistic microscopic dynamics must be based on quantum mechanics.
In quantum equilibrium, the dynamics is given by quantum dynamical 
semigroups. We outline the ideas involved, in order to emphasize some 
issues that are usually swept under the carpet.

Even when described at the microscopic level, thermal systems
of sizes handled in a laboratory are in contact with their environment, 
via containing walls, emitted or absorbed radiation, etc.. 
We therefore embed the system of interest into a bigger, completely 
isolated system and assume that the quantum state of the big system is 
described at a fixed time by a value map that assigns to a linear 
operator $g$ in the big system the value $\<g\>$ and satisfies the rules
(R1)--(R4) for a state.
The small system is defined by a Euclidean $*$-algebra $\Ez$ of linear 
operators densely defined on $\widehat \Hz$, composed of all meaningful
expressions in field operators at arguments in the region of interest; 
the integral  is given by the trace in the big system.
Since the value map restricted to $g\in \Ez$ also satisfies the rules
(R1)--(R4) for a state, the big system induces on the system of
interest a state. By standard theorems (see, e.g.,
\sca{Thirring} \cite{Thi}), there is a unique \bfi{density operator}
$\rho\in\Ez$ such that $\<g\> = \sint \rho g$ for all $g\in \Ez$ with 
finite value. Moreover, $\rho$ is Hermitian and positive semidefinite. 
If $0$ is not an eigenvalue of $\rho$ then $\<\cdot\>$ is a Gibbs state 
with entropy $S=-\kbar \log \rho$. Note that the entropy defined in
this way depends on the choice of $\Ez$, hence on the set of quantities
found to be relevant. (In contrast, if the big system that includes the 
environment is in an approximately pure state, as is often assumed, 
the value of the entropy of the big system is approximately zero.)

To put quantum equilibrium into the thermal setting, we simply choose 
a set of extensive variables spanning the algebra $\Ez$; then $S$ 
can be written in the form \gzit{3-2}. (A thermal description is not 
possible if $0$ is an eigenvalue of $\rho$, an exceptional situation 
that can be realized experimentally only for systems with extremely 
few quantum levels.
This happens, e.g., when the state is pure, $\rho=\psi\psi^*$.)

Of course, $\psi$ and hence the state $\<\cdot\>$ depend on time.
The time evolution is now quite different from the conservative
dynamics usually assumed for the big system that includs the 
environment. The system of interest does {\em not} inherit a 
Hamiltonian dynamics from the isolated big system; instead, 
the dynamics of $\rho$ is given by an integro-differential equation 
with a complicated memory term, defined by the so-called projector 
operator formalism described in detail in \sca{Grabert} \cite{Gra}; 
for summaries, see 
\sca{Rau \& M\"uller} \cite{RauM} and \sca{Balian} \cite{Bal2}. 
In particular, one can say nothing specific about the dynamics of
$\ol S$. (In contrast, were the reduced system governed by a 
Hamiltonian dynamics, $\rho$ would evolve by means of a unitary 
evolution; in particular, $\ol S = \<S\> = -\kbar \tr \rho \log \rho$
would be time-independent.)
A suitable starting point for a fundamental derivation, based on 
quantum field theory, are provided by the so-called exact 
renormalization group equations (see, e.g.,
\sca{Polonyi \& Sailer} \cite{PolS}, \sca{Berges} \cite{Ber}).

In typical treatments of reduced descriptions, one assumes that
the memory decays sufficiently fast; this so-called \bfi{Markov 
assumption} can be justified in a weak coupling limit 
(\sca{Davies} \cite{Dav}, \sca{Spohn} \cite{Spo}), corresponding to
a system of interest that is only weakly interacting with the 
environment.
But a typical thermal system, such as a glass of water on a desk
is held in place by the container. Considered as a nearly independent 
system, the water would behave very differently, probably diffusing 
into space. Thus, it is questionable whether the Markov assumption
is satisfied; a detailed investigation of the situation would be
highly desirable. Apparently there are only few discussions of the 
problem how containers modify the dynamics of a large quantum system;
see, e.g., \sca{ Lebowitz \& Frisch} \cite{LebF},
\sca{Blatt} \cite{Bla} and \sca{Ridderbos} \cite{Rid}.
One should expect a decoherence effect (\sca{Brune} et al. \cite{BruHD})
of the environment on the system that, for large quantum systems, 
is extremely strong (\sca{Zurek} \cite{Zur}).

However, simply assuming the Markov assumption as the condition 
for regarding the system of interest to be \bfi{effectively isolated} 
allows one to deduce for the resulting \bfi{Markov approximation} 
a deterministic differential equation for the density operator. 
The dynamics then describes a linear quantum dynamical semigroup. 
For all known linear quantum dynamical semigroups 
(cf. \sca{Davies} \cite{Dav}) on a Hilbert space, the
dynamics takes the form of a \bfi{Lindblad equation} 
\lbeq{e.lind}
\dot\rho= \frac{i}{\hbar} (\rho H-H^*\rho)+P^*\rho
\eeq
(\sca{Lindblad} \cite{Lin}, \sca{Gorini} et al. \cite{GorKS}),
where the \bfi{effective Hamiltonian} $H$ is a not necessarily 
Hermitian operator and $P^*$ is the dual of a completely positive map 
$P$ \at{explain defining properties} of the form 
\[
P(f) = Q^*J(f)Q \Forall f\in\Ez,
\]
with some linear operator $Q$ from $\Ez$ to a second $*$-algebra $\Ez'$ 
and some $*$-algebra homomorphism $J$ from $\Ez$ to $\Ez'$.
(\sca{Stinespring} \cite{Sti}, \sca{Davies} \cite[Theorem 2.1]{Dav}).
The resulting dynamics is inherently dissipative; for time 
$t\to\infty$, $P^*\rho$ can be shown to tend to zero, which implies 
under a natural nondegeneracy assumption that the limiting state is a 
global equilibrium state.
 
No matter how large we make the system, it is necessary to take account 
of an unobserved environment, since all our observations are done in a 
limited region of space, which, however, interacts with the remainder
of the universe. As a consequence, the time evolution of any system of 
signifcant size is irreversible. In particular, the prevalence here on 
earth of matter in approximate equilibrium could possibly be 
explained by the fact that the earth is extremely old.

\bigskip 
We now consider relations within the hierarchy of the four levels. 
The quantum equilibrium entropy $S_\fns{qu}$, the microlocal 
equilibrium entropy $S_\fns{ml}$, the local equilibrium 
entropy $S_\fns{lc}$, and the global equilibrium entropy $S_\fns{gl}$ 
denote the values of the entropy in a thermal description of the 
corresponding equilibrium levels. The four levels have an increasingly 
restricted set of extensive quantities, and the relative entropy 
argument of Theorem \ref{t4.5} can be applied at each level. Therefore
\lbeq{e.hi1}
S_\fns{qu} \le S_\fns{ml} \le S_\fns{lc} \le S_\fns{gl}.
\eeq
In general, the four entropies might have completely different
values. We discussfour essentially different possibilities,

 (i) $S_\fns{qu}\approx S_\fns{ml}\approx S_\fns{lc}\approx S_\fns{gl}$,

 (ii) $S_\fns{qu}\approx S_\fns{ml}\approx  S_\fns{lc} \ll S_\fns{gl}$,

 (iii) $S_\fns{qu} \approx S_\fns{ml} \ll S_\fns{lc} \le S_\fns{gl}$,

 (iv) $ S_\fns{qu} \ll S_\fns{ml} \le S_\fns{lc} \le S_\fns{gl}$,

with different physical interpretations. As we have seen in 
Section \ref{s.detail}, a thermal description is valid only if the 
entropy in this description approximates the true entropy sufficiently 
well. All other entropies, when significantly different, do not 
correspond to a correct description of the system; their disagreement 
simply means failure of the coarser description to match reality.
Thus which of the cases (i)--(iv) occurs decides upon which 
descriptions are valid.
(i) says that the state is in global equilibrium, and all four 
descriptions are valid.
(ii) that the state is in local, but not in global equilibrium,
and only the three remaining descriptions are valid.
(iii) says that the state is in microlocal, but not in local 
equilibrium, and in particular not in global equilibrium. 
Only the quantum and the microlocal descriptions are valid.
Finally, (iv) says that the state is not even in microlocal equilibrium,
and only the quantum description is valid.

Assuming that the fundamental limitations in observability are 
correctly treated on the quantum level, the entropy is an objective 
quantity, independent of the level of accuracy with which we are able 
to describe the system. The precise value it gets in a model 
depends, however, on the model used and its accuracy. The observation 
(by \sca{Grad} \cite{Grad}, \sca{Balian} \cite{Bal2}, and others) 
that entropy may depend significantly on the description level 
is explained by two facts that hold for variables in models of any 
kind, not just for the entropy, namely\\
(i) that if two models disagree in their observable predictions, 
at most one of them can be correct, and\\
(ii) that if a coarse model and a refined model agree in their 
observable predictions, the more detailed model has unobservable 
details.\\
Since unobservable details cannot be put to an experimental test, 
the more detailed model in case (ii) is questionable unless dictated
by fundamental considerations, such as symmetry or formal simplicity.

\section{Statistics and probability}\label{s.stat}

Recall from Section \ref{s.limit} that a quantity $g$ is considered to 
be significant if its resolution $\res(g)$ is much smaller than one, 
while it is considered as noise if it is much larger than one. 
If $g$ is a quantity and $\widetilde g$ is a good 
approximation of its value then $\Delta g:=g-\widetilde g$ is 
noise. Sufficiently significant quantities can be treated as 
deterministic; the analysis of noise is the subject of 
statistics.

Statistics is based on the idea of obtaining information about noisy
quantities of a system by repeated \bfi{sampling} from a 
\bfi{population}\footnote{
Physicists usually speak of an \bfi{ensemble} in place of a population; 
but since in connection with the microcanonical, canonical, or grand 
canonical ensemble we use the term ensemble synonymous with state, 
we prefer the statistical term population to keep the discussion 
unambiguous.
} 
of independent systems with identical preparation, but differing in 
noisy details not controllable by the preparation. In the present 
context, such systems are described by the same Euclidean 
$*$-algebra $\Ez_0$, the same set of quantities to be sampled, and the 
same state $\<\cdot\>_0$. 

More precisely, the systems may be regarded as subsystems of a bigger
system (e.g., the laboratory) whose set of quantities is given by a
big Euclidean $*$-algebra $\Ez$. To model identically prepared 
subsystems we consider injective homomorphisms from $\Ez_0$ into $\Ez$
mapping each reference quantity $f\in\Ez_0$ to the quantity 
$f_l \in\Ez$ of the $l$th subsystem considered to be `identical' 
with $f$. Of course, in terms of the big system, the $f_l$ are not 
really identical; they refer to quantities distinguished by position 
and/or time. That the subsystems are \bfi{identically prepared} 
is instead modelled by the assumption 
\lbeq{e.identical}
\<f_l\> = \<f_0\> \Forall f \in\Ez_0,
\eeq
and that they are \bfi{independent} by the assumption 
\lbeq{e.independent}
\<f_kg_l\> = \<f_k\>\<g_l\>  \Forall f,g \in\Ez_0 \mbox{~and~} k\ne l.
\eeq
The following result is fundamental for statistical considerations:

\begin{thm}\label{t.weaklaw} \bfi{(Weak law of large numbers)}\\
For a family of quantities $f_l$ $(l=1, \ldots , N)$ satisfying
\gzit{e.identical} and \gzit{e.independent}, the \bfi{mean} quantity
\[
  \widehat f := \frac{1}{N} \D \sum ^N _{l=1} f_l
\]
(which again is a quantity)
satisfies
\[
  \< \widehat f\, \> =\<f_0\>,
\]
\lbeq{e.sigN}
\sigma (\widehat f\,) = \sigma(f_0)/\sqrt{N},
\eeq
\end{thm}

\bepf 
Writing $\mu:=\<f\>_0$ and $\sigma:=\sigma(f_0)$, we have
\[
\<\widehat f\,\> =\frac{1}{N}(\<f_1\>+\dots+\<f_N\> )
=\frac{1}{N}(\mu+\dots+\mu)=\mu
\] 
and
\lbeq{e.ffsum}
\<\widehat f\,^*\widehat f\,\>
=\frac{1}{N^2}\Big\<\Big(\sum_jf_j\Big)^*\Big(\sum_kf_k\Big)\Big\>
=N^{-2}\sum_{j,k}\<f_j^*f_k\>.
\eeq
Now
\[
\<f_j^*f_j\>=\<f_j\>^*\<f_j\>+\sigma(f_j)^2=|\mu|^2+\sigma^2,
\]
and by \gzit{e.independent} for $j\neq k$,
\[
\<f_j^*f_k+f_k^*f_j\>=2\re \<f_j^*f_k\>
=2\re \<f_j\>^*\<f_k\>=2\re \mu^*\mu=2|\mu|^2.
\]
In the sum in \gzit{e.ffsum}, this leads to a contribution of 
$|\mu|^2+\sigma^2$ for each of the $N$ diagonal elements, and of 
$2|\mu|^2$ for each of the ${N \choose 2}$ pairs of off-diagonal 
elements. Therefore 
\[
\<\widehat f\,^*\widehat f\,\>
=N^{-2}\Big(N(|\mu|^2+\sigma^2)+{N \choose 2}2|\mu|^2\Big)
=N^{-1}\sigma^2 + |\mu|^2,
\]
so that 
\[
\sigma(\widehat f\,)^2 
= \<\widehat f\,^*\widehat f\,\>-\<\widehat f\,\>^*\<\widehat f\,\>
=N^{-1}\sigma^2,
\]
and the assertions follow.
\epf

As a significant body of work in probability theory shows, the
conditions under which $\sigma(\widehat f\,)\to 0$ as $N\to\infty$ can
be significantly relaxed; thus in practice, it is sufficient if 
\gzit{e.identical} and \gzit{e.independent} are approximately valid.

The significance of the weak law of large numbers lies in the fact that
\gzit{e.sigN} becomes arbitrarily small as $N$ becomes sufficiently 
large. Thus the uncertainty of quantities when averaged over a large 
population of identically prepared systems becomes arbitrarily small 
while the mean value reproduces the value of each quantity.
Thus quantities averaged over a large population of identically 
prepared systems become highly significant when their value is nonzero,
even when no single quantity is significant. 

This explains the success of statistical mechanics to provide an 
effectively deterministic description of ideal gases, where all 
particles may be assumed to be independent and identically prepared. 
In real, nonideal gases, the independence assumption is only 
approximately valid because of possible interactions, and in liquids,
the independence is completely lost. The power of the abstract 
theory discussed in the preceding chapters lies in the fact that
it allows to replace simple statistical reasoning based on independence
by more sophisticated algebraic techniques that give answers even in
extremely complex interacting cases.
 
The weak law of large numbers also implies that, in a context where 
many repeated experiments are feasible, states can be given a 
\bfi{frequentist} interpretation, in which $\<g\>$ is the 
\bfi{expectation} of $g$, empirically defined as an average over 
many realizations. In this case (and only in this case), $\res(g)$ 
becomes the standard deviation of $g$, divided by the absolute value 
of the expectation; therefore, it measures the relative accuracy of 
the individual realizations.

On the other hand, in equilibrium thermodynamics, where a tiny number 
of macroscopic observations on a single system completely determine 
its state to engineering accuracy, such a frequentist interpretation
is inappropriate. Indeed, as discussed by \sca{Sklar} \cite{Skl}, 
a frequentist interpretation of statistical mechanics 
has significant foundational problems, already in the framework of 
classical physics. 

Thus, the present framework captures correctly the experimental 
practice, and determines the conditions under which  
deterministic and statistical reasoning are justified:

Deterministic reasoning is sufficient for all quantities whose limit
resolution is below the relative accuracy desired for a given 
description level.

Statistical reasoning is necessary precisely when the limit resolution 
of certain quantities is larger than the desired relative accuracy,
and these quantities are sufficiently identical and independent 
to ensure that the limit resolution of their mean is below this 
accuracy.

\at{adapt}
In this way, we delegate statistics to its role as {\em the art of 
interpreting measurements}, as in classical physics. 
Indeed, to have
a consistent interpretation, real experiments must be designed such that
they allow one to determine approximately the properties of the state 
under study, hence the values of all quantities of interest.
The uncertainties in the experiments imply approximations, which, 
if treated probabilistically, need an {\em additional} probabilistic 
layer accounting for measurement errors. 
Expectations from this secondary layer, which involve probabilistic
 statements about situations that are uncertain due to neglected 
but in principle observable details (cf. \sca{Peres} \cite{Per}), 
happen to have the 
same formal properties as the values on the primary layer, though 
their physical origin and meaning is completely different.

\bigskip
\bfi{Classical probability.}
Apart from the traditional axiomatic foundation of probability theory 
by \sca{Kolmogorov} \cite{Kol} in terms of measure theory there is 
a less well-known axiomatic treatment by \sca{Whittle} \cite{Whi}
in terms of expectations, which is essentially the commutative
case of the present setting. 
The exposition in \sca{Whittle} \cite{Whi} (or, in more abstract 
terms, already in \sca{Gelfand \& Naimark} \cite{GelN}) shows that, 
if the $X_j$ are pairwise commuting, it is possible to define
for any Gibbs state in the present sense, random variables $X_j$ 
in Kolmogorov's sense such that the expectation of all sufficiently
regular functions $f(X)$ defined on the joint spectrum of $(X)$
agrees with the value of $f$. It follows that in the pairwise commuting 
case, it is always possible to construct a probability interpretation 
for the quantities, completely independent of any assumed microscopic 
reality. 

The details (which the reader unfamiliar with measure theory may 
simply skip) are as follows. We may associate with every vector $X$ 
of quantities with commuting components a time-dependent, monotone 
linear functional $\<\cdot\>_t$ defining the \bfi{expectation}
\[
      \<f(X)\>_t:=\sint \rho(t) f(X)
\]
at time $t$ of arbitrary bounded continuous functions $f$ of $X$. 
These functions define a commutative $*$-algebra $\Ez(X)$.
The \bfi{spectrum} $\spec X$ of $X$ is the set of all $*$-homomorphisms 
(often called \bfi{characters}) from $\Ez(X)$ to $\Cz$, and has
the structure of a Hausdorff space, with the \bfi{weak-$*$ topology}
obtained by calling a subset $S$ of $\spec X$ closed if, for any
pointwise convergent sequence (or net) contained in $S$, its limit is 
also in $S$. Now a monotone linear functional turns out to be 
equivalent to a multivariate probability measure $d\mu_t(X)$ 
(on the sigma algebra of Borel subsets of the spectrum $\Omega$ of $X$) 
defined by
\[
   \int d\mu_t(X) f(X) := \sint \rho(t) f(X) =\<f(X)\>_t.
\]
Conversely, classical probability theory may be discussed in 
terms of the Euclidean $*$-algebra of \bfi{ random variables},
i.e., Borel measurable complex-valued 
functions on a Hausdorff space $\Omega$ where bounded continuous 
functions are strongly integrable and the integral is given by 
$ \sint f := \int d\mu(X) f(X)$ for some distinguished measure $\mu$.

If -- as in quantum systems -- the extensive quantities do 
not commute, a probabilistic interpretation in the Kolmogorov sense 
is no longer possible. In Section \ref{s.qprob}, we discuss what may 
take its place.

\section{Classical measurements}\label{s.measurement}

A measuring instrument measures properties of a system of interest.
However, the measured value is read off from the instrument, and 
hence is primarily a property of the measuring instrument and not
one of the measured system. On the other hand, properties of the 
system are encoded in the state of the system and its dynamics.
This state and what can be deduced from it are the only objective 
properties of the system. 

In order that a measurement on a system deserves its name there must be
a quantitative relation between the state of the system and the 
measured values. This relation may be deterministic or stochastic, 
depending on what is measured.

Measurements are therefore possible only if the microscopic laws imply 
relations between properties of the measured system and the values 
read off from the measuring instrument.
These relations may be either deduced from a theoretical analysis, 
or they may be guessed from experimental evidence.
In general, the theoretical analysis leads to difficult many-particle 
problems that can be solved only in a stochastic approximation by
idealized situations; from such idealizations one then transfers 
insight to make educated guesses in cases where an analysis is too 
difficult.

The behavior required in the following discussion for a classical or a 
statistical instrument guarantees \bfi{reproducibility} of
measurements, a basic requirement of natural sciences, in the
sense that systems prepared in the same state will behave alike
when measured. Here `alike' is interpreteted for classical instruments 
in the deterministic sense of `approximately equal within the 
specified accuracy', and for statistical instruments in the 
sense of `reproducing in the long run approximately the same 
probabilities and mean values'.

\bigskip
When measuring classical or quantum systems that are \bfi{macroscopic},
i.e., large enough to be described sufficiently well by the methods of 
statistical mechanics, one measures more or less accurately extensive 
or intensive variables of the system and one obtains essentially
deterministic results. A \bfi{classical instrument} is a measuring 
instrument that measures such deterministic values within some known
margin of accuracy. Note that this gives an operational meaning 
to the term {\em classical}, although every classical instrument is, 
of course, a quantum mechanical many-particle system when modelled 
in full detail. Whether a particular
instrument behaves classically can in principle be found out by an 
analysis of the measurement process considered as a many-particle 
system, although the calculations can be done in practice only under 
simplifying assumptions. For some concrete models, see, e.g.,
\sca{Allahverdyan} et al. \cite{AllBM}.
Thus there is no split between the classical 
and the quantum world but a gradual change from quantum to classical 
as the system gets larger and the limit resolution improves.

It is interesting to discover the nature of thermodynamic 
observables\footnote{
We use the term observable with its common-sense meaning. In quantum 
mechanics, the term has also a technical meaning that we do not use, 
denoting there a self-adjoint linear operator on a Hilbert space. 
}. 
We encountered intensive variables, which are parameters
characterizing the state of the system, extensive variables, 
values that are functions of the intensive variables and of the
parameters (if there are any) in the Hamiltonian, and limit 
resolutions, which, as functions of values, are also functions of the 
intensive variables. Thus all thermodynamic observables of practical 
interest are functions of the parameters defining the thermal 
state or the Hamiltonian of the system.
Which parameters these are depends of course on the assumed model.

For an arbitrary model of an arbitrary system we perform a natural 
step of extrapolation, substantiated later (in Section 
\ref{s.quantdyn}) by the Dirac-Frenkel variational principle,
\at{check Section reference}
and take the parameters characterizing a family of Hamiltonians 
and a family of states that describe the possible states of a system.
as the basic variables. We call these parameters the 
\bfi{model parameters}; the values of the model parameters completely
characterize a particular system described by the model.
An \bfi{observable} of the model is then a function of these basic 
variables.

Thus we may say that a classical instrument is characterized by
the fact that upon measurement the measurement result approximates 
with a certain accuracy the value of a function $F$ of the model
parameters.
As customary, one writes the result of a measurement as an 
\bfi{uncertain number} $F_0 \pm\Delta F$ consisting of a main 
value $F_0$ and a deviation $\Delta F$, with the meaning that the 
error $|F_0 - F|$ is at most a small multiple of $\Delta F$.
Because of possible systematic errors, it is generally not possible 
to interpret $F_0$ as mean value and $\Delta F$ as standard 
deviation. Such an interpretation is valid only if the instrument
is calibrated to satisfy the implied statistical relation.

In particular, since $\<f\>$ is a function of the model parameters,
a measurement may yield the value $\<f\>$ of a quantity 
$f$, and is then said to be a classical instrument for \bfi{measuring} 
$f$. As an important special case, all readings from a photographic 
image or from the scale of a measuring instrument, done by an observer,
are of this nature when considered as measurements of the instrument 
by the observer. Indeed, what is measured by the eye is the particle 
density of blackened silver on a photographic plate or of iron of the
tip of the pointer on the scale, and these are extensive variables 
in a continuum mechanical local equilibrium description of the
instrument. 

\bigskip
The measurement of a tiny, \bfi{microscopic} system, often consisting 
of only a single particle, is of a completely different nature. 
Now the limit resolutions do not benefit from the law of large numbers, 
and the relevant quantities often are no longer significant.
Then the necessary quantitative relations between properties of the 
measured system and the values read off from the measuring instrument
are only visible as stochastic correlations. In a single measurement 
of a microscopic system, one can only glean very little information
about the state of a system; conversely, from the state of the system 
one can predict only probabilities for the results of a single 
measurement. 
The results of single measurements are no longer reproducably 
observable numbers; reproducably observable -- and hence the
carrier of scientific information -- are only probabilities and
statistical mean values.

To obtain comprehensive information about the state of a single 
microscopic system is therefore impossible. To collect enough 
information about the prepared state and hence the state of each 
system measured, one needs either 
time-resolved measurements on a single system (available, e.g., for 
atoms in ion traps or for electrons in quantum dots), or a population
of identically prepared systems.

Extrapolating from the macroscopic case, it is natural to consider
again the parameters  characterizing a family of states that describe 
the possible states of a system as the basic numbers whose functions 
define observables in the present, nontechnical sense. 
This is now a less well-founded assumption based
only on the lack of a definite boundary between the macroscopic and 
the microscopic regime, and an application of Ockham's razor to
minimize the needed assumptions. 

Measurements in the form of clicks, flashes or events (particle tracks) 
in scattering experiments may be described in terms of a 
\bfi{statistical instrument} characterized by a discrete family of 
possible measurement results  $a_1,a_2,\dots$ that may be real or 
complex numbers, vectors, or fields, and nonnegative Hermitan 
quantities  $P_1,P_2,\dots$ satisfying
\lbeq{e.Psum} 
      P_1 + P_2 + \dots  = 1
\eeq
such that the instrument gives the result $a_k$ with probability
\lbeq{e.Pprob}
      p_k = \<P_k\>
\eeq
if the measured system is in the state $\<\cdot\>$.
The nonnegativity of the $P_k$ implies that all probabilities are 
nonnegative, and \gzit{e.Psum} guarantees that the probabilities always 
add up to 1.

\bigskip
An instructive example is the \bfi{photoelectric effect}, 
the measurement of a classical free electromagnetic field by means 
of a photomultiplier. A detailed discussion is given in Chapter 9 
of \sca{Mandel \& Wolf} \cite{ManW}; here we only give an informal
summary of their account. 

Classical input to a quantum system is conventionally represented in 
the Hamiltonian of the quantum system by an interaction term 
containing the classical source as an external field or potential.
In the semiclassical analysis of the photoelectric effect, the
detector is modelled as a many-electron quantum system, while the
incident light triggering the detector is modelled as an external
electromagnetic field. The result of the analysis is that if the
classical field consists of electromagnetic waves (light) with a 
frequency exceeding some threshold then the detector emits a random 
stream of photoelectrons with a rate that, for not too strong light, 
is proportional to the intensity of the incident light. The
predictions are quantitatively correct for normal light.
 
The response of the detector to the light is statistical, 
and only the rate (a short time mean) with which the electrons are 
emitted bears a quantitative relation with the intensity.
Thus the emitted photoelectrons form a statistical measurement
of the intensity of the incident light.

The results on this analysis are somewhat surprising: The discrete
nature of the electron emissions imply that a photodetector responds 
to classical light as if it were composed of randomly arriving photons
(the explanation of the photoeffect for which Einstein received the 
Nobel prize), although the semiclassical model used to derive the 
quantitatively correct predictions does not involve photons at all!

This shows the importance of differentiating between prepared states 
of the system (here of classical light) and measured events in the
instrument (here the amplified emitted electrons). The measurement
results are primarily a property of the instrument, and their 
interpretation as a property of the system needs theoretical analysis 
to be conclusive.

\section{Quantum probability}\label{s.qprob}

Although quantum mechanics generally counts as an intrinsically
statistical theory, it is important to realize that it not only
makes assertions about probabilities but also makes many
deterministic predictions verifiable by experiment.

These deterministic predictions fall into two classes:

(i) Predictions of numerical values believed to have a precise value 
in nature:
\begin{itemize}
\item The most impressive proof of the correctness of quantum field 
theory in microphysics is the \bfi{magnetic moment} of the electron, 
predicted by
quantum electrodynamics (QED) to the phenomenal accuracy of 12 
significant digit agreement with the experimental value. 
It is a universal constant, determined solely by the two parameters
in QED, the electron mass and the fine structure constant.

\item QED also predicts correctly emission and absorption spectra
of atoms and molecules, both the spectral positions 
and the corresponding line widths.

\item Quantum hadrodynamics allows the prediction of the masses of all 
isotopes of the chemical elements in terms of models with only a 
limited number of parameters.
\end{itemize}

(ii)  Predictions of qualitative properties, or of numerical values
believed to be not exactly determined but which are accurate with a 
high, computable limit resolution. 
\begin{itemize}
\item QED predicts correctly the color of gold, the liquidity of 
mercury at room temperature, and the hardness of diamond.

\item Quantum mechanics enables the computation of thermodynamic
state equations for a huge number of materials. Equations of states
are used in engineering in a deterministic manner.

\item From quantum mechanics one may also compute transport 
coefficients for deterministic kinetic equations used in a variety
of applications.
\end{itemize}

Thus quantum mechanics makes both deterministic and stochastic 
assertions, depending on which system it is applied to and on the state 
or the variables to be determined. 
Statistical mechanics, as discussed in Chapters \ref{c.quants} and 
\ref{c.lawtherm}, is mainly concerned with deterministic
prediction of class (ii) in the above classification.

Interestingly, our definition of classical instruments also covers 
joint position-momentum measurements of \bfi{coherent states}, the 
quantum states discussed in Section \ref{sec-coh-state}. 
They are parameterized
by position and momentum, and describe single quantum particles with 
essentially classical trajectories, such as they can be seen as 
particle tracks on photographic plates or in bubble chambers.
The deterministic nature of the recorded tracks is due to the 
interaction of such a particle with the many-particle system formed
by the recording device.

Predictions of class (i) are partly related to spectral properties 
of the Hamiltonian of a quantum system, which we shall discuss in  
Chapter \ref{c.spec}, and partly to properties deduced from form 
factors, which are deterministic byproducts of scattering calculations.
In both cases, classical measurements account adequately for the 
experimental record.

Particle scattering itself, however, is a typical stochastic phenomenon.
The same holds for radioactive decay, when modelled on the level of 
individual particles; it needs a stochastic 
description as a branching process, similar to classical birth and 
death processes in biological population dynamics.
In the remainder of this section, we consider the fundamental aspects 
of this stochastic part of quantum mechanics.

A \bfi{statistical instrument} in the quantum case is mathematically
equivalent to what is called in the literature a \bfi{positive 
operator-valued measure}, short \bfi{POVM}, defined as a 
family $P_1,P_2,\dots$ of Hermitian, positive semidefinite operators 
satsifying \gzit{e.Psum} (or a continuous generalization of this).
They originated around 1975 in work by \bfi{Helstrom} \cite{Hel} 
on quantum detection and estimation theory and are discussed in some 
detail in \sca{Peres} \cite{Per}. They describe the most 
general quantum measurement of interest in quantum information theory.
Which operators $P_k$ correctly describe a statistical instrument 
can in principle be found out by suitable \bfi{calibration 
measurements}. Indeed, if we feed the instrument with enough systems 
prepared in known states $\<\cdot\>_j$, we can measure approximate 
probabilities $p_{jk}\approx\<P_k\>_j$. By choosing the states diverse 
enough, one may approximately reconstruct $P_k$ from this information 
by a process called \bfi{quantum tomography}. In quantum information 
theory, the Hilbert spaces are finite-dimensional, hence the quantities 
form some algebra $\Ez=\Cz^{N\times N}$; then $N^2$ values $\<P_k\>_j$
for linearly independent states suffice for this reconstruction.
The optimal reconstruction using a minimal number of individual 
measurements is the subject of \bfi{quantum estimation theory},
still an active frontier of research.

Before 1975, quantum measurements used to be described in terms of
\bfi{ideal} statistical measurements, the special case of POVMs where 
the $P_k$ form a family of \bfi{orthogonal projectors}, i.e., linear 
operators satisfying
\[
P_k^2=P_k=P_k^*,~~~P_jP_k=0\for j\ne k,
\]
on the eigenspaces of a self-adjoint 
operator $A$ (or the components of a vector $A$ of commuting, 
self-adjoint operators) with discrete spectrum given by $a_1,a_2,\dots$.
In this case, the statistical instrument is said to perform an
\bfi{ideal measurement} of $A$, and the rule \gzit{e.Pprob} defining 
the probabilities is called \bfi{Born's rule}. The rule is named after 
Max {\sc Born} \cite{Bor}, who derived it in 1926 in the special case 
of pure states (defined in \gzit{e.pureex}) and was rewarded in 1954 
with the Nobel prize for this at that time crucial insight into the 
nature of quantum mechanics.

Ideal measurements of $A$ have quite strong properties since under the
stated assumptions, the instrument-based statistical average 
\[
      \ol {f(A)} = p_1 f(a_1) + p_2 f(a_2) + \dots 
\]
agrees for all functions $f$ defined on the spectrum of $A$ with the
model-based value $\<f(A)\>$. On the other hand, these strong 
properties are bought at the price of idealization, since they result in
effects incompatible with real measurements. For example, according
to Born's rule, the ideal measurement of the energy of a system whose 
Hamiltonian $H$ is discrete always yields an exact eigenvalue of $H$,
the only statistical component is the question which of the eigenvalues
is obtained. This is impossible in a real measurement; the precise 
measurement of the Lamb shift, a difference of eigenvalues of the 
Hamiltonian of the hydrogen atom, was even worth a Nobel prize 
(1955 for Willis Lamb).

In general, the correspondence between values and eigenvalues is
only approximate, and the quality of the approximation improves with 
improved resolution. The correspondence is perfect only at resolution 
zero, i.e., for completely sharp measurements.
To discuss this in detail, we need some results from functional 
analysis. The \bfi{spectrum} $\spec f$\index{$\spec f$, spectrum} of a 
linear operator on a Euclidean space $\Hz$ is the set of all 
$\lambda\in\Cz$ for which no linear operator $R(\lambda)$ from the 
completion $\ol \Hz$ of $\Hz$ to $\Hz$ exists such that 
$(\lambda-f)R(\lambda)$ is the identity. $\spec f$ is always a closed 
set. A linear operator $f\in\Lin\Hz$ is called 
\bfi{essentially self-adjoint} if it is 
Hermitian and its spectrum is real (i.e., a subset of $\Rz$).
For $N$-level systems, where $\Hz$ is finite-dimensional, the spectrum
coincides with the set of eigenvalues, and every
Hermitian operator is essentially self-adjoint. In infinite dimensions,
The spectrum contains the eigenvalues, but not every number in the
spectrum must be an eigenvalue; and whether a Hermitian operator is 
essentially self-adjoint is a question of correct boundary conditions. 

\begin{thm}
Let $f$ be essentially self-adjoint, with value
$\ol f :=\<f\>$ and standard deviation $\sigma(f)$ in a given state. 
Then the spectrum of $f$ contains some real number $\lambda$ with 
\lbeq{e.specabs}
|\lambda-\ol f|\le \sigma(f).
\eeq
Moreover, if $\ol f \ne 0$ then the spectrum of $f$ contains some 
real number $\lambda$ with 
\lbeq{e.specrel}
\frac{\lambda-\ol f|}{|\ol f|} \le \res(f).
\eeq
\end{thm}

\bepf
The linear operator $g=(f-\ol f)^2-\sigma(f)^2$ is a quadratic function
of $f$, hence its spectrum consists of all 
$\lambda':=(\lambda-\ol f)^2-\sigma(f)^2$ with $\lambda\in \spec f$;
in particular, it is real. Put $\lambda_0:=\inf\spec g>0$. Then
$g-\lambda_0$ is a Hermitian operator with a real, nonnegative spectrum,
hence positive semidefinite. (In infinite dimensions, this requires 
the use of the spectral theorem.) Thus $g-\eps\ge 0$ and 
$0\le \<g-\lambda_0\>=\<(f-\ol f)^2\>-\sigma(f)^2-\lambda_0=-\lambda_0$.
Therefore $\lambda_0\le 0$. Since $\spec g$ is closed, $\lambda_0$ is 
in the spectrum, hence has the form $(\lambda-\ol f)^2-\sigma(f)^2$
with $\lambda\in \spec f$. This $\lambda$ satisfies \gzit{e.specabs},
and, if $\ol f \ne 0$, also \gzit{e.specrel}.
\epf

In particular, if one can make a sharp measurement (with $\sigma(f)=0$)
then the value $\<f\>$ belongs to the spectrum. In practice, this is 
the case only for quantities $f$ whose spectrum (set of sharp values)
consists of small integers. 

\bigskip
\bfi{Binary tests.}
There is, however, an important special case of Born's rule that 
frequently applies essentially exactly. 
An \bfi{ideal binary statistical measurement}, e.g., the click of a 
detector, is described by a single orthogonal projector $P$; the POVM 
is then given by $P_1 = P$ for the measurement result $a_1= 1$ (click) 
and by $P_2 = 1-P$ for the measurement result $a_2= 0$ (no click). 
In particular, a \bfi{test for a state $\varphi$} with 
$\varphi^*\varphi = 1$ is an ideal  binary statistical measurement 
with orthogonal projector 
$P=\varphi\varphi^*$; the reader should check that indeed $P^2=P=P^*$.
By the above, such a test turns out positive with probability 
$p = \<\varphi\varphi^*\>$. In particular, if the system is in a 
pure state $\psi$ then  \gzit{e.pureex} implies that 
\[
p = \<\varphi\varphi^*\>=\psi^*\varphi\varphi^*\psi 
= |\varphi^*\psi|^2.
\]
This is the well-known \bfi{squared probability amplitude} formula,
the original form of Born's rule. As a consequence, the test for 
$\varphi$ always turns out positively if the measured system is in 
the pure state $\varphi$. However, it also turns out positively with 
a positive probability if the measured system is in a pure state 
different from $\varphi$, as long as it is not orthogonal to it.

By a suitable sequence of binary tests, it is possible
in principle to determine with arbitrary accuracy the state in which
a stationary source of particles prepares the particles.
Indeed, this can be done again with quantum tomography. In case of
$N$-level systems represented by $\Ez=\Cz^{N\times N}$, a general state
is characterized by its \bfi{density matrix} $\rho$, a complex
Hermitian $N\times N$-matrix with trace one, together with the 
trace formula
\[
\<f\>= \tr \rho f.
\]
This implies that a set of $N^2-1$ tests for specific states, repeated 
often enough, suffices for the state determination. Indeed,
it is easy to see that repeated tests for the states $e^j$, the unit 
vectors with just one entry one and other entries zero, tests the
diagonal elements of the density matrix, and since the trace is one,
one of these diagonal elements can be computed from the knowledge of 
all others. Tests for $e^j+e^k$ and $e^j+ie^k$ for all $j<k$ then 
allow the determination of the $(j,k)$ and $(k,j)$ entries.
Thus frequent  repetition of a total of $N-1+2{N\choose 2} = N^2-1$ 
particular tests determines the full state.
The optimal reconstruction to a given accuracy, using a minimal number 
of individual measurements, is again a nontrivial problem of quantum 
estimation theory.

\at{Discuss collapse as a subjective phenomenon caused by a a change 
of the model, and the transition to conditional expectations.\\
Mott on particle tracks from spherical waves -- an example of a 
collapse as change of model and going over to conditional probabilities:
The first ionization localizes the particle and puts it in an 
approximately coherent state.}

\section{Entropy and information theory} 
\label{s.complexity}

The concept of entropy also plays an important role in \bfi{information 
theory}.
To connect the information theoretical notion of entropy with the 
present setting, 
we present in this section an informal example of a simple 
stochastic model in which the entropy has a natural information 
theoretical interpretation. We then discuss what this may
teach us about a non-stochastic macroscopic view of the situation.

We assume that we have a simple stationary device
that, in regular intervals, delivers a reading $n$ from a countable
set $\cal N$ of possible readings. For example, the device might
count the number of events of a certain kind in fixed periods of
time; then ${\cal N}=\{0,1,2,\dots \}$.

We suppose that, by observing the device in action for some time, we
are led to some conjecture about the (expected) relative frequencies
$p _n$ of readings $n\in \cal N$; since the device is 
stationary, these relative frequencies are independent of time. 
If $\cal N$ is finite and not too
large, we might take averages and wait until these stabilize to a
satisfactory degree; if $\cal N$ is large or infinite, most 
$n\in \cal N$ will not have been observed, and our conjecture must 
depend on educated guesses. (The appropriateness of the conjecture,
the relation to the knowledge of the guesser, and how to improve
a conjecture when new information arrives are the subject of 
\bfi{Bayesian statistics}; cf. Section \ref{s.maxent}.)

\at{review the use of conjecture, knowledge, subjective etc. in this 
and the next Section, and whether the narrative is coherent.
The point is, things depend on the actual situation, not on what we 
know about it.}

Clearly, in order to have a consistent interpretation of the $p _n$ 
as relative frequencies, we need to assume that {\em each} reading is 
possible:
\lbeq{1-1}
p_n > 0 \mbox { for all }n\in \cal N,
\eeq
and {\em some} reading occurs with certainty:
\lbeq{1-2}
\sum _{n\in \cal N}p _n=1.
\eeq
For reasons of economy, we shall not allow $p_n=0$ in \gzit{1-1},
which would correspond to readings that are either impossible,
or occur too rarely to have a scientific meaning. Clearly, this is no
loss of generality.

Knowing relative frequencies only means that (when ${\cal N}>1$) we only
have incomplete information about future readings of the device.
We want to calculate the information deficit by counting the expected
number of questions needed to identify a particular reading unknown
to us, but known to someone else who may answer our questions with
yes or no.

Consider arbitrary strategies $s$ for asking questions, and denote by
$s_n$ the number of questions needed to determine the reading $n$
with strategy $s$. Since there are two possible answers for each 
question, we can distinguish with $q$ questions at most $2^q$ different 
cases. However, since reading $n$ is assumed to be determined after 
$s_n$ questions, the answers to the later questions do not matter, and 
reading $n$ is obtained in $2^{q-s_n}$ of the $2^q$ cases when 
$s_n\le q$. Thus, no matter which strategy is used,
\[
\sum_{s_n\le q}2^{q-s_n}\le 2^q.
\]
If we divide by $2^q$ and then make $q$ arbitrarily large we find that
\lbeq{1-3}
\sum _{n\in \cal N}2^{-s_n}\le 1.
\eeq
Since we do not know in advance the reading,
we cannot determine the precise number of questions needed in a
particular unknown case. However, knowledge of the relative frequencies
allows us to compute the average number of questions needed, namely
\lbeq{1-4}
\ol{s}=\sum _{n\in \cal N}p_n s_n.
\eeq
To simplify notation, we introduce the abbreviation
\lbeq{1-5}
\sint  f:=\sum _{n\in \cal N}f_n
\eeq
for every quantity $f$ indexed by the elements from $\cal N$, and we 
use the convention that inequalities, operations and functions of such 
quantities are understood componentwise. Then we can rewrite 
\gzit{1-1}--\gzit{1-4} as
\lbeq{1-6}
p >0,\quad \sint p=1\ ,
\eeq
\lbeq{1-7}
\ol{s}=\sint p s,\quad \sint 2^{-s}\le 1\ ,
\eeq
and
\lbeq{1-8}
\ol{f}=\<f\>:=\sint p f
\eeq
is the average of an arbitrary quantity $f$ indexed by $\cal N$.

\bigskip
It is not difficult to construct a strategy $s$ realizing given 
integral values $s_n$ ($n\in\cal N$) whenever \gzit{1-3} holds. 
\at{give the general recipe}
We now idealize the situation a little by allowing the $s_n$
to be arbitrary nonnegative real numbers instead of integers only.
This is justified when the size of $\cal N$ is large or infinite
since then most $s_n$ will be large numbers that can be approximated 
by integers with a tiny relative error.

Thus we redefine a \bfi{strategy} to be a quantity $s\ge 0$ satisfying
$\sint 2^{-s}\le 1$. We now ask for a strategy $s$ (in the new, 
generalized sense) that makes the number $\ol{s}$ as small as possible.

\begin{thm} \label{1.1.}

The \bfi{entropy} $S$, defined as the quantity
\lbeq{1-9}
S:=-\kbar \log p, ~~~\mbox{where}~~\kbar=\frac {1} {\log 2},
\eeq
satisfies $\ol{S}\le \ol{s}$, with equality if and only if $s=S$.
(One also needs $\sint 2^{-s}=1$, but this holds for $s=S$.)
\end{thm}

\bepf
\gzit{1-9} implies $\log p =-S\log 2$, hence $p =2^{-S}$.
Therefore
\[
2^{-s}=p 2^{S-s}=p e^{(S-s)\log 2}\ge p (1+(S-s)\log 2),
\]
with equality iff $S=s$. Thus
\[
p (S-s)\le \frac {1} {\log 2}(2^{-s}-p )=\kbar (2^{-s}-p )
\]
and
\[ 
\begin{array}{lll} 
\ol{S}-\ol{s}&=\sint p (S-s)\le \sint 
\kbar (2^{-s}-p ) \\
 &=\kbar~\sint 2^{-s}-\kbar ~ \sint p 
\le \kbar -\kbar =0.
\end{array} 
\]
Hence $\ol{s}\ge \ol{S}$, and equality holds iff $s=S$.
\epf

Since \gzit{1-9} implies the relation 
\[
p =e^{-S/\kbar }, 
\]
we have
$\<f\>=\sint p f= \sint e^{-S/\kbar }f$. Thus, the expectation mapping 
is a Gibbs state with entropy $S$, explaining the name.
Note that $s=S$ defines an admissible strategy since
\[
\sum _{n\in \cal N}2^{-S_n}=\sint 2^{-S}=\sint p =1, 
\]
hence $2^{-S_n}\le 1$, $S_n\ge 0$ for all $n\in \cal N$. 
Thus, the entropy $S$ is \bfi{the unique optimal decision strategy.}
The \bfi{expected entropy}, i.e., the mean number
\lbeq{1-10}
\ol{S}=\<S\>=\sint pS =-\kbar~ \sint p \log p 
\eeq
of questions needed in an optimal decision strategy, is nonnegative,
\lbeq{1-11}
\ol{S}\ge 0.
\eeq
$\ol{S}$ measures the \bfi{information deficit} of the device with 
respect to our conjecture about relative frequencies. Traditionally, 
the expected entropy is simply called the {\em entropy}, while we 
reserve this word for the random variable \gzit{1-9}.
Also commonly used is the name {\em information} for $\ol{S}$, 
which invites linguistic paradoxes since ordinary language associates 
with information a connotation of relevance or quality that is absent 
here. 
The classical book on information theory by \sca{Brillouin} \cite{Bri} 
emphasizes this very carefully, by distinguishing absolute information 
from its human value or meaning. \sca{Katz} \cite{Kat} uses the 
phrase {\em missing information}.

The information deficit says nothing at all about the quality
of the information contained in the summary $p $ of our past 
observations.
An inappropriate $p $ can have arbitrarily small information
deficit and still give a false account of reality.
For example, if for some small $\eps>0$,
\lbeq{1-10a}
 p _n=\eps^{n-1}(1-\eps) \quad \mbox{for}~ n=1,2,\dots,
\eeq
expressing that the reading is expected to be
nearly always 1 ($p_1=1-\eps$) and hardly ever large, then
\[
 \ol{S}=\kbar \Big(\log(1-\eps )+\frac{\eps}{1-\eps}\log\eps\Big) 
\to 0 ~\mbox{as} ~\eps \to 0. 
\]
Thus the information deficit
can be made very small by the choice \gzit{1-10a} with small $\eps$, 
independent of whether this choice corresponds to the known facts. The
real information value of $p $ depends instead on the care 
with which the past observations were interpreted, which is a matter of
data analysis and not of our model of the device. If the data analysis 
is done poorly, the resulting expectations will simply not be matched 
by reality.
This shows that the entropy reflects objective properties of the 
stochastic process, and -- contrary to claims in the literature -- 
has nothing to do with our knowledge of the system, 
a subjective, ill-defined notion. \at{A proper formalization of 
``knowledge'' would have to identify it with the quantity $p$.}

\bigskip
\bfi{Relations to thermodynamics.}
Now suppose that the above setting happens at a very fast, unobservable 
time scale, so that we can actually observe only short time 
averages (\ref{1-8}) of quantities of interest. Then $\ol f=\<f\>$ 
simply has the interpretation of the time-independent observed value 
of the quantity $f$. The information deficit
simply becomes the observed value of the entropy $S$. 
Since the information deficit counts the number of optimal 
decisions needed to completely specify a (microscopic)
situation of which we know only (macroscopic) observed values,
the observed value of the entropy quantifies the \bfi{intrinsic} 
(microscopic) \bfi{complexity} present in the system.

However, the unobservable high frequency fluctuations of the device 
do not completely disappear from the picture. 
They show up in the fact that generally $\ol{g^2}\ne\ol g^2$, 
leading to a nonzero limit resolution \gzit{e.res} of Hermitian 
quantities.
This is precisely the situation characteristic of the traditional 
treatment of thermodynamics within classical equilibrium statistical 
mechanics, if we assume that the system is \bfi{ergodic}, i.e., that 
population averages equal time averages.  Then, all observed values are 
time-independent, described by equilibrium thermal variables.
But the underlying high-frequency motions of the atoms making up 
a macroscopic substance are revealed by nonzero limit resolutions.
However, the assumption that all systems for which thermodynamics works
are ergodic is problematic; see, e.g., the discussion in 
\sca{Sklar} \cite{Skl}.

Note that even a deterministic but chaotic high frequency dynamics, 
viewed at longer time scales,
looks stochastic, and exactly the same remarks about the unobservable
complexity and the observable consequences of fluctuations apply. 
Even if fluctuations are observable directly, these observations are 
intrinsically limited by the necessary crudity of any actual 
measurement protocol. For the best possible measurements (and only for
these), the resolution of $f$ in the experiment is given by the limit
resolution $\res(f)$, the size of the unavoidable fluctuations.

Due to the quantum structure of high frequency phenomena on an atomic 
or subatomic scale, it seems problematic to interpret 
thermodynamic limit resolutions in terms of a simple short time average 
of some underlying microscopic reality. Thus an information theoretic 
interpretation of the physical entropy seems questionable.

\section{Subjective probability}\label{s.maxent}

The formalism of statistical mechanics is closely related to that 
used in statistics for random phenomena expressible in terms of 
exponential families; cf. Remark \ref{r7.2}(viii). 
Exponential families play an important role in Bayesian statistics.
Therefore a Bayesian, subjective probability interpretation 
to statistical mechanics is possible in terms of the knowledge of 
an observer, using an information theoretic approach.
See, e.g., \sca{Balian} \cite{Bal3} for a recent exposition in terms 
of physics, and \sca{Barndorff-Nielsen} \cite{BarN,BarN2} for a formal 
mathematical treatment. In such a treatment, the present integral 
plays the role of a \bfi{noninformative prior}, i.e., of the state 
considered to be least informative. This noninformative prior is often 
improper, i.e., not a probability distribution, since $\sint 1$ 
need not be defined. 

Motivated by the subjective, information theoretic approach to 
probability, \sca{Jaynes} \cite{Jay1,Jay2} used the maximum 
entropy principle to derive the thermodynamic formalism. 
The \bfi{maximum entropy principle} asserts that one should
model a system with the statistical distribution that maximizes 
the expected entropy subject to the known information about certain 
expectation values. This principle is sometimes considered as a 
rational, 
unprejudiced way of accounting for available information in 
incompletely known statistical models.
Based on Theorem \ref{t4.5},
it is not difficult to show that when the known information is given 
by the expectations of the quantities $X_1,\dots,X_n$,
the optimal state in the sense of the maximum entropy principle
is a Gibbs state whose entropy is a linear combination of 1 and the 
$X_k$.

However, the maximum entropy principle is an unreliable general 
purpose tool, and gives an appropriate distribution only under quite 
specific circumstances.

\begin{expl} 
If we have information in the form of a large but finite sample of 
realizations $x(\omega_k)$ of a random variable $x$ in $N$ independent
experiments $\omega_k$ ($k=1,\dots,n$),
we can obtain approximate information about all moments $\<x^n\>$
$\<x^n\>\approx N^{-1}\sum x(\omega_k)^n/N$ ($n=0,1,2,\dots$)
by taking the appropriate sample means, 
\[
\<x^n\>\approx \sum w_k x(\omega_k)^n/\sum w_k~~~ (n=0,1,2,\dots),
\]
where the $w_k$ are appropriate positive weights (typically chosen 
such that the experimental errors in $w_k x(\omega_k)$ is approximately 
constant.
It is not difficult to see that the maximum entropy principle would 
infer that the distribution of $x$ is discrete, namely that of the
sample distribution.

If we take as uninformative prior for a real-valued random 
variable $x$ the Lebesgue measure, $\sint f(x):=\int f(x) dx$,
and only know that the mean of $x$ is 1, say, 
the maximum entropy principle does not produce a sensible 
probability distribution. If we add the knowledge of the second 
moment $\<x^2\>=1$, say, we get a Gaussian distribution with mean 1 and 
standard deviation $1/\sqrt{2}$. Adding the further knowledge of 
$\<x^3\>$, the maximum entropy principle fails again to produce a 
sensible distribution. If, on the other hand, after knowing that 
$\<x\>=1$ we learn that the random variable is in fact nonnegative 
and integer-valued, this cannot be accounted for by the principle, 
and the probability of obtaining a negative value remains large. 
But if we take as prior the discrete measure on nonnegative 
integers defined by $\sint f(x):=\sum_{x=0}^\infty f(x)/x!$,
the supposedly noninformative prior has become much more informative,
the knowledge of the mean produces via the maximum entropy principle 
a Poisson distribution.

If we know that a random variable $x$ is nonnegative and has 
$\<x^2\>=1$; the  Lebesgue measure on $\Rz_+$ as noninformative prior
gives for $x$ a distribution with density $\sqrt{2/\pi}e^{-x^2/2}$.
But we can consider instead our knowledge about $y=x^2$,
which is nonnegative and has $\<y\>=1$; the same noninformative
prior now gives for $y$ a distribution with density $e^{-y}$.
The distribution of $x=\sqrt{y}$ resulting from this has density
$2xe^{-x^2/2}$. Thus the result depends on whether we regard $x$
or $y$ as the relevant variable.
\end{expl} 

We see that the
choice of expectations to be used as constraints reflects prior 
assumptions about which expectations are likely to be relevant. 
Moreover, the prior, far from being uninformative,
reflects the prejudice assumed in the complete absence of knowledge. 
The prior that must be assumed to describe the state of complete 
ignorance significantly affects the results of the maximum entropy 
principle, and hence makes the application of the principle ambiguous.

The application of the maximum entropy principle becomes reliable
only if the information is available in form of the expectation values 
of a sufficient statistics of the true model; see, e.g., 
\sca{Barndorff-Nielsen} \cite{BarN}. Which statistical model may be 
considered sufficient depends on the true situation and is difficult 
to assess in advance.

In particular, a Bayesian interpretation of statistical mechanics
in the manner of Jaynes is appropriate if and only if
\begin{itemize}
\item
correct, complete and sufficiently accurate information about 
the expectation of all relevant quantities is assumed to be known, 
and
\item
the noninformative prior is fixed by the constructions of
Example \ref{e3.1}, 
namely as the correctly weighted Liouville measure in classical physics
and as the microcanonical ensemble  (the trace) in quantum physics.
\end{itemize}
Only this guarantees that the knowledge assumed and hence the 
results obtained are completely impersonal and objective, as required 
for scientfic results, and agree with standard thermodynamics, as 
required for agreement with nature.
However, this kind of knowledge is clearly completely hypothetical 
and has nothing to do with the real, partial and imprecise knowledge 
of real observers.

    \part{Lie algebras and Poisson algebras}\label{p.liepoisson}
\chapter{Lie algebras}\label{c.lie}

Part \ref{p.liepoisson} introduces the basics about Lie algebras and 
\at{...} Lie groups,
with an emphasis on the concepts most relevant to the conceptual side
of physics.

\bigskip
This chapter introduces Lie algebras together with the slightly richer 
structure of a Lie $*$-algebra usually encountered in the mechanical
applications. We introduce tools for verifying
the Jacobi identity, and establish the latter both for the Poisson 
bracket of a classical harmonic oscillator and, for quantum systems, 
for the commutator of linear operators.

Further Lie algebras arise as algebras of matrices closed under 
commutation, as algebras of derivations in associative algebras, 
as centralizers or quotient algebras, and by complexification.
An overview over semisimple Lie algebras and their classification
concludes the chapter.

\bigskip
\at{adapt}
In finite dimensions, the relation is almost one-to-one, the ''almost'' 
being due to the fact that the so-called universal covering group of 
a finite-dimensional Lie algebra (defined in Section \ref{univ.cover}) 
may have a nontrivial discrete normal subgroup.

Many finite-dimensional Lie groups arise as groups of square 
invertible matrices, and we discuss the most important families, 
in particular the unitary and the orthogonal groups. We introduce
group representations, which relate groups of matrices (or linear 
operators) to abstract Lie groups, and will turn out to be most 
important for understanding the spectrum of quantum systems.

Of particular importance for systems of oscillators are the Heisenberg 
groups, the universal covering groups of the Heisenberg algebras.
Their product law is given by the famous Weyl relations, which are 
an exactly representable case of the Baker--Campbell--Hausdorff
formula valid for many other Lie groups, in particular for 
arbitrary finite-dimensional ones.
We also discuss the Poincar\'e group. This is the symmetry group of 
space-time, and forms the basis for relativity theory.

\section{Basic definitions}\label{s.lie}

We start with the definition of a Lie algebra over a field
$\Kz$, usually implicitly given by the context. 
In our course, $\Kz$ is either the field $\Cz$ of complex
numbers, occasionally the field $\Rz$ of real numbers. Lie algebras 
over other fields, such as the rationals $\Qz$ or finite fields $\Zz_p$ 
for $p$ prime, also have interesting applications in mathematics,
physics and engineering, but these are outside the scope of this book. 
To denote the Lie product, we use the symbol $\lp$ introduced at the 
end of Section \ref{s.thstat}. (This replaces other, bracket-based 
notations common in the literature.)

\begin{definition}~\\
(i) A {\bfi{Lie product}} on a vector space $\Lz$ over $\Kz$ is a 
bilinear operation\index{$\lp$} on $\Lz$ satisfying

(L1) $f\lp f = 0$,

(L2) $f\lp (g\lp h) + g \lp (h \lp f)+ h \lp (f \lp g)=0$
for all $f,g,h\in \Lz$.

Equation (L2) is called the {\bfi{Jacobi identity}}.

(ii) For subsets $A,B$ of $\Lz$, we write
\[
A \lp B :=\{f\lp g\mid f\in A,g\in B\},
\]
and for $f,g\in\Lz$,
\[
A\lp g :=A\lp \{g\},~~~ f\lp B:=\{f\}\lp B.
\]
(iii) 
A {\bfi{Lie algebra}} over $\Kz$ is a vector space $\Lz$ over $\Kz$ 
with a distinguished Lie product. Elements $f\in\Lz$ with $f\lp \Lz=0$ 
are called \bfi{(Lie) central}\index{central}\index{Lie central}; 
the set $Z(\Lz)$ of all these elements is called the {\bfi{center}} 
of $\Lz$. A \bfi{real (complex)} Lie algebra is a Lie algebra over 
$\Kz=\Rz$ (resp. $\Kz=\Cz$).
Unless confusion is possible, we use the same symbol $\lp$ for the Lie 
product in different Lie algebras.
\end{definition}

Clearly, if $f\lp g$ defines a Lie product of $f$ and $g$, so does 
$f {\lp\!}_\iota g := \iota (f\lp g)$ for all $\iota\in \Kz$.
Thus the same vector space may be a Lie algebra in different ways.

In physics, finite-dimensional Lie algebras are often defined in terms 
of basis elements $X_k$ called \bfi{generators}\index{generator} and
{\bfi{structure constants}} $c_{jkl}$, such that
\lbeq{phys.struct}
X_j\lp X_k =\sum_l c_{jkl}X_l\,.
\eeq
By taking linear combinations and using the bilinearity of the Lie 
product, the structure constants determine the Lie product completely.
Conversely, since the generators form a basis, the structure constants 
are determined uniquely by the basis. They depend, however, on the 
basis chosen. Frequently, there are distinguished bases with a physical 
interpretation in which the structure constants are particularly simple,
and most of them vanish. If a basis and the structure constants are 
given, many Lie algebra computations can be done automatically; 
important software packages include LIE (\sca{van Leeuwen} et al. 
\cite{vLeuCL}) and LTP (\sca{Torres-Torriti} \cite{TorT}).
In this book, we usually prefer a basis-free approach, resorting to 
basis-dependent formulas only to make connections with traditional 
physics notation.

As a consequence of (L1) (and in fact equivalent to it), we have the 
following antisymmetry property:
\[
f\lp g = - g\lp f \,.
\]
This follows from observing that $f\lp 0 = 0 \lp f = 0$ and
\beqar
0&=& (f+g)\lp (f+g)= f\lp
f
+ f\lp g + g\lp f + g\lp g \nonumber\\
&=& f\lp g + g\lp f\,. \nonumber
\eeqar
Using the antisymmetry property of the Lie product one can write
the Jacobi identity in two other important forms, each equivalent 
with the Jacobi identity:
\lbeq{jac.der}
 f  \lp (g\lp h)  = (f \lp g)\lp h + g\lp (f\lp h)\,,
\eeq
\lbeq{jac.der2}
 (f  \lp g)\lp h  = (f \lp h)\lp g+ f\lp (g\lp h)\,.
\eeq
These formulas say that one can apply the Lie product to a compound 
expression in a manner familiar from the product rule for 
differentiation.

An important but somewhat trivial class of Lie algebras are the
{\bfi{abelian} Lie algebras}, where $f\lp g=0$ for all $f,g\in \Lz$.
It is trivial to check that (L1) and (L2) are satisfied.
Clearly, every vector space can be turned into an abelian Lie algebra
by defining $f\lp g=0$ for all vectors $f$ and $g$.
In particular, the field $\Kz$ itself and the center of any Lie 
algebra are abelian Lie algebras. 

A subspace $\Lz'$ of a Lie algebra $\Lz$ is a {\bfi{Lie
subalgebra}} if it is closed under the Lie product, i.e., if 
$f\lp g \in\Lz'$ for all $f,g\in\Lz'$. In this case, 
 the restriction of the Lie product $\lp$ of $\Lz$ to
$\Lz'$ turns $\Lz'$ into a Lie algebra. That is, a Lie subalgebra is a
subspace that is a Lie algebra with the same Lie product.
(For example, the subspace $\Kz f$ spanned by an arbitrary element $f$ 
of a Lie algebra is an abelian Lie subalgebra.)
A Lie subalgebra is \bfi{nontrivial} if it is  not the whole Lie algebra
and contains a nonzero element.

\bigskip
The property $(L1)$ is usually easy to check.
It is harder to check the Jacobi identity $(L2)$ for a
proposed Lie product; direct calculations can be quite messy
when many terms have to be calculated before one finds that they
all cancel. Since we will encounter many Lie products that must be
verified to satisfy the Jacobi identity, we first develop some
technical machinery to make life easier, or at least more structured.
For a given binary bilinear operation $\circ$ on $\Lz$, we define
the {\bfi{associator}} of $f,g,h\in\Lz$ as
\lbeq{e.assoc}
[f,g,h]:= (f\circ g)\circ h - f\circ (g\circ h)\,.
\eeq

\begin{prop}\label{assoclie}
If the associator of a bilinear operator $\circ$ on $\Lz$
satisfies
\lbeq{assoid}
[f,g,h]+[g,h,f]+[h,f,g]-[f,h,g]-[h,g,f]-[g,f,h]=0\,,
\eeq
then
\[
f\lp g := f\circ g - g\circ f\,
\]
defines a Lie product on $\Lz$.
\end{prop}

\bepf
Define
\[
J(f,g,h):= f\lp (g\lp h) + g \lp (h \lp f)+ h \lp (f \lp g)\,,
\]
and define
\[
S(f,g,h):= [f,g,h]+[g,h,f]+[h,f,g]-[f,h,g]-[h,g,f]-[g,f,h]\,.
\]
Writing out $S(f,g,h)$ and $J(f,g,h)$ with
$f \lp g :=f\circ g - g\circ f$, one sees $J(f,g,h)= - S(f,g,h)$
and hence if $S(f,g,h)=0$ for all $f,g$ and $h$, then the Jacobi
identity is satisfied for all $f,g$ and
$h$. The antisymmetry property $f\lp f= 0$ is trivial.
\epf

\begin{thm}\label{t.1dpoisson}
The  binary operation $ \lp$ defined on the vector space
$C^\infty(\Rz\times\Rz)$ by
\[
f\lp g := f_p g_q - g_p f_q\,,
\]
where $f_p=\partial f/\partial p$ and  $f_q=\partial f/\partial q$,
is a Lie product.
\end{thm}

\bepf
We calculate the associator for the bilinear operator
$f\circ g = f_p g_q$. We have
 \beqar
 [f,g,h]&=& (f\circ g)_p h_q -f_p(g\circ h)_q \nonumber \\
 &=& (f_pg_q)_p h_q - f_p(g_ph_q)_q \nonumber\\
 &=& f_{pp}g_qh_q + f_pg_{qp}h_q - f_pg_{pq}h_q - f_p g_p h_{qq}
\nonumber\\
 &=& f_{pp}g_qh_q  - f_p g_p h_{qq}\nonumber
 \eeqar
Writing the cyclic permutations we get
 \beqar
 [f,g,h]+[g,h,f]+[h,f,g]&=& f_{pp}g_qh_q +g_{pp}h_qf_q + h_{pp}f_q
 g_q
 \nonumber\\
 &&-f_pg_p h_{qq} - g_ph_p f_{qq} - h_p f_p g_{qq}\,,\nonumber
 \eeqar
which is symmetric in $f,g$; hence the identity
\gzit{assoid} is satisfied. Proposition \ref{assoclie} therefore
implies that $\lp$ is a Lie product. 

The reader is invited to prove this result also by a direct calculation.
\epf

We end this section by introducing some concepts needed at various 
later points but collected here for convenience. 
If $\Lz$ and $\Lz'$ are Lie algebras we call a linear map $\phi:\Lz\to
\Lz'$ a {\bfi{homomorphism}} (of Lie algebras) if
\[
\phi(f\lp g)=\phi(f)\lp \phi(g)
\]
for all $f,g\in\Lz$. Note that the left-hand side involves the Lie 
product in $\Lz$, whereas the right-hand side involves the Lie product 
in $\Lz'$. An injective homomorphism is called an {\bf\idx{embedding}}
of $\Lz$ into $\Lz'$.
We call two Lie algebras $\Lz$ and $\Lz'$ {\bfi{isomorphic}}
if there is a homomorphism $\phi:\Lz\to \Lz'$ and a homomorphism
$\psi:\Lz'\to\Lz$ such that $\psi\circ\phi$ is the identity on $\Lz$
and $\phi\circ\psi$ is the identity on $\Lz'$. Then $\phi$ is called
an {\bfi{isomorphism}}, and $\psi$ is the inverse isomorphism.

Given a Lie algebra $\Lz$ and a subalgebra $\Lz'$, the 
{\bfi{centralizer}} \idx{$C_{\Lz'}(S)$} in $\Lz'$ of a subset 
$S\subset\Lz$ is defined by
\[
C_{\Lz'}(S) = \left\{ f\in \Lz' \mid f\lp g =0
\Forall g\in S\right\}\,.
\]
In words, $C_{\Lz'}(S)$ consists of all the elements in $\Lz'$ that Lie
commute with all elements in $S$. One may use the Jacobi identity to see
that $C_{\Lz'}(S)$ is a Lie subalgebra of $\Lz'$. 

An {\bfi{ideal}} of $\Lz$ is a subspace $I\subseteq \Lz$ such that
$f\lp g\in I$ for all $f\in \Lz$ and for all $g\in I$. 
In other notation $\Lz \lp I=I\lp \Lz \subseteq I$. 
Note that $0$ and $\Lz$ itself are always ideals;
they are called the trivial ideals. Also, the center of a Lie algebra
is always an ideal. A less trivial ideal is the
{\bfi{derived Lie algebra}\index{Lie algebra!derived} $\Lz^{(1)}$ 
of $\Lz$} consisting of all elements that can be
written as a finite sum of elements of $\Lz\lp\Lz$.
If $I\subseteq \Lz$ is an ideal in $\Lz$ one may form the 
\idx{quotient Lie algebra}\index{Lie algebra!quotient} $\Lz/I$, 
whose elements are 
the equivalence classes $[f]$ of all $g\in\Lz$ such that $f-g\in I$,
with addition, scalar multiplication, and Lie product given by
\[
\alpha [g]:=[\alpha g]\,,~~~
[f] + [g]:=[f + g]\,,~~~
[f]\lp [g]:=[f\lp g]\,.
\]
It is well-known that the vector space operations are well-defined.
The Lie product is well-defined since $f'\in[f]$ implies $f'-f\in I$,
hence $(f-f')\lp g\in I$ and 
$[f']\lp [g]=[f'\lp g]=[f \lp g + (f'-f)\lp g]=[f\lp g]$.

If $\Lz$ and $\Lz'$ are Lie algebras, their {\bfi{direct sum}} 
$\Lz\oplus \Lz'$ is the direct sum of the vector spaces equipped with
the Lie product defined by
\[
(x+x')\lp (y+y')=x\lp y + x'\lp y'
\]
for all $ x,y\in\Lz$ and all $x',y'\in\Lz'$. It is easily verified that
the axioms are satisfied.

\section{Lie algebras from derivations}\label{sec.calc.lie}

Equation \gzit{jac.der},
\[
 f  \lp (g\lp h)  = (f \lp g)\lp h + g\lp (f\lp h)\,.
\]
 resembles the product rule for (partial) differentiation;
\[
\frac{\partial}{\partial x}(gh) = \frac{\partial g}{\partial x} h
+ g\frac{\partial h}{\partial x}\,.
\]
To make the similarity more apparent we introduce for every
element $f\in \Lz$ a linear operator $\ad_f:\Lz\rightarrow \Lin\Lz$,
the {\bfi{derivative in direction} $f$}, 
given by\index{$\ad_f$}
\[
\ad_f g := f\lp g\,.
\]
The notation reflects the fact that the operator $\ad:\Lz\to\Lin\Lz$ 
defined by 
\[
\ad f:=\ad_f
\]
is the {\bfi{adjoint representation}}\index{$\ad$} of $\Lz$; see 
Sections \ref{fin-dim} and \ref{sec-fin-dim}. \at{check placement}

Note that an element $f\in\Lz$ is in the center $Z(\Lz) = C_{\Lz}(\Lz)$ 
of $\Lz$ if and only if the linear operator $\ad_f$ is zero.

\begin{expl}
For $f$ in the Lie algebra $C^\infty(\Rz\times\Rz)$ constructed in
Theorem \ref{t.1dpoisson}, we have
\lbeq{e.Xf}
 \ad_f g= f\lp g= f_pg_q - f_qg_p
= \left( f_p\frac{\partial}{\partial q} -
f_q\frac{\partial}{\partial p}\right) g\,.
\eeq
The vector field \idx{$X_f$} on $\Rz\times \Rz$ defined by the 
coefficients of $\ad_f$ \at{.} is
called the {\bfi{Hamiltonian vector field}} defined
by $f$; cf. Chapter 12. \at{.}
In particular, the Hamiltonian derivative operators with
respect to $p$ and $q$ take the explicit form
\[
X_p = \frac{\partial}{\partial q}\,, \ \ \
X_q = -\frac{\partial}{\partial p}\,,
\]
and we have
\[
\ad_f = f_p X_p +f_q X_q.
\]
\end{expl}

With the convention that operators bind stronger than the Lie product,
the Jacobi identity can be written in the form
\[
\ad_f (g\lp h) = \ad_f g \lp h + g\lp \ad_f h\,.
\]
The Jacobi identity is thus equivalent to saying that the operator
$\ad_f$ defines for every $f$ a derivation of the Lie algebra.

\begin{dfn}~\\
(i) A {\bfi{derivation}} of a vector space $\Az$ with a 
bilinear product $\circ$ is a linear map $\delta:\Az\rightarrow\Az$ 
satisfying the product rule (or \bfi{Leibniz identity})
\[
\delta (f\circ g) = \delta f\circ g + f \circ \delta g\,,
\]
for all $f,g\in\Az$. We denote by \idx{$\der\Az$} the
set of all derivations of $\Az$. (In the cases of interest,
$\Az$ is an associative algebra with the associative product as $\circ$,
or a Lie algebra with the Lie product as $\circ$.

(ii) If $\Ez$ is an associative algebra $\Ez$, a
\bfi{(left) $\Ez$-module} is an additive abelian group $\Vz$ 
together with a multiplication mapping which assigns to
$f\in\Ez$ and $x\in\Vz$ a product $fx\in\Vz$ such that
\[
f(x+y)=fx+fy,~~~(f+g)x=fx+gx,~~~f(gx)=(fg)x
\]
for all $f,g\in\Ez$ and all $x,y\in\Vz$.
\end{dfn}

\begin{prop} \label{p.der}
The commutator of two derivations is a derivation. In particular,
$\der \Az$ is a Lie subalgebra of $\Lin \Az$ with Lie product
\[
\delta\lp \delta' := [\delta,\delta'].
\]
Moreover, if $\Az$ is commutative and associative then the \bfi{product}
$f\delta$ of $\delta\in\der\Ez$ and $f\in\Ez$ defined by
\[
(f\delta)g:=f(\delta g)
\]
is a derivation, and turns $\der  \Ez$ into an $\Ez$-module.
\end{prop}
\bepf
Since $\der \Ez$ is a linear vector space, and since
the antisymmetry property and the Jacobi identity are already
satisfied in $\Lin\Ez$, we only need to check that the Lie
product of two derivations is again a derivation. We have:
 \beqar
 (\delta\lp\delta')(fg) &=& (\delta\delta')(fg) -
 (\delta'\delta)(fg) \nonumber\\
 &=& \delta ( (\delta'f)g + f(\delta'g) ) - \delta' ( (\delta f)g +
 f(\delta g) ) \nonumber\\
 &=& (\delta \delta' f)g +f\delta\delta' g - (\delta'\delta f)g -
 f\delta\delta' g \nonumber\\
 &=& (\delta \lp \delta' f)g + f(\delta\lp\delta' g)\nonumber
 \eeqar
This proves the first part. The second part is straightforward.
\epf

\begin{prop}\label{lem.der.cent}
The {\bfi{centralizer}} of a subset $S$ of $\Lin \Ez$, defined as
\[
C(S):=
\left\{ \delta \in\der \Ez \mid A\lp \delta=0\,\Forall A\in S\right\},
\]
is a Lie subalgebra of $\der\Ez$.
\end{prop}
\bepf As before, we only need to prove that the Lie product closes
within $C(S)$. If $\delta,\delta'\in C(S)$, the
Jacobi identity in the form \gzit{jac.der} implies
\[
A\lp (\delta\lp \delta') = (A\lp \delta)\lp \delta' +
\delta \lp (A\lp \delta') = 0\,.
\]
\epf

\section{Linear groups and their Lie algebras}\label{s.LinLie}

In quantum mechanics, linear operators play a central role; 
they appear in two essentially different ways:
Operators describing time evolution and canonical transformations
are linear operators $U$ on a Hilbert space, that are \bfi{unitary} 
in the sense that $U^*U=UU^*=1$, and hence bounded\footnote{
The bounded operators on a Hilbert space a so-called $C^*$-algebra;
see for example \sca{Rickart} \cite{Ric}, \sca{Baggett} \cite{baggett}, 
or \sca{Werner} \cite{werner}. But we do not use this fact. 
}. 
The unitary operators form a group, which in many cases of interest is 
a so-called Lie group.

On the other hand, many important quantities in quantum mechanics are 
described in terms of unbounded linear operators that are 
defined not on the whole Hilbert space but only on a dense subspace.
Usually, the linear operators of interest have a common dense domain 
$\Hz$ on which they are defined and which they map into itself. 
$\Hz$ inherits from the Hilbert space the Hermitian inner product,
hence is a complex \bfi{Euclidean space}, and the Hilbert space can
be reconstructed from $\Hz$ as the \bfi{completion} $\ol\Hz$ of $\Hz$ 
by equivalence classes of Cauchy sequences, in the way familiar from 
the construction of the real numbers from rationals. We therefore
consider the algebra \idx{$\Lin\Hz$} of continuous linear operators 
on a Euclidean space $\Hz$, with composition as associative 
product. 

In this section, we define the basic concepts relevant for a study of
groups and Lie algebras inside algebras of operators. Since for these
concepts neither the operator structure nor the coefficient field 
matters in most cases - as long as the characteristic is not two -, 
we provide a slightly more general framework.
In the next section, we apply the general framework to the algebra  
$\Cz^{n\times n} = \Lin \Cz^n$ of complex $n\times n$-matrices,
considered in the standard way as linear operators on the space
$\Cz^n$ of column vectors with $n$ complex entries.
Many of the Lie groups and Lie algebras arising in the applications
are naturally defined as subgroups or subspaces of this algebra.

\bigskip
An {\bfi{(associative) algebra}} over a field $\Kz$ 
is a vector space $\Ez$ over $\Kz$ with a bilinear, associative
multiplication. For example, every $*$-algebra is an associative algebra
over $\Cz$. As traditional, the product of an associative algebra
(and in particular that of $\Lin\Hz$ and  $\Kz^{n\times n}$) is
written by juxtaposition. An associative algebra $\Ez$ is called 
\bfi{commutative} if $fg=gf$ for all $f,g\in\Ez$, and  
\bfi{noncommutative} otherwise. In many cases we assume that such an 
algebra has a unit element $1$ with respect to multiplication;
after the identification of the multiples of $1$ with the elements 
of $\Kz$, this is equivalent to assuming that $\Kz\subseteq \Ez$.
If $\Ez$ and $\Ez'$ are associative algebras over $\Kz$ with $1$, then 
a $\Kz$-linear map $\phi:\Ez\to \Ez'$ is an 
\idx{algebra homomorphism} if $\phi(fg)=\phi(f)\phi(g)$ and
$\phi(1)=1$. Often we omit the reference to the ground field $\Kz$ and
assume a ground field has been chosen. 

We now show that every associative algebra has many Lie products, 
and thus can be made in many ways into a Lie algebra.
For commutative algebras, the construction is uninteresting since it 
only leads to abelian Lie algebras.

\begin{thm}\label{ass.lie.J}
Let $\Ez$ be an associative algebra. Then, for every $J\in\Ez$, the
binary operation ${\lp\!\!}_J$ defined on $\Ez$ by
\[
f{\lp\!\!}_J \, g := fJg-gJf\,
\]
is a Lie product. In particular ($J=1$), the binary operation
$\lp$ defined on $\Ez$ by
\[
f\lp g := [f,g]\,
\]
where
\[
[f,g]:=fg-gf
\]
denotes the {\bfi{commutator}} of $f$ and $g$, is a Lie product.
\end{thm}

\bepf
We compute the associator \gzit{e.assoc} for the bilinear  
operation $f\circ g := fJg$:
\[
[f,g,h]= (f\circ g)Jh - fJ(g\circ h) = fJgJh - fJgJh =0\,,
\]
by associativity. Hence the associator of $\circ$ satisfies 
\gzit{assoid}, and we conclude that ${\lp\!\!}_J$ is a Lie product.
\epf

Note that $Jf \lp Jg =
J(f{\lp\!\!}_J g)$. Hence the corresponding  
Lie algebras are isomorphic when $J$ is invertible.

If $\Ez$ and $\Ez'$ are two associative algebras with unity, 
we may turn them into Lie algebras by putting $f\lp
g =[f,g]$ in both $\Ez$ and $\Ez'$. We denote by $\Lz$ and $\Lz'$ the
Lie algebra associated to $\Ez$ and $\Ez'$, respectively. 
If $\phi$ is an algebra homomorphism
from $\Ez$ to $\Ez'$ then $\phi$ induces a Lie algebra
homomorphism between the Lie algebras $\Lz$ and $\Lz'$. Indeed
$\phi(f\lp g)=\phi(fg-gf)=\phi(f)\phi(g)-\phi(g)\phi(f)
=\phi(f)\lp \phi(g)$.

Theorem \gzit{ass.lie.J} applies in particular to $\Ez=\Kz^{n\times n}$.
The Lie algebra $\Kz^{n\times n}$ with Lie product $f\lp g:=[f,g]$ is 
called the {\bfi{general linear algebra}} $gl(n,\Kz)$ over $\Kz$. 
If $\Kz=\Cz$, we simply write $gl(n)=gl(n,\Cz)$;
similar abbreviations apply without notice for the names of other
Lie algebras introduced later.

\begin{dfn}\label{d.llie}~\\
(i) A \bfi{Hausdorff  $*$-algebra} is a $*$-algebra $\Ez$ with a 
Hausdorff topology in which addition, multiplication, and conjugation 
are continuous. An element $f\in\Ez$ is called \bfi{complete} if
the initial-value problem  
\lbeq{e.complete}
\frac{d}{dt} U(t) = f U(t),~~~U(0)=1
\eeq
has a unique solution $U:\Rz\to\Ez$. Then the mapping $U$ is called a
\bfi{one-parameter group} with \bfi{infinitesimal generator} $f$,
and we write $e^{tf}:=U(t)$; this notation is unambiguous since
it is easily checked that $e^{t(sf)}=e^{(ts)f}$ for $s,t\in\Rz$.
An element $f\in\Ez$ is called \bfi{self-adjoint} if $f^*=f$ and
the product $if$ with the imaginary unit is complete.
We call an element $g \in \Ez$ \bfi{exponential} if it is of the form
$g=e^f$ for some complete $f\in\Ez$. 
We call a Hausdorff $*$-algebra $\Ez$ an \bfi{exponential algebra} 
if the set of exponential elements in $\Ez$ is a neighborhood of $1$.
\at{Is it enough to require that there is a path-connected 
neighborhood of 1?}

(ii) A \bfi{linear group} is a set $\Gz$ of invertible elements of some 
associative algebra $\Ez$ such that $1\in\Gz$ and 
\[
g,g'\in\Gz\implies g^{-1},gg'\in \Gz.
\]
If $\Ez$ is given with a topology in which its operations are 
continuous, we consider $\Gz$ as a topological group with the
topology induced by calling a subset of $\Gz$ open or closed if it 
is the intersection of an open or closed set of $\Ez$ with $\Gz$. 

(iii) A \bfi{linear Lie group} is a closed subgroup of the group 
$\Ez^\times$ of all invertible elements of an exponential algebra $\Ez$.
A \bfi{Lie group} is a group $\widetilde\Gz$ with a 
Hausdorff topology that is \bfi{isomorphic} to some linear Lie group 
$\Gz$, i.e., for which there is a continuous, invertible mapping 
$\phi:\widetilde\Gz\to\Gz$ such that $\phi$ and $\phi^{-1}$ are 
continuous and $\phi(1)=1$,  $\phi(gg')=\phi(g)\phi(g')$ for all 
$g,g'\in\widetilde\Gz$.
\end{dfn}

For all exponential algebras $\Ez$, the group $\Ez^\times$ is a linear 
Lie group. Note that the law $e^fe^{f'}=e^{f+f'}$ holds if $f$ and $f'$ 
commute but not in general. In particular, 
\[
e^fe^{-f}=e^{0}=1.
\]

If $\Ez$ is a \bfi{Banach algebra}, i.e., if the topology of $\Ez$ is 
induced by a norm $\|\cdot\|$ satisfying $\|fg\|\le \|f\|\,\|g\|$,
it is not very difficult to show that every $f\in\Ez$ is complete, and 
we have for all $f\in\Ez$ the absolute convergent series expansion
\[
e^f = \sum_{k=0}^\infty \frac{f^k}{k!},
\] 
and that $f=\log g$, where
\[
\log g := -\sum_{k=1}^\infty \frac{(1-g)^k}{k}\for \|1-g\|<1,
\]
provides an $f$ such that $g=e^f$.
Therefore every Banach algebra is exponential.
Note that the \bfi{exponential mapping}, which maps a matrix $f\in\Ez$ 
to $e^f\in\Ez^\times$, is usually not surjective. \at{counterexample?} 

The above applies to the case $\Ez=\Cz^{n\times n}$ with the maximum 
norm, which is a Banach algebra, which covers all finite-dimensional
Lie groups.
In infinite dimensions, however, many interesting linear Lie groups
are not definable over Banach algebras (see, e.g., 
\sca{Neeb} \cite{Nee}).

\section{Classical Lie groups and their Lie algebras}\label{s.Liega}

This section is not yet in a good form.
\at{this section needs some polishing -- some things are explained 
twice}

\bigskip
A \bfi{matrix group} is a linear group in an algebra $\Kz^{n\times n}$.
In this section, we define the most important matrix groups and 
the corresponding Lie algebras. Although these are defined no matter
which field is involved, the Lie algebras in quantum physics have $\Kz$ 
is the field of real numbers of the field of complex numbers.
Because exponentials can be defined for the real and complex fields, 
the groups have a natural differential geometric 
structure as differentiable manifolds; cf. Section \ref{s.Liegroups}. 
\footnote{For general fields, there are no exponentials, and one needs 
to replace the differential geometric structure inherent in Lie groups 
by an algebraic geometry structure, and may then interpret general 
matrix groups as so-called \bfi{groups of Lie type}. 
In particular, for finite fields, one gets the \bfi{Chevalley groups}, 
which figure prominently in the classification of finite simple groups. 
}. 

\begin{expl}
The group 
$GL(n,\Kz)$\index{$GL(n,\Kz)$}\index{$gl(n,\Kz)$} of all invertible 
$n\times n$-matrices over $\Kz=\Rz$ or $\Kz=\Cz$ is a linear group,
The subgroup of $GL(n,\Kz)$
consisting of the matrices with unit determinant is denoted by
\idx{$SL(n,\Kz)$}. In other words, $SL(n,\Kz)$ is the kernel of the map
$\det:GL(n,\Kz)\to \Kz^*$, where \idx{$\Kz^*$} is the group of 
invertible elements in $\Kz$. The Lie algebra of $SL(n,\Kz)$ is 
denoted by \idx{$sl(n,\Kz)$} and consists of the traceless $n\times n$ 
matrices with entries in $\Kz$.
\at{show $\log GL(n,\Kz)=gl(n,\Kz)$.}
By Theorem \ref{ass.lie.J}, the algebra of $n\times n$-matrices 
with entries in $\Kz$ is a Lie algebra the commutator as Lie product; 
this Lie algebra is denoted by \idx{$gl(n,\Kz)$}. 
The center of $gl(n,\Kz)$ is easily seen to be the
$1$-dimensional subalgebra spanned by the identity matrix,
$Z(gl(n,\Kz))=\Kz 1 = \Kz$.
\end{expl}

Every subspace of a Lie algebra closed under the Lie product is again
a Lie algebra. This simple recipe provides a large number of useful
Lie algebras defined as Lie subalgebras of some $gl(n,\Kz)$. 
Conversely, the (nontrivial) {\bfi{theorem of Ado}}, 
\at{ref? e.g., Jacobsen}
not proven here but see e.g. \sca{Jacobsen} \cite{jacobsen}, states 
that every finite-dimensional Lie algebra 
is isomorphic to a  Lie subalgebra of some $gl(n,\Rz)$.

The group $GL(n,\Kz)$ is one of the most important finite-dimensional
linear groups and all finite-dimensional linear groups are isomorphic 
tosubgroups of $GL(n,\Kz)$ for some $n$. 
If $\Kz=\Rz$ or $\Kz=\Cz$ then every closed subgroup $\Gz$ of 
$GL(n,\Kz)$ is a Lie group. These Lie groups have associated 
Lie algebras $\Lz=\log\Gz$  of infinitesimal generators.
For any Lie subgroup $G$ of $GL(n,\Kz)$ one gets the Lie
algebra by looking at the vector space of those elements $X$ of
$gl(n,\Kz)$ such that
$e^{\eps X}$ is in $G$ for $\eps$ small enough. This criterion
is very useful since we can take $\epsilon$ so small that we only have
to look at the terms linear in $\epsilon$ so that we don't have to
expand the exponential series completely. If the subgroup $\Gz\subset
GL(n,\Kz)$ is connected and either compact or nilpotent, \at{defined?}
then the exponential map can be shown to be surjective, see
e.g. \sca{Knapp} \cite{knapp}.


The Lie algebra $sl(n,\Kz)$ is the Lie subalgebra of $gl(n,\Kz)$ given
by the traceless matrices. The dimension is $n^2-1$ and we have
\[
sl(n,\Kz) \cong gl(n,\Kz) / \Kz\,.
\]
The quotient is well defined and is a Lie algebra because $\Kz$
is the center and thus in particular an ideal.

If $\Lz$ is a Lie algebra over $\Rz$ then by taking the tensor product
with $\Cz$ and extending the Lie bracket in a $\Cz$-linear way, one
obtains the \bfi{complexification} of $\Lz$, denoted $\Lz^\Cz$. The
process of complexification is also called \bfi{extension of scalars}.
In particular, if we write $\Lz^\Cz=\Cz\otimes_\Rz \Lz$ then in
$\Lz^\Cz$ the Lie bracket is given by
$(\alpha\otimes x)\lp (\beta\otimes y)=\alpha\beta\otimes (x\lp y)$.
The reverse process is called \bfi{realization} or \bfi{restriction of
scalars}; we clarify the process of restriction of scalars by an
example.

\begin{example}\label{ex.real.sl}
Consider $\Lz=sl(2,\Cz)$. We wish to calculate \idx{$sl(2,\Cz)^\Rz$}. 
A basis of $sl(2,\Cz)$ is given by the elements
\[
\pmatrix{1& 0\cr 0&-1}\,,~~\pmatrix{0&1\cr0&0}\,,~~\pmatrix{0&0\cr
  1&0}\,.
\]
This basis is as well a basis for $sl(2,\Rz)$; therefore we see
$sl(2,\Cz)^\Rz\cong sl(2,\Rz) \oplus_\Rz i\,sl(2,\Rz)$ as real vector
spaces. The Lie product of $f+ig$ and $f'+ig'$ for $f,f'\in sl(2,\Rz)$ 
and $ig,ig'\in i\,sl(2,\Rz)$ is given by
\[
(f+g)\lp (f'+g')= f\lp f' - g\lp g' + i(f\lp g' +f'\lp g)\,.
\]
The reader who has already some experience with Lie algebras is
encouraged to verify the isomorphism $sl(2,\Cz)^\Rz\cong so(3,1)$.
\end{example}


\begin{expl}\label{ex.ortho}
Suppose we have a symmetric bilinear form $B$ on
$\Kz^n$. The Lie algebra $so(n,B;\Kz)$ is the subspace of all
$f\in sl(n,\Kz)$ satisfying 
\lbeq{lie.sopq}
B(fv,w)=-B(v,fw).
\eeq
We leave it to the reader to show that if $f$ and $g$ satisfy
\gzit{lie.sopq}, then so does $fg-gf$; thus we have indeed a Lie
algebra. In the special case where $B(v,w)=v^Tw$, the
Lie algebra $so(n,B;\Kz)$ is called the 
{\bfi{complex orthogonal Lie algebra}}\index{orthogonal Lie
algebra}\index{Lie algebra!orthogonal}  \idx{$so(n,\Kz)$}. In
matrix language, $so(n,\Kz)$ is the Lie algebra of antisymmetric 
matrices with entries in $\Kz$ and has dimension $n(n-1)/2$.

An {\bfi{orthogonal matrix}} is a matrix $Q$ satisfying 
\lbeq{e.orth}
Q^TQ=1.
\eeq
The orthogonal $n\times n$-matrices with coefficients in a field $\Kz$
form a subgroup of the group $GL(n,\Kz)$, the 
{\bfi{orthogonal group}} $O(n,\Kz)$\index{$O(3)$}. 
\at{this overlaps with the next example}
Since \gzit{e.orth} implies that $(\det Q)^2=1$, orthogonal matrices 
have determinant $\pm 1$.
The orthogonal matrices of determinant one form a
subgroup of $O(n,\Kz)$, the {\bfi{special  orthogonal group}}
\idx{$SO(n,\Kz)$}.
The corresponding Lie algebra is \at{why?} $so(n,\Kz)=\log O(n,\Kz) =
\log SO(n,\Kz)$, the Lie algebra of antisymmetric $n\times n$-matrices.
In particular, the group $SO(3)=SO(3,\Rz)$\index{$SO(3)$} consists of 
the rotations in 3-space and was discussed in some detail in 
Section \ref{s.rot3}.
\end{expl}

For a \bfi{nondegenerate}\index{bilinear form!nondegenerate} $B$ 
(i.e., one where $B(v,w)=0$ for all $v$ implies $w=0$) and 
$\Kz=\Cz$ (or any algebraically closed field), we can always choose 
a basis in which the bilinear form is represented as the identity 
matrix. Therefore all $so(n,B;\Kz)$ with nondegenerate $B$ are
isomorphic to  $so(n,\Kz)$.

Over $\Kz=\Rz$, symmetric bilinear forms
are classified by their signature, i.e., the triple $(p,q,r)$
consisting of the number $p$ of positive, $q$ of negative, and $r$ of
zero eigenvalues of the symmetric matrix $A$ representing the bilinear
form $B$; $B(v,w)=v^TAw$. The form $B$ is nondegenerate if and only if $r=0$.
Bilinear forms with the same signature lead to
isomorphic Lie algebras. In particular, $so(p,q)$ denotes a
Lie algebra $so(p+q,B,\Rz)$ where $B$ is a nondegenerate symmetric
bilinear form $B$ on $\Rz^n$ of signature $(p,q,0)$.
The basis can always be chosen such that the representing matrix $A$
is
\[
I_{p,q}= \pmatrix{ 1_p & 0 \cr 0&- 1_q }\,,
\]
where $1_p$ and $1_q$ are the $p\times p$ and $q\times q$ identity
matrix, respectively. In this basis, the Lie algebra \idx{$so(p,q)$} is
the subalgebra of $gl(n,\Rz)$ consisting of elements $f$ satisfying
\[
f^T I_{p,q}+I_{p,q}f=0\,.
\]
Note that if $f\in so(p,q)$ then
\[
0= \tr\left((f^T
  I_{p,q}+I_{p,q}f)I_{p,q}\right)=2\tr(fI_{p,q}^2)=2\tr(f)
\]
and hence $so(p,q)\subset sl(n,\Rz)$.

\begin{expl}\label{iso-ortho}
Let $V$ be a vector
space over a field $\Kz$. Suppose $V$ is equipped
with a symmetric or antisymmetric nondegenerate bilinear form $B$.
There is a symmetry group associated to the bilinear form consisting
of the linear transformations $Q:V\to V$ such that
\[
B(Qv,Qw)=B(v,w)
\]
for all $v,w$ in $V$. If $B$ is symmetric one calls the group of these
linear transformations an {\bfi{orthogonal group}} and denotes it by
\idx{$O(B,\Kz)$}. The associated Lie algebra is $o(B,\Kz)$. Indeed,
$e^{tf}$ transforms $x,y$ into
\[
\bary{lll}
B(e^{tf}x,e^{tf}y) &=& B(e^{tf}x,e^{tf}y)\\
&=& B((1+tf)x,(1+tf)y)+ O(t^2)\\
&=& B(x,y) +t B(fx,fy) + O(t^2).
\eary
\]
\at{but in the Lie algebra section, we used $x^TJy$ in place of 
$B(x,y)$; adapt!}
\end{expl}

\begin{expl}\label{ex.SOpq}
When $\Kz=\Rz$, one has for symmetric bilinear forms another
 subdivision, since $B$
can have a definite \idx{signature} $(p,q)$ where $p+q$ is the dimension
of $V$. If $B$ is of signature $(p,q)$, this means that there exists
a basis of $V$ in which $B$ can be represented as
\[
B(v,w) = v^T A w\,,~~~\textrm{where}~~~A =
\textrm{diag}(\underbrace{-1,\dots, -1}_{p \mbox{\scriptsize ~times}},
\underbrace{1,\dots, 1}_{q\mbox{\scriptsize~times}})\,.
\]
The group of all linear
transformations that leaves $B$ invariant is denoted by \idx{$O(p,q)$}. 
The subgroup of $O(p,q)$ of transformations with determinant one is 
the so-called \bfi{special orthogonal group}\index{orthogonal 
group!special} and is denoted by \idx{$SO(p,q)$}.
The associated real Lie algebra is denoted \idx{$so(p,q)$} and its
elements are linear transformations $A:V\to V$ such that
for all $v,w\in V$ we have $B(Av,w)+B(v,Aw)=0$. The Lie product
is given by the commutator of matrices.

The group of all translations in $V$ generates together with $SO(p,q)$
the group of {\bfi{inhomogeneous special orthogonal
transformations}}, which is denoted \idx{$ISO(p,q)$}. One can obtain
$ISO(p,q)$ from $SO(p,q+1)$ by performing a contraction; that is,
by rescaling some generators with some parameter $\epsilon$ and then
choosing a singular limit $\epsilon\to 0$ or $\epsilon \to \infty$.
\at{describe this}
The group $ISO(p,q)$ can also be seen as the group of
$(p+q+1)\times (p+q+1)$-matrices of the form
\[
\pmatrix{ Q & b \cr 0 & 1 }  ~~~ \textrm{with }~ Q\in SO(p,q)\,,~ b\in
V\,.
\]
The Lie algebra of $ISO(p,q)$ is denoted \idx{$iso(p,q)$} and can be
described as the
Lie algebra of $(p+q+1)\times (p+q+1)$-matrices of the form
\[
\pmatrix{ A & b \cr 0 & 0 }  ~~~ \textrm{with }~ A\in so(p,q)\,,~ b\in
V\,.
\]
Again, the Lie product in $iso(p,q)$ is the commutator of matrices.
\end{expl}

We define the 
{\bfi{symplectic Lie algebra}}\index{Lie algebra!symplectic}
\idx{$sp(2n,\Kz)$} as the
Lie subalgebra of $gl(2n,\Kz)$ given by the elements $f$ satisfying
\lbeq{symp.alg}
f^TJ+Jf=0\,,
\eeq
where $J$ is the $2n\times 2n$-matrix given by
\[
J=\pmatrix{ 0 & -1_n \cr 1_n& 0}\,.
\]
We leave it to the reader to verify that if $f$ and $g$ satisfy 
\gzit{symp.alg},
then so does $fg-gf$. Another useful exercise is to prove
$sl(2,\Kz)\cong sp(2,\Kz)$. (Caution: The reader is warned that
in the literature there are different notational conventions 
concerning the symplectic Lie algebras. For example, some people write 
$sp(n,\Kz)$ for what we and many others call $sp(2n,\Kz)$.)

If $B$ is antisymmetric in the example \ref{iso-ortho}, the group is 
called a {\bfi{symplectic group}} and one writes \idx{$Sp(B,\Kz)$}. 
The associated Lie algebras is $sp(B,\Kz)$.
If $V$ is of finite dimension $m$ one writes $Sp(B,\Kz)=Sp(m,\Kz)$.
Note that $m$ is necessarily even.

Other real Lie algebras that play a major role in many areas of
physics are the 
{\bfi{unitary Lie algebras}}\index{Lie algebra!unitary} and the
{\bfi{special unitary Lie algebras}}\index{Lie algebra!special 
unitary} -- called so because they are the
generating algebras of the groups of (special) unitary matrices,
a term that will be explained in Section \ref{s.Liegroups}.
The unitary Lie algebra \idx{$u(n)$} consists of all antihermitian
complex $n\times n$ matrices. The special unitary Lie algebra is
defined as the antihermitian $n\times n$ complex traceless matrices
and is denoted \idx{$su(n)$}. It is clear that $su(n)\subset u(n)$.
It might seem weird to
call a Lie algebra real if it consists of complex-valued
matrices. However, as a vector space the antihermitian complex
$n\times n$ matrices form a real vector space. If $f$ is a
antihermitian matrix, then $if$ is Hermitian. The dimension (as a
real vector space) of
$su(n)$ is $n^2-1$, and the dimension of $u(n)$ is $n^2$.
It is a good exercise to check that $so(3)\cong su(2)$ since these 
two Lie algebras will return very often. A hint: $so(3)$
consists of anti-symmetric real $3\times 3$ matrices, so there are
only three. Choosing an obvious basis for both $su(2)$ and $so(3)$
will do the job.

\begin{example}
A complex matrix $U$ is {\bfi{unitary}} if it satisfies 
\[
UU^*=1,
\]
where $(U^*)_{ij}=\bar U_{ji}$. Since the inverse of a matrix is
unique, it follows that also $U^* U=1$. By splitting all the
matrix entries into a real and imaginary part $U_{ij}=A_{ij}+iB_{ij}$
we see that the set of $n\times n$ unitary matrices makes up a
submanifold of $\Rz^{2n^2}$ of dimension $n^2 $. The linear group of
unitary $n\times n$ matrices is denoted \idx{$U(n)$}. 
\at{adapt text; part of it also holds for $SL(n)$.}
\[
U=e^A=\sum_{k=0}^{\infty}\frac{1}{k!}A^k\,.
\]
Then multiply $A$ with a parameter $t$, take $t\to 0$ and keep only
the linear terms: $U=1+tA+O(t^2)$. Then
since $U$ has to be unitary, we obtain
\[
1=(1+tA+O(t^2))(1+tA+O(t)^2)^*=1+t(A+A^*)+O(t)^2\,,
\]
implying that $A$ has to be antihermitian. 
Thus the Lie algebra of infinitesimal generators of $U(n)$ is 
\idx{$u(n)$}. 

The subgroup of $U(n)$ of all elements with determinant 1 is denoted 
by \idx{$SU(n)$} and is called the {\bfi{special unitary group}}. 
The dimension of $SU(n)$ is $n^2-1$. For the determinant we get
\[
\det  (1+tA+O(t^2))=1+\tr tA+O(t)^2\,,
\]
and thus the trace of infinitesimal generators of $SU(n)$ has to 
vanish, and we see that the corresponding Lie algebra is $su(n)$.
Note that the Lie algebra \idx{$u(n)$} 
contains all real multiples of $i\cdot 1$, which commutes with all
other elements. Hence $u(n)$ has a center, whereas $su(n)$ does not.

In the case $n=2$ it is a nice exercise to show that each special
unitary matrix $U$ can be written as
\[
U=\pmatrix{ x& y\cr -\bar y & \bar x }\,,~~~x,y\in \Cz\,,~
|x|^2+|y|^2=1\,.
\]
Writing $x=a+ib$ and $y=c+id$ for $a,b,c,d\in\Rz$ we see that 
$a^2+b^2+c^2+d^2=1$. This implies that there is
a one-to-one correspondence between $SU(2)$ and the set of points 
on the unit sphere $S^3$ in $\Rz^4$.
Thus $SU(2)$ is as a manifold homeomorphic to $S^3$. 
In particular $SU(2)$ is compact.
Hence every element $U$ of $SU(2)$ can be
written as the exponent of a matrix $A$.
\end{example}

Physicists prefer to work with Lie algebras defined by Hermitian 
matrices, corresponding to Lie $*$-algebras. \at{clarify} 
In the applications, distinguished real generators typically represent 
important real-valued observables. Therefore they tend
to replace the matrix $A$ by $iA$ for a Hermitian matrix $A$. This is
one of the reasons why the structure constants for real algebras
appear in the physics literature with an $i$, as alluded at the end of
Section \ref{sec.calc.lie}.

\section{Heisenberg algebras and Heisenberg groups}

A {\bfi{Heisenberg algebra}} is a Lie algebra $\Lz$ with a
1-dimensional center and a distinguished Lie central element 1 
called {\bfi{one}} or {\bfi{identity}}, such that every 
$f\lp g$ is a multiple of 1 for all $f,g\in\Lz$. There is an 
embedding of $\Kz$ into $\Lz$ given by $\alpha\mapsto \alpha 1$
which can be used to identify the multiples of 1 with the multipliers 
from the field, so that $\Kz=Z(\Lz)\subset \Lz$. 

When we divide out the center of a Heisenberg algebra we obtain an 
abelian Lie algebra. More generally, let $\Lz$ be any Lie algebra and
let $\Lz'$ be another Lie algebra with a subalgebra $Z$ contained in the
center of $\Lz'$. If $\Lz'/Z$ is isomorphic to $\Lz$, one calls $\Lz'$
a {\bfi{central extension}} of $\Lz$.\footnote{
In more abstract terms, central extensions are conveniently described
by short exact sequences. Let $A_i$ be a set of Lie algebras and suppose
that there are maps $d_i:A_i\to A_{i+1}$;
\begin{equation}
\begin{diagram}[heads=LaTeX]
\ldots  &\rTo & A_{i-1}&  \rTo^{d_{i-1}} & A_{i} & \rTo^{d_i}&
A_{i+1} & \rTo\\
\end{diagram}\,.
\end{equation}
We call the sequence {\bfi{exact}} if
$\textrm{Ker~} d_i = \textrm{Im~} d_{i-1}$ for all $i$ where
there are $d_{i-1}$ and $d_i$.
As an exercise, the reader is invited to verify the following
assertion: The sequence $0\to A\to B\to 0$ is exact if and only if
$A\cong B$ and the isomorphism is the map from $A$ to $B$. A
short exact sequence is a sequence of maps of the form
\[
0\rightarrow A \rightarrow B \rightarrow C \rightarrow 0\,.
\]
A {\bfi{central extension}} of $\Lz$ is then a Lie algebra $\Lz'$ 
such that there
is an exact sequence $0\to Z\to \Lz' \to \Lz \to 0$ with $Z$ abelian.
}  

Corresponding to any Heisenberg algebra there is an alternating
bilinear form $\omega:\Lz\times\Lz\rightarrow \Kz$ given by
\[
f\lp g = \omega(f,g)\,.
\]
Conversely, given such a form on an arbitrary vector space $\Vz$
not containing 1, this formula turns $\Lz:=\Kz\oplus\Vz$ into a
Heisenberg algebra. If $\omega$ is nondegenerate on $\Vz$ it defines
a \bfi{symplectic form} on $\Vz$. 

The Heisenberg algebra \idx{$h(n)$} is the special case where 
$\Kz=\Cz$, $\Vz=\Cz^{2n}$, and $\omega$ is nondegenerate. Thus $h(n)$ 
is a central extension of the abelian Lie algebra $\Cz^{2n}$ and has 
dimension $2n+1$. We can find a
basis of $\Vz$ consisting of vectors $p_k$ and $q_l$ for $1\leq k,l
\leq n$ such that $\omega(p_k,p_l)=\omega(q_k,q_l)=0$ for all $k,l$ and
$\omega(p_k,q_l)=\delta_{kl}$; that is, $\omega$ is then the standard
symplectic form on $\Kz^{2n}$ represented by the matrix
$\pmatrix{ 0 & -1 \cr 1 & 0}$. Thus Heisenberg algebras encode 
symplectic vector spaces in a Lie algebra setting. Everything done here
extends with appropriate definitions to general symplectic manifolds,
and, indeed, much of classical mechanics can be phrased in terms of
symplectic geometry, the geometry of such manifolds -- we refer the
reader to the exposition by \sca{Arnold} \cite{arnold} on 
classical mechanics and symplectic geometry. 

\begin{expl}
Let us write \idx{$t(n,\Kz)$} for the Lie subalgebra of
$gl(n,\Kz)$ consisting of upper-triangular matrices and
$n(n,\Kz)$ as the Lie subalgebra of $gl(n,\Kz)$
consisting of strictly upper-triangular matrices, which have zeros 
on the diagonal.

The Lie algebra $t(3,\Kz)$ of strictly upper triangular
$3\times 3$-matrices is a Heisenberg algebra with 
\[
1=\pmatrix{ 0&0&1\cr 0&0&0\cr 0&0&0 }\,,
\]
since
\[
\pmatrix{ 0 & \alpha & \gamma \cr 0 & 0 & \beta \cr0&0&0 } \lp
\pmatrix{ 0 & \alpha' & \gamma' \cr 0 & 0 & \beta'\cr 0&0&0 } =
\pmatrix{ 0 & 0 & \alpha\gamma'-\gamma\alpha' \cr 0 & 0 & 0 \cr
0&0&0 } = \alpha\gamma'-\gamma\alpha'.
\]
The Lie algebra $t(3,\Cz)$ is called \bfi{the \idx{Heisenberg algebra}};
thus if one talks about ''the'' (rather than ''a'') Heisenberg algebra,
this Lie algebra is meant and is denoted $h(1)$. 
Introducing names for the special matrices
\[
p:=\pmatrix{ 0&1&0\cr 0&0&0\cr 0&0&0},
 ~~~
q:=\pmatrix{ 0&0&0\cr 0&0&1\cr 0&0&0 },
\]
we find that $p,q$ and 1 form a basis of $t(3,\Cz)$, and we can
express the Lie product in the more compact form
\lbeq{e.heislp}
(\alpha p+\beta q+\gamma) \lp (\alpha' p+\beta' q+\gamma')
= \alpha\beta' - \beta\alpha'.
\eeq
Defining 
\[
(\alpha p+\beta q+\gamma)^*:=\ol\alpha p+\ol\beta q+\ol\gamma
\]
turns the Heisenberg algebra into a Lie $*$-algebra in which 
$p$ and $q$ are Hermitian. 
\at{definition of Hermitian in a Lie $*$-algebra? Real is better!}
Note that here $*$ is {\em not} the conjugate transposition of matrices!

\gzit{e.heislp} implies that $p$ and $q$ satisfy the so-called 
{\bfi{canonical commutation relations}}
\lbeq{e.CCR} 
p \lp q =1\,,~~~ p \lp p=q\lp q=0. 
\eeq
Since $f \lp 1 =0$ when $1$ is Lie central, \gzit{e.CCR}
completely specifies the Lie product. The canonical commutation
relations are frequently found in textbooks on quantum mechanics,
but we see that they just characterize the Heisenberg algebra.

The notation $q$ and $p$ is chosen to remind of position of momentum.
Indeed, the canonical commutation relations arise naturally in 
classical mechanics. In the Lie algebra $C^\infty(\Rz\times \Rz)$ 
constructed in Theorem \ref{t.1dpoisson},
we consider the set of {\bfi{affine functions}}, that is, those that
are of the form $f(p,q) = \alpha_f p + \beta_f q+\gamma_f$, with
$\alpha_f,\beta_f,\gamma_f\in\Cz$. In particular, the constant functions
are included with $\alpha_f=\beta_f=0$, and we identify them with
the constants $\gamma_f\in\Cz$. Given another affine function
$g(p,q)=\alpha_gp+\beta_gq+\gamma_g$, we find
\[
f\lp g = \alpha_f\beta_g - \beta_f\alpha_g \in\Cz\,.
\]
Since $f\lp g$ is just a complex number times the function that is
$1$ everywhere, it is a central element, that is, it Lie
commutes with all other algebra elements. Thus the affine functions
form a Heisenberg subalgebra of $C^\infty(\Rz\times \Rz)$, and $p$ and 
$q$ satisfy the canonical commutation relations.
\end{expl}

\at{move to Poisson algebras!}
Suppose that a commutative Poisson algebra $\Ez$ 
contains two elements 
$p$ and $q$ satisfying the canonical commutation relations \gzit{e.CCR}.
Then $\Ez$ contains a copy of the Heisenberg algebra. 
The algebra of polynomials
in $p$ and $q$ is then a Poisson subalgebra of $\Ez$ in which
\gzit{e.Xf} is valid. This follows from Proposition \ref{der.bracket}.
Thus the canonical commutation relations capture the essence of the
commutative Poisson algebra $C^\infty(\Rz\times \Rz)$.
But getting the bigger algebra requires taking limits which
need not exist in $\Ez$, since with polynomials alone, one does not get
all functions.

\at{Define the \bfi{oscillator algebra} $os(n)$.}

\at{interlace this with the algebra part}

\begin{expl}
An upper triangular $n\times n$-matrix
is called {\bfi{unit upper triangular}} if its elements on
the diagonal are $1$, and {\bfi{strictly upper triangular}} if its
elements on the diagonal are zero.  It is straightforward to check
that the unit upper triangular $n\times n$-matrices form a subgroup
\idx{$T(n,\Kz)$} of the group $GL(n,\Kz)$, and the strictly
upper triangular $n\times n$-matrices form a Lie subalgebra of
$gl(n,\Kz)$, which we denote by \idx{$t(n,\Kz)$}.
We have $t(n,\Kz)=\log T(n,\Kz)$. In the following we shall look more
closely at the case $n=3$ which is especially important.
\end{expl}

\at{we need $H(n)$; otherwise the section should be called 
''The....''}

The {\bfi{Heisenberg group}} is the group
\lbeq{heis.grp}
  T(3,\Cz)=\Big\{ \pmatrix{ 1 & a & c \cr 0 & 1 & b \cr 0&0&1 } \,
\Big|\,  a,b,c\in \Cz \Big\}
\eeq
of unit upper triangular matrices in
$\Cz^{3\times 3}$; its corresponding Lie algebra is the Heisenberg
algebra $t(3,\Cz)$. Since the Heisenberg group is defined in terms of
matrices, it comes immediately with a representation, the defining
representation. Note that the defining representation is not unitary.
\at{group representations not yet defined.}

The relation between the Heisenberg algebra and the Heisenberg
group is particularly simple since the exponential map
has a simple form. Indeed, if $A\in \Cz^{n\times n}$ then
\lbeq{exp.mat}
e^A = \sum_{k=0}^{\infty}\frac{A^k}{k!}\,,
\eeq
where $A^0=1$ is the identity matrix and the series \gzit{exp.mat}
is absolutely convergent. A note on the infinite-dimensional case: 
For linear operators $A$ on a Hilbert space $\Hz$, the series 
converges absolutely only when $A$ is bounded (and hence
everywhere defined); for unbounded but self-adjoint $A$ (which are
only densely defined), convergence holds in a weaker sense giving
\lbeq{exp.mat2}
e^A \psi= \sum_{k=0}^{\infty}\frac{A^k}{k!} \psi
\eeq
for a dense set of vectors $\psi\in\Hz$ that are analytic for $A$.

If $A\in t(3,\Cz)$, then a direct calculation shows that $A^2$ is of
the form
\[
\pmatrix{ 0&0&c\cr 0&0&0\cr 0&0&0}
\]
for some $c\in \Cz$. Hence $A^3=0$ and the exponential of $A$ is
simply given by $ e^A = 1+A+\shalf A^2 $. Thus if $A$ is given by
\[
A=\pmatrix{ 0 & \alpha & \gamma \cr 0 & 0 & \beta \cr 0&0&0 }\,,
\]
the exponential is given by
\[
e^A = \pmatrix{ 1 & \alpha & \gamma+\shalf \alpha\beta \cr 0 & 1 &
\beta \cr 0&0&1  }\,.
\]
The map $A\to e^A$ is clearly bijective. The inverse map is given
 by the logarithm, which is
for matrices defined by
\lbeq{log-taylor}
\log(1 + X) =
\sum_{k=1}^{\infty}\frac{(-1)^{k-1}}{k}X^k\,,
\eeq
so that for the Heisenberg group $\Gz$ we have
\[
\log(X) = (X-1) - \shalf (X-1)^2 = -2+ 2X - \shalf X^2 \,.
\]
We are thus in the situation that both $T(3,\Cz) = \exp t(3,\Cz)$
and $t(3,\Cz) =\log T(3,\Cz)$. This is not special to the Heisenberg
group, neither does
it hold in general. But there is a class of groups for which
this holds. For example, the exponential map is surjective for all
connected Lie groups that are compact or nilpotent (see below), 
see, e.g., \sca{Helgason} \cite{helgason} or
\sca{Knapp} \cite{knapp}. 
The Heisenberg group is a noncompact but nilpotent Lie group.

Let us shortly repeat what it means when a group is \idx{nilpotent}. 
Given any group $G$, we can form the \idx{commutator subgroup} 
$G^{(1)}$, which is generated by all elements of the form 
$aba^{-1}b^{-1}$ for all $a,b\in G$. We can also consider 
the commutator subgroup of $G^{(1)}$ and denote it by $G^{(2)}$. 
Repeating this procedure we get a sequence of groups
\[
G\supseteq G^{(1)}\supseteq G^{(2)}\supseteq ...
\]
A group is nilpotent if the procedure ends in a finite number of
steps with the trivial group $G^{(n)}=1$. It is easy to see that the 
Heisenberg group is two-step nilpotent since $G^{(2)}=1$.

Since the exponential map is bijective for the Heisenberg group,
there exists a binary operation $\oplus$ on $t(3,\Cz)$, where 
$A\oplus B$ is the element with
\lbeq{e.oplus}
e^A e^{B} = e^{A\oplus B}\,.
\eeq
It is not difficult to give an explicit formula for $A\oplus B$.
Since $A$ and $B$ are strictly upper triangular, we have $A^p B^q=0$
for $p+q\geq 3$. We thus have
\[
e^Ae^B = (1+A+\shalf A^2) (1+B+\shalf B^2) 
= 1+A+B +\shalf (A^2+B^2+2AB)\,.
\]
Applying \gzit{log-taylor} we find
\beqar A\oplus B &=& \log\left(
1+A+B +\shalf
(A^2+B^2+2AB)\right)\nonumber\\
&=& A+B +\shalf (AB-BA)\,,\nonumber
\eeqar
hence
\lbeq{e.weyl0}
 A\oplus B=A+B +\shalf A\lp B \,.
\eeq
Thus we get from \gzit{e.oplus} the formula
$e^Ae^B=e^{A+B +\shalf A\lp B}$. Since $A\lp B$ is central, it behaves 
just like a complex number, and we find
the {\bfi{Weyl relations}}
\lbeq{e.weyl1}
e^{A+B}=e^{-\half A\slp B}e^Ae^B\,.
\eeq
In fact this result is also a direct consequence of the famous
(but much less elementary) \bfi{Baker--Campbell--Hausdorff (BCH)
formula}\index{Baker--Campbell--Hausdorff formula}
that gives for general matrix Lie groups a series expansion of
$A\oplus B$ when $A$ and $B$ are not too large.
Even more generally, the \idx{Baker--Campbell--Hausdorff
formula} applies to abstract finite-dimensional Lie groups\footnote{
In infinite dimensions, additional assumptions are needed for the
BCH-formula to hold.
}
that are not necessarily matrix groups and says that for two fixed
Lie algebra elements $A$ and $B$ and for small enough real numbers $s$
and $t$ there is a function $C$ from $\Rz\times \Rz$ to the Lie algebra
such that we have
\[
e^{sA}e^{tB} = e^{C(s,t)}\,.
\]
The function $C(s,t)$ is given by a (for small $s,t$ absolutely
convergent) infinite power series, the first terms of
which are given by
\[
C(s,t)= sA+tB +\frac{st}{2} A\lp B + \frac{st}{12}\left(s A\lp
(A\lp B) - t B\lp (A\lp B)\right)+\ldots.
\]
In fact, this series expansion may be derived from a closed form 
integral expression.

The Baker--Campbell--Hausdorff formula is of great importance in
both pure and applied mathematics. It gives (where it applies; in
particular in finite dimensions)
the relation of a Lie group with the associated Lie algebra. It for
example says that the product of $e^A$ and $e^B$ for some $A$ and $B$
in the Lie algebra is again an element of the form $e^C$ with $C$ in
the Lie algebra. Hence the exponents of the Lie algebra generate a
subgroup of the corresponding Lie group.

For infinite-dimensional Lie algebras and groups, one has to
use a refined argument centering around the Hille--Yosida theorem.
Let $U(t)$ denote a \at{merge with linear Lie group part. = motion?}
one-parameter group of linear operators on a Hilbert space $\Hz$
such that $t\to U(t)$ is {\bfi{strongly continuous}}, which
means that $t\to U(t)\varphi$ is continuous for all
$\varphi\in\Hz$. Then we can differentiate $U(t)$ to obtain the
strong limit
\[
A= \lim_{t\rightarrow 0}\frac{U(t)-U(0)}{t}\,.
\]
The object $A$ is called {\bfi{infinitesimal
generator}}\index{generator!infinitesimal} of the
one-parameter group $U(t)$. It turns out that $A$ is a closed linear
operator that is defined on a dense subspace in $\Hz$. The Hille--Yosida
theorem gives a necessary and sufficient condition \at{state these
and give a reference! Define selff-adjoint}
for a closed linear
operator $A$ to be the infinitesimal generator of some strongly
continuous one-parameter semigroup
\[
U(t) =  e^{tA} \,,
\]
since in general one might not get a group.
The Hille--Yosida theorem is very useful for analyzing the solvability
of linear differential equations
\[
\frac{d}{dt}\psi(t)= A\psi(t)\,, \ \ \ \psi(0)=\psi_0\,,
\]
examples of which are the Schr\"odinger equation or the heat equation.
If the conditions of the Hille--Yosida theorem hold
for $A$, the solution to this initial value problem takes the form
\[
\psi(t)= e^{tA}\psi(0)\,.
\]
For the (hyperbolic, conservative) Schr\"odinger equation,
$A=-\frac{i}{\hbar}H$ with a self-adjoint Hamiltonian $H$,
the solution exists for all $t$, and the $U(t)$ form a one-parameter
group.
For the (parabolic, dissipative) heat equation, $A=k\Delta$ is a
positive multiple of the Laplacian 
$\Delta = \partial_x^2+\partial_y^2+\partial_z^2$, the solution exists 
only for
$t\ge 0$, and we only get a semigroup.

\section{Lie $*$-algebras}

Many Lie algebras of interest in physics have an additional structure:
an adjoint mapping compatible with the Lie product.

\begin{definition}\label{d.liester}
A {\bfi{Lie $*$-algebra}} is a Lie algebra $\Lz$ over $\Cz$
with a distinguished
element $1\ne 0$ called {\bfi{one}} and a mapping
$*$ that assigns to every $f\in\Lz$ an {\bfi{adjoint}} $f^*\in\Lz$
such that
\[
f\lp 1 =0,
\]
\[
(f+g)^*= f^*+g^*,~~~
(f\lp g)^* = f^*\lp g^*,
\]
\[
f^{**}=f,~~~
(\lambda f)^* = \bar \lambda f^*,~~~
1^*=1
\]
for all $f,g\in \Lz$ and $\lambda\in \Cz$. We identify the multiples
of 1 with the corresponding complex numbers.
\end{definition}

The reason why we include the 1 into the definition of a Lie 
$*$-algebra is that many physically relevant
Lie algebras are equipped with a distinguished central
element\footnote{Many such Lie algebras are realized most naturally
as central extensions of semisimple Lie algebras, corresponding to
projective representations of semisimple Lie algebras.
By including the 1 automatically we work directly in the central
extension, and avoid the cohomological technicalities
associated with the formal discussion of central extensions and
projective representations.}.
But the presence of $1$ is not a restriction, since one can always 
adjoin a central element $1$ to a Lie algebra $\Lz'$ without nonzero 
central element and form the direct sum $\Lz=\Lz'\oplus \Kz$.

An important Lie $*$-algebra for nonrelativistic quantum mechanics
is the algebra $\Ez=\Lin\Hz$ of linear operators of a Euclidean space 
$\Hz$ (usually a dense subspace of a Hilbert space $\ol \Hz$).
The relevant Lie product is defined by Theorem \ref{ass.lie.J} with 
the choice
\[
J:=\frac{i}{\hbar}=\frac{i}{\hbar} 1_\Hz \in\Lin\Hz,
\]
where $1_\Hz$ is the identity operator on $\Hz$, and the conjugate 
of $f\in \Ez$ is given by the \bfi{adjoint} of $f$, defined as the
linear mapping $f^*$ satisfying $\phi^* f^*\psi=(f\phi)^*\psi$
for all $\phi,\psi\in\Hz$. Dropping the index $J$ in the Lie product
of Theorem \ref{ass.lie.J}, we get the {\bfi{quantum Lie product}}
\lbeq{e.liequant}
f\lp g = \frac{i}{\hbar}(fg-gf) =\frac{i}{\hbar}[f,g]
\eeq
of $f,g\in\Lin \Hz$, already familiar from \gzit{e.lp}.
Note that the axioms require the purely imaginary factor in this
formula, whereas the value of Planck's constant $\hbar$ is arbitrary
from a purely mathematical point of view.
In quantum field theory, a different choice of $J$
is sometimes more appropriate.

For any Lie $*$-algebra, the set
\[
\re \Lz :=\{f\in\Lz\mid f^*=f\}
\]
is a Lie algebra over $\Rz$. When describing symmetries, physicists
often work with Lie algebras over the reals; the present Lie 
$*$-algebras are then the complexifications of these real algebras, 
with a central element 1 adjoined if necessary.

The {\bfi{complexification}}  of a real Lie algebra $\Lz$ is the 
Lie $*$-algebra \idx{$\Cz\Lz$} defined as follows. In case that 
a complex scalar multiplication is already defined on $\Lz$, one first 
replaces $\Lz$ by an isomorphic Lie algebra in which $if\not\in \Lz$ 
if $f\in\Lz$ is nonzero. Then one defines 
\[
\Cz\Lz=\Lz\oplus i\Lz, 
\]
extending scalar multiplication in a natural way to the complex field. 
That is, any element $f\in\Cz\Lz$ is of the form
\[
f= f_1 + i f_2
\]
with $f_1,f_2\in\Lz$, and one defines 
\[
\alpha(\beta f):=(\alpha\beta)f,~~~
\alpha f\lp \beta g:= (\alpha\beta)f\lp g
\]
for all $f,g\in \Lz$ and $\alpha,\beta\in\Cz$. Conjugation is
defined as 
\[
(f_1 + i f_2)^*:=f_1 - i f_2\for f_1,f_2\in\Lz;
\]
The axioms for a Lie $*$-algebra are easily established if $1\in\Lz$.
Note that the real dimension of $\Lz$ equals the complex dimension of
$\Cz\Lz$. It is easy to check that 
\[ 
\re \Cz\Lz \cong \Lz.
\]
Conversely, for a Lie $*$-algebra $\Lz$,
\[ 
\Cz \re \Lz \cong \Lz.
\]
If a complex Lie algebra $\Lz'$ is isomorphic to $\Cz\Lz$ as a Lie 
algebra, one says that $\Lz$ is a {\bfi{real form}} of the complex 
Lie algebra $\Lz'$. 

We leave it as an exercise to verify $\Cz su(n)\cong
sl(n,\Cz)$ and $\Cz so(p,q) =so(p+q,\Cz)$. 
\at{describe more generally all classical $*$-algebras!}
In general, a complex Lie
algebra has more than one real form as we can see since for $p\neq
q,n-q$ the Lie algebras $so(p,n-p)$ and $so(q,n-q)$ are not isomorphic.

An {\bfi{involutive Lie algebra}}\index{Lie algebra!involutive} 
(\sca{Neeb}
\cite{neeb}) is a Lie algebra
$\tilde \Lz$ with Lie product $[\cdot,\cdot]$ and with an involutive,
antilinear anti-automorphism $\sigma$, i.e., a mapping 
$\sigma:\tilde \Lz\to\tilde \Lz$ satisfying 
\[
\sigma(\alpha f) = \alpha^* \sigma f,~~~
\sigma(f\lp g) = \sigma g \lp \sigma f
\]
for $\alpha\in\Cz,f,g\in\tilde \Lz$.
Associated to an involutive Lie algebra $\tilde \Lz$ is
the \bfi{real form} $\tilde \Lz_\Rz=\Big\{ x\in \tilde \Lz\mid \sigma x
  = -x\Big\}$. Our definition of a Lie $*$-algebra is closely
related and obtained as follows, after adjoining to $\tilde \Lz$
a central element 1 if necessary. We define $\Lz$ as the vector space
$\tilde \Lz$ equipped with the Lie product $\lp$ defined by 
$x\lp y = \frac{i}{\hbar} [x,y]$. 
Then the mapping $x\mapsto i\hbar x$, with $\hbar$ a positive
real constant (in physical applications Planck's constant) is an
isomorphism of Lie algebras. The map $\sigma$ induces the
conjugation $x^* = -\sigma x$ and $\re \Lz = \tilde \Lz_\Rz$.

\begin{rems}
~\at{give here the example of $so(3)$}\\
(i) The nomenclature of Lie $*$-algebras is a bit tricky. If $\Lz$ is a
Lie $*$-algebra, we therefore denote it (usually) with the name of the 
real Lie algebra $\re \Lz$. To avoid confusion, 
it is important to keep track of whether we are discussing real Lie
algebras, complex Lie algebras or Lie $*$-algebras.

(ii) In the physics literature, one often sees the defining relations 
\gzit{phys.struct} for real Lie algebras written in terms of complex 
structure constants,
\[
X_j \lp X_k =\sum_l ic_{jkl}X_l\,.
\]
where $i=\sqrt{-1}$ and the $c_{jkl}$ are real. That is, the Lie 
product takes values outside of the real Lie algebra! 
What is done by the physicists is that -- as in the above definition 
of a Lie $*$-algebra from an involutive Lie algebra -- they
multiply all elements in the Lie algebra by $i$. The reasons for
making this seemingly difficult construction mainly has historical
reasons. One is that in some real algebras the elements are
antihermitian matrices. By multiplying with $i$ one obtains Hermitian 
matrices and in quantum mechanics, observable quantities are represented
as Hermitian operators. 
\end{rems}

The converse process of complexification is {\bfi{realization}}. 
Given a complex Lie algebra $\Lz$, one regards it as a real Lie 
algebra  \idx{$\Lz^\Rz$} by restricting scalar multiplication to  
real factors. Since $f$ and the imaginary scalar multiple $if$ are 
linearly independent over $\Rz$,
the real dimension of $\Lz^\Rz$ is twice the complex dimension of $\Lz$.
In the finite-dimensional case, a convenient way to obtain the 
realization is as follows: 
Choose a basis $t_1,\ldots,t_n$ of $\Lz$ and then form
the elements $s_j=it_j$ for all $j$. All real linear combinations of
$s_j$ and $t_j$ make up $\Lz^\Rz$. Given two elements $f,g$ in
$\Lz^\Rz$ one calculates their Lie product as if they were elements 
of $\Lz$; the result can be written as
\[
f\lp g =\sum (\alpha_j +i\beta_j)t_j\,.
\]
The Lie product of $f$ and $g$ in $\Lz$ is then defined as
\[
f\lp g =\sum \alpha_jt_j +i \sum\beta_j s_j\,.
\]
See also Example \ref{ex.real.sl}. \at{\and/or the $so(3)$ example}

\chapter{Mechanics in Poisson algebras}\label{c.ppoisson}

This chapter brings more physics into play by introducing Poisson
algebras, i.e., associative algebras with a compatible Lie algebra 
structure. These are 
the algebras in which it is possible to define Hamiltonian
mechanics. Poisson algebras abstract the algebraic features of both
Poisson brackets and commutators, and hence serve as a unifying tool
relating classical and quantum mechanics.
In particular, we discuss classical Poisson algebras for oscillating 
and rotating systems. 

\at{Although you've defined Poisson algebras earlier in this chapter,
you still given no hint of how/why this extra structure is useful.\\
Based on a Poisson algebra, a general physical system can be
characterized in terms of a Poisson representation 
\at{put definition here?}
of the kinematical Lie algebra of distinguished
quantities of interest, a Hamiltonian, a distinguished Hermitian
quantity in the Poisson algebra defining the dynamics, and a state
defining a particular system at a particular time. The same abstract
concept of a state has different realization in the classical
(commutative) and quantum (noncommutative) case.}

\section{Poisson algebras}\label{s.poisson}

Many algebras that we will encounter have both an associative product
and a Lie product, which are compatible in a certain sense. Such
algebras are \bfi{Poisson algebras}, our definition of which is
the noncommutative version discussed, e.g., in 
\sca{Farkas \& G. Letzter} \cite{farkasletzter}. (In contrast, in 
classical mechanics on Poisson manifolds, one usually assumes Poisson 
algebras to be always commutative.)

\begin{definition}
A {\bfi{Poisson algebra}} $\Ez$ is a Lie algebra with an associative
and distributive \bfi{multiplication} which associates with $f,g\in\Ez$ 
its \bfi{product} $fg$, and an \bfi{identity} 1 with respect to 
multiplication, such that the compatibility condition
\lbeq{poissonalg}
f \lp (gh) = (f\lp g)h + g(f\lp h)
\eeq
holds.
Equation \gzit{poissonalg} is also called the {\bfi{Leibniz
identity}}.

In expressions involving the associative product and the
Lie product, the binding of the associative product is stronger than
the Lie product, i.e., $f\lp gh$ is interpreted as $f\lp (gh)$,
and $fg \lp h$ as $(fg)\lp h$.
\end{definition}

\begin{rems}
Since Poisson algebras have two products, neither of which is
assumed to be commutative, we reserve the notation \idx{$[f,g]$} 
for the {\bfi{commutator}} 
\[
[f,g]:=fg-gf
\] 
with respect to the associative product.
If $[f,g]=0$ we say that $f$ and $g$ {\bfi{commute}}. If $f\lp g=0$ 
we say that $f$ and $g$ {\bfi{Lie commute}}. An element which 
commutes (Lie commutes) with every element in $\Ez$ is called 
\bfi{central (Lie central)}\index{central}.
\end{rems}

\at{add domains and ranges of $f,g,...$ in examples, definitions, 
etc.}

\begin{example}
We take $C^\infty(\Rz\times \Rz)$ where the associative product is
given by ordinary multiplication of functions, and where the Lie
product is given by $f\lp g = f_p g_q - f_q g_p$. To see that the
Leibniz condition is satisfied we write
\beqar f\lp gh
&=& f_p (gh)_q - f_q (gh)_p \nonumber \\
&=& f_p g_q h + f_p g h_q - f_q g_p h - f_q g h_p
\nonumber\\
&=& (f\lp g)h + g(f\lp h)\nonumber\,.\eeqar Thus
$C^\infty(\Rz\times\Rz)$ is a commutative Poisson algebra.
\end{example}

\begin{example}
For a Euclidean space $\Hz$ we consider the space $\Lin \Hz$ of
continuous linear operators on $\Hz$. The Lie product is given by
\[
f\lp g = \frac{i}{\hbar}[ f,g] = \frac{i}{\hbar}(fg -gf)\,.
\]
We have \beqar f\lp gh &=& \frac{i}{\hbar}\Big(fgh -
ghf\Big)\nonumber\\
&=&\frac{i}{\hbar}\Big( fgh - gfh +gfh - ghf\Big)\nonumber \\
&=&\frac{i}{\hbar}\Big( [f,g]h + g[f,h]\Big)\nonumber\,.
 \eeqar
Hence $\Lin \Hz$ is a non-commutative Poisson algebra. In particular, 
taking $\Hz=\Cz^n$, we find that $\Cz^{n\times n}$ is a non-commutative 
Poisson algebra.
\end{example}

These examples are prototypical for the application in physics. Indeed,
the Poisson algebras relevant for classical mechanics are commutative 
and are defined via differential operators, while the Poisson algebras 
relevant for qunatum mechanics are noncommutative and typically 
defined via a commutator. 
\at{or its generalization in Theorem \ref{ass.lie.J}?}

We note some immediated consequences of the axioms, which show that 
the Lie product has close similarities with differentiation.

\begin{prop}\label{der.bracket}
Let $\Ez$ be a Poisson algebra. Then
\[
f\lp 1 = 0,
\]
and
\[
f\lp g^n = n g^{n-1}(f\lp g) ~~~\mbox{if~~}[f\lp g,g]=0\,.
\]
\end{prop}
\bepf We first take $n=0$ and calculate
\[
f\lp 1 = f\lp (1\cdot 1) = (f\lp 1)1 + 1(f\lp 1)= 2(f\lp 1)\,,
\]
from which it follows that $f\lp 1 = 0$. Let us therefore suppose that
the proposition is true
for all $k$ with $0\leq k\leq n$, then for $k=n+1$ we have
\[
f\lp (g^{n+1}) = (f\lp g^n)g+g^n(f\lp g) 
= ng^{n-1}(f\lp g)g + g^n (f\lp g)
= ng^{n}(f\lp g)\,.
\]
\epf

\begin{definition}
A {\bfi{Poisson $*$-algebra}} is a Poisson algebra that as a Lie
algebra is a Lie $*$-algebra (defined in Definition \ref{d.liester})
satisfying the additional rule
\[
(fg)^* = g^* f^*\,.
\]
\end{definition}

Note the change of order in $(fg)^* = g^*f^*$, while the ordering of 
the Lie product is preserved under the involution $*$.

\begin{example}
The commutative Poisson algebra $C^\infty(\Rz\times \Rz)$ is made into a
Poisson $*$-algebra by defining 
\[
f^*(p,q):=\ol{f(p,q)}. 
\]
We have $(f^*)_p(p,q) = \ol{f_p(p,q)}$,
and
\[
(fg)^*(p,q)=\ol{fg}(p,q)=\ol{f(p,q)}\;
\ol{g(p,q)}=(f^*g^*)(p,q),
\]
hence $(fg)^*=f^*g^*=g^*f^*$ since the algebra is commutative.
From these considerations it follows immediately that
$C^\infty(\Rz\times \Rz)$ is a Poisson $*$-algebra.
\end{example}

\begin{example}
We make $\Lin\Hz$ with the quantum Lie product \gzit{e.liequant}
into a non-commutative Poisson $*$-algebra by defining $A^*$
to be the adjoint conjugate transpose of $A$, which is defined as the
linear operator $A^*$ such that
\[
\langle A\phi,\psi\rangle=\langle\phi,A^*\psi\rangle\,, \quad
\Forall \phi,\psi\in \Hz\,,
\]
where $\langle\cdot,\cdot\rangle$ denotes the inner product on $\Hz$.
In particular, if $\Hz=\Cz^n$
then $\<\phi,\psi\>=\phi^*\psi$ and $A^*$ is the conjugate transpose
of the matrix $A\in\Cz^{n\times n}$.
For general $\Hz$, we have
\[
\langle AB \phi,\psi\rangle = \langle B\phi,A^*\psi\rangle =
\langle \phi, B^*A^* \psi\rangle\,,
\]
from which we read off that $(AB)^* = B^*A^*$. Then it follows
that
\beqar
A\lp B &=& \Big(\frac{i}{\hbar}[A,B]\Big)^*
= -\frac{i}{\hbar}((AB)^* - (BA)^*)=-\frac{i}{\hbar}(B^*A^*-A^*B^*)
\nonumber\\
&=& \frac{i}{\hbar}(A^*B^* - B^*A^* )= \frac{i}{\hbar}[A^*,B^*]
= A^*\lp B^*
\nonumber\,.
\eeqar
Hence $\Lin\Hz$ is a Poisson $*$-algebra.
\end{example}

\at{add the notion of a 
\bfi{canonical transformation}, a 1-parameter group of linear mappings 
$U(t)$ satisfying 
\[
U(0)=1,~~~\frac{d}{dt}\Big(U(t)f\Big) = H(t) \lp U(t)f
\]
for 
some generating family $H(t)$ of Hamiltonians, or also the element 
$U(1)$ of such a 1-parameter group. Mention the invariance of 
Hamiltonian dynamics under canonical transformations, as a consequence 
of the fact that these are Lie automorphisms.}

\section{Rotating rigid bodies}\label{s.rigid}

The spinning top is the classical model of a spinning particle.
Like a football, the top can be slightly deformed but when the
external force is released it jumps quickly back to its
equilibrium state. Molecular versions of a football are the
fullerenes, the most football-like fullerene being a molecule with 60
carbon atoms arranged in precisely the same manner as the vertices
that can be seen in the corners between the patches on the surface of
an official football. In a reasonable approximation, the deformability 
can be neglected; the spinning top, and also the fullerene soccer ball, 
is most often treated as a rigid body.

The spinning top is treated in most undergraduate courses in mechanics;
hence there is a rich literature on the topic.\footnote{
Good accounts of the standard approach can be found, e.g., in 
\sca{Arnold} \cite{arnold}, \sca{Marion \& Thornton} \cite{thornton},
 or \sca{Goldstein} \cite{goldstein}.}
Due to the abundance of classical treatments of the spinning top we 
pursue here a nonstandard approach based on Poisson algebras, which 
shows how it is a special prototypical case of a uniform algebraic 
approach to mechanical systems.

\bigskip
A rigid body can be moving as a whole, that is, its center of mass
can have a nonzero velocity, but changing to comoving coordinates
via a time-dependent translation, one may assume that the center of
mass is not moving. The coordinate system in which the center of
mass of the rigid body is fixed is in physics literature called
the {\bfi{center of mass coordinate system}}. Without loss of 
generality we then assume the center of mass is at the origin $(0,0,0)$.

Having fixed the center of mass the rigid body can still rotate,
but after rotating the coordinate system to the body-fixed one,
no freedom is left. This means that the {\bfi{pose}} of a rigid 
body with fixed center of mass is completely described by a rotation 
$Q(t) \in SO(3)$\index{$SO(3)$}.

\at{some material appeared already earlier; 
the following is now icn Chapter 1}

Thus $Q(t)$ satisfies $Q(t)Q(t)^T = Q(t)^TQ(t) = 1$ and $\det Q(t)=1$.
Differentiating we get
\[
\dot Q(t) Q(t)^T + Q(t)\dot Q(t)^T = 0\,.
\]
Calling $\Omega(t) = \dot Q(t)Q(t)^T=\dot Q(t)Q(t)^{-1}$ we thus have
\[
\Omega(t)^T = -\Omega(t)\,,
\]
that is $\Omega$ is antisymmetric. We can therefore parameterize
$\Omega$ as
\[
\Omega = \pmatrix{ 0 & -\omega_3 & \omega_2 \cr \omega_3 & 0 &
-\omega_1 \cr -\omega_2 & \omega_1 & 0 }\,.
\]
We then have $\Omega \vv = \omega\times \vv$, where $\omega$ is the
vector $(\omega_1,\omega_2,\omega_3)^T$. We view $\Omega(t)$ as a
matrix $X(\omega(t))$ depending on the vector $\omega(t)$, called
the {\bfi{angular velocity}}.

A rigid body in a conserved system has an energy that can depend
on the position determined by $Q(t)$ and the velocity $\dot Q(t)$.
Since $\dot Q = \Omega Q^{-T}=\Omega Q = X(\omega)Q$, the energy
thus depends on $Q$ and $\omega$: the Hamiltonian $H=H(Q,\omega)$
is a function of $Q$ and $\omega$.

For a {\bfi{freely rotating body}}, the Hamiltonian only depends on
the kinetic energy and is quadratic in the angular velocity;
\[
H =H(\omega) = \frac{1}{2}\omega^T I\omega\,,
\]
and we can always take $I$ symmetric, $I = I^T$.
The $3\times 3$-matrix $I$, called the
{\bfi{tensor of moments of inertia}}\index{moment of inertia}, 
or just {\bfi{inertia tensor}}, has the meaning of an 
{\bfi{angular mass matrix}} analogous to the mass
matrix $M$ given in Chapter \ref{c.oscillating} for the case of an oscillating
particle, where the kinetic energy was given by
$H=\frac{1}{2}\vv^TM\vv$.
The reason why it is called a tensor and not a matrix is because $I$
is in fact a bilinear form.\footnote{The same holds for the mass matrix 
-- but there the terminology has become traditional.}
Under a coordinate change $I$ does
transform as a bilinear form and not as a matrix.  Indeed, 
under the change of coordinates $\omega \mapsto \tilde Q\omega$ for some
$\tilde Q\in SO(3)$, the Hamiltonian is invariant and thus
$I$ transforms as $I\mapsto Q^{-T}IQ^{-1}$, that is, by a congruence
transformation. In contrast, a matrix $A$ transforms as
$A\mapsto QAQ^{-1}$, which is a similarity transformation.
By a coordinate change $I$ can be made diagonal, so that we may assume 
that
\[
H = \frac{1}{2} \sum_{k=1}^{3} I_k \omega_{k}^{2}\,.
\]
The coefficients $I_k$ are called the {\bfi{principal moments of
inertia}}. To have a Hamiltonian that is bounded from below we
require $I_i\geq 0$. In practice one has $I_k>0$ for all $k=1,2,3$;
then $I$ is invertible.

In analogy to the linear momentum
$p=Mv=\frac{\partial H}{\partial v}$ for an oscillating particle
with kinetic energy $H=\frac{1}{2}v^TMv$, we define the
{\bfi{angular momentum}} $\J$ by
\[
\J:=\frac{\partial H}{\partial \omega} = I\omega\,.
\]
We rewrite the Hamiltonian as a function of $\J$;
\lbeq{hamilton.spin}
H = \frac{1}{2} \J^T I^{-1} \J \,,
\eeq
in analogy to the formula $H =\frac{1}{2}\p^TM^{-1}\p$ for the
oscillating particle. We have
\lbeq{J.ham}
\frac{\partial H}{\partial \J} = I^{-1}\J=\omega,
\eeq
in analogy with $\vv=\frac{\partial H}{\partial \p}=M^{-1}\p$.

\section{Rotations and angular momentum}\label{s.obs.rot}

\at{As a warm-up example we discuss the group $SO(3)$ of 3-dimensional
rotations, the corresponding Lie algebra $so(3)$ of infinitesimal
rotations, generated by the components of the angular momentum,
and the Lie product induced on the algebra of smooth functions of the
angular momentum. It also shows how the quantities of a classical
Poisson algebra are naturally interpreted as physical observables.
This turns out to be a simple instance of Lie--Poisson algebras,
a class of classical (i.e., commutative) Poisson algebras canonically
associated with any Lie algebra. }

In Section \ref{s.rotSU2}, we used the $J_k$ as generators of the 
rotations; they
are basis elements of the Lie algebra $\Lz=so(3)$. The $J_k$
correspond to the angular momenta of a spinning particle
(see Section \ref{s.rigid}). Thus there is a more physical 
interpretation;
the $J_k$ correspond to measurable quantities, the components of the
angular momentum.
We denote the observable that corresponds to $J_k$ with the
same symbol $J_k$. Purely classical, the state of a rigid rotating
body in its rest frame is defined by specifying a numerical value for
$\J=(J_1,J_2,J_3)^T$, called the \bfi{angular momentum} of the 
rigid body.

The dynamics of a rigid body is determined by the equation 
$\dot \J = \J\times \omega$, where $\omega=I^{-1}\J$ is the 
angular velocity of the rigid body and $I$ is the constant
inertia tensor.

Thus the state at a given time determines uniquely its value at any 
time, and therefore the value of every classical observable
$f(\J)$, i.e., every function of the angular momentum, such as the
angular velocity or the \bfi{total angular momentum} $\J^2$.
In analogy with the case of a single particle, we therefore consider 
the manifold $\Rz^3$ of possible states $\J$ to be the phase space of 
the rotating rigid body. 

To study the observables, i.e., functions of $\J$, we begin with 
polynomials. We write $\Pol\Lz$ for the \bfi{polynomial algebra}
generated by $1$ and the $J_i$, and give this algebra the structure of 
a Poisson algebra. The recipe obtained will then be
further generalized to cover arbitrary $C^\infty$-functions of $\J$.

Motivated by the $so(3)$ structure we define a product $\lp$
recursively, starting with the commutation relations of $so(3)$ with
1 adjoined, 
\[
1\lp J_k =0\,,~~~ J_k\lp J_l = \sum_m \epsilon_{klm}J_m\,.
\]
With the abbreviation $a_1J_1+a_2J_2+a_3J_3=a\cdot \J$, this gives
\[
1\lp a\cdot \J =0\,,~~~ a\cdot \J\lp b\cdot \J =(a\times b)\cdot \J \,.
\]
Having given the product on the generators of $\Pol\Lz$, the
product is completely determined by the Leibniz rule
\[
a\cdot \J f(\J)\lp b\cdot \J = (a\cdot \J\lp b\cdot \J)f(\J)+a\cdot
\J(f(\J)\lp b\cdot \J)
\]
for $f\in\Pol\Lz$.

\begin{lem}\label{l.lp1}
We have the identity
\[
f(\J)\lp b\cdot \J
= (b\times \J) \cdot \frac{\partial f(\J)}{\partial \J}\,,
\]
where $\J$ as a vector in $(\Pol\Lz)^3$ means $(J_1,J_2,J_3)^T$, with 
components $J_k\in \Pol\Lz$.
\end{lem}
\bepf
The proof is by induction. For degree of $f$ zero the statement is
trivial. For degree $1$ we have
\[
a\cdot \J\lp b\cdot \J = (a\times b) \cdot \J = (b\times \J)\cdot a
= (b\times\J) \cdot \frac{\partial a\cdot \J}{\partial \J}\,.
\]
Here we use a vector notation, that is, we consider $\Pol\Lz$. Now
suppose the statement is true for some $n\geq 1$, then we consider
next a homogeneous polynomial of degree $n+1$ and write it as $a\cdot
\J f(\J)$ (or a linear sum of such). Next we consider on the one hand
\[
a\cdot \J f(\J) \lp b\cdot \J = (a\times b)\cdot \J f(\J) + a\cdot \J
b\times \J\cdot \frac{\partial f(\J)}{\partial \J}\,,
\]
and on the other hand
\beqar
(b\times \J)\cdot \frac{\partial }{\partial \J}
\Big( a\cdot \J f(\J)\Big)
&=& (b\times \J) \cdot a f(\J) + a\cdot \J (b\times \J)\cdot
\frac{\partial}{\partial \J}f(\J)\nn\\
&=& (a\times b) \cdot \J f(\J)
+a\cdot \J (b\times \J)\cdot \frac{\partial}{\partial \J}f(\J)\nn\,,
\eeqar
and by inspection the two expressions are the same.
\epf

\begin{lemma}\label{l.lp2}
The product $\lp$ satisfies
\lbeq{e.lprot}
f(\J)\lp g(\J)
= \Big( \frac{\partial g(\J)}{\partial \J}\times
  \J\Big)\cdot \frac{\partial f(\J)}{\partial \J}
= \J\cdot\Big(
\frac{\partial f(\J)}
{\partial \J}\times\frac{\partial g(\J)}{\partial \J}
\Big)\,.
\eeq
\end{lemma}
\bepf
We again proceed by induction, this time on the degree of $g$. For
degree $\leq 1$ of $g$, the previous lemma gives the result. Now
suppose the result holds for polynomials up to degree $n\geq 1$. Now
consider the polynomials of degree $n+1$ and write such a
polynomial as a sum of terms $g(\J)h(\J)$ where $g$ and $h$ both have
degree $\leq n$. Then for each such term we have
\beqar
f(\J)\lp \Big( g(\J)h(\J)\Big)
&=& \Big(f(\J)\lp g(\J)\Big)h(\J) +
g(\J)\Big(f(\J)\lp h(\J)\Big)\nn\\
&=& \Big( \frac{\partial g}{\partial \J}\times \J\Big)\cdot
\frac{\partial f}{\partial \J} h(\J) +
 g(\J)\Big( \frac{\partial h}{\partial \J}\times \J\Big)\cdot
\frac{\partial f}{\partial \J}\nn\\
&=&  \Big( \frac{\partial (gh)}{\partial \J}\times \J\Big)\cdot
\frac{\partial f}{\partial \J}
= \J\cdot\Big(
\frac{\partial f}{\partial \J}\times\frac{\partial (gh)}{\partial \J}
\Big)\,.\nn
\eeqar
\epf

Note that although \gzit{e.lprot} was derived only for polynomials, its 
right hand side makes sense for arbitrary $C^\infty$ functions of $\J$
Thus we take it as the definition of a Lie product on $C^\infty(\Rz^3)$:

\begin{prop}
The algebra $\Ez=C^\infty(\Rz^3)$ and its subalgebra $Pol(\Lz)$ are
Poisson algebras. That is, the product \gzit{e.lprot} is a Lie product
satisfying the Leibniz identity.
\end{prop}

\bepf
The antisymmetry of the product $\lp$ is obvious on the generators,
for the other cases we use Lemma \ref{l.lp1} and Lemma \ref{l.lp2}
together with the observation that
$u\cdot \J\times w = w\cdot u\times \J = -w\cdot \J\times u$.
The Leibniz identity is a direct consequence of the product rule for
partial derivatives. The Jacobi identity is a bit tedious to check.
Using the notation $f_k=\frac{\partial f}{\partial J_k}$
and the Levi--Civit\`a symbol, one writes the outer product for vectors
as $(u\times v)_k = \sum_{lm}\epsilon_{klm}u_lv_m$. Then we find for
the Lie product
\[
f\lp g = \sum_{klm} \epsilon_{klm}J_lf_mg_k\,,
\]
from which the antisymmetry follows immediately.
Using the identity
\[
\sum_{m}
\epsilon_{klm}\epsilon_{mnp}
=\delta_{kn}\delta_{lp}-\delta_{kp}\delta_{ln}\,,
\]
one obtains after some algebraic calculations
\[
(f\lp g)\lp h = \sum h_kf_k J_k g_l - h_k g_k J_lf_l
+ \sum \epsilon_{klm}\epsilon_{abc}J_mJ_ch_l(f_ag_{bk}+f_{ak}g_b)\,,
\]
where the summations are over all present indices. When summing over
the cyclic permutations of $f,g$ and $h$ the first summation is easily
seen to give zero. We write the second sum as
\[
\epsilon_{klm}\epsilon_{abc}J_mJ_c \Big(f_ag_{bk}h_l+f_{ak}g_bh_l+
  g_ah_{bk}f_l+g_{ak}h_bf_l +h_af_{bk}g_l+h_{ak}f_bg_l \Big),
\]
and focus on the term with two derivatives on $f$
\beqar
\epsilon_{klm}\epsilon_{abc}J_mJ_c
\Big(f_{ak}g_bh_l+h_af_{bk}g_l\Big)&=&
\epsilon_{klm}\epsilon_{abc}J_mJ_c
\Big(f_{ak}g_bh_l+h_kf_{la}g_b\Big)\nn\\
&=& \epsilon_{klm}\epsilon_{abc}J_mJ_c \Big(f_{ak}g_bh_l -
  h_lf_{ka}g_b\Big)\nn\\
&=&0\,.\nn
\eeqar
The other terms cancel similarly.
\epf

\section{Classical rigid body dynamics}\label{sec-clas-rig}

Many books on classical mechanics, see for example
\sca{Marion and Thornton} \cite{thornton}, \sca{Arnold} \cite{arnold} 
or \sca{Goldstein} \cite{goldstein}, present the standard 
approach to the dynamics of a spinning rigid body, resulting in the
{\bfi{Euler equations}}. We take an alternative route, exploiting the 
Lie algebra structure corresponding to the rotation group. 
We write down the Lie product that determines the mechanics.
We then derive the Euler equations and reproduce the same equations
of motion. Thus we are giving an equivalent description.

The motivation for the form of the Lie product is determined by symmetry
considerations. We have seen that the algebra of infinitesimal
rotations -- which must be involved in the differential equations
describing the state of the spinning object -- is $so(3)$, the 
Lie algebra of real, antisymmetric $3\times 3$-matrices. 
In Section \ref{lie--poiss}, we shall see that we can obtain a
Lie--Poisson algebra out of any Lie algebra; in particular, we 
construct the Lie--Poisson algebra of $so(3)$ in Example
\ref{ex.so3.pois}.\at{Check and adapt}
Since the  
dynamical observables of a physical system form a Poisson algebra, we
consider the Lie--Poisson algebra of arbitrarily often differentiable 
functions on $\Rz^3$, with
coordinates $J_1$, $J_2$ and $J_3$, equipped with the Lie product
given in Section \ref{s.obs.rot}
\[
f\lp g := \left( \J\times \frac{\partial f}{\partial \J} \right) \cdot
\frac{\partial g}{\partial \J} 
\]
for $f,g\in C^\infty(\Rz^3)$.

Now that we have the Poisson algebra and the Hamiltonian
\gzit{hamilton.spin} for the
classical mechanics of the spinning top, we can apply the
usual recipe. For an observable $f$ the time-evolution is given by
\[
\dot f = H \lp f \,.
\]
In particular, for the angular momentum we have from \gzit{J.ham}
\[
\dot J_k = H\lp J_k = \left(\J\times\frac{\partial H}{\partial
\J} \right)\cdot \hat e_k =  (\J\times\omega)_k\,,
\]
where $\hat e_k$ is the unit vector in the direction $k$, and where
we use $\partial H / \partial \J = I^{-1}\J = \omega$. We thus have
\[
\dot \J = \J\times\omega \,.
\]
Further, since $\J= I\omega$ we find
$I\dot \omega = I \omega\times \omega$.
\at{shouldn't all these $\omega$'s (here and
in previous sections) be in bold, since they're 3-vectors like $\J$ ?
You could use the bm (bold math) macro, like this: ${\bm \omega}$.
And what about $I$? Maybe it's simpler just to drop the emboldening 
of J?}
Writing this out in components we find
\lbeq{euler}
\pmatrix{I_1\dot\omega_1\cr I_2 \dot\omega_2 \cr I_3
\dot\omega_3}= \pmatrix{I_1\omega_1\cr I_2\omega_2\cr I_3\omega_3}\times
\pmatrix{\omega_1\cr \omega_2\cr \omega_3}=
\pmatrix{\omega_2\omega_3(I_2-I_3) \cr \omega_1\omega_3(I_2 - I_1)
\cr \omega_1\omega_2(I_1 - I_2)} %
\eeq
The equations \gzit{euler} are the {\bfi{Euler equations}} for the
spinning rigid body. The spinning direction is given by the vector
$\n:=\omega/|\omega|$ and the spinning speed is given by $|\omega|$.
Thus knowing the trajectory of $\omega(t)$ in the phase space $\Rz^3$
at all times implies knowing everything about the direction and speed
of the spinning motion.

We claim that $\J^2=\J\cdot\J$ is a Casimir \at{where defined?}
of the Lie algebra $so(3)$. Indeed,
from \gzit{so.3.first} we have $J_1\lp J_2 = J_3$ and the other
commutation relations can be obtained by cyclic permutation. But then 
\[
J_1 \lp \J^2 = J_1\lp J_{2}^2+  J_1\lp
J_{3}^2=J_3J_2+J_2J_3-J_2J_3-J_3J_2=0\,,
\]
and for the other generators the results are similar.
Since $\J^2$ is a Casimir of the Lie algebra, it is
conserved by the dynamics. Indeed, calculating the time-derivative of
$\J^2$ we find
\[
(\J^2)^\pdot=2\J\cdot \dot\J = 2\J \cdot \J\times\omega =0\,.
\]
Hence the motion preserves surfaces of constant $\J^2$, which are
spheres. The radius of the sphere is determined by the initial
conditions.

Note that the angular momentum phase space $\Rz^3$ cannot be symplectic
since it is not even-dimensional. However, since we have a
Poisson algebra, it is a Poisson manifold as described in Section
\ref{s.sympl}.

\at{The rest of this paragraph and the next one assume far too
much about things yet to come later in the book.}
In the present case, the symplectic leaves (co-adjoint
orbits) are the surfaces where the Casimir $\J^2$ has a constant
value; hence they are the spheres on which the motion takes place.
Indeed, 2-dimensional spheres in $\Rz^3$ have a natural symplectic
manifold structure, on which the rotation group $SO(3)$ acts as a
group of symplectic transformations. 

Since the Hamiltonian is conserved (although for completely
different reasons), the motion also preserves surfaces of constant
$E  = \shalf \J^T I^{-1}\J$, which are ellipsoids. If $I$ is not a
multiple of the identity, this forces the motion to be on curves
of constant $\J^T I^{-1}\J$ on the co-adjoint orbit, i.e.,
on the intersection of the sphere defining the co-adjoint orbit
with the ellipsoid $\J^T I^{-1}\J=2E$,
where $E$ is again determined by the initial conditions.
Then only the speed along these curves needs to be determined to
specify the motion. Thus the free spinning rigid body motion is
exactly solvable.

\at{The following paragraph seems misplaced}
Let us consider affine functions on the Poisson algebra of the
classical spinning top. We calculate (for $a,b\in\Cz^3$ and 
$\alpha,\beta\in\Cz$)
\[
(\alpha + a\cdot \J)\lp (\beta + b\cdot \J) = (a\times \J)\cdot b 
= (b\times a)\cdot \J\,,
\]
which describes the Lie algebra $u(2)$. Looking only at linear
functions, that is, the linear subspace spanned by $\J$, we find the
Lie algebra $so(3)$. The two Lie algebras only differ by the center of
$u(2)$ and thus $u(2)\cong su(2)\oplus \Rz\cong \Rz\oplus so(3)$. This
coincidence is due to the 
sporadic isomorphism $so(3)\cong su(2)$.

\section{Lie--Poisson algebras}\label{lie--poiss}

In the above section we started from a the Lie algebra structure 
of $so(3)$ to construct an associated Poisson algebra. 
This program can be repeated for arbitrary real Lie algebras.

The formulation closest to the physical applications is in terms of a
Lie $*$-algebra $\Lz$. It applies to arbitrary real Lie algebras such as
so(3) by taking their complexification and adding, if necessary, a
central element 1, thus extending the dimension of the Lie algebra by
one. As usual, we write $\Cz$ for the complex linear subspace spanned 
by the element $1$. In case that $\Lz$ is
infinite-dimensional, we assume $\Lz$ to be equipped with a topology
in which all operations are continuous and that $\Lz$ is reflexive
(see below);
in finite dimensions this is automatic.

We consider the dual space $\Lz^{*}$ of continuous linear maps from
$\Lz$ to $\Cz$, and the bidual space $\Lz^{**}$ of continuous linear
maps from the dual space $\Lz^*$ to $\Cz$. For finite-dimensional
vector spaces we have canonically $\Lz^{**}=\Lz$, for
infinite-dimensional vector spaces in general only
$\Lz\subseteq \Lz^{**}$; in both cases we have an injective map $\Lz\to
\Lz^{**}$ given by
\[
\delta \in\Lz\,,~~\xi\in\Lz^*:~~~ \delta(\xi):=\xi(\delta)\,.
\]
A normed vector space is called reflexive if $\Lz^{**}\cong\Lz$. We need
$\Lz$ to be reflexive for the construction
that follows. We thus assume $\Lz^{**}\cong\Lz$ in the following.

For any real number $\lambda$ (we shall need $\lambda=0$ and
$\lambda=1$),
we define the family of parallel affine hyperplanes
\[
M_\lambda:=\{ \xi\in \Lz^* \mid \xi(f^*)
=\xi(f)^*~\Forall f\in\Lz,~\xi(1)=\lambda\}\,.
\]
One should note that $M_0$ is a real linear
subspace in $\Lz^*$. The affine hyperplane $M_1$ carries the structure
of a real submanifold, with the tangent space \at{avoid this} 
at each point being isomorphic to $M_0$.

If $\Lz$ is the complexification of a real Lie algebra $\Lz'$, so that
we have $\Lz=\Lz'\otimes_\Rz \Cz$, then the elements of $M_0$ are the
linear functionals $\xi$ on $\Lz'$ that are zero on the element $1$,
and are extended to linear forms on $\Lz$ by linearity:
$\xi(a+bi)=\xi(a)+i\xi(b)$ for $a,b\in\Lz'$. So we can
identify $M_0$ in this case with the dual of the quotient Lie algebra
$\Lz'/\Rz$, where $\Rz$ denotes the real subspace spanned by the
distinguished central element $1$. Therefore the dual of $M_0$ is
again $\Lz'/\Rz$. \at{why?} 
In the general case $M_0$ is a real subspace in $(\Lz/\Cz)^*$, so 
that $M_0\Cz=M_0+iM_0$ satisfies $(M_0\Cz)^*\cong \Lz/\Cz$.

We consider for a non-empty open subset $M$ of $M_1$ the
commutative algebra $\Ez=C^\infty(M)$. We define
for every $f\in\Ez$ and $\xi\in M$ a linear map $df(\xi):M_0 \to \Cz$ by
\[
df(\xi) v= \lim_{t\to 0} \frac{f(\xi+tv)-f(\xi)}{t}\Forall v \in M_0\,.
\]
So we have $df(\xi)\in\Lin(M_0,\Cz)$. Extending by $\Cz$-linearity we
can view $df(\xi)$ as an element of $\Lin(M_0 \Cz,\Cz)$. Hence
$df(\xi)$ defines an element in $(M_0 \Cz)^*\cong
\Lz/\Cz$. We can find an element $Df(\xi)$ in $\Lz$ such that
under the projection $\Lz\to\Lz/\Cz$ the element $Df(\xi)$ goes to
$df(\xi)$. The choice of $Df(\xi)$ is not unique, but another choice
$D'f(\xi)$ differs from $Df(\xi)$ by an element in $\Cz$, which is
contained in the center.

We now show how the object $Df(\xi)$  can be chosen. We choose an
arbitrary element $\omega\in \Lz^*$ with $\omega(1)=1$.
Then we can write $\Lz^*$ as a
direct sum $\Lz^*=M_0\Cz \oplus W$ (as a complex vector space),
where $W = \Cz \omega := \{ \alpha \omega \mid \alpha \in\Cz\}$
is the $1$-dimensional span of $\omega$. Indeed, for an arbitrary
element of $\xi\in \Lz^*$, the element $\xi':=\xi-\xi(1)\omega $ 
satisfies $\xi'(1)=0$. Now $\xi'$ can be written
as a linear combination $u+iv$ of two elements $u,v\in M_0$.
Thus $\xi=u+iv+\xi(1)\omega\in  M_0\Cz \oplus W$.
For any fixed choice of $\omega$ we define $Df(\xi)$ by
\lbeq{choice.w}
Df(\xi)u := df(\xi)(u-u(1)\omega)\,.
\eeq
Note that $u-u(1)\in M_0\Cz$.
The extended $Df(\xi)$ lies thus in $\Lz^{**}$. But $\Lz$ was
assumed to be reflexive, hence we have $Df(\xi)\in \Lz$.

We are now in a position to define a Lie product on $\Ez$ by
\[
(f\lp g)(\xi) = (Df(\xi)\lp Dg(\xi))(\xi)\Forall \xi\in M\subset M_1\,,
\]
where the Lie product on the right-hand side is that of $\Lz$.
The left-hand side above is the complex number obtained by
evaluating the function $h:=f\lp g$ for the argument
$\xi\in M \subseteq M_1 \subset \Lz^*$. The right-hand side is the
complex number obtained from the bilinear pairing between
$df(\xi)\lp dg(\xi)\in\Lz$ and the same $\xi$. Since the derivative of
a smooth function is again smooth, $f\lp g$ is again an element of
$\Ez$.

We see that the Lie product $f\lp g$ is independent of the choice of
$Df(\xi)$ and $Dg(\xi)$, or equivalently, of the choice of
$\omega$ in \gzit{choice.w}. Indeed, any other choice would differ
only by an element in the center. But taking the Lie product in $\Lz$
the dependence of the central element drops out.

We have the following theorem:

\begin{thm}
The algebra $\Ez$ with the Lie product $\lp$ defined above is a Poisson
algebra, called the \bfi{Lie--Poisson algebra} over $\Lz$.
The restriction of the Lie product of $\Ez$ to affine functions
coincides with the Lie product of $\Lz$.
\end{thm}
\at{Complete the  proof using the Lie $*$-algebra workfile}
\bepf
(Sketch): The definition of $\lp$ is independent of $\omega$. The
antisymmetry of $\lp$ is clear, and the Jacobi identity follows from
that of $\Lz$, using the fact that partial derivatives commute.
 The Leibniz identity follows from the Leibniz property
of differentiation. The injection $\Lz\to
\Lz^{**}$ gives a map from the Lie algebra to the affine functions. 
We therefore regard the Lie algebra as a subalgebra of the affine 
functions. Since we assumed the Lie algebra $\Lz$ to be reflexive 
the affine
functions represent elements of the Lie algebra. Indeed, for an affine
function $f$ we obtain a linear function by subtracting $f(0)$ and
thus defines an element $f'$of $\Lz$. But $f(0)$ is a multiple of $1$
and thus also an element of the Lie algebra, therefore 
$f=f'+f(0)\in\Lz$.
\epf

We give two important examples.

\begin{example} \label{ex.heis.pois}
Consider $\Lz =h(1)$, the Heisenberg algebra, which is spanned by
generators $p$, $q$ and $1$, with $p\lp q=1$ and all other Lie
products between the generators vanishing. We identify the dual
$\Lz^*$ with $\Cz^3$ as follows,
\[
\pmatrix{x\cr y\cr z}(\alpha p+\beta q + \gamma)
=x\alpha+y\beta+ z\gamma\,,
\]
for any choice $\alpha$, $\beta$, and $\gamma$ in $\Cz$. The affine
hyperplane $M_0$ is in this case given by
\[
M_0=\Big\{ \pmatrix{x\cr y \cr 0} ~\Big|~ x,y\in \Rz\Big\}\,,
\]
and similarly, for $M_1$ we find
\[
M_1=\Big\{ \pmatrix{x\cr y \cr 1} ~\Big|~ x,y\in \Rz\Big\}\,.
\]
If $f$ is a smooth function on $M_1$, it is a smooth function
$\Rz^2\to \Rz$. For $df(\xi)$, $\xi=(x,y,1)^T$ and $v=(a,b,0)^T$ we find
\[
df(\xi)v= \frac{\partial f(x,y)}{\partial x}a + \frac{\partial
f(x,y)}{\partial y}b\,.
\]
The simplest choice for $Df(\xi)$ corresponding to writing $h(1)=\Cz
p\oplus \Cz q\oplus \Cz$, is
\[
Df(\xi) =  \frac{\partial f(x,y)}{\partial x}p +  \frac{\partial
  f(x,y)}{\partial y} q\,.
\]
If $g$ is another smooth function $\Rz^2\to \Rz$ we have
\[
Df(\xi)\lp Dg(\xi) = \frac{\partial f(x,y)}{\partial x}\frac{\partial
  g(x,y)}{\partial y} - \frac{\partial f(x,y)}{\partial
  y}\frac{\partial g(x,y)}{\partial x} \in \Cz\subset h(1)\,,
\]
and thus
\[
(f\lp g) (\xi) = \frac{\partial f(x,y)}{\partial x}\frac{\partial
  g(x,y)}{\partial y} - \frac{\partial f(x,y)}{\partial
  y}\frac{\partial g(x,y)}{\partial x}\,,
\]
which precisely corresponds to the Lie product associated to the
dynamics of a single particle in one dimension. \at{add equation no}

More generally, an arbitrary Heisenberg algebra leads to general 
symplectic Poisson algebras on convenient vector spaces. 
\at{where is convenient needed?}
\end{example}

\begin{example}\label{ex.so3.pois}
We now show that for the choice $so(3)$ we recover the Lie product
\gzit{e.lprot}. We identify the real Lie algebra $so(3)$ with $\Rz^3$
equipped with the vector product. We adjoin a central element to
obtain $so(3)\oplus 1$ and call $\Lz$ the complexification of
$so(3)\oplus 1$. We write an element of $\Lz$ as $(v, a)$ where
$v\in \Cz^3$ and $a\in \Cz$, so that the Lie product is given by
\[
(v,a)\lp (w,b)= (v \times w,0)\,.
\]
Of course, $v\times w$ is defined by extending the vector product on
$\Rz^3$ by $\Cz$-linearity. We identify $\Lz^*$ with $\Cz^4$ as
follows
\[
\pmatrix{x \cr y\cr z \cr t}(v,a)=xv_1+yv_2+zv_3+ta\,,~~v=\pmatrix{
v_1\cr v_2\cr v_3}\,.
\]
Thus we find that $M_\lambda $ consists of the vectors
$(x,y,z,\lambda)^T$ with $x$, $y$ and $z$ real numbers. A smooth
function on $M_1$ is just a smooth function $\Rz^3 \to \Rz$.
For any smooth
$f:\Rz^3\to \Rz$, and $\xi=(x,y,z,1)$ we define
\[
\Nabla f(\xi) = \Big(\frac{\partial f(\xi)}{\partial x},
  \frac{\partial f(\xi)}{\partial y},\frac{\partial f(\xi)}{\partial
    z}\Big)^T \,,
\]
where we identify the vector $\xi = (x,y,z,1)$ in $M_1$ with the vector
$(x,y,z)$ in $\Rz^3$. We see that we can choose $Df(\xi)=(\Nabla
f(\xi),0)$ and the Lie product on $\Ez$ is then given by
\[
f\lp g (\xi) = \xi \cdot (\Nabla f(\xi)\times \Nabla g(\xi) )\,,
\]
which is precisely \gzit{e.lprot}; $(x,y,z)^T$ corresponds to
$(J_1,J_2,J_3)^T$.
\end{example}

The attentive reader might have noticed that in 
Example \ref{ex.so3.pois},
the central element $1$ played no role at all.
\at{Mike wondered what value there is in keeping the central
element distinct in the $h(1)$ case (Example \ref{ex.heis.pois}).
Why not keep $\partial_z$ on the same footing as $\partial_x$ and
$\partial_y$? After all, the central element is part of $h(1)$ just
like $p,q$.}
As mentioned before, when
a Lie algebra has no distinguished central element one can always add
one. However, in this case one can also proceed directly as
follows. For a real Lie algebra $\Lz$, we consider the dual $\Lz^*$ and
the algebra $\Ez$ of real-valued smooth functions on $\Lz^*$. Let
$f\in \Ez$ and $\xi\in\Lz^*$. The
1-form $df(\xi)$ is an element of the dual of the tangent space at
$\xi$. \at{not yet defined}
Since $\Lz^*$ is a vector space and $\Lz$ is assumed to be
reflexive, the dual of the tangent space at $\xi$ is again
$\Lz$. Hence $df(\xi)$ defines an element of the Lie algebra, which we
also denote by $df(\xi)$. Then we define the Lie product on $\Ez$ for
$f,g\in \Ez$ as follows $(f\lp g)(\xi)= \xi(df(\xi)\lp dg(\xi)) $,
that is, to get $(f\lp g)(\xi)$ the function $\xi$ is evaluated at 
the Lie algebra element $df(\xi)\lp dg(\xi)$. 
We leave it as an exercise that this
gives the same result for real Lie algebras that do not have a
distinguished central element.

It turns out that the majority of commutative
Poisson algebras relevant in physics are Lie--Poisson algebras
constructible from a suitable Lie algebra, or natural quotients
of such algebras.
In particular, this holds for the Poisson algebra of classical
symplectic geometry in $\Rz^N$, which come from general Heisenberg
algebras, and for all but one of the Poisson algebras
for nonequilibrium thermodynamics constructed in 
\sca{Beris and Edwards} \cite{BerE}.

\section{Classical symplectic mechanics}\label{s.csymp}

A conservative \at{define this somewhere!}
physical system is completely characterized by three
main ingredients: the kinematical algebra, the Hamiltonian, and the
state. The {\bfi{kinematical algebra}} of the system is a 
Lie $*$-algebra $\Lz$ which defines the kinematics, i.e., 
the structure of the quantities whose knowledge determines the system.
The {\bfi{Hamiltonian}} $H$ defines the dynamics. It is a Hermitian
quantity in an associative algebra $\Ez$ carrying a particular
representation of the kinematical algebra, a Poisson representation
in the classical case, and a unitary representation in the quantum case.
The state encodes all properties of the physical state of
an individual realization of the system at a fixed time.

The kinematical algebra determines the kinematical symmetries of
a whole class of systems which differ in Hamiltonian and state.
This means that applying a transformation of the corresponding
symmetry group transforms a system of this class into another system
of the same class, usually with a different Hamiltonian.
Those (often few) symmetries which preserve a given Hamiltonian
are called symmetries of the system; applying a symmetry of the
system changes possible state space trajectories of the system into
other possible trajectories, usually affecting the states.
Those (even fewer) symmetries which preserve the Hamiltonian and
the state are symmetries of the particular realization of the system,
and hence directly measurable.

The kinematical algebra may admit (up to isomorphism) one or many
Poisson representations \at{where defined?}
for classical systems, and one or many unitary
representations for the corresponding quantum systems. For example,
a Heisenberg algebra with finitely many degrees of freedom
admits only one unitary representation, which is
the content of the Stone--Von Neumann theorem. \at{refer to Chapter x}

In the nonrelativistic case, the Hamiltonian is an element of the
Poisson algebra for classical systems, and for quantum systems the
Hamiltonian is an element of the universal enveloping algebra of the
Lie $*$-algebra. 

Let $\Ez$ be the algebra determined by the physical system, that is,
either $\Ez$ is the Lie--Poisson algebra of the classical system, or
$\Ez=\Lin\Hz$ for a Euclidean space $\Hz$ whose closure is a Hilbert
space. Both the Lie algebra $\Lz$ and the space-time symmetry group
are represented inside $\Ez$.
\at{relate with symmetries from Chapter 1?}

\at{The remainder of this section seems quite mis-sequenced and even
misplaced. You consider the special case of N-particle systems
(molecular motion) and talk about characters, evaluations, etc, which
seems to have little or no connection to other stuff, then you break
and talk about mixed classical states, densities and Liouville measure,
and then you resume talking about N-particle stuff. (I note in the next
section on Molecular mechanics you have an "attention" about merging it
with stuff here.)}

We now consider the special case of classical $N$-particle systems 
describing the motions of a molecule. \at{what is the phase space?
realte to Part II}]
The algebra $\Ez$ consists of the complex-valued functions on
phase space $M$ and each point $z\in M$ in phase space determines a
state $\<\cdot\>_z$ by
\[
\<f\>_z :=f(z),
\]
called a {\bfi{classical pure state}}. Note that
evaluation at a point $z\in M$ is more than a linear functional;
an evaluation gives
an algebra homomorphism $C^\infty(M;\Rz)\to \Rz$ since
$(fg)(z)=f(z)g(z)$; hence we have a character of the commutative 
algebra $\Ez$. \at{where defined?}
If the phase space $M$ is an open subset of
$\Rz^n$, the evaluations are the only characters of $\Ez$. This can be
seen as follows. Take any algebra homomorphism $\varphi:C^\infty
(M)\to \Rz$. Let $x_1,\ldots,x_n$ be coordinates on $M$ and denote by
$a_i=\varphi(x_i)$ the images of the coordinate functions. The
homomorphism $\varphi$ thus determines a point $z=(a_1,\ldots,a_n)$ in
$\Rz^n$. We have to show $z\in M$. Suppose $z\notin M$, then
\[
f(x_1,\ldots,x_n) = \sum_j (x_j-a_j)^2
\]
is a function that does not vanish on $M$, and thus is an invertible
element of $C^\infty(M)$. If an element $x$ is invertible, then so is
its image under any homomorphism. Indeed, if $xy=1$, then
$\varphi(xy)=\varphi(x)\varphi(y)=\varphi(1)=1$. But the function $f$
is mapped to zero under $\varphi$ and hence cannot be
invertible. Hence we arrive at a contradiction and the assumption
$z\notin M$ is false.

A mixed classical state is a weighted mixture of pure classical states.
That is, there is a real-valued function $\rho$ on the phase space $M$,
called the {\bfi{density}}, taking nonnegative values and 
integrating to one
\[
\int_M \rho(z) d\mu(z) =1\,,
\]
such that
\lbeq{state-cl}
\langle f \rangle = \int_M \rho(z)f(z)d\mu(z)\,.
\eeq
The integration measure $d\mu$ depends on the application. \at{adapt}
In symplectic mechanics, the symplectic Poisson 
bracket determines the Lie product, one uses the 
\bfi{Liouville measure}, defined in local coordinates $(q,p)$ by
\[
d\mu(z) = dq_1\cdots dq_n dp_1\cdots dp_n.
\]

Consider a system containing $N$ particles. Then each particle has a
momentum and a position. Hence phase space is $6N$-dimensional. The
Lie algebra is given by the relations
$p_{ia}\lp q_{jb}=\delta_{ij}\delta_{ab}$ where $p_{ia}$ is the $a$th
component of the momentum of the $i$th particle and $q_{jb}$ is the
$b$th component of the position of the $j$th particle. The obtained
Lie algebra is the Heisenberg algebra $h(N)$.  

In molecular mechanics, the Hamiltonian is of the simple 
form\footnote{
The more general form
$H= \frac{1}{2}\sum_{i,j=1}^{N}G_{ij}(\q_1,\ldots,\q_N)\p_i\p_j +
V(\q_1,\ldots,\q_N)$,
where $G$ is a configuration-dependent inverse mass matrix,
appears at various places in physics. When the
potential $V$ is constant (so that we can put it to zero), the
physical system is sometimes called a \bfi{$\sigma$-model}. Such models
play an important role in modern high-energy physics and cosmology.
Some authors prefer to include a potential into the definition of a
$\sigma$-model.
} 
\[
H= \sum_{i=1}^{N}\frac{\p^{2}_{i}}{2m_i} + V(\q_1,\ldots,\q_N)\,,
\]
where the \bfi{potential} $V(\q_1,\ldots,\q_N)$ describes the potential
energy of the configuration with positions $(\q_1,\ldots,\q_N)$.

The states in symplectic mechanics are precisely the states of the
form \gzit{state-cl}. \at{move this and that together!}
If the system is such that we can measure
at one instant of time all positions and momenta exactly (obviously
an idealization), the configuration is precisely given by the point
$z=(\q_1,\ldots,\q_N,\p_1,\ldots,\p_N)$ in phase space, and 
$\<f\> = f(z)$ for all $f\in\Ez$. Thus the density degenerates to a 
product of delta functions of each phase space coordinate. Thus 
classical pure states are equivalent to points $z$ in phase space,
marking position and momentum of each point of interest, such as the 
centers of mass of the stars, planets, and moons making up a celestial 
system.

\section{Molecular mechanics}\label{s.molmech}

\at{merge with preceding}
Consider a molecule consisting of $N$ atoms. The molecule is chemically
described by assigning bonds between certain pairs of atoms, reflecting
the presence of chemical forces that -- in the absence of chemical
reactions which may break bonds -- hold these atoms close together.
Thus a molecule may be thought of as a graph embedded in
3-dimensional space, in which some but usually not all atoms are
connected by a bond. The chemical structure of the molecule is thus
described by a connected graph, the \bfi{formula} of the molecule.
(In the following, we ignore multiple bonds, which are just a way to
indicate stronger binding than for single bonds, reflected in the
interaction potential.)
We write $i\sim j$ if there is a bond between atom
$i$ and atom $j$ and similarly we write $i\sim j\sim k$ if there is
a bond connecting $i$ and $j$ and there is a bond connecting $j$
and $k$. The
notation is extended to longer chains: $i\sim j \sim k\sim l\sim \dots$.

The interactions between the atoms in a molecule are primarily
through the bonds, and to a much smaller extent through forces
described by a pair potential and through multibody forces for
joint influences of several adjacent bonds.

The geometry is captured mathematically by assigning to the $j$th atom a
3-dimensional coordinate vector
\[ q_j = \pmatrix{q_{j1}\cr q_{j2}\cr q_{j3} }
\]
specifying the position of the atom in space. If two
atoms with labels $j$ and $k$ are joined by a chemical bond, we consider
the corresponding {\bfi{bond vector}} $q_j-q_k$, with {\bfi{bond
length}}
$\| q_j - q_k \|$. At room temperature, the bonds between adjacent
atoms $i$ and $j$ are quite rigid,
meaning that the deviation from the average distance $r_{ij}$ is
generally small and the
force that tries to maintain the atoms at distance $r_{ij}$ is strong.
In chemistry this is modeled by a term
\[
V_\fns{bond}(q_1,\ldots,q_N)
= \sum_{i\sim j}\frac{a_{ij}}{2}(\|q_i-q_j\|
- r_{ij})^2
\]
in the Hamiltonian, where the $a_{ij}$ are \bfi{stiffness constants},
parameters determined by the particular chemical structure.

\begin{figure}
\centerline{\psfig{figure=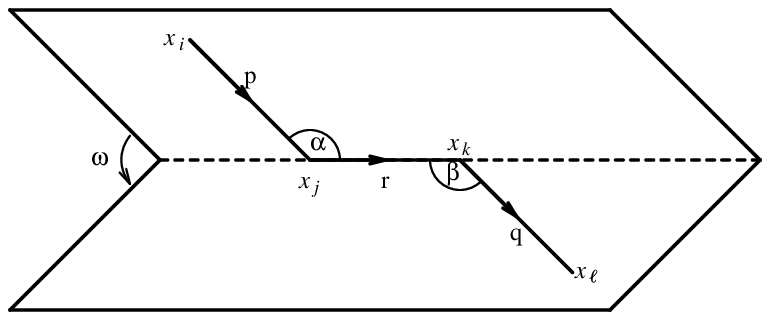,height=5cm}}
\caption{Bond vectors, bond angles, and the dihedral angle}
\label{dihedral}
\end{figure}

\at{figure \ref{dihedral} fails to display.\\
make the following consistent with the figure: $i \sim j \sim k$}
Consider two adjacent bonds $i\sim j$ and $k\sim l$. The {\bfi{bond
angle}} $\alpha$ is the angle between the bond
vectors $q_j-q_i$ and $q_l-q_k$. The bond angle $\alpha$ can then be
computed  from the formulas
\[
\cos \alpha
=  \frac{(q_i-q_j)\cdot
  (q_k-q_j)}{\|q_i-q_j\|\|q_k-q_j\|}\,,~~~
\sin\alpha = \frac{\|(q_i-q_j)\times
  (q_k-q_j)\|}{\|q_i-q_j\|\|q_k-q_j\|}\,,
\]
and is thus invariant under the simultaneous action of the group
$ISO(3)$ \at{defined?}
on all $3$ vectors. In most molecules the bond angles are
determined from the interaction between the atoms in the molecule.
There is thus an $ISO(3)$-invariant term
\[
V_\fns{angle}(q_1,\ldots,q_N)
=\sum_{i\sim j \sim k}a_{ijk}\Phi(q_i,q_j,q_k)
\]
in the potential with $\Phi:(\Rz^3)^3¸\to\Rz$ an $ISO(3)$-invariant
function, and $a_{ijk}$ are some parameters.

Finally, the {\bfi{dihedral angle}}
$\omega = \winkel (i\sim j\sim k \sim l)$
(or the complementary {\bfi{torsion angle}} $2\pi-\omega$) measures
the relative orientation of two adjacent angles in a chain
$i \sim j \sim k\sim l$ of atoms. It is defined as
the angle between the normals through the planes determined
by the atoms $i,j,k$ and $j,k,l$, respectively, and can be calculated
from
\[ \cos \omega =
\frac{(q_i-q_j) \times (q_k-q_j) \cdot (q_j-q_k) \times
  (q_l-q_k)}{\|(q_i-q_j) \times (q_k-q_j) \| \| (q_j-q_k) \times
  (q_l-q_k)\|}\,,
\]
and \[
\sin \omega =
\frac{|(q_i-q_j) \times (q_l-q_k) \cdot(q_k-q_j)| \,\|  q_k-q_j\|}{\|(q_i-q_j) \times (q_k-q_j) \| \| (q_j-q_k) \times
  (q_l-q_k)\|}\,. \]
Again, the angle between the planes is $ISO(3)$-invariant and
therefore described by an
$ISO(3)$-invariant function $\Psi:(\Rz^3)^4\to \Rz$ of the positions
of the four atoms. Hence to model the molecule there is a term
\[
V_\fns{dihedral}(q_1,\ldots,q_N)=\sum_{i\sim j\sim k \sim
  l}a_{ijkl}\Psi(q_i,q_j,q_k,q_l)
\]
in the Hamiltonian, with again $a_{ijkl}$ parameters. 
The total Hamiltonian is then taken to be
\[
H=\sum_i \Big(\frac{p_{i}^{2}}{2m_i}
+V_\fns{bond}(q_1,\ldots,q_N)
+V_\fns{angle}(q_1,\ldots,q_N)
+V_\fns{dihedral}(q_1,\ldots,q_N)\Big)\,.
\]
The above Hamiltonian is of a special type; it is a member of the
family of Hamiltonians of the form
\[
H=\sum_i \frac{p_{i}^{2}}{2m_i} +V(q_1,\ldots,q_N)\,.
\]
This family of Hamiltonians is favorable since there are no mixed terms
between the momenta and the positions. Therefore, in the quantum
theory there is no ambiguity in how the quantum mechanical Hamiltonian
has to be written, since the momenta commute among themselves and the
positions commute among themselves, too.

\at{add context, check dependence on later things,
$so(m,n)$ form of $iso(3)$ invariants?}

The group $ISO(3)$ \at{all groups here not yet defined}
plays an important role here purely on symmetry
grounds; how and where a molecule is located in $\Rz^3$ does not
determine the chemical properties. Hence the Hamiltonian should depend
on $ISO(3)$-invariant quantities only. 

From the above we see \at{?} 
that the one in practice is given a representation $j:G\to GL(V)$
\at{why not $J$?}
of some group in a vector space. The construction of a suitable
Hamiltonian can be facilitated by knowing the invariants in the tensor
representations $V\otimes V$, $V\otimes V\otimes V$ and so on. If $V$
is irreducible, $V$ might contain one or more one-dimensional
subrepresentations \at{not defined}
on which $G$ acts trivially; these are precisely
the invariants \at{not defined} in $V$. Hence knowing the
irreducible representations of $G$ is of great importance. 

Now suppose that $G$ has an irreducible representation $j$ on $V$. 
Then $g\in G$ acts on $V\otimes V$ as follows: 
$g: v\otimes w\mapsto j(g)v\otimes j(g)w$, and similarly for higher
order tensor products.
It is almost never the case that the representation of $G$ in tensor 
products of irreducible representations is again irreducible. 
But, in many cases, the decomposition of 
$V\otimes V$, $V\otimes V\otimes V$, etc. 
into irreducible representations is known. 

In the case of $ISO(3)$ the representation is in $\Rz^3$, 
\at{it is not a linear representation, however!}
which 
contains no invariants since all points of $\Rz^3$ form a single orbit.
But, as we have shown in Section \ref{s.molmech}, 
$V\otimes V$, $V\otimes V \otimes V$ and 
$V\otimes V\otimes V\otimes V$ do have invariants: distances and angles.

\section{An outlook to quantum field theory}

\at{is this still at the right place? we have not yet introduced 
operators
...}

Quantum field theory is the area in physics
where fields are treated by quantum mechanics. The way
physicists think of this is more or less as follows.
As we have seen in Chapter \ref{c.oscillating}, classical linear field
equations, such as the Maxwell equations, can be seen as describing
a family of harmonic oscillators labeled by a continuum of pairs
$(\p,s)$ of momenta $\p$ and spin or helicity $s$. \at{both undefined}
Therefore,
what has been treated above is nice, but for quantum field theory
it is not enough. One needs an infinite number of
oscillators.
Treating such a system becomes mathematically sophisticated, because 
topological details start playing a dominant role. A way to deal with
this heuristically, often employed by physicists, is by discretizing 
space-time in a box.
On each point of the lattice one places a harmonic oscillator; then
there are just a finite number of oscillators. To get
the quantum field theory, one considers the limit in which the size of
the box goes to infinity and the spacing of the lattice goes to
zero. Then the oscillators are not described by operators $a_k$
and $a_{k}^{+}$ that are labeled by vectors $k$, but by operators
$a(x)$ and $a^+(x)$ that are labeled by the continuous four-vector
index $x$. The limit might not exist\dots. 

In two space-time dimensions
the limit is well-defined for interacting field theories, that is, for
field theories where the different fields can interact. 
In case of four
dimensions, the correct limit is only known for non-interacting field
theories. From experience we know that there is interaction, of
course, so our description shows serious shortcomings. After the
preceding description of representations, it is interesting to
note that,- in the field theory limits, the metaplectic representations
\at{this is now out of context. Move the remark!}
still exist.

For 2-dimensional field theories with one space and one time dimension,
this leads to satisfactory quantum field theories (such as conformal
field theory) \at{refs from my FAQ, and more metaplectic/spin stuff}.
But for 4-dimensional field theories, the metaplectic representation
is restricted to a class of operators not flexible enough for
capturing the physics. This is the main mathematical obstacle for
formulating a consistent framework for  4-dimensional quantum field
theories. \at{fill in ref's from FAQ}

\chapter{Representation and classification}\label{c.repclass}

\section{Poisson representations}\label{sec-quad-rep}

\at{Lie algebras in Lie--Poisson algebras are almost the same 
as Poisson representations of Lie algebras}

Consider the Heisenberg algebra $h(n)$ with the usual generators 
$1$, $p_i$, and $q_i$, and the corresponding Lie--Poisson algebra
$\Ez(h(n))$. The subalgebra of all polynomials in $q_i,p_j$ 
is closed under the Lie product, and hence a Poisson subalgebra.
More interestingly, there are several Lie subalgebras of low degree 
polynomials, which we shall now explore. We
write $z$ for the $2n$-tuple
\[
z=\pmatrix{ p \cr q \cr}
\]
of all the generators except $1$. All linear polynomials without
constant term can be written as $a\cdot z$ for some $a\in
\Cz^{2n}$. On $\Cz^{2n}$ we introduce the antisymmetric bilinear form
$\omega$, represented in the given basis by the matrix $J$:
\[
\omega(a,b)= a^TJb\,,~~~J=\pmatrix{0&-1\cr 1&0\cr}\,,
\]
where the entries in $J$ are $n\times n$-matrices, i.e., $1=1_n$, etc.. 
The bilinear form $\omega$ is nondegenerate and antisymmetric,
and we have
\lbeq{sympl.comm}
a\cdot z \lp b\cdot z = \omega(a,b)\,.
\eeq
Any quadratic expression in $\Ez(h(n))$ is a linear sum of
expressions of the form
\[
(a\cdot z)( b\cdot z)\,.
\]
We consider two such expressions and calculate their Lie product, 
using the Leibniz rule twice. \at{keep the terms in
the second line in a sequence matching the first?}
\beqar
(a\cdot z)( b\cdot z)\lp (c\cdot z)( d\cdot z)
&=&a\cdot z(( b\cdot z)\lp (c\cdot z)( d\cdot z))
+ (a\cdot z\lp (c\cdot z)(d\cdot z)) b\cdot z \nn\\
&=& (b\cdot z) (c\cdot z) \omega(a,d)
+(b\cdot z)( d\cdot z) \omega(a,c) \nn\\
&&+(a\cdot z)( c\cdot z) \omega(b,d)
+(a\cdot z)( d\cdot z) \omega(b,c)\nn\,,
\eeqar
which is a quadratic expression. Hence the homogeneous quadratic
polynomials form a Lie subalgebra of $\Ez(h(n))$.
We show below that this Lie algebra is related to $sp(2n,\Cz)$.
We proceed in physicist's fashion by looking at a conveniently chosen 
basis.
In Section \ref{s.meta} we give a second derivation in a coordinate
independent fashion, which generalizes to the fermionic case and
gives Lie algebras related to the real orthogonal groups.

The generators of $h(n)$ are $1$, $p_i$ and $q_i$. Consider the
elements
\[
Q_{ij}=q_iq_j\,,~~~P_{ij}=
p_{i}p_{j}\,,~~~E_{ij}=\frac{1}{2}(q_ip_j+p_jq_i)\,,
\]
of the universal enveloping algebra. We have $Q_{ij}=Q_{ji}$ and 
$P_{ij}=P_{ji}$. We find the commutation relations:
\beqar
Q_{ij}\lp Q_{kl}&=&0\,,~~~P_{ij}\lp P_{kl}=0\nn\,,\\
E_{ij}\lp E_{kl}&=& -\delta_{il}E_{kj}+\delta_{jk}E_{il}\nn\,,\\
E_{ij}\lp Q_{kl}&=& \delta_{jl}Q_{ik}+\delta_{jk}Q_{il}\nn\,,\\
E_{ij}\lp P_{kl}&=& -\delta_{il}P_{jk}-\delta_{ik}P_{jl}\nn\,,\\
Q_{ij}\lp P_{kl}&=&
-\delta_{ik}E_{jl}-\delta_{jk}E_{il}-\delta_{jl}E_{ik}
-\delta_{il}E_{jk}\nn\,.
\eeqar
The Lie algebra $sp(2n,\Cz)$ is given by the complex $2n\times
2n$-matrices that preserve the above given $J$:
\[
X\in sp(2n,\Cz)~ \Leftrightarrow~X^TJ+JX=0\,.
\]
Taking $X$ in block form as
\[
X=\pmatrix{A&B\cr C&D\cr}\,,
\]
we find $X\in sp(2n,\Cz)$ if and only if $C=C^T$, $B=B^T$ and $D=-A^T$.
If we introduce the $n\times n$-matrices $e_{ij}$ that are $1$ on the
$ij$-entry and zero elsewhere, we have
$e_{ij}e_{kl}=\delta_{jk}e_{il}$ and the matrices
\[
A_{ij}=\pmatrix{ e_{ij}&0\cr 0&
  -e_{ji}\cr}\,,~~~B_{ij}=\pmatrix{0&e_{ij}+e_{ji}\cr 0 &
  0}\,,~~~C_{ij}=\pmatrix{0&0\cr e_{ij}+e_{ji} & 0}\,,
\] form a basis for $sp(n,\Cz)$. We find the commutation rules
\beqar
B_{ij}\lp B_{kl}&=&0\,,~~~C_{ij}\lp C_{kl}=0\nn\,,\\
A_{ij}\lp A_{kl}&=& -\delta_{il}A_{kj}+\delta_{jk}A_{il}\nn\,,\\
E_{ij}\lp B_{kl}&=& \delta_{jl}B_{ik}+\delta_{jk}B_{il}\nn\,,\\
A_{ij}\lp C_{kl}&=& -\delta_{il}C_{jk}-\delta_{ik}C_{jl}\nn\,,\\
B_{ij}\lp C_{kl}&=&
\delta_{ik}E_{jl}+\delta_{jk}E_{il}+\delta_{jl}E_{ik}
+\delta_{il}E_{jk}\nn\,.
\eeqar
Sending $Q_{ij}$ to $-B_{ij}$, $P_{ij}$ to $C_{ij}$ and $E_{ij}$ to
$A_{ij}$ we have an isomorphism between the algebras.

We now allow for inhomogeneous quadratic polynomials by adjoining
the linear forms of the algebra $\Ez(h(n))$ to this Lie algebra. 
Everything commutes with the central element $1$, so we will not write 
down the commutation relations with $1$. The commutation relations of 
the other basis elements are found to be
\beqar
Q_{ij}\lp q_k&=&P_{ij}\lp p_k=0\nn\,,\\
Q_{ij}\lp p_k&=& -\delta_{ik}q_j-\delta_{jk}q_i\nn\,,\\
P_{ij}\lp q_k&=& -\delta_{ik}p_j-\delta_{jk}p_i\nn\,,\\
E_{ij}\lp q_k &=& \delta_{ik}q_j\nn\,,\\
E_{ij}\lp p_k &=& -\delta_{jk}p_i\nn\,.
\eeqar
We define the Lie subalgebra $\Lz'$ of $\Ez(h(n))$ as the Lie
subalgebra of quadratic expressions in the generators and we define
$\Lz=\Lz'/\Cz $, so that in
$\Lz$ we have $q_i\lp p_j=0$. Using the previously established
isomorphism with $sp(2n,\Cz)$ it is not too
hard to see that $\Lz$ is isomorphic to the Lie algebra 
\idx{$isp(2n,\Cz)$},
which is defined as the Lie algebra of all 
$(2n+1)\times (2n+1)$-matrices of the form
\[
\pmatrix{ A & r \cr 0 & 0\cr}\,,
\]
with $A$ a $2n\times 2n$-matrix in $sp(2n,\Cz)$ and $r$ a $2n$-vector.
We have thus shown that $\Lz'$ is a central extension of $isp(2n,\Cz)$.

\section{Linear representations}

Of great interest in quantum mechanics are certain realizations of
Lie algebras and of Lie groups \at{not yet defined}
by means of operators on vector spaces.
We therefore address the concept of a representation of a Lie algebra. 
In the previous chapter we have already given a short discussion of
 finite-dimensional representations of finite-dimensional
Lie algebras. \at{merge with here?}

\begin{definition}\label{d.lie-rep}~\\
(i) A \bfi{(linear) \idx{representation}}\index{representation!linear}
of a Lie algebra $\Lz$ in an associative algebra $\Ez$ is a linear map 
$J:\Lz\rightarrow \Ez$ such that
\[
J(f\lp g)=J(f)J(g)-J(g)J(f) \Forall f,g\in\Lz.
\]
The representation is called {\bfi{faithful}} if $J$ is injective.
A {\bfi{linear representation}}\index{representation!linear} on a
(finite- or infinite-dimensional)
vector space $\Hz$ is a representation in the algebra $\Ez=\Lin\Hz$.
In the case that $\Ez$ is the algebra of $n\times n$ matrices with
entries in $\Kz$ one obtains the definition of Section \ref{fin-dim}.
A linear representation is called {\bf
\idx{irreducible}}\index{representation!irreducible} when
the only subspaces closed under multiplication by linear mappings
of the form $J(f)$ are $0$ and $\Hz$.

(ii) A {\bfi{unitary representation}}\index{representation!unitary} 
of a Lie $*$-algebra $\Lz$ is a linear map $J:\Lz\to \Ez$ in the 
$*$-algebra $\Ez=\Lin\Hz$ of continuous linear operators of a 
Euclidean space $\Hz$ (with $*$ being the adjoint), satisfying
\[
J(1)=1\,,~~~ J(f^*)=J(f)^*\,,~~~ 
J(f\lp g)=\frac{i}{\hbar}\Bigl(J(f)J(g)-J(g)J(f)\Bigr).
\]
\end{definition}

Note that by Proposition \ref{p.der},
an associative algebra $\Ez$ becomes in a natural way a
Lie algebra by defining $f\lp g= [f,g]=fg-fg$.
Hence a representation of a
Lie algebra $\Lz$ in an algebra $\Ez$ is a Lie algebra
homomorphism from $\Lz$ to $\Ez$, with $\Ez$ regarded as a Lie
algebra. If the representation is faithful, the image of $\Lz$ is
a Lie subalgebra of $\Ez$ isomorphic to $\Lz$. 
In this case, one often identifies the elements of $\Lz$ with their
images, and then speaks of an {\bfi{embedding}} of $\Lz$ into $\Ez$.
\at{simile for Lie $*$-algebras}
By the Theorem of Ado mentioned in Section \ref{s.Liega}, every 
finite-dimensional real Lie algebra has a faithful representation.

\bigskip
\bfi{The enveloping algebra.}
In a representation, the elements of $\Lz$ are represented by 
matrices or linear operators. From a
given set of matrices we can form the algebra that these matrices
generate, containing the unit matrix, all finite products and
their linear combinations. This motivates us to consider an
object that already encompasses this algebra for all representations:
the universal enveloping algebra of a Lie algebra $\Lz$. In general
it is constructed by considering the \idx{tensor algebra} 
\idx{$T(\Lz)$}, which is given by
\[
T(\Lz) = \Kz\oplus \Lz\oplus (\Lz\otimes \Lz) \oplus
(\Lz\otimes\Lz\otimes \Lz) \oplus\ldots = \bigoplus_{i=0}^{\infty}
\Lz^{\otimes i}\,.
\]
One makes $T(\Lz)$ into an associative noncommutative algebra over the
complex numbers by defining the product $ab$ to be the tensor product 
$a\otimes b$. \at{more elementary definition?}

Within $T(\Lz)$ we consider the ideal $\mathcal{J}$ generated by all
elements of the form
\[
x\otimes y - y\otimes x -[x,y]
\]
for all $x,y$ in $\Lz$. Thus an element in $\mathcal{J}$ is a sum of
elements of the
form
\[
a\otimes(x\otimes y - y\otimes x -[x,y])\otimes b\,,
\]
for some $a,b\in T(\Lz)$. The {\bfi{universal enveloping algebra}}
of $\Lz$ is then
defined as the associative noncommutative algebra  
\idx{$\mathcal{U}(\Lz)$} over the complex numbers given by
\[
\mathcal{U}(\Lz)= T(\Lz)/\mathcal{J}\,.
\]
Another view on the universal enveloping algebra
$\mathcal{U}(\Lz)$ would be as follows. One chooses a basis
$\left\{t_i\right\}$ for $\Lz$
and considers the associative noncommutative polynomial algebra in the
generators while imposing the relation
\[
t_it_j - t_jt_i = [t_i,t_j]\,.
\]
Thus we consider the associative algebra generated by $1$ and by the
generators of $\Lz$ and impose the Lie product, which in this case is
the commutator, by hand.
The algebra we obtain in this way is canonically isomorphic to the
universal enveloping algebra $\mathcal{U}(\Lz)$.

The universal enveloping algebra thus contains the Lie algebra,
i.e. envelopes the Lie algebra. This approach is very practical and 
therefore often used by physicists. 
There exists a more sophisticated definition, using a
so-called universal property. One then proves that such an object is
unique and that the given
definition above has this universal property. We do not expand on
the definition using the universal property but refer to the
literature, see, e.g., 
\sca{Jacobsen} \cite{jacobsen},
\sca{Knapp} \cite{knapp}, or
\sca{Fuchs \& Schweigert} \cite{fuchsandschweigert}.
It is because of this
universal property that $\mathcal{U}(\Lz)$ is usually called the 
universal enveloping algebra, and not just the enveloping algebra.

The main reason to define the universal enveloping algebra is to
study the representations of the Lie algebra. Every representation of
the Lie algebra induces a unique representation of the {\it
  associative} universal enveloping algebra, and conversely, every
representation of the universal enveloping algebra induces a
representation of the Lie algebra itself. 
\at{this is the universal property; clasical analogue?}
In a sense, \at{.} all
finite-dimensional representations are maps of the associative
universal enveloping algebra to the associative algebra of
$n\times n$-matrices for some $n$.

\bigskip
{\bfi{Casimir elements}.}
An element $C\in \mathcal{U}(\Lz)$ in the center of the universal 
enveloping algebra, i.e. that commutes with all other elements of 
$\mathcal{U}(\Lz)$, is called a 
{\bfi{Casimir element}}\index{Casimir operator}, or just Casimir 
and sometimes also Casimir operator. If $\Lz$ has a representation 
in a vector space $V$, then for any $c\in\Kz$ the subspace 
$V_c =\left\{ v\in V | Cv=cv\right\}$ is invariant under the action 
of $\Lz$, precisely because $C$ is in the center of $\mathcal{U}(\Lz)$. 
Hence if the representation $V$ is
irreducible, $V_c$ must be the whole of $V$ for some $c$ and the other
$V_c$ are zero. That means, $C$ acts diagonally in irreducible 
representations.

The classical analogue of the universal enveloping algebra is the
Lie--Poisson algebra discussed in Chapter \ref{lie--poiss}.

\section{Finite-dimensional representations}\label{fin-dim}

We have already seen in Section \ref{s.Liega}
that the Lie algebra $gl(n,\Kz)$ has many interesting Lie subalgebras.
Given an arbitrary Lie algebra $\Lz$ it
is interesting to see how we can represent $\Lz$ as a Lie algebra of
matrices. In this section we consider finite-dimensional Lie
algebras and finite-dimensional representations in more detail.

For any vector space $V$ over $\Kz$ we denote $gl(V)$ the Lie algebra
of linear maps from $V$ to $V$ with the Lie product given by the
commutator $f\lp g = fg-gf$. If $V$ is identified with $\Kz^n$ we
write $gl(V)=gl(n,\Kz)$ (see Section \ref{s.Liega}).
A Lie algebra homomorphism $\phi: \Lz\to gl(V)$ is called a
\bfi{finite-dimensional 
representation}\index{representation!finite-dimensional} of $\Lz$; 
the vector space $V$ is then called an \bfi{$\Lz$-\idx{module}}.
We call the representation complex if $\Kz=\Cz$ and real
if $\Kz=\Rz$. We have already seen that $su(n)$ has a complex
representation, since it is defined as a (real) subalgebra of
$gl(n,\Cz)$.

Given a representation $\phi:\Lz\to gl(V)$ we call $W$ an
{\bfi{invariant subspace}} of $V$ if $\phi(f)w\in W$ for all
$f\in \Lz$ and all $w\in W$. The representation is called
\bfi{irreducible} \at{alredy defined}
if the only invariant subspaces are $0$ and $V$.
We call the representation {\bfi{decomposable}} or {\bfi{fully
reducible}}\index{representation!decomposable}\index{representation!fully
reducible},
if for any invariant subspace $W$ there is a complementary invariant
subspace $W'$ such that $V=W\oplus W'$.

If $\phi_1:\Lz\to gl(V_1)$ and $\phi_2:\Lz\to gl(V_2)$ are
representations of $\Lz$ we can form the direct sum representation
$\phi_{1\oplus 2}:\Lz\to gl(V_1\oplus V_2,\Kz)$ by defining
\[
\phi_{1\oplus
  2}(f)(v_1+v_2)=\phi_1(f)(v_1)+\phi_2(f)(v_2)
\]
for $f\in \Lz$ and $v_1\in V_1$, $v_2\in V_2$. It is easy to check 
the representation property. In terms of matrices,
the direct sum representation corresponds to the map given by
\[
f\mapsto \pmatrix{\phi_1(f) & 0 \cr 0 & \phi_2(f)}\
\]
in block matrices.

If $\phi_1:\Lz\to gl(V)$ and $\phi_2:\Lz\to gl(W)$ are
representations of $\Lz$ we can form the tensor product representation
$\phi_{1\otimes 2}$ as follows: Each element $f$ in $\Lz$ is sent to
the linear map
\lbeq{tens.repr}
\phi_{1\otimes 2}(f) (v\otimes w)= (\phi_1(f)v)\otimes w + v\otimes
(\phi_2(f)w)
\eeq
for all $v\in V$ and $w\in V$. It is easy to check that
\gzit{tens.repr} defines a representation.

In Section \ref{sec-fin-dim} we have already mentioned the adjoint
representation $\ad: \Lz\to gl(\Lz)$ defined by
\[
\ad_f:g\mapsto [f,g]\,.
\]
The map $\ad$ is clearly linear, and from
the Jacobi identity we see
\[
\ad_x \ad_y (z) -\ad_y\ad_x(z) = \ad_{[x,y]}(z)\,,
\]
hence
\[
[\ad_x,\ad_y] = \ad_{[x,y]}.
\]
We can now rephrase the definition of the ideal (see Section
\ref{sec-fin-dim}) as follows: $I\subset \Lz$ is
an ideal if and only if $I$ is an invariant subspace of the adjoint
representation.

\section{Representations of Lie groups}\label{univ.cover}

Lie group representations have a similar definition as Lie algebra
representations.

\begin{definition}
A \idx{representation of a Lie group} $\Gz$ in an associative algebra
 $\Ez$ with identity $1$ is a map $U:\Gz \rightarrow\Ez$ such that
\[
U(fg)=U(f)U(g), ~~~U(1)=1.
\]
The representation is called {\bfi{faithful}} if $U$ is injective.
If $\Ez=\Lin\Hz$, one speaks again of a {\bf\idx{linear representation}}
on $\Hz$. A linear representation is called {\bfi{irreducible}} if
the only subspaces closed under multiplication by linear mappings
of the form $U(f)$ are $0$ and $\Hz$.
A {\bfi{unitary representation}} of a Lie group $\Gz$ is a linear
 representation in the $*$-algebra $\Ez=\Lin\Hz$ of continuous
linear operators of a Euclidean space $\Hz$, satisfying
\[
U(f)^*U(f)=1.
\]
\end{definition}

It is easy to see that $U(f^{-1})=U(f)^{-1}$, and in the unitary case,
$U(f^{-1})=U(f)^{-1}=U(f)^*$.

Note that the invertible elements of $\Ez$ form a group and a Lie
group representation of $\Gz$ in $\Ez$ is a group homomorphism of
$\Gz$ into this group. Again, if the representation is faithful,
one may identify group elements with their images under the
representation, and then has an {\bfi{embedding}} of $\Gz$ into the
algebra $\Ez$. Thus if $\Ez$ is the algebra of $n\times n$ matrices
with entries in $\Kz$ we get a group homomorphism of $\Gz$ into
$GL(n,\Kz)$. For $\Kz=\Cz$ the representation is unitary if the image
of $\Gz$ lies inside $U(n)$.

If a Lie algebra representation $J:\Lz \rightarrow \Ez$ is an
embedding, we can get something that is close to a representation 
of the Lie group by exponentiation, i.e., by defining
\[
U(e^f):= e^{J(f)}= \sum_{k=0}^\infty \frac{J(f)^k}{k!}\,,
\]
provided this converges for all $f\in\Lz$ in the topology of
$\Ez$. In Subsection \ref{univ.cover} we go deeper into the question
of how to get a Lie group representation from a Lie algebra
representation and the problems one encounters.
On the other hand, given a representation $U$ of a Lie group $\Gz$
with Lie algebra $\Lz$
we can get a representation $J$ of the Lie algebra
by differentiation, i.e., by defining
\[
J(X):= \frac{d}{dt} U(e^{tX})\Big|_{t=0}\,,
\]
provided the derivative always exists. In finite dimensions, both
constructions work generally; in infinite dimensions, suitable
assumptions are needed to make the constructions work.

The group $\Gz$ acts on the Lie algebra $\Lz$. We will discuss this
shortly for groups of matrices. For every element $g\in \Gz\subset
GL(n,\Kz)$ we define $\Ad(g)$\index{$\Ad$} which is a linear 
transformation of $\Lz$ given by
\[
\Ad (g):X\mapsto gXg^{-1}\,.
\]
It holds that $\Ad(g)X\in\Lz$, which we will not prove. The interested
reader is referred to 
\sca{Knapp} \cite{knapp},
\sca{Helgason} \cite{helgason},
\sca{Frankel} \cite{frankel}, or
\sca{Kirillov} \cite{kirillovLiegroup}. 
For all the
examples discussed so far, the reader can check it by hand. The map
$\Ad:g\to \Ad(g)$ clearly satisfies $\Ad(gh)=\Ad(g)\Ad(h)$ and is thus
a representation, which is called the {\bf
\idx{adjoint}}\index{representation!adjoint} representation of
the group $\Gz$.

\bigskip
\bfi{Universal covering group.}
For Lie algebra representations an important construct is the
universal enveloping algebra. For Lie groups there is an
analogue. Above we mentioned that by differentiating a representation
of a Lie group, one obtains a representation for the corresponding Lie
algebra. By exponentiating a representation of the Lie algebra one
gets a representation for those group elements that can be written as
exponents. If a group is not connected, one does not obtain a
representation of the group in this way. 

Other problems arise when the
group is not simply connected. \at{not yet defined}
For example $SO(3)$ is not simply connected \at{why?}
and therefore certain representations of the Lie algebra
cannot be lifted to representations of the Lie group; the spin
representations become multivalued. Even other problems arise when
two Lie groups that are fundamentally different have isomorphic Lie
algebras. Consider for example the group $U(1)$ of complex numbers of
absolute value $1$. As a manifold $U(1)$ is just the circle $S^1$. 
The Lie
group of $U(1)$ is the one-dimensional abelian Lie algebra (there is
only one). Now consider the Lie group $\Rz$ where the group operation
is addition $a\cdot b = a+b$. Then $\Rz$ is a one-dimensional abelian
Lie group with a one-dimensional Lie algebra. The Lie algebras of
$U(1)$ and $\Rz$ are isomorphic, but the Lie groups are totally
different. When we want to lift a Lie algebra representation of either
of them to a Lie group representation, which group do we choose then?

These topological considerations
lead one to the question whether there is a unique simply connected
Lie group for a given Lie algebra. The answer is positive: for every
real finite-dimensional Lie algebra $\Lz$ there is a unique simply 
connected
Lie group $\Gz$ with Lie algebra $\Lz$. So given a Lie group $\Hz$
with Lie algebra $\Lz$ one
can construct a unique simply connected Lie group $\Gz$ with Lie
algebra $\Lz$. The group $\Gz$ is called the 
{\bfi{universal covering group}} of $\Hz$. 
Then $\Hz$ and $\Gz$ are locally isomorphic; there
are small neighborhoods of the origin in both groups on which $\Hz$
and $\Gz$ are diffeomorphic to each other. The Lie group $\Hz$ is then
a quotient of $\Gz$; $\Hz\cong \Gz / D$ for some discrete normal
subgroup $D$ of $\Gz$.

The exponential map $\Lz\to \Gz$ is in general not surjective,
however, the image of the exponential map generates an interesting
subgroup of $\Gz$, the connected component of $\Gz$, denoted
$\Gz_0$. If $g\in\Gz$
lies in the connected component, we can write $g=e^{f_1}\cdots e^{f_r}$
for some Lie algebra elements $f_1,\ldots,f_r$. \at{explain why}
Given a Lie algebra
representation we can uniquely lift it to a representation of the
connected component of the Lie group $\Gz_0$ if $\Gz_0$ is simply
connected. Therefore, in this case, the representations of the Lie 
algebra $\Lz$ are in a one-to-one correspondence with the 
representations of the universal covering group  corresponding to $\Lz$.

Now let $\Hz$ be a Lie
group with Lie algebra $\Lz$ and with universal covering group $\Gz$
such that $\Hz\cong \Gz/K$ for some normal subgroup $K$ of
$\Gz$. \at{which will be discrete. Why?}
Given a Lie algebra representation of $\Lz$, we get a Lie
group representation of $\Gz$. If the normal subgroup $K$ is in the
kernel of the representation, we get a well-defined representation of
$\Hz$ as well. Conversely, given a representation of $\Hz$, we get
a representation of $\Gz$ by first projecting to $\Gz/K$, so that $K$
is in the kernel. Hence, representations of $\Hz$ are in one-to-one
correspondence with representations of $\Gz$ that map $K$ to the unit
matrix.

\section{Finite-dimensional semisimple Lie algebras}\label{sec-fin-dim}

For finite-dimensional Lie algebras a lot is known about the general
structure; here we give an overview over the results most useful in 
physics. Since no details are given, this section may be skipped on 
first reading.

Classifying all finite-dimensional Lie algebras is in a
certain sense possible; all finite-dimensional Lie algebras are a
semidirect product of a semisimple and a solvable Lie algebra (to be
defined below). 
The classification of all semisimple real and complex Lie algebras is
completely understood. It turns out that the semisimple complex Lie 
algebras can be classified by studying certain root systems. 
The semisimple real Lie algebras are obtained by applying the 
classification of complex Lie algebras to the complexified Lie algebras 
and then finding all ways of turning the resulting complex Lie algebras
into a Lie $*$-algebra; their real parts then give all semisimple real 
Lie algebras.

In the semisimple case, every representation is faithful; hence
a representation is nothing more than an embedding into a matrix 
Lie algebra $gl(n,\Cz)$, realizing the Lie algebra elements by matrices.
Every Lie algebra $\Lz$ comes with a canonical representation, the
{\bfi{adjoint representation}}\index{representation!adjoint}, 
denoted \idx{$ad$}, \at{defined several times?}
which maps an element $f$
to the Hamiltonian derivative $\ad_f$ in direction $f$, introduced in
Section \ref{sec.calc.lie}. Thus to each Lie algebra element $f$ we
assign a linear operator on a vector space. The vector space is the
Lie algebra itself and an element $f$ of the Lie algebra is represented
by the linear transformation $\ad_f$ that maps an element $g\in\Lz$
to $f\lp g$. In the mathematical literature, one often writes the 
Lie product as a commutator. Then the definition takes the form
\[
ad_x(y) = [x,y]\,.
\]
Due to the Jacobi identity this indeed defines a representation.
For finite-dimensional Lie algebras, there is a canonical symmetric
bilinear form called the 
\bfi{(Cartan--)Killing form}\index{Cartan--Killing form}\index{Killing 
form}, which we write as $B_{CK}$\index{$B_{CK}$, 
(Cartan--)Killing form} and 
defined by \at{change notation to $(x,y)_K = \iint xy$!}
\[
B_{CK}(x,y)=\tr (ad_x ad_y)\,.
\]
Due to the Jacobi identity, the Killing form is invariant,
\[
B_{CK}([x,z],y)=B_{CK}(x,[z,y])\,.
\]
Recall that an ideal of a Lie algebra $\Lz$ is a subspace $I$ in $\Lz$ 
such that  $\Lz\lp I\subseteq I$. Thus an ideal is an 
invariant subspace under the adjoint action of the Lie algebra on 
itself. 
A Lie algebra $\Lz$ is called {\bfi{simple}} if it is not
one-dimensional and has no nontrivial ideals (distinct from $0$ and 
$\Lz$). Thus the adjoint action of $\Lz$ on itself has no nontrivial 
invariant subspace. A Lie algebra is {\bf\idx{semisimple}} if it is a 
direct sum of simple Lie algebras. 
There is a convenient criterion for a Lie algebra to be semisimple:

\begin{thm}\label{lem.cartan}\bfi{(Lemma of Cartan)}\\
A Lie algebra is semisimple if and only if its Killing form is 
nondegenerate.
\end{thm}
\bepf
The proof can be found in many Lie algebra textbooks such as
\sca{Jacobsen} \cite{jacobsen}, 
\sca{Humphreys} \cite{humphreys},
\sca{Knapp} \cite{knapp}, or 
\sca{Fulton \& Harris} \cite{fultonharris}.
\epf

A finite-dimensional real Lie algebra $\Lz$ is called \bfi{compact} 
if its Killing form is negative definite. In this case, 
the Lemma of Cartan implies that $\Lz$ is semisimple.
For example, the Lie algebra $so(3)$ is compact, whereas
$so(2,1)$ is noncompact.
\at{where remarked?}
However, note that Lie algebras are vector spaces and therefore not
compact as topological spaces in the usual topology.

For a given Lie algebra one may form the so-called 
{\bfi{lower central series}} (or {\bfi{derived series}})
of ideals:
\[
\Lz_0 = \Lz\,,~~~ \Lz_{n+1} = \Lz_n\lp \Lz_n\,, ~~n\geq 0\,.
\]
The Lie algebra $\Lz$ is called {\bfi{solvable}} if there is an 
$n$ such
that $\Lz_n =0$. A theorem of Levi \at{ref} 
says that every Lie algebra is a semidirect sum \at{undefined} 
of a semisimple part $P$ and a solvable ideal $S$
(that is, $S$ is a solvable Lie subalgebra that is an ideal in $\Lz$),
such that $P\cong \Lz /S$. It follows that an important part of the
classification of all Lie algebras is the classification of the simple
Lie algebras.

The classification of the finite-dimensional complex simple Lie
algebras can be done by classifying certain objects called finite
{\bfi{root systems}}, associated to a choice of maximal
commutative subalgebras called {\bfi{Cartan subalgebras}}.
Associated to each root system is a finite {\bfi{reflection group}},
i.e., a group generated by elements whose square is 1. The finite
reflection groups (also called {\bfi{Coxeter groups}}) which are not
direct products of nontrivial smaller reflection groups arise as
symmetry groups of regular polytopes. They have all been
classified by Coxeter, and fall into five infinite families denoted
 by $A_n$ (simplices), $B_n$, $C_n$, $D_n$ (all three related to cubes
and crosspolytopes), and $I_n$ (polygons), and a few sporadic cases
denoted by $E_6, E_7, E_8, F_4, H_3$, and $H_4$ ($H_3$ is the symmetry
group of the dodecahedron and the icosahedron).

Most of the finite reflection groups are also realized as symmetry
groups of a root system. All root systems give rise to semisimple
Lie algebras, and irreducible root systems lead to simple Lie algebras.
The classification says there are four infinite series of
Lie algebras denoted $A_n$, $B_n$, $C_n$ for $n\geq 1$ and $D_n$ for
$n\geq 4$ and five exceptional Lie algebras called $E_6$, $E_7$, $E_8$,
$G_2$ and $F_4$. The corresponding reflection groups have the same
labels, except for $G_2$ which corresponds to the hexagon $I_6$.
It is a highly nontrivial result -- and one of the most beautiful pieces
of mathematics -- that this gives a complete classification of the
finite-dimensional semisimple complex Lie algebras.

The four infinite series of Lie algebras, called the {\bfi{classical
Lie algebras}}\index{Lie algebra!classical}, are realized geometrically
as infinitesimal symmetry groups of certain bilinear forms, i.e.,
Lie algebras of linear transformations with zero trace whose
exponentials leave the form invariant. The Lie algebras
$A_n$ are isomorphic to the \bfi{special linear} Lie algebras\index{Lie
algebra!special linear} \at{already defined}
$sl(n+1,\Cz)$ of $(n+1)\times (n+1)$-matrices with complex entries and
trace zero. The Lie algebras $B_n$ and $D_n$ are the odd and even
\bfi{special orthogonal} Lie algebras\index{Lie algebra!special
orthogonal} $so(m,\Cz)$ ($m=2n+1$ and
$m=2n$, respectively), consisting of complex antisymmetric
$m\times m$-matrices. For the $C$-series we have
$C_n = sp(2n,\Cz)$, where the \bfi{symplectic} Lie algebras\index{Lie
algebra!symplectic}
$sp(2n,\Cz)$ \at{already defined}
are given by the complex $2n\times 2n$-matrices $X$ satisfying
$X^TJ + JX=0$ where $J$ is the antisymmetric $2n\times 2n$-matrix
given in block form by
\[
J = \pmatrix{ 0 & -1 \cr 1 & 0 \cr}\,.
\]
For each complex Lie algebra $\Lz$ of the $A$-, $B$-, $C$- or
$D$-series, there is an associated (simply connected) Lie group 
\at{not yet introduced}
denoted by the same, but capitalized letters, whose complexified 
tangent space at the identity
coincides with the Lie algebra $\Lz$.

There is some redundancy in the nomenclature for low-dimensional Lie
algebras: $A_1\cong B_1\cong C_1$, $ B_2\cong C_2$.
It is easy to check that $sp(2,\Cz) \cong sl(2,\Cz)\cong so(3,\Cz)$.
The Lie algebra $so(2)$ is one-dimensional (and hence abelian) and
therefore not simple. The Lie algebra $so(4,\Cz)$ is in fact
semisimple,
$so(4,\Cz)\cong so(3,\Cz)\oplus so(3,\Cz)$, and not simple since each
$so(3,\Cz)$-factor is a nontrivial ideal. The Lie algebra $so(6,\Cz)$ is
isomorphic to $sl(4,\Cz)$. For the just mentioned reasons, one starts
the $D$-series for $n\geq 4$; $so(8,\Cz)$ is the first in the series
that is not isomorphic to any other. In fact, $so(8,\Cz)$ is very
special in that it has a large automorphism group (related to
{\bfi{triality}}). \at{both undefined}
For the exceptional simple Lie algebras $E_6,E_7,E_8,F_4$ and $G_2$,
there is no simple geometric description as for those in the
$A$-, $B$-, $C$- and $D$-series.
However, the exceptional simple Lie algebras can be realized as
infinitesimal symmetry groups of some algebraic structure. And to each
exceptional Lie algebra $\Lz$ one can associate a Lie group, such that
$\Lz$ is the complexification of the tangent space at the identity.

It is important to keep in mind over which field the Lie algebra is
considered. For example, over the real numbers the Lie algebras
$so(p,q)\equiv so(p,q;\Rz)$ are non-isomorphic, apart from the trivial
isomorphism
$so(p,q)\cong so(q,p)$. Over the complex numbers we have
$so(p,q;\Cz) \cong so(p+q,\Cz)$, since 
over the complex numbers the sign of a
nondegenerate symmetric bilinear form is not invariant.
Even more severe things are dependent of the field;
the real Lie algebra $so(1,3)$, which is extremely important in
physics, is simple, but extending the field to the complex numbers we
have $so(1,3;\Cz)\cong so(4,\Cz) \cong so(3,\Cz)\oplus so(3,\Cz)$,
which is not simple. (For applications to physics, this is actually
an advantage.)
However, this is as bad as it can get from the structural point of
view; if a Lie algebra is semisimple over some field $\Kz$, then it is
semisimple over all fields containing $\Kz$. 
This follows from the Lemma 
of Cartan \ref{lem.cartan}: If the Killing form is
nondegenerate over
some field, then extending the field does not change this property.

The (semi-)simple real Lie algebras can also be classified, albeit the
classification is a bit more complicated. See for example the books of
\sca{Gilmore} \cite{gilmore} (or, for the more mathematically minded,
\sca{Helgason} \cite{helgason} or \sca{Knapp} \cite{knapp}).
If a real Lie algebra $\Lz$ is simple, the complex extension --
letting the scalars be complex -- is either simple or of the form
$S\oplus S$ for a simple complex Lie algebra $S$. Hence the
classification of the real simple Lie algebras is still `close' to the
classification of the simple complex Lie algebras in the sense that no
completely new structures appear. It is an amusing historical fact
that \'Elie Cartan provided the classification of the complex simple
Lie algebras and his son, Henri Cartan, finished the project so to say
by classifying the real simple Lie algebras.

As we shall see in Chapter \ref{c.harmonic}, the unitary representations
\at{not yet defined}
of different real forms of the same complex Lie algebra can be quite
different.
The Lie algebra $so(2,1)$ does not admit a finite-dimensional unitary
representation, whereas $so(3)$ does. 
All compact Lie algebras admit a unitary
representation, and in fact, the adjoint representation is already
unitary. The main difficulty in the proof of this lies in establishing
that all compact Lie algebras admit a Lie $*$-algebra structure;
this requires more theory and will not be discussed here.
Since finite-dimensional unitary groups are compact,
noncompact semisimple Lie algebras cannot have finite-dimensional
unitary representations, apart from the trivial one which maps
everything to zero.

\section{Automorphisms and coadjoint orbits}

\at{adapt notation and move to the right place!}

The adjoint and coadjoint representations of a Lie algebra $\Lz$ 
extend to elements $g\in\Aut\Lz$ \at{this seems to require 
$\Aut\Lz \subseteq \Lin\Lz$} by defining
\[
 \delta^g:=g\delta g^{-1}=\Ad_{g}\delta \for \delta \in\Lz
\]
\[
 \omega^g=\Ad^{*}_{g^{-1}}\omega\for \omega \in\Lz^*
\]
with the properties
\[
 (\delta\lp\eps)^g=\delta^g\lp\eps^g,
\]
\[
 (\delta\lp\omega)^g=\delta^g\lp\omega^g,~~~(\omega\lp\delta)^g=\omega^g
 \lp\delta^g,
\]
\[
 \delta^g(\omega^g)=\delta(\omega)
\]
and for continuous motions $g\in C^1([0,1],\Aut\Lz)$,
\[
 \frac{d}{d\tau}\delta^{g(\tau)}=\frac{dg}{d\tau}(\tau)\lp\delta^
 {g(\tau)},
\]
\[
 \frac{d}{d\tau}\omega^{g(\tau)}=\frac{dg}{d\tau}(\tau)\lp\omega^
 {g(\tau)};
\]
in short,
\[
 (\delta^g)^\pdot=\dot g\lp\delta^g,~~~(\omega^g)^\pdot=\dot g\lp
 \omega^g.
\]
A set $\Omega\subseteq V$ is called \bfi{$\Lz$-invariant} (in a given
representation $Q$ of $\Lz$ on $V$) if,
\at{rather phrase it as equivalence relation; 
then orbits = equivalence classes and invariant = union of orbits}
for all $\delta \in C([0,1],\Lz)$ and all $\omega_0\in\Omega$ there is 
a unique $\omega \in C^1([0,1],\Omega)$ such that
\lbeq{eli13}
  \dot \omega(\tau)=Q(\delta(\tau))\omega(\tau),~~~\omega(0)=\omega_0.
\eeq
The set of points $\omega(1)$ reachable from a fixed $\omega_0$ in 
this way is called the \bfi{orbit} $\orb(\omega_0)$ of $\omega_0$.
The orbits partition $V$, and $\Omega$ is invariant iff it is a union of
orbits. The \bfi{coadjoint orbits} are the orbits in the coadjoint
representation on $\Lz^*$. Apparently, $\Omega=\orb(\omega)$ is a 
manifold homeomorphic to $\Aut\Lz/\stab(\omega)$, and the tangent space
at $\omega$ is 
\[
 T_\omega\Omega=\{Q(\delta)\omega\mid\delta\in\Lz\}.
\]
The coadjoint orbits correspond to maximal subgroups and are symplectic
manifolds with closed 2-form $\omega(f,g) := \tr \rho (f\lp g)$
\at{or so} The set of 
all $\omega\in\Lz^*$ for which a fixed set of casimirs takes fixed 
values is always invariant. 
\at{relate to irreducible representations, 
and give the Poincar\'e classification}

    \part{Nonequilibrium thermodynamics}\label{p.noneq}

\chapter{Markov Processes}\label{c.markov}

Part \ref{p.noneq} discusses the dynamics of nonequilibrium phenomena,
i.e., processes where the expectation changes with time, in as far as 
no fields are involved. 

It should be complemented (in a later stage of the book) by a treatment 
of space-time dependent thermodynamics, and its derivation from 
quantum field theory.

\bigskip
We first develop a formal mathematical language for representing the 
physical concepts related to experiments with quantum systems in an 
unambiguous way, such that the relations between the mathematical 
concepts precisely mirror the relations between the corresponding 
physical concepts. In particular, we define sources, activities, 
processes, observers, protocols and observables. 

In this way, phenomenological quantum physics gets a formal 
representation in the Platonic world of precise ideas, in the same way
as it has been custumary for centuries for mathematics.
 
A general formal framework for phenomenological quantum mechanics is 
given that allows a concise formulation of the problems of observation,
in a way close to real life.

It makes the ideas developed in quantum measurement theory, and in
particular the theory of positive operator valued measures 
(POVM's \at{crossref}) intuitive and useful for actual 
modeling. 

\at{New problems -- identifying protocols that approximate given 
observables -- are posed that cry for an answer. 
Relate to Helmstrom and Holevo}

\at{Everything will be illustrated by showing how linear classical and 
quantum optics fits into the framework. In particular,
we discuss quantum jump experiments.}

\bigskip
We then take the continuum limit of the present framework; it results 
in the traditional Lindblad theory of dissipative quantum processes.
However, we shall put the latter in the broader framework of Markov 
processes.

The most important class of nonequilibrium processes are the  
memory-less Markov processes. But by disregarding some variables in a 
Markov process, \at{to be done} one also finds a natural dynamics for
processes with memory. Since it can be argued that the memory in any 
physical process is due to hidden variables, Markov processes can be 
regarded as the fundamental processes, and we shall concentrate on the 
latter.

A Markov process is characterized in our set-up by a linear operator 
with properties resembling those of a derivation, and hence called a 
\bfi{forward derivation}. We shall discuss forward derivations in 
Section \ref{s.forward}, general Markov processes in Section 
\ref{s.markov}. 
\at{Then we discuss their simplest case, so-called Markov 
chains, in Section ???, and memory in Section ???, both to be written.
Positivity preservation is missing; so swap with M2.tex?}

\at{Also treat time-dependent Markov processes. 
Is every process with memory a contraction of a Markov process on 
equivalence classes of histories?}

\bigskip
The building blocks of phenomenological (classical or quantum) objects 
are {\em sources}, i.e., physical objects producing a definite state. 
For example, in optical experiments, a source is typically an object 
or arrangement that produces one or several light beams of a certain 
kind. On the formal level, sources are represented by certain monotone 
linear functionals.

Part of experimental physics consists in the art of 
devising real arrangements that {\em prepare} a source, i.e., that
produce output whose ensemble properties agree with that of a
formal source.

Sources are further modified by {\em conditioning}, i.e., subjecting 
them to one or several {\em activities} that change the output of a 
source. For example, in optical experiments, an activity may be passing 
a light beam through a beam splitter or an optical filter. On the formal
level, activities are represented by certain monotone linear operators.
Activities and how they condition sources are discussed in 
Section \ref{activities}.

Informally, a {\em process} is a description of everything that may
happen to the output of a source while passing through an arrangement
of physical equipment. Since experiments are not completely 
reproducible, they need a stochastic description; thus, we describe
processes by a (classical) probability distribution on the possible 
activities that characterize the corresponding possible changes. 

The relation between activities and real-life observations is 
established by an {\em observer} who classifies the activities 
according to more or less objective principles, resulting in classical 
{\em records}. Further processing of the records according to
established scientific standards yields {\em protocols}
that can be communicated by classical means. Associated to each 
protocol is a set of {\em observables} defined by the protocol.
Processes, observers, and protocols are discussed in 
Section \ref{processes}.

\section{Activities}\label{activities}

The building blocks of phenomenological (classical or quantum) objects 
are {\em sources}, i.e., physical objects producing a state. 
For example, in optical experiments, a source is typically an object 
or arrangement that produces one or several light beams of a certain 
kind. On the formal level, sources are represented by certain monotone 
linear functionals.

A \bfi{source} is a monotone $*$-linear functional $E$ on the space 
$\Ez$ of quantities, i.e., a mapping 
$E : \Ez \to \Cz$ 
\at{in principle we might use $\Pol \Ez$ in place of $\Ez$ everywhere}
satisfying
\[
   E(f+g)=E(f)+E(g), ~~~E(\lambda f) = \lambda E(f),
\]
\[
   E(f^*f) \geq 0, ~~~E(f^*)=E(f)^*.
\]
A source is \bfi{proper} if it has a finite and positive 
\bfi{partition function}
\[
   Z:=E(1) \in~] 0, \infty [.
\]
(This is a function $Z(\theta)$ only when the source $E=E_\theta$ 
is parameterized by one or several control variables, collected in the 
vector $\theta$.)

Typical examples of sources are:
\[
   E(f) = f(\omega)~~~\mbox{(pure classical source)}
\]
if $f$ is a quantity from an algebra $\Ez$ of functions of a set 
$\Omega$ and $\omega \in \Omega$,
\[
   E(f) = \psi^*f \psi~~~\mbox{(pure quantum source)}
\]
if $f$ is a quantity from the algebra $\Ez$ of linear operators on
a Hilbert space $\Hz$ and $\psi \in \Hz$,
\[
   E(f) = \sint e^{- \beta H} f~~~\mbox{(canonical thermal source)}
\]
for an arbitrary Hermitian Hamiltonian $H\in\Ez$ and an inverse 
temperature $\beta$ such that $Z=\int e^{- \beta H}$ is finite.

The \bfi{ensemble} associated with a proper source is defined by the 
expectation functional $\<E \cdot\>$ that associates with a quantity 
$f$ the \bfi{expectation}
\[
   \<Ef\>: = E(f) /E(1).
\]
Sources that are multiples of each other are equivalent and define the 
same ensemble. Sources form a closed convex set:

\begin{prop}
Let $E_\alpha$ be a family of sources.

(i) For any directed set order on the $\alpha$, the limit
$E=\lim_{\alpha} E_\alpha$ defined by
\[
   E(f) = \lim_{\alpha} E_\alpha (f),
\]
is again a source. 

(ii) For any probability measure 
$d \mu (\alpha)$ (with $\int d \mu (\alpha)=1$), the
\bfi{convex combination} $E= \int d_\mu (\alpha)E_\alpha$ of the 
$E_\alpha$, defined by 
\[
   E(f) = \int d \mu (\alpha) E_\alpha (f),
\]
is again a source. In particular, if $p_\alpha$ are nonnegative
numbers with sum $1$ then the weighted sum $E= \sum p_\alpha E_\alpha$
of sources is a source.
\end{prop}

\bepf
Straightforward.
\epf

Part of experimental physics consists in the art of 
devising real arrangements that {\em prepare} a source, i.e., that
produce output whose ensemble properties agree with that of a
formal source $E$. If a family of sources $E_\alpha$ is already 
available, arbitrary convex 
combinations of these sources can be realized in practice by 
randomizing the selection of sources in sufficiently narrow 
time intervals. 

\bigskip
Sources are further modified by {\em conditioning}, i.e., subjecting 
them to one or several {\em activities} that change the output of a 
source. For example, in optical experiments, an activity may be passing 
a light beam through a beam splitter or an optical filter such as a
polarizer or an absorber. On the formal
level, activities are represented by certain monotone linear operators.

An \bfi{activity} is a monotone *-linear mapping $A:\Ez \to \Ez$ 
\[
   A(1) \leq 1;
\]
$A(1)$ is called the \bfi{effect} of $A$. An activity is called 
\bfi{conservative} if it is invertible and 
\[
A(fg)=A(f)A(g) \forall f,g\in\Ez;
\]
i.e., if it is a $*$-automorphism of $\Ez$.
An activity is called \bfi{autonomous} if
\lbeq{ac.1}
   A(f) = E_A(f)g_A,~~~0 \leq g_A \leq E_A(1)^{-1}
\eeq
for some source $E_A$ and some $g_A \in \Ez$, and \bfi{primitive} if 
\lbeq{ac.2}
   A(f) = L^* f L
\eeq
for some quantity  $L \in \Ez$ with
\[
L^*L \leq 1, 
\]
called the \bfi{Lindblad operator}
of the activity. (Nonprimitive activities have no associated
Lindblad operator.)
A \bfi{von-Neumann activity} is an activity with $A^2 = A$, e.g., a
primitive activity whose Lindblad operator is a projector $L$ 
(satisfying $L^2 = L$). 

A primitive activity with Lindblad operator $L$ is conservative iff 
$L$ is unitary, and a von-Neumann activity iff $L$ is idempotent, 
$L^2 = L$. 

Primitive activities with Lindblad operators 
\[
   L=e^{-\iota tH}
\]
with a Hermitian \bfi{Hamiltonian} $H$ describe conservative unitary 
evolution (simply passing time). 

A \bfi{screen} is an orthogonal projector 
to an invariant subspace of the position operator, i.e., to the closed
subspace spanned by some Borel set of position eigenstates.
For classical algebras, the corresponding Lindblad operator is a 
characteristic function, for quantum algebras an orthogonal projector. 
Thus (displaying on) a screen is a von-Neumann activity.

In real applications to quantum systems (for quantum optical devices,
see, e.g., {\sc Leonhardt \& Neumaier} \cite{LeoN}), we typically have 
dissipative systems described by \bfi{dissipative Lindblad operators} 
of the form
\[
   L = e^a~~~\mbox{where}~~~a \in \Ez,~\re a \leq 0.
\]
Note that $\re a \leq 0$ implies $L^*L \leq 1$: From
\[
\frac{d}{dt} (e^{ta})^*e^{ta} =(e^{ta})^*(a+a^*)e^{ta} \le 0,
\]
we see that $b(t):=(e^{ta})^*e^{ta}$ is monotone decreasing; hence 
$L^*L=b(1)\le b(0)=1$.
 
\at{Note that we may have $L^*L=1$ but $LL^* \neq 1$; 
e.g., classically, ``dropping the first index''.} 

\at{Classical systems require $A(f)=e^{tD}(f)$ with a forward 
derivative $D$ since conservative motion is described by {\em outer} 
automorphisms.}

\begin{thm} ~

(i) If $A$ is an activity then its \bfi{transpose} $A^T$, defined by
$A^T(f)=A(f)^T$, is an activity.

(ii) The product $A=A_1 \dots A_n$ of a sequence $A_1,\dots,A_n$ of 
activities is again an activity. 

(iii) Activities form a closed convex set.

(iv) When $\Ez$ is the algebra of bounded linear operators on a 
Hilbert space then all activities are convex combinations of primitive 
activities, transposed primitive activities and autonomous activities.
\end{thm}

\bepf
(i)--(iii) are straightforward, and (iv) follows from a well-known 
classification theorem; see, e.g.,  
{\sc Davies} \cite[Section 3.1]{Dav}.
\epf

(For classical $\Ez$, the analogue of (iv) is wrong.)

\bigskip
For any source $E$ and any activity $A$, the product $EA$ is again a 
source, the source $E$ \bfi{conditioned} by $A$. The associated 
ensemble has the \bfi{conditional expectation}
\[
   \<EAf\> = EA(f)/EA(1).
\]
For an autonomous activity \gzit{ac.1}, $\<EAf\> = \<E_A(f)\>$ 
independent of the source $E$. Thus autonomous activities wipe 
out the information in the input.

A source $E$ can be conditioned by a sequence $A_1,\dots, A_n$ of 
activities, resulting in the \bfi{conditioned source} $EA_1 \dots A_n$.
Except for von-Neumann activities, conditioning by repeating an activity
generally gives results different from conditioning by the single 
activity, $EAA \neq EA$ (e.g., when applying several absorbers in turn
to an optical system).

\at{adapt this -- von Neumann is already the activity picture!
$E^*$ must be defined! Relations to time correlations?
Heisenberg is an automorphism only for conservative systems, and is not
very useful in the dissipative case. Perhaps delete this!}
Activities can be considered in the pictures of 
Heisenberg and von Neumann: \at{explain connection}
\[
  f \to L^*fL = E(f),~~~\psi \to L \psi,~~~
  \rho \to L \rho L^* = E^*(\rho).
\]
Primitive activities (which preserve or reduce the rank) can also be 
considered in the Schr\"o\-dinger picture; nonprimitive activities 
have no associated Schr\"odinger picture.

\bigskip
In physics, one frequently passes from a fundamental description in
terms of microscopic quantities to coarse-grained description in terms
of certain effective quantities of interest. The effective quantities
form a subalgebra, and we may look at the consequences of restricting
attention to such a subalgebra.

A \bfi{restriction} is an idempotent $*$-linear mapping $\Pi $ from 
$\Ez$ onto a subalgebra $E_\Pi $  of $\Ez$. Typical examples are
\[
   \Pi (f \otimes g) = f
\]
(projecting away so-called {\em heat bath} quantities $g$),
\[
   \Pi (f) = \Diag~(f)
\]
(projecting quantum observables to corresponding classical observables
in the maximal commuting subalgebra of diagonal operators). The latter
may be combined with a basis change, giving
\[
   \Pi (f) = P  \Diag~(P ^*fP )P ^*~~~
\mbox{for}~ P : \Hz' \to \Hz,~~~P ^*P =1.
\]
To each source $E$ of $\Ez$ we associate the \bfi{restricted source} 
$E_\Pi $ on $\Ez_\Pi $ with 
\[
   E_\Pi  (f) = E\Pi (f) = E(f) ~~~\mbox{for}~f \in \Ez_\Pi ,
\]
and to each activity on $\Ez$ the \bfi{restricted activity} $A_\Pi $ 
on $\Ez_\Pi $ with 
\[
   A_\Pi  (f) = \Pi A\Pi (f) = \Pi A(f)~~~\mbox{for}~ f \in \Ez_\Pi .
\]
Note that $(EA)_\Pi \neq E_\Pi  A_\Pi $ in general, whence the 
composition of activities (and sources) depends on the context in 
which they are described!

Over an algebra $\Ez = \Ez_\Pi \oplus \Ez_\env$, 
\bfi{separable} activities are those satisfying 
\[
  A(f \otimes g) = A_\Pi f \otimes  A_\env g.
\]
They behave well under restriction,
\[
   (EA)_\Pi (f) = EA (f \otimes 1) = E (A_\Pi  f \otimes 1) 
  = EA_\Pi (f) =  E_\Pi A_\Pi (f).
\]
\at{For classical algebras, separability corresponds to a block 
diagonal structure.}
Nonseparable activities are {\em entangled} and their restriction has 
no simple description.

However, {\em primitivity is not preserved} by restriction, whence 
restriction generally cannot produce truly pure sources.  
\at{Indeed, this makes questionable the  traditional assumption 
of pure states in quantum mechanics.}

\section{Processes}\label{processes}

Informally, a {\em process} is a description of everything that may
happen to the output of a source while passing through an arrangement
of physical equipment. Since (as far as human observations are 
concerned) nature is not deterministic, we describe
processes by a (classical) probability distribution on the possible 
activities that characterize the corresponding possible changes. 

\begin{figure}[htb]
\caption {Conditioning a source by a process}
\label{f.1}~
\begin{center}
\setlength{\unitlength}{1cm}
\begin{picture}(4,1)
\put(1,0.5){$E$}
\put(2.5,0.5){\fbox{$\mathcal{A}$}}
\put(4.2,0.5){$EA_\alpha$}
\multiput(1.4,0.65)(1.75,0){2}{\vector(1,0){1}}
\end{picture}
\end{center}
\end {figure}  

A \bfi{process} ${\cal A}$ changes a source $E$ into the conditioned 
source $EA_\alpha$ with 
probability density $d \mu(\alpha) EA_\alpha (1)/ E \ol{A} (1)$, 
cf. Figure \ref{f.1}. Here the $A_\alpha$ are activities  
indexed by \bfi{labels} $\alpha$, and $d\mu(\alpha)$ is a probability 
measure  on the set of labels, and the expression
\lbeq{e.ac3}
   \ol{A}:= \int d \mu (\alpha) A_\alpha
\eeq
is the \bfi{mean activity} of the process. 
A process is called \bfi{complete} if $\ol{A} = 1$. 
(Traditionally, a complete process is referred to as a 
{\em positive operator-valued measure}, or POVM; see, e.g., 
{\sc Davies} \cite{Dav}. We prefer to use the above more intuitive 
terminology.)

\at{Check the known examples whether we need 
$\ol A<1$. If needed, specialize the above to complete processes, 
otherwise replace. This also gives a list of examples.}

\at{add classical examples. I want to have Markov chains as a 
particular case! see notes on ``arrangements''}

\at{add the quantum jump experiment!}

\bigskip
The relation between activities and real-life observations is 
established by an {\em observer} who classifies the activities 
according to more or less objective principles, resulting in classical 
{\em records} that can be objectively processed.

An (ideal) \bfi{observer} is a mapping $r$ that associates with each 
label $\alpha$  a \bfi{record} $r(\alpha) \in R$; here $R$ is an 
arbitrary set containing the possible records. \at{Nonideal observers 
would associate records according to some probability distribution.}
The expression
\[
   \ol{A}_r:= \int d \mu (\alpha) \CHI_{r(\alpha)=r}A_\alpha
\]
is called the \bfi{mean activity} corresponding to record $r$. 
\at{is $E\ol A_r$ meaningful? Is the contracted process useful?}

\begin{prop}
Given a source $E$ and a process $\cal A$, the distribution of the 
observed records is described by the expectation
\lbeq{e.ac4}
   \<f(r)\> = \int d \mu (\alpha) EA_\alpha (f(r(\alpha))) /E\ol{A}(1)
\eeq
for {\em classical} functions $f:R \to \Cz$, hence is a classical 
probability distribution. 
\end{prop}
\gzit{e.ac4} follows from the definition of the change a process
causes, and defines an expectation since $\<1\>=1$ by \gzit{e.ac3}. 
\bepf

\epf

\at{can I define conditional expectation relative to \bfi{records} 
instead of activities? would be useful since we never know the precise 
activities}

\begin{figure}[htb]
\caption {History generation by observing a composite process}
\label{f.2}~

\begin{center}
\setlength{\unitlength}{1cm}
\begin{picture}(12,2.5)
\put(1.2,1.5){\fbox{$\mathcal{A}_1$}}
\put(4.4,1.5){\fbox{$\mathcal{A}_2$}}
\put(9.2,1.5){\fbox{$\mathcal{A}_n$}}
\put(0,1.6){\vector(1,0){1.2}}
\multiput(1.9,1.6)(3.2,0){2}{\vector(1,0){2.5}}
\put(7.8,1.57){$\dots$}
\put(8.4,1.6){\vector(1,0){0.8}}
\put(9.95,1.6){\vector(1,0){2.5}}
\multiput(1.55,1.325)(3.2,0){2}{\vector(0,-1){0.8}}
\put(9.6,1.325){\vector(0,-1){0.8}}
\put(0.4,1.75){$E$}
\put(2.6,1.75){$EA_{\alpha_1}$}
\put(5.5,1.75){$EA_{\alpha_1}A_{\alpha_2}$}
\put(10,1.75){$EA_{\alpha_1}\dots A_{\alpha_n}$}
\put(0.9,0.2){$r_1 = r(\alpha_1)$}
\put(4.1,0.2){$r_2 = r(\alpha_2)$}
\put(8.9,0.2){$r_n = r(\alpha_n)$}
\end{picture}
\end{center}

\end {figure}
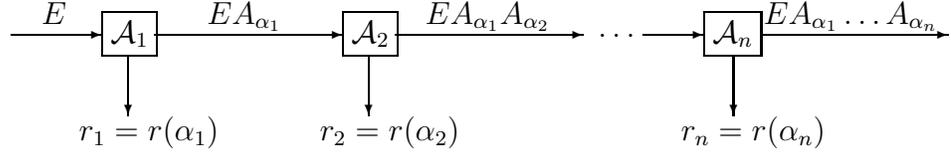

A \bfi{composite process} with \bfi{subprocesses} 
${\cal A}_1, \dots, {\cal A}_n$ is the family of activities 
\[
A_{(\alpha_1 \dots, \alpha_n)}: =  
A_{\alpha_1} \dots A_{\alpha_n}(A_{\alpha_s} \in {\cal A}_s)
\]
with associated product measure $d\mu(\alpha_1,\dots,\alpha_n) = 
d \mu_1 (\alpha_1) \dots d \mu_n ( \alpha_n)$. A corresponding 
sequence $(r_1, \dots,r_n) = (r (\alpha_1) \dots, r(\alpha_n))$ 
is called a \bfi{history}, cf. Figure \ref{f.2}.
\at{There may also be times $t_s$ associated with the subprocesses, 
with $t_1<\dots<t_n$, or space-time positions $x_s$ with 
$x_1<\dots<x_n$ in the causal partial order.}

Note that things happening at the same time (e.g., in quantum 
entanglement experiments) are considered to be part of a {\em single} 
activity of a composite process. This is an expression of the 
nonlocality of quantum mechanics, and assumes a preferred time axis 
(given in the relativistic case by the 4-momentum of the observer).

\at{unitary embedding -- ancilla = Squark's toy universe}

\bigskip
Further processing of the records according to
established scientific standards yields {\em protocols}
that can be communicated by classical means. Associated to each 
protocol is a set of {\em observables} defined by the protocol.

A \bfi{protocol} is a mapping $v:R \to \Cz^n$ that assigns to each 
record $r$ a vector $v(r)$, and hence (given an observer), to each
label $\alpha$ the vector $v_\alpha = v (r(\alpha))$. For any protocol 
$v$ with values in $D\subseteq \Rz^n$ and any function 
$\phi:D\to \Rz^m$, we define the protocol $\phi(v)$ with 
\[
\phi(v)_\alpha:=\phi(v_\alpha).
\]
Each protocol $v$ defines a vector of \bfi{observables}
\[
   \widehat v := \int d \mu(\alpha) A_\alpha (v_\alpha).
\]

\begin{thm}
For an arbitrary source $E$, the (observable) record expectation 
$\<v(r)\>$ is related to the (computable) ensemble expectation 
$\<E\widehat v \>$ by 
\[
   \<v(r)\> = \<E\widehat v\> / \<E\widehat 1\>.
\]
\end{thm} 

\begin{proof}
We have
\[
   E\widehat v  = E\int d\mu (\alpha) A_\alpha (v_\alpha) = \int d \mu 
   (\alpha) EA_\alpha (v(r(\alpha)))= \<v(r)\> E\ol{A}(1)
\]
by \gzit{e.ac4}, hence
\[
   \<E\widehat v \>=E\widehat v /E1=\<v(r)\> E\ol{A}(1)/E1
  =\<v(r)\>\<E\ol{A}(1)\>.
\]
Since $\ol{A}(1)=\widehat 1$, the result follows.
\end{proof}

Thus $\<E\widehat 1\> = \<E\ol{A}(1)\>\in[0,1]$ is the \bfi{efficiency} 
of a source $E$ for a given process. If the process is complete, all 
sources are 100\% efficient.

Note that the $A_\alpha(1)$ need not commute (randomize the decision 
of what to measure); thus we can jointly measure noncommuting 
quantities. \at{But how well?}

\bigskip
Scientific or industrial standards carefully define protocols for 
objectively observing key observables. The art of experimental 
design consists in finding protocols whose associated observables 
approximate a desired vector quantity $f \in \Ez^n$ as closely as 
possible. Note that generally $\widehat{\phi(v)}\neq \phi(\widehat v)$, 
so that operations on protocols are only approximately matched by the 
corresponding operations on the associated observables. 
\at{error estimates in terms of the uncertainty? 
Does $\widehat{v^*v}=\widehat v^*\widehat v$ imply 
$\widehat{\phi(v)} = \phi(\widehat v)$?}

\bfi{Conjecture.}
(in the 2-norm w.r. to $E$ or in the $\infty$-norm? Is there a related 
result in terms of protocols?)
\[
   [\tilde{f}, \tilde{g}]=0~~~\implies 
   \|f-\tilde{f}\| ~ \|g-\tilde{g}\| \geq \unc (f,g)
\]

One way to measure the quality of the approximation of a quantity $f$ 
by an observable $\widehat v$ is in terms of the \bfi{maximal deviation}
\[
   \|\widehat v -f\|
\]
which is defined independent of sources. Numerically, this leads to a 
semidefinite programming problem that can be approximately solved 
with high efficiency (see, e.g., \cite{Ali,Hel,VanB,WolSV}), 
namely
\[
  \bary{ll}
   \min & \lambda\\
   \mbox{s.t.} & - \lambda \leq {\D\sum_r} v(r)f_r-f\leq \lambda,
  \eary
\]
where
\[
   f_r = \int d_\mu(\alpha) \CHI_{r(\alpha)=r} A_\alpha(1).
\]
If the source is 
known, we may instead minimize the empirical expectation of the 
\bfi{surprise} ({\sc Neumaier} \cite{Neu.surprise})
\[
   s = \frac{1}{n} (v(r)-f)^* \Cov (f) (v(r)-f)
\]
where
\[
   \ol{f}=\<Ef\>,~~~\Cov (f) = \<E(f-\ol{f})(f-\ol{f})^*\>.
\]
This defines a least squares problem. Note that $\<s\>\geq1$, with 
equality iff $\widehat v =f$. \at{why?}

\at{we might want to minimize some distance involving 
$\widehat{\phi(v)}$ and $\phi(f)$ for some class of $\phi$'s.}

\section{Forward morphisms and quantum dynamical semigroups}
\label{a.semi}

This section is still \at{quite} incomplete.

Dissipative quantum systems are described by quantum dynamical 
semigroups and their associated evolution equations (often called 
Lindblad equations). The former are characterized in terms of what we 
shall call \bfi{forward morphisms}, satisfying a relaxed version of the
laws for a homomorphisms;
the latter are defined in terms of what we shall call \bfi{forward 
derivations}, which relax the property of a derivation, and are 
discussed in the next section.

\at{the following definition is motivated by my core interpretation!
So this section should preced the one on forward derivations.}

\begin{dfn}
Let $\Ez$ be a Euclidean *-algebra and $\Hz$ a Euclidean space. 
A *-linear map $A\in\Lin(\Ez,\Lin\Hz)$ is called \bfi{completely 
positive} if a \bfi{Stinespring factorization}
 \[
A(f)=L^*U(f)L \for f\in\Ez
\]
holds for some Euclidean space $\Fz$, some linear mapping $L:\Hz\to\Fz$,
and some homomorphism $U:\Ez\to\Lin\Fz$ (satisfying $U(1)=1$, 
$U(f^*)=U(f)^*$, and $U(fg)=U(f)U(g)$). 
\end{dfn}

\begin{prop}
If $A$ is completely positive then, for every $n=1,2,\dots$ and every
positive semidefinite $F\in\Ez^{n\times n}$, the matrix 
$A(F)\in\Lin(\Hz)^{n\times n}$ with component $A(F)_{jk}=A(F_{jk})$ is 
positive semidefinite, in the sense that for all $u\in\Cz^n$, the
operator $\sum_{j,k} u_j^* A(F_{jk})u_k^*$ is positive semidefinite.
\end{prop}

\at{this needs explanations -- we only need the case $n=2$.}

If all $f\in\Ez$ are bounded, these conditions are equivalent 
(Stinespring's theorem); then the latter property may be taken to be 
the definition of complete positivity.

\begin{dfn}
A \bfi{forward morphism} on a Euclidean *-algebra $\Ez$ is a *-linear 
functional $E:\Ez\to\Cz$ such that
\lbeq{e.forMor1}
E(f^*f)\ge E(f)^*E(f) 
\eeq
\lbeq{e.forMor2}
E(f^*f)\ge E(f)^*f+f^*E(f)-f^*E(1)f
\eeq
hold for all bounded $f\in\Ez$.
\end{dfn}

\at{We need the first condition for the positivity proof. The first 
condition implies $E(1)^2\le E(1)=E(1)^*$, hence $0\le E(1)\le 1$. 
We need the second condition for the generic Lindblad form, without 
any restriction on $E(1)$. Both follow from complete positivity. }

\begin{prop}
(i) The identity mapping is a forward morphism.

(ii) Convex combinations of forward morphisms are forward morphisms.

(iii) If $E$ is a forward morphism on $\Ez$ and $P\in\Lin\Ez$ satisfies 
$PP^*\le 1$ then the mapping $E_P:\Ez\to\Cz$ defined by 
$E_P(f):=P^*E(f)P$ is a forward morphism.
\end{prop}

Apparently, the set of forward morphisms is the smallest set with the 
properties stated in this proposition. This can be proved in case 
$\Ez$ is finite-dimensional.

\begin{prop}
If $E$ is a forward morphism on $\Ez$ and $P\in\Lin\Ez$ satisfies 
$PP^*\le 1$ then the mapping $E_P:\Ez\to\Cz$ defined by 
$E_P(f):=P^*E(f)P$ is a forward morphism.
\end{prop}

\at{To be done -- combine with activities.tex!!!}

\section{Forward derivations}\label{s.forward}

Let $\Ez $ be an Euclidean *-algebra.

\begin{definition} A mapping $D: \Ez \rightarrow \Ez$ is called a
\bfi{derivation} if it satisfies\\
(D1)~~$D(\alpha f + \beta g) = \alpha D f + \beta D g$ 
    for $\alpha, \beta \in \Cz$ , $f, g, \in \Ez$,\\
(D2)~~$D(f^*) = (Df)^*$ for $f \in \Ez $, \\
(D3a)~~$D (fg) = (Df) g + f(Dg)$ for $f, g \in \Ez $,\\
and a \bfi{forward derivation} if $(D1), (D2)$ hold and\\
(D3)~~$D(f^*f) \ge (Df)^* f + f^* (Df)$ for $f \in \Ez $\\
\at{We may also need continuity!?}
\end{definition}

\begin{proposition}
For a forward derivation, $ D1 \le 0$, and for a derivation, $D1 = 0$.
\end{proposition}

\bepf
Insert $f = 1$ into $(D3)$ and $f = g = 1 $ into $(D3_0)$.
\epf

\begin{examples} 
~\at{Rewrite -- verify $D(f^*f)$! 
In place of (iii) and (iv) consider $Df=\sum_j A_jfB_j$.
Note that (i) (ii) also hold for algebra valued $f$ and noncommutative 
$x,h$: $Df=(e^{-isH}fe^{isH}-f)/s$, a special case of (iii)}

(i) If $\Ez  = C^\infty (\Rz )$ then $D f (x) := f'(x)$ defines
a derivation. This example is responsible for the name.

(ii) If $\Ez  = C (\Rz ^n)$, $h \in \Rz ^n$ and $\eps > 0$ then the
\bfi{coarse-grained directional derivative}
\[
  D_hf(x):= \frac{ f(x + \eps h) -f (x)} { \eps }
\]
defines a forward derivation. Indeed, $(D1)$ and $(D2)$ are trivial, 
and since $D_h(ff^*) \fct{maps} x$ to
\[
  \bary{lll}
  D_h(ff^*) (x) &=& 
  \D \frac{ f (x+ \eps h) f^* (x + \eps h) - f(x) f^* (x)} {\eps} \\
  ~\\
  &=& \D \frac{ (f (x) + \eps D_h f (x)) ( f^* (x) + \eps D_hf^* (x)) - f(x) f^* (x)}{\eps}\\
  &=& \D D_hf (x) f^* (x) + f (x) D_hf^* (x) + \eps D_hf (x) (D_hf (x))^*,  
\eary
\]
we find
\[
  D_h(ff^*) - (D_hf) f^* - f(D_hf^*) = \eps (D_hf) (D_hf)^* \ge 0 
\]
since $\eps > 0$. Therefore, $(D3)$ holds.
\at{the central deriviative approximation $\half(D_h+D_{-h})$ is also 
a forward derivation!}

(iii) If $B$ is an arbitrary quantity then
\[
  Df:= 2B fB^* - BB^* f -f BB^*
\]
defines a forward derivation; $(D3)$ follows from [details?]
\[
  D(ff^*) - (Df) f^* - f (Df^*) = 2 [ B,f] [B,f]^* \ge 0.
\]
(iv) If $B$ is a quantity such that $B + B^* \ge 0$ then
\[
  Df := - Bf - fB^*
\]
defines a forward derivation. Again, only $(D3)$ is nontrivial and follows from
\[
  \bary{l}
  D(ff^*) -(Df) f^* - f(Df^*) \\
  = - B ff^* -ff^* B^* + (Bf + fB^*) f^* + f (Bf^* +fB^*) \\
  = f(B + B^*) f^* \ge 0.
  \eary
\]
(v) If $H$ is a Hermitian quantity then \at{change sign, and introduce 
$\iota$}
\[
  Df := i [H,f] = i(Hf -fH)
\]
defines a derivation.
Indeed, $(D1)$ is trivial, $(D2)$ follows from
\[
  (Df)^* =
   -i (Hf - fH)^* = -i (f^* H - Hf^*) = i(Hf^* - f^* H) = D f^*,
\]
and $(D3)$ from
\[
  \bary{l}
   D(H^*) = i(H ff^* -ff^* H) 
   = i((Hf - fH) f^* + f(Hf^* - f^* H))  \\
  = (Df) f^* + f(Df^*).
  \eary
\]
(vi) If $D$ is a derivation then $D^2$ is a forward derivation. Indeed, $(D1)$
and $(D2)$ are trivial, and
\[
  D^2 (ff^*) = D(D(ff^*)) = D \big( (Df)f^* + f (Df^*) \big)
               (D^2 f) f^* +_2 (Df) (Df)^* + f (D^2 f^*)
\]
so that
\[
  D^2 (ff^*) - (D^2 f) f^* - f(D^2 f^*) = 2(Df) (Df)^* \ge 0.
\]
In particular, $Df(x) := f'' (x)$ is a forward derivation.

(vii) Clearly, a nonnegative linear combinaton of forward derivations is 
again a forward derivation.

The fact that the derivation (v) is the special case $B=-iH$ of (iv), where
$B + B^* = 0$ can be extended to the following result.
\end{examples}

\at{prop: If $D_k$ are forward derivations and $G$ is positive semidefinite then $D=\sum_{j,k} G_{jk}D_jD_k$ is a forward derivation. Also give diagonal normal from and partial converses -- there is none for jump processes?} 

\begin{proposition}
If $D$ and $-D$ are forward derivations then $D$ is a derivation.
\end{proposition}

\bepf
Under our assumptions, $(D3)$ holds with equality. Therefore
\[
  \bary{l}
  D(ff^*) + \alpha D (gf^*) + \alpha^* D (fg^*) + \alpha \alpha^* D (gg^*) \\
  = D \big( (f + \alpha g) (f + \alpha g)^* \big) = \big( D (f + \alpha g) \big)
   (f + \alpha g)^* + (f + \alpha g) D(f + \alpha g)^* \\
  = (Df) f^* + \alpha (Dg) f^* + \alpha ^* (Df)g^* + \alpha \alpha^* (Dg)g^*  \\
  ~~~+ f(Df^*) + \alpha g (Df^*) + \alpha ^* f(Dg^*) + \alpha \alpha^* g(D g^*).
  \eary
\]
Again, since $(D3)$ holds with equality, this simplifies to 
\[
  \alpha D(gf^*) +\alpha^* D(fg^*) = \alpha \big( (Dg)f^* + g (Df^*) \big) + 
  \alpha^* \big( (Df) g^* + f(Dg^*) \big),
\]
If we add this equation for $\alpha = 1$ and $i$ times for $\alpha = i$,
we find
\[
  (1 + i) D(fg^*) = (1+i) (Df) g^* + (1+i)f(Dg^*),
\]
and this implies $(D3_0)$. 
\epf

\begin{dfn}
A quantity  $c \in \Ez $ is called \bfi{conserved} if
\lbeq{e2}
   D c = 0
\eeq
and
\lbeq{e3}  
  D(cf) = c(Df),~~~ D (fc) = (Df) c~~~\mbox{ for all } f \in \Ez .
\eeq
The forward derivation $D$ is called \bfi{dissipative} if every 
quantity $c$ satisfying
\lbeq{e4} 
   D(c^*c) = (Dc)^*c + c^*(Dc)
\eeq
is conserved, and \bfi{primitive} if it is dissipative and only 
constants are conserved.
\end{dfn}

The conserved quantities form a *-subalgebra of $\Ez$ containing $\Cz $.
Indeed, $D \alpha = 0$ for $\alpha \in C$ follows from \gzit{e3} for 
$f = \alpha$, and closure under addition, multiplication and 
conjugation is straightforward. 
\at{does this also hold if we impose \gzit{e4}?}
Therefore any algebraic
expression $f$ involving only conserved quantities is conserved, too.
In particular, we automatically have \gzit{e4} for conserved quantities.

\section{Single-time, autonomous Markov processes}\label{s.markov}

To motivate our abstract concept of a (single-time, autonomous) Markov 
process, we first consider two special cases: a classical deterministic 
dynamics and the quantum Schr\"odinger dynamics.

\begin{example}
(\bfi{Deterministic processes})
Let $x: \Rz  \rightarrow \Rz ^n$ be a solution of the differential
equation
\lbeq{e6}
   \dot{x} (t) = F (x(t),t),
\eeq
and
\[
   E = C^1 (\Rz ^n), ~~~ \fct{tr} f = \int dx^n f(x).
\]
If we define
\lbeq{e7}
   \< f \>_t : = f(x(t),t)
\eeq
then
\[
  \bary{rcl}
   \frac{d}{dt} \< f \> _t = \frac{d}{dt} f(x(t),t) 
   &=& \partial_x f(x(t),t) \cdot \dot{x} (t) + \partial_t f(x(t),t) \\
   &=& \partial_x f(x(t),t) \cdot F (x(t),t) + \partial_t f(x(t),t) \\
   &=& \< \partial_x f \cdot F + \partial_t f \>_t = \< D_F f + f \>_t,
  \eary
\]
where $D_F$, defined by 
\lbeq{e8}
   D_F f = F \cdot \partial_x f
\eeq 
is the \bfi{Lie derivative} with respect to $F$. Clearly, $D_F$ is a 
derivation on $\Ez $. Thus \gzit{e1} holds with $D = D_F$, and we have 
a reversible Markov process. \at{rewrite without referring forward!}
The general solution of \gzit{e1} with initial condition
\lbeq{e9x}
   \< f \> _{t_0} =  \int dz  \rho_0 (z)f (z,t_0)
\eeq
is easily seen to be
\lbeq{e10}
   \< f \> _t = \int dz  \rho_0 (z)f(x(z,t),t),
\eeq
where $x(z,t)$ is the solution of \gzit{e6} with $x(z,t_0) =z$. The particular
solution \gzit{e7} corresponds to the density $\rho_0 (z) = \delta (z-x (t_0))$,
and has vanishing variance
\[
  \< (f - \< f \> _t )^2 \>_t = \big( f (x(t),t) - \< f \>_t \big)^2 = 0,
\]
as one would expect from a deterministic process. However, unsharp initial
conditions \gzit{e9x} lead to solutions \gzit{e10} with, in general, nonzero,
variance.
\end{example}

\begin{example}(\bfi{Reversible classical mechanics})\label{ex.revC}
If we specialize the preceding example to an autonomous Hamiltonian system,
defined by
\lbeq{e11}
  \dot{q} = \frac{\partial}{\partial p} H(p,q),~~~~~\dot{p} = - \frac{\partial}
  {\partial q} H (p,q),
\eeq
for a Hamiltonian $H: \Rz ^n \times \Rz ^n \rightarrow \Rz $,
we find (for time-independent $f$) the reversible dynamics 
\lbeq{e1}
   \frac{d}{dt} \< f \>_t = \< Df \>_t ~~~~\mbox{ for } t \in \Rz , f \in
   \Ez ,
\eeq
with the \bfi{Poisson bracket}
\lbeq{e12} 
   Df = \{ f,H \} := \sum_\nu \left( \frac{\partial f}{\partial q_\nu}
   \frac{\partial H}{\partial p_\nu} - 
   \frac{\partial f}{\partial p_\nu}
   \frac{\partial H}{\partial q_\nu} \right)
\eeq
as associated derivation.
\end{example}

\begin{example}
\bfi{(Reversible quantum mechanics) }\label{ex.revQ}

Let $\psi$ be a solution of the time-dependent Schr\"odinger equation,
\lbeq{e13}
   i \hbar \frac{d}{dt} \psi = H \psi,
\eeq
defined in a Hilbert space $\Hz $ for a Hermitian Hamilton operator $H$, and
let $\Ez $ be the algebra of linear operators on $\Hz $, with standard 
trace. If we define
\lbeq{e14}
   \< f \>_t := \psi ^* f \psi
\eeq
then
\[
  \bary{rcl}
  \frac{d}{dt} \< f \>_t &=& \psi ^* f \psi) = \dot{\psi}^* \dot{f} \psi 
  + \psi^* \dot{f} \psi + \psi^* f \dot{\psi} \\
  &=& \left( \frac{H \psi}{i \hbar} \right) ^* f \psi + \psi^* \dot{f} \psi
  + f \frac{H}{i \hbar} \psi \\
  &=& \psi^* \left( - \frac{H}{i \hbar} f + \dot{f} + f \frac{H}{i \hbar}
  \right) \psi = \< \frac{i}{\hbar} [H,f] + \dot{f} \> _t \\
  &=& \< Df + \dot{f} \> _t,
  \eary
\]
where
\lbeq{e15}
   Df = \frac{i}{\hbar} [H,f]
\eeq
is a commutator. As we have seen, $D$ is a derivation on $\Ez $, and 
\gzit{e1} holds again. The general solution of \gzit{e1} can be 
expressed in terms of solutions of the Schr\"odinger equation, see
below \at{}.
\end{example}

We now discuss an axiomatic irreversible dynamics, of which our 
reversible examples are particular cases.
\at{adapt text; add later $D(t)$ and the multitime version.
It seems that the time-dependent case is covered by going from $\Ez$ 
to $C^1(\Rz,\Ez)$, but the latter lacks the right integral...}

\begin{definition}
A (single-time, autonomous) \bfi{Markov process} is a flow 
\at{explain} on the set of states of a Euclidean *-algebra $\Ez $ such 
that \gzit{e1} holds with a forward derivation $D$ on $\Ez $. 
We use a dot to denote differentiation with respect to time.
The process is called \bfi{reversible} if $D$ is a derivation, and 
 \bfi{dissipative} if $D$ is dissipative.

A \bfi{stationary state} of the process \gzit{e1} is a state with 
$\< Df \> = 0.$
\end{definition}

\begin{remarks}
(i) In \gzit{e1}, it is assumed that $f$ is independent of $t$; 
otherwise, the correct dynamics is given instead by
\lbeq{e5}
   \frac{d}{dt} \< f(t) \>_t = \< (Df)(t) + \frac{d}{dt}f(t) \> _t~~~~~
  \mbox{for } t \in  \Rz ,~ f \in C^1 (\Rz , \Ez ).
\eeq

(ii) For a reversible Markov process, the backward dynamics 
\at{yet undefined} is also a Markov process.

(iii) A stationary state is invariant under the dynamics \gzit{e1}.
The expectation of conserved quantities satisfies 
$\frac{d}{dt} \< f\> _t = 0$, and hence is time-invariant also for 
nonstationary states.
\end{remarks}

In reversible quantum mechanics (Example \ref{ex.revQ}), the conserved 
quantities are precisely the quantities commuting with $H$.

\begin{proposition} (\bfi{Schr\"odinger picture and Liouville equation})
\[
  \dot{\rho} = D^* \rho.
\]
\end{proposition}

 \at{Liouville version for the reversible examples}

\section{to be done}
\begin{itemize}
\item Markov chains. 
This is the case $\Ez = \Cz^n$ with pointwise operations
\item Conservative classical and quantum mechanics.
Here  $\Ez$ is a Poisson algebra and $Df=H\lp f$ for some 
Hamiltonian $H$.
\item General properties of Markov processes (the second law)
\item Low noise approximation (drift and diffusion)
\end{itemize}

\chapter{Diffusion processes}\label{c.diffusion}

In this chapter we describe an important class of classical primitive 
Markov processes, characterized by their drift vector and their 
diffusion matrix. 
As particular cases we obtain \bfi{damped Hamiltonian systems}, which 
are shown to converge towards the canonical ensemble, and another 
important case, the exactly solvable \bfi{Ornstein-Uhlenbeck processes},
which describe coupled damped harmonic oscillators.

Our Euclidean *-algebra is $\Ez  = \Ez _0 = {\cal S} (\Rz^n)$, the 
Schwartz space of rapidly decaying $C^\infty$ functions, and the trace 
is integration over $\Rz ^n$ 
\at{certain process dependent boundary conditions are admissible, too; 
we need $(\rho, D(\nabla) f) = (D(- \nabla) \rho , f)$.}
Partial integration then gives $\nabla^* = - \nabla.$

\section{Stochastic differential equations}

\begin{theorem}
Let $v$ be a vector field and let $G$ be a symmetric tensor field on 
$\Rz^n$, $G(x)$ positive semidefinite ($v$ and $G$ may be time 
dependent). Then
\[
  Df := v \cdot \nabla f + \half G : \nabla ^2 f
\]
defines a forward derivation on $E_0 = C^\infty (\Rz ^n)$, with drift 
vector $v$ and diffusion matrix $G$.

Moreover, if $G$ is definite then $D$ is primitive.
\end{theorem}

\bepf
*-Linearity is clear, and
\begin{eqnarray*}
  D(f,g) : &=& D(fg^*) - (Df)g^* - f (Dg^*) \\
           &=& u \cdot (\nabla (fg^*) -(\nabla f)g^* - f \nabla g^*) \\
           &~& + \half G: \nabla^2(fg^*) - \half (G: \nabla^2) g^* 
               - \half f (G:\nabla^2 g) \\
           &=& \half G: (\nabla^2 (fg^*) - (\nabla^2 f)g^* 
               - f \nabla^2 g^*) \\
           &=& G :(\nabla f)(\nabla g) =  (\nabla f)^T G (\nabla g). 
\end{eqnarray*}
In particular, $D(f,f^*) \ge 0$. Moreover if $G$ is positive definite 
then $D(f,f^*) > 0$ unless $\nabla f = 0$, i.e., unless $f$ is a 
constant. Hence in this case $D$ is primitive.

Finally, the drift vector is $Dx= v \cdot \nabla x = v$
and the diffusion matrix is $D(x,x^*) = (\nabla x)^T G(\nabla x) = G$.
\epf

\begin{definition}
The Markov process
\lbeq{e6.1.2.1}
   \frac{d}{dt} \< f \> = \< v \cdot \nabla f + \half G : \nabla^2 f \>
\eeq
is called the \bfi{diffusion process} with drift vector $v$ and diffusion 
matrix $G$.

We note that the low noise approximation of any Markov process is determined by
drift vector and diffusion matrix, hence any Markov process can be approximated
by a diffusion process when the noise is sufficiently small.
(This is the first approximation of the \bfi{Kramers-Moyal}, or \bfi{system 
size expansion} and gives a more accurate approximation than the low noise 
approximation.
The diffusion approximation treats slow time scales as infinitely slow; hence 
metastable states appear to be stable.) This accounts for
the importance of diffusion processes as approximations of more complex
Markov processes.
\end{definition}

\begin{proposition}
The Liouville equation for a diffusion process is the {\bf Fokker-Planck
equation} 
\[
  \dot{\rho} = D^* \rho 
   = -\nabla \cdot (v\rho) + \half \nabla ^2 : (G \rho).
\]
\end{proposition}
\bepf
By integration by parts of \gzit{e6.1.2.1}.
\epf

\begin{Remark}
Associated with the diffusion process \gzit{e6.1.2.1} is the 
\bfi{stochastic Ito differential equation}
\lbeq{e6.1.2.2a}
  dx = v(t,x) dt + B(t,x) dW(t),
\eeq
where $BB^T$ is an arbitrary factorization of $G$. Here $W(t)$ denotes 
the \bfi{Wiener process}, i.e., the special process \gzit{e6.1.2.1} 
with $v = 0$, $G = 1$, and $x$ is defined by \gzit{e6.1.2.2a} via 
stochastic integration. We shall not use this, and refer \at{} to 
Gardiner (4.3.18) and (4.3.3) for the equivalence of \gzit{e6.1.2.1} 
and \gzit{e6.1.2.2a}. To avoid the arbitrariness in the choice of $B$ 
we shall write \gzit{e6.1.2.2a} as
\lbeq{e6.1.2.2}
   dx = vdt + (Gdt)^\half = vdt + d \eps, ~~~d \eps \sim N (0, Gdt),
\eeq
which is suggestive in view of
\begin{eqnarray*}
  \< f(x (t + dt)) \> = \< f(x + dx)\> 
  = \< f + \nabla f \cdot dx + \half \nabla^2 f : (dx)^2 \> \\
  = \< f \> + \< \nabla f \cdot v + \half \nabla^2
  f : G \> dt = \< f\> + d \< f \>
\end{eqnarray*}
if we note that $\< ud \eps \> = 0$, $\< A: (d \eps)^2 \> = A: G$,
and $d \eps^3 = d \eps dt = dt^2 = 0$. We shall see that the definition 
\gzit{e6.1.2.1} is very satisfactory, and we need no integrals; these 
are needed for questions of existence and other, more mathematical 
developments.
\end{Remark}

\begin{Remark} 
Stochastic integrals from diffusion processes:
In $\hat{\Ez }$ (the extended algebra of Heisenberg operators, \at{} 
with fixed $ \< \cdot \> = \< \cdot \>_0)$, we \bfi{define} the 
relation \gzit{e6.1.2.2} with \bfi{arbitrary} operators 
$v,G$ (possibly unrelated to the operator $x$ to be equivalent with
\[
  \frac{d}{dt} \< f(t,x)\> 
  = \< \partial_t f(t,x)+v\cdot\nabla f(t,x)+\half G:\nabla^2 f(t,x) \>
\]
for all expolynomials $f$ (together with a multitime Markov property
\at{}). Then it is meaningful to ask for solutions of stochastic 
differential equations (where now a relation between $x,v,G$ is 
\bfi{prescribed}), and we can consider an existence and uniqueness 
theory. Clearly things are linear, and 
\[
  dx = v(t) dt,~ x(0) = x_0
\]
is solved by 
\[
  x = x_0 + \int_0 ^t v(t) v(t) dt
\]
so one has to give a meaning to  
$\int_0^t (G dt) ^\half$ which is an Ito-stochastic integral. 
With this as \bfi{definition}, it is easy to show that the \bfi{Ito 
transformations formula} holds:
\[
  df(x) = (v \cdot \nabla f + \half G : \nabla^2 f) dt 
  + \left( ( \nabla f)^T  G(\nabla f) dt \right) ^\half~~~~~
  \mbox{where } \nabla f = (f')^T.
\]
\at{This might give a \bfi{neat} introduction to the Ito integral.
Can I \bfi{define} it via path integrals constructively? (all in 
\bfi{one} $\hat{\Ez }$ ?) $Q:$ What about the det term? $\< \>$ seems 
to change, too.}
\end{Remark}

\begin{proposition}
(Ito transformation formula) 

If functions of $x$ satisfy the Markov process
\gzit{e6.1.2.1} then functions of $\bar{x} = \xi(x)$ satisfy the Markov
process $(\bar{1})$ with
\lbeq{e6.1.3.3}
  \bar{v} = \xi ' v + \half \xi '' : G,~~~ \bar{G} = \xi ' G \xi^{'T}.
\eeq
Thus, the Ito differential equation transforms according to
\lbeq{e6.1.3.4}
  d \xi (x) = (\xi ' v + \half \xi '' :G) dt + d \eta,~~d \eta \sim N(0, \xi '
  G \xi '^Tdt).
\eeq
\bepf Differentiation of $\bar{f}(\bar{x}) = \bar{f} (\xi(x)) = f(x)$
gives
\lbeq{e6.1.3.4a}
  (\nabla f)^T =
  (\overline{\nabla} \bar{f})^T \xi ' \Rightarrow \nabla f = \xi '^T 
  \overline{\nabla} \bar{f}
\eeq
and hence
\[
v \cdot \nabla f = v \cdot \xi^{'T} \overline{\nabla} \bar{f} 
= \xi ' v \cdot \overline{\nabla} \bar{f}. 
\]
Differentiating \gzit{e6.1.3.4a}
again gives
\[
  \nabla^2 f = \nabla (\xi^{'T} \overline{\nabla} \bar{f}) = \xi '' 
  \overline{\nabla} \bar{f} + \xi^{'T} \nabla (\overline{\nabla} \bar{f}) 
  = \xi'' \overline{\nabla} \bar{f}
  + \xi^{'T} \xi^{'T} \overline{\nabla}^2 \bar{f}.
\]
Therefore, 
$G:\nabla^2f=(\xi'':G)\ol{\nabla}\bar{f}+\xi'G\xi^{'T}:\ol{\nabla}^2f$,
so that
\[
  v \cdot \nabla f + \half G : \nabla^2 f = (\xi ' v + \half
  \xi '' : G) \overline{\nabla} \bar{f} + \half \xi ' G \xi^{'T} : 
  \overline{\nabla} ^2 f.
\]
\epf
\end{proposition}   

\begin{corollary} (Covariant form)

We may rewrite the Markov process \gzit{e6.1.2.1} as
\lbeq{e6.1.4.5}
   \frac{d}{dt} \< f \> = \< u \cdot \nabla f + \half \nabla \cdot 
   (G \nabla f) \>
\eeq
with the {\bf covariant drift}
\lbeq{e6.1.4.6}
   u^T = v^T - \half \nabla \cdot G;
\eeq
then $u$ transforms under $\bar{x} = \xi (x)$ according to
\lbeq{e6.1.4.7}
  \bar{u} = \xi ' u.
\eeq
[but $\< dx - udt \> \neq 0$ in general.]
\end{corollary}

\at{Can be find a condition under which $\bar{v}$ (or $\bar{u})$ 
becomes linear and $\bar{G}$ constant? This would solve the 
construction problem, and is simpler than the path integral approach, 
or even equivalent.}

\bepf
\begin{eqnarray*}
   \bar{u} ^T 
   &=& \bar{v}^T - \half \overline{\nabla} \cdot \bar{G} = v^T \xi^{'T}
   + \half G: \xi '' - \half \overline{\nabla} \cdot \xi ' G \xi^{'T} \\
   &=& v^T \xi^{'T} + \half G : \xi '' - \half \nabla \cdot (G \xi^{'T}) =
   ( v^T - \half \nabla \cdot G) \xi^{'T} = u ^T \xi^{'T}.
\end{eqnarray*}
\epf

One can also derive the \bfi{Stratonovic transformation formula}.
The \bfi{Stratonovic drift}
\[
  w^T : = u^T + \half (\nabla \cdot B) B^T , \mbox{ where } G = BB^T,
\]
transforms by
\begin{eqnarray*}
   \bar{w}^T &=& ( \xi' u)^T + \half (\nabla \cdot \xi' B) (\xi' B)^T \\
             &=& u^T \xi^{'T} + \half (\xi^{'T} \nabla \cdot B)( B^T 
                 \xi^{'T}) \\
             &=& (u^T + \half (\overline{\nabla} \cdot B) B^T) 
                 \xi^{' T} = w \xi^{'T},
\end{eqnarray*}
hence according to standard rules.
However, it depends on the factorization of $G$ and hence is less useful
than the covariant form \gzit{e6.1.4.5}.

\bfi{Note:} The equation $w^T = v^T -\half B : (\nabla B^T)$ gives the 
traditional translation between the Ito version and the Stratonovic 
version of a stochastic process.

\begin{proposition}
The Liouville equation for a diffusion process is a {\bf continuity 
equation}
\lbeq{e6.1.5.8}
   \dot{\rho} + \nabla \cdot j = 0,
\eeq
with the current
\lbeq{e6.1.5.9}
   j = u \rho - \tilde{G} (\nabla \rho),
\eeq
where $\tilde{G}$ is an arbitrary (possibly $x$-dependent) matrix with 
$G = \tilde{G} + \tilde{G}^T$.
\end{proposition}
\bepf 
We have $\dot{\rho} = D^* \rho$ and by symmetry,
$D = u \cdot \nabla + \half \nabla \cdot\tilde{G} \nabla$.
\epf

In particular, for the Wiener process $(u=0, \tilde{G} = \half G = \kappa,$
constant) we get the \bfi{diffusion equation}
\[
  \dot{\rho} = \half \nabla \cdot G \nabla \rho = \kappa \Delta \rho.
\]
In general, the current is composed of a streaming part (velocity $u$ 
times density $\rho$) and a diffusion part.

\begin{proposition}
The relative entropy increases, and takes its maximum at the 
equilibrium.
\at{max = $0$ ?, need not be attained as $t \rightarrow \infty$ ? 
Equilibrium need not be unique. Is the residual dynamics on the 
equilibrium states reversible?}
\end{proposition}
\bepf
\at{adapt. This must be copied to Chapter \ref{c.coll}, since the 
argument works for quasimonotone operators with a diffusion and jump 
part. 
In terms of Thirring IV $(2.1.5)$, $\rho (t)$ gets more \bfi{mixed}.}
For any convex $\Phi$ which is bounded below, and
\[ 
  f := \rho ^{-1} \rho_0,
\] 
we have
\begin{eqnarray*}
  \frac{d}{dt} \fct{tr} (\rho \Phi f )) 
   &=& \fct{tr} (\dot{\rho} \Phi (f) + \rho \dot{f} \Phi' (f)) 
   = \fct{tr} (\dot{\rho} [\Phi (f) - f \Phi' (f)]+ \dot{\rho}_0 \Phi' (f)) \\
   &=& \fct{tr} (\rho D [\Phi (f) - f \Phi' (f)] + \rho_0 D [\Phi' (f)]) \\
   &=& \fct{tr} (\rho \{ D[\Phi (f) - f \Phi ' (f)] + f D [\Phi ' (f)] \} ).
\end{eqnarray*}
Now the part in curly brackets simplifies to 
\[
  \bary{rcl}
- \Phi '' (f) (\nabla f)^T G (\nabla f) \\
- \int d \nu w (\nu,t,x) [\Phi (f(t, x + \nu)) 
       - \Phi (f(t,x)) - (f(t,x+ \nu) - f(t,x)) \Phi' (f(t,x + \nu))].
  \eary
\]
For convex $\Phi$, this is nonpositive. In particular,
and
\[
  \left.{
  \bary{l}
    \Phi (s) = s-1- \log s \\
    \Phi (s) = 1-s + s \log s 
\eary} \right\}
  \rightarrow \mbox{~standard relative entropy}
\left \{ \bary{l}
  \rho~ \log (\rho / \rho_0) \\
  \rho_0 \log (\rho _0  / \rho )
\eary \right.
\]
\at{and for $\Phi (s) = s^2$~~~$\rightarrow$, we get the previous 
proof of Chapter \ref{c.markov}}
\epf

\section{Closed diffusion processes}
We consider \bfi{closed systems} at constant temperature $T$. Closed systems 
are characterized by the fact that there is a current which vanishes at 
equilibrium. Thus any equilibrium state $\rho _{eq}$ is given by the solution 
if
\lbeq{e6.2.1}
   u \rho_{eq} = \tilde{G} ( \nabla \rho_{eq}), ~~\fct{tr} \rho _{eq} = 1.
\eeq
It is traditional to write some positive equilibrium state in terms of a
\bfi{thermodynamic potential} $\Phi$ (with units of energy) as
\[
  \rho _{eq} = Z^{-1} e^{- \beta \Phi},~~~~~~~~~\beta = (\kbar T)^{-1}
\]
with a constant $Z$, the \bfi{partition function} determined by the 
normalization condition to 
\[
  Z = \fct{tr} e^{-\beta \Phi}.
\]
Equation \gzit{e6.2.1} then becomes 
\[
  u = - \beta \tilde{G} \nabla \Phi.
\]
In terms of the \bfi{transport matrix}
\[
  L:= \beta \tilde{G}
\]
and the \bfi{thermodynamics force} 
\[
  F:= \nabla \Phi
\]
(named so in analogy with the ordinary force and a mechanical potential)
we find as covariant drift 
\lbeq{e6.2.1a}
  u = -LF,
\eeq
and the diffusion matrix becomes 
\lbeq{e6.2.1b}
  G = \kbar T (L + L^T).
\eeq
\at{Use $L^T$ in place of $L$ to get similar canonical forms for jump 
and diffusion? In any case check consistency of using $L$ or $L^T$.}

Two cases are particularly important:

(i) For \bfi{closed systems with fixed boundary} (i.e., at constant 
$T,V$), the thermodynamic potential is called the \bfi{Helmholtz 
potential}, and $Z = : e^{-\beta U}$ determines the \bfi{internal 
energy} $ U = \Phi - TS = - \kbar T \log Z$.

(ii) For \bfi{closed systems with free boundary} (i.e.,  at constant 
$T,P$), the thermodynamic potential is the \bfi{Gibbs potential}, and
$Z = :e^{- \beta H}$ determines the \bfi{enthalpy} $H = \Phi -TS = 
  - \kbar T \log Z$.
(Actually, since we ignore space coordinates, we describe only 
well-mixed systems, and ``diffusion'' is here in the space $\Rz ^n$ of 
species densities, not in physical space.)

\gzit{e6.2.1a}--\gzit{e6.2.1b} allow one to obtain the parameters 
of a diffusion process from macroscopic measurements:
The thermodynamical potential can be obtained from measurements at 
equilibrium, and the transport matrix from measuring the response of 
the system to small external forces. \at{see later}

\begin{theorem}
Let the transport matrix $L$ be positive semidefinite. Then the 
diffussion process with covariant drift $u=-LF$ and diffusion matrix 
$G= \kbar T(L+L^T)$ can be written in the {\bf canonical form}
\lbeq{e6.2.1.2}
   \frac{d}{dt} \< f \> =  \< (\kbar T \nabla - F) \cdot L^T \nabla f \>.
\eeq
For {\bf closed} systems, i.e., if the thermodynamic force $F$ has the 
form
\lbeq{e6.2.1.3}
   F = \nabla \Phi
\eeq
with a thermodynamical potential $\Phi$, then the process 
\gzit{e6.2.1.2} has an equilibrium state
\lbeq{e6.2.1.4}
  \rho _{eq} = Z^{-1} e^{-\beta \Phi},~~~~~\beta = (\kbar T)^{-1},
\eeq
with the partition function $Z= \fct{tr} e^{\beta \Phi}$. 
The equilibrium state is unique if $L$ is positive definite.
\end{theorem}

\bepf Since
\[
  u \cdot \nabla = - LF \cdot \nabla = - F \cdot L^T \nabla,
\]
\[
   \nabla \cdot G \nabla = \kbar T \nabla \cdot (L + L^T)
   \nabla = 2 \kbar T \nabla \cdot L^T \nabla,
\]
we can write the Markov equation as
\[
  \frac{d}{dt} \< f \> = \< (\kbar T \nabla - F) \cdot L^T \nabla f \>.
\]
\epf

\begin{remarks}
(i) In terms of $\rho_{eq}$ we can write \gzit{e6.2.1.2} as
\lbeq{e6.2.1.5}
  \frac{d}{dt} \< f \> = \kbar T \< \rho_{eq}^{-1} \nabla \cdot (\rho_{eq} L^T
  \nabla f) \>
\eeq
since $\rho_{eq} ^{-1} \nabla \cdot (\rho_{eq} g) = e^{\beta \Phi} \nabla 
(e^{\beta \Phi} g) = 
e^{\beta \Phi} (e^{- \beta \Phi} \nabla g - \beta (\nabla \Phi) e^{\beta \Phi} 
g) =(\nabla - \beta F) g$. 

This shows directly that $\rho_{eq}$ is a stationary solution.

(ii) For {\bf open} systems, the canonical form extends by adding to $F$ the
external force, i.e., we replace \gzit{e6.2.1.3} by
\lbeq{e6.2.1.3open}
  F= \nabla \Phi + F_{\fct{ext}}.
\eeq
(iii) In the zero temperature limit, $T \rightarrow 0$, noise can be neglected
$(G \rightarrow 0)$, and the equilibrium state approches (saddle point 
approximation!) $\rho_{eq} \rightarrow \delta (x - x_{eq}),$ where $x_{eq}$ is
the global minimizer of $\Phi$ (assumed unique). The motion becomes 
deterministic.

We now consider the deterministic approximation. According to Chapter $5$ \at{},
this is given by the differential equation.
\[
  \dot{x} = v(x)
\]
with the drift 
\[
  v = u + \half \nabla \cdot G = - LF + \frac{\kbar T}{2} \nabla 
  \cdot (L + L^T).
\]
However, because of covariance and the next result, it is 
more appropriate to use in place of $v$ the covariant drift $u=-LF$.
The two are the same when $L + L^T$ is constant, a very common case;
in general they differ at low noise ($\kbar T$ small) only in higher order
terms.

\at{I must explain why I use $\< v (x) \> \approx u ( \< x \> ) $! 
This is strange when $v$ is linear but $G$ is not constant, but this 
is unnatural.}
\end{remarks}

\begin{theorem}
For the {\bf covariant deterministic approximation} of the closed 
diffusion process \gzit{e6.2.1.2}, given by 
\lbeq{e6.2.2.6}
  \dot{x} = - L(x) F(x), \mbox{ where } F(x) = \nabla \Phi (x),
\eeq
the potential $\Phi (x)$ decreases with time (``energy dissipates''):
\lbeq{e6.2.2.7}
   \frac{d}{dt} \Phi (x) = - F(x)^T L(x) F(x) \le 0.
\eeq
If $L$ is definite and $\Phi$ is coercive and below then any limit point of $x(t)$ for
$t \rightarrow \infty$ is a stationary point of the thermodynamic potential, 
and the only stable equilibria are the local minima of $\Phi$. 
\end{theorem}
\bepf
\gzit{e6.2.2.7} holds since $ \frac{d}{dt} \Phi (x) = \nabla \Phi (x) \cdot 
\dot{x}$.
If $\Phi$ is coercive and bounded below then $\D \lim_{t \rightarrow \infty} 
\Phi (x)$ exists, $\frac{d}{dt} \Phi (x) \rightarrow 0$, whence $F^T LF = 0$ 
at any limit point. If $L$ is definite, this implies $F=0$.
\epf

By \gzit{e6.2.2.7}, $L$ describes the amount of energy \bfi{dissipation};
and since
\[
  \<G\> = \< (x - \bar{x}) (x - \bar{x})^T \>
\]
describes the size of \bfi{fluctuations},
the relation \gzit{e6.2.1b} between $G$ and $L$ is called a 
\bfi{fluctuation-dissipation theorem}. 

In particular, \bfi{conservative} systems have

\centerline{$L^T = -L \Leftrightarrow$ no dissipation $\Leftrightarrow$ no 
fluctuation.}

In this case, we can transform the system into \bfi{symplectic form} 
\at{does this also work when $L$ depends on $x$?}:
\[
  L = MJM^T,~J =  
  \left( \bary{rrr}
    0 & I & 0 \\
   -I & 0 & 0 \\
    0 & 0 & 0 
  \eary \right), 
\]
\[ 
  H(z) := \Phi (Mz), \nabla H = M^T \nabla M
\]
\[
  \Rightarrow x = Mz,~ \dot{z} = J \nabla H (z).
\]

\begin{remark}
Other instances of \bfi{fluctuation - dissipation theorems}:

(i) Huang: ``susceptibilites'' or ``response functions''
are expressible as covariances (fluctuations); $cf$ therm.tex.
\[
  (7.14)~~~\kbar T^2 C_V = \< H^2 \>_{ph} - \< H \> _{ph}^2
\]
\[
  (7.43)~~~N \kbar T K_T/V = \< N^2 \>_{ph} - \< N \> _{ph}^2
\]
\[
  ch.16(?):~~~ \frac{\kbar T}{V} \frac{\partial M}{\partial H_{magn}} = 
  \< \Gamma (r) \> _\Omega = \kbar T \chi
\]
where

$M =$ Magnetization $= \< m(r) \> _\Omega$~~~~~~~(extensive, \bfi{an order
parameter}) ($15.275$)

$H_{magn} =$ conjugate magnetic field (Hamiltonian $H_0 - H_{magn} M$)

$ \Gamma (r) = \< m(r) m(0)\>_{ph} - \< m(r) \>_{ph} \< m(0) \>_{ph} =$
space correlation (measures persistence of memory)

(ii) Reichl p. $545, 554-6, 573$ $(15.67)$:

$\chi (\omega)$ is the Fourier transform of $K (\tau)$ (Reichl p. $550/l.1)$

defined by $(15.19)$ \bfi{response matrix} (in $1D =$ response function)
 
for its interpretation see also $15.5.$~$3+4$ (and following lines)

Note that due to $(15.56)$ this is based on the \bfi{linear} model.

so we should have to derive this from the general canonical form.

(iii) Kreuzer [Ho Stat Me $96$]~~p.$11/12$:

\[
  (1.15)~~ \< (dU)^2 \> = \kbar T^2 n C_V
\]
\[
  (1.32)~~ \< (dT)^2 \> = \kbar T^2/nC_V
\]
\[
  \Rightarrow \bigtriangleup U \bigtriangleup T = \kbar T^2 = 
  \frac{RT^2}{N_A} = O (N_A^{-1})
\]
uncertainty relation between time and energy.

(iv) relation between equilibrium covariance and damping term in a linear
stochastic differential equation

(v) response matrix (to past forces) $\leftrightarrow$ time correlations;
generalize $C_V$ expression?

(vi) The differential equation for the Ornstein-Uhlenbeck process.
\end{remark}

\section{Ornstein-Uhlenbeck processes} \label{s8.3}

The \bfi{driven Ornstein-Uhlenbeck process} is defined by the diffusion 
process
\lbeq{e8.3.1}
   \frac{d}{dt} \<f\> =  \< v \cdot \nabla f + \half G : \nabla^2 f \> ,
\eeq
where
\lbeq{e8.3.2}
   v(x) = - L (\Sigma x - F_{ext} (t))
\eeq
\lbeq{e8.3.3}
   G = \kbar T (L + L^T)
\eeq
and $L, \Sigma$ are positive semidefinite matrices, $\Sigma$ symmetric. 
Usually they are nonsingular 9and hence definite).

The most important feature of these processes is that they preserve 
Gaussian distributions. \at{a fact which I do not want to prove 
here} In particular, this implies that their statistical behavior is
completely determined by the mean
\[
  \bar{x} (t) := \< x \>_t
\]
and the covariance
\[
  C(t) := \< (x- \bar{x})(x-\bar{x})^T \>_t.
\]
From Chapter $5$ \at{}, we have
\lbeq{e8.3.4}
  \frac{d}{dt}\bar{x}(t) = \<v\> = -L(\Sigma \bar{x} (t) - F_{ext} (t)),
\eeq
\lbeq{e8.3.5}
   \frac{d}{dt} C(t) = \< G + v (x - \bar{x})^* + (x - \bar{x}) v^* \> =
   G - L \Sigma C (t) - C (t) \Sigma L^T,
\eeq    
and these equations are \bfi{exact} consequences of \gzit{e8.3.1} - 
\gzit{e8.3.3}. [i.e., all approximations made are already in the model 
formulation.] \gzit{e8.3.4} describes a deterministic system of 
coupled and damped driven harmonic oscillators, while \gzit{e8.3.1} is 
the corresponding stochastic version.

The mean equation \gzit{e8.3.4} and the covariance equations 
\gzit{e8.3.5} are decoupled, and can be solved explicitly:
\lbeq{e8.3.6}
  \bar{x} (t) = e^{-tL\Sigma} \bar{x} (0) + \int^t_0 e^{- (t -  \tau) L 
  \Sigma} L F_{ext} (\tau) d \tau, 
\eeq
\lbeq{e8.3.7}
  C(t) = \kbar T \Sigma^{-1} + e^{-tL \Sigma } B (e^{-tL \Sigma})^T,
\eeq
where
\lbeq{e8.3.7a}
   B = C(0) - \kbar T \Sigma ^{-1}.
\eeq 
We now dicuss the solution \gzit{e8.3.6}, \gzit{e8.3.7}, assuming that 
$\Sigma$ and $L$ are \bfi{definite}. The \bfi{dissipation matrix} 
\[
  A := -L \Sigma
\] 
has its spectrum in the negative half plane $\Cz _-$,
\[
  A x = \lambda x, x \ne 0 ~~ \Rightarrow ~~\fct{Re} \lambda = 
  - \frac{(\Sigma x)^* L (\Sigma x)}{x^* \Sigma x} < 0.
\]
\at{Is $\fct{spec} A$ also the spectrum of the associated forward 
derivation?}

Thus the initial state $\bar{x} (0)$ and ``old'' forces get 
exponentially damped, and \bfi{after long times,} the system behaves
 ike the special solution
\lbeq{e8.3.8} 
  \bar{x} (t) = \int^t_{- \infty} e^{-(t - \tau) L \Sigma} LF_{ext}
  (\tau) d \tau  = \int_0^\infty e^{-sL \Sigma} L F_{ext} (t-s) ds.
\eeq
The equilibrium of an Ornstein-Uhlenbeck process with positive definite 
$G$ is unique; since the potential is quadratic, the equilibrium is 
\bfi{Gaussian}, with mean $0$ and covariance $C= \kbar T \Sigma ^{-1}$. 
The covariance matrix $C$ is related to the dissipation matrix $A$ and 
the diffusion matrix $G$ by
\lbeq{e8.3.9}
  G = AC + CA^T
\eeq
which is another expression of the \bfi{fluctuation-dissipation
theorem}.

In particular, since \gzit{e8.3.9} implies that $C(t) = C$ is a 
solution of \gzit{e8.3.5}, uniqueness of the equilibrium implies unique 
solvability of \gzit{e8.3.9} when $A$ and $G$ are given. This is a 
linear system for $C$; the solution gives also the transport matrix 
\[
  L = (\kbar T)^{-1} AC.
\]
  
\section{Linear processes with memory}

\gzit{e8.3.8} is a special case of a \bfi{deterministic 
dissipative linear process with memory}, defined by
\lbeq{e8.4.10}
   \bar{x} (t) := \int^\infty _0 R(s) F_{ext} (t-s) ds,
\eeq
where the \bfi{response function} $R(s)$ decays to $0$ exponentially as 
$s \rightarrow \infty$.

In particular, for a harmonic external force 
\[
  F_{ext} (t) = F_0 \cos \omega t = \fct{Re} (F_0 e^{-i \omega t}),
\]
the response is given by
\lbeq{e8.4.11}
 \bary{rl} 
   \bar{x}(t)=& \D\fct{Re} \int_0^\infty R(s)F_0 e^{-i \omega (t-s)} ds 
   = \fct{Re} e^{-i \omega t} \hat{R} ( \omega ) F_0\\
   = & (\cos( \omega t) \fct{Re} \hat{R} (w) + \sin ( \omega t) \fct{Im}
     \hat{R} ( \omega )) F_0
 \eary
\eeq
with the \bfi{transfer matrix}
\lbeq{e8.4.12}
  \hat{R} (\omega) = \int_0^\infty e^{i \omega s} R(s) ds = 
  \int^\infty _{- \infty} e^{i \omega s} R (s) ds
\eeq
if one extends the response function to $s < 0$ by setting $R(s) = 0$ 
for $s < 0$. 
The energy dissipated in a period is \at{complete...}
\lbeq{e8.4.12a}
  E = \int F_{ext}(t)\cdot d\bar{x}(t) 
    = \int_0^{2\pi / \omega} F_{ext}(t) \cdot\dot{\bar{x}} (t) dt 
    =\ldots= \pi F_0^T (\fct{Im} \hat{R} (\omega))F_0 .
\eeq
Since $R(t)$ can be recovered form $\fct{Im} \hat{R} (w)$ as
\[
  R(t) = \frac{2}{\pi}\int^\infty_0\fct{Im}\hat{R}(\omega)\sin(\omega t)
  d \omega \] 

(from Fourier inversion formula since $R$ is real \at{check ... causal
Fourier transform}), this implies that (the symmetric part of) the
response function and the transfer matrix can be obtained very accuratly
by measuring the power absorption $dE/dt$ (averaged over $N$ periods of
a harmonic driving force with angular frequency $\omega$. This gives
$\frac{\omega N}{2\pi} E$; hence (the symmetric part of) $\fct{Im}
\hat{R} (\omega)$ is directly computable from \gzit{e8.4.12a}. Then one
can calculate $R(t)$. Finally, $\fct{Re} \hat{R} (\omega)$ is
reconstructed either by Fourier transform of $R (t)$, or by the
Kramers-Kronig relations. \at{}

The system responds by forced oscillations of the same frequences, but
the force is weighted by the transfer matrix. In particular, {\em
resonances} occur at frequencies where $\hat{R} (\omega)$ is large.

For the Ornstein-Uhlenbeck process, (\ref{s8.3}.\ref{e8.3.8}) shows 
that the response function is
\lbeq{e8.4.13}
     R(s) = 
   \left\{
   \bary{ll}
     e^{-sL \Sigma} L & \mbox{if~~} s > 0, \\
     0                & \mbox{if~~} s < 0.
   \eary
   \right.
\eeq
The transfer matrix becomes \at{check sign}
\[
  \hat{R} (\omega) = \int_0^\infty e^{-s(L \Sigma - i \omega)} L ds = 
  (L\Sigma-i\omega)^{-1} e^{-s(L \Sigma - i \omega)} L \Big|_0^\infty ;
\]
hence
\lbeq{e8.4.14}
  \hat{R} (\omega) = - (i \omega - L \Sigma)^{-1} L
\eeq
is a rational function of $\omega$. 
The dissipated energy \gzit{e8.4.12a} involves \at{check sign}
\[
 \bary{rl}
   \fct{Im} \hat{R} (\omega) = & \fct{Im} (i \omega - L \Sigma)^{-1} L 
   = [\omega^2 + (L\Sigma)^2] ^{-1} \fct{Im} (-i \omega - L \Sigma) L \\
   = & [\omega ^2 + (L \Sigma)^2] ^{-1} \omega L.
 \eary \] 
In particular, $\hat{R} (\omega)$ will be large when $~\omega~$ is close
to the imaginary part of an isolated eigenvalue of $L \Sigma$ with small
real part. This is the reason why one defines \bfi{resonances}
mathematically by the poles of the transfer matrix in the half plane
$\omega \in \Cz _+$ where \gzit{e8.4.12} makes sense.

For numerical calculation, we use a Cholesky factorization 
\[
  \Sigma = R ^T R
\]
and a spectral factorization
\[
  RL R^T = Q \Lambda Q^{-1},~~~~\Lambda \mbox{~diagonal,~~~~} \fct{Re}
  \Lambda \ge 0.
\]
The columns of $R^{-1} Q$ are eigenvectors of $L\Sigma$, and we have 
(for $s > 0$)
\[
  R(s) = e^{-sL\Sigma} L = R^{-1} Q e^{-s \Lambda} \Lambda Q^{-1} R^{-T}
  = R^{-1} Q,
\]
\[
  e^{-tL \Sigma} B(e^{-tL \Sigma})^T = R^{-1} Q e^{-s\Lambda} Q^{-1} R B
  (Q^{-1} R)^* e^{-s \Lambda^*}.
\]
For symmetric $L$, we rather factor 
\[
  L=R^T R
\]
and
\[
  R \Sigma R^T = Q \Lambda Q^T ,
\]
with orthogonal $Q$ and diagonal $\Lambda \geq 0$. Now the columns of 
$R^T Z$ are eigenvectors of $LZ$, and we get
\[
  e^{-sL \Sigma} L = R^T Q e^{-s \Lambda} Q^T R,
\]
\[      
  e^{-tL \Sigma} B (e^{-tL \Sigma})^T = R^T Q E^{-t \Lambda} \bar{B} e^
  {- t \Lambda} Q ^T R,
\]
where 
\[
  \bar{B} = Q^T (R^{-1} B R^{-1}) Q.
\]
\at{add transfer matrix. Then we'd need to discuss:}

\begin{itemize}
\item \bfi{spectral density} 
\item time correlations. why $|t|$?~~~~~``two kinds of order''
\item relation to $B(t) = $ return to equilibrium response after a 
  force switch

~~~~~~~~~~~~~~~~~~~~~This is another $F/D$ version

\item $B(t) = \frac{2}{\pi} \int_0^\infty \fct{Im} \hat{R} (\omega)
\cos \omega (t) \frac{d \omega}{\omega} $ \at{Reichl 15.67}
($B$ is the fluctuation = time correlation, $\fct{Im} \hat{R}$ the 
dissipation.)

\item  Quantum version \at{Reichl p. $571$ ff}

\item The \at{ex. in Reichl, p.$557$ ff} Brownian particle correponds 
to linearized damped Hamiltonian systems. Force only applies to the 
second order term. \at{Details?}
\end{itemize}

\bfi{driven continous-time linear state space model} = driven 
Ornstein-Uhlenbeck process with derived measurable quantities $y$.
\[
  \dot{x} (t) = - L (\Sigma x(t) - F_{ext} (t)) ~~+ ~~\mbox{noise,} 
\]
\[
  y(t) = Yx(t)~~+~~\mbox{ noise,}       
\]
\[
  \bar{x} (t) = \int^\infty _0 e^{-sL \Sigma} L F_{ext} (t - s )ds.
\]
For a constant pulse of length $T$,
\[
  F_{ext} (t) = 
  \left\{ \bary{ll}
    F_0 & \mbox{ for } t \in [-T,0], \\
    0   & \mbox{ otherwise,}
   \eary \right.
\]
the system responds for times $t>0$ with
\[ 
  \bary{rl}
    \bar{x} (t) &= \D \int^0_{-T} e^{-(t - \tau) L \Sigma} L F_0 d \tau 
  = e^{-tL\Sigma}\Big( \int^0_{-T} e ^{\tau L \Sigma} d\tau\Big) LF_0 \\
           &= \D e^{-tL \Sigma}(1-e^{-TL\Sigma}) (L \Sigma)^{-1} L F_0 
            = (e^{-tL \Sigma} -e^{-(t+ T)L \Sigma}) \Sigma^{-1} F_0.
    \eary
\]
In particular, if $ L \Sigma Q = Q \Lambda~ (\Lambda $ diagonal) then
\[
   \bar{y} (t) = Y Q (e^{-t \Lambda} -e^{-(t + T) \Lambda}) Q ^{-1} 
    \Sigma^{-1} F_0 = Q e^{-t \Lambda} (1 - e ^{-T \Lambda}) 
    Q ^{-1} \Sigma^{-1} F_0 .
\] 
So the \bfi{pulse response} $y(t)$ is a linear combination of 
exponentials. More generally, when the spectral factorzation is not 
possible because of defective $L \Sigma, y(t)$ is still expolynomial.

\at{ can we model a \bfi{delayed response} of the form of a peaked hat 
function by expolynomials $\Sigma a_i (1-e^{-\lambda t})$ ? 
All $\lambda_i$ must be positive since decay at $\infty$. 
Try two real eigenvalues
\[
  y(t) = a(1-e^{- \lambda t}) - b(1-e^{-\mu t}),~~ \lambda > \mu .
\]
Unique extremum at $t = \tau$, where $\frac{\lambda a}{\mu b} =
e^{(\lambda - \mu) \tau}$. Thus
\[
  y(t) = \fct{const} [ \frac{e^{\lambda \tau}}{\lambda} (1-e^
  {- \lambda t}) - \frac{e^{\mu \tau}}{\mu} (\Lambda - e ^{- \mu t})], 
\]
and
\[
\Rightarrow \dot{y} (t) = \fct{const} [e^{\lambda (\tau - t)} 
  -e ^{\mu (\tau -t)}]
       ~~~ \cases{\ge 0 \mbox{ for } t \in [0, \tau], \cr
                 \le 0 \mbox{ for } t \in [\tau, \infty] 
                 \mbox{ when } \fct{const} > 0,}
\]
\[
    \ddot{y} (t) 
   = \fct{const} [ - \lambda e^{\lambda (\tau - t)} + \mu e 
    ^{\mu (\tau - t)}]. 
\]
So we have only \bfi{one} inflection point which is not good enough.
} 

\section{Dissipative Hamiltonian Systems}

\at{Reformulate this section as theorem: If the conservative equation
\gzit{e6.3.1} is the covariant deterministic approximation of a 
canonical form then $\Phi = H$ and \gzit{e6.3.2} holds.}

A conservative Hamiltonian system is an energy-conserving process 
characterized by a \bfi{Hamiltonian function} $H(p,q)$ in the position 
vector $q$ and the conjugate momentum vector $p$ and defined by the 
differential equations
\lbeq{e6.3.1}
  \dot{q} = \frac{\partial H}{\partial p},~~~
  \dot{p} = - \frac{\partial H}{\partial q}.
\eeq
Energy conservation follows since
\[
  \frac{d}{dt} H (p,q) = H_p \dot{p} + H_q \dot{q} = 0,
\]
so that $H(p,q)$ must be constant. But energy conservation for 
macroscopic systems is an idealization, and more realistic models 
include energy loss due to \bfi{friction}. Friction is energy 
dissipation caused by random microscopic forces producing \bfi{heat}. 
We use the canonical form to derive an equation for the phase vector 
\[
  x= {q \choose p}.
\] 
Friction is experienced as a restraining force proportional to the 
velocity $\dot{q}$; in the covariant approximation, Newton's law gives 
the modified differential equation
\lbeq{e6.3.2}
  \dot{q} = \frac{\partial H}{\partial p}, ~~ 
  \dot{p} = - \frac{\partial H}{\partial q} - C (q) \dot{q} 
  = - \frac{\partial H}{\partial q}- C (q) \frac{\partial H}{\partial p}
\eeq
describing a \bfi{dissipative Hamiltonian system}.

With this dynamics, the energy change is
\[
  \frac{d}{dt} H (p,q) 
  = \frac{\partial H}{\partial p} \dot{p} 
    + \frac{\partial H}{\partial q} \dot{q} 
  = - \dot{q} ^T C(q) \dot{q},    
\]
and we see that the damping matrix $C(q)$ must be assumed to be 
\bfi{positive definite} (but not necessarily symmetric) in order to 
have dissipation except at rest. \at{Onsager: $C$ symmetric}

In terms of the potential (total energy)
\[
  \Phi (x) = H (p,q),
\]
the thermodynamic force becomes 
\[
    \D 
      F =
    \left( \bary{c}
      \frac{\partial H}{\partial p} \\
      ~ \\
      \frac{\partial H}{\partial q}
    \eary \right);
\]
and we infer from \gzit{e6.3.2} that the covariant drift has the form
\[
    u = 
  \left( \bary{c}
    \frac{\partial H}{\partial p}  \\
    ~  \\
    -\frac{\partial H}{\partial q} - C (q) \frac{\partial H}{\partial p} 
  \eary \right)  
    = - 
  \left( \bary{cc}
    0 & - I \\
    ~ & ~ \\
    I & C(q)
  \eary \right)
  \left( \bary{c}
    \frac{\partial H}{\partial p} \\
    ~ \\
    \frac{\partial H}{\partial q}
  \eary \right)
    = - LF,
\]
where
\lbeq{e6.3.3}
     L = 
   \left( \bary{cc}
     0 & -I \\
     I & C(q)
   \eary \right).
\eeq
The fluctuation-dissipation theorem now shows that the correct diffusion 
matrix is
\lbeq{e6.3.4}
     G = \kbar T(L + L^T) = \kbar T 
   \left( \bary{cc}
     0 & 0 \\
     0 & C(q) + C (q)^T
   \eary \right)
\eeq
Thus we end up with the canonical diffusion process
\lbeq{e6.3.5}
  \displaystyle
  \frac{d}{dt} \< f(p,q) \> = \left< \frac{\partial H}{\partial p} \cdot \frac{
  \partial f}{\partial q} - \left( \frac{\partial H}{\partial q} + C 
  \frac{\partial H}{\partial p} \right) \cdot \frac{\partial f}{\partial p}
  + \kbar T \frac{\partial}{\partial p} \cdot \left( C \cdot \frac{\partial f}
  {\partial p} \right) \right>.
\eeq
The stationary density \at{unique?} turns out to be the 
\bfi{canonical ensemble}
\lbeq{e6.3.6}
  \rho _{eq} (p,q) = Z^{-1} e ^{- \beta H (p,q)},~~~ Z = \fct{tr} 
  e^{-\beta H (p,q)}.
\eeq
\at{In special cases (Kramers equation), this gives the 
Maxwell-Boltzmann distribution.}

For constant $C$ we get the associated stochastic differential equation
\[
  dq = \frac{\partial H}{\partial p} dt, ~~ dp = 
  \left( - \frac{\partial H}{\partial q} - C \frac{\partial H}{\partial p}
  \right) dt + d \eps ,
\]
\[
  d \eps \sim N (0, \kbar T(C + C^T)).
\]
In the special case of Cartesian coordinates, where
\[
  H = \half p^T M^{-1} p + V(q)
\]
we can write $q = x,~ p = M \dot{x}$ (from $dq = M^{-1} pdt$) and get
\[
  M \ddot{x} + C \dot{x} + \nabla V (x) = B \dot{W} (t)
\]
where
\[
  BB^T = \kbar^T (C + C^T)
\]
and $B,C$ are constant.

\chapter{Collective Processes}\label{c.coll}

\section{The master equation} \label{s7.1}

In this section we discuss the general set-up of a system in which a 
large number of individuals interact through private communication in 
an environment where collective forces govern the frequency of 
communication events.

Our communication model has the following ingredients.

$(C 1)$ there are $q$ \bfi{species} $X_j (j = 1,\ldots,q)$ describing 
\bfi{individuals}. The number of individuals of species $X_j$ in a 
system is written as $N_j$; these numbers define the \bfi{population 
vector} $N \in \Zz  ^q$;
\bfi{The density} $\rho_t(N)$ describes the likelihood to have at time 
$t$ a population number $N$, and the integral is 
\[
  \sint f  = \sum_{N \in \Zz ^q} f(N).
\]
$\Ez $ consists of all geometrically decaying functions of $N$.

In practice, species may be elementary particles, molecules of fixed 
chemical composition, biologigal species (of animals or plants), 
diseases of people, 
professions of people, spectral classes of stars, etc. We take the 
number of different species as finite, although everything extends to
$q = \infty$, with examples e.g. in polymer physics, where $j$ is the 
chain length).

$(C 2)$ There are $r$ kinds of \bfi{events}
\lbeq{e7.1.1}
  \sum_{j=1} ^q \nu _{lj}^+ X_j \rightleftharpoons \sum _{j = 1} ^q \nu
  _{lj} ^- X_j~~~~~~(l=1,\ldots,r)
\eeq
with nonnegative numbers $\nu _{lj}^\pm~~~(l=1,\ldots,r; j=1, \ldots,q)$.

Events model the elementary units of communtication; an event consists in
the meeting of a collection of $ \nu_{lj}^+$ individuals of kind $X_j
(j=1,\ldots,q)$ changing during the meeting into a collection of kind 
$\nu_{lj}^-$ 
individuals of kind $X_j (j=1, \ldots, q);$ or reversed. If the reverse process
is impossible we write $\rightarrow$ in place of $\rightleftharpoons$.
Events are considered as black boxes about which no details are available.
Again we assume $r$ to be finite.

Usually $\sum v_{lj}^{+}$ and $\sum v_{lj}^-$ are very small; typically
$\le 3$. We may illustrate an event $A + B \rightleftharpoons C + D $
as follows. \at{figure missing}

\bigskip
\vspace{3cm}  

Typical instances are:
\[
e^+ e^- \rightleftharpoons 2 \gamma
\]
\[
  H^+ + OH^- \rightleftharpoons H_2 O
\]
\[
  \mbox{[hungry bird] $+$ [caterpillar] $\rightarrow$ [satisfied bird]}
\]
\[
  \mbox{[healtly person] $+$ [ill person] $\rightarrow 2$ [ill person]}
\]
\[ 
  \left\{ \bary{l}
  \mbox{[husband]} + \mbox{[wife]} \rightarrow \mbox{[husband]} + 
  \mbox{ [pregnant wife]}\\
  \mbox{[pregnant wife]} \rightarrow \mbox{[wife]} + \mbox{[child]}  
  \eary \right.
\]
\[
  \mbox{[buyer} \& \mbox{money]} + \mbox{[seller} \& \mbox{object]} 
  \rightleftharpoons \mbox{[buyer} \& \mbox{object]} + \mbox{[seller} 
  \& \mbox{money]}
\]
$(C 3)$ There are \bfi{transition rates} $u_l ^{\pm} : \Zz ^n \rightarrow
\Rz _+ (l = 1,\ldots, r); u_l^+ (N)$ and $u_l ^-(N)$
specify the likelihood that event $l$ happens in forward or backward direction
in a population described by $N; u_l^- = 0$ specifies an event which only
oceurs in the forward direction. 

The transiton rates are global, collective, properties of the system; they
account for nonlocal, long range interaction between individuals, and for
\bfi{limitations of freedom} due to overpopulation, mutual attraction, and
mutual repulsion. A very common Ansatz for the transmition rates is that
of \bfi{combinatorial kinetics}, where
\lbeq{e7.1.2}
  u_l ^\pm (N) = K_l ^\pm \pi^q _{j=1} {N_j \choose \nu^\pm _{lj}} =
  \left \{ \bary{ll}
    0     &  \mbox{if } N_j < \nu _{lj} ^\pm \mbox{ for some } j, \\
    \D k_l ^\pm \prod^q_{j=1} \frac{N_j!}{(N_j - v_{lj} ^\pm)!} & 
    \mbox{ otherwise},
  \eary \right.
\eeq
with constants $k_l^\pm = k_l^\pm \prod_j (\nu_{lj}^\pm)!$; one writes 
$\rightleftharpoons_{k_l-} ^{k_l^+}$ (or ${k_l^+ \atop \rightarrow}$ if 
$k_l^- = 0)$.
This models the assumption that individuals are completely independent
and meet only by chance, so that the transition rates are proportional
to the number of ways to assemble the required collection of individuals in
a population described by $N$. Combinatorial kinetics describes correctly 
the chemistry of ideal gases and ideal solutions; it is also used for most 
systems in biology, medicine, economy, and social sciences, mainly because
of simplicity and lack of more detailed knowledge. 

$(C 4)$ The \bfi{probability current} produced by the event $l$ is defined as
\lbeq{e7.1.3}
  j_l (N) := u_l^- (N + \nu_l^-) \rho(N+ \nu_l^-) -u_l^+(N+ \nu_l^+) \rho(n+
  \nu_l^+).
\eeq
Here $N$ is interpreted as the part of the population \bfi{not} involved in 
the event, and the current consists of a positive contribution due to the 
outcome of the event and a negative contribution due to the input of the 
event. If we interpret the probability currents as the rate of change of the
density due to single events, and count forward events positively, backward
events negatively, we end up with the \bfi{master equation}
\lbeq{e7.1.4}
  \frac{d}{dt} \rho_t (N) =  \sum_{l=1} ^r \left(j_l(N- \nu_l ^+)- j_l
  (N-\nu_l^-)\right)
  = \sum_{l=1}^r \left( [u_l^- \rho_t]_N^{N + \eps_l} 
  +[ u_l ^+ \rho_t]_N^{N-\eps_l} \right)                
\eeq
with the \bfi{jump vectors}
\lbeq{e7.1.5}
  \eps_l := \nu_l^- - \nu_l^+.
\eeq
The master equation is a system of infinitely many ordinary differential
equations. 
\at{Note that this is an idealisation of reality: it assumes that 
events happen instantaneously, and neglects time scales much shorter 
than the time scale of interest.}

$(C 5)$ Finally we shall assume the restrictions
\beq
   N_j \ge \nu_{lj} ^+ \mbox{~~~~for all~} j \Rightarrow \eps_l^+(N) > 0
\eeq
and
\beq
  n_j < \nu_{lj}^\pm \mbox{~~~~for some~} j \Rightarrow u_l ^\pm (N) = 0;
\eeq
the latter implies that the dynamics preserves the natural condition
\beq
  \rho_t (N) = 0 \mbox{~~~if some } N_j < 0
\eeq
at all times if it holds at $t = 0$.

A system satisfying (C1)--(C5) is called a \bfi{collective process}.  
   
\begin{theorem} Any collective process is a Markov process, whose 
forward derivation $D$ is given by
\lbeq{e7.1.1.6}
  Df: N \rightarrow \sum_l \left( u_l^- (N) (f(N-\eps_l) - f(N)) 
  + u_l ^+ (N) (f(N + \eps_l) - f(N)) \right).  
\eeq
For $x=N$, the drift vector $v$ and the diffusion matrix $G$ are given by
\lbeq{e7.1.1.7} \label{7.1.7}
  v(N) = \sum_l (u_l ^+ (N) -u_l ^-(N)) \eps_l,
\eeq
\lbeq{e7.1.1.8}
  G(N) = \sum_l (u_l^+(N) + u_l ^- (N)) \eps_l \eps_l^*.
\eeq
\end{theorem}

\bepf
We first note that the sum over $N \in \Zz ^q$ is translation invariant.
Using the notation
\[
  [f]_{N_1}^{N_2} = f(N_2) - f(N_1),
\]
we find  
\begin{eqnarray*}
  \sum_N [g]_N ^{N+\eps} f(N) &=& \sum_N g(N + \eps) f(N) -\sum_N g(N) f(N)\\
      &=& \sum_N g(N) f(N - \eps) - \sum_N g (N) f(N) = 
          \sum_N g(N) [f]_N ^{N - \eps}.
\end{eqnarray*}
Hence 
\begin{eqnarray*}
   \frac{d}{dt} \<f\> &=& \sum_N \frac{d}{dt} \dot{\rho}_t (N) f(N) 
   = \sum_N \sum_l ( [ u_l^- \rho]_N ^{N + \eps_l} + [ u_l^+ \rho]_N^{N-\eps_l}
   ) f(N)\\
   &=& \sum_N \sum_l ( u_l^- (N) \rho (N)_N ^{N-\eps_l} + u_l^+ (N) \rho (N)
   [f]_N ^{N+\eps_l})\\
   &=& \sum_N \rho (N) (Df) (N) = \< Df\>,
\end{eqnarray*}
with $D$ defined by \gzit{e7.1.1.6}. Clearly, $D$ is $*$-linear and $D1 = 0$. 
Now 
\lbeq{e7.1.1.8a}
  Q(f,g) := D(fg^*) - (Df)g^* - f(Dg^*)
\eeq
maps $N$ to
\begin{eqnarray*}
    \sum_\pm \sum_l & & \D u_l ^\pm (N) \left\{ \left( f(N \pm \eps_l) g^* 
    (N \pm \eps_l) -f(N) g^*(N) \right) \right. \\
    & & \left. \D - \left( f(N \pm \eps_l) g^* (N) -f(N) g^* (N) \right)
    - \left( f(N) g^* (N \pm \eps - f(N) g^*(N) \right) \right\}, 
\end{eqnarray*}
hence
\lbeq{e7.1.1.9}
    Q(f,g)(N) = \sum_\pm \sum_l \D u_l^\pm (N) (f (N \pm \eps_l)-f(N)) 
    (g(N \pm \eps_l)-g(N))^*. 
\eeq
For $g=f$, this is clearly nonnegative; hence $D$ is a forward derivation
and we have a Markov process. For $f(N)=N_j,~ g(N)=N_k$ we find the 
$(j,k)$-component of the diffusion matrix \gzit{e7.1.1.8}, and the drift
vector follows from \gzit{e7.1.1.6} for $f(N)=N_j$.
\epf
   
\begin{remarks} \begin{enumerate}
\item Higher jump momemts are found similarly, with alternating signs of
      $u_l^-$.
\item Any Markov process in $\Ez  = \Cz ^n$ can be brought to this form:
      Introduce for each unordered pair $~(j,k)$ an event $X_j 
      \rightleftharpoons X_k$ and note that $\sum N_j=1$ (or $=N_{\fct{tot}}$,
      with frequency representation) is conserved.
\end{enumerate}
\end{remarks}

\begin{examples} 
(i) (\bfi{Poisson process}). The simplest \bfi{birth process} 
(e.g., electrons in $\beta$-decay, change of composition of the source due to 
the decay) is given by a single event for a simple species
\[
  0
  \bary{c}
   ~ \\
   \to \\
   \lambda
   \eary
   X.
\]
With combinational kinetics, we have $u_0^+ (N) = \lambda,~ u_0^- (N) = 0$, 
hence the master equation
\lbeq{e7.1.2.10}
  \dot{\rho}_t (N) = \lambda (\rho_t (N-1) - \rho_t(N)).
\eeq
With the initial condition $\rho_0 (N) = \delta_{N0}$ (no individual at 
$t = 0$), the solution of \gzit{e7.1.2.10} is 
\lbeq{e7.1.2.11}
  \rho_t (N) = \frac{(\lambda t)^N e^{-\lambda t}}{N!},
\eeq
describing a Poisson distribution with 
\lbeq{e7.1.2.11a}
  \<N\>_t = \lambda t, ~~\mbox{$\fct{var}$}_t (N) = \lambda t.
\eeq
\gzit{e7.1.2.11a} follows more directly from the moment equations
\[
  \frac{d}{dt} \<N\> = \<v\> = \< \lambda \> = \lambda,
\]
\[
  \frac{d}{dt} \sigma^2 = \<G\> = \< \lambda \> = \lambda.
\]
For $t \rightarrow \infty$ the variance diverges; hence no stationary solution
exists.

(ii) The simplest \bfi{birth and death process} (e.g., fog formation) is given
by a single event for a single species
\[
      0 
    \bary{c}
      \alpha \gamma \\  
      \rightleftharpoons \\ 
      \gamma 
    \eary
      X.
\]
with combinational kinetics. We get from $u^+ (N) = \alpha \gamma,~ u^- (N) = 
\gamma N$ the master equation
\lbeq{e7.1.2.12}
  \D \dot{\rho}_t(N) = \alpha \gamma \big( \rho_t (N-1) -\rho_t(N) \big) + 
  \gamma \big((N + 1) \rho_t (N + 1) - N \rho_t (N)\big). 
\eeq
This can be solved exactly, with a complicated solution (see e.g.,
{\sc Gardiner} \cite{Gar}). 
Rather we use the moment equations for mean and variance,
\[
  \frac{d}{dt} \<N\> = \<v\> =  \< \gamma (N-\alpha \gamma)\> = \gamma (\<N\>
  - \alpha),
\]
\[
  \frac{d}{dt} \sigma^2 = \<G\> = \<\gamma N + \alpha \gamma \> = \gamma
  (\<N\> + \alpha).
\]
With initial condition $N=N_0$ (a number) at $t=0$, we get the solution
\lbeq{e7.1.2.13}
  \left.{ \bary{l}
    \<N\>_t = \alpha (1-e^{-\gamma t}) + N_0 e^{-\gamma t,}\\
    \sigma_t ^2 = (\alpha + N_0 e^{-\gamma t}) (1-e^{- \gamma t}).
  \eary} \right\}
\eeq
The equilibrium solution is, characterized by $\dot{\rho} (N) = 0$ for all $N$,
can be shown to be given by a Poisson distribution
\[
  \rho(N) = \frac{\alpha^N e^{-\alpha}}{N!},
\]
and indeed, \gzit{e7.1.2.13} converges in the limit $t \rightarrow \infty$
to mean and variance of this distribution.

(iii) More generally, systems consisting of solitary individuals, where all 
events are of the form 
$X_j \rightleftharpoons X_k$, $X_j \leftrightarrow 0$, $X_j \rightarrow$ 
anything, or $0 \rightarrow$ anything, have explicitely solvable master 
equations and linear, triangular moment equations which can be solved 
recursively.
\end{examples}

\at{{\sc Gardiner} \cite{Gar}, $7.2.30/42/52$ also has a system size expansion
\[
  x = \Omega \alpha + \Omega ^{1/2} z, \alpha = \Omega \sigma,
\]
giving approximately a Gaussian
\[
  \tilde{\rho} (z) = \fct{const} e^{-z^2/2 \sigma^2}, \tilde{\rho}(x) = 
  \fct{const} e^{-(x- \Omega \alpha)^2/2 \Omega \sigma^2}.
\]
The above example corresponds \cite{Gar}, to $(1.4.4-8)$ (i) and p.$238$ ff (ii)}

More interesting cases can only be solved numerically, using the low noise 
approximation, a discrete Galerkin method, or Monte Carlo simulations.

We say that the linear combination $a^T N = \sum a_k N_k$ is \bfi{conserved}
in a collective process if 
\lbeq{e7.1.2.14} 
  a^T \nu_l^+ = a^T \nu_l^- \mbox{~~~ for } l = 1, \ldots, r.
\eeq
The typical reason for such a relation is that there is a family 
$Y_i(i=1,\ldots,p)$
of \bfi{invariants} (conserved quantities, e.g., charge, lepton number, atoms,
functional groups, dollars) which, in every event, are exchanged in full 
units and don't get lost. If each individual of species $X_j$ contains $A_{ij}$
invariants $Y_j$ then 
\lbeq{e7.1.2.15}
  A \nu_l^+ = A \nu_l^- \mbox{~~~~ for~} l = 1, \ldots, r,
\eeq
since the $i$th component of both sides counts the total number of invariants
$Y_i$ involved in event $l$. Hence, if $a_k = \sum b_j A_{jk}$ with arbitrary
constants 
$b_j$ then $a^T \nu_l^+ = b^T A \nu_l^+ =b^T A \nu _l^- = a^T \nu_l^-,$
i.e., $a^TN$ is conserved. We say that the invariants $Y_i (i=1,\ldots,p)$ 
form a \bfi{full set of invariants} if every conserved linear combination
$a^T N$ is a \at{modular} linear combination of the components of $AN$, i.e. if
\lbeq{e7.1.2.16}
  a^T \nu_l^+ =  a^T \nu_l^-(l = 1, \ldots, r)~~ \Rightarrow~~ a^T \equiv b^TA
  \mbox{~~~for some } b \in \Rz ^P.  
\eeq
The matrix $A$ is referred to as the composition matrix of the process with 
respect to $Y_1,\ldots,Y_p$.

\begin{proposition} If $\rho_t(N)$ is a solution of a master equation
\gzit{e7.1.4} satisfying \gzit{e7.1.2.15} then, for all functions $g$,
\[
  \tilde{\rho}_t (N) := \rho_t(N) g (AN)
\]
is also a solution of \gzit{e7.1.4}.
\end{proposition}  

\bepf \gzit{e7.1.2.15} implies $A \eps_l = 0$, hence $g(A(N \pm 
\eps_l)) = g(AN)$, where we can cancel in $(\tilde{ \ref{e7.1.4}})$ a common 
factor $g(AN)$. 
\epf

The proposition reflects the fact that \bfi{any} initial distribution of $AN$
is fixed by the dynamics. Usually one fixes the distribution by assuming 
deterministic values for the components of $AN$ (i.e. numbers instead of
quantities); this reduces the Euclidean *-algebra and turns $g(AN)$ into
a number, which cancels under normalization of $\tilde{\rho}.$

A \bfi{component} of collective process is a set $N$ consisty of all population
vectors $N \in \Zz ^q$ which are reachable from some fixed $N_0$ by a
sequence of events. \at{explain! positive rate} Clearly, \gzit{e7.1.2.15} 
implies that all $N \in {\cal N}$ have the same value of $AN$, and typically 
components are characterized by the common value of $AN$ for $N \in {\cal N}$. 
However, there are processes like
\[
  2 X_1 \rightleftharpoons 0,
\]
where the parity of $N_1$ is conserved, too, and there are processes like
\[
  2 X_1 \rightleftharpoons 2 X_2 \rightleftharpoons X_1
\]
which have no invariant but several components, here 
$\left\{ {0 \choose 0} \right\}$ and $\{N \mid N_1, N_2 \ge 0 \} \setminus
\left\{ {0 \choose 0} \right\}$.

Clearly, the dynamics in different components is completely independent;
so we may restrict $\Ez _0$ (and hence the density) to functions of $N$
which vanish outside some fixed component.

\begin{theorem} If $\Ez _0$ consists of the functions of $N$ which vanish
outside some fixed component of a collective process, the forward derivation 
of the process is primitive. In particular, if a positive equilibrium state 
exists, it is unique, and is reached from any initial state as $t \rightarrow 
\infty$.
\end{theorem}

\bepf 
By \gzit{e7.1.1.8a} and \gzit{e7.1.1.9}, $D(ff^*)= 
(Df)f^* + f(Df^*) \Rightarrow f(N \pm \eps_l) = f(N)$ whenever 
$u_l^\pm (N) > 0.$
Thus $f$ is constant on each component. Since by assumption only one 
component is nontrivial, $D$ is primitive.
\epf

\section{Canonical form and thermody\-namic limit}
A generalization of combinatorial kinetics is given by collective processes
in which the transition rates have the form
\lbeq{e7.2.1}
  u_l^\pm(N) = \omega_l^\pm \frac{\rho_\Omega(N-\nu_l^\pm)}{\rho_\Omega(N)}, 
\eeq
where $\omega_l ^\pm \ge 0$ and $\rho_\Omega$ is a density which is positive when 
all $N_j \ge 0$, and vanishes otherwise. We call processes satisfying 
\gzit{e7.2.1} \bfi{canonical}. Combinatorial kinetics is special case of 
\gzit{e7.2.1} where $\rho_\Omega$ is a \bfi{multivariate Poisson distribution},
defined by 
\lbeq{e7.2.2}
  \rho_\Omega (N) = \prod^q_{j=1} \frac{\alpha_j ^{N_j} e^{-\alpha_j}}{N_j !};
\eeq
substitution into \gzit{e7.2.1} and comparison  with (\ref{s7.1}.\ref{e7.1.2}),
shows that the rate constants are related to the $\omega_l^\pm$ by the equations
\lbeq{e7.2.3}
   k_l^\pm = \omega_l^\pm \prod^q _{j=1} \alpha_j^{-\nu_{lj}^\pm}~~(l=1,\ldots,r).
\eeq
$[$Apparently all processes considered in applicatons are in canonical form.
The reason is unclear to me.$]$

\begin{theorem} For a canonical collective process with
\lbeq{e7.2.1.4}
  \omega_l^+ = \omega_l^-~~~(l=1,\ldots,r),
\eeq
we have:

(i)~ At equilibrum, all probability currents $j_l (N)$ vanish.

(ii) On each component of the process, $\rho_{eq}(N)$ is a constant multiple of
$\rho_\Omega(N)$.
\end{theorem}

\bepf 
The density $\rho = \rho_\Omega$ produces the current
\[
  j_l(N) = (\omega_l^+ - \omega_l^-) \rho_\Omega (N).
\]
$j_l$ vanishes when $\omega_l^+ =  \omega_l^-$. By the master equation, this implies 
that $\rho_\Omega$ is time invariant, and by Theorem $7.1.4$, this shows that
$\rho_{eq}$ is a multiple of $\rho_\Omega$.
\epf

If \gzit{e7.2.1.4} holds we say the system satisfies {\bf detailed balance}. 
Each event then satisfies a separate balance equation $j_l (N) = 0$, whereas 
in general at equilibrium only the total (signed) sum of currents vanishes. 
Thus (i) and (ii) characterize {\bf closed systems}.

For a noncanonical system, detailed balance does not say too much, only 
\[
  u_l^\pm (N) = \frac{\omega_l(N-\nu_l^\pm)}{\rho_{eq}(N)}
\]
 for suitable $\omega_l$.
Note that microreversibility only gives detailed balance 
\at{\cite{Gar} argues that all variables are even since there are no 
continuous indices}, but not 
the canonical form, which is an independent axiom.

To discover the equilibrium of a combinatorial process one can try to
solve \gzit{e7.2.3} with $\omega_l^+ = \omega_l^-$ for the $\omega_l$ and $\alpha_j$ ($2r$
equations for $q + r$ unknowns); if these equatons are consistent, 
\gzit{e7.2.2} provides the equilibrium solution (upto a constant factor 
which depends on the initial distribution of the conserved quantities).
If the equations are inconsistent, the system cannot be closed. 
\at{whereas,
sometimes, open systems happen to satisfy detailed balance, too; 
e.g., $X \leftrightarrow 0$ (from Example $7.1.2$) would not be considered 
closed.}

The analogy to the canonical form for diffusion processes is seen by 
introducing the discrete forward derivations $\delta_l$ with
\[
 \delta_l f: N \rightarrow f(N + \nu_l^-) - f (N + \nu_l^+);
\]
one easily sees that the adjoint $\delta _l^*$ is given by
\[
  \delta_l^* f:N \rightarrow f(N - \nu_l^-) - f(N-\nu_l^+).
\]
Thus we can write
\begin{eqnarray*}
  Df: N \rightarrow 
  & & \sum_l \left(u_l^- (N) (\delta_l f)(N - \nu_l ^-) -u_l^+
  (N) (\delta _l f)(N -\nu _l^+) \right) \\
  = & & \sum _c \omega_l \rho_{eq} (N)^{-1} \left( (-\rho_{eq} \delta_l f) 
  (N-\nu_l^-) + (\rho_{eq} \delta_l f) (N-\nu_l^+) \right) \\
  = -  & & \sum_l \rho_{eq}^{-1} \delta_l^* (\omega_l (\rho_{eq} \delta_l f)).
\end{eqnarray*}
With the vector $\delta = (\delta_l)_{l=1}^r$ and $\Omega = \fct{diag}
(\omega_l)$ we get
\[
  Df = - \rho_{eq}^{-1} \delta^* (\Omega \rho_{eq} \delta f)
\]
in close analogy to the diffusion case
\[
  Df = \rho_{eq}^{-1} \nabla (L \rho_{eq} \nabla f) = - \rho_{eq} ^{-1} \nabla
  ^* (L \rho_{eq} \nabla f).
\]
The latter formula is obtained if we use the approximations
\[
  \delta_l f \approx \eps_l \cdot \nabla f, ~~ \delta_l^* f \approx -\eps_l 
  \cdot \nabla f
\]
and put
\[
  L:=  \sum _{\| l \| \le r} \omega_l \eps_l \eps_l^T.  
\]
\at{this is too crude, ok only to $O(1, \Omega F)$; note  $\omega_l = 
O(\Omega)$.}

In practice, collective processes are often studied when $N$ is very large;
then a very useful approximation is the consideration of the so-called
\bfi{thermodynamic limit} $N \rightarrow \infty$.
The interesting quantities are the relative sizes of the $N_j$. Thus we shall
write
\lbeq{e7.2.1.5} 
  N =  \Omega x,~~~~\omega_l ^\pm = \Omega \kappa_l^\pm,
\eeq
and for combinatorial kinetics also \at{??}
\lbeq{e.7.2.1.6}
  \alpha _j = \Omega a_j  
\eeq
with $x, \kappa, a_j$ of order $1$ and a number $\Omega$ which becomes very 
large.
$\Omega$ could be the total number of individuals, or any other extensive 
quantity (total volume, total mass, etc.).
Our next text theorem justifies \bfi{deterministic physics} for 
\bfi{macroscopic objects}.

\begin{theorem} Suppose the {\bf thermodynamic potential}
\lbeq{e7.2.2.5}
  \Phi (x) := -k T \lim _{\Omega \rightarrow \infty} \frac{\log \rho_\Omega
  (\Omega x)}{\Omega} 
\eeq
exists and is continuously differentiable. Then:

(i) $\Phi$ is homogeneous of degree one, $\Phi (\lambda x) = \lambda \Phi (x)$,
and the {\bf thermodynamic force}
\lbeq{e7.2.2.6}
  F(x):= \nabla \Phi (x)
\eeq
is scaling invariant, $F(\lambda x) = F(x)$. Moreover,
the {\bf Euler equation}
\lbeq{e7.2.2.7}
  \Phi (x) = F \cdot x
\eeq
holds.

(ii) The dynamics becomes deterministic in the thermodynamic limit $\Omega 
\rightarrow \infty$, and is given by the differential equation
\lbeq{e7.2.2.8}
  \dot{x} = u(x) := \sum_l \left(\kappa_l^+ e^{\beta \nu_l^+ \cdot F(x)} -
  \kappa_l ^- e ^{\beta \nu^- _l \cdot F(x)} \right) (\nu_l^- - \nu_l ^+),
\eeq
and for finite $\Omega$ we have the relation
\lbeq{e7.2.2.9}
  \frac{d}{dt} \< f(x) \> = \< u (x) \cdot \nabla f (x) \> + O(\Omega ^{-1}
  \| f \|).
\eeq
\end{theorem}

\bepf 
From the definition \gzit{e7.2.2.5}, we find 
\[
  \Phi (\lambda x) = - k T \D \lim_{\Omega \rightarrow \infty} \log 
  \rho_\Omega (\Omega\lambda x
  / \Omega) \Omega = - \lambda \kbar T \D \lim_{\lambda \Omega \rightarrow 
  \infty} \log \rho_\Omega (\lambda \Omega x) / \lambda \Omega = \lambda \Phi 
  (x),
\]
whence $\Phi$ is homogeneous. Differentiation with respect to $\lambda$ 
yields $\nabla \Phi (x) \cdot x = \Phi (x)$, and for $\lambda=1$ we obtain the 
Euler equation \gzit{e7.2.2.7}.

To derive \gzit{e7.2.2.8}, we assume \at{} more specifically that
\lbeq{e7.2.2.10}
  -\kbar T \log \rho_\Omega (\Omega x) = \Omega \Phi (x) + \psi(x) + 
  O(\Omega ^{-1}).
\eeq
\at{better: we can proceed more generally, using directly the argument 
on p.$10a$ with $v$ in place of $u$, and then approximate $v$ by $u$. 
This shows that the diffusion approximation is also valid, hence is 
more informative. Also use the correct expansion with 
$+ \psi (x)  \log \Omega + O(\Omega^{- 1} \log \Omega)$ which gives 
the correct error term.} 

Thus
\[ \bary{rcl}
   \kbar T \log \rho_\Omega (N-\nu) 
   & {= \atop \gzit{e7.2.2.10} } &  -\Omega \Phi (x-
   \nu/\Omega) - \psi ( x - \nu/ \Omega) + O(\Omega ^{-1})\\
   ~~ & {= \atop \mbox{Taylor}} & - \Omega \Phi (x) + \nabla \Phi (x) \cdot 
  \nu -  \psi(x) + O(\Omega^{-1}),
\eary
\]
hence
\[
  \bary{rcl}
     \log (\rho_\Omega (N - \nu) / \rho_\Omega(N))
     & = & \log \rho_\Omega (N-\nu) -\log \rho_\Omega (N)\\
     & = & \beta \nabla \Phi (x) \cdot \nu + O(\Omega^{-1}) =  \beta \nu \cdot
     F(x) + O(\Omega ^{-1}).
  \eary
\]
The transition rates \gzit{e7.2.1} become
\[
  u_l^\pm (N) = \Omega \kappa_l^\pm e^{\beta \nu_l^\pm \cdot F(x)}
  (1 + O(\Omega ^{-1})).
\]
For any function $f(x)$ and $\tilde{f} (N) := f(N/\Omega)$ we obtain the
forward derivation

$ \bary{rl}
  D \tilde{f} (N) & = \D \sum_\pm \sum_l u_l^\pm (N) \left(\tilde{f} (N\pm 
  \eps_l) - \tilde{f} (N)\right)\\  
  & = \D \sum \sum u_l ^\pm (N) \left(f(x \pm \eps_l/\Omega) -f(x)\right) \\
  & = \D \sum \sum \Omega \kappa_l e^{\beta \nu_l^\pm \cdot F(x)} 
  \left(1 + O(\Omega ^{-1})\right) \left(\nabla f(x) (\pm \eps _l/ \Omega ) 
  + O(\Omega ^{-2})\right)\\
  & = \sum \sum \kappa_l e^{\beta \nu_l ^\pm \cdot F(x)} (\pm \eps_l \cdot
  \nabla f(x) + O(\Omega ^{-1})) \\
  & = u(x) \cdot \nabla f(x) + O(\Omega ^{-1}).
\eary $   

Thus, with $u(x)$ defined by \gzit{e7.2.2.8}, we find
\[
  \frac{d}{dt} \< f(x \> = \frac{d}{dt} \< \tilde{f} (N) \> = \< u(x) \cdot
  \nabla f(x) \> + O(\Omega ^{-1}),
\]
which is \gzit{e7.2.2.9}. In the thermodynamic limit, we obtain the 
deterministic dynamics
\[
  \frac{d}{dt} \< f(x) \> = \< u(x) \cdot \nabla f(x) \>
\]
belonging to the ordinary differential equation \gzit{e7.2.2.8}.
\epf

We call a deterministic process \gzit{e7.2.2.8} with a thermodynamic force 
$F= \nabla \Phi$ determined by a homogeneous thermodynamic potential $\Phi$ of
degree $1$ and corresponding a \bfi{macroscopic collective process}.

\begin{example}[Combinatorial kinetics]

For combinatorial kinetics \gzit{e7.2.2}, Stirlings formula 
$\log n ! = n \log n-n + o(n)$
gives
\[
   \D \log \rho_\Omega = \sum_j (N_j \log \alpha_j - \alpha_j - \log N_j!)   
   = - \sum_j N_j (\log (N_j/ \alpha_j) -1) +o(\sum N_j),
\]
and by \gzit{e7.2.1.5} - \gzit{e7.2.2.5} we find
\lbeq{e7.2.3.10}
  \beta \Phi (x)= \sum_j x_j(\log(x_j/ a_j) -1),  
\eeq
as the negative entropy contribution to an ideal mixture.
Noting that in the multivariate Poisson distribution
\[
  \< u_l ^\pm (N) \>_{\mbox{Poisson}} = 
  k_l ^\pm \prod_j \< N_j \> ^{\nu_{lj}^\pm}
  _{\mbox{Poisson}},
\]
it is more appropriate to replace \at{} expectation by Poisson expectation 
(= equilibrium expectation!) and get from 
\[
    \frac{d}{dt} \< N \> = \< v(N)\> 
  \bary{c}
    ~ \\    
    = \\
    (\ref{7.1.7})
  \eary 
    \sum_l \left( \< u_l^+ (N) \> - \<_l^- (N) \> \right) \eps_l
\]
the \bfi{Poisson approximation}
\[
  \frac{d}{dt} N (t) = \sum_l \left(k_l^+ \prod_j N_j (t)^{\nu_{lj}^+} - k_l^- 
  \prod_j N_j (t)^{\nu_{lj}^-}\right) (\nu_l^- - \nu_l^+)
\]
which is precisely the same. \at{ as what?} 
In the thermodynamic limit, we find
\lbeq{e7.2.3.11}  
  \dot{x} = \sum_l \left(\kappa_l^+ \prod_j \left(\frac{x_j}{a_j}
  \right)^{\nu_{lj}^+} - \kappa_l^- \prod_j \left(\frac{x_j}{a_j}\right) 
  ^{\nu_{lj}^-}\right) (\nu_l^- - \nu_l^+)
\eeq
\[
  = \sum_l (\bar{k}_l^+ \prod x_j^{\nu_{lj}^+} - \bar{k}_l^- \prod x_j 
 ^{\nu_{lj}^-}) (\nu_l^- - \nu_l^+).
\]
\end{example}

For a macroscopic, closed system, we can again prove a dissipation theorem.
   
\begin{theorem}
For a macroscopic colllective process \gzit{e7.2.2.6} - \gzit{e7.2.2.9}
in a closed system, the potential $\Phi (x(t))$ decreases with time. If $A$
is the composition matrix of the process with respect to a full set of 
invariants then $Ax(t)$ is time invariant. 

Moreover, if $\Phi$ is coercive \at{} and bounded 
below then any limit point $x^*$ of $x(t)$ for $t \rightarrow \infty$ is
a stationary point of $\Phi (x)$ on the affine subspace $A x = A x(0)$, and
the only stable equilibria are the local minimizers of  $\Phi (x)$ on some 
subspace $Ax= b$. Moreover, at equilibrium we have detailed balance:
\[
  \nu_l^+ \cdot F = \nu_l^- \cdot F ~~ \mbox{ for all } l.
\]
\end{theorem}

\bepf
We assume $\kappa_l^+ = \kappa _l^- = : \kappa_l$ and write 
$f_l^\pm := \nu_l^\pm \cdot F(x)$. Then
\[
   \bary{rcl}
     \frac{d}{dt} \Phi (x(t)) & = & \nabla \Phi (x(t)) \cdot \dot{x} (t) 
     {= \atop \gzit{e7.2.3.10}} F(x) \cdot U (x)\\
     ~ & {= \atop \gzit{e7.2.3.11}} &  \sum_l \kappa_l (e^{\beta f _l^+} -
     e^{\beta f_l^-}) (f_l^- - f_l^+).
   \eary
\]
By the mean value theorem, there are $f_l \in \overline{f_l^+ f_l}$ such that 
this equals
\[
   \sum_l \kappa_l \beta e^{\beta f_l} (f _l ^+ - f_l^-) ( f_l^- -
   f_l^+).
\]  
Therefore,
\lbeq{e7.2.4.13}
  \frac{d}{dt} \Phi (x(t)) = - \beta \sum_l \kappa_l e^{\beta f_l} (f_l^+
  - f_l ^-)^2 \le 0, 
\eeq
and $\Phi (x(t)$ decreases.

Since $A \nu_l^+ = A\nu_l^+$, \gzit{e7.2.2.8} implies $A \dot{x} = 0$,
so that $Ax(t) = A x(0)$ for all $t$. If $\Phi$ is coercive and bounded 
below, then $\D \lim_{t \rightarrow \infty} \Phi (x(t))$ exists, so
$\frac{d}{dt} \Phi (x(t)) \rightarrow 0$. Therefore \gzit{e7.2.4.13} 
gives $f_l ^+ - f_l^- = 0$ at any limit point $x^*$ of $x(t)$ for 
$t \rightarrow \infty$. By definition of $f_l ^\pm$, this implies that
$F(x^*) \cdot N$ is conserved, and hence that 
\lbeq{e7.2.4.14}
  F(x^*) ^T = \lambda^T A \mbox{ for some }\lambda \in \Rz ^p. 
\eeq
Now we note that the stationary points $x^*$ of $\Phi (x)$ on the
affine subspace $Ax = Ax(0)$ are the stationary points of the Lagrangian
\[
  L(x) = \Phi (x) - \lambda ^T(Ax -  Ax (0)),
\] 
and since $L'(x) = (\nabla \Phi)^T - \lambda^T A = F^T - \lambda^T A$, 
this is just the condition \gzit{e7.2.4.14}.
\epf

\section{Stirred chemical reactions}
Here the $X_i$ are \bfi{substances} (molecule species),
\[
  [X_j] := x_i = N_i / \Omega \mbox{~\bfi{concentrations,}}
\]
\[
  \nu_j := F; \mbox{~\bfi{chemical potentials,}}
\]
\[
  \mbox{transition rates } = : \mbox{ \bfi{reaction rates,}}
\]
\[
  \mbox{rate constants } = : \mbox{ \bfi{reaction constants.}}  
\]
The constants are normalized with respect to so-called \bfi{standard states} 
defined by fixed reference chemical potentials $\mu_j^0$.

The rate constants at standard state are then defined as
\lbeq{e7.3.1}
  k_l^\pm : = \kappa_l^\pm e^{\beta \nu _l^\pm \cdot \mu^0}
  \mbox{ (with } \kappa_l^+ = \kappa_l^- \mbox{ for closed systems),}
\eeq
and we get
\[
  \kappa_l^\pm e^{\beta \nu _l^\pm \cdot F(x)} = f_l^\pm e
  ^{\sum_j \beta \nu _{lj} ^\pm (F_j (x) - \mu_j^0)} =
  k_l^\pm \D \prod_j z_j (x)^{\nu_{lj} ^\pm},
\]
where
\lbeq{e7.3.2}
  z_j (x) := e^{\beta (F_j(x) - \mu_j^0)}
\eeq
is the \bfi{activity} of the $j$th species.
The \bfi{macroscopic} reaction process therefore takes the form of the system 
of differential equations 
\lbeq{e7.3.3}
  \dot{x} = u(x) := \sum_{l=1}^r \left(k_l + \prod_{j=1}^q z_j (x)^{\nu_{lj}
  ^+}
  - k_l ^- \prod_{j = 1} ^q z_j(x) ^{\nu_{lj}^-}\right)
  (\nu_l^- - \nu_l^+).
\eeq
This looks like combinatorial kinetics, which is the special case
$z_j(x) = x_j =[X_i]$ corresponding to an ideal mixture.

Note: \at{} 
For conservation of nonnegativity we need $z_j(x) = 0$ if some
$x_j = 0$; thus $F_j$ must contain a $\log x_j$ term, i.e., $\rho_0$ is an
analytic multiple of the Poisson $-\rho_0$. Thus the $N!$ is perhaps best
moved into the trace?

Chemical reactions are most commonly described at constant temperature $T$
and constant pressure $P$; 
then the appropriate thermodynamic potential is $\Phi= G,$ the \bfi{Gibbs
potential}. A useful phenomenological form is the so-called \bfi{NRTL model}
which describes the potential by a correction to the ideal mixture potential,
\lbeq{e7.3.4}
   \beta G = \sum_j \beta G_j (x_j) + \sum x_j \log x_j - \left(\sum x_j
   \right) \log \left(\sum x_j\right) + \sum_j \frac{(Ax)_j x_j}{x_j + (Bx)_j}.
\eeq
where $A,B$ are matrices with $A_{jj} = B_{jj} = 0, B_j n \ge 0$.
[the $\sum \log \sum$ term vanishes when $\sum x_j = 1 ~(\le \Omega = 
\sum N_j)$ but preserves the homogeneity of $\beta G$ in the general case.]

This model
has the advantage that the $G_j(x_j)$ can be determind form pure substances, 
and the coefficients $A_{j k}, B_{j k}$ can be determined from experiments
with binary mixtures.

For a closed system, the equilibium is characterized by detailed balance,
which says that in \gzit{e7.3.3} the  contribution of each reaction 
vanishes separately. This gives the \bfi{law of mass action},
\lbeq{e7.3.5}
  k_l^+ \prod_{j=1}^q z_j (x) ^{\nu_{lj}^+} = k_l^- \prod_{j=1} ^q z_j (x)
  ^{\nu_{lj}^-},
\eeq
which, in the case of ideal mixing, reduces to the more familiar form
\[
  k_l^+ \prod_{j=1}^q [ X_j] ^{\nu_{lj}^+} = k_l^- \prod _{j=1}^q [X_j] 
  ^{\nu_{lj}^-}.
\]
In particular, for a binary reaction
\[
    A + B 
  \bary{c}
    \alpha \\
    \rightleftharpoons \\
    \beta 
  \eary
    C+D~~~~~~~~~~~~~~~~~+ = \mbox{ input}
\]
we get
\[
  \alpha [A] [B] = \beta [C] [D].
\]
(Traditionally, this is derived by probabilistic hard sphere arguments.)

However, for practical calculation of nonideal cases if is preferable to solve
the constrained optimization problem
\lbeq{e7.3.6}
  \bary{rl}
    & \D \min_N G(T,P,N)\\
    & \D \mbox{s.t. } AN = AN^0,~~ N \ge 0,
  \eary
\eeq
where $N_0$ is the initial composition.

If the Helmholtz potential $A(T,V,N)$ is given as the thermodynamic function, 
then we must also consider
variation of volume by considering volume elements as separte species 
$X_{\fct{vol}}$, and specifying the change of volume in each reaction.

The Gibbs potential is now
\[
  G = PV + A 
\]
(with $V$ corresponding to $N_{\fct{Vol}}$),
and the optimization problem becomes
\lbeq{e7.3.7}
  \bary{l}  
    \D \min_{N,V} (PV +A (T,V,N)\\
    \mbox{s.t. } AN= AN^0,~ N \ge 0.
  \eary
\eeq
($V$ does not contribute to conservation laws.)    
 
If the entropy $S(U,V,N)$ is given as thermodynamic function, we must also 
consider variation of internal energy by considering energy elements as 
separate species $X_{ie}$
and specifying the change of energy in each reaction. (This is typically 
expressded in terms of change of enthalpy $H_{\fct{ent}} = U +PV.)$
\at{at equilibrum $dH = dG+TdS = TdS$ is the change of heat} The Gibbs 
potential is now
\[
  G=PV + U -TS 
\]
with $U \hat{=} N_{jl})$, and the optimization problem becomes
\lbeq{e7.3.8}
  \bary{rl}
  & \D \min_{N,U,V} PV +V -TS (U,V,N) \\
  & \mbox{s.t. } AN = AN^0, ~~ N \ge 0.
  \eary
\eeq

\section{Linear response theory}

We assume a macroscopic situtation ($\Omega$ large but not infinite)
so that the concept of a time-dependent \bfi{external} thermodynamic 
force $F_{ext} (t)$ makes sense. As can be seen from the thermodynamic 
limit
\[ 
  \dot{x} 
= \sum_l \left( \kappa_l^+ e ^{\beta \nu_i^+ \cdot F(x)} -\kappa_l^- 
  e^{\beta \nu_l^- F(x)} \right) (\nu_l^- - \nu_l^+),
\] 
changing $F(x)$ to $F(x) - F_{ext} (t)$ amounts the replacement of 
$\kappa_l ^\pm = \kappa_l$ for a closed system by time dependent rate 
constants
\lbeq{e8.2.1}
  \kappa_l^\pm = \kappa_l e^{-\beta \nu_{\pm l} \cdot F_{ext}(t)}
\eeq
for an open system. \gzit{e8.2.1} is valid for arbitrary systems.

For small forces $F, F_{ext} = O(\Omega ^{- 1/2})$ we can use 
$\omega_l^\pm = \Omega \kappa e^\pm$ to express drift and diffusion in 
terms of
\lbeq{e8.2.2}
  L: = \beta \sum \omega_l \varepsilon_l \varepsilon_l^*
\eeq	
(of order $O(\Omega)$, symmetric) as
\[
  v = -L (F(N) - F_{ext}(t)) + O (1),
\]
\[
  G = 2 \kbar T L+ 0(\Omega^{-1/2}).
\]
An analysis similar to that for the thermodynamic limit reveals that 
higher than second derivatives in $Df$ can be neglected, so that we 
can approximate the process by a diffusion process. Ignoring also the 
error terms in drift and diffusion we find the following \bfi{canonical 
form of linear response theory}, 
\lbeq{e8.2.3}
   \frac{d}{dt}\<f\> =  \<v\cdot\nabla f + \half G : \nabla ^2 f \> ,
\eeq
valid for open macroscopic systems with \bfi{small thermodynamic 
forces}, where
\lbeq{e8.2.4}
   v = - L (\nabla \Phi (N) - F_{ext} (t))
\eeq
\lbeq{e8.2.5}
   G = 2 \kbar TL
\eeq
since, by \gzit{e8.2.2}, the \bfi{transport matrix} $L$ is constant and 
symmetric positive semidefinite.

\begin{remarks}

$1.$ In the absense of external forces, (\ref{e8.2.3} - \ref{e8.2.5}) 
describes a diffusion process in canonical form; since $L$ is constant, 
the covariant drift $u$ agrees with $v$.

$2.$ The entries of $L$ are called the \bfi{transport coefficients}; the
symmetry relations $L_{ik} = L_{ki}$ are called the \bfi{Onsager 
relations}. 

$3.$ The rate constants $\omega_l$ often grow nearly linear with $T$ so 
that $\beta \omega_l$ and hence $L$ only depends weakly on temperature.

$4.$ Written as stochastic differential equation we have
\lbeq{e8.2.5a}
  dN = L (-\nabla \Phi (N) + F_{ext} (t)) dt + d \eps,~~
  d\eps \sim N (0,  2 \kbar T L dt).
\eeq
$5.$ In the space-dependent case, the Onsager relations must be
modified for variables like velocities which are not time-reversal 
invariant; then $L$ is no longer symmetric (i.e., self-adjoint) and 
\gzit{e8.2.5} reads
\lbeq{e8.2.5b}
  G= \kbar T(L+L^T).
\eeq
Linear response theory is used for an impressively large collection of
applications.
\end{remarks}

A particular case where thermodynamical forces are small is when a 
system operates \bfi{close to equilibrium}. In this case the 
potential can be expanded in powers of deviations $x:= N-N^*$ from a 
minimizer $N^*$ of $\Phi (N)$, and
sufficiently close to equilibrium, a quadratic expansion is sufficient.
The Hessian $\Sigma := \Phi'' (N^*)$ at the minimizer is symmetric and 
positive semidefinite, and we get 
\[
  \Phi (N) = \Phi (N^*) + \half x ^T \Sigma x + O(\| x \| ^3).
\]
Ignoring the error term, the substitution into linear response theory 
yields the driven Ornstein-Uhlenbeck process discussed in Section 
\ref{s8.3}.

\section{Open system}\label{s.open}

All interesting phenomena in our world are \bfi{alive} in a more or less
complex way, and this is due to the fact that the systems involved are 
not closed but \bfi{open}, interacting with the \bfi{environment}. 
\bfi{Life is dependent on communication; a closed system is doomed to 
death}, by the second law of thermodynamics which moves the system to 
equilibrium where nothing happens anymore. Such a system can be brought 
back to life only by exerting external influence.
 
Now it is a very remarkable fact that the same thermodynamic laws which 
force closed systems towards death operate on open systems in such a way
that an enormously rich variety of living structures appear, evolve and
change. Indeed, we shall see that the universe is teleological and 
comprehensible precisely because of dissipation: \bfi{Life forces and 
death forces are identical}.

Modern science has just started to understand some details of this 
fascinating vision of the world, and like concepts 
\bfi{self-organization, evolution, synergetics, chaos} created new 
\bfi{paradigms} whose further unfolding will enrich and change our 
scientific understanding of the world.

Mathematically, open systems are characterized by the occurence of
(in general time dependent) \bfi{external flows or forces} 
$\lambda = \lambda(t)$. Corresponding to each value $\lambda$ there is 
a forward derivation $D_\lambda$ which specifies the dynamics at 
constant external conditions.
$\lambda = 0$ describes a closed system with detailed balance, but for 
$\lambda \neq 0$, detailed balance is usually violated. A general open
 system is described by the forward derivation $D$ defined by
\lbeq{e8.1.1}
   D f (\lambda,x) := \frac{\partial f}{\partial \lambda} (\lambda,x) 
   \dot{\lambda} + D_{\lambda} f(\lambda,x)
\eeq
or a time-dependent version of it.
When decay to equilibrium is \bfi{extremely slow}, on time scales 
accessible to experiment, the behavior of a system may, for practical 
purposes, remain different from the steady state; this situation is 
usually approximated by setting tiny reaction rates $\kappa_l^-$ to 
zero and huge concentrations to infinity. The resulting reduced system 
then accounts for the behavior on \bfi{shorter time scales} which are 
hardly affected by these approximations, but detailed balance
is now already lost for $\lambda = 0$.

If \bfi{$\lambda$ is fixed} in \gzit{e8.1.1}, the corresponding Markov 
process is \bfi{autonomous} (i.e., has a time-independent dynamics) 
and we expect a unique limit behavior as $t \rightarrow \infty$. 
If $\lambda$ is small, the system is nearly closed and we expect decay 
to a steady state near equilibrium. When $\lambda$ becomes larger one 
expects a repetitive ``mechanical'' behavior restricted to a 
low-dimensional \bfi{attractor} parametrized by a few microscopically
conserved quantities; depending on the details, the system
may be periodic or chaotic. (This is impossible for dissipative systems 
with detailed balance.)

The fact that attractors are usually of much lower finite-dimensional 
dimension than the system itself accounts to a large extend for the 
\bfi{comprehensibility of our world}, since our awareness can only 
process a limited amount of information.

In reality, the external conditions are usually \bfi{not constant}. 
As $\lambda$ changes slightly the attractor changes its position but 
``generically'' preserves its qualitative (topological) features. 
However, typically, when $\lambda$ crosses certain surfaces (points if 
$\lambda \in \Rz $, curves if $\lambda \in \Rz ^2$), the topology of 
the attractor (fixed point, number of independent periods,
chaos) changes. This situation is described mathematically as 
\bfi{bifurcation}, physically as \bfi{(nonequilibrium) phase 
transition}, and in general language as \bfi{structural change}, 
\bfi{catastrophe}, \bfi{self-organization}, or \bfi{miracle}, depending 
on the context of observer and observation 
(see books \at{} on dynamical systems or synergetics).

The term \bfi{self-organization is somewhat misleading}, since ``self'' 
only refers to the collective response of the system to the 
\bfi{external} stimulus $\lambda (t)$; any life pattern depends
in its existence on the presence of the appropriate environment.
The external stimulus may be viewed as a \bfi{constraint} (canceling 
internal forces) or as \bfi{enticement} (reinforcing internal forces). 
\at{Do constraints lead to higher dissipation of energy?}

The individuals try to \bfi{optimize their owns interest}
\at{is this local equilibrium? relate to competitive equilibrium!}, 
and this collectively leads --- because of dissipation --- to a 
decrease of the free energy of the system. Thus the system remains in 
a state of \bfi{nearly minimal free energy} (subject to the external 
contraints) and hence preserves (``self-organizes'')
its structure. At certain thresholds for 
$\lambda$, the system's free energy surface changes its global minimum,
and in a \bfi{crisis}, a new structure (more alive or more dead)
forms. It is clear that both the external stimulus and the physical law 
(defined by potential and rate constants) are needed to 
``self-organize'' something \bfi{new}.

\vspace*{3cm} 

\bfi{Figure:} A crisis of a collection of noninteracting individuals. 
When (second diagram) \at{} the global minimum is not unique, two 
equilibrium states coexist.

\section{Some philosophical afterthoughts}

Related to the dynamics of open systems is the \bfi{teleological} 
(i.e., goal-directed) nature of our world. In contrast to a widely 
held view, physical laws have a natural teleological interpretation as 
\bfi{democracy of forces} in collaboration and conflict:

Forces are teleological, their goal is trying to move particles along 
the field lines, in a way similar to the way we try to earn our 
livings, make a career, win a game, etc.. The \bfi{laws of physics are 
constraints} which resolve conflicts between competing forces in a 
democratic way (forces are additive).
As in society, if many individual forces are present the 
\bfi{collective behavior is often different from what the individuals 
hope for}. The analogy to human affairs is close, and indeed one can 
model sociological systems by the same mathematics as chemical systems, 
say, though much less accurately.

The \bfi{mind-matter problem} is located on this level, and perhaps one 
is not too far away from modeling mind-matter interaction by open 
collective  processes involving \bfi{mind fields} expressing feeling, 
awareness and will -- and on the society level, mass media).

In a \bfi{local} perspective, our mind is able to set some external 
stimuli to the working of our physical body; further external stimuli 
come through our senses (and perhaps further through inspiration, 
telepathy, etc.).
We all know the lack of self-organization in learning due to wrong 
circumstances --- distracting thoughts (mind stimuli), talking 
neighbours (physical stimuli), missing information (lack of stimuli),
and the phase transitions induced by the presentation of strange new
information --- after a period of intermittend chaos the formation of 
understanding: ``it dawned upon him, she catched on''.  
Reaching a stable equilibrium corresponds to the death of doubts and 
questions.

In a \bfi{global} perspective, \bfi{God's mind} sets the conditions for 
a world created by Him to serve this purpose. Some people think of God 
as the mind of the universe, and in this view one might consider the 
universe as the body of God; but, like with all images of God, this 
view is only partially appropriate.

    \part{Mechanics and differential geometry}\label{p.diffgeom}
\chapter{Fields, forms, and derivatives}\label{c.manifolds}

Part \ref{p.diffgeom} introduces the relevant background from 
differential geometry and applies it to classical Hamiltonian and 
Lagrangian mechanics, to a symplectic formulation of quantum mechanics,
and to Lie groups.

In this chapter we introduce basic material on manifolds,
the associated commutative algebra of scalar fields, and the Lie
algebra of vector fields. All manifolds
used in this book are arbitrarily often differentiable, real manifolds
whose dimension need not be finite. However, we are very brief and
sometimes incomplete in the technical details that need attention in
the infinite-dimensional case; on first reading, the reader may restrict
everything to the finite-dimensional case, where these details are not
required.

We first recall some basics from differential geometry. 
Our approach differs from standard introductions to differential
geometry since, consistent with the theme of the book, all definitions 
are given in an algebraic way. As a side benefit, this prepares the 
reader to noncommutative geometry, only briefly touched in this book,
where a manifold structure is no longer available and all geometry
enters in an algebraic way.
Among other applications, noncommutative geometry gives an interesting 
geometric perspective to the quantum field theory of the standard model.
\at{ref Connes}

Vector fields on a manifold $\Mz $ are essentially equivalent to 
derivations on the commutative algebra $C^\infty(\Mz )$ of scalar 
fields. However, to be able to use the traditional terminology,
where vector fields and the corresponding derivations (Lie derivatives)
are distinguished, we introduce an abstract set $\Wz=\vect \Mz $
of vector fields, whose elements are put into correspondence with 
derivations by means of a mapping $d:\vect \Mz  \to \der \Mz $ which is 
applied at the right. In this way, the calculus on manifolds can be 
formulated in a purely algebraic way, without any reference to the 
manifold.

We therefore formulate everything in terms of an arbitrary topological 
commutative algebra $\Ez$ in place of $C^\infty(\Mz )$, and an 
arbitrary set $\Wz$ in place of $\vect \Mz $. 
However, the main situation that the reader should have in mind 
is where $\Ez$ is an algebra of complex-valued, arbitrarily often 
differentiable functions on a finite-dimensional manifold,
for example $C^\infty(\Rz^n)$. But $\Ez$ could also be the Schwartz 
space of arbitrarily often differentiable functions all of whose 
derivatives decay faster than polynomially at infinity.

As a result, our presentation is completely coordinate-free, except in
some examples. For readers accustomed to differential geometry in 
index notation but not to the coordinate-free Cartan notation, 
we suggest that they translate the definitions and main results into 
coordinates to understand their meaning, but to treat proofs as if the 
concepts introduce new abstract algebraic notions.

\section{Scalar fields and vector fields}\label{s.scalar}

We introduce the objects, operators, and 
operations needed for presenting the traditional differential calculus
in a purely algebraic framework: Lie derivatives applied to multilinear 
forms, and exterior products and the exterior derivative of alternating
forms. As the most important special case, we consider manifolds and
associated geometric notions, in particular diffeomorphisms.

Before giving the definitions, we discuss the letter conventions 
and priority rules used in the formulas.

We typically (i.e., when not forced by conflicts or tradition to do
otherwise) use lower case letters from the middle of the alphabet, such
as $f,g,h$, to denote scalar fields, capital letters from the end of
the alphabet, such as $X,Y,Z$ to denote vector fields, capital letters
from the beginning of the alphabet, such as $A$, for general
multilinear forms, but $z$, for linear forms, $\omega$ for alternating
bilinear forms, and $\eta$ for symmetric bilinear forms.

We use the convention that a Lie derivative acts on the shortest
following expression which is syntactically a vector field or a
multilinear form. Similarly, the exterior derivative operator $d$ acts
from the right on a vector field $X$, giving $Xd$, or from the left
on the shortest following expression that is syntactically an
alternating form $\omega$, giving $d\omega$.

The wedge product $\wedge$ has lower priority than
the operations written as juxtaposition, but higher priority than $+$
and $-$.

\begin{dfn}~\\
(i) A \bfi{differential geometry} consists of a  commutative algebra 
$\Ez$ containing $\Cz$, a left $\Ez$-module $\Wz$ with an additional
Lie product $\lp$, both equipped with a topology such that all 
operations are continuous, 
and a continuous mapping $d$ (written on the right),
which maps $X\in\Wz$ to $Xd\in \der \Ez$, such that
\lbeq{e.dgeom}
(X+Y)d= Xd+Yd,~~~ (f X)d=f(Xd),~~~(X\lp Y)d = [Xd,Yd],
\eeq
for all $X,Y\in \Wz,f\in\Ez$.
The differential geometry is called \bfi{(non-)commutative} if
the multiplication in $\Ez$ is (non-)commutative.

(ii) We refer to the elements of $\Ez$ as \bfi{scalar fields}, and to 
the elements of $\Wz$ as \bfi{vector fields}.
The \bfi{Lie derivative} of a vector field $X$ is the linear 
mapping $L_X$ which maps a scalar field $ f$ to\footnote{
As will become apparent in Section \ref{s.ext} (cf. Theorem \ref{t6.2}),
we may read the term
$Xd~f$ also as product of the vector field $X$ with the 
exact linear form $df$. Until then, we shall write an explicit space 
after $d$ to remind the reader of the correct way to group the
letters. 
} 
\lbeq{e.ld6s}
L_X f:=Xd~ f,
\eeq
and a vector field $Y$ to 
\lbeq{e.ld6}
L_XY ~:= ~ X \lp Y ~=~ -L_Y X ~.
\eeq
The scalar field $L_X f$ (resp. the vector field $L_XY$) 
is called the \bfi{directional derivative} of the scalar field $ f$ 
(resp. the vector field $Y$) in the direction of the vector field $X$.
\end{dfn}

The following example is responsible for the naming.
Interpreting the set $\Mz $ in the example as the 
domain of a chart of a finite-dimensional manifold, one can translate 
everything said here to general finite-dimensional manifolds by a 
process described in all books on differential geometry.  
Thus the example gives essentially the full intuition for our 
constructions, except for the complications that may arise in 
infinite dimensions. 

\begin{expl}\label{ex.Rn}
\bfi{(Differential geometry of open subsets in $\Rz^{\times n}$)}\\
Let $\Rz^{\times n}$\index{$\Rz^{\times n}$} denote the vector space 
of row vectors\footnote{
The index notation corresponds to standard differential 
geometric practice when working in a chart of a manifold (which is 
essentially the situation we are discussing here).
The interpretation in terms of rows (row vectors = \bfi{rovectors}, 
indexed by upper indices = \bfi{roindices}) and columns
(column vectors = \bfi{covectors}, indexed by lower indices = 
\bfi{coindices}) makes the transition to 
standard linear algebra transparent.
} 
$x=(x^1,\dots,x^n)$ with $n$ real components $x^j$,
let $\Mz $ be a nonempty, open subset of  $\Rz^{\times n}$, and let
$\Ez=C^\infty(\Mz )$ and $\Wz=C^\infty(\Mz ,\Cz^{\times n})$, equipped 
with the weak topology. \at{define}
Thus scalar fields are real-valued functions, while vector fields are
row vector valued functions. In terms of the partial differential
operators $\partial_j$ defined by 
\[
\partial_jf(x):=\partial f(x)/\partial x^j,
\]
we define the \bfi{gradient} $\partial f$ of a scalar field as the
column vector
with $n$ entries 
\[
(\partial f)_j=\partial_j f.
\]
It is not difficult to show that an arbitrary derivation $\delta$ on t
he algebra of scalar fields can be uniquely expressed as a linear 
partial differential operator of the form
\[
\delta = X\partial = \sum_{j=1}^n X^j \partial_j,
\]
with a vector field $X$. This derivation then acts on scalar fields $f$ 
as
\[
\delta f = X\partial f = \sum_{j=1}^n X^j \partial_j f.
\]
Thus the mapping $d$ which maps the vector field $X$ to the 
differential operator $X\partial$ is a bijection of the type required 
in the previous example. Thus we have a canonical differential geometry;
it is clearly commutative.
The reader is invited to check that the Lie derivative takes the form
\lbeq{e.ldc}
L_X f   = X \partial  f,~~~L_X Y = X\partial Y - Y \partial X.
\eeq
A second, noncanonical differential geometry results by using in the
above construction in place of $C^\infty(\Mz )$ the subalgebra
\idx{$C^\infty_0(\Mz )$} of scalar fields with compact support, 
and in place of $C^\infty(\Mz ,\Cz^{\times n})$
the subspace \idx{$C^\infty_0(\Mz ,\Cz^{\times n})$} of vector fields
with compact support.
\end{expl}

\begin{expl}
\bfi{(Canonical differential geometries)}\\
Let $\Ez$ be an arbitrary topological algebra containing $\Cz$. 
We may give
$\Ez$ the structure of a differential geometry by picking an arbitrary
set $\Wz$ with the same cardinality as $\der \Ez$, and choosing an
arbitrary bijection $d$ from $\Wz$ to $\der \Ez$.
$\Wz$ inherits all properties of $\der\Ez$ by means of
the bijection $d$: We turn $\Wz$ into a Lie algebra and a
topological $\Ez$-module by defining
\[
X+Y:= (Xd+Yd)d^{-1},~~~ f X:=( f(Xd))d^{-1}\,,
\]
\[
X\lp Y := [Xd,Yd]d^{-1}.
\]
for $X,Y\in\Wz$ and $f\in\Ez$, and by calling a set $S\in\Wz$
closed if its image under $d$ is closed in the
topology of $\der \Ez$ induced by that of $\Ez$.
The result is a differential geometry.
We call differential geometries constructed in this way 
\bfi{canonical}. 
\end{expl}

\begin{prop}
The \bfi{product rule}
\lbeq{e.ld3v}
L_X( f g)=(L_X f )g+ f (L_Xg),~~~
L_X( f Y)=(L_X f )Y+ f (L_XY),
\eeq
the \bfi{commutation rule}
\lbeq{e.ld4v}
[L_X \,, L_Y]=L_{X\!\lp Y},
\eeq
and the equations
\lbeq{e.ld27a}
L_{ f X}\, g = f L_X \, g,
\eeq
\lbeq{e.ld27}
L_{ f X}Y= f L_XY -  XYd~f
\eeq
hold for $ f,g \in\Ez$ and $X,Y\in\Wz$,
\end{prop}

\bepf
The first part of \gzit{e.ld3v} is trivial since in this case
$L_X = Xd$ is a derivation on $\Ez$. For the second part of 
\gzit{e.ld3v}, we note that
\[
\bary{rcl}
\Big(L_X(f Y)\Big)d\;g &=& (X \lp fY)d\;g ~=~ [Xd \,, fYd\,] g
	~=~ Xd\Big(f(Yd~g)\Big) - fYd\;(Xd~g) \\
&=& (Xd~f)(Yd~g) + f Xd\;(Yd~g) -fYd\;(Xd~g) ~.
\eary
\]
Also,
\[
((L_X f)Y)d~g ~=~ ((Xd~f)Y)d~g ~=~ (Xd~f)(Yd~g)
\]
and
\[
\bary{rcl}
\Big(f (L_XY)\Big)d~g &=& \Big(f (X \lp Y)\Big)d~g ~=~ f[Xd \,, Yd]g \\
&=& f\Big( Xd\;(Yd~g) - Yd\;(Xd~g) \Big) ~=~ fXd\;(Yd~g) - fYd\;(Xd~g)
\eary
\]
Putting these three pieces together proves the second part of
\gzit{e.ld3v}.

To prove \gzit{e.ld4v}, note that $L_Y f = Yd~f$ is in $\Ez$,
so $L_X L_Y f = L_X(Yd~f) = Xd\;(Yd~f)$. Interchanging $X,Y$ we find
for the commutator:
\[
[L_X , L_Y]f ~=~ Xd~(Yd~f) - Yd~(Xd~f) ~=~ [Xd, Yd]f ~=~ (X\lp Y)d~f
	~=~ L_{X\slp Y} \, f ~.
\]
Formula \gzit{e.ld27a} is immediate from the definition, and
\gzit{e.ld27} follows from the product rule
$X\lp  f Y = (Xd~ f )Y+ f (X\lp Y)$ by swapping $X$ and $Y$, using the
anticommutativity of $\lp$.
\epf

In the following, we develop the differential calculus for
commutative differential geometries only; thus, with exception of the 
remarks on noncommutative geometry in Section \ref{s.mani},
\bfi{the algebra $\Ez$ of scalar fields is always assumed to be 
commutative.} In this case, we extend the left module structure on 
vector fields to a bimodule structure by putting
\[
Xf:=fX
\]
for $f\in\Ez$ and $X\in\Wz$.
Note that some authors treat vector fields as synonymous with 
derivations and therefore write $X(f)$ for $Xd~f$. This should not
be confused with the present notation $Xf$ for multiplying the vector 
field $X$ with the scalar field $f$.

\section{Multilinear forms}\label{s.mforms}

Apart from scalar and vector fields, differential geometry makes
heavy use of multilinear forms and tensors, which we define 
next.

\begin{dfns}~\\
(i) A \bfi{linear form} $\zeta$ is a continuous, $\Ez$-linear mapping
$\zeta:\Wz\to\Ez$ (written on the right\footnote{Strictly speaking, 
they should be called $\Ez$-linear forms, and a similar remark applies 
later to multilinear forms.  Talking about a {\em form} rather than a 
mapping implies the assumption of continuity.\\
The standard notation for $X\zeta$ is $i_X\zeta=\zeta(X)$; the present
notation simply replaces $i_X$ by $X$. This way of writing the mapping
generalizes standard matrix calculus if we use the intuition gained
from Example \ref{ex.Rn} and think of vector fields as row vectors and
of linear forms as column vectors, an intuition that extends to matrix
fields. Since in the general situation, linear forms are often called
\bfi{covectors}, we shall occasionally use the analogous word {\bf
rovector} to denote a vector field, although, strictly speaking, one
should talk about covector fields and rovector fields. The same
ambiguity is traditionally maintained for multilinear forms on
manifolds, which refer both to the corresponding fields and to their
values at a particular point.
}\label{fn.iX}) 
which maps the vector field $X$ to the scalar field $X\zeta$. We
write $\Wz^*$ for the $\Ez$-module consisting of all linear forms,
(sometimes called the "$\Ez$-dual" of $\Wz$), with scalar
multiplication of $\zeta\in\Wz$ by $f\in\Ez$ defined via
\[
X( f \zeta):= f(X\zeta).
\]
(ii) A \bfi{$c$-linear} form $\phi$ is a mapping
$\phi : \Wz\times\dots\times\Wz \to \Ez$ (with $c$ factors of $\Wz$
in the Cartesian product) such that the image\footnote{
The traditional notation for $X_1\dots X_c \phi$ is
$i_{X_1}\dots i_{ X_c}\phi = \phi(X_c,\dots,X_1)$; as for linear forms, 
the present notation simply replaces the $i_X$ by $X$.
Note the reverse order resulting in the arguments written in the 
traditional way, needed 
in order that \gzit{e.XA} together with our definition \gzit{e.ld8} 
of insertion is consistent with the traditional definition
$(i_X\phi)(X_1,\dots,X_{c-1})=\phi(X,X_1,\dots,X_{c-1})$. In our
notation this translates into
$X_{c-1}\dots X_1(i_X\phi)=X_{c-1}\dots X_1X\phi$.
} 
$X_1\dots X_c \phi$ of $(X_c,\dots,X_1)\in \Wz\times\dots\times\Wz$
depends $\Ez$-linearly on each argument $X_k$, i.e., if, for all
$X_j,Y,Z\in\Wz$ and $ f, g \in\Ez$,
\[
X_1\dots ( f Y+ g Z)\dots X_c \phi  
~=~   f X_1\dots Y\dots X_c \phi  
~+~ g X_1\dots Z\dots X_c \phi ~.
\]
Here the unindexed argument between the dots replaces the $k$th 
argument $X_k$, for some $k$ in $1,\dots,c$.
In the degenerate case $c=0$, we consider the $0$-linear mappings 
to be the scalar fields.

(iii) We write\footnote{
Writing the ``$c$" in $\Wz_c$ as a subscript serves
as a reminder that when an element $\phi$ of $\Wz_3$ (say) is written
in index notation, it has 3 {\em lower} indices: ``$\phi_{ijk}$".
Indeed the ``$c$" is intended to be suggestive of ``covector". Note
that in terms of the direct products of $c$ factors $\Wz$ or $\Wz^*$, 
there is a canonical isomorphism $\Wz^*\times\dots\times\Wz^* 
\cong (\Wz\times\dots\times\Wz)^* =\Wz_c^*$.
} 
$\Wz_c$ for the $\Ez$-module of continuous $c$-linear 
mappings on $\Wz$. Scalar multiplication of $\phi\in\Wz_c$ by 
$f\in\Ez$ is defined via
\[
X_1\dots X_c( f \phi) ~:=~ f(X_1\dots X_c \phi) ~.
\]
The elements of $\Wz_c$ are called \bfi{multilinear forms}
or \bfi{$c$-linear forms}; for $c=2$ also \bfi{bilinear forms}.
Note that $\Wz_0=\Ez$ consists of scalar fields (or 0-forms), and 
$\Wz_1=\Wz^*$ consists of linear forms (or 1-forms).

(iv) The \bfi{product} of a vector field $X\in\Wz$ and a $c$-linear
form $\phi\in\Wz_c$ is for $c=0$ the vector field $X \phi$ defined by
scalar multiplication with the scalar $\phi$, and for $c>0$ the
$(c-1)$-linear form $X \phi$ defined by
\lbeq{e.XA}
X_1\dots X_{c-1}(X \phi) ~:=~ X_1\dots X_{c-1}X \phi
~,~\Forall X_1,\dots,X_{c-1}\in\Wz.
\eeq
The operator $i_X$ defined on multilinear forms $\phi$ by
\lbeq{e.ld8}
i_X\phi ~:=~ X\phi
\eeq
is traditionally called the \bfi{insertion} of $X\in\Wz$; cf.
footnote \ref{fn.iX}. \at{footnote reference comes out wrong}

(v) A $c$-linear form $\phi$ is called \bfi{alternating} (or 
a $c${\bf-form}) if either $c\le 1$ or $X\phi$ is alternating
and $XX\phi=0$ for all vector fields $X$. 
$\phi$ is called \bfi{symmetric} if either $c\le 1$ or $X\phi$ 
is symmetric and $XY\phi=YX\phi$ for all vector fields $X,Y$.
We write $\Az_c$ and $\Sz_c$ for the space of alternating and 
symmetric $c$-linear forms, respectively. In particular,
\[
\Ez_0=\Wz_0=\Ez,~~~ \Ez_1=\Wz_1,~~~ \Ez_2\oplus \Sz_2 = \Wz_2 ~.
\]
(vi) The \bfi{transpose} of a bilinear form $\phi$ is the bilinear form
$\phi^T$ defined by
\lbeq{e.transpose}
XY\phi^T:=YX\phi
\eeq
for all vector fields $X,Y$.
In particular, a bilinear form $\phi$ is symmetric iff $\phi^T=\phi$ and
alternating iff $\phi^T=-\phi$. 
A bilinear form $ \phi$ is called \bfi{nondegenerate} if every
linear form $\zeta\in\Wz^*$ can be written as $\zeta=X\phi$ for a unique
vector field $X$; otherwise \bfi{degenerate}.
A bilinear form $\phi$ may be considered as a linear 
mapping from $\Wz$ to $\Wz^*$ that maps the vector field $X$ to the 
linear form $X\phi$. If $\phi$ is nondegenerate, this mapping is 
invertible, and the inverse $\phi^{-1}$ is a linear 
mapping from $\Wz^*$ to $\Wz$, which maps a linear form $\zeta$
to the vector field $\zeta\phi^{-1}$ in such a way that 
\lbeq{e.Ainv}
\zeta\phi^{-1}\phi=\zeta.
\eeq
A nondegenerate bilinear form is called a \bfi{symplectic form}
if it is alternating, and a \bfi{metric} if it is symmetric.

(vii) A $[c,r]${\bf-tensor field}\footnote{ 
An $[c,r]$-tensor is also called a tensor of $[{r \atop c}]$-valence
(\sca{Penrose \& Rindler} \cite{PenRI}).
With traditional index notation, a $c$-linear form is written with
$c$ lower (co)indices, and 
a $[c,r]$-tensor is written with $r$ upper (ro)indices and
$c$ lower (co)indices. E.g., a $[3,2]$-tensor $T$ is written
${T_{ijk}}^{mn}$, and the image $T\phi$ of a bilinear form is written
$(T\phi)_{ijk}={T_{ijk}}^{mn}\phi_{mn}$, using the traditional {\bf
Einstein summation convention} (which deletes the explicit indication
of the sum over $m$ and $n$ so as not to unnecessarily inflate the
formulas without conveying any more information). In the more modern
\bfi{abstract index notation} of \sca{Penrose \& Rindler} \cite{PenRI}, 
such repeated indices
denote instead an insertion (dual-pairing) {\em without} any implied
connotation of summation over basis-dependent components, and such
indices may be used to keep explicit track of the types of complicated
objects.
} 
$T$ is a continuous $\Ez$-linear mapping $T:\Wz_r \to \Wz_c$.
The space of $[c,r]$-tensor fields
is denoted by\footnote{ 
For the differential geometry of open subsets $\Mz $ of $\Rz^n$
(Example \ref{ex.Rn}), $\Wz[c,r] =\Wz_c^r$ is, in the traditional
terminology, the space of sections of the tensor bundle $T_c^r\Mz $.
} 
$\Wz[c,r] =\Wz_c^r=\Lin(\Wz_r,\Wz_c)$. A $[1,1]$-tensor
field is called a \bfi{matrix field}.

\end{dfns}

\begin{rems}~\\
(i) Multilinearity implies
\[
XY\omega=-YX\omega 
\mbox{~~~for alternating $c$-forms $\omega$ with $c>1$}.
\] 
Thus, for $c\ge 2$,
\lbeq{e.ld10b}
 i_X^2=0,~~i_Xi_Y =-i_Yi_X \mbox{~~~on alternating $c$-linear forms},
\eeq
whereas
\lbeq{e.ld10c}
i_Xi_Y =i_Yi_X \mbox{~~~on symmetric $c$-linear forms}.
\eeq
(ii) Note that there is a canonical identification
of $\Wz[c,0]$ with $\Wz_c$, and a canonical embedding of $\Wz$ into
$\Wz[0,1]$. In the case of finite-dimensional manifolds, we may also
identify $\Wz[0,1]$ with $\Wz$.

(iii) The ordinary operator product of a $[c',c]$-tensor and an
$[c,r]$-tensor is well-defined, and is a $[c',r]$-tensor:
$\Wz[c',c]\Wz[c,s]\subseteq\Wz[c',s]$. In particular,
$\Wz[1,1]=\Lin \Wz_1$ is an algebra of matrix fields.
\end{rems}

\begin{thm}\label{t.GenLieDeriv}
For every vector field $X$, the Lie derivative can be extended uniquely
to a linear operator $L_X$ mapping\footnote{
The Lie derivative can also be extended to tensors $T\in\Wz[c,r]$ by 
defining
\[
(L_XT)B:=L_X(TB)-TL_XB \for B \in \Wz_r.
\]
We do not need such an extension for our limited 
applications; it would be needed, however, in a treatment of general 
relativity. The reader is invited to verify that $L_XT\in\Wz[c,r]$
and to formulate and prove the analogues to \gzit{e.ld3} and  
\gzit{e.ld4}.
} 
vector fields to vector fields and $c$-linear forms to $c$-linear 
forms, and satisfying the \bfi{product rule}
\lbeq{e.ld3}
L_X( f \phi)=(L_X f )\phi+ f (L_X\phi),~~~
L_X(Y\phi)=(L_XY)\phi+Y(L_X\phi).
\eeq
for $ f \in\Ez$, $Y\in\Wz$, and $\phi\in \Wz$ or $\phi\in \Wz_c$ .
The extended Lie derivative satisfies the commutation rules
\lbeq{e.ld4}
[L_X,L_Y]=L_{X\lp Y},
\eeq
\lbeq{e.ld9}
[L_X,i_Y]=i_{X\lp Y}
\eeq
for $X,Y\in\Wz$.
\end{thm}

\bepf 
We first assume that the product rule holds, and show that this
fixes the operation of $L_X$ on all multinear forms.
By the product rule  \gzit{e.ld3},
\lbeq{e.ld21}
YL_X\phi = L_X(Y\phi)-(L_XY)\phi, 
\eeq
This formula shows that $L_X$ is determined on $c$-linear forms
by its action on  $(c-1)$-linear forms, and since it is given on
scalar fields, it is unique if it exists at all. 

Conversely, to show existence of the extension, we define $L_X$ 
recursively by \gzit{e.ld21},
starting with the known action of $L_X$ on scalar fields. 

Since
$( f Y)L_X\phi = L_X( f Y\phi)-L_X( f Y)\phi 
= (L_X f )Y\phi+ f L_X(Y\phi)-(L_X f )Y\phi- f (L_XY)\phi
=  f (L_X(Y\phi) -  f  (L_XY)\phi 
=  f (YL_X\phi)$, we see inductively that $YL_X\phi$ is
$\Ez$-linear in $Y$, so that $L_X\phi$ is indeed a tensor.

The first part of the product rule holds since, by \gzit{e.ld21},
the equation
$YL_X(f\phi)=L_X(Yf\phi)-(L_XY)(f\phi)=L_X(fY\phi)-(L_XY)(f\phi)
= (L_Xf)Y\phi+fL_X(Y\phi)-(L_XY)f)\phi-f(L_XY)\phi
=Y(L_Xf)\phi+Yf(L_X\phi)$
holds for all vector fields $Y$.
The second part of the product rule follows directly from \gzit{e.ld21}.

To prove \gzit{e.ld4}, we first note that by \gzit{e.ld21}, we have
$ZL_XL_Y\phi = L_X(ZL_Y\phi)-(L_XZ)L_Y\phi
=L_X(L_Y(Z\phi)-(L_YZ)\phi) - (L_XZ)L_Y\phi
=L_XL_Y(Z\phi) - (L_XL_YZ)\phi - (L_YZ)L_X\phi - (L_XZ)L_Y\phi$.
The last two terms are symmetric in $X,Y$, hence cancel when taking 
the difference with $ZL_YL_X\phi$ in 
$Z[L_X,L_Y]\phi=ZL_XL_Y\phi-ZL_YL_X\phi
=(L_XL_Y-L_YL_X)(Z\phi) - (L_XL_YZ-L_YL_XZ)\phi
=[L_X,L_Y](Z\phi)-([L_X,L_Y]Z)\phi$.
Since \gzit{e.ld4} is already known by \gzit{e.ld4v} to hold on vector 
fields and on scalar fields (0-linear forms),
we assume that we know its validity for the action on $c$-linear forms. 
Taking for $\phi$ a $(c+1)$-linear form, we may conclude that
$Z[L_X,L_Y]\phi=L_{X\lp Y}(Z\phi)-(L_{X\lp Y}Z)\phi = ZL_{X\lp Y}\phi$
by \gzit{e.ld21}. Since $Z$ was arbitrary, we conclude that
$[L_X,L_Y]\phi=L_{X\lp Y}\phi$ for $(c+1)$-linear forms $\phi$. 
By induction, \gzit{e.ld4} holds in general.

\gzit{e.ld9} follows from the product rule \gzit{e.ld3} since 
\[
[L_X,i_Y]\phi=L_X(Y\phi)-Y(L_X\phi)=(L_XY)\phi=(X\lp Y)\phi
	=i_{X\lp Y}\phi.
\]
\epf

The reader may wish to prove inductively that, for $c'$-linear forms 
$\phi$ with $c'\ge c$, 
\[
X_1\dots X_cL_X\phi 
= L_X(X_1\dots X_c\phi)-\sum_{k=1}^c X_1\dots X\lp X_k\dots X_c\phi.
\]

\begin{prop}
The Lie derivative of an alternating $c$-form is again an alternating 
$c$-form. 
\end{prop}

\bepf
Indeed, for any alternating form $\omega$, 
\[
\bary{rcl}
YYL_X\omega
&=& Y(L_X(Y\omega)-(L_XY)\omega) ~=~ Y L_X(Y\omega)-Y(L_XY)\omega \\
&=& Y L_X(Y\omega)+(L_XY)Y\omega = L_X(YY\omega) = 0 ~.
\eary
\]
\at{proof that $YL_X\omega$ is alternating is missing.}
\epf

\section{Exterior calculus}\label{s.ext}

In the case of differential geometry in $\Rz^{\times n}$ (Example
\ref{ex.Rn}), the gradient operator $d=\partial$ behaves symbolically
similarly to a covector, except for the nontrivial behavior implied by
the Leibniz product rule. However, the gradient operator $d$, defined
there on scalar fields only, cannot be extended to a gradient operator
that associates with a general $c$-linear form $\phi$ a $(c+1)$-linear
form $d\phi$ such that $L_X\phi=Xd\phi$ for all vector fields $X$. 
The existence of such an
extension would imply that $L_{ f X}\phi= f L_X\phi$. While this holds
by \gzit{e.ld27a} when $\phi$ is a scalar field, it fails already when
$\phi$ is a linear form. For a linear form $\zeta$, we have instead
\[
L_{ f X}\zeta= f L_X\zeta + d f X\zeta,
\]
which follows as a special case of \gzit{e.extb} below, or from 
\[
L_X\zeta=X(\partial \zeta)^T + \partial X\zeta
\]
by substituting $fX$ for $X$. However, the
gradient can be generalized in a different way to alternating forms, 
leading to the exterior derivative. 

The generalization is valid not only for Example \ref{ex.Rn}, but in 
full generality. To define the  exterior derivative we need some 
preparations.

\begin{thm}

For every linear form $\zeta$, there is a unique $\Ez$-linear mapping 
$\zeta\wedge$ mapping alternating $c$-forms to alternating 
$(c+1)$-forms for $c=0,1,2,\dots$,
and vector fields to zero, such that
\lbeq{e.ld26}
(X\zeta)\omega = X(\zeta\wedge\omega) + \zeta\wedge X\omega
\eeq
for all vector fields $X$ and all alternating $c$-forms $\omega$.
$\zeta\wedge\omega$ is called the \bfi{exterior product} or 
\bfi{wedge product}\footnote{
One can define an exterior product $\omega'\wedge\omega$
for arbitrary alternating forms $\omega,\omega'$,
but we do not need it.
} 
of $\zeta$ and 
$\omega$. The exterior product satisfies the rules
\lbeq{e.exta}
\zeta\wedge \zeta' = - \zeta'\wedge \zeta,
\eeq
\lbeq{e.extb}
L_X(\zeta\wedge\omega) = L_X\zeta\wedge\omega +\zeta \wedge L_X\omega
\eeq
for alternating $c$-forms $\omega$, linear forms $\zeta,\zeta'$, 
and vector fields $X$.
\end{thm}

\bepf 
A necessary and sufficient condition for \gzit{e.ld26} to hold is
that
\lbeq{e.ld26b}
X(\zeta\wedge\omega)=(X\zeta)\omega - \zeta\wedge X\omega;
\eeq
\MMt{
($p=2$)
\[
\bary{rcl}
X^k \zeta_{[k} \omega_{ij]}
  &=& \frac{1}{2} X^k \Big(\zeta_k\omega_{ij} - \zeta_i\omega_{kj}
  		- \zeta_j\omega_{ik} - \zeta_k\omega_{ji} \Big)
  ~=~ \frac{1}{2} X^k \Big(2 \zeta_k\omega_{ij}
		- \zeta_i\omega_{kj} - \zeta_j\omega_{ik} \Big)\\
  &=& X^k \zeta_k\omega_{ij}
	  - \frac{1}{2}X^k(\zeta_i\omega_{kj} + \zeta_j\omega_{ik})
  ~=~ (X\zeta)\omega_{ij}
      - \frac{1}{2}(\zeta_i X^k\omega_{kj} - \zeta_j X^k\omega_{ki})\\
  &=& (X\zeta)\omega_{ij}
      - \frac{1}{2}(\zeta_i (X\omega)_j - \zeta_j(X\omega)_i)
  ~=~ (X\zeta)\omega_{ij} - \zeta_{[i} (X\omega)_{j]} \\
  &=& \Big((X\zeta)\omega\Big)_{ij}
  	- \Big(\zeta\wedge (X\omega)\Big)_{ij} \\
\eary
\]
} 
in particular,
\lbeq{e.ld26a}
\zeta\wedge \omega =\zeta\omega \mbox{~~~for a 0-form~} \omega,
\eeq
This completely specifies the exterior product of a linear form 
$\zeta$ and an alternating $(c+1)$-form $\omega$, given the exterior 
product with an alternating $c$-form. 
Therefore, if the exterior product exists, it is unique.

To prove the existence of the exterior product, we have to
define the exterior product of a linear form $\zeta$ and an 
alternating $c$-form $\omega$ to be the expression $\zeta\wedge\omega$
defined for $c=0$ by \gzit{e.ld26a}
and for $c>0$ recursively by \gzit{e.ld26b}.
To show that we really get an alternating $(c+1)$-form, we need to 
show that $X(\zeta\wedge\omega)$ is alternating for $c>0$ and any 
vector field $X$, and verify
\lbeq{e.ld23a}
( f X)(\zeta\wedge\omega) =  f (X(\zeta\wedge\omega))
\eeq
and
\lbeq{e.ld23b}
XX(\zeta\wedge\omega) =0.
\eeq
\at{proof missing, also proofs of \gzit{e.exta} and \gzit{e.extb}}

\epf

\at{add sum formula for the general case?
The reader may wish to prove inductively that, for alternating 
$c'$-forms $\omega$ with $c'\ge c$, ...}

\begin{thm}\label{t6.2}
There is a unique linear mapping $d$ mapping vector fields to zero
and alternating $c$-forms to alternating $(c+1)$-forms 
(for $c=0,1,2,\dots$) such that\footnote{
In particular, the relation $L_X=Xd$ valid on scalar fields
fails to hold for the extension of $L_X$ and $d$ to alternating forms.
} 
\lbeq{e.ld22}
L_X\omega = Xd\omega + d(X\omega)
\eeq
for all alternating $c$-forms $\omega$ and vector fields $X$. 
The alternating form $d\omega$ is called
the \bfi{exterior derivative} of $\omega$, and satisfies the
\bfi{exactness relation}
\lbeq{e.ld25}
dd\omega = 0
\eeq
and the \bfi{product rules}
\lbeq{e.ld28}
d( f\omega)= f d\omega+d f \wedge\omega,
\eeq
\lbeq{e.ld28a}
d(\zeta\wedge\omega)=d\zeta\wedge\omega-\zeta\wedge d\omega,
\eeq
\lbeq{e.ld29}
L_{ f X}\omega =  f L_X\omega +d f \wedge X\omega,
\eeq
for all alternating froms $\omega$, scalar fields $f$, linear forms 
$\zeta$ and vector fields $X$.
Note the minus sign in \gzit{e.ld28a}!
\end{thm}

\bepf
A necessary and sufficient condition for \gzit{e.ld22} to hold is
that
\lbeq{e.ld22b}
X(d\omega)=L_X\omega - d(X\omega);
\eeq
in particular,
\lbeq{e.ld22a}
X(d\omega)= Xd~\omega =L_X\omega  \mbox{~~~for a 0-form~} \omega.
\eeq
This completely specifies the exterior derivative of an alternating 
$c$-form.
Therefore, if the exterior derivative exists, it is unique.
To prove the existence of the exterior derivative, we have to
define the exterior derivative of an alternating $c$-form $\omega$ 
to be the expression $d\omega$ determined for $c=0$ by 
\gzit{e.ld22a} and for $c>0$ recursively by \gzit{e.ld22b}.

To show that we really get an alternating $(c+1)$-form, we need to 
show that $X(d\omega)$ is alternating for $c>0$ and any 
vector field $X$, and verify
\lbeq{e.ld30}
( f X)d\omega =  f (Xd\omega)
\eeq
which shows that $Xd\omega$ is $\Ez$-linear 
in $X$, so that $d\omega$ is a $(c+1)$-linear form,
and
\lbeq{e.ld23}
XXd\omega =0
\eeq
which proves antisymmetry.
The proof of \gzit{e.ld30} is based on \gzit{e.ld29}.
To prove \gzit{e.ld29}, we \at{start missing} get inductively
$Y(L_{ f X}\omega- f L_X\omega-d f \wedge X\omega) 
=YL_{ f X}\omega- f YL_X\omega-Yd f \wedge X\omega
=L_{ f X}(Y\omega)-(L_{ f x}Y)\omega
 - f (L_X(Y\omega)-(L_XY)\omega)
 -((Yd f ) X\omega-d f \wedge YX\omega)$.
Using the induction hypothesis on the first term and \gzit{e.ld28}
on the second term, one finds that all terms cancel.
From \gzit{e.ld29} one obtains
$( f X)d\omega = L_{ f X}\omega-d( f X\omega) 
=  f L_X\omega+d f  \wedge X\omega -( f d(X\omega)+d f \wedge X\omega)
= f (L_X\omega-d(X\omega) = f (Xd\omega)$,
showing that \gzit{e.ld30} holds.
\gzit{e.ld23} follows inductively \at{check initial value} from
\[
XXd\omega = XL_X\omega - Xd(X\omega)
=XL_X\omega-L_X(X\omega)-d(XX\omega)
=-(L_XX)\omega-d(XX\omega)=0-0=0.
\]
To prove the product rule \gzit{e.ld28}, we 
\at{initialize}, and then inductively
$Xd( f \omega)=L_X( f \omega)-d( f X\omega)
=(L_X f )\omega+ f L_X\omega-( f d(X\omega)+d f \wedge X\omega)
= f (L_X\omega-d(X\omega))+(Xd f )\omega-d f \wedge X\omega
= f Xd\omega+X(d f \wedge\omega)=X( f d\omega+d f \wedge\omega)$,
completing the induction.

\at{proof of the other product rule?}

To prove the exactness relation \gzit{e.ld25}, we need the formula
\lbeq{e.ld24}
d(L_X\omega)=L_X(d\omega).
\eeq
Indeed,
$Yd(L_X\omega)=L_Y(L_X\omega)-d(YL_X\omega)
=LYL_X\omega-d(L_X(Y\omega)-(L_XY)\omega)$, whereas
$YL_X(d\omega)=L_X(Yd\omega)-(L_XY)d\omega
=L_X(L_Y\omega-d(Y\omega))-(L_{L_XY}\omega-d(L_XY)\omega)
=L_YL_X\omega+d((L_XY)\omega)-L_Xd(Y\omega)$
since $L_{X\lp Y}=L_{X\lp Y}=[L_X,L_Y]=L_XL_Y-L_YL_X$.
Comparing the two expressions, one finds that
$Yd(L_X\omega)-YL_X(d\omega)  = d(L_X(Y\omega)-L_Xd(Y\omega)$,
which vanishes inductively.

Now \gzit{e.ld25} follows inductively from the relation
$Xd(d\omega)=L_Xd\omega-d(Xd\omega) 
=L_Xd_\omega-d(L_X\omega-d(X\omega))
=L_Xd\omega-d(L_X\omega)+dd(X\omega)=-dd(X\omega)$,
obtained by using \gzit{e.ld24}. 
(For 0-forms, the $dd$-term is absent, which starts the induction.)

\epf

In particular,
the exterior derivative of the linear form $\zeta$ is the alternating
bilinear form $d\zeta$ with
\[
YXd\zeta = YL_X\zeta-XL_Y\zeta+(X\lp Y)\zeta,
\]
since $YXd\zeta = Y(L_X\zeta-d(X\zeta)) = YL_X\zeta-L_Y(X\zeta)
= YL_X\zeta-((L_YX)\zeta+XL_Y\zeta) 
= YL_X\zeta-XL_Y\zeta-(Y\lp X)\zeta$.

\at{add sum formula for the general case?
The reader may wish to prove inductively that, for alternating 
$c'$-forms $\omega$ with $c'\ge c$, ...}

An alternating $c$-form $\omega$ is called \bfi{closed} if $d\omega=0$, 
and \bfi{exact} if it can be written in the form $\omega=d\theta$
for some $(c-1)$-form $\theta$.
In particular, a linear form $\zeta$ is exact if it has the form
$\zeta=d f$ 
for some scalar field $ f$.

By \gzit{e.ld25}, every exact $c$-form is closed.
The converse is not generally valid but holds in simple cases,
e.g., by the \bfi{Poincar\'e Lemma}, when $\Ez=C^\infty(\Mz )$, where $\Mz $ 
is a nonempty, open and convex subset of $\Rz^n$).

\section{Manifolds as differential geometries}\label{s.dmani}

A central notion for analysis on infinite-dimensional spaces is that 
of a convenient vector space. This notion is discussed in detail in
\sca{Kriegl \& Michor} \cite{KriM}, and refines the notion of a
Hausdorff vector space, which is a vector space with the minimal amount 
of topological structure to allow the definition of a meaningful limit.
A convenient vector space has in addition a meaningful notion
of differentiability of paths, a property essential for differential
geometry on manifolds.
(For more details on basic notions from topology and functional 
analysis; see, for example, \sca{Rudin} \cite{rudin}.)

\begin{dfn}
 A vector space $\Fz$ over $\Rz$ is called \bfi{locally convex} if 
there is a family $S$ of \bfi{seminorms}, i.e., mappings $s:\Fz\to\Rz$ 
such that
\[
s(\alpha x + \beta y)\le |\alpha| s(x) + |\beta| s(y) 
\]
for all $x,y\in\Fz$ and all $\alpha,\beta\in\Rz$, with the property 
that 
\[
s(x)=0 ~~\mbox{for all } s\in S \implies x=0.
\]
A locally convex vector space becomes a Hausdorff space by defining 
a \bfi{neighborhood} of $x\in\Fz$ to be a set containing for
each $s\in S$ some set of the form $\{y\in\Fz\mid s(y-x)< r\}$ 
for some real number $r>0$. Thus a sequence $x_l$ ($l=0,1,2,\dots$)
in $\Fz$ converges to $x\in\Fz$ iff $s(x_l-x)\to 0$ for all $s\in S$.
A path $\pi$, i.e., a continuous mapping $\pi :\Rz\to\Fz$, is called
\bfi{smooth} or 
\bfi{arbitrarily often differentiable} if there are paths  
$\pi^{(k)} :\Rz\to\Fz$ ($k=0,1,2,\dots$) such that 
\[
\pi(t+h)=\sum_{k=0}^n \frac{h^k}{k!}\pi^{(k)}(t) + O(h^{k+1})
\]
for all $t,h\in\Rz$ and all natural numbers $n$.
Clearly, $\pi^{(0)}=\pi$, and we write $\dot\pi:=\pi^{()}$.
\end{dfn}

The reader should verify that, for any seminorm $s$ and 
$\alpha\in\Rz$, $x\in\Fz$, 
\[
s(0)=0,~~~s(\alpha x)=|\alpha|s(x)\ge 0,
\]
that in any locally convex vector space, addition and scalar 
multiplication are continuous, and that differentiation satisfies
the traditional rules.

\begin{dfn}
A \bfi{convenient vector space} is a locally convex vector space
$\Fz$ over $\Rz$ such that every smooth path in $\Fz$ is the 
derivative of another path in $\Fz$. A complex-valued function $f$ on 
a nonempty and open subset $\Mz $ of $\Fz$
is called \bfi{smooth} or \bfi{arbitrarily often differentiable} if 
the complex-valued function $f\circ \pi $ is smooth for every 
smooth path $\pi :\Rz\to\Fz$. The space of smooth functions 
$f$ from $\Mz$ to a topological vector space $\Vz$ is denoted by 
$C^\infty(\Mz,\Vz)$, and the space of smooth functions 
$f:\Mz \to\Cz$ is denoted by $C^\infty(\Mz )$.
\end{dfn}

The reader should verify that in a convenient vector space there is
a mapping $\partial\in\Lin(C^\infty(\Mz ),C^\infty(\Mz,\Fz^*)$
called the \bfi{gradient} such that 
\[
\frac{d}{dt} f(\pi(t)) = \dot\pi(t) \partial f(\pi(t))
\]
for all $f\in :C^\infty(\Mz )$, all  arbitrarily often differentiable
paths $\pi:\Rz\to \Mz $ and all $t\in\Rz$.

\begin{expl}
The space $\Fz=\Rz^n$ is convenient, with $S$ consisting of 
the Euclidean norm only. Other
examples of convenient vector spaces are Hilbert spaces and Schwartz
spaces; see \sca{Kriegl \& Michor} \cite{kriegl97convenient}.
\end{expl}

\at{extend Example \ref{ex.Rn} to nonempty open subsets of a 
convenient vector space, and give formulas for $L_X,\zeta\wedge \omega$ 
and $d\omega$ in terms of the gradient.}

\begin{dfn}\label{d.pcm}
Let $\Ez$ be a differential geometry with Lie algebra $\Wz$ of 
vector fields.

(i) A \bfi{point} is a algebra homomorphism from $\Ez$ to 
$\Cz$ which maps $1$ to $1$.
We write $\Mz (\Ez)$ for the set of points, and say that the point 
$\xi$ maps the scalar field $f$ to the \bfi{value} $f(\xi)$ of $f$ at 
$\xi$.

(ii) If $\Fz$ is a convenient vector space, an \bfi{$\Fz$-chart}
is a homomorphism $C$ from $\Ez$ to $C^\infty(U(C),\Cz)$ for some 
nonempty, open subset $U(C)$ of $\Fz$. 

(iii) An \bfi{$\Fz$-manifold} is a differential geometry satisfying
the axioms

(M1)~~ If $f(\xi)=0$ for all $\xi\in \Mz (\Ez)$ then $f=0$.

(M2)~~ For all charts $C$ and all $x\in U(C)$, there is a unique 
$\xi=\xi_C(x)\in \Mz (\Ez)$ such that
\[
f(\xi_C(x)) = (Cf)(x) \Forall f \in\Ez.
\]
(M3)~~  For all charts $C$ and all $\delta\in\der C^\infty_0(U(C))$,
there is a unique $\widehat \delta \in\der \Ez$ such that
\[
(\widehat \delta f)(\xi)=0 \Forall \xi\not\in \xi_C(U),
\]
\[
(\widehat \delta f)(\xi_C(x))=(\delta Cf)(x) \Forall x \in U.
\]
\end{dfn}

Informally, property (M1) says that there are sufficiently many 
points to separate scalar fields. It implies not only that $\Ez$
is commutative, since $(fg-gf)(x)=f(x)g(x)-g(x)f(x)=0$ for all points 
$x$, but also excludes many other commutative algebras, such as
nontrivial quotients of the algebra of polynomials in a single variable.

(M2) expresses that charts are
sufficiently large to represent scalar fields locally, and (M3) says 
that there are sufficiently many derivations to reduce differentiation
locally to charts.

\at{Questions: How much of (M1)--(M3) must be assumed and what can 
already be proved?}

\section{Manifolds as topological spaces}\label{s.mani}

\begin{dfn}\label{d.mani} ~\\
(i) Let $\Fz$ be a convenient vector space.
A {\bfi{manifold} modeled on $\Fz$} 
(short {\bfi{$\Fz$-manifold}} or simply \bfi{manifold}\footnote{
More precisely, this defines arbitrarily often differentiable, real 
manifolds whose dimension need not be finite. There are a number of 
other notions of a manifold which make somewhat different assumptions.
} 
if $\Fz$ is apparent from the context)
is a set $\Mz $ whose elements are called \bfi{points} together with a 
family $\mathcal{C}$ of maps $\xi:U\to \Mz $ from a ($\xi$-dependent) 
nonempty open subset $U$ of $\Fz$ to $\Mz $ called {\bfi{charts}}, 
with the properties

(SM1) Every point of $\Mz $ is in the {\bfi{range}} $\xi[U]$ of some 
chart $\xi:U\to \Mz $;

(SM2) A map $\xi:U\to \Mz $ is in $\mathcal{C}$ if and only if $\xi$ is 
injective and,
for every nonempty open subset $V$ of $U$ and every chart
$\xi':U'\to \Mz $ in $\mathcal{C}$ with $\xi[V]\subseteq \xi'[U']$,
\[
\xi'^{-1}\xi\big|_V \in C^\infty(V,\Fz).
\]
The manifold is canonically a topological space by declaring as
open sets arbitrary unions of finite intersections of ranges of
charts.

(ii) The inverse of a chart is called a 
{\bfi{local coordinate system}}.
An {\bfi{atlas}} is a family of charts whose ranges cover $\Mz $;
the family $\cal C$ of all charts is the \bfi{universal atlas}.

(iii) The dimension of $\Fz$ is called the {\bfi{dimension}} of $\Mz $.
In particular, $\Mz $ is called \bfi{finite-dimensional ($d$-dimensional)}
if $\dim \Fz<\infty$ (resp. $\dim\Fz =d$).

(iv) A mapping $F$ from $\Mz $ to convenient some vector space $\Uz$ is
 called {\bfi{smooth}} (or \bfi{infinitely differentiable}) 
if $F(\xi)\in
C^\infty(U,\Uz)$ for every chart $\xi:U\to \Mz $. A {\bfi{scalar
field}} on $\Mz $ is a smooth complex-valued function on $\Mz $; the algebra
of all scalar fields on $\Mz $ with pointwise operations is denoted by 
$C^\infty(\Mz )$\index{$C^\infty(\Mz )$}. A {\bfi{derivation}} on $\Mz $ 
is a mapping $\delta\in\Lin C^\infty(\Mz )$ satisfying
\[
\delta(fg) = (\delta f)g+f(\delta g)\Forall f,g\in C^\infty(\Mz ).
\]
Thus a derivation on $\Mz $ is an element of
$\der C^\infty(\Mz )$, which we also denote by
$\der \Mz $\index{$\der \Mz $}.

(v) A canonical differential geometry whose scalar
fields form the algebra $\Ez=C^\infty(\Mz )$ of a manifold $\Mz $
is called a \bfi{differential geometry} of $\Mz $.
\end{dfn}

Note that a chart $\xi$ is injective, hence its inverse
$\xi^{-1}$, the corresponding local coordinate system is
well-defined on the range of the chart, and maps a nonempty, open
subset of $\Mz $ to an open subset of $\Fz$. In many treatments of
differential geometry, the local coordinate system $\xi^{-1}$
rather than $\xi$ is called the chart.

\begin{prop}~\\
(i) The set $C^\infty(\Mz )$ of all scalar fields on $\Mz $ is a
commutative $*$-algebra under pointwise multiplication.

(ii) The set $\Lie \Mz :=\Der C^\infty(\Mz )$ of all derivations on
$\Mz $ with the commutator of derivations as Lie product is a Lie 
algebra.
\end{prop}

\bepf This is left to the reader as a straightforward exercise.
\epf

\at{add left module structure and topology for 
$\Der C^\infty(\Mz )$, if possible in a way 
that works for arbitrary $\der \Az$!}

\begin{thm}
$\Fz$-Manifolds in the sense of Definition \ref{d.pcm}(iii) and
$\Fz$-manifolds in the sense of Definition \ref{d.mani}(i) 
are equivalent 
concepts. 
\end{thm}

\bepf
~\at{give a more precise formulation of the theorem, and give a proof}
\epf

The motivating example defining the terminology is the surface of the
earth, the \bfi{globe}, which  may be regarded as a 2-dimensional
manifold $\Mz $ with $\Fz=\Rz^{1\times 2}$, the vector space of 
2-dimensional row vectors\footnote{
It is convenient to think of points as row
vectors; then tangent vectors are row vectors, too, and 
gradients of scalar fields are naturally column vectors.
Thus later expressions like the directional derivative $Xdf$ of a 
scalar field $f$ in the direction of a vector field $X$ have a 
natural interpretation as ''scalar = row times colums'' in terms of 
ordinary matrix algebra.
}. 
Here the domain $U$ of a
chart $\xi:U\to \Mz $ may be viewed as the paper on which
the chart (a road map, say) is printed. The important points
$q=(x,y)\in U$ correspond to Cartesian coordinates of marks on the
road map labeled by towns. $\xi(q)$ denotes the location of the
corresponding town on the globe. Note that our charts may have 
domains which are not neatly cut and may be disconnected or unbounded.

The simplest examples of manifolds are open, nonempty subsets of 
$\Rz^n$. \at{find its general chart!}

\begin{expl}
We consider the concrete case where smooth manifolds
modeled on the vector spaces $\Rz^d$ for some $d\in \Nz$
are embedded into a bigger vector space $\Rz^n$, and where membership
in the manifold is characterized by $m$ equations $F_k( x)=0$
($k=1,\dots,m$). For example, a {\bfi{$d$-sphere}} is the set of
 points $ x\in \Rz^{d+1}$ satisfying the single ($m=1$) equation
$F( x):= x^T x-1=0$, where the superscript $T$ denotes the transpose. 

Let $\Mz _0$ be an open subset of $\Rz^n$ and let $F\in
C^\infty(\Mz _0,\Rz^m)$. The {\bfi{gradient}} at $ x$ of $F$ is 
given by
\[
d F( x) = F'( x)^T \in \Rz^{n\times m}\,.
\]
This generalizes
the traditional terminology for the case where $m=1$.
The implicit function theorem implies that if the gradient has
constant rank $m$, i.e.,
$\rk dF( x)=m$ for all $ x\in \Mz _0$, then the set $\Mz $ given by
\[
\Mz = \left\{  x\in \Mz _0 \mid F( x)=0\right\}\,,
\]
is a $d$-dimensional manifold with $d=n-m$. If $\Mz $ defines a
$d$-dimensional manifold given by an equation $F( x)=0$, then the
tangent space at a point $ x\in \Mz $ is given by\index{$T_ x \Mz $}
\[
T_ x \Mz  = \left\{ X \in \Rz^{1\times n} \mid X\cdot dF ( x) = 0
\right\}\,.
\]
\at{prove this?}
Thus the tangent space consists of those vectors perpendicular to
the gradient, that is, the tangent vectors at $ x$ are tangent
to $\Mz $ at $ x$. Hence the name tangent space. The vector fields of 
$\Mz $ are similarly given by\index{$\vect \Mz $} 
\[
\vect \Mz  = \left\{ X\in C^\infty(\Mz ,\Rz^{1\times n}) \mid X( x)\cdot
dF( x) = 0, \, \Forall  x\in \Mz  \right\}\,.
\]
~\at{find its general chart!}
\end{expl}

\bigskip
Given an $\Fz$-manifold $\Mz $ and an $\Fz'$-manifold $\Nz$, we define 
$C^\infty(\Mz ,\Nz)$ as the set of maps $A:\Mz \to \Nz$ such that if
$\xi:U\to \Mz $ is a chart on $\Mz $ and $\xi':V\to \Nz$ is a chart on 
$\Nz$ such that $A(\xi(U))\subseteq \xi'(V)$ implies  
$(\xi')^{-1}\circ A \circ \xi\in C^\infty(U,V)$.
A {\bfi{diffeomorphism}} of $\Mz $ is an invertible
mapping in $C^\infty(\Mz ,\Mz )$ with an inverse in $C^\infty(\Mz,\Mz)$;
we write $A x$ for the image of a point $ x\in \Mz $ under a
diffeomorphism $A$.

We assume that the identity map
on $\Mz $ is in $C^\infty(\Mz ,\Mz )$ and that the composition of $f\in
C^\infty(\Mz ,\Nz)$ and $g\in C^\infty(\Nz,\Nz')$ is in 
$C^\infty(\Mz,\Nz')$, a condition\footnote{
In technical terms this says that the modeling vector spaces should 
admit a category of smooth manifolds.
}
automatically satisfied in finite dimensions.
\at{and now in general?}
Then the set $\Diff \Mz $\index{$\Diff \Mz $} of all diffeomorphisms is 
a group under
composition of maps. Additional conditions are needed to ensure that
$\Diff \Mz $ is a $\vect \Mz $-manifold and hence (in the terminology
of Section \ref{s.Liegroups} below) a Lie group; 
see, e.g., \sca{Neeb} \cite{Nee},
where one can find a detailed discussion of pathologies that can
arise in infinite-dimensional Lie groups.

We define a {\bfi{motion}} on $\Mz $ as a mapping $A\in
C^\infty([0,1],\fct{Diff}(\Mz ))$ such that $A(0)=1$ is the identity.
The intuition is that the points $ x=A(0) x$ of an object
(subset of $\Mz $) at time $t=0$, the start of the motion, is moved by
the motion to the point $A(t) x$ at time
$t\in[0,1]$, ending up in $A(1) x$ at the end of the motion.
For every motion $A$ and for all
$t\in[0,1]$ we define a vector field $\dot A(t)$ on $\Mz $ by
\[
\dot A(t)d f :  x \to
\frac{d}{ds} f(A(s)A(t)^{-1} x)\Big|_{s=t}
\]
for all scalar fields $ f$ and all $ x\in \Mz $. Since the product rule 
holds for
smooth functions in $C^\infty(\Mz )$, the
object $\dot A(t)d$ is indeed a derivation on $\Mz $, and hence
$\dot A(t)$ is a vector field. From the definition of $\dot A(t)$
we get the {\bfi{chain rule}}:
\[
\frac{d}{dt} f(A(t) x) = \dot A(t)d  f(A(t) x)\,.
\]
If we are only interested in what happens in an infinitesimal
neighborhood of a point $ x\in \Mz $, the vector fields in
\[
N( x) := \left\{ X_0 \in \vect \Mz  \mid X_0d f ( x) = 0
\Forall  f\in C^\infty(\Mz )\right\}
\]
have no effect at $ x$. Since $N( x)$ is a vector space,
we can form the quotient space
\[
T_ x \Mz  = \vect \Mz  / N( x)\,,
\]
called the {\bfi{tangent space}} or 
{\bfi{tangent (hyper-)plane}} at
$ x$.
We denote with
\[
X( x):=X+N( x),
\]
the equivalence class of $X$ with respect to the equivalence relation
$X\sim Y \Leftrightarrow X-Y\in N( x)$. We call the equivalence
class that contains the vector
field $X$ the {\bfi{tangent vector} of $X$ at the
point $ x$}.

The union $T\Mz $ of all $T_ x\Mz$ is naturally a manifold called the 
{\bfi{tangent bundle}} of $\Mz$. 
\at{also in infinite dimensions? 
give charts or say we don't explain it. 
Define $T^*\Mz ,T_s\Mz ,T^r_s\Mz $}

\at{Question: is there an algebraic construction of $C^\infty(T\Mz )$ 
and $C^\infty(T^*\Mz )$ which does not mention points?}

\bigskip
\bfi{The Lie derivative in the traditional approach.}
In the special case where $\Ez=C^\infty(\Mz )$ for some 
finite-dimensional manifold $\Mz $, there is
an alternative, traditional route to the calculus on manifolds,  
using the following traditional definition of the Lie derivative. 

For any vector field $X$, the initial value problem
\lbeq{e.ivp}
\xi(0)=x,~~~\frac{d}{d\tau} \xi(\tau)= X(\xi(\tau))
\eeq
is solvable for every $x\in \Mz $, for $\tau$ in some $x$-dependent 
neighborhood of zero.
\at{But I have not yet defined the tangent vector 
$\frac{d}{d\tau} \xi(\tau)\in T_{\xi(\tau)} \Mz $.}
This follows from the standard theory of ordinary differential 
equations, since differentiable vector fields are locally Lipschitz.

We denote by $e^{\tau X}$ the local diffeomorphism which maps $x$ 
into the value $\xi(\tau)$ of the solution $\xi$ of \gzit{e.ivp}. 
Clearly, $e^0=1$ is the identity, but for fixed $\tau$, the map 
$e^{\tau X}$ need not be defined everywhere. The latter is the case 
only when \gzit{e.ivp} is solvable for all $\tau\in\Rz$; in this case,
the vector field is called \bfi{complete}, and the $e^{\tau X}$ 
form a 1-parameter group of diffeomorphisms. In general, we have,
on the domain of definition, 
\[
e^{\tau X}e^{\tau' X}=e^{(\tau+\tau') X},
\]
\[
\frac{d}{d\tau} e^{\tau X}x = X(e^{\tau X}x).
\] 
We define the \bfi{directional derivative} $L_X\phi$ of a tensor field 
$\phi$ with respect to the complete vector field $X$ by
\[
(L_X\phi)(x):= \frac{d}{d\tau} \phi(e^{\tau X}x)\Big|_{\tau=0}.
\]
This defines a linear differential operator $L_X$ mapping tensor 
fields to tensor fields of the same type $[c,r]$, called the  
\bfi{Lie derivative} of $X$. Clearly,
\[
(e^{\tau L_X}\phi)(x) = \phi(e^{\tau X}x).
\]
It is not difficult to show the \bfi{chain rule}
\[
\frac{d}{d\tau} \xi(\tau) =X(\tau)\xi(\tau) \implies
\frac{d}{d\tau} \phi(\xi(\tau)) = L_{X(\tau)}\phi(\xi(\tau))
\]
for every smooth path $x:[0,1]\to  \Mz $.
Here $\dot \xi(\tau)$ is the tangent vector of the path at $\xi(\tau)$.
The chain rule implies the product rule \gzit{e.ld3} for the Lie 
derivative.
Therefore Theorem \ref{t.GenLieDeriv} implies that the traditional 
concept coincides with our algebraic concept when $\Ez$ is the algebra 
of scalar fields of a finite-dimensional manifold.

In the infinite-dimensional case, this approach can also be carried 
through, although it requires considerable technicalities to account
for the fact that initial-value problems for differential equations 
in infinite-dimensional spaces are not always solvable. For details,
see \sca{Kriegl \& Michor} \cite{KriM}.

\bigskip
\section{Noncommutative geometry}

In this short section, we indicate how things generalize
to \idx{noncommutative geometry}, without giving details; the reader 
not familiar with the notions used may simply skip the section.

In noncommutative geometry, position measurements are
limited by uncertainty relations. The notion of a point therefore loses
its meaning, and the evaluation of functions and vectors at a point 
is no longer well-defined. 
Thus, in noncommutative geometry, a manifold of points no
longer exists, but in place of $C^\infty(\Mz )$ one has a noncommutative
algebra $\Ez$ whose elements behave in a way analogous to scalar fields.
All constructions based only on this algebra rather than a manifold
generalize in an appropriate way to the noncommutative situation.
Thus most geometric notions extend formally, but they can be matched
with true geometric concepts only in certain commutative subalgebras.
\at{match this with earlier stuff}
The basic observation is that a point evaluation is a 
 $*$-homomorphism of $C^\infty(\Mz )$ to $\Cz$, and conversely,
all such homomorphisms of $C^\infty(\Mz )$ are obtained as point 
evaluations.
Now, if $\Ez_0$ is a commutative normed $*$-subalgebra of $\Ez$ whose
completion is a {\bfi{$B^*$-algebra}} (a term we shall not 
further use, and hence not introduce formally \at{but give ref's!}) 
then one can reconstruct on $\Ez_0$
a topological space $\Mz _0$ by calling the characters of $\Ez_0$
points; the $B^*$-algebra is then canonically isomorphic to
the algebra of bounded continuous functions on
$\Mz _0$. If $\Ez_0$ admits sufficiently many derivations then $\Mz _0$
is a (smooth) manifold. When $\Ez_1$ and $\Ez_2$ are two such
commutative subalgebras that do not commute, then, in contrast to the
commutative situation, the  corresponding manifolds $\Mz _1$ and 
$\Mz _2$ are not naturally embedded into a bigger manifold. 
Thus there may be many maximal manifolds embedded in a single 
noncommutative geometry.

\at{add here the procedure for constructing an algebra from an 
atlas of algebras, and invite the reader experienced in differential 
geometry to verify that this reduces to the traditional construction
 of manifolds by means of an atlas of charts.}

\at{relate linear forms and Part II}

\section{Lie groups as manifolds}\label{s.Liegroups}

This section defines Lie groups in full generality.
Differential equations defining the flow along a vector field 
naturally produce Lie groups and the exponential map, which relates 
Lie groups and Lie algebras.  \at{polish the definition}

\begin{dfn}~\\
(i) A {\bfi{Lie group}} is a group $\Gz$ which is at the same time 
a manifold, such
that multiplication and inversion are arbitrarily often differentiable.
A Lie group is both a manifold and a group and the two structures
are compatible. The identity element in a Lie group will always be
written as 1.

(ii) We canonically embed $\Gz$ into $\fct{Diff}(\Gz)$ by associating
to $A\in\Gz$ the map $B\to AB$, which is a diffeomorphism.
For the definition of the Lie algebra associated with a Lie group,
it is important to know that the group $\Gz$ acts on $C^\infty(\Gz)$ by
right multiplication, that is, to every $A\in\Gz$ we associate the
map $R_A: C^\infty(\Gz)\to C^\infty(\Gz)$ given by
\[
(R_A\varphi)(B) := \varphi(BA)
\]
for all $B\in \Gz$, $\varphi\in C^\infty(\Gz)$. Of
course, the group also acts by left-multiplication on
$C^\infty(\Gz)$ but this action is not directly related to the Lie
algebra.

(iii) The Lie algebra $\vect \Gz$ contains the set
\[
\log \Gz =\left\{ X\in \vect \Gz \mid R_B\lp Xd = 0 \Forall
B\in \Gz\right\}
\]
of {\bfi{invariant vector fields}}.
\end{dfn}

It is not difficult to show that every Lie group in the above sense 
is a Lie group in the sense of Definition \ref{d.llie}, since $\Gz$ 
is canonically embedded into $\Lin C^\infty(\Gz)$. The converse is
also valid but a bit more difficult to establish. \at{give details!}

\begin{prop}
The invariant vector fields $\log \Gz$ form a Lie algebra. 
\end{prop} 
\bepf
To check the statement, we only need to show that
the Lie product $X\lp Y$ of two invariant vector fields $X$ and $Y$
is invariant. But this follows from  Proposition \ref{lem.der.cent}
since
the invariant vector fields form the centralizer of the
set $\left\{ R_A \mid A\in\Gz\right\}$.
\epf

\begin{prop}
For any smooth motion $A\in C^\infty([0,1],\Gz)$, where $\Gz$
is identified with a subset of $\fct{Diff}(\Gz)$, the vector field
$\dot A(t)$ is an invariant vector field.
\end{prop}
\bepf We know that $\dot A(t)$ is a vector field. Hence we need to
check that $\dot A(t)$ Lie commutes with $R_B$ for all $B\in\Gz$.
For arbitrary $B\in\Gz$ and $\varphi\in C^{\infty}(\Gz)$ we have
 \beqar
 (R_B\lp \dot A(t)d)\varphi (A(t)\zeta) &=& R_B (\dot A(t)
 d\varphi)(A(t)\zeta) - \dot A(t) d (R_B \varphi)(A(t)\zeta) \nonumber\\
 &=&\dot A(t)d \varphi (A(t)\zeta B ) - \frac{d}{dt}(R_B
 \varphi)(A(t)\zeta) \nonumber \\
&=& \frac{d}{dt}\varphi(A(t)\zeta B) - \frac{d}{dt}
 \varphi (A(t)\zeta B) =0\nonumber\,.
 \eeqar
\epf
\begin{rems}
Note that an essential ingredient in the above proof is that the
action of $\Gz$ on $C^\infty(\Gz)$ is defined from the right and
the action of the vector field $\dot A(t)$ from the left.
\end{rems}

\begin{definition}\label{d.exp}
A motion $A(t)$ is called a {\bfi{uniform motion}} if there exists a
unique $f\in \log\Gz$ such that
\lbeq{e.exp}
\dot A (t) = f A(t) \Forall t\in [0,1].
\eeq
In this case we write $e^f$ for the group element $A(1)$ and call it
the {\bfi{exponential}} of $f$. Conversely, $f$ is called the
{\bfi{infinitesimal generator}}\index{generator!infinitesimal} of 
the motion.
\end{definition}

Formula \gzit{e.exp} is a linear differential equation with 
constant
coefficients; the initial condition $A(0)=1$ is already part of the
definition of a motion. In finite dimensions, such initial value
problems are uniquely solvable; in infinite dimensions, unique
solvability depends on additional conditions. It is easy to check that
a uniform motion with infinitesimal generator $f\in\log\Gz$ is given
by $A(t)=e^{tf}$.

\begin{expl}
In any associative algebra, the set of invertible elements is a group.
In many cases, the group of invertible elements is a Lie group.
\at{adapt}
In particular, the group 
$GL(n,\Kz)$\index{$GL(n,\Kz)$}\index{$gl(n,\Kz)$} of all invertible 
$n\times n$-matrices over $\Kz=\Rz$ or
$\Kz=\Cz$ is a Lie group, since it is the open set of points in
$\Kz^{n\times n}$ where the determinant does not vanish, so that
any point has an open neighborhood on which the identity is a
chart. We can choose coordinates $x_{ij}$ for $1\leq i,j\leq n$ and
then $GL(n,\Kz)$ is the open set where $\det (x_{ij})\neq 0$. Any
derivation is of the form
\[
(Xf)(x_{ij}) = \sum_{1\leq i,j\leq n} X^{ij}\frac{\partial}{\partial
  x_{ij}}f(x_{ij})\,,
\]
for all $f\in C^{\infty}(GL(n,\Kz))$ and for some $X^{ij}\in \Kz$. One
finds that
$\log GL(n,\Kz)=gl(n,\Kz)$ is the Lie algebra of all
$n\times n$-matrices over $\Kz$. It is easy to verify these properties
by describing everything with matrices. The subgroup of $GL(n,\Lz)$
consisting of the matrices with unit determinant is denoted by
\idx{$SL(n,\Kz)$}. In other words, $SL(n,\Kz)$ is the kernel of the map
$\det:GL(n,\Kz)\to \Kz^*$, where \idx{$\Kz^*$} is the group of 
invertible elements in $\Kz$. The Lie algebra of $SL(n,\Kz)$ is 
denoted by \idx{$sl(n,\Kz)$} and consists of the traceless $n\times n$ 
matrices with entries in $\Kz$.
\end{expl}

\chapter{Conservative mechanics on manifolds}\label{c.pmani}

\at{adapt this introduction, and also the summary in Section 1.5 
of QML}

We consider closed 2-forms in manifolds and their associated Poisson 
algebras. This naturally leads to symplectic geometry and a symplectic 
formulation of the dynamics of quantum mechanics.
It also leads to classical Hamiltonian and Lagrangian mechanics,
including constraints.

\at{{\bf phys/qft contains a workfile mech.tex improving on this 
chapter!!!}}

\section{Poisson algebras from closed 2-forms}\label{s.sympl}

In general, a classical, conservative dynamical system is
described in terms of motion on a manifold $\Mz $, called the 
{\bfi{phase space}}, such that some algebra of 
functions on it has a Poisson algebra structure; more precisely,
$C^\infty(\Mz )$ is
equipped with a Lie product $\lp$ that is antisymmetric and satisfies
the Jacobi identity and the Leibniz identity. Such a manifold is
called a {\bfi{Poisson manifold}}; see, e.g., 
\sca{Vaisman} \cite{Vai} or \sca{da Silva \& Weinstein} \cite{daSW} 
Poisson manifolds provide a
general setting for the study of the dynamics of classical
conservative mechanical systems by differential geometric methods;
for a more comprehensive discussion of different aspects see 
\sca{Marsden  \& Ratiu} \cite{marsdenratiu}, 
\sca{Ratiu} \cite{courseratiu} and
\sca{Morrison} \cite{morrison98}.

Every symplectic manifold is a Poisson manifold since the symplectic
structure gives rise to a natural Poisson bracket. 
In the symplectic case, a Hamiltonian is a function of some
coordinates $q_i$ and the conjugated momenta $p_i$. In such cases, the
phase space is even-dimensional. In more general cases described, 
e.g., by Lie-Poisson algebras, the phase space need not be a symplectic 
manifold. Indeed, symplectic manifolds are always even-dimensional 
while the manifold $SO(3)$ of the spinning rigid body (see Section 
\ref{s.rigid}) has dimension 3.

\bigskip
Many Poisson algebras of relevance in classical mechanics may be 
constructed via a uniform construction based on a 
closed 2-form characterizing the kinematics of the system of interest.
The description is then completed by specifying the dynamics through
a Hamiltonian in the resulting Poisson algebra, and by selecting an 
initial state describing the preparation of the system.
In this section, we discuss the general construction principle.

Let $\omega$ be a closed 2-form on a differential geometry $\Ez$.
We call a scalar field $f\in \Ez$ \bfi{compatible} with
$\omega$ if there is a vector field $X_f$ such that\MMt{\footnote{
Equation \gzit{e.obs} can be
understood better via the "stick" tensor type notation. The left hand
side $df$ corresponds to "$|\cdot$" (covector times scalar), and the
right hand side $X_f \omega$ corresponds to "\_$||$" (rovector
contracted with a bilinear form). Both side are therefore a covector = 
linear form.
} 
} 
\lbeq{e.obs}
   df ~=~ X_f \, \omega ~;
\eeq
any such $X_f$ is called a \bfi{Hamiltonian vector field} 
associated with $f$.
\at{Note somewhere that $\ad_f=X_fd$.}
We write $\Ez(\omega)$ for the set of all scalar fields $f\in \Ez$ 
which are compatible with $\omega$.
In general, $X_f$ need not exist for all $f$, and if it 
exists, it need not be unique. Thus $\Ez(\omega)$ may be a 
proper subspace of $\Ez$; this situation is typical for examples
arising from constrained Hamiltonian mechanics.

\begin{prop}
Let $\omega$ be a symplectic form. Then every scalar field $f$ is 
compatible with $\omega$, 
\lbeq{e.comp}
X_f=df\omega^{-1},
\eeq
and $\Ez(\omega)=\Ez$.

\end{prop}

\bepf
Since $\omega$ is a symplectic form, $\omega$ is 
nondegenerate and has an inverse satisfying \gzit{e.Ainv}.
The defining condition for $X_f$ can therefore be solved uniquely 
for $X_f$, for all $f\in \Ez$, resulting in \gzit{e.comp}.
\epf

A vector field $ X$ is called \bfi{locally Hamiltonian} 
(with respect to $\omega$) if the linear form $ X\omega$ is closed, 
and \bfi{Hamiltonian} (with respect to $\omega$) if $ X\omega$ is 
exact (and hence closed). Thus, for any $f\in\Ez(\omega)$, the vector 
field $X_f$ is Hamiltonian with respect to $\omega$,

\begin{prop}\label{p4.1}
If $X,Y$ are locally Hamiltonian vector fields with respect to
the closed 2-form $\omega$ then $X\lp Y$ is Hamiltonian,
and
\lbeq{e.locham}
(X \lp Y)\omega = d(XY\omega).
\eeq
In particular, the locally Hamiltonian vector fields and the
Hamiltonian vector fields form Lie subalgebras of $\Wz$.
\end{prop}

\bepf
Since $\omega$ and $X\omega$ are closed, \gzit{e.ld22} implies
that $L_X\omega=Xd\omega+d(X\omega)=0$. Again by \gzit{e.ld22},
$d(XY\omega) = L_X(Y\omega)-Xd(Y\omega) =  L_X(Y\omega)
= (L_XY)\omega+YL_X\omega=(L_XY)\omega=(X\lp Y)\omega$,
using the closedness of $Y\omega$ and the product rule \gzit{e.ld3}.
This proves \gzit{e.locham}. The concluding statement is an immediate 
consequence.
\epf

\begin{thm} \label{t2.2}
For every closed 2-form $\omega$ over the manifold $ \Mz $, the set
$\Ez(\omega)$ is a Poisson algebra, with Lie product given by
\lbeq{e.liep}
   f \lp g := X_f dg = X_fX_g\omega  = -X_gX_f\omega  = -X_g df.
\eeq
A Hamiltonian vector field associated with $f\lp g$ is given by 
\lbeq{e.hampl}
  X_{f \slp g}:=X_f\lp X_g.
\eeq
In particular, if $\omega$ is a symplectic form then 
\lbeq{e.lies}
f\lp g =df\omega^{-1}dg.
\eeq
\end{thm}

\bepf
We first show that $\Ez(\omega)$ is a subalgebra of the algebra 
$\Ez$. 
If $f,g\in\Ez(\omega)$ and $\lambda\in \Cz$ then $\lambda f, f\pm g,
fg \in\Ez(\omega)$ since we may take
\[
   X_{\lambda f} = \lambda X_f,~~~X_{f \pm g} = X_f \pm X_g,~~~
 X_{fg} = f X_g + gX_f.
\]
We next show that $f \lp g$ is well-defined. Indeed, if $X_f,X_f'$ 
are two Hamiltonian vector fields associated with $f$ then
$\omega(X_f'-X_f,Y)=0$, hence $f \lp g$ does not depend on the
choice of the Hamiltonian vector fields associated with $f$ and $g$.

Proposition \ref{p4.1} implies that \gzit{e.hampl}
is a Hamiltonian vector field for $f\lp g$; therefore
$f\lp g\in\Ez(\omega)$.

The operation $\lp$ defined by \gzit{e.liep} is bilinear, 
antisymmetric, and satisfies the Leibniz identity.
To conclude that $\Ez(\omega)$ is a Poisson algebra it therefore 
suffices to show that the Jacobi identity holds. This follows 
since, with $X:=X_f,Y:=X_g$,
\[
\bary{lll}
(f \lp g) \lp h &=& X_{f\slp g} d h = (X_f \lp X_g) d h
= (X\lp Y) d h = L_{X\slp Y} h \\
&=& [L_X,L_Y] h
= L_X L_Y h - L_Y L_X h 
= X_f d (X_g d h) -  X_g d (X_f d h)\\
&=& f\lp (g \lp h) - g\lp (f \lp h)
= (f\lp h)\lp g + f\lp (g \lp h).
\eary
\]
Finally, if $\omega$ is a symplectic form, \gzit{e.comp} implies that 
the Lie product \gzit{e.liep} can be rewritten in the form 
\gzit{e.lies}.
\epf

Note that the Lie product can be extended to the case where one argument
is in $\Ez(\omega)$ and the other may be an arbitrary quantity from 
$\Ez$:
\[
f \lp g = X_f dg \for f\in \Ez(\omega),~g\in\Ez,
\]
\[
f \lp g = - X_g df \for f\in \Ez,~g\in\Ez(\omega).
\]
Thus if $f$ is compatible with $\omega$, the Lie product is defined 
even when $g$ is not compatible with $\omega$.

\bigskip
In the manifold case, the above theorem defines, for each
closed 2-form $\omega$ on an $\Fz$-manifold $\Mz $, a Poisson algebra 
$\Ez(\omega)$ which is the set of functions $f \in \Ez= C^\infty(\Mz )$ 
which are compatible with $\omega$.
In the special case where $\omega$ is symplectic, we have seen that
$\Ez(\omega)=\Ez$; thus we may define the \bfi{Poisson bracket} 
\lbeq{e.pbracket}
\{f,g\}:= g \lp f = dg\omega^{-1}df
\eeq
of $f,g\in \Ez$. This is the traditional Poisson bracket
associated with the \bfi{symplectic space} $(\Mz ,\omega)$.

The affine functions, which map $\xi\in \Mz $ to $\xi u+\gamma$
for some $u\in \Fz$ and some $\gamma\in\Cz$, satisfy
\[
(\xi u +\gamma) \lp (\xi v +\gamma') = u \omega{-1} v \in \Cz,
\]
hence form a Lie subalgebra, which is a Heisenberg algebra.
This provides a faithful classical Poisson representation of general
Heisenberg algebras.

\begin{expl}~
We continue the discussion of Example \ref{ex.Rn}, where scalar fields
(resp. vector fields) are the smooth 
complex-valued (resp. row vector valued) functions
on a nonempty, open subset $\Mz $ of the space $\Rz^{\times n}$ 
of rovectors of length $n$. In this case, it is natural to identify
$\Wz^*$ with the vector space $C^\infty(\Mz ,\Cz^n)$ of covector-valued 
fields via
\[
(X\zeta)(x)=X(x)\zeta(x)
\]
for $X\in \Wz= C^\infty(\Mz ,\Cz^{\times n})$ and 
$\zeta\in\Wz^*= C^\infty(\Mz ,\Cz^n)$. In particular, the gradient 
$df=\partial f$ appears naturally as an element of $\Wz^*$, consistent 
with our abstract development. 

Now let $\theta$ be a distinguished linear form. Then we can define its 
\bfi{Jacobian}, 
the $x$-dependent square array\MMt{\footnote{
an object of type $||$ in the stick notation
}
} 
$\partial \theta$ whose entries are the partial derivatives 
\[
\partial_j\theta_k(x) ~=~ \Pdrv{\theta_k(x)}{x^j} ~.
\]
We now consider the exact 2-form $\omega=-d\theta$ (the minus
sign is traditional). We have
\lbeq{e.dtheta}
YX\omega = Yd(X\theta)-Xd(Y\theta)+(X\lp Y)\theta \for \omega=-d\theta
\eeq
since $YX\omega = -YXd\theta=-Y(L_X\theta-d(X\theta))
=-YL_X\theta+Yd(X\theta) = -L_X(Y\theta)+(L_XY)\theta + Yd(X\theta)$
by \gzit{e.ld22b} and \gzit{e.ld21}.
It is not difficult to show that now
\lbeq{e.omega1}
(X\omega)(x) = \sum X^j(x)\omega_{jk}(x),
\eeq
where 
\lbeq{e.omega2}
\omega_{jk}(x) = \partial_k\theta_j(x)-\partial_j\theta_k(x)
\eeq
are the components of the antisymmetric expression 
$(\partial \theta)^T-\partial \theta$ in the Jacobian of $\theta$.

As a consequence, $\omega=-d\theta$ is nondegenerate precisely when 
$(\partial \theta)^T-\partial \theta$ is nonsingular.
If this holds, $\omega$ is a symplectic form, and all our results apply.
This matches the present development with that found in standard 
treatises such as \sca{Marsden \& Ratiu} \cite{MarR}.

\end{expl}

\section{Conservative Hamiltonian dynamics}\label{s.hamphase}

We now apply the results of Section \ref{s.sympl} to 
classical Hamiltonian mechanics of conservative systems. 
The phase space of a classical system 
is the set of all states that may be attained in some realization of 
the system. We begin with the unconstrained case, 
where the phase space is a cotangent bundle over a manifold $\Mz $, 
and then extend the discussion to the constrained case, where the 
phase space has a more complicated structure.

To avoid technicalities, we only treat the case where the manifold
can be described by a single chart, so that it can be treated as an
open subset of some topological vector space. However, using standard
techniques from differential geometry, it is not difficult 
to lift the discussion to arbitrary manifolds. Thus, in the following,
the \bfi{configuration space}
$\Mz _c$ is a nonempty, open subset of a convenient vector space $\Fz$
over $\Rz$. Thinking of $\Mz _c$ as a chart of a general manifold, 
everything we say here extends in a standard way to arbitrary 
$\Fz$-manifolds in place of $\Mz _c$.

We write the bilinear pairing between elements $q$ from $\Fz$ and 
elements $p$ from the dual space $\Fz^*$ as product 
$p\cdot q = q \cdot p$.
We extend this product linearly to the compexifications $\Cz\Fz$
of $\Fz$ and $\Cz\Fz^*$ of $\Fz^*$, and extend it further pointwise to 
$\Cz\Fz$-valued or $\Cz\Fz^*$-valued functions.

\at{
Earlier you started with $\Mz_c$, then said it is a convenient vector
space $\Fz$, then told the reader to think of $\Mz_c$ as a chart. But
in the next paragraph you use plain $\Mz$, and then define it as the
cotangent bundle $\Mz_c\times\Fz^*$ of $\Mz_c$ -- which is puzzling
because if $\Mz_c$ is an ordinary vector space, why bother with
bundles?}

In this section, we consider the case of unconstrained dynamics
Here $\Ez=C^\infty(\Mz )$ and $\Wz=C^\infty(\Mz ,\Cz\Fz\times\Cz\Fz^*)$ 
are the spaces of scalar fields and vector fields, respectively, on the 
\bfi{cotangent bundle} $\Mz  = T^*\Mz _c:=\Mz _c\times\Fz^*$ of $\Mz_c$.

The reader may think of the Euclidean space $\Fz=\Fz^*=\Rz^n$ of 
vectors with $n$ real components and bilinear pairing 
$p\cdot q = \sum_k p_k q_k$.  As discussed in Section 
\ref{s.cao}, this accounts for the mechanics of point particles.
For field theories, $\Fz$ is an infinite-dimensional function space.

A classical, conservative, \bfi{unconstrained mechanical system} 
is defined by a \bfi{Hamiltonian} $H\in\Ez$ and considering the 
full cotangent bundle $\Mz $ as the \bfi{phase space} of the system.
The point $x =(q,p)\in \Mz $ is called the \bfi{state} 
with \bfi{position} $q\in \Mz _c$ and \bfi{momentum} $p\in\Fz^*$.
The \bfi{energy} of the system in the state $(q,p)$ is 
the value $H(q,p)$ of the Hamiltonian at $(q,p)$. 
\at{The Hamiltonian of a physically realistic 
system must be bounded below.}

The state of the system varies with \bfi{time} $t$, which we 
consider to be a number in the interval $[\ul t,\ol t]$,
where $\ul t$ is the \bfi{initial time} and $\ol t>\ul t$ is the 
\bfi{final time} for which the system is considered. 
The time dependence is modeled by a \bfi{trajectory}, a state-valued, 
arbitrarily often differentiable function of time, mapping 
$t\in[\ul t,\ol t]$ to $(p(t),q(t))\in \Mz $. The position $q(t)$ 
and the momentum $p(t)$ at time $t$ are constrained by the
\bfi{Hamilton equations} in \bfi{state form},
\lbeq{e.hamc}
  \dot q = \frac{dq}{dt} = \partial_p H,~~~ 
  \dot p = \frac{dp}{dt} = -\partial_q H.
\eeq
Here $\partial_p = \partial/\partial p$ and 
$\partial_q = \partial/\partial q$ denote the gradient with respect 
to momentum $p$ and position $q$, respectively. Note that if $f$ is a 
scalar field then $\partial_p f(q,p)\in\Cz\Fz$ and 
$\partial_q  f(q,p)\in \Cz\Fz^*$.

The Hamiltonian equations automatically imply
the \bfi{conservation of energy}:
$\frac{d}{dt} H(q,p)
=\partial_q H \cdot \dot q + \partial_p H \cdot \dot p = 0$.

The Hamiltonian equations may be derived from a \bfi{variational 
principle}. We define the \bfi{action} as the functional on smooth 
paths in $\Mz$ defined by 
\lbeq{e.action}
I(q,p) := \int_{\ul t}^{\ol t} dt~ 
\Big(p(t)\cdot \dot q(t) - H(q(t),p(t))\Big),
\eeq
and consider small variations $\delta q$ and $\delta p$ of the arguments
$q$ and $p$, respectively. Since we do not make further 
use of the principle, we assume without the discussion that the 
integral can be manipulated as accustomed from the finite-dimensional 
case, where $\Fz=\Rz^n$.
\at{The action needs a Banach space setting. Treat only the flat case
since this is only motivation!
later add a boundary term 
$I_b(q(\ul t),q(\ol t),\dot q(\ul t),\dot q(\ol t))$
and discuss the resulting boundary conditions?
In the case without boundary conditions, apparently the 
variation $h$ must be required to vanish at the end points?}
For variations vanishing
at $t=\ul t$ and $t=\ol t$, we have, up to higher order terms,
\[
\bary{lll}
I(q+\delta q,p) - I(q,p) 
&\approx& \D\int_{\ul t}^{\ol t} dt~ 
\Big(p(t)\cdot \delta \dot q(t)
-\partial_q H(q(t),p(t))\cdot \delta q\Big)\\
&=& \D\int_{\ul t}^{\ol t} dt~ 
\Big(-\dot p(t)\cdot \delta q(t)
-\partial_q H(q(t),p(t))\cdot\delta q\Big),
\eary
\]
\[
I(q,p+\delta p) - I(q,p) \approx  \int_{\ul t}^{\ol t} dt~ 
\Big(\delta  p(t)\cdot \dot q(t)
-\partial_p H(q(t),p(t))\cdot\delta p\Big),
\]
so that the path $(q,p)$ is a stationary point of the action if
and only if the extended Hamiltonian equations \gzit{e.hamc} hold.

\bigskip
A vector field $X\in\Wz$ is a pair of functions $X=(X^q,X^p)\in 
C^\infty(\Mz ,\Cz\Fz)\times C^\infty(\Mz ,\Cz\Fz^*)$; 
its value at the state $(q,p)$ is $X(q,p)=(X^q(q,p),X^p(q,p))$. 
Associated with each vector field $X$ is the derivation $Xd$ defined by
\[
Xdf := X^q\cdot \partial_q f + X^p \cdot \partial_p f.
\]
Using the mapping $d$ defined in this way, it is easily checked that
we have a commutative differential geometry.
In particular, a general linear form $\zeta$
is described by a pair of functions $(\zeta_q,\zeta_p)\in 
C^\infty(\Mz ,\Cz\Fz^*)\times C^\infty(\Mz ,\Cz\Fz)$
such that
\lbeq{e.zeta}
X\zeta = X^q\cdot\zeta_q  + X^p \cdot \zeta_p.
\eeq

\begin{thm}
Let $\theta = (p,0)$ be the linear form defined by
\lbeq{e.thetap}
(X\theta)(q,p):=X^q(q,p) \cdot p ~.
\eeq
Then $\omega=-d\theta = \pmatrix{0 & -1 \cr 1 & 0}$ 
is an exact symplectic form satisfying
\lbeq{e.omegap}
(YX\omega)(q,p):=X^q(q,p)\cdot Y^p(q,p)-Y^q(q,p)\cdot X^p(q,p)
\eeq
for arbitrary vector fields $X,Y$. Its inverse 
satisfies
\lbeq{e.omegaq}
X=\zeta\omega^{-1} \iff X^q=\zeta_p,~~X^p=-\zeta_q.
\eeq
for arbitrary linear forms $\zeta$. With the Lie product
\lbeq{e.canpoisson}
f\lp g := df \omega^{-1} dg 
= \partial_p f\cdot \partial_q g - \partial_p g \cdot  \partial_q f,
\eeq
the algebra $\Ez=C^\infty(\Mz )$ of scalar fields on phase space $\Mz $ 
is a Poisson algebra. $\lp$, $\omega$, and $\theta$ are called the 
\bfi{canonical Lie product}, the \bfi{canonical symplectic form},
and the \bfi{canonical linear form}\footnote{
In the notation using components and the Einstein 
summation convention, we have $\theta = p_j dq^j$
and $\omega = dq^j \wedge dp_j$. Here the linear forms $dq^j$ and
$dp_j$, given by $Xdq^j:=(X^q)^j$ and $Xdp_j:=(X^p)_j$, are the
gradients of the functions $q^j$ and $p_j$ mapping a general state
$(q,p)$ to the indicated components.
} 
on phase space $\Mz $.
\end{thm}
\bepf
$\omega$ is an exact 2-form since $\omega=d(-\theta)$. To prove 
\gzit{e.omegap}, \at{check the following} 
we use  \gzit{e.dtheta} to work out 
$(YX\omega)(q,p)=Yd(X\theta)-Xd(Y\theta)+(X\lp Y)\theta
=Yd(X^qp)-Xd(Y^qp)+(X\lp Y)^qp
=Y^q\partial_q(X^qp)+Y^p\partial_p(X^qp)
-X^q\partial_q(Y^qp)-X^p\partial_p(Y^qp)+(X\lp Y)^qp$. 
Using $(X\lp Y)^q=X\partial Y^q-Y \partial X^q 
=X_q\partial_qY^q+X_p\partial_pY^q-Y_q\partial_q X^q-Y_p \partial_pX^q$,
which follows from \gzit{e.ldc}, 
the product rule, and $\partial_p p =1$, everything cancels except 
for $Y^pX^q-X^pY^q$. This proves \gzit{e.omegap}.
\MMt{
Write $X\theta$ as $X^i\theta_i = X^{q^{\mu}}p_\mu$.
\[
\bary{rcl}
Y^jX^i\omega_{ij}
 &=& Y^jX^i d_{[j}\theta_{i]}
     ~=~ Y^jX^i d_j\theta_i - Y^jX^i d_i\theta_j \\
 &=& Y^j X^{q^\mu} d_j(p_\mu) - Y^{q^\mu} X^i d_i(p_{\mu})
 	~=~  Y^{p_{\mu}}X^{q^{\mu}} - Y^{q^{\mu}}X^{p_{\mu}}
\eary
\]
} 
By comparing \gzit{e.zeta} with \gzit{e.omegap}, we see that
\lbeq{e.zetaX}
\zeta=X\omega \iff \zeta_q=-X^p,~~\zeta_p=X^q.
\eeq
Thus $\omega$ maps $X=(X^q,X^p)$ to $(-X^p,X^q)$, corresponding to 
right multiplication by the matrix 
$\pmatrix{0 & -1 \cr 1 & 0}$.
Since the equations \gzit{e.zetaX} are uniquely solvable for $X$
by \gzit{e.omegaq}, we conclude that $\omega$ is nondegenerate, 
hence symplectic.

Since every exact 2-form is closed, Theorem \ref{t2.2} applies and 
gives the final assertion.
\epf

Using the chain rule, the dynamics \gzit{e.hamc} is easily seen to be 
equivalent to the \bfi{Hamilton equations} in \bfi{general form}, 
\lbeq{e.heg}
\dot f = \frac{df}{dt} = H \lp f;
\eeq
cf. Chapter \ref{s.cao}.

\section{Constrained Hamiltonian dynamics}\label{s.hamcon}

In the constrained case, additional parameters (e.g., Lagrange 
multipliers) are needed to describe the possible states of the system.
Therefore we take $\Ez=C^\infty(\Mz \times \Uz)$ and 
$\Wz=C^\infty(\Mz \times \Uz,\Cz\Fz\times\Cz\Fz^*\times\Cz\Uz)$, 
the spaces of scalar fields and vector fields, respectively, on an 
\bfi{augmented cotangent 
bundle} $\Mz \times \Uz$ of $\Mz _c$, where, as before, the phase space 
is $\Mz =\Mz _c\times \Fz$, and $\Uz$ is a convenient vector space.

A classical, conservative, \bfi{constrained mechanical system} 
is again defined by a \bfi{Hamiltonian} $H\in\Ez$. 
The point $x =(q,p,u)\in \Mz \times \Uz$ is called the \bfi{state} 
with \bfi{position} $q\in \Mz _c$, \bfi{momentum} $p\in\Fz$,
and \bfi{parameter} $u\in \Uz$; however, due to the constraints 
derived below from $H$, not all points in $\Mz \times \Uz$
are physical. As we shall see, the accessible phase space may also be 
smaller than $\Mz $.
The \bfi{energy} of the system in the state $(q,p,u)$ is 
the value $H(q,p,u)$ of the Hamiltonian at $(q,p,u)$. 

The state of the system again varies with time $t\in [\ul t,\ol t]$.
The time dependence is modeled by a \bfi{trajectory}, a state-valued, 
arbitrarily often differentiable function of time, now defining 
position $q(t)$, momentum $p(t)$, and parameter $u(t)$ at time $t$.
These are constrained by the \bfi{extended Hamiltonian equations},
\lbeq{e.hamcu}
  \dot q = \frac{dq}{dt} = \partial_p H,~~~ 
  \dot p = \frac{dp}{dt} = -\partial_q H,~~~
  0 = \partial_u H.
\eeq
Here $\partial_u = \partial/\partial u$ denotes the gradient operator 
with respect to the parameter $u$. Thus, in place of a system of 
ordinary differential equations in the unconstrained case we now have
a system of  
\bfi{differential-algebraic equations} (\bfi{DAE}) involving the 
\bfi{holonomic constraints} 
\lbeq{e.constr}
0 = \partial_u H(q,p,u).
\eeq
Again the extended Hamiltonian equations automatically imply
the \bfi{conservation of energy}:
$\frac{d}{dt} H(q,p,u)
=\partial_q H \cdot \dot q + \partial_p H \cdot \dot p
+ \partial_u H \cdot \dot u = 0$. 

The case where the symmetric \bfi{Hessian matrix}
\[
G := \partial_u^2H(q,p,u)
\]
is invertible is referred to as the \bfi{regular} case.
Then, by the implicit function theorem, \gzit{e.constr} can be solved
locally uniquely for $u=u(q,p)$, which implies that \gzit{e.hamcu}
may be viewed as an ordinary differential equation in $q$ and $p$ alone.
In the \bfi{singular} case where the Hessian $G$ is not invertible, 
the constraints imply restrictions on $p$. Thus, not the whole phase 
space is dynamically accessible, and the analysis of solvability of 
the DAE is more involved. The details depend on the so-called 
\bfi{index} of a DAE, index 1 corresponding to the regular case, 
index $>1$ to the singular case, and are beyond our treatment.

\begin{expl}\label{ex.gauge}
We consider the constrained Hamiltonian system with $\Fz=\Rz^3$ and 
$\Uz=\Rz$, defined by the Hamiltonian
\[
H(\q,\p,u):= \half \p^2 + V(\k\times\q)-(\k\cdot\p)u,
\]
where $V(\E)$ is a potential energy function.
The special case $V(\E):=\half \E^2$, describes the dynamics of a 
single Fourier mode with wave vector $\k$ of the free electromagnetic 
field. A straightforward calculation gives the dynamics
\[
\dot\q = \p-\k u,~~~\dot\p = \k \times \Nabla V(\k\times\q),~~~
0=\k\cdot\p.
\]
Since $G=\partial_u^2H=0$, this is a singular case. Indeed, the 
dynamically relevant part of the phase space is characterized by the 
\bfi{transversality condition} $0=\k\cdot\p$, whereas the multiplier
$u$ is completely undetermined by the dynamics. This implies that the 
dynamics of $q$ is determined only up to an arbitrary multiple of $\k$;
in other words, only $\k\times\q$ is determined at all times by the 
initial conditions.

Note that $H\lp \k\cdot\p = \k\cdot(H\lp \p)=\k\cdot\p=0$,
hence the constraint $0=\k\cdot\p$ is automatically satisfied at all
times if it is satisfied at some time. Thus, in the terminology of 
constrained mechanics, it is called a \bfi{first class constraint},
and gives rise to \bfi{gauge symmetries}. 
A \bfi{gauge transform} replaces $\q$ by $\q+\k s(\q)$ with an 
arbitrary scalar field $s(\q)$, and leaves everything of dynamical 
interest invariant. The \bfi{gauge invariant} quantities are those
in the centralizer $C(\k\cdot\p)$ of the constraint. $f$ belongs to the
centralizer iff it Lie commutes with $\k\cdot\p$, which is the case 
iff $\k\cdot\partial_q f=0$, hence iff $f$ depends only on $p$ and 
$\k\times\q$. Thus, the centralizer consists of all smooth functions of 
\[
\B:=\k\times \q,~~~\E:=-\p,
\]
and the dynamics of the gauge invariant quantities is determined by
\lbeq{e.BE}
\dot \B = - \k \times \E,~~~\dot\E=\k \times \nabla V(\B),~~~
0 = \k\cdot\E.
\eeq
\end{expl}

The extended Hamiltonian equations may also be derived from a 
variational principle. Now the \bfi{action} is defined on smooth 
paths in $\Mz\times \Uz$,
\lbeq{e.actionu}
I(q,p,u) := \int_{\ul t}^{\ol t} dt~ 
\Big(p(t)\cdot \dot q(t) - H(q(t),p(t),u(t))\Big).
\eeq
Variations of the arguments show as before that 
the path $(q,p,u)$ is a stationary point of the action if
and only if the extended Hamiltonian equations \gzit{e.hamcu} hold;
the constraint equations derive from
\[
I(q,p,u+\delta u) - I(q,p,u) \approx  \int_{\ul t}^{\ol t} dt~ 
\Big(-\partial_u H(q(t),p(t),u(t))\cdot\delta u\Big).
\]
A vector field $X\in\Wz$ is now a triple of functions 
\[
X=(X^q,X^p,X^u)\in C^\infty(\Mz ,\Cz\Fz)\times 
C^\infty(\Mz ,\Cz\Fz^*)\times C^\infty(\Mz ,\Cz\Uz); 
\]
its value at the state $(q,p,u)$ is 
$X(q,p,u)=(X^q(q,p,u),X^p(q,p,u),X^u(q,p,u))$. 
Associated with each vector field $X$ is the derivation $Xd$ defined by
\[
Xdf := X^q\cdot \partial_q f + X^p \cdot \partial_p f 
+ X^u \cdot \partial_u f.
\]
It is again easy to check that this defines a commutative differential 
geometry.
In particular, a general linear form $\zeta$
is described by a triple of functions $(\zeta_q,\zeta_p,\zeta_u)\in 
C^\infty(\Mz ,\Cz\Fz^*)\times C^\infty(\Mz ,\Cz\Fz^*)
\times C^\infty(\Mz ,\Cz\Uz)$
such that
\lbeq{e.czeta}
X\zeta = X^q\cdot\zeta_q  + X^p \cdot \zeta_p  + X^u \cdot \zeta_u.
\eeq
In analogy to the unconstrained case, we define the linear form 
$\theta$ by
\[
(X\theta)(q,p,u):=X^q(q,p,u) \cdot p ~.
\]
Thus $\theta=(p,0,0)$ and a similar calculation as before gives
the exact 2-form
\[
\omega:=-d\theta = \pmatrix{0 & -1 & 0 \cr 1 & 0 & 0 \cr 0 & 0 & 0},
\] 
and 
\[
(YX\omega)(q,p,u)
:=X^q(q,p,u)\cdot Y^p(q,p,u)-Y^q(q,p,u)\cdot X^p(q,p,u).
\]
Since no differentiation by the parameters $u$ is involved, 
the 2-form $\omega$ is now degenerate and hence no longer 
symplectic. As a result, $\Ez(\omega)$ is strictly smaller than $\Ez$;
a scalar field $f$ is found to be compatible with $\omega$ and hence in 
$\Ez(\omega)$ only if $\partial_u f = 0$, i.e., $f$ is independent of 
$u$. Thus $\Ez(\omega)=C^\infty(\Mz )$ is again the Poisson algebra of 
scalar fields on phase space, with Lie product \gzit{e.canpoisson}.

The Hamilton equations \gzit{e.heg} remain valid, too; note that
by the general theory, $ H \lp f\in\Ez(\omega)$, although $H$ depends
on $u$.

If the Hamiltonian $H(q,p,u)=H(q,p)$ is independent of $u$, everything
reduces to what we said about unconstrained Hamiltonian mechanics.
Constrained Hamiltonian mechanics with unconstrained Hamiltonian 
$H_0(q,p)$ and $u$-independent holonomic constraints $C(q,p)=0$
are obtained by introducing a vector $u$ of Lagrange multipliers for 
the constraints and defining $H(q,p,u)=H_0(q,p) +C(q,p) \cdot u$.
Note that $\partial_u H(q,p,u) =  C(q,p)$ simply recovers the 
holonomic constraints. Thus, we see that the components of $u$ which 
occur only linearly in $H$ behave as multipliers of $u$-independent 
holonomic constraints.
\at{But what happened with the constraint 
equation? check this on examples! Treat the regular case and the
particular case of a regular quadratic term. Discuss the gauge case.}

\bigskip
Note that there is another class of models for conservative Hamiltonian 
dynamics, defined by so-called \bfi{nonholonomic constraints}. 
\at{refs} There 
the constrained dynamics is not given by \gzit{e.hamcu} but instead by
\[
  \dot q = \frac{dq}{dt} = \partial_p H(q,p),~~~ 
  \dot p = \frac{dp}{dt} = -\partial_q H(q,p)+A(q) u,~~~
  0 = \partial_p H(q,p)\cdot A(q),
\]
where $A(q)$ maps a multiplier vector $u\in\Uz$ to an element from 
$\Fz$, and the Hamiltonian $H$ again defines the energy.
The energy is conserved since $\frac{d}{dt} H(q,p) 
= \partial_q H(q,p)\cdot \dot q+\partial_p H(q,p)\cdot \dot p
=\partial_p H(q,p)\cdot A(q) u =0$. However, now the dynamics can 
usually no longer be written in terms of a variational principle. 
Only the special \bfi{integrable} case where $A(q) = \partial_q C(q)$ 
corresponds to holonomic constraints of the form $C(q)=0$ and a
modified Hamiltonian $\widetilde H(q,p,u):=H(q,p)-C(q) \cdot u$,
which agrees on the space of trajectories with $H$.
The most general conservative Hamiltonian system may have both 
holonomic and nonholonomic constraints; the reader may wish to write
down the defining equations and generalize the above discussion 
accordingly.

\section{Lagrangian mechanics}\label{s.lagr}

Frequently, and especially in relativistic field theory, a classical
system is defined in terms of the Lagrangian approach to mechanics.
We consider here the \bfi{autonomous} case only, where the Lagrangian 
is time-independent.

The basic object is now a \bfi{Lagrangian} $L\in\C^\infty(T\Mz_c)$,
a function of points in the tangent space $T\Mz_c$ of a configuration
manifold $\Mz_c$. As in the Hamiltonian case, we restrict our attention
to the case where $\Mz_c$ is a nonempty, open subset of a convenient
vector space $\Fz$ over $\Rz$. Then the tangent space is 
$T\Mz_c=\Mz_c\times\Fz$, points in $T\Mz_c$ are pairs $(q,v)$
consisting of a configuration point $q\in\Mz_c$ and a tangent vector 
$v\in\Fz$ at $q$, referred to as \bfi{velocity}, and the Lagrangian is 
a function with function values $L(q,v)$.

The Lagrangian approach to mechanics can be represented
in the framework of constrained Hamiltonian dynamics by taking
$\Uz=\Fz$, and $u=v$. Then the choice 
\lbeq{e.Lham}
H(q,p,v):=up-L(q,v)
\eeq
for the Hamiltonian gives unconstrained Lagrangian mechanics.
Constrained Lagrangian mechanics with holonomic constraints $C(q,v)=0$ 
is similarly obtained by taking $\Uz=\Fz\times\Uz_0$ and $u=(v,u_0)$ 
and $H(q,p,v,u_0)=pv - L(q,v) +u_0^TC(q,v)$, where $u_0$ is a 
Lagrange multiplier. However, in the following, we only discuss the
unconstrained Lagrangian case.

Applying the general machinery of Section \ref{s.hamcon} to
\gzit{e.Lham}, we find as dynamical equations the 
\bfi{Euler-Lagrange equations}
\lbeq{po.2}
   \dot q = v,~~~\dot p = \partial_q L(q,v),~~~p=\partial_v L(q,v);
\eeq
and the action \gzit{e.action} reduces on the submanifold defined by 
$\dot q = v$ to 
\lbeq{e.Laction}
I(q) := \int_{\ul t}^{\ol t} dt~  L(q(t),\dot q(t)).
\eeq
The Hamiltonian \gzit{e.Lham} is time invariant since
\[
   (p\cdot\dot q)^\pdot = \dot p\cdot\dot q+p\cdot \ddot{q} = 
L_q\cdot \dot q+ L\dot q\cdot \ddot{q} = L(q,\dot q)^\pdot = \dot L.
\]
It is easily verified directly that the condition for $I(q)$ to be 
stationary at the path $q$ gives again the Euler-Lagrange equations 
\gzit{po.2}; this is usually taken as the starting point of the
Lagrangian approach.

\begin{expl}
The Lagrangian $L(q,v)=  \half m v^2 - \half k q^2$ defines the 
harmonic oscillator, as can be seen by writing down the
Euler-Lagrange equations. Note that the action need not be bounded 
below, as can be seen from the path $q(t)=s(1-t^2)$ in
 $[\ul t,\ol t]=[-1,1]$, where $I(q)=(4m-\frac{8}{15}k)s^2$ diverges 
to $-\infty$ when $k>7.5m$ and $s\to\infty$. Thus, it is inappropriate
to refer to the stationary action principle as \idx{principle of least 
action}, as often done for historical reasons.
\at{Check that at the harmonic solution we only have a saddle point!}
\end{expl}

If we change a Lagrangian $L(q,v)$ to
\[
\widetilde L(q,v):=L(q,v)+ v \dot \partial_q \phi(q)
\]
for some smooth function $\phi$, the action $I(q)$ remains
unchanged apart from a boundary term arising through integration by 
parts.  As a result, the new equations of motions and the old ones
are equivalent. On the other hand, the momentum changes from $p$
to $\widetilde p = p +  \partial_q \phi(q)$. This does not affect the
equation of motion in the form \gzit{e.heg} since the transformation
from $p$ to $\tilde p$ is a canonical transformation leaving the 
Lie product invariant. Indeed, it is not difficult to see that the
more general substitution of $p'=p+\chi(q)$ for $p$ preserves the
Lie product iff the Jacobian $\partial_q\chi(q)$ is a symmetric matrix.
Necessity follows since for constants $a,b\in \Fz$,
\[
a \cdot p' \lp b \cdot p' = a\cdot \chi(q) b -b \cdot \chi(q) a
\]
must vanish, and sufficiency can be established by a more involved 
computation.

\at{reconsider the Hamiltonian example;
mention the field theory case, Lagrangian density, and covariant 
actions.}

\at{{\bf *********** Ignore the remainder of this section ***********}
It treats the constraint Lie product, and is 
not yet adapted to the uniform notation used throughout the book.}

We may work directly in the tangent manifold \at{define $\Ez$ etc.}
and define the linear form
\[
\theta_L = p = \partial_{\dot q} L, 
\]
\lbeq{e.66a}
 \theta_L(X):=Xp,
\eeq
and the canonical 2-form
\[
\omega_L = -d\theta_L.
\]
Then $d\omega_L = -ddp = 0$ implies that $\omega_L$ is closed, 
and hence Theorem \ref{t2.2} applies.
If $\omega_L$ is non-degenerate, we can solve for $\dot q$ in terms
of $p,q$ and get the Hamiltonian picture in the traditional way. The
Poisson algebra becomes the standard Poisson algebra on the cotangent
bundle. If $\omega_L$ is degenerate, we cannot solve for $\dot q$
and compatibility restricts the space $\Ez(\omega_L)$ of quantities.
\[
   p = L_{\dot q} \in \Ez \otimes \Cz\Fz 
\]
is the \bfi{canonical momentum}.

\at{check notation; how are the Lie products related?
Is $f\in\Ez(\omega_l)$ iff it is a function of $q$ and 
$\partial_v L(q,v)$? If so, just make a remark that not every $p$ 
has the right form in the singular case.}
On $\Ez=C^\infty(T  \Mz  )$, any Lagrangian $L=L(q,\dot q)$ defines
a Lie product on $\Ez(\omega_L)$
which induces the Euler-Lagrange dynamics defined by the action 
$I=\int dt L$. 

We rearrange the canonical 2-form as
\[
   \omega_L = dq \wedge dp = dq \wedge (p_q dq+ p_{\dot q} d \dot q).
\]
The condition for $f\in \Ez(\omega_L)$ to be compatible with $\omega$ 
requires the 
existence of a Hamiltonian vector field $X_f$ with 
\lbeq{e.sing}
   \frac{\partial f}{\partial \dot q} = G dq (X_f),~~~
   \frac{\partial f}{\partial q} = -G d \dot q (X_f),
\eeq
with the symmetric \bfi{Hessian matrix}
\lbeq{po.6a}
G:=\partial_q^2 L = 
\partial_q \partial_v L=\frac{\partial p}{\partial  \dot q}
\eeq

{\sc Case 1.} In the {\em regular} case, i.e., if the Hessian matrix
is invertible, we can solve the constraint equation $p=\partial_v L$ 
at least locally for $v$, getting an equation $\dot q = v(q,p)$.
In this case, we find from \gzit{e.sing} that
\lbeq{po.7}
   f \lp g = \frac{\partial f}{\partial q} \cdot G^{-1} 
   \frac{\partial g}{\partial \dot q}  - \frac{\partial g}
   {\partial q} \cdot G^{-1} \frac{\partial f}{\partial \dot q},
\eeq
where $f$, $g$ are functions of $q$ and $\dot q$. 
Note that
\[
   L_v(q,v(q,p))=p,
\]
and 
\[
   H(q,p) = pv(q,p)-L(q,v(q,p))
\]
has derivatives 
\[
   H_p=v(q,p)+p v_p(q,p) - L_v(q,v(q,p)) v_p (q,p) = v(q,p) = \dot q,
\]
\[
   H_q=pv_q (q,p) - L_q(q,v(q,p))-L_v(q,v(q,p)) v_q(q,p)
      = -L_q(q, \dot q)  = - \dot p,
\]
so that
\[
   \frac{d}{dt} f(q,p) = f_p \dot p + f_q \dot q = -f_p H_q+f_q H_p
   = H \lp f,
\]
with the canonical Lie product on phase space.

\at{is this just a repeat?}
Since $H = (p| \dot q)-L$, we have for solutions $q$
\[
\frac{\partial H}{\partial \dot q} 
  = \Big( \frac{\partial p}{\partial\dot q}\Big|\dot q\Big)+p-L_{\dot q}
  =\frac {\partial p}{\partial \dot q} \dot q,
\]
\[
  \bary{ll}
    \D\frac{\partial H}{\partial q} & = \D\Big( \frac{\partial p}
    {\partial q} \Big | \dot q \Big) - \frac{\partial L}{\partial q} = 
    \Big( \frac{\partial p}{\partial q}\Big| \dot q \Big) - \dot p\\
    & = \D\Big( \frac{\partial p}{\partial q} \Big| \dot q\Big) - 
    \Big( \Big( \frac{\partial p}{\partial q} \Big| \dot q\Big) +
    \Big( \frac{\partial p}{\partial \dot q}\Big|\ddot{q}\Big) \Big) =
   - \Big( \frac{\partial p}{\partial \dot q} \Big| \ddot{q} \Big).
  \eary
\]
Hence \at{why?}
\[
   X_H = \Big( \frac{d}{dt} q \Big| \partial_q \Big) +
   \Big( \frac{d}{dt} \dot q \Big| \partial_{\dot q} \Big)
\]
and 
\[
   H \lp g = dg (X_H) = \Big( \frac{\partial g}{\partial q} \Big| 
   \dot q \Big) + \Big( \frac{\partial g}{\partial \dot q} \Big|
   \ddot{q} \Big) = (g(q, \dot q))^\pdot = \dot g,
\]
so that $H$ generates the dynamics.

\at{but the dynamics is already assumed in the derivation!?}

{\sc Case 2.} 
In the {\em singular} case, i.e., when the Hessian matrix \gzit{po.6a} 
is not invertible,  condition \gzit{e.obs} is nontrivial, 
not all $f(q,\dot q)$ are compatible with $\omega$ and hence in the 
Poisson algebra. Then 
\gzit{po.7} only holds for the generalized inverse and \gzit{e.sing}
requires that the partial derivatives are in the range of $G$.
\at{These should be mappings 
of $T \Mz $ which leave both $H$ from \gzit{e.Lham} and $\theta_L$ 
from \gzit{e.66a}
invariant. These form a group. Typical transformations should have 
the form $\ol q = Q(q), \ol v = Q'(q)v+w(q)$.
Clarify! Only first class constraints 
have undetermined multipliers and hence an associated 
gauge freedom -- see 
[R. Jackiw,
(Constrained) Quantization Without Tears,
arXiv:hep-th/9306075}
The Poisson manifold (or orbifold?) 
is the set of orbits of the gauge group;
 cf. M/R p. 325. Restrict $\Ez$ accordingly, as in the symplectic case:]

The resulting Lie product (cf. \gzit{e.liep}) is 
\lbeq{po.lie}
   f \lp g = d g (X_f) 
   = \frac{\partial g}{\partial q} dq (X_f)
   + \frac{\partial g}{\partial \dot q} d \dot q (X_f).
\eeq

Note that the standard treatment in terms of symplectic 
manifolds requires regularity. In the singular case, complicated 
additional assumptions and arguments are needed to bring theories
with gauge symmetries (which are always singular) into the 
framework of symplectic geometry.

\chapter{Hamiltonian quantum mechanics}\label{c.hquant}

\at{adapt the summary in Section 1.5 of QML}

In this chapter, Hamiltonian quantum mechanics is described in 
differential geometric, classical terms. In particular, this
enables one to formulate dynamics for mixed quantum-classical systems
in which -- as in the Born-Oppenheimer approximation in quantum 
chemistry -- slow degrees of freedom are modelled classically, while
the fast motion (typically of electrons) is modelled by quantum
mechanics. 

Also discussed is the relation between classical mechanics and quantum 
mechanics in terms of quantization procedures.

\section{Quantum dynamics as symplectic motion}\label{s.quantdyn} 

As a particular case of dynamics in the Poisson algebra of
a symplectic form we discuss here the dynamics of wave functions and 
expectations in quantum mechanics.

We consider the special case of the unconstrained setting of 
Section \ref{s.hamphase} where $\Mz_c=\Fz$. Then $\Hz=\Cz\Fz\cong\Mz$ 
is a complex Euclidean space in which we may do quantum mechanics.
The isomorphism between $\Hz$ and $\Mz$ as real vector spaces is
made explicit by writing
\lbeq{e.qs1}
\psi=q+\iota p \in\Cz\Fz,
\eeq
where
\lbeq{e.qs1a}
\iota:=\frac{i}{\hbar}.
\eeq
Then
\lbeq{e.qs2}
\ol\psi=q-\iota p,
\eeq
and the Hermitian inner product in $\Hz$ is 
\lbeq{e.qs3}
\phi^*\psi=\ol\phi\cdot\psi.
\eeq
We may regard arbitrary smooth functions of $\psi$ and $\ol\psi$ as
functions of $q$ and $p$ by writing (with slight abuse of notation) 
\lbeq{e.qs4}
f(\psi,\ol\psi)=f(q+\iota p,q-\iota p).
\eeq
The chain rule then implies the relations 
\lbeq{e.qs5}
\partial_\psi+\partial_{\ol\psi} = \partial_q,
\eeq
\lbeq{e.qs6}
\iota(\partial_\psi-\partial_{\ol\psi}) = \partial_p
\eeq
for the partial derivatives. Using these, it is an easy matter to 
rewrite the Lie product \gzit{e.canpoisson} in the form
\lbeq{e.qs7}
f \lp g = \iota(f_\psi\cdot g_{\ol\psi} - g_\psi\cdot f_{\ol\psi}).
\eeq
Now we consider the classical Hamiltonian 
\[
H_c(\psi,\ol\psi):= \psi^* H\psi,
\]
where $H\in\Lin \Hz$ is a quantum Hamiltonian. Then we find
\lbeq{e.qs8}
\dot\psi = H_c\lp\psi = -\iota H\psi,
\eeq
giving the \bfi{Schr\"odinger equation}
\lbeq{e.qs8a}
i\hbar\dot\psi = H\psi
\eeq
as classical Hamiltonian equation of motion for the state vector 
$\psi\in\Hz$. Thus, quantum mechanics may be discussed in a classical 
framework. 
The variational principle for classical Hamiltonian systems discussed
in the context of \gzit{e.action}, rewritten for the present situation,
is called the \bfi{Dirac-Frenkel variational principle}. It was first
used by \sca{Dirac} \cite{Dir.var} and \sca{Frenkel} \cite{Fre},
and found numerous applications; a geometric treatment is given in
\sca{Kramer \& Saraceno} \cite{KraS}. The action takes the form
\lbeq{e.DFaction}
I(\psi,\ol\psi) := \int_{\ul t}^{\ol t} dt~ 
\psi^*(t)\Big(i\hbar\frac{d}{dt}  - H)\Big)\psi(t);
\eeq
setting its variation to zero indeed recovers \gzit{e.qs8a}.
The Dirac-Frenkel variational principle plays an important role
in approximation schemes for the dynamics of quantum systems.
In many cases, a viable approximation is obtained by restricting the
state vectors $\psi(t)$ to a linear or nonlinear manifold of
easily manageable states $|z\>$ (for example \bfi{coherent states}) 
parameterized by classical parameters $z$ which can often be given a 
physical meaning. Inserting the ansatz $\psi(t)=|z(t)\>$ into the
action \gzit{e.DFaction} gives an action for the path $z(t)$, and
the variational principle for this action defines an approximate
classical Lagrangian (and hence conservative) dynamics for the 
parameter vector $z(t)$.
Thus, the Dirac-Frenkel variational principle fits in naturally with
the interpretation in Section \ref{s.measurement} of the parameter 
vectors characterizing a state as the natural observables.
An important application of this situation are the 
\bfi{time-dependent Hartree-Fock equations} which are at the heart of 
dynamical simulations in quantum chemistry. 

We note that $\psi^*\psi$ is a constant of the motion,
hence we may restrict the dynamics \gzit{e.qs8a} to \bfi{normalized 
state vectors} $\psi$ satisfying $\psi^*\psi=1$. In this case, we may
interpret the function $A_c\in\Ez$ defined for $A\in\Lin\Hz$ by 
\[
A_c(\psi,\ol\psi):= \psi^* A\psi = \<A\>
\]
as the classical \bfi{value} of the quantity $A$ in the pure state
defined by the normalized state vector $\psi$, or, equivalently, by the
rank one density matrix 
\lbeq{e.prho}
\rho = \psi\psi^*.
\eeq
\at{discuss gauge invariance and functions of values as the gauge 
invariant classical quantities.}
The Lie product 
of two values is again a value, since one easily calculates
\lbeq{e.qs9}
\<A\>\lp\<B\>= \<\iota [A,B]\> = \<A\lp B\>,
\eeq
where the Lie product on the right hand side is the quantum bracket.
In particular, the dynamics of the values is given by the
\bfi{Ehrenfest equation} 
\lbeq{e.qs10}
   \frac{d}{dt}\<A\> = \<H\>\lp\<A\> = \<H\lp A\> 
   = \frac{i}{\hbar}\<[H,A]\>.
\eeq
In the special case, where $H=T(p) +V(q)$ is expressible as a sum
of a kinetic energy operator $T(p)$ depending on a momentum vector $p$
and of a potential energy operator $V(q)$ depending on a position 
vector $q$, whose components are operators satisfying 
the traditional \bfi{canonical commutation rules}
\[
q_j\lp q_k = p_j\lp p_k =0,~~~p_j\lp q_k=\delta_{jk},
\]
the special cases of the Ehrenfest equation,
\[
\frac{d}{dt}\<q\> = \<\partial_p H(q,p)\> = \<\partial_pT(p)\>,~~~
\frac{d}{dt}\<p\> = -\<\partial_q H(q,p)\>= -\<\partial_q V(q)\>,
\]
often called the \bfi{Ehrenfest theorem},
are due to \sca{Ehrenfest} \cite{Ehr}.
The Ehrenfest equation, here derived in the Schr\"odinger 
picture, is valid also in the Heisenberg picture (or even more general 
interaction pictures); 
the dynamical objects of physical interest are neither
the states nor the quantities, but the values. We may also compute
the dynamics of the density matrix \gzit{e.prho}, and find the
\bfi{Liouville equation}
\lbeq{e.qliouv}
    i \hbar \dot\rho = [H(p,q), \rho].
\eeq
More generally, it is not difficult to check that taking \gzit{e.qs9}
as a definition of the Lie product of values in arbitrary states
(not necessarily pure states as in the above derivation) indeed turns
the family of $\<A\>$ with $\<\cdot\>$ ranging over states defined by
\lbeq{e.erho}
\<A\>=\tr \rho A
\eeq
for some strongly integrable density matrix $\rho$ and $A$ ranging 
over the elements of $\Lin\Hz$ into a Lie algebra. Therefore, the
Ehrenfest equation is valid for arbitrary states, not only for pure 
states. By inserting \gzit{e.erho} into the Ehrenfest equation and
comparing coefficients, one also sees that the Liouville equation
\gzit{e.qliouv} remains valid.

\section{Quantum-classical dynamics} \label{s.qc}

There are many systems of practical interest which are treated
in a hybrid quantum-classical fashion. The most important example
is the Born-Oppenheimer approximation, where nuclei are treated 
classically, while electrons remain quantized. Another truly
quantum-classical system is a quantum Boltzmann equation with spin;
here the spin is still an operator, represented by $4\times 4$ matrices 
parameterized by classical phase space variables. On the other hand,
the quantum-Boltzmann equation for spin zero is already a purely 
classical equation, since its dynamical variables are all commuting.

In the Liouville picture, where the density matrices are the dynamical 
variables, the basic equations for a large class 
of quantum-classical models are the \bfi{generalized Liouville equation}
\[
    i \hbar \dot\rho = [H(p,q), \rho],  
\]
and the \bfi{generalized Hamilton equations} 
\[
    \dot q =  \tr \rho \partial_p H(p,q),
\]
\[
    \dot p=  - \tr \rho \partial_q H(p,q).
\]
Here $H\in C^\infty(\Mz,\Lin\Hz)$ is an operator valued function
on a classical phase space $\Mz$. Thus $H(p,q)$ is, for any fixed
vectors $p,q$, a linear operator on some Euclidean space $\Hz$, the
\bfi{density matrix} $\rho=\rho(t)$ is a time-dependent trace-class 
operator on $\Hz$, and $q=q(t), p=p(t)$ are classical, time-dependent 
vectors, not quantum objects. The classical quantities are the 
functions of the values
\[
   \<f(p,q)\> = \tr \rho f(p,q)
\]
where $f$ is a $(p,q)$-dependent operator on $\Hz$.
Expressed in terms of values, we have
\[
   \dot q = \Big\<\partial_p H(p,q)\Big\>, ~~~
\dot p= -\Big\<\partial_q H(p,q)\Big\>,
\]
which looks like the Ehrenfest theorem,
except that on the left hand 
side we have classical variables and no expectations.
The equations are conservative equations for the evolution of
values (the value $\<H\>$ of the energy is conserved);
dissipative systems and stochastic systems can be also modelled,
but this is beyond the scope of the present exposition.

The quantum-classical dynamics preserves the rank of the density $\rho$.
In particular, if $\rho$ has the rank 1 form  
\lbeq{esc5}
   \rho = \psi \psi^*   
\eeq
at some time, it has at any time the form \gzit{esc5} 
with time-dependent $\psi$.
The fact that $\rho$ has trace 1 translates into the statement that
the state vector $\psi$ is normalized to $\psi^*\psi=1$.
One easily checks that the Liouville equation holds iff the state vector
psi, determined by  \gzit{esc5} up to a phase, satisfies the 
Schr\"odinger equation 
\[
   i \hbar \dot\psi = H(p,q) \psi.
\]
In terms of the state vector,  values take the familiar form
\[
   \<f(p,q)\> = \psi^* f(p,q) \psi.
\]
The reader is invited to formulate a Hamiltonian description
of quantum-classical systems, by starting with a symplectic 
dynamics in which only a part of the position and momentum variables 
are complexified into a quantum state vector, and to derive the
 corresponding Poisson algebra. Now the Lie product is the tensor 
product of that of the classical subsystem and that of the quantum 
subsystem treated as a classical Hamiltonian system. The 
\bfi{Ehrenfest equation} still has the form
\[
   \frac{d}{dt} \<A\> = \<H\> \lp \<A\>,
\]
but the right hand side no longer simplifies to the value of a 
commutator; instead, one gets a nonlinear dependence on values. 
Such nonlinearities are common for reduced descriptions coming 
from a pure quantum theory by coarse graining. 
Usually, quantum-classical systems are regarded as reduced 
descriptions, and the same phenomenon occurs.
There are plenty of other examples of practical importance,
the primary one being the \bfi{Schr\"odinger-Poisson equations}
in semiconductor modeling.

\begin{expls}
We mention two important examples, 
molecular quantum chemistry and a spinning electron.

(i) The \bfi{Born-Oppenheimer approximation} of the dynamics of 
molecules, widely used in quantum chemistry, is a typical 
quantum-classical system of the above kind.
The nuclei are described by classical phase space variables, 
while the electrons are described quantum mechanically by means of a
state vector $\psi$ in a Hilbert space of antisymmetrized electron wave 
functions. \at{give $H$}

(ii) A \bfi{spinning electron}, while having no purely classical 
description, can be modelled quantum-classically by classical phase 
space variables $p,q$ and a quantum 4-component spin. 
Then, with $\alpha,\beta$ as in the Dirac equation, 
\at{define these in section \ref{s.beta}!}
\lbeq{esc6}
    H(p,q) = \alpha \cdot p + \beta m + e V(q)  
\eeq
is a $4\times 4$ matrix parameterized by classical 3-vectors $p=p(t)$ 
and $q=q(t)$,
$\rho =\rho(t)$ is a positive semidefinite $4\times 4$ matrix of 
trace 1, and the trace in the above equation is just the trace of a 
$4\times 4$ matrix.

One gets the equations from Dirac's equation and Ehrenfest's theorem by
an approximation involving coherent states for position and momentum.
Note that this is just a toy example. More useful field theoretic 
quantum-classical versions lead to \bfi{Vlasov equations} for 
$(p,q)$-dependent $4\times 4$ densities, describing a fluid of 
independent classical electrons of the form \gzit{esc6}. With even 
more realism, one needs to add also a collision term accounting
for interactions, resulting in a \bfi{quantum Boltzmann equation};
and for even more accurate modeling, \gzit{esc6} is no longer adequate 
but needs additional dissipative terms.
\end{expls}

The quantum-classical dynamics, given in 
the Schr\"odinger picture, can also be written in the Heisenberg 
picture. The equivalent Heisenberg dynamics is 
\[
  \frac{d}{dt} f = \partial_q f \<\partial_p H\> 
                  -\partial_p f \<\partial_q H\> + \frac{i}{\hbar} [H,f]
\]
where now $\<\pdot\>$ is the fixed Heisenberg state.
From this, one can immediately see that everything depends only on 
 values by applying $\<\pdot\>$ to this equation:
\[
 \frac{d}{dt} \<f\> = \<\partial_q f\> \<\partial_p H\> 
      - \<\partial_p f\> \<\partial_q H\> 
      + \<\frac{i}{\hbar} [H,f]\>.
\]
This is now a fully classical equation for classical values
of the quantum-classical hybrid model considered.

In the interpretation given in Chapter \ref{c.models}, densities are
irreducible objects describing a single quantum system,
not stochastic entities that make sense only under repetition.
(This is analogous to the way phase space densities appear in the
Boltzmann equation, though the analogy is not very deep.)

\bigskip
In general,  values in the quantum-classical dynamics
are to be interpreted as objects characterizing 
a single quantum system, in the sense of the consistent experiment
interpretation, and not as the result of averaging over many 
realizations. 

By design, in the Heisenberg picture, the state does not take part in 
the dynamics. What is new, however, compared to pure quantum 
dynamics is that the Heisenberg state occurs explicitly in the
differential equation. In practical applications, the Heisenberg 
state is fixed by the experimental setting; hence 
this state dependence of the dynamics is harmless. However, 
because the dynamics depends on the Heisenberg state, calculating
results by splitting a density at time $t=0$ into a mixture of pure
states no longer makes sense. One gets different evolutions of the
operators in different pure states, and there is no reason why
their combination should at the end give the correct dynamics
of the original density. (And indeed, this will usually fail.)
This splitting is already artificial in pure quantum mechanics
since there is no natural way to tell of which pure states a mixed
state is composed of. But there the splitting happens to be valid
and useful as a calculational tool since the dynamics in the
Heisenberg picture is state independent.

In contrast to the pure quantum case, there is now a difference
between averaging results of two experiments $\rho_1, \rho_2$ and
the results of a single experiment $\rho$ given by $(\rho_1+\rho_2)/2$.
That, in ordinary quantum theory, the two are indistinguishable
in their statistical properties is a coincidental consequence
of the linearity of the Schr\"odinger equation, and the resulting
state independence of the Heisenberg equation;
it does no longer hold in effective quantum theories
where nonlinearities appear due to a reduced description.

\at{apply to test particles and their response to a quantum field.
$H=(\p-e\A)^2/2m + e A_0$
This is consistent with the Copenhagen interpretation!}

\at{
It is important to have an interpretation in which these can be 
consistently interpreted. This is, however, impossible in the 
traditional statistical interpretation; there are several theorems
in the literature documenting this \cite{Bou,Sal}.
On the other hand, the consistent experiment interpretation can 
cope successfully with this challenge.\\
Any hybrid theory must be interpreted in terms of concepts that have 
identical form in classical and in quantum mechanics; 
otherwise there are inevitable conflicts. The consistent experiment 
interpretation does this -- it is a theory which contains
the classical and the quantum case as two special cases of the
same conceptual framework. In this framework one can therefore discuss
things consistently that lead to puzzles if interpreted either on
a pure classical or on a pure quantum basis, or in some ill-defined
in-between limbo.\\
It is in principle conceivable (though not desirable
on the basis of simplicity) that the most fundamental description 
of nature is truly quantum-classical and not purely quantum. 
In the absence of an interpretation which a consistent quantum-classical
setting, this would have been unacceptable, but now there are no
fundamental reasons apart from elegance that would forbid it.
} 

\section{Deformation quantization}\label{s.defq}

There are many ways to \bfi{quantize} a classical system, i.e., to 
relate to a dynamical description of a classical system a corresponding 
quantum version. This process is far from unique, but there are a 
number of well-explored (and only sometimes equivalent) routes for 
doing this. Whether a particular quantization is useful depends on how 
well the resulting quantum system describes the intended application 
-- something outside the scope of our discussion. 

An important algebraic approach to quantization is 
{\bfi{Berezin quantization}}, 
also called the {\bfi{method of orbits}}. Here classical 
Poisson representations of Lie algebras are lifted to unitary
representations. We only hint at the constructions, and refer 
for details to \sca{Berezin} \cite{berezin74},
\sca{Bar-Moshe \& Marinov} \cite{bar-moshemarinov}, \sca{Landsman}
\cite{landsman-1998}, and \sca{Kirillov} \cite{kirillov04}.
The construction of the Lie--Poisson algebra in Section \ref{lie--poiss}
from a Lie $*$-algebra $\Lz$ implies that the dual of $\Lz$ becomes in a
natural way a Poisson manifold; the corresponding symplectic leaves
are the so-called {\bfi{co-adjoint orbits}}, the orbits of the 
universal covering group \at{not yet defined} corresponding to $\Lz$ 
in its co-adjoint action on $\Lz^*$. The canonical Poisson algebras 
on the co-adjoint orbits carry an \bfi{irreducible Poisson 
representation} of $\Lz$, and any irreducible Poisson representation 
of $\Lz$ arises in this way (up to equivalence).
Thus, classifying the co-adjoint orbits is the classical analogue of
classifying irreducible unitary representations.
The quantization constructions mentioned above rely on close
relations between co-adjoint orbits, coherent states over Lie groups,
and irreducible unitary representations.
These relations can even be generalized further, replacing the
Lie algebra structure by a purely geometric setting, which then leads
to the framework of {\bfi{geometric quantization}}, 
cf.\ \sca{Woodhouse} \cite{woodhouse}.

\at{add backreferences to definition of Poisson algebra}
Another possibility is \bfi{deformation quantization} which deforms a 
commutative product into a so-called \bfi{Moyal product}; for
definitions and details, see, e.g., \sca{Rieffel} \cite{Rie}. 
Alternatively, deformation quantization may be viewed as a deformation
of the quantities in a Poisson algebra $\Ez$. This is the procedure we 
shall discuss in more detail.

The deformation can be 
obtained by embedding $\Ez$ into the algebra $\Lin \Ez$, identifying 
$f \in \Ez$ with the multiplication mapping $M_f$ which maps $g$
to 
\[
M_f\{g\}:=fg,
\]
writing for emphasis the arguments (in $\Ez$) of operators from 
$\Lin\Ez$ (often referred to as \bfi{superoperators}, to distinguish 
them from operators acting on $\Ez$ itself) in curly braces.

Recall the linear operator $\ad_f\in\Lin \Ez$ defined by
\lbeq{equ1}
  \ad_f\{g\}:=f \lp g,
\eeq
For $f\in\Ez$, we define the \bfi{quantization} $\widehat{f}$ of $f$ by
\lbeq{equ2}
  \widehat{f} :=f-\frac{i \hbar}{2} \ad_f \in\Lin\Ez.
\eeq
Note that the quantization preserves nonlinear operations (product and 
Lie product) only up to terms of formal order $O(\hbar)$. 
This reflects the ordering ambiguity in traditional quantization
procedures. 

\at{example. what
precisely does "traditional" mean here? Does it include Berezin
quantization? Just the standard poisson-brackets-to-commutators rule?
What about Dirac-Bergman constrained quantization?}

For an arbitrary Gibbs state on $\Lin\Ez$, the expectation
\[
  \< \widehat f \> = \< f   \> - \frac{i \hbar }{2} \< \ad_f \> 
\]
differs from those of $f$ by a term of numerical order $O(\hbar )$, 
justifying an interpretation in terms of deformation. 

\begin{prop}          
For $f,g$ in a not necessarily commutative Poisson algebra $\Ez$, 
\lbeq{equ3}
  [\ad_f,g]=[f,\ad_g]=f \lp g,
\eeq
\lbeq{equ4}
  [\ad_f,\ad_g]=\ad_{f \slp g}.
\eeq
\lbeq{equ6}
  [\widehat{f},g]=[f,\widehat{g}]=[f,g]-\frac{i \hbar}{2} f \lp g.
\eeq
\lbeq{equ5}
  [ \widehat{f} ,\widehat{g}]=[f,g]-i \hbar f \lp g - \frac{\hbar ^2}{4}
  \ad_{f\lp g},
\eeq
\end{prop}

\bepf  
\at{in qft.tex, there is a version of this for super Poisson algebras.} 
We have, for all $h \in \Ez$,
\[
 \bary{rcl}
  [\ad_f,g]\{h\}&=&\ad_f\{gh\}-g \ad_f\{h\} \\
  ~&=& f \lp gh - g(f \lp h)=(f \lp g)h,
 \eary
\]
hence $[\ad_f,g]=f\lp g$. Therefore, also $[f,\ad_g]=-[\ad_g,f]
=-g \lp f =f \lp g$, so that \gzit{equ3} holds. Similarly, 
\[
 \bary{rcl}
  [\ad_f,\ad_g]\{h\}&=&\ad_f\{\ad_g\{h\}\}- \ad_g\{\ad_f\{h\}\} \\
  ~&=& f \lp (g \lp h)- g \lp (f \lp h) \\
  ~&=& (f \lp g) \lp h= \ad_{f \lp g}\{h\}
 \eary
\]
by (S5), hence \gzit{equ4} holds. \gzit{equ6} follows from
\[
  [\widehat{f},g]=\Big[ f-\frac{i\hbar}{2} \ad_f,g \Big] 
  =[f,g]-\frac{i \hbar}{2} [\ad_f,g]=
  [f,g]- \frac{i \hbar}{2} f \lp g ,
\]
and
\[
  [f,\widehat{g}]=-[\widehat{g},f]=-[g,f]+\frac{i \hbar}{2} g \lp f =
  [f,g]-\frac{i\hbar }{2} g \lp f.
\]
Finally, since
\[
  \bary{rcl}
   [\widehat{f} ,\widehat{g} ]
    &=&\D \Big[ f-\frac{i \hbar }{2} \ad_f, g - \frac{i \hbar }{2}\ad_g \Big]
     \\
    &=&\D [f,g]-\frac{i \hbar }{2} [\ad_f,g]- \frac{i \hbar}{2} [f,\ad_g]+
    \Big( \frac{i \hbar }{2} \Big) ^2 [\ad_f, \ad_g],
   \eary
\]
\gzit{equ5} follows from \gzit{equ3} and \gzit{equ4}.
\epf

To actually quantize a classical theory, one may choose a 
Lie algebra of relevant quantities generating the Poisson algebra, 
quantize its elements by the above rule, express the classical
action as a suitably ordered polynomial expression in the 
generators, and use as quantum action this expression with all
generators replaced by their quantizations.

In general, the above recipe for phase space quantization gives an
approximate Poisson isomorphism, up to $O(\hbar)$ terms.

\at{example. A Poisson isomorphism is an isomorphism that preserves 
the Poisson algebra structure}

We now show that, however, Lie subalgebras are mapped into 
(perhaps slightly bigger) Lie algebras defining an abelian extension 
\at{explain!}, and that one gets a true isomorphism for all embedded 
Heisenberg Lie algebras and all embedded abelian Lie algebras.

\begin{thm}~\bfi{(Quantization Theorem)}\\    
If $\Ez$ is commutative then, with
\[
M_f\{g\}:=fg,~~~
Q_f\{g\}:= \widehat{f}\{g\} =fg-\frac{i \hbar}{2} f\lp g,
\]
the quantum Lie product 
\[
A\lp B = \iota[A,B]~~~\mbox{for } A,B\in\Lin \Ez
\]
satisfies, for $f,g\in\Ez$,
\lbeq{equ7}
  Q_{f} \lp Q_{g}=M_{f \slp g}- \frac{i \hbar }{4} \ad_{f \slp g}
  = \half (M_{f \slp g} + Q_{f \slp g}),
\eeq
\lbeq{equ8}
  Q_{f} \lp M_g =M_f \lp Q_{g} = \half M_{f \slp g}.
\eeq
Any Lie subalgebra $\Lz$ of $\Ez$ defines a Lie algebra 
\lbeq{equ9}
  \widehat{\Lz}= \{ M_{f \slp g} + Q_{h} \mid f,g,h \in \Lz \}
\eeq
under the quantum Lie product. If $\Lz$ is an abelian Lie algebra or 
a Heisenberg Lie algebra then $Q_{}$ is a Lie isomorphism between 
$\Lz$ and $\widehat\Lz$.
\end{thm}

\bepf                 
Since $\Ez$ is commutative, the first term in \gzit{equ5} and 
\gzit{equ6} vanishes, and multiplication by $\iota =i/ \hbar$ gives 
\gzit{equ7} and \gzit{equ8}. The final statement is immediate from 
\gzit{equ7} and \gzit{equ8}.
\epf

\at{explain how
the naive impression one gets from elementary QM books that every
self-adjoint operator on a Hilbert space corresponds to a physically
meaningful observable property is not tenable in general.}

Note that by the so-called \bfi{Groenewold-van Hove Theorem}
(\sca{Groenewold} \cite{GrovH}, \sca{van Hove} \cite{vHov}, 
\sca{Gotay} et al. \cite{GotGT}), 
no quantization procedure can exist which possesses all features 
desirable from a naive point of view. The present quantization 
procedure sacrifices the exact preservation of commutation rules.

\at{If there is a Euclidean structure,  
$Q_{f^*}=Q_ f^*$ follows from partial integration,
so that we have a unitary representation! Why is 
$Q_ H$ bounded below? This relates to \bfi{fans}. 
It requires a Euclidean structure 
and hence an integral, which isn't defined yet. 
Relate this to Chapter 5.
relate to Clebsch representation!}

\bigskip
\at{adapt this; treat only the Lie-Poisson case.
We can restrict any classical Poisson algebra $\Ez=C^\infty(\Omega)$ to
any open subset $\Oc$ of an invariant set, and get a restricted Poisson
algebra $C^\infty(\Oc)$ with
\lbeq{eli14}
  f|_\Oc\lp g|_\Oc=(f\lp g)|_\Oc.
\eeq
For example, in the Heisenberg Lie-Poisson algebra, we take $\Oc=\{
\omega\mid1(\omega)=1\}$ and get the standard symplectic Poisson 
algebra. The same works for any projective Lie algebra and produces the
corresponding projective Lie--Poisson algebra.\\
If $\Ez$ is the Lie-Poisson algebra over $\Lz$ and $\Omega$ is a 
coadjoint orbit, these results can be strengthened. We have for $\omega
\in\Omega$
\lbeq{eli15}
  T_\omega\Omega=\{\delta\lp\omega\mid\delta\in\Lz\}~~~\subseteq 
  \{\nu\mid\lp\omega=0~\Rightarrow~\delta(\nu)=0\}, 
\eeq
with equality if $\dim V<\infty$. The first equality implies that
there is a unique symplectic form $\Pi_\omega\in(T_\omega\Omega \times 
T_\omega\Omega)^*$ with 
\lbeq{eli16}
  \Pi_\omega(\delta\lp\omega,\delta'\lp\omega)=(\delta\lp\delta')
  (\omega);
\eeq
for details see M/R p. 475. This is a special case of the construction
of symplectic leaves (M/R p. 360) for general Poisson algebras. The 
latter are defined as orbits under $\Aut\Ez$, and the above has a 
curved generalization to this situation; now $\delta\lp\omega\in T_
\omega\Omega$ instead of $V$.}

\section{The Wigner transform}\label{s.wigner}

We now specialize the preceding to the standard symplectic Poisson 
algebra $\Ez=C^\infty (\Rz^n \times \Rz^n)$. Thus, $\Ez$ is a 
commutative Poisson algebra of phase space functions as discussed in 
Section \ref{s.hamphase}. in this very important case, which covers
$N$-particle quantum mechanics, the embedding discussed in 
Section \ref{s.defq} turns out to be equivalent to standard
quantization.

The equivalence is given in terms of the so-called Wigner transform.
The Wigner transform relates kernels of linear integral operators over 
$\Rz^n$ to corresponding phase space functions. It therefore mediates 
between a quantum (operator) ansd a classical (phase space) description 
of the same situation, and is heavily used in semiclassical 
approximations of quantum mechanics. For example (though this is 
outside the scope of the present book), they turn the Heisenberg 
equation of motion for quantum field expectations (combined with 
certain approximations) into quantum kinetic equations on phase space. 

The basic idea is to rewrite a kernel $K(x,y)$ as a function of the
mean coordinate $q=(x+y)/2$ and the difference $q'=x-y$, and then 
Fourier transform with respect to $q'$ to get a function of $q$ and a 
momentum vector $p$. 

\at{relate to Moyal quantization}

\at{check $\iota,j,\hbar$}
In the present special case, phase space quantization
amounts to using the reducible representation
\[
 \widehat p = p- \frac{i \hbar }{2} \partial_q,~~~
 \widehat q =q+ \frac{i \hbar }{2} \partial_p
\]
of the canonical commutation rules on the Hilbert space of square 
integrable functions on phase space 
instead of the traditional irreducible \bfi{position representation} by
\[
  \widetilde p=-i \hbar \partial_x,~~~\widetilde q=x
\]
on the Hilbert space of square integrable functions of configuration 
space, or of the irreducible 
\bfi{momentum representation} by
\[
  p'=p,~~~q'=i \hbar \partial_p
\]
on the Hilbert space of square integrable functions of 
momentum space. Since the momentum representation is 
obtained from the position representation by the simple canonical 
transformation which interchanges $x$ and $p$ and then writing 
$q'=-\widetilde p$, $p'=\widetilde q$, it is enough to discuss
in the following the transformation to the position representation.

By quantizing in phase space, one gives up irreducibility
(and hence the description of a state by a {\em unique} density)
but gains in simplicity. This may be compared to the situation
in gauge theory, where the description by gauge potentials introduces
some arbitrariness with which one pays for the more elegant 
formulation of the field equations but which does not affect the
observable consequences.

\bigskip
We now show that these representations are related by a 
\bfi{Wigner transform} (cf. \sca{Wigner} \cite{Wig}).

We consider the quantization of the commutative Poisson algebra 
$\Ez=C^\infty (\Rz^n \times \Rz^n)$ with standard Poisson bracket 
\gzit{e.canpoisson}. Since for $f=f(p,q)$ we have 
\lbeq{ewi2}
  p \lp f= \partial _qf,~~~q \lp f=- \partial_pf,
\eeq
the quantization rule amounts to 
\lbeq{ewi3}
 \widehat p = p- \frac{i \hbar }{2} \partial_q,~~~
\widehat q =q+ \frac{i \hbar } {2} \partial_p.
\eeq
By \gzit{equ7},
\lbeq{ewi4}
  \widehat p _\mu \lp \widehat q _\nu = p_\mu \lp q _\nu 
  =\delta _{\mu \nu}.
\eeq
Thus we have a unitary representation of the Heisenberg
algebra $H(n)$ by linear operators on the phase space 
$\Omega=\Rz^n \times \Rz^n$, equipped with the standard inner product.
To relate this representation to the traditional 
position representation given by 
\lbeq{ewi5}
  \widetilde p=-i \hbar \partial_x,~~~\widetilde q=x,
\eeq

\at{explain that $\widetilde p =-i\hbar\partial_x$ is Hermitian, 
due to integration by parts}

we introduce the \bfi{Wigner transform}
\lbeq{ewi6}
  \widetilde f(x,y)
:=\D \int dpe^{\iota p^T(x-y)}f \Big(p,\frac{x+y}{2}\Big)
\eeq
of a function $f \in C^ \infty(\Rz^n \times \Rz^n)$.

\at{I must put the factor $h^n$ into \gzit{ewi6} and start with 
\gzit{ewi7} as definition since $\widetilde f(x,y)$ is independent 
of $\hbar$ and I need to investigate the limit $\hbar\to 0$ of 
traditional quantum field theory.}

\begin{thm}
The Wigner transform has the inverse transform
\lbeq{ewi7}
  f(p,q)
=h^{-n} \D \int d \xi e^{-2 \iota p^T \xi}\widetilde f(q+ \xi,q- \xi),
\eeq
where
\lbeq{ewi8}
  h=2 \pi \hbar,~~~n= \dim p= \dim q,
\eeq
and satisfies the rules
\lbeq{ewi9}
  \widetilde {\widehat pf}= \widetilde p \widetilde f,~~~
\widetilde {\widehat qf}= 
\widetilde q\widetilde f.
\eeq
\end{thm}

\bepf
We have
\[
\bary{rcl}
  \D \int d \xi e^{-2 \iota p^T \xi}\widetilde f(q+ \xi,q- \xi)
  &=& \D \int d \xi e^{-2 \iota p^T \xi} \D \int dke^{2 \iota k^T\xi}
  f(k,q)\\
  ~&=& \D \int dp \Big( \D \int d \xi e^{2 \iota(k-p)^T \xi } \Big)
  f(k,q)\\
  ~&=& \D \int dp~h^n \delta(k-p)f(k,q)=h^nf(p,q),
\eary
\]
proving \gzit{ewi7}. \gzit{ewi9} follows from
\[
\bary{rcl}
  \widetilde p \widetilde f(x,y) 
  &=& -i \hbar \partial_x \D \int dpe^{\iota p^T(x-y)}f \Big(p, 
  \frac {x+y}{2}\Big)\\
  ~&=& -i \hbar \D \int dpe^{\iota p^T(x-y)}\Big( \iota p+ \half 
  \partial _q \Big) f \Big( p,\frac{x+y}{2}\Big) \\
  ~&=& \D \int dp e^{\iota p^T(x-y)}(p-\frac{i\hbar}{2} \partial_q)f 
\Big(p, \frac{x+y}{2}\Big)\\
  ~&=& \D \int dp e^{\iota p^T(x-y)}\widehat p f
 \Big(p,\frac{x+y}{2}\Big)
   =\widetilde {\widehat pf}(x,y)
\eary
\]
and
\[
\bary{rcl}
  \widetilde {\widehat qf}(x,y)
  &=& \D \int dpe^{\iota p^T(x-y)} \Big(\frac{x+y}{2}+
  \frac{i \hbar}{2}\partial p\Big) f \Big( p,\frac{x+y}{2}\Big)\\
  ~&=& \D \int dpe^{\iota p^T(x-y)} \Big(\frac{z+y}{2}-\frac{i \hbar}{2}
  \iota(x-y)\Big)f \Big(p,\frac{x+y}{2}\Big)\\
  ~&=& \D \int dpe^{\iota p^T(x-y)}xf \Big(p,\frac{x+y}{2}\Big)\\
  ~&=& \D x\int dpe^{\iota p^T(x-y)}f \Big(p,\frac{x+y}{2}\Big)=
  x \widetilde f(x,y)=\widetilde q \widetilde f(x,y).
\eary
\]
\epf

Thus the Wigner transform provides an isomorphism between the two
representations. Note that the phase space representation is highly
redundant since the position representation does not act at all on the  
$y$-coordinate. The redundancy is apparent from the fact that the
algebra generated by $\widehat p$ and $\widehat q$ is much smaller than 
$\Lin \Ez$, and in fact isomorphic (modulo convergence issues) to $\Lin
C^ \infty(\Rz ^n)$ via the Wigner transform.
However, this redundancy is very helpful since it makes
the classical limit and the approximation by semiclassical techniques
much simpler. 

The Wigner transform can be applied to all nonlinear PDEs of 
Schr\"odinger or Dirac type. These have the form
\lbeq{ewi10}
  I(\psi \psi^*) \psi=0,~~~\psi \in C^\infty(\Rz^n),
\eeq
where $I$ is an operator-valued function of the density matrix
\lbeq{ewi11}
   \widetilde \rho:=\psi \psi^*
\eeq
and can be rewritten in terms of it as the equation
\[
 I(\widetilde \rho)\widetilde \rho=0
\]
which after an inverse Wigner transform becomes an equation
\lbeq{ewi12}
  \bar I(\rho)\rho=0
\eeq
in phase space. However, \gzit{ewi11} loses the rank $1$ condition
implicit in \gzit{ewi10}, hence corresponds to a ``mixing'' of pure 
states. 
\at{In practical applications, it is probably \gzit{ewi12} which is
of true physical significance, while \gzit{ewi11} is only an irrelevant
restriction. (\gzit{ewi10} preserves the rank if $L$ is gauge 
invariant.)}

\begin{prop}
The bilinear inner product
\lbeq{ewi17}
  (f|g)= \D \int dpdq~f(p,q)g(p,q)
\eeq
satisfies
\lbeq{ewi18}
  (f|g)=\Big( \frac{2}{h}\Big)^n \D \int dxdy ~
\widetilde f(x,y)\widetilde g(y,x).
\eeq
\end{prop}

\bepf
This follows from
\[
\bary{rcl}
  \D \int dxdy \widetilde f(x,y) \widetilde g(y,x)
  &=& \D \int dxdy \D \int dpe^
  {\iota p^T(x-y)}f \Big(p,\frac{x+y}{2}\Big)\D \int dke^{\iota k^T
  (y-x)}g \Big(k,\frac{y+x}{2}\Big)\\
  ~&=& \D \int dpdkdxdye^{\iota(p-k)^T(x-y)}f \Big(p,\frac{x+y}{2}\Big)g
  \Big(k,\frac{x+y}{2}\Big)\\
  ~&=& \D \int dpdkdqdwe^{\iota(p-k)^Tw}f(p,q)g(k,q)\\
  ~&=& \D \int dpdkdq \Big(\frac{h}{2}\Big)^n \delta (p-k)f(p,q)g(k,q)\\
  ~&=&\D \Big(\frac{h}{2}\Big)^n \D \int dpdq~f(p,q)g(p,q)=\Big(\frac{h}
  {2}
  \Big)^n(f|g).
\eary
\]
\epf

Under conjugation, we have directly from \gzit{ewi6},
\lbeq{ewi13}
  \widetilde f^*(x,y)= \ol{\widetilde f(y,x)}\iff f^*(p,q)=\ol{f(-p,q)},
\eeq
indicating that the complex combination
\[
 z=q+ip
\]
behaves naturally.

\begin{prop}
The \bfi{conditional expectation} of an operator $A \in S(\Omega, \Lz)$
at fixed $x$ or $k$, respectively, defined by
\lbeq{e8a}
  \< A \> _x := \int dk \rho (k,x) A ,~~~
  \< A \>_k := \int dx \rho (k,x) A,
\eeq
satisfies
\lbeq{e9}
  \< A(\p)\>_k = A(k) \< 1 \> _k,~~~~ \< A (\q) \>_x =A(x) \<1\>_x
\eeq
and
\lbeq{e9a}
\< A \>:= \int dk dx A \rho (k,x)  
= \int dk \< A \> _x   = \int dx \< A \> _k.
\eeq
Thus $\p$ is the momentum operator and $\q$ the position operator.
\end{prop}

\bepf
The proof is straightforward.
\epf

\begin{thm}
With the pointwise convolution
\lbeq{ewi14}
  (f*g)(p,q)= \D \int dkf(k,q)g(p-k,q),
\eeq
we have for the pointwise product
\lbeq{ewi15}
  \widetilde f \widetilde g= \widetilde{f*g}.
\eeq
\end{thm}

\bepf
By \gzit{ewi7}, we have $\widetilde f \widetilde g=\widetilde e$, where
\[
\bary{rcl}
  h^ne(p,q)
  &=& \D \int d \xi e^{-2 \iota p^T \xi}\widetilde f(q+\xi,q- \xi)
  \widetilde g(q+\xi,q-\xi)\\
  ~&=& \D \int d \xi e^{-2 \iota p^T \xi} \D \int dke^{2 \iota k^T \xi}
  f(k,q) \D \int dle^{2 \iota l^T \xi}g(l,q)\\
  ~&=& \D \int dkdl \Big(\D \int d \xi e^{2 \iota(k+l-p)^T \xi}\Big)
  f(k,q)g(l,q)\\
  ~&=& \D \int dkdl~h^n \delta (k+l-p)f(k,q)g(l,q)\\
  ~&=& h^n \D \int dk~f(k,q)g(p-k,q)=h^n(f*g)(p,q).
\eary
\]
\epf

A comparison with \gzit{ewi9} shows that, formally,
\[
 p*f= \widehat{p}f,~~~q*f= \widehat{q}f.
\]
For a fully localized phase space function, \gzit{ewi6} implies 
directly 
\lbeq{ewi16}
  f(p,q)=f(q)\iff \widetilde f(x,y)=h^nf(x) \delta (x-y).
\eeq

\bigskip
\bfi{The symbol formulation.}
To extend the quantization rule to arbitrary smooth functions we need
the symbol formulation of $\widehat p$ and $\widehat q$.

Let $\Omega = \Rz ^m \times \Rz ^m$
the classical phase space, and let 
\lbeq{ecc1}
   j= i \hbar / 2.
\eeq

\at{$1/j=-2\iota$ Don't use $j$ in exponents, and change sign of $j$ 
everywhere to conform to above text.}

\begin{prop}
For every distribution $\rho$ on $\Omega$,
\lbeq{ecc2}
   \rho (k,x) = \int \widehat{\rho} (k+p, x+q) e^{-p \cdot q/j}dp dq,
\eeq
where
\lbeq{ecc3}
   \widehat{\rho} (p,q)= (\pi \hbar)^{-2m} \int \rho (k+p, x+q) e^{k \cdot x/j}
   dk dx.
\eeq
\end{prop}

\bepf  
We use the formula
\lbeq{ecc4}
  \int f(q) e^{-p \cdot q/j} dp dq = (\pi \hbar)^m f(0),
\eeq
and find for $\widehat{\rho}$ defined by \gzit{ecc3}:
\[
  \bary{rl}
    ~ & \D \int \widehat{\rho} (k+p,x+q) e^{-p \cdot q/j} dp dq \\
    = & (\pi \hbar)^{-2m} \D \int \rho (k' +k+p,x'+x+q) e^{k' \cdot 
    x' -p \cdot q)/j} dk'dx'dp dq \\
    = & (\pi \hbar)^{-2m} \D \int \rho (k+p',x+q') e^{p' \cdot q' /j} 
    e^{-p \cdot q'/j} e^{-p'\cdot q/j}  dp dq'dq dp' \\
    = & (\pi \hbar)^{-m} \D \int \rho (k,x +q') e^{-p \cdot q' /j} dp 
    dq' \\
    = & \rho (k,x ).
  \eary
\] 
We used the substitution $k' =p'-p,~x'=q'-q$.               
\epf

\begin{prop}
For every Schwartz function $A$ on phase space $\Omega$, the symbol $A(\p,\q)$ 
defined by
\lbeq{ecc5}
   A(\p, \q) \rho (k,x):= \int_\Omega A(k +p,x-q) \widehat{\rho} (k+p,
   x+q) e^{-p \cdot q /j} dp dq
\eeq
is consistent with the interpretation
\lbeq{ecc6}
   \p = k +j \partial_x, ~~~~\q = x-j \partial _k
\eeq
when $A$ is a normally ordered polynomial acting on $\rho(k,x)$ 
where all $\p_\mu$ are to the right of all $\q_\nu$. 
Moreover, we have the canonical commutation rules $(CCR)$
\lbeq{ecc7}
   \iota[\p _\mu, \q_{\nu}] = \delta _{\mu \nu}.
\eeq
\end{prop}

\bepf
We can rewrite \gzit{ecc2} as
\[
  \rho (k,x) = \int \widehat{\rho} (k', x')e ^{-(k' -k) \cdot (x' - x)/j}
  dk' dx',
\]
and find inductively for a monomial $\q ^{\alpha} \p ^{\beta}$  (with
multiexponents)
\[
  \bary{rcl}
    \q ^{\alpha} \p ^{\beta} \rho (k,x) &=& \D \int (k')^{\alpha} 
    (2x -x')
    ^\beta \widehat{\rho} (k', x') e^{-(k'-k) \cdot (x' -x)/j} dk' dx' \\
    &=& \D \int (k+p) ^{\alpha} (x-q)^{\beta} \widehat{\rho} (k+p,x+q) e^{-p
    \cdot q/j} dp dq.
  \eary
\]
Now take linear combinations and limits to find \gzit{ecc5}.

The CCR follow since (with $f_u=\partial f/\partial u$)
\[
  \bary{rcl}
    p_\mu q_\nu f & = & p_\mu (x_\nu f-jf_{k_\nu})=k_\mu (x_\nu
    f-jf_{k_\nu}) +j (x_\nu f - j f_{k_\nu})_{x_\nu} \\
   ~&=& k_\mu x_\nu f+j (\delta_{\mu \nu} f +x_\nu f_{x_\mu} -k_\mu 
    f_{k_\nu}) -j^2 f_{k_\nu} x_\mu, 
  \eary
\]
\[
  \bary{rcl}
    q_\nu p_\mu f &=& q_\nu (k_\mu f +j f_{x_\mu}) = x_\nu (k_\mu f +
    j f_{x_\mu})-j (k_\mu f+jf_{x_\mu}) k_\nu \\
    &=& k_\mu x_\nu f +j (- \delta _{\mu_\nu}f + x_\nu f_{x_\mu} -k_\mu
    f_{k_\nu}) -j^2 f_{k_\nu} x_\mu,
  \eary
\]
hence
\[
  [p_\mu, q_\nu]f= p_\mu q_\nu f-q_\nu p_\mu f = 2j \delta_{\mu \nu}f =
  i \hbar \delta _{\mu\nu} f.
\]       
This also implies \gzit{ecc6}.       
\epf

Write $\Jm(f):=\iota(f-f^*)$, so that
\[
  \Jm (a+jb):=b~~~\mbox{if } a , b \in \Ez_0 \mbox{~ are Hermitian.}
\]

\begin{prop}
If
\lbeq{ecc10}
   \rho (-k,x) = \overline{\rho (k,x)}~~~\mbox{for all }(k,x) \in \Omega
\eeq
then the dynamics
\lbeq{ecc11}
  \dot{\rho} = \{ H, \rho\},
\eeq
where
\lbeq{ecc12}
  \{ H, \rho \} = \frac{1}{2j} \left( H \rho - (H \rho)^* \right) = \Jm
  (H \rho)
\eeq
leaves the total density $\< 1 \>$ invariant. Moreover, if for all 
$A$ we have $\< A^* A \> \ge 0$ at time $t=0$ then $\< A^* A \>
\ge 0$ at all times $t \ge 0$.
\end{prop}

\at{proof?}

\begin{prop} \bfi{(Classical limit)} 

In the limit $j \rightarrow 0$,
\[
  A(\p, \q) = A(k,x) + O( \hbar ),
\]
and
\[
  \{ H(\p,\q), \rho \} (p,q)= \partial_p H \partial_q \rho - \partial
  _q H \partial_p \rho + O(\hbar)
\]
reduces to the classical Poisson bracket.
\end{prop}

\at{proof? get precise error estimates from \sca{Mauser} GM-MP!}

Note that \at{check signs!}
\[
  \{ p, \rho \} = \rho _q,~\{ q, \rho \} =  - \rho _p
\]
so this is consistent with 
\[
  \dot{\rho} = (\dot{q} \partial_q + \dot{p} \partial _p) \rho 
  = H_p \rho_q -H_q \rho_p
\]
from Hamiltonian dynamics.

To get the density in a position representation, write
\[
  \underline{\rho} (x,x')=\int \underline{\widehat{\rho}} (p,x') e^{p \cdot 
  x /i \hbar} dp 
\]
where
\[
  \underline{\widehat{\rho}} (p,x')=(2 \pi \hbar)^{-m} \int \underline{\rho}
  (x,y)e^{-p \cdot x/ i \hbar}dx
\]
For normal ordered $H$, we have
\[
  H(\underline{\p}, \underline{\q}) \underline{\rho} (x,y)=(2\pi \hbar)
  ^{-m} \int H(p,x) \underline{\widehat{\rho}} (p,y) e^{p \cdot x /i \hbar}
  dp.
\]
\at{show directly that this holds}.

Similarly,
\[
  \bary{l}
   \underline{\rho} (k+p,k-p):= \D \int \rho (k,x) e^{p\cdot x/j} dx, \\
   \rho (k,x)= (\pi \hbar) ^{-m} \D \int \underline{\rho} (k+p,k-p)
   e^{-p \cdot x /j}dx
  \eary
\]
defines a CCR isomorphism with the momentum representation.

Note that in the position and momentum representation,
\[
  \underline{\rho} ^* (x,x')=\overline{\underline{\rho}(x',x)},~~~
  \underline{\rho} ^* (k,k')= \overline{ \underline{\rho} (k',k) } 
\]
\at{check signs of $k$!} while in the phase repesentation,
\[
  \rho ^* (k,x)=\overline{\rho (-k,x)}.
\]
This becomes natural in an analytic representation $z=x+ik$.
\bigskip

For numerical calculation, approximate $H \in S (\Omega , \Ez_0)$ by
trigonometric functions (complex exponentials). This gives finite 
difference formulas. The initial density $\rho$ can be just given,
or it can be regularized (using a Husimi function? GM-MP 1.24)

    \part{Representations and spectroscopy}\label{p.spectra}
\chapter{Harmonic oscillators and coherent states}\label{c.harmonic}

Part \ref{p.spectra} applies the concepts introduced so far to the 
study of the 
dominant kinds of elementary motion in a bound system, vibrations of
oscillators (described by Poisson representations of the Heisenberg 
group), rotations of rigid bodies (described by Poisson representations 
of the rotation group), and their interaction. On the quantum level, 
quantum oscillators are always bosonic systems, while spinning systems 
may be bosonic or fermionic depending on whether or not the spin is 
integral. The analysis of experimental spectra, concentrating on 
the mathematical contents of the subject, concludes our discussion.

\bigskip
This chapter is a detailed study of harmonic oscillators (bosons, 
elementary vibrations), both from the classical and the quantum point 
of view. We introduce raising and lowering operators in the symplectic 
Poisson algebra, and show that
the classical case is the limit $\hbar\to 0$ of the quantum harmonic 
oscillator. 

The representation theory of the single-mode Heisenberg 
algebra is particularly simple since by the Stone--von Neumann theorem, 
all unitary representations are equivalent. We find that the quantum 
spectrum of a harmonic oscillator is discrete and consists of the 
classical frequency (multiplied by $\hbar)$ and its nonnegative 
integral multiples (overtones, excited states).

We shall work in the representation where the harmonic oscillator 
Hamiltonian is diagonal, which gives rise to the ladder operators 
mediating between neighboring eigenstates. 
We introduce Dirac's bra-ket notation, and
deduce the basic properties of the bosonic Fock spaces, first for
a single harmonic oscillator and then for a system of finitely many 
harmonic modes. 

We then introduce coherent states, an overcomplete basis representation
in which not only the Heisenberg algebra, but the action of the 
Heisenberg group is explicitly visible. 
Coherent states are quantum states that behave as classically as
possible, thereby making a bridge between the quantum system and
classical systems.
The coherent state 
representation is particularly relevant for the study of quantum 
optics, but we only indicate its connection to the modes of the 
electromagnetic field.

\section{The classical harmonic oscillator}\label{s.charm}

The classical one-dimensional harmonic oscillator without damping,
introduced in Section \ref{s.cao}, is defined by the Hamiltonian
\lbeq{ch5.ham}
H = \frac{p^2}{2m}+ V(q)\,,
\eeq
where $V(q)$ is quadratic and bounded from below, so that there are
constants $q_0$, $V_0$ and $k>0$ with
\[
V(q) = V_0 + \frac{k}{2}(q - q_0)^2\,.
\]
The number $k$ is called the {\bfi{stiffness}}; the greater
the constant $k$, the more difficult is it to move away from
equilibrium. The Hamilton equations are:
\[
 \dot q = \frac{p}{m}\,, ~~~ \dot p = - V'(q) = - k (q- q_0)\,.
\]
A complex exponential ansatz shows that the solution of the Hamilton
equations is:
\[
q(t) = q_0 + 2 \re (e^{i\omega t}x)\,, ~~~
p(t) = \re (i\omega m e^{i\omega t}x) \,,
\]
where $x$ is a complex number depending on the initial
conditions, and
\[
\omega = \sqrt{\frac{k}{m}},
\]
is the {\bfi{frequency}} of the harmonic oscillator.
It is convenient to express the variables in terms of a so-called
complex {\bfi{normal mode}}, the function $a(t)$ defined by
\[
a(t) := \sqrt{\frac{k}{2\omega}}(q(t)-q_0)
       + i\frac{p(t)}{\sqrt{2m\omega}}\,.
\]
One can recover $q$ and $p$ through
\lbeq{ch5.pq}%
 q(t) = q_0 +
\frac{1}{2}\sqrt{\frac{2\omega}{k}}(a(t)+a^*(t))\,, ~~~ p(t) =
\frac{1}{2i}\sqrt{2m\omega}(a(t)-a^*(t))\,,%
\eeq
hence the description by a normal mode is equivalent to the
original description. Differentiating $a(t)$ and using $\dot q = p/m$,
we obtain
\[
\dot a(t) = \sqrt{\frac{k}{2\omega}}\frac{p(t)}{m}
               -  \frac{i}{\sqrt{2m\omega}}k(q(t)-q_0)
          = -i\omega a(t)\,.
\]
We conclude that $a(t)$ has to obey
\[
a(t)= a(0) e^{i\omega t}\,.
\]
We calculate the Lie product of $a$ and $a^*$ and find
\beqar
a \lp a^* &=& \frac{\partial a }{\partial p}\frac{\partial a^* }{\partial q} - \frac{\partial a^* }{\partial p} \frac{\partial a }{\partial q}\nonumber\\
&=& 2i\sqrt{\frac{k}{2\omega}}\frac{1}{\sqrt{2m\omega}} \nonumber\\
&=& i\,,\nonumber
\eeqar
that is, we obtain the relation
\lbeq{ch5.ccr}
a\lp a^* = i\,.
\eeq
The relation \gzit{ch5.ccr} is called the
{\bfi{canonical commutation relation} (CCR)} for the harmonic 
oscillator.
More generally, one finds for the Lie product of general functions
$f,g$ of $a$ and $a^*$ the formula
\lbeq{ch5.pois}
f\lp g = i \frac{\partial f}{\partial a}\frac{\partial g}{\partial a^*}
        -i \frac{\partial f}{\partial a^*}\frac{\partial g}{\partial a}.
\eeq
This will be seen later as a special case of a general principle for
constructing so-called Lie--Poisson algebras from a Lie algebra.

\section{Quantizing the harmonic oscillator}\label{s.qharm}

For a classical harmonic oscillator, the Lie product in the CCR
\gzit{ch5.ccr} is defined via the Poisson bracket.
To quantize the harmonic oscillator, all we do is replace the Lie
product in the CCR by its quantum analogue. Thus we postulate the
existence of an operator $a$ and its conjugate $a^*$ with the relation
\[
\frac{i}{\hbar} [a,a^*] = i\,,
\]
equivalently
\lbeq{ch5.qccr}
[a,a^*]=\hbar\,.
\eeq
Note that equation \gzit{ch5.qccr} has the right behavior under
$\hbar \to 0$, since in the limit that $\hbar$ goes to zero, we have
to end up in the classical regime, where the operators $a$ and $a^*$
become functions on phase space and hence commute.

Equation \gzit{ch5.qccr} defines a $*$-algebra, i.e., an associative
algebra with unity and an involution $*$, generated by $a$ with the
relation $aa^* - a^*a=\hbar$.
Later we look for representations in a Hilbert space, where the
involution then corresponds to Hermitian conjugation. But already at
this level, we call expressions in $a$ and $a^*$ operators.

The quantum mechanical Hamiltonian for the harmonic oscillator
is the operator given by direct substitution of the $p$ and $q$
from \gzit{ch5.pq}:
\beqar
H &=& \omega \Bigl(\frac{a-a^*}{2i}\Bigr)^2 
+ \omega \Bigl(\frac{a+a^*}{2}\Bigr)^2+ V_0 \nonumber\\
&=& \shalf \omega (aa^* + a^*a) + V_0 \nonumber\\
&=& \omega a^* a + \shalf \omega \hbar + V_0 \,.
\eeqar
Since only differences in energy are important, one often chooses 
$V_0= -\shalf \hbar \omega$ to get the simple formula $H=\omega a^* a$.

In the classical theory we have commuting variables $a$ and $a^*$ with
a Lie product $a\lp a^* = i$. That is, we have a commutative $*$-Poisson
algebra. In the quantum theory we have an
associative algebra generated by $a$ and $a^*$ with the relation
$aa^*-a^*a=\hbar$. Since in the quantum theory two seemingly different
polynomial expressions (such as $aa^*$ and $a^*a+\hbar$) can be the
same, there is a need for a preferred ordering of $a$ and $a^*$ in
monomials. The {\bfi{normal ordering}} is that ordering of $a$ 
and $a^*$
in monomials where all $a^*$'s are moved to the left of the $a$'s.
It is easy to see that every noncommutative polynomial in $a$ and
$a^*$ can be normally ordered by repeated use of the relation
$aa^*=a^*a+\hbar$; in the process of normal ordering, lower degree
monomials are generated with higher powers of $\hbar$. We give the
following proposition that guarantees that taking $\hbar \to 0$ we
recover the classical theory:

\begin{prop}
Let $f$ and $g$ be noncommutative polynomials in $a$, $a^*$ and
$\hbar$. Viewing $f$ and $g$ as polynomials in commuting variables
$a$ and $a^*$, one can calculate $f\lp g$ using \gzit{ch5.pois}.
As noncommutative polynomials one can calculate the commutator
$[f,g]= fg-gf$. The two results are related by:
\lbeq{e.classlim}
\frac{i}{\hbar} [f,g] = f \lp g + O(\hbar)\,.
\eeq
One expresses this relation by saying that the quantum Lie product is
a {\bfi{deformation}} of the classical Lie product.
\end{prop}
\bepf
The order of the $a$ and $a^*$ does not matter since changing the
order we generate powers of $\hbar$. We use induction on the degree
of the polynomials. For degree zero and one, \gzit{e.classlim} holds. 
Suppose it holds
for degree of $f$ smaller than $n$ and degree of $g$ one. If we write
$f= a^* S +  Ta $ for some normally ordered polynomials $S$ and $T$
with degrees smaller than $n$, we see that for $g=a^*$:
\beqar
i[a^* S + Ta,a^*] &=&  ia^* [S,a^*]+i\hbar T + i[T,a^*]a\nonumber\\
&=&\hbar a^*S\lp a^* + i\hbar T +\hbar T\lp a^* a +O(\hbar^2)\nonumber\\
&=& \hbar (a^* S)\lp a^* + \hbar (Ta)\lp a^* +O(\hbar^2)\nonumber\,,
\eeqar
and the result holds for $f$ arbitrary and $g=a^*$. For $g=a$ it goes
similar. Suppose the claim holds for all $g$ with degree $k$, with
$0\leq k\leq n$. Then for degree $n+1$ let us write
$g = a P+a^* Q + R$, where $P$, $Q$ and $R$ are polynomials of degree
strictly less than $n+1$. Then we have:
\beqar
\frac{i}{\hbar}[f,g] &=& \frac{i}{\hbar}[f, aP+a^*Q+R]\nonumber\\
&=& \frac{i}{\hbar}[f,a]P+\frac{i}{\hbar}a[f,P]+\frac{i}{\hbar}[f,a^*]Q
    +\frac{i}{\hbar}a^*[f,Q]+\frac{i}{\hbar}[f,R]\nonumber\\
&=& f\lp aP+af\lp P+f\lp a^*Q+a^*f\lp Q+f\lp R + O(\hbar)\nonumber\\
&=& f\lp (aP) + f\lp (a^*Q) + f\lp R +O(\hbar)\nonumber\\
&=& f\lp (aP+a^* Q+R)+O(\hbar)\,.
\eeqar
And the proof is complete.
\epf

\bigskip
\bfi{Extension to the anharmonic case.}
The anharmonic oscillator can in principle be treated in a similar
fashion. Since the classical Lie product (the Poisson bracket) is
the same, we may proceed exactly as before, except that the formulas
involving the Hamiltonian are no longer valid. In particular, since
the frequency $\omega$ was determined by the Hamiltonian, it is now
an arbitrary constant. Thus there are multiple, inequivalent ways
of defining the quantities $a(t)$. Indeed, there is even more freedom
since the only important property to be preserved is the canonical
commutation relation.

Generalizing the affine form of $a(t)$ in the harmonic case,
we choose it as an arbitrary affine combination of $q(t)$ and $p(t)$,
\lbeq{e.affine}
a = \lambda+ \mu q + i\nu p
\eeq
for suitable complex numbers $\mu$, $\nu$ and $\lambda$. As can be 
easily verified, the canonical commutation relations \gzit{ch5.ccr}
are reproduced, so that the classical Lie product takes again the
form \gzit{ch5.pois}, exactly when the restriction 
\[
2\re \mu\ol \nu =1
\]
holds.
Having made a choice, we obtain a classical Hamiltonian $H=H(a,a^*)$ in
terms of $a$ and $a^*$. Using the Heisenberg dynamics and
\gzit{ch5.pois}, we obtain
\lbeq{e.normaldyn}
\dot a(t) = H\lp a(t) = -i \frac{\partial H}{\partial a^*}\,.
\eeq
We remark that if \gzit{ch5.ccr} holds for
$t=0$ then it holds for all $t$. Indeed, the derivative
of the left-hand side of \gzit{ch5.ccr} vanishes identically.

Using a different choice of the parameters defining $a$ we get a
different variable $a'$, which is affinely related to the original $a$,
\lbeq{ch5.bog}
a' = \alpha + \beta a + \gamma a^*\,.
\eeq
The requirement that $a'$ satisfies the same commutation relations as
$a$ leads to the restriction
\lbeq{ch5.bog2}
|\beta|^2-|\gamma|^2 = 1.
\eeq
A transformation of the form \gzit{ch5.bog} satisfying \gzit{ch5.bog2}
is called a {\bfi{Bogoliubov transformation}}. Bogoliubov
transformations have important applications; for example, they were at 
the heart of Hawkings' proof that black holes radiate. 
\at{add references}
The generalization
of Bogoliubov transformations to systems of oscillating electron pairs
in metals is an important ingredient for the theory of Cooper pairs,
which explains superconductivity effects in metals at low temperature.
\at{add references}

Different choices of the coefficients in the definition of $a$ lead
of course to different forms of $H(a,a^*)$; this means that different
Hamiltonians $H(a,a^*)$ can describe the same oscillator.
The particular choice above for the harmonic oscillator is the one
leading to $H(a,a^*)=E_0+\omega a^*a$, for which the dynamics
\gzit{e.normaldyn} takes the simple form $\dot a = -i\omega a$. In
theoretical physics there are different operators $a_k$ and
$a^{*}_{k}$ labeled by some parameter $k$.
One tries to find by means of Bogoliubov
transformations the simplest form of the Hamiltonian. The preferred
form is the form where $H$ is diagonalized: $H=\sum_k a^{*}_{k}a_k +
\ldots$, where the dots contain terms of higher order in the operators
$a_k$ and $a^{*}_{k}$.

\bigskip
The quantization of an anharmonic oscillator is done as in the
harmonic case.
For each classical Hamiltonian polynomial in $a$ and $a^*$,
there is a unique normally ordered quantum version.
However, when modeling the same system both in a classical and in
a quantum setting, the coefficients of the quantum system in a normal
ordering of the operators must be taken to depend on $\hbar$,
and the form of this dependence is not determined by the
quantum-classical correspondence. Therefore, the best
fit of coefficients of $H$ to experimental data will generally produce
different optimal values in the classical and the quantum case.
In a quantum field theory, the coefficients will also be dependent on
the scale at which frequencies remain unresolved, giving so-called
{\bfi{running coupling constants}} which play an important role in
renormalization techniques.

\section{Representations of the Heisenberg algebra}
\label{section-reps-heis}

We saw at the end of Section \ref{s.charm}
that the Heisenberg algebra $t(3,\Cz)$ can be considered
as being generated by $1,a$, and $a^*$ where $a$ and $a^*$ satisfy
the CCR $a\lp a^*=i$.

In the classical case, we know a realization of these
commutation relations in terms of a Poisson bracket. In the quantum
case, we must find a representation in terms of operators in a Hilbert
space. The representations of physical interest are the
unitary representations, which represent the one as identity and
behave properly under the $*$-operation.
In this section we construct a unitary representation of the Heisenberg
algebra.

In the quantized version of a classical theory the functions on
phase space become elements of some associative algebra $\Ez$.
For a representation we want to realize the algebra $\Ez$ as a
subalgebra of an algebra of linear operators.

The approach of Schr\"odinger (1926) to this
problem was to take as Hilbert space the space of square integrable
complex-valued functions $\psi$ on $\Rz^3$; then the Schr\"odinger
equation for the dynamics of a pure state takes the form of a wave
equation, which was familiar to physicists at that time and hence
came to dominate quantum mechanics. The approach taken by Schr\"odinger
proved to be very successful and is also presented in many quantum
physics textbooks, since (for a single particle) the real-valued 
function $|\psi(x)|^2$ has an intuitive semiclassical probability 
interpretation (discussed in Section {s.motQM}).
For multiparticle systems, the intuitive advantages of Schr\"odinger's
representation is no longer given, as the wave functions are no longer
in physical space $\Rz^3$ but, for $n$ particles, in an abstract
$3n$-dimensional configuration space. For systems involving an
unconserved number of particles, in particular for interactions with 
light, and for systems in the thermodynamic limit, things are even 
more complicated since the configuration space becomes 
infinite-dimensional, and the wave function representation becomes
unwieldy -- instead one usually resorts to the techniques of quantum 
field theory. Nevertheless, there are
interesting papers using the resulting functional Schr\"odinger
equation to illuminate the relations between classical solitons and
quantum bound states (see, e.g., \sca{Jackiw} \cite{jackiw77}). 

One year earlier than Schr\"odinger, Heisenberg invented his
infinite-dimensional
matrix algebra. We present Heisenberg's approach since it generalizes
easily to the most complex quantum systems, including the universe as
a whole.

\bigskip
We now look at an arbitrary unitary representation
$J:\Lz\to \Lin\Hz$ in a Euclidean space $\Hz$
satisfying
\[
J(a^*)=J(a)^*, ~~~J(1)=1.
\]
We shall write the
operators corresponding to $a$ and $a^*$ in the representation again
by $a$ and $a^*$ (rather than using $J(a)$, etc.), in order to
avoid clumsy notation. This will not cause problems since the
representation turns out to be faithful. Then the operator
\[
n:=\frac{1}{\hbar}a^*a,
\]
for reasons that will soon be apparent, is called the {\bfi{number
operator}}, satisfies the commutation relations
\[
[a,n]= a\,, ~~~ [a^*,n]=-a^*\,,
\]
as is easily checked. This implies that the vector space generated by
$1$, $a$, $a^*$ and $n$ is closed under the commutator, and hence
forms a Lie $*$-algebra $\Lz$ with the quantum Lie product,
called the {\bfi{oscillator algebra}} $os(1)$. In this
section (as always when classifying unitary representations),
it will be more convenient to work directly with commutators.

\bigskip
We now illustrate an important technique in representation theory,
which in many cases of interest provides all irreducible representations
of a certain kind. See Section \ref{s.highest} for some other
applications.

We define the {\bfi{Verma module}} 
\at{check adequateness of the name}
corresponding to a complex number $\lambda$ by
\[
V_\lambda = \{ \psi \in \ol\Hz \mid n\psi = \lambda \psi\} \,,
\]
where the Hilbert space $\ol \Hz$ is the closure of $\Hz$.
If $V_\lambda$ is nontrivial, it contains a nonzero vector,
$\lambda$ is an eigenvalue of $n$, and
any nonzero $\psi\in V_\lambda$ is a corresponding eigenvector.
Thus the nonzero Verma modules are just the eigenspaces of the
eigenvalues of $n$.
Since we consider here only unitary representations where * is the
adjoint, this implies that
\[
\lambda=\frac{\psi^*
  n\psi}{\psi^*\psi}=\frac{\|a\psi\|^2}{\hbar \|\psi\|^2}\geq 0
\]
is real and nonnegative. Noting that in general $n$ is Hermitian, we
now make the slightly stronger assumption that $n$ is self-adjoint
as a densely defined operator of the Hilbert space $\ol\Hz$. Then
the spectral theorem implies that the infimum
\[
\widehat\lambda = \inf_{\psi\ne 0} \frac{\psi^* n\psi}{\psi^*\psi}
\]
is a real and nonnegative number, attained for some $\widehat\psi\ne
0$, and $\widehat\psi$ is an eigenvector of $n$ corresponding to the
eigenvalue $\widehat\lambda$. Thus $V_{\widehat\lambda} \neq 0$.
Now consider
an arbitrary $\lambda$ with $V_\lambda \neq 0$ and a nonzero
$\psi\in V_\lambda$. Then \at{why?}
$na\psi = (\lambda - 1)a\psi$, hence
$a\psi \in V_{\lambda-1}$. If $a\psi =0$ then
\[
\lambda\psi^*\psi=\psi^*n\psi = \frac{1}{\hbar } \psi^* a^* a \psi = 0,
\]
hence $\lambda=0$; and if $a\psi \ne 0$ then $V_{\lambda-1}\ne 0$, and
$\lambda-1$ is an eigenvalue of $n$. In the latter case, we can repeat
the step once, or more often. But since all eigenvalues are nonnegative,
this can happen only a finite number of times, and ultimately we must
end up with the other alternative. Hence zero is an eigenvalue
(in particular $\widehat\lambda=0$) and $\lambda-n=0$ for some
nonnegative
integer $n$. Thus the only possible eigenvalues are
nonnegative integers. That all these actually are eigenvalues follows
by a similar argument. Indeed, with $\psi$ as before, we have
\[
na^*\psi = ([n,a^*]+a^*n)\psi = (n+1)a^*\psi = (\lambda+1)a^*\psi,
\]
hence $a^*\psi\in V_{\lambda+1}$. Since
\[
\|a^*\psi\|^2=\psi^* a a^* \psi = \psi^* (\hbar+a^*a)\psi =
\hbar\|\psi\|^2+\|a\psi\|^2\ge \hbar\|\psi\|^2>0,
\]
$V_{\lambda+1}\ne 0$ and $a^*\psi$ is an eigenvector for the
eigenvalue $\lambda+1$. By induction, we reach all positive integers
from $V_{\widehat\lambda}=V_0$.
Thus we have proved the following theorem:

\begin{thm}
In a representation in which $n$ is self-adjoint, 
a Verma module $V_\lambda$ of the oscillator algebra $\Lz$ is nonzero 
if and only if $\lambda$ is a nonnegative integer.

In particular, the spectrum of the Hamiltonian $H=\omega a^*a$ of a
quantum harmonic oscillator consists of the nonnegative integral
multiples of $\omega\hbar$.
\end{thm}

\at{The self-adjointness requirement is just the requirement that 
we actually represent the group! mention this somewhere in an earlier 
chapter!}

The results obtained justify the following terminology. The operator
$a$ is called a {\bfi{lowering operator}}, since its application 
to an eigenstate of the number operator $n$ lowers the associated 
eigenvalue by one. The operator $a^*$ is called a 
{\bfi{raising operator}}, since
its application to an eigenstate of the number operator $n$ raises
the associated eigenvalue by one. Together, the operators $a$
and $a^*$ are called {\bfi{ladder operators}}. A unit vector in the
Verma module $V_0$ is called a {\bfi{ground state}} (in the second
quantized language of quantum field theory a {\bfi{vacuum vector}}).
For a {\bfi{ground state}}, i.e., a nonzero vector  $\psi\in V_0$, 
we have $n\psi=0$,
hence  $\|a\psi\|^2=\psi^*a^*a\psi=\hbar \psi^*n\psi=0$ and therefore
$a\psi=0$. Thus the ground state is annihilated by the lowering
operator. Therefore $a$ is also called an {\bfi{annihilation 
operator}}; if this term is used then $a^*$ is called a 
{\bfi{creation operator}}.

\section{Bras and Kets}\label{s.braket}

In his groundbreaking work on quantum mechanics, Dirac introduced a
notation for vectors and operators that is widely used by physicists
but is quite different from what mathematicians are used to. Dirac's
bra-ket calculus is not very well defined in the way actually used
by physicists, since the basis vectors considered in the calculus
do not necessarily lie in the Hilbert space in which everything
should happen from a strictly axiomatic point of view.

We define here a precise version of Dirac's bra-ket calculus,
which can also satisfy mathematicians. Instead of working in a
Hilbert space we consider a fixed dense subspace which we denote by
$\Hz$. Thus $\Hz$ is a vector space with a Hermitian inner product
$\<\cdot|\cdot\>$, antilinear in the first argument and linear in
the second, such that $\<\psi|\psi\>$ is always real and nonnegative,
and the relation
\lbeq{e.sym}
\<\phi|\psi\>^*=\<\psi|\phi\>
\eeq
holds, where $\alpha^*$ denotes the complex conjugate of a number
$\alpha\in\Cz$. The inner product defines a Euclidean norm
$\|\psi\|:=\sqrt{\<\psi|\psi\>}$, and the Hilbert space is the
closure $\ol\Hz$ of $\Hz$ in the topology induced by this norm.
We refer to the elements of $\Hz$ as {\bfi{smooth vectors}} since
they correspond in the important special case $\Hz=C^\infty(\Rz)$
to arbitrarily often differentiable functions.

Every smooth vector $\psi$ defines a continuous linear functional,
denoted by $\psi^*$, which maps $\phi\in\Hz$ to the complex number
\lbeq{e.dual}
\psi^*(\phi):=\<\psi|\phi\>.
\eeq
Dirac's idea was to turn this formula into a more suggestive
form by splitting the bracket $\<\psi,\phi\>$ into a
{\bfi{bra}} $\<\psi|$, standing for $\psi^*$, and a {\bfi{ket}}
$|\phi\>$, standing for $\phi$, and deleting the now superfluous
parentheses. Then the formula becomes
\lbeq{e.bra-ket}
\<\psi|\,|\phi\>:=\<\psi|\phi\>,
\eeq
which just asks us to replace two adjacent vertical bars by a single
one.

If $\Hz$ is itself a Hilbert space (and in particular, if the
dimension of $\Hz$ is finite) then it is not difficult to see
that all continuous linear functionals arise in this way.
However, in many interesting infinite-dimensional vector spaces $\Hz$,
the situation is different.
For example, if $\Hz=C^\infty(\Rz)$ and $z\in\Rz$ then the
mapping $\delta_z$ which maps $\psi\in\Hz$ to
\[
\delta_z(\psi):=\psi(z)
\]
is a continuous linear functional which cannot be obtained as
$\psi^*$ for some smooth vector $\psi$.

We can accommodate this in the bra-ket calculus by allowing
as bras {\em all} continuous linear functionals rather than only those
which have the form $\psi^*$ with $\psi\in\Hz$. We simply need to
label the continuous linear functional as bras $\<\psi|$ with symbols
$\psi$ from a set $\Hz^*$ such that the functionals of the form
$\psi^*$ with $\psi\in\Hz$ get the label $\psi$. The set $\Hz^*$ can
be made canonically into a vector space containing $\Hz$ as a subspace
by requiring the mapping $*:\psi\to \psi^*:=\<\psi|$ to be antilinear.
Then $\Hz^*=\Hz$ in case $\Hz$ is a Hilbert space, but in general
$\Hz$ may be a proper subspace of $\Hz^*$. Since the inner product
extends continuously from $\Hz$ to the Hilbert space completion
$\ol\Hz$, every element of $\ol\Hz$ defines a continuous linear
function. Thus, in general, the Hilbert space $\ol\Hz$ sits somewhere
in between $\Hz$ and $\Hz^*$,
\lbeq{e.triple}
\Hz \subseteq \ol\Hz \subseteq \Hz^*.
\eeq
Frequently, some extra ''nuclear'' structure on $\Hz$ is assumed
which turns \gzit{e.triple} into a so-called \bfi{Gelfand triple}
or \bfi{rigged Hilbert space}
(see, e.g., \sca{Maurin} \cite{Mau}, \sca{Bohm \& Gadella} \cite{BohG}, 
and for applications to resonances \sca{Kukulin} et al. 
\cite{KukKH}); however, on the 
level of our discussion, we don't need this extra structure.

If $\Hz=C^\infty(\Rz)$, physicists call the vectors $\psi\in\Hz^*$
\bfi{wave functions}\index{wave function} -- well being aware that 
they are not always functions in the standard sense --,
and write them with a dummy argument $x$ as
$\psi(x)$. For example, they consider $\delta_z$
to be a \bfi{shifted delta function}\index{delta function}, and write 
it as $\delta(x-z)$.

A wave function which is in the Hilbert space $\ol\Hz\subseteq \Hz^*$
is called {\bfi{normalizable}}, the remaining wave functions
are called {\bfi{non-normalizable}}. In mathematical terms, the
normalizable wave functions are equivalence classes of square integrable
functions, with two functions being regarded as equivalent when they
differ only on a set of measure zero. The shifted delta functions are
examples of non-normalizable wave functions.

For a general Euclidean space $\Hz$, we refer to the elements of
$\Hz^*$ as {\bfi{rough vectors}} since they correspond in the 
special case $\Hz=C^\infty(\Rz)$ to functions that are less smooth, 
possibly not even continuous, and possibly (as in case of the 
$\delta_z$) not functions at all.

Having extended the bra-ket notation to allow rough vectors as labels
in bras, the symmetry property \gzit{e.sym} is lost. To restore that,
we simply extend the inner product to enforce the validity of
\gzit{e.sym} by defining $\<\psi|\phi\>:=\<\phi|\psi\>^*$ if
$\phi\in\Hz^*$ and $\psi\in\Hz$. This can be done consistently,
and implies that now kets can be labeled by rough vectors, too.
But now the formula \gzit{e.bra-ket} makes trouble. What is
$\<\psi|\phi\>$ when both $\phi$ and $\psi$ are rough vectors?
In general, there is no solution; this product cannot be always
defined. However, one can consistently define it in certain cases,
namely when $\phi$ is in some subspace $\tilde\Hz$ of $\Hz^*$ and
the linear functional $\psi^*$ defined at first only on $\Hz$
can be extended to $\tilde\Hz$ by some limiting procedure.
We won't list here the various possibilities; our usage of bras
and kets will be restricted to cases where at least one of the
two labels in an inner product is smooth.

The main use of Dirac's notation is for the specification of
vectors and matrices in a particular representation of the
algebra of quantities. We first review the notation in the case
where a countable orthonormal basis of smooth states is available.
In this case there is a countable set $K$ of {\bfi{labels}}
such that the basis consists of the kets $|k\>$ with $k\in K$,
and orthogonality implies that
\[
\<j|k\> = \delta_{jk}\,,
\]
and the {\bfi{resolution of unity}}
\[
\sum_k |k\>\<k| = 1\,.
\]
In the finite-dimensional case, there is a close correspondence to
the notation of linear algebra if we take $\<k|$ to be the $k$th unit
row vector with a 1 in position $k$ and zeros elsewhere, and
$|k\>$ to be its transpose, the $k$th unit column vector.
\[
x = \sum_k x_k|k\>
\mbox{~~~ represents the vector $x=(x_k)$},
\]
and
\[
\<k|x = x_k
\]
gives the components of $x$.
\[
A = \sum_{jk} |j\>A_{jk}\<k|
\mbox{~~~ represents the matrix $A=(A_{jk})$},
\]
\[
\<j|A = \sum_{k} A_{jk}\<k|
\mbox{~~~ represents $A_{j:}$, the $j$th row of $A$},
\]
\[
A|k\> = \sum_{j} |j\>A_{jk}
\mbox{~~~ represents $A_{:k}$, the $k$th column of $A$},
\]
and
\[
\<j|A|k\>=A_{jk}
\]
gives the matrix entries of $A$. Compared to the standard linear
algebra notation there is no gain.

The situation is different when $K$ is a structured set, for example
a set of pairs $(k,s)$ where $k$ is a momentum label and $s$ a spin
label, or other such sets arising naturally in the dynamical symmetry
approach of Section \ref{s.chains}. Then the index notation becomes
somewhat cumbersome to comprehend, and the more lengthy bra-ket notation
is superior.

\at{continuous ''basis'' case missing}

\section{Boson Fock space}\label{s.bfock}

As we have seen in Section \ref{section-reps-heis}, every nice unitary
representation of the oscillator algebra contains a ground state
$\widehat\psi$ of norm 1, and hence the representation contains the
vectors $(a^*)^k \widehat\psi$ ($k=0,1,2,\dots$). Their span defines a 
Euclidean vector space \idx{$\Fz_+$} whose closure $\ol\Fz_+$
is a Hilbert space, called the {\bfi{single mode bosonic Fock 
space}}, or simply Fock space. Clearly, $\Fz_+$ is closed
under the action of $\Lz$, hence we have a unitary representation of
$\Lz$ on $\Fz_+$.
It is not difficult to see that different choices of the ground
state either define the same Fock space (if the ground states
differ only by a phase) or orthogonal Fock spaces. 
Indeed, if $\Fz\subset \Fz_+$
is an invariant submodule, it needs to have a vector $\tilde\psi_0$,
which necessarily coincides with the ground state of $\Fz_+$ up to a
complex number. Thus an arbitrary unitary representation is a direct
sum of Fock spaces. Thus the representations on a Fock space
are {\em irreducible} representations. We shall show in
a moment that the unitary representation on a Fock space is
essentially unique. This is the content of the celebrated {\bf
\idx{Stone--Von Neumann theorem}}, which actually is about the
representation of the Heisenberg group.

Bosonic Fock spaces with more degrees of freedom are obtained by taking
tensor products of the Fock space with one degree of freedom, and
describe systems of quantum oscillators. As we shall see in 
Chapter \ref{c.spin}, there is also a fermionic counterpart of Fock 
spaces, which are related to so-called
Clifford algebras. The single mode case describes a so-called
{\bfi{qubit}} and is simply the vector space $\Cz^2$; the general
case is a tensor product of these, and describes systems of qubits.

We now study the structure of $\Fz_+$ for a given ground state
$\widehat\psi$ of norm 1 in more detail. The properties found will
lead to a construction of a Hilbert space which actually contains a
representation of the Heisenberg algebra (which, so far, we simply
had assumed).

\begin{prop}\label{prop-5.5}
The vectors
\lbeq{e.kket}
|k\> := \frac{1}{k!} (a^{*})^{k} \widehat \psi~~~(k=0,1,2,\dots)
\eeq
satisfy the relations
\[
a^*|k-1\> = k |k\> \nn,~~~ a|k\> = \hbar |k-1\> \nn,~~~
n|k\> = k |k\>,
\]
\[
\<k|k'\> = \frac{\hbar^k}{k!}\delta_{kk'}.
\]
\end{prop}
\bepf
The first relation is just definition. For the second observe that
$a\widehat\psi = 0$ and $[a,(a^{*})^k]=k(a^{*})^{k-1}$. For the third,
just combine $n=\frac{1}{\hbar}a^*a$ and the first and second
relation. For the fourth relation we have $\<k| =
\frac{1}{k!}(\widehat\psi)^* a^k$ and $\<k|k'\>=0$ if $k\neq
k'$ since eigenvectors of a Hermitian operator corresponding to
different eigenvalues are orthogonal.
So only the normalization needs to be checked:
\[
\< k| k\> = \frac{1}{k^2 }\<k-1|a a^*|k-1\> = \frac{1}{k^2}\<k-1 |
\hbar +\hbar n |k-1\> = \frac{\hbar}{k}\<k-1|k-1\>\, .
\]
Using induction the fourth equality follows.
\epf

In the Fock space $\Fz_+$, the vectors are by definition the linear
combinations
\[
\psi = \sum_{k=0}^\infty \psi_k |k\>\,,~~~\psi_k\in \Cz\,.
\]
Proposition \ref{prop-5.5} gives us the relations \lbeq{fock-op}
(a\psi)_k = \hbar \psi_{k+1}\,,~~~(a^*\psi)_k = k\psi_{k-1}\,,~~~
(n\psi)_k = k\psi_k\,, \eeq and $ \varphi^* \psi = (\sum \varphi_k
|k\> )^*  \sum \psi_l |l\> = \sum \frac{\hbar^k}{k!}\bar\varphi_k
\psi_k$, hence \lbeq{fock-ip}
 \varphi^* \psi =  \sum \frac{\hbar^k}{k!}\bar\varphi_k \psi_k\,.
\eeq
Equations \gzit{fock-op} and \gzit{fock-ip} are an equivalent
description of the equations of Proposition \ref{prop-5.5}.

We now define $\Hz$ as the closure of $\Fz_+$. This makes $\Fz_+$
a dense subspace of $\Hz$, and we say that the operators
$a,a^*$ and $n$ are \bfi{densely defined} in $\Hz$ (meaning that they 
are defined on a dense subspace). Previously we have
seen that if the canonical commutation relations admit an
irreducible representation, then it has to be of the form as
described by Proposition \ref{prop-5.5}. But now we can say more:

The set $\Hz$ of vectors $\psi\in\Cz^\infty$ with finite norm
\[
\|\psi\|:=\sqrt{\sum_{k=0}^\infty \frac{\hbar^k}{k!}|\psi_k|^2}
\]
is a Hilbert space with inner product \gzit{fock-ip}, on which
the definitions \ref{fock-op} give densely defined operators
$a,a^*,n$. The components of $a^* \psi$ grow significantly
faster than those of $\psi$, so that $a^* \psi\in \Hz$ only for
$\psi$ in a proper subspace of $\Hz$. This subspace is dense,
since it contains the dense subset of $\psi$ with only finitely many
nonzero entries. Note that the operators $1$, $a$, $a^*$ and $n$,
and hence all elements of $\Lz$ are represented by infinite tridiagonal 
matrices, where only matrix elements in which the indices differ by at
most one are nonzero. This is the representation of the
quantum harmonic oscillator discovered by Heisenberg in his
groundbreaking paper \cite{Hei}.

It is now easy to check that the operators
$1$, $a$, $a^*$ and $n$ satisfy the canonical commutation relations,
that $a^*$ is the Hermitian conjugate of $a$, and that $n=a^*a$.
Thus we have a representation of $\Lz$. The representation is
irreducible since acting repeatedly with $a^*\in\Lz$ on the vector
$\widehat\psi$ with entries $\widehat\psi_k=\delta_{k0}$ (the ground
state) gives a basis of $\Hz$. Combining this with the uniqueness
statement obtained before, we arrive at the
following theorem of Stone and Von Neumann (but essentially already
obtained in \cite{Hei}): \at{repeat why unique}

\begin{thm}
The canonical commutation relations admit an (up to equivalence)
unique irreducible unitary representation on a Hilbert space such that
the action of $a$, $a^*$ and $n=a^*a$ is defined on a dense subspace
and $n$ is self-adjoint.
\end{thm}

The theorem holds with a similar proof for arbitrary finite-dimensional
Heisenberg algebras coming from a nondegenerate alternating form.
It fails spectacularly in infinite dimensions. In this case there are
uncountably many inequivalent representations; see, e.g., 
\sca{Barton} \cite{Bar} for an (in spite of the title of the book) 
elementary discussion of these. Their existence
is one of the main stumbling blocks for extending quantum mechanics
to quantum field theory.

\section{Bargmann--Fock representation}\label{s.bargmann}

We present an important but easy representation of the Heisenberg
algebra $h(n)$, which will be useful to us when we study coherent
states in Section \ref{sec-coh-state}. 
\at{where is $os(n)$ defined? Not needed here, but it should be 
defined somewhere, for completeness.}
Consider the vector space of
complex polynomials in $n$ variables $\Cz[z_1,\ldots,z_n]$. We then
identify $a_k$ and $a^{*}_{k}$  with the operators defined
by\footnote{Remember the transformations
$a_k=\frac{p_k-iq_k}{\sqrt{2}}$ and
$a_{k}^{*}=\frac{p_k+iq_k}{\sqrt{2}}$.}
\[
(a_k p)(z_1,\ldots,z_n):= z_kp(z_1,\ldots,z_n)\,,
\]
and
\[
(a^{*}_{k}p)(z_1,\ldots,z_n):=
\frac{\partial}{\partial z_k}p(z_1,\ldots,z_n)\,.
\]
It is easy to check that this indeed defines a representation. We can
even make a unitary representation out of this. For that purpose we
consider the vector space $\Hz$ of all entire functions on $\Cz^n$ with
finite norm with respect to the inner product
\[
\< f|g\> = \int_{\Cz^n} \overline{f(z)}g(z)e^{-\bar z\cdot
  z}\,.
\]
The space $\Hz$ with the above inner product is a Euclidean space; its
closure is a Hilbert space, the multi-dimensional version of the
Bargmann--Fock space described in Section \ref{s.bfock}.
\at{check name. It should be the Bargmann-Fock representation only.
The space is called Fock space almost universally}

The operators $a_k$ and $a^{*}_{k}$ are adjoints of each
other. An orthogonal
basis is given by the monomials:
\[
\< z_{1}^{k_1}\ldots  z_{n}^{k_n}|z_{1}^{l_1}\ldots
z_{n}^{l_n}\> = \prod_{i=1}^{n}k_i!\delta_{k_i,l_i}\,.
\]
From the discussion in Section \ref{sec-quad-rep} it follows that the
quadratic expressions
modulo the linear expressions in the elements $z_i$ and
$\frac{\partial}{\partial z_k}$ form the Lie algebra $sp(2n,\Cz)$.
Taking
all quadratic expressions (so not modding out by the linear
polynomials) in the elements $z_i$ and
$\frac{\partial}{\partial z_k}$ one obtains a central extension of
$isp(2n)$.

The above representation is irreducible (one sees rather quickly that
starting with $1$, acting with $p_k$ gives all entire functions) and
is called the Bargmann--Fock representation. By the Stone--Von Neumann
theorem, which says that there is only one irreducible representation
of the Heisenberg algebra, the Bargmann--Fock representation is up to
isomorphism the only irreducible representation of the Heisenberg
algebra.

\at{The rest of this section following below seems too terse for
a book of this kind.}

We have seen in Section \ref{sec-quad-rep}
that the quadratic expressions (modulo linear terms) in
the $q_i$ and the $p_k$ rotate the generators $q_i$ and $p_k$ into
each other under the action of the Lie product. In other words,
the action of the quadratic expressions
builds a representation of $sp(2n,\Cz)$ inside the Bargmann--Fock
representation. That this happens is not so strange. Let us consider
the automorphism group of the Heisenberg algebra, consisting of all
the invertible maps $h(n)\to h(n)$ preserving the Lie product. But
from equation \gzit{sympl.comm} we see that the automorphism group
contains the group $Sp(2n,\Cz)$. Now let us denote the above given
Bargmann--Fock representation by $U: h(n)\to \Lin(H)$, then using
$Sp(2n,\Cz)$ we get a new representation of the Heisenberg algebra
as follows. For each $g\in Sp(2n,\Cz)$ we consider the representation
\[
U_g: h(n)\to \Lin(H)\,,~~~ U_g(x) = U(gx)\,.
\]
Since $Sp(2n,\Cz)\subset \Aut(h(n))$ the $U_g$ are indeed
representations. But the unitary irreducible representation of $h(n)$
is unique, up to isomorphism, and hence there must be a unitary
operator $R(g)$ such that
\[
U_g = R(g)UR(g)^{-1}\,.
\]
It is clear that the $R(g)$ are
determined up to a sign. Thus $R(g)R(h)=\pm R(gh)$ and we say that the
$R(g)$ form a 
{\bfi{projective representation}}\index{representation!projective} 
of the group $Sp(2n,\Cz)$. 
This representation is called the {\bfi{metaplectic 
representation}}\index{representation!metaplectic}. 
The operators $R(g)$ themselves form a group,
closely related to the {\bfi{metaplectic group}} $Mp(2n,\Rz)$,
the universal covering group of the Lie
algebra $sp(2n,\Rz)$. The metaplectic group is a two-fold cover of
$Sp(2n,\Rz)$, hence has a center of order 2, while our group has
the multiplicative group of the reals as center. Factoring out the
positive reals leaves the metaplectic group.

\section{Coherent states for the harmonic oscillator}
\label{sec-coh-state}

Coherent states were introduced in 1963 by \sca{Glauber} \cite{glauber},
who recognized their importance in quantum optics; he received in 2005  
the Nobel prize for his work in this direction. But the notion
of a coherent state (without the name) was already
introduced by Erwin Schr\"odinger \cite{schroedinger} in 1926 when
he was looking for solutions to the Schr\"odinger equation that
satisfy the {\bfi{Heisenberg uncertainty relation}} 
\lbeq{e.qmunc}
\Delta p\Delta q \geq \frac{\hbar}{2}, 
\eeq
\at{derive this in Chapter 5} where $\Delta x$ denotes the
variance of a quantity $x$. Schr\"odinger was looking for states
that were as classical as possible, having equality $\Delta p\Delta q =
\frac{\hbar}{2}$. The coherent states, and only these satisfy 
equality; they therefore build a connection between classical physics
and quantum physics that grew stronger as the notion of coherent states 
was extended to more general situations. 

\at{remark that the Schr\"odinger equation is easy to solve either 
in a spectral representation, or in a coherent state representation.
Also prove the above minimal uncertainty property!}

To introduce Glauber's coherent states, we remind the reader 
that for the harmonic oscillator we constructed the Fock
space $\Hz$ of $\psi=(\psi_k)_{k\geq 0}$ satisfying
\lbeq{7.norm}
\sum_{k=0}^{\infty} \frac{\hbar^k}{k!} \ol \psi_k \psi_k <
\infty\,.
\eeq
One may regard $\psi$ either as a vector with infinitely many
components, or as an infinite sequence.
Equivalently, in Dirac's bra-ket notation, the $\psi_k$'s are the
complex coefficients in the expansion of $\psi$ with respect to an
eigenbasis $|k\>$ of the number operator,
$\psi = \sum \psi_k|k\>$.
The inner product is given by
\[
 \varphi^* \psi = \sum_{k=0}^{\infty} \frac{\hbar^k}{k!} \ol
\varphi_k \psi_k\,.
\]
The operators $a$, $a^*$ and $n$ act as $(a\psi)_k = \hbar
\psi_{k+1}$, $(a^* \psi)_k = k \psi_{k-1}$ and $(n\psi)_k =
k\psi_k$. We now define a {\bfi{coherent state}} for the harmonic
oscillator to be a vector of the form
\[
|\lambda,z\> := (\ol \lambda, \ol \lambda \ol z, \ol \lambda
\ol z^2,\ldots)\,;  ~~~(\lambda,z\in\Cz)
\]
in other words, a state $\psi$ with coefficients 
$\psi_k = \ol \lambda \ol z^k$. By \gzit{e.kket},
we can write
\[
|\lambda,z\> = \sum_{k=0}^{\infty} \ol \lambda \frac{\ol
z^k}{k!}(a^{*})^{k}
 \hat\psi\,,
\]
where $\hat\psi$ is the ground state. Even more, we have
\lbeq{exp.in-n}
|\lambda,z\> = \sum_{k\geq 0} \bar \lambda  \bar z^k |k\>\,,
\eeq
and we see that $\psi\in\Hz$ since
\[
\sum_{k\geq 0} \hbar^k  |\lambda|^2 \frac{|z|^{2k}}{k!} =|\lambda|^2
e^{\hbar |z|^2}<\infty\,.
\]
The inner
product between two coherent states is given by
\[
\< \lambda',z'|\lambda,z\> = \sum_{k=0}^{\infty} \frac{\hbar^k
z'^{k}\ol z^{k}}{k!}\lambda'\ol \lambda = \lambda'\ol \lambda
e^{\hbar z'\ol z}\,.
\]
It is easy to see that
\[
\< \lambda,z | n\> =\frac{\lambda \hbar^nz^n}{n!}\,.
\]
Suppose $\psi$ is an element of $\Hz$, then
\lbeq{7.coh.fun}  %
\<\lambda,z|\psi\> = \lambda \sum_{k=0}^{\infty}
\frac{(\hbar z)^k}{k!} \psi_k \equiv \lambda \psi(z)\,,%
\eeq%
which defines the function $\psi(z)$ corresponding to
$\psi$. Conversely, given an analytic function $g$
\[
g(z) = \sum_{k\geq 0} \frac{(\hbar z)^k}{k!}g_k\,,
\]
with $\sum_{k\geq 0}\frac{\hbar^2|g_k|^2}{k!}<\infty$ we assign to
$g$ the
element $\psi_g= (g_k)_{k\geq 0}$ in $\Hz$. We claim that
$\psi \mapsto \psi(z)$ is a map from $\Hz$ to the set of analytic
functions. In order to prove the claim we have to prove that the
power series \gzit{7.coh.fun} converges everywhere. We calculate
the radius of convergence $R$
\[
R = \limsup_{k\to \infty}
\frac{1}{\sqrt[k]{\frac{\hbar^k\psi_k}{k!}}} = \limsup_{k\to
\infty}\sqrt[k]{\frac{k!}{\hbar^k \psi_k}} \to \infty
\]
since $\psi_k$ satisfies \gzit{7.norm}. Hence the function
$\psi(z)$ is analytic everywhere. The state $\psi$ is uniquely
described by the function $\psi(z)$ in the sense that
$\psi(z) = 0 \Leftrightarrow \psi =0$, since
\[
\frac{1}{\lambda \hbar^k}\frac{d^k}{dz^k} \psi(z)\Big|_{z=0} =
\psi_k\,.
\]
The inner product between $\varphi$ and $\psi$ now becomes
\[
\varphi^* \psi = \sum_{k=0}^{\infty}
\frac{1}{k!|\lambda|^2\hbar^k} \Big[\frac{d^k}{dz^k}
\ol\varphi(z) \frac{d^k}{dz^k} \psi(z)\Big]_{z=0}\,.
\]
So we can use the powerful theorems of complex analysis to deal
with the states in the Hilbert space $\Hz$. For the relations between
complex analysis and coherent states, including important
generalizations to coherent states associated with other Lie groups,
see \sca{Perelomov} \cite{perelomov}, \sca{Upmeier} \cite{upmeier},
\sca{Faraut \& Koranyi} \cite{faraut}.

Every element in $\Hz$ is a linear combination
of coherent states, but the combination is in general not unique.
For the harmonic oscillator a set of finitely many coherent states
$|\lambda,z\>$ with different $z$ is linearly independent, since
suppose
\[
|v\>:=\sum_{i=1}^{n} |\lambda_i,z_i \> = 0\,, ~~~ \lambda_i\neq
0\,,
\]
then it follows that
\[
\<\mu,w |v\> = \sum_{i=1}^{\infty} \mu\lambda_{i}e^{\hbar
wz_i}=0\,,
\]
for all $w$. But a finite set of exponential functions is
linearly independent. Hence it follows that a finite set of
coherent states $|\lambda,z\>$ with different $z$ is linearly
independent. The set of linear combinations of finitely many
coherent states is dense in $\Hz$. The coherent states form a
kind of a ``basis", but an overcomplete set. Such a set is called a
{\bfi{frame}}. Frames are widely used in wavelet analysis.

We now show that coherent states of unit norm have a basis-like 
property, expressed through a so-called resolution of the identity. 
To simplify the notation we put $\hbar=1$. Next we define
\[
| z \> = | e^{-\shalf |z|^2}, z\> \,.
\]
The vectors $|z\>$ have a unit norm; $\< z|z\>=1$.
We calculate for an element $|f\> = \sum_k f_k|k\>$ the following
\[
|z\> \< z | f\> = \sum_{k,n} e^{-|z|^2} \bar z^n z^k
\frac{f_k}{k!}|n\>\,.
\]
The coefficient of each component $|n\>$ equals
\[
\varphi_n=e^{-|z|^2}f(z)\bar z^n\,,
\]
where $f(z)$ is defined as
\[
f(z)=\sum_n \frac{z^n}{n!}f_n.
\]
From the above discussion we know that $f(z)$ is analytic
everywhere. Thus the vector $\sum_n\varphi|n\>$ is of finite norm;
\[
\sum_n \frac{|\varphi_n|^2}{n!}=\sum_n
e^{-2|z|^2}|f(z)|^2\frac{|z|^{2n}}{n!}=e^{-|z|^2}|f(z)|^2<\infty\,.
\]
Hence $ |z\> \< z | f\>$ represents an element in $\Hz$
and we can integrate each component to get
\lbeq{res.id1}
\frac{1}{\pi}\int_{\Cz} |z\> \< z | f\> d^2 z =\frac{1}{\pi}\int_{\Cz}
|z\> f(z) e^{-\shalf |z|^2}dz = |f\>\,,
\eeq
where the integration measure is $dz= d({\rm Re} z)d({\rm Im} z)$
and where we used
\lbeq{nice.prop}
\int_\Cz \bar z^n z^m e^{-|z|^2} dz  =\pi n!\delta_{n,m}\,.
\eeq
In physics literature one writes the result \gzit{res.id1} as
\[
\frac{1}{\pi}\int |z\> \< z | dz =1\,.
\]
In mathematics, such an expression is called a \bfi{resolution of the
identity}\index{resolution of unity}\index{resolution of the identity}. 
The fact that the coherent states admit a resolution of
the identity makes them useful. We now wish to show that
the expansion of $|f\>$ in coherent states is unique, thereby proving
that the coherent states make up a tight frame. We use \gzit{res.id1}
to compute the inner product of $|f\>$ with a coherent state $\< w|$
\[
\< w| f\> = \frac{1}{\pi}e^{-\shalf |w|^2}\int_\Cz e^{w\bar z - |z|^2}
f(z) dz\,.
\]
But $f$ is an analytic function, so we first try $f=z^n$. Using
\gzit{nice.prop} we obtain the identity
\[
\frac{1}{\pi}\int_\Cz e^{-|z|^2}e^{w\bar z}z^m dz = w^m\,.
\]
Hence we derive the more general identity for analytic functions
\[
\frac{1}{\pi}\int_\Cz e^{-|z|^2}e^{w\bar z}f(z) dz = f(w)\,,
\]
from which we obtain
\[
f(w)=e^{\shalf |w|^2}\< w| f\> \,.
\]
Hence the expansion of $f$ in coherent states is unique, since if
$f=0$, then all the $\< z | f\>=0$ and the expansion vanishes
identically. Note that the above discussion only works for analytic
functions $f$. If we admit a non-analytic $f$ we get for example
\[
\int_\Cz |z\> \bar z^n e^{-\shalf |z|^2} dz =0\,,
\]
for all $n>0$. There is a relation between coherent states and the
Hilbert space $\Hz_\Cz$ of analytic functions $f:\Cz\to \Cz$ such that
\[
\int_\Cz |f(z)|^2 e^{-|z|^2}dz<\infty\,,
\]
for which the $(z^n)_{n\geq 0}$ form a basis. We refer the interested
reader to \sca{Glauber} \cite{glauber}, 
\sca{Segal} \cite{segal}, \sca{Bargmann} \cite{bargmann1,bargmann2}. 
We just remark
that if $f\in\Hz_\Cz$ and expand $f$ as
\[
f(z)=\sum_{n\geq 0} \frac{f_nz^n}{n!}
\]
then
\beqar
\int_\Cz |f(z)|^2 e^{-|z|^2} dz &=& \int_\Cz \sum_{n,m\geq 0}\frac{\bar
  z^nz^m \bar f_n f_m}{n!m!}e^{-|z|^2}dz\nn\\&=& \pi \sum_{n\geq
  0}\frac{|f_n|^2}{n!}\,.\nn
\eeqar
Hence $f$ defines an element in the Fock space $\Hz$. (We have assumed
one can change the order of integration and summation, but that can be
made rigorous, see \sca{Bargmann} \cite{bargmann1} for a readable 
explanation.)
The above
discussion on the uniqueness of the expansion of an analytic function
in terms of coherent states was taken from \sca{Glauber} \cite{glauber},
which is a
very readable account on coherent states and the physics and
mathematics behind them. 

We now reinsert the constant $\hbar$ to see some of the behavior of the
coherent states. Remember the formula for $q$
\[
q={2\omega m}^{-\shalf}(a+a^*)\,.
\]
We see easily that 
\lbeq{e.anneig}
a|\lambda,z\> = \hbar \ol z |\lambda,z\>.
\eeq
\at{use this eigenstate property as a definition, and derive the
other formula and the minimum uncertainty property from that!}

With a bit more work we see that
\[
\<\lambda,z|a^* |\lambda,z\> = \sum_{k,l} |\lambda|^2
\frac{z^i\ol z^j}{k!l!} \hat\psi^* a^k (a^{*})^{l+1}\hat\psi = z
|\lambda|^2 \sum_{k=0}^{\infty}\frac{|z|^{2k}}{k!}\,,
\]
from which it follows that
\[
\<\lambda,z|a^*|\lambda,z\> = z\<\lambda,z|\lambda,z\>\,.
\]
We thus see that we can associate the real part of $z$ with the
position;
\[
\< \lambda, z| q|\lambda,z \>
= {2\omega m}^{-\shalf}(z+z^*) \<\lambda, z|\lambda,z \>   \,.
\]
In a similar fashion the imaginary part of $z$ is related to the
momentum $p$. Let us pause for a while to see what the above
means. The harmonic oscillator has a very symmetric shape. One can
show that the wave functions which are eigenvectors of the number
operator $n$ respect the symmetry $q\to -q$ in
the sense that if $q\to -q$ then they change with a factor $(-1)^n$ (see
any introductory book on quantum mechanics,
e.g., \sca{Griffiths} \cite{griffiths}). 
This means that for all wave functions that are
eigenfunctions of the number operator the
average position is precisely in the middle, at $q_0$. Therefore the
momentum has zero expectation value. The coherent states represent
shifted states; their position is not in the middle. Note that we have
used the Heisenberg picture where the states are time-independent.
To see the time-dependent behavior of the coherent states, we consider
the product
\[
e^{-iHt}|\lambda,z\>\,.
\]
If we now shift the lowest energy to zero (that is, we choose
$V_0=\shalf \hbar \omega $), and use \gzit{exp.in-n} and
\[
e^{-i\frac{H}{\hbar} t}|n\>=e^{-i\omega n t}|n\>\,,
\]
we see
\[
e^{-i\frac{H}{\hbar}t}|\lambda,z\>= |\lambda,ze^{-i \omega t}\>\,.
\]
The coherent states thus swing from left to right in the potential
with a frequency $\omega$ and with amplitude $|z|/2m\omega$.
\at{mention explicitly anywhere that coherent states evolve
into coherent states, and the
corollary that the minimum-uncertainty property is preserved under time
evolution.}
In order to see the action of the Heisenberg group on the coherent
state, we calculate
\[
(e^{\alpha n}\psi )_k = e^{\alpha k} \psi_k~~\Rightarrow e^{\alpha
n} |\lambda,z\> = |\lambda,e^\alpha z\>\,.
\]
From $a|\lambda,z\> = \hbar \ol z |\lambda,z\>$ it follows that
\[
e^{\alpha a}|\lambda,z\> =e^{\hbar \alpha \ol z}|\lambda,z\> =
|e^{\hbar \alpha \ol z}\lambda,z\>\,.
\]
Further, we have
\[
e^{\ol \alpha a^*} |\lambda,z\> = \sum_{k,l} \frac{\ol \alpha^k
(a^*)^{k}}{k!} \ol \lambda \frac{\ol z^l (a^*)^{l}}{l!}\hat \psi
= \sum_k \ol\lambda \frac{\ol{\alpha+z}^k}{k!}(a^*)^{k}\hat\psi =
|\lambda,z+\alpha\>\,,
\]
and also we have
\[
e^\alpha |\lambda,z\> = |e^\alpha\lambda,z\>\,.
\]
We summarize this and write
\beqar\label{7.group}
e^{\alpha n } |\lambda,z\> &=& |\lambda,e^\alpha z\>\,, ~~~
e^{\alpha a}|\lambda,z\> = |e^{\alpha\hbar z}\lambda,z\> \\
e^{\ol \alpha a^*}|\lambda,z\> &=&
|\lambda,z+\alpha\>\,,~~~e^\alpha |\lambda,z\> =
|e^\alpha\lambda,z\>\,.%
\eeqar%
We can apply arbitrary group elements by taking products.

\bigskip
The {\bfi{Glauber coherent states}}\index{coherent state!Glauber} 
introduced in the present section
for the Heisenberg group, can be generalized. Indeed, the concept of
coherent states extends to a large class of
Lie groups acting on so-called co-adjoint orbits of the group.
In each case, the co-adjoint orbit provides a manifold of labels
for the coherent states on which the group acts, and the
coherent states span in an overcomplete fashion a Hilbert space
on which the group acts as an irreducible highest weight
representation. We shall discuss highest weight representations 
in Chapter \ref{c.highest}, but cannot give details for the general 
case mentioned here. Instead, we refer the reader to the book by
\sca{Perelomov} \cite{perelomov} and to the extensive survey by
\sca{Zhang} et al. \cite{ZhaFG}.

\section{Monochromatic beams and coherent states}

As indicated in Section \ref{s.gamma}
for the case of a beam of monochromatic light, the modes
of the electromagnetic field play the role of the annihilation and
creation operator of the quantum field. Classically the observables
are functions on phase space, hence specified to a certain observable
we have an operator on the configuration space. Namely a physical
configuration is specified by giving the values of the observables,
and to any observable we assign the operator that reads off the value
of that observable. Thus if a configuration of a laser beam, which we
suggestively denote $|\E\>$, is
specified by an electric field $\E(x,y,z,t)$, then the operator
$\mathcal{E}(x,y,z,t)$ reads off the values of the components of the
electric field at the space-time point $(x,y,z,t)$:
\[
\mathcal{E}(x,y,z,t)|\E\> = \E(x,y,z,t)|E\>\,.
\]

In the transition from classical mechanics to quantum mechanics the
role of the operator $\mathcal{E}(x,y,z,t)$ is played by the operator's
positive frequency part of the electromagnetic field. 
The above equation then
tells us that $|\E\>$ is an eigenvalue of the annihilation operator.

In a classical system there are many photons and the number of photons
need not be constant, due to absorption and due to the constant photon
production of the laser. Hence, from a micromechanical point of view,
the quantum number $n$ is no longer a good quantum number to assign to
a system resembling a laser. However we know that the electric field is
nearly perfectly constant, and if the beam goes in one direction we
can take the expression
\[
\E (x,y,z,t) =  a^*\e^{i\omega_k t }\mathbf{u}(\x)\,,
\]
for the electric field, where $a^*$ is an annihilator  operator.
\at{this is not clear...}
Since the classical state of the laser has a
well-defined value of the electric field, the quantum state
$|\E\>$ that
mimics the classical state the most is the one where
\[
 a^*\e^{i\omega_k t }\mathbf{u}(\x)|\E\>
=\E (x,y,z,t) |\E\> \,.
\]
But then $|\E\>$ is an eigenvector of $a^*$. This similarity
between coherent states and classical states is what motivated Roy
Glauber to investigate coherent states and apply his analysis to the
(quantum and semiclassical) theory of light.

All this extends with suitable modifications to the other wave
equations described in Sections \ref{s.beta} and \ref{s.alpha}.
In each case, there are families of coherent states describing
nearly classical ray-like behavior, and there are more exotic
quantum states which behave quite unlike any classical system.

\at{describe these coherent states?}

\chapter{Spin and fermions}\label{c.spin}

This chapter discusses the quantum mechnaics of spinning systems, 
where the only relevant degrees of freedom correspond to rotation. 

The quantum version of the classical rotator discussed in 
Section \ref{sec-clas-rig} can be obtained by looking for canonical 
anticommutation relations, which naturally produce the Lie algebra
of a spinning top. As for oscillators, the canonical 
anticommutation relations have a unique irreducible unitary 
representation, which corresponds to a spin $1/2$ representation of 
the rotation group. The multimode version gives rise to fermionic
Fock spaces; in contrast to the bosonic case, these are 
finite-dimensional when the number of modes is finite. In particular,
the single mode fermionic Fock space is 2-dimensional.

Many constructions for bosons and fermions only differ in the signs 
of certain terms, such as commutators versus anticommutators.
For example, quadratic expressions in bosonic or fermionic Fock spaces 
form Lie algebras, which give natural representations of the universal 
covering groups of the Lie algebras $so(n)$ in the fermionic case and
$sp(2n,\Rz)$ in the bosonic case, the so-called spin groups and 
metaplectic groups, respectively.
In fact, the analogies apart from sign lead to a common generalization 
of bosonic and fermionic objects in form of super Lie algebras, which,
however are outside the scope of the book. 

Apart from the Fock representation, the rotation group has a unique 
irreducible unitary representation of each finite dimension. We derive 
these spinor representations by restriction of corresponding nonunitary 
representations of the general linear group $GL(2,\Cz)$ on homogeneous 
polynomials in two variables, and find corresponding spin coherent 
states.

\section{Fermion Fock space}\label{s.qanti}

As we have seen in Section \ref{sec-clas-rig} the affine functions in 
the Poisson algebra of the
spinning top make up the Lie algebra $u(2)$. One can thus expect
that the quantization of the spinning top boils down to
representation theory of $su(2)$ and $u(2)$ and indeed it does. In the
following sections the representations of $su(2)$ and $u(2)$ play an
important role. See for example \sca{Humphreys} \cite{humphreys} or
\sca{Jacobsen} \cite{jacobsen} for a comparison of the methods used.
 
In this section, however, we look at a particular representation,
the Fock representation of $u(2)$. It behaves in many respects
like the Fock representation of the Heisenberg algebra, and gives 
the right generalization to the case of many fermionic modes, and in 
particular to quantum field theory. 

In fact, there are many analogies between bosonic and fermionic 
systems -- many formulas look alike, apart for the occurrence of 
additional minus signs in certain places.\footnote{
To see how this leads to the vast mathematical area of superalgebras
and supergeometry we refer the interested reader to for example {\sc
  Varadarajan} \cite{varadarajan}, \sca{Scheunert} \cite{scheunert},
\sca{Tuynman} \cite{tuynman}, \sca{Deligne} et al. \cite{delignemorgan}
and references.}
Although very similar in many respects, there is a fundamental 
difference with basic representation theory of bosons and fermions. 
While bosons are characterized by canonical commutation relations, 
fermions are quantized using canonical anticommutation relations. 
We shall see in a moment that this naturally reproduces the Lie algebra
$u(2)$ of a spinning top, and -- just like for canonical commutation 
relations -- uniquely fixes the representation.

We define a {\bfi{signed commutator}}\index{$[f,g]_\pm$}
\[
[f,g]_\pm := fg \mp gf\,;
\]
the upper sign applies to `bosonic' quantities $f,g$, and reproduces the
ordinary commutator, $[f,g]_+=[f,g]$,
while the lower sign applies to `fermionic' quantities $f,g$,
and reproduces the {\bfi{anticommutator}}
\[
[f,g]_-=fg+gf\,.
\]
Often, the anticommutator is written instead as $\{f,g\}$, which looks 
like a Poisson bracket, so that we don't recommend this notation. 
In the theory of Lie superalgebras, the sign at the commutator is
not written at all, since the context already determines the nature
of the arguments, and hence implies the commutator sign.

To understand how anticommutators give rise to the $u(2)$ Lie algebra 
governing a spinning top, we impose the {\bfi{canonical 
anticommutation relations}} on operators $a$ and $a^*=(a)^*$ in some 
Hilbert space
\[
[a , a^*]_- =\hbar\,,~~~
[a,a]_- =0\,,~~~[a^*,a^*]_- = 0\,.
\]
In particular we have $a^2=(a^{*})^{2}=0$. The algebra $\Ez$
spanned by $1,a$ and $a^*$ is four-dimensional since these generators 
together with $aa^*-a^*a$ already span $\Ez$. Hence $\Ez$ is
isomorphic to the algebra of complex $2\times 2$-matrices; an explicit 
isomorphism is obtained by identifying $a$ and $a^*$ with the matrices
\[
\pmatrix{0&1\cr 0 &0}\,,~~~{\rm and}~~~\pmatrix{0&0\cr 1&0}\,,
\]
respectively and $aa^*-a^*a$ with $\sigma_3$. The Lie $*$-algebra $\Lz$
described by $1$, $a$, $a^*$ and $[a,a^*]$ is thus $u(2)$.
Thus the anticommutation relation automatically produce the right 
Lie algebra for a spinning rigid body, and $u(2)$ is the fermionic 
analogue of the oscillator algebra $os(1)$.

We can get the same result more formally on the quantum level in a way 
which is completely analogous to the bosonic case, by considering an 
arbitrary unitary representations of the canonical anticommutation 
relations, i.e., for linear operators $a$ and $a^*$ satisfying these 
relations. We introduce the operator
\[
n := \hbar^{-1} a^*a\,,
\]
and we let $\psi\in V_\lambda$ for some \idx{Verma module} $V_\lambda$;
as in Section \ref{section-reps-heis}, this means 
$n\psi = \lambda\psi$. We obtain
\[
n a\psi = \hbar^{-1}a^*aa\psi = 0\,,
\]
and hence $a\psi \in V_0$. Further,
\[
na^*\psi = \hbar^{-1} a^*aa^* \psi =a^*\psi \,,
\]
and therefore $a^*\psi \in V_1$. To compute $\lambda$ we proceed
\[
\hbar^2n^2\psi = a^*aa^*a \psi = \hbar a^*a\psi = \hbar^2n\psi
\]
and hence $\lambda^2-\lambda= 0$, from which we deduce $\lambda=0,1$.
Thus we have arrived at the remarkable conclusion that the canonical
anticommutation relations lead to two-dimensional Hilbert spaces.

We take a basis vector $|0\>$ in $V_0$ and we define $|1\> =
a^*|0\>$. The Lie $*$-algebra $\Lz$ acts on the space spanned by $|0\>$ 
and $|1\>$, which is isomorphic to $\Cz^2$. This representation is 
called the {\bfi{Pauli representation}}. 
In the above we have thus shown 
that the Pauli representation is the unique irreducible unitary 
representation of the canonical anticommutation relations. This is the
fermionic analogue of the \idx{Stone--von Neumann theorem}.

In analogy with the boson case, the representation space
$\Cz^2$ is called the \bfi{single mode fermion Fock space}. 

We shall see in Section \ref{s.spinor} that the irreducible 
representations of $su(2)$ are in one-to-one correspondence with the 
(finite) dimension of the representation space. For historical reasons,
this dimension is usually denoted by $2j+1$, and $j$ is called the 
{\bfi{spin}} of the representation. Clearly, the spin $j$ is half 
 a nonnegative integer. In particular, the single mode fermionic Fock 
space has dimension 2 and hence spin $j=1/2$. 

In general, cf. Section \ref{s.poincare}, elementary particles are 
associated with an irreducible representation of the Poincar\'e 
algebra (or in the nonrelativistic limit the Galileo algebra),
which is characterized by mass and spin. The spin assigment in these
representations is such that, in the massive case, the restriction to 
a center of mass frame at a fixed time gives an irreducible 
representation of the Lie algebra $so(3)=su(2)$ of the same spin.
(The massless case is not related to $u(2)$.) 

Elementary particles of integral spin (bosons) are represented by a 
bosonic Fock space, those of nonintegral spin (fermions) by a 
fermionic Fock space. This fact is a consequence of the so-called
{\bfi{spin-statistics theorem}} which holds under certain
causality assumptions related to Poincar\'e invariance of a field 
theory. Fermionic particles obey the Pauli exclusion principle 
(\sca{Pauli} \cite{pauli-exclusion}, \sca{Schwinger}
\cite{schwinger1958}, \sca{Streater} \cite{streater}).

\section{Extension to many degrees of freedom}

Suppose that the algebra of linear operators on some vector space
$\Hz$ contains, for some linearly ordered set\footnote{A set $M$ is a
  linearly ordered if there is a binary relation $\leq$ such that for
  all $m,n,p\in M$:
  $(1)$ $m\leq m $, $(2)$ $m\leq n$ and $n\leq m$ then
  $m=n$, $(3)$ $m\leq n$ and $n\leq p$ then $ m\leq p$, $(4)$ either
  $m\leq n$ or $n\leq m$.} $M$ of labels,
quantities $a_j$ and $a_{j}^{*}$ ($j\in M)$ satisfying the relations
\lbeq{e.ccarn}
[a_j,a_k]_\pm = [a_j^*,a_k^*]_\pm = 0,~~~
[a_j,a^{*}_{k}]_\pm=\delta_{jk}\,,~~~(j,k\in M).
\eeq
For the upper sign (the bosonic case), these are just the canonical
commutation relations defining a Heisenberg algebra corresponding
to harmonic oscillators with finitely many degrees of freedom.
For the lower sign (the fermionic case), the relations \gzit{e.ccarn}
generalize the canonical anticommutation relations which we have met
for the spinning top; we thus expect to get an analogue of the
spinning top with $n$ degrees of freedom.

In this section, we consider the fermionic case. We first assume that 
we have a unitary faithful representation and deduce
enough properties that determine the representation uniquely.
Then we use the properties deduced to construct the representation.

The canonical anticommutation relations imply that
\lbeq{e.carn}
a_j a_k = - a_ka_j\,,~~~a_j^* a_k^* = - a_k^*a_j^*\,,~~~
a_j a_k^* = \delta_{jk} - a_k^*a_j\,,
\eeq
and again we have in particular $a_j^2= (a_j^*)^2=0$.
To find the unitary representations of physical interest, we assume -- 
in analogy to the bosonic case of the canonical commutation relations --
the existence of a nonzero vector $\psi_0$, the {\bf\idx{ground state}},
such that
\lbeq{e.vac}
a_j\psi_0 =0 \Forall j\in M\,.
\eeq
We next define for any finite set $J=\left\{j_1,\ldots,j_l\right\}$
of distinct labels $j_1<\ldots<j_l$ from $M$ the vectors
\lbeq{e.I}
|J\> :=|j_1 \ldots j_l\>:= a^{*}_{j_1}\cdots a^{*}_{j_l}\psi_0\,~~~
|\emptyset\>=\psi_0\,.
\eeq
Since we want a faithful representation, we may assume that $|J\> \neq
0$. Indeed suppose $|J\>=0$ for $J=\{j_1,\ldots,j_l\}$, then acting on
$|J\>$ with $a_{j_{l-1}}\cdots a_{j_1}$ we see that $a_{j_l}^{*}$ acts 
as $0$. Because of \gzit{e.carn}, we have
\lbeq{e.aI}
\protect{ 
a_{j}|J\> =
\cases{
\eps_j(J)|J\setminus \{j\}\>& if $j\in J$,\cr
0& if $j\not\in J$,
}~~~~~~
}
a_{j}^{*}|J\> =
\cases{
0& if $j\in J$,\cr
\eps_j(J)|J\cup \{j\}\>&  if $j\notin J$,
}
\eeq
where the sign $\eps_j(J)$ is defined to be $+1$ if there is an even
number of indices in $J$ that are smaller than $j$ and $-1$ otherwise.

\begin{prop}
We have the following identities:
\lbeq{e.eps1a}
\eps_j(J\setminus \{j\})=\eps_j(J) \for j\in J\,,
\eeq
\lbeq{e.eps1b}
\eps_j(J\cup\{j\})=\eps_j(J) \for j\notin J\,,
\eeq
\lbeq{e.eps2}
\eps_{j}(J)\eps_{k}(J\cup\{j\})
=-\eps_{k}(J)\eps_{j}(J\cup\{k\})
\for j,k\not\in J\,,~ j\ne k\,,
\eeq
\lbeq{e.eps3}
\eps_{j}(J)\eps_{k}(J\setminus\{j\})
=-\eps_{k}(J)\eps_{j}(J\setminus\{k\})
\for j,k \in J\,,~ j\neq k\,,
\eeq
\lbeq{e.eps4}
\eps_j(J)\eps_k(J\cup \{j\}) = - \eps_k(J)\eps_j(J\setminus \{k\})
\for j\notin J\,, k\in J\,.
\eeq
\end{prop}

\bepf
This is a straightforward consequence of the definition, taking into
account when $\eps_{j}(J)$ and $\eps_{k}(J)$ change sign if an index
is removed or added to $J$.
\epf

We define $\Fz_-(M)$ to be the vector space spanned by the $|J\>$. By
definition, $\Fz_-(M)$ consists of the finite linear sums of the
elements $|J\> $.

\begin{prop}~\\
(i) The vectors $|J\>$ are linearly independent.

(ii) The vector space $\Fz_-(M)$ is an irreducible representation 
space for the canonical anticommutation relations.
\end{prop}

\bepf
(i) Suppose that we have $\sum_J c_J |J\> =0$
with finitely many nonzero coefficients.
and let $J=\{j_1, \ldots ,j_l\}$ ($j_1<\ldots<j_l$) be a set of 
maximal size among the sets with $c_J\ne 0$. In view of \gzit{e.aI}, 
multiplication by
$a_{j_l}\cdots a_{j_1}$ leaves as only nonzero term
$\pm c_J|\emptyset\>=0$. Since the ground state $\psi_0$
is nonzero, we conclude that $c_J=0$, contradiction. Therefore, the
vectors \gzit{e.I} are linearly independent and form a basis of
$\Fz_-(M)$.

(ii) Equations \gzit{e.aI} imply that $a_j$ and
$a_j^*$ map $\Fz_-(M)$ into itself. Irreducibility of the
representations follows since the same argument used in (i) implies
that any invariant subspace of $\Fz_-(M)$ containing a nonzero
element contains the ground state, hence all $|J\>$, and hence all
elements of $\Fz_-(M)$.
\epf

Since the $|J\>$ form a basis of $\Fz_-(M)$, we may identify a vector
$\psi\in \Fz_-(M)$ with the {\bfi{fermion wave function}} $\psi$
defined on the finite subsets of $M$ whose value at $J\subseteq M$ is
the coefficient $ \psi(J)$ in the basis expansion
\[
\psi = \sum_{J\subseteq M} \psi(J)|J\>\,,
\]
where the summation is over all finite subsets $J$ of $M$. Note that
only finitely many coefficients $\psi(J)$ are nonzero.

\begin{prop}
Under the assumption \gzit{e.vac}, the anticommutation relations
\gzit{e.carn} imply that the linear operators
\[
a(u):=\sum_{j\in M} u_j a_j, ~~~a^*(u):=\sum_{j\in M} u_ja^{*}_{j},
\]
defined for all vectors $u$ indexed by $M$ which have only finitely
many nonzero entries, act on fermion wave functions according to
\lbeq{e.afermi1}
(a(u)\psi)(J)=\sum_{j\notin J}\eps_j(J)u_j\psi(J\cup\{ j\}) \,,~~~
(a^*(u)\psi)(J)=\sum_{j\in J} \eps_j(J)u_j\psi(J\setminus \{j\})\,.
\eeq
\end{prop}

\bepf
We have
\[
a_j\psi = \sum_J \psi(J) a_{j}|J\>
= \sum_{J\ni j}  \psi(J) \eps_j(J)|J\setminus \left\{j\right\}\>
= \sum_{J\not\ni j}  \psi(J\cup \left\{ j\right\}) \eps_j(J\cup
\left\{ j\right\})|J\>,
\]
and using \gzit{e.eps1b}, we find
\[
(a_j\psi)(J)=\cases{
0 & if $j\in J$,\cr
\eps_j(J)\psi(J\cup \{j\}) & if $j \not\in J$.
}
\]
Taking linear combinations proves the first assertion.
Similarly,
\[
a_{j}^{*}\psi = \sum \psi(J) a_{j}^{*}|J\>
= \sum_{J\not\ni j}  \psi(J) \eps_j(J)|J\cup \left\{ j\right\}\>
= \sum_{J\ni j} \psi(J\setminus \left\{j\right\})
\eps_j(J\setminus \left\{j\right\})|J\>\nn\,.
\]
Using \gzit{e.eps1a}, we find
\[
(a_{j}^{*}\psi)(J)=\cases{
\eps_j(J)\psi(J\setminus\{j\}) & if $j\in J$,\cr
0 & if $j\notin J$\,.
}
\]
Taking linear combinations proves the second assertion.
\epf

For a unitary representation, we need that $a_j$ is the Hermitian 
conjugate of $a_{j}^{*}$, which is the case if and only if the $|J\>$ 
are orthonormal. Indeed, suppose $J\neq J'$, then we may assume there is
$j\in J$ that is not in $J'$ (else turn the role of $J$ and $J'$
around). But then $\< J|J'\> = \< J\setminus\{j\}|a_j|J'\> =0$. Hence
we may assume that $\Fz_-(M)$ has the inner product
\lbeq{e.afermi2}
\phi^*\psi = \sum_J \overline{\phi(J)}\psi(J)\,.
\eeq
To show that unitary representations with the desired conjugation and
anticommutation relations actually exist, we start with the space
$\Fz_-(M)$ of complex valued functions $\psi$ defined on finite subsets
of an arbitrary set $M$ such that only finitely many values $\psi(J)$
are nonzero. Then \gzit{e.afermi2} defines an inner product
on $\Fz_-(M)$, and the completion $\ol{\Fz_-(M)}$ of $\Fz_-(M)$ in the
associated norm is a Hilbert space, called the
{\bfi{fermion Fock space}} over $M$.

For a concise formulation of the result, we use a slightly more
abstract notation. We introduce the Euclidean space $\Hz$ of vectors
indexed by $M$ with finite support, equipped with the bilinear form
\[
u^Tv:=\sum_{k\in M} u_k v_k\,,
\]
and write
$\Fz_-\Hz:=\Fz_-(M)$.
In applications to quantum field theory, $\Hz$ becomes the
infinite-dimensional single-particle Hilbert space, and the sums
become integrals over momentum vectors, but the formulas
below remain valid with an appropriate interpretation.

\begin{thm}
The relations \gzit{e.afermi1} define two linear
mappings $a,a^*$ from $\Hz$  to the algebra $\Lin(\Fz_-,\Hz)$,
and we have
\lbeq{e.a*}
(a(u))^*=a^*(\ol u)\,,
\eeq
\lbeq{e.caruv1}
a(u)a(v)+a(v)a(u)=0\,,~~~~
a^*(u)a^*(v)+a^*(v)a^*(u)=0\,,
\eeq
\lbeq{e.caruv2}
a(u)a^*(v)+a^*(v)a(u)=u^Tv.
\eeq
In particular, taking for $u,v$ vectors with a single nonzero entry,
we find the canonical anticommutation relations \gzit{e.carn}.
\end{thm}

\bepf
From the definitions \gzit{e.afermi1} and \gzit{e.afermi2}, we find
\[
\phi^*a^*(\ol u)\psi =\sum_J\ol{\phi(J)}\sum_{j\in J} \eps_{j}(J)\ol u_j
\psi(J\setminus \{j\})\,.
\]
Renaming the $J\setminus\left\{j\right\}$ to $J$, we get in view of
\gzit{e.eps1a} and \gzit{e.eps1b}
\[
\bary{lll}
\phi^*a^*(\ol u)\psi &=& \D\sum_J\sum_{j\notin J} \eps_{j}
(J\cup\left\{j\right\}) \ol u_j
\ol{\phi ( J \cup\left\{j\right\})} \psi(J)\\
&=& \D\sum_J\sum_{j\notin J} \eps_{j}(J) \ol u_j
\ol{\phi ( J \cup\left\{j\right\})} \psi(J)
= (a(u)\phi)^*\psi\,.
\eary
\]
This implies \gzit{e.a*}.
To prove \gzit{e.caruv1}, we note that in view of \gzit{e.eps2},
\[
\bary{lll}
(a(u)a(v)\psi)(J)
&=& \D\sum_{j\not \in J} \sum_{k\not\in J\cup\{j\}}
\eps_j(J)\eps_k(J\cup \{ j\})u_j v_k \psi(J\cup\{j,k\})\\
&=& \D-\sum_{k\not \in J} \sum_{j\not\in J\cup\{k\}}
\eps_k(J)\eps_j(J\cup \{ k\})u_j v_k \psi(J\cup\{j,k\})\\
&=& -(a(v)a(u)\psi)(J)\,.
\eary
\]
This proves the first formula in \gzit{e.caruv1}, and
the second formula follows with \gzit{e.a*}.
Finally, to prove \gzit{e.caruv2}, we note that
\lbeq{e.aa*}
(a(u)a^*(v)\psi)(J)
 = \sum_{j\not \in J} \sum_{k\in J\cup\{j\}}
\eps_j(J)\eps_k(J\cup \{j\})u_j v_k \psi(J\cup\{j\}\setminus\{k\})\,,
\eeq
and
\lbeq{e.a*a}
(a^*(v)a(u)\psi)(J)
=\sum_{k\in J} \sum_{j\notin J\setminus\{k\}}
\eps_k(J)\eps_j(J\setminus\{k\})u_jv_k\psi(J\setminus\{k\}\cup\{j\})\,.
\eeq
The sets of pairs $(j,k)$ over which the summation in the two
equations is taken, contain the pairs with $j\not\in J,k\in J$, for which
the signs are opposite by \gzit{e.eps4}. Thus the corresponding terms
in the sums cancel when the two equations are added. The remaining
terms consist in \gzit{e.aa*} of the terms with $k=j\notin J$ and in
\gzit{e.a*a} of the terms with $j=k\in J$; for the corresponding terms
in the sums, all signs are $+1$. Therefore, adding the two equations
results in
\[
(a(u)a^*(v)\psi)(J)+(a^*(v)a(u)\psi)(J)
=\sum_{j\not\in J} u_j v_j \psi(J) + \sum_{k\in J} u_k v_k \psi(J)
= u^Tv\psi(J)\,.
\]
\epf

\bigskip
\section{Exterior algebra representation}

We now show another important realization of the
anticommutation relations, equivalent to that of the preceding
section but phrased in a different language familiar from differential
geometry.

For any vector space $\Hz$ we consider the tensor algebra
$\bigoplus\Hz =\Cz\oplus \Hz \oplus (\Hz \otimes \Hz )\ldots$,
which is an associative
algebra with unity where multiplication is the tensor product. We
define the ideal $\Jz_\mp$ to be the ideal generated by the elements 
$v\otimes w \mp w\otimes v$ for $v,w\in \Hz $. 
The quotients\index{$\bigvee \Hz$}\index{$\bigwedge\Hz$}
\[
\Hz / \Jz_- = \bigvee \Hz\,, ~~~ \Hz/\Jz_+=\bigwedge \Hz\,,
\]
that we obtain by dividing out by the ideals $\Jz_\mp$ are equipped with
a natural algebra structure, since we divided out by ideals. We call
$\bigvee \Hz$ the {\bfi{symmetric algebra}} and $\bigwedge \Hz$
the {\bfi{exterior algebra}}. The product in the exterior algebra
is written as $v \wedge w$ in place of 
$vw$, and is then called the {\bfi{exterior product}}
or {\bfi{wedge product}}. The product $\wedge$
satisfies the anticommutative law
\[
v\wedge w = -w\wedge v
\]
for any two vectors $v,w\in\Hz$, but not for general elements of
$\bigwedge \Hz$; $u\wedge v\wedge w = w\wedge u \wedge v$.

The symmetric algebra leads to a representation of the canonical
commutation relations (see Section \ref{s.bargmann}); the exterior
algebra to one of the canonical anticommutation relations. 

We concentrate on the latter, and restrict to finite-dimensional
vector spaces. If $\Hz$ is a vector
space of finite dimension $n$ we may
choose a basis $e_1,\ldots,e_n$. Using the anticommutation
relations one easily verifies that the exterior algebra $\bigwedge
\Hz$ has a basis consisting of the elements
\[
1\,;~~~e_i\,;~~~e_i\wedge e_j\,~(i<j)\,;~~~e_i\wedge e_j\wedge
e_k\,~(i<j<k)\,;~~\ldots; ~~ e_1\wedge e_2\wedge\cdots\wedge e_n\,,
\]
making a total of $2^n$ basis vectors. Thus the dimension of
$\bigwedge \Hz $ is $2^n$. We now introduce operators $a_k$
given by
\[
a_k(\omega)= e_k \wedge\omega \for \omega\in\bigwedge
\Hz \,;
\]
in particular, $a_k(1)=e_k$.
Similarly we define operators $a_{k}^{*}$ as follows:
On elements of the form $\omega = e_k\wedge \omega'$ we put
\[
a_{k}^{*}(\omega) := a_{k}^{*}(e_k\wedge \omega')=\omega',
\]
but if we cannot write $\omega $ into the form $\omega=e_k\wedge
\omega'$ we put
\[
a_{k}^{*}(\omega):=0\,.
\]
We now show that we have
\lbeq{ferm.circ}
a_ka_{l}^{*}+ a_{l}^{*} a_k = \delta_{kl}\,.
\eeq
\at{check this calculation}
First we assume that we cannot write $\omega$ in the form
$e_k\wedge\omega'$ and neither in the form $e_l\wedge
\omega''$. In this case we have
\[
a_k\circ a_{l}^{*}(\omega)+ a_{l}^{*}\circ a_k(\omega)=
a_{l}^{*}(e_k\wedge\omega)=\delta_{kl}\omega\,.
\]
If we can write $\omega$ as $e_k\wedge\omega'$ but not as $e_l\wedge
\omega''$ then we have $k\neq l$ and
\[a_k\circ a_{l}^{*}(\omega)+ a_{l}^{*}\circ a_k(\omega)=
a_{l}^{*}\circ a_k(e_k\wedge\omega')=0\,.
\]
\at{Is the 2nd step above correct? I would have thought:
$
a_k(a_{l}^{*}(\omega)) + a_{l}^{*}(a_k(\omega))
= a_k(0) + a_{l}^{*}(e_k \wedge \omega')
= 0 + a_{l}^{*}(e_k \wedge \omega')
$
and the last term is also zero because $\omega'$ cannot be written
as $e_l\wedge\omega''$.}

If $k=l$ and $\omega=e_k\wedge\omega'$ we have
\[
a_k\circ a_{k}^{*}(\omega)+a_{k}^{*}\circ a_k(e_k\wedge\omega')
=a_k\circ a_{k}^{*}(e_k\wedge\omega')=\omega\,.
\]
In the last case, when $k\neq l$ but we can write $\omega= e_k\wedge
e_l\wedge \omega$ we have
\[
a_k\circ a_{l}^{*}(\omega)
= a_k\circ a_{l}^{*}(-e_l\wedge e_k\wedge\omega')=0
\]
and
\[
a_{l}^{*}\circ a_k (\omega)=a_{l}^{*}\circ a_k(e_k\wedge
e_l\wedge \omega)= 0\,.
\]
Putting it all together we indeed have \gzit{ferm.circ}.

Generalized to infinite dimensions, \gzit{ferm.circ} is the basis for 
the description of fermion fields in quantum field theory.

\section{Spin and metaplectic representation}\label{s.meta}

In analogy to the bosonic case treated in Section \ref{sec-quad-rep},
we now show that quadratic expressions in anticommuting operators $a$
and $a^*$ make up well-known finite-dimensional Lie algebras, in this
case the orthogonal algebras $so(2n)$ and $so(2n+1)$.

The method of derivation is different however.
It works for bosons and fermions simultaneously,
with differences only in certain signs, and gives in the bosonic case
a construction of the metaplectic representation of
$sp(2n,\Rz)$ and the central extension of $isp(2n,\Rz)$. In the sequel, 
the upper signs apply for the bosonic case, and the lower signs apply 
for the fermionic case. 
We use coordinate-independent notation, so that the 
method can be taken over almost literally to the infinite-dimensional
case. 

We assume that we have a linear mapping $a:\Hz\to \Ez$ that assigns
to each $\alpha$ from some vector space $\Hz$ an element $a(\alpha)$
in an associative algebra $\Ez$ with identity $1$ such that
\lbeq{eq.spin.meta.def}
[a(\alpha),a(\beta)]_\pm =
a(\alpha)a(\beta)\mp a(\beta)a(\alpha)\in \Cz\,.
\eeq
For example, with the standard generators $a_k,a_k^*$ in a bosonic or
fermionic Fock space, we can take
\[
a(\alpha)=\sum_k (\alpha_k a_k + \alpha_{-k}a_{k}^*).
\]
For the bosonic case, \gzit{eq.spin.meta.def} means that the Lie 
algebra that is obtained by equipping $\Ez$ with the commutator as 
Lie product, contains a central extension of a commutative algebra.

The ground state on $\Ez$ is a positive linear functional $\<\cdot\>$ 
that satisfies $\langle 1\rangle =1$. Linearity implies that there is
a linear operator $G$ satisfying
\[
\langle a(\alpha)a(\beta) \rangle = \alpha^T G \beta\,;
\]
\gzit{eq.spin.meta.def} then implies that 
\[
a(\alpha)a(\beta)\mp a(\beta)a(\alpha)=\alpha^T J \beta\,.
\]
where 
\[
J:=G\mp G^T,~~~J^T=\mp J\,.
\] 
In Section \ref{sec-quad-rep}, the bilinear form $\omega$ was
represented by an antisymmetric nondegenerate $2n\times 2n$-matrix.
This is the most interesting case, although in the first part of the 
discussion below $J$ can be degenerate.

Any quadratic expression in the $a_k$ and $a_{k}^*$ is a sum of
terms of the form $a(\alpha)a(\beta)$; this is nothing else than the
statement that any matrix is a sum of matrices
of the form $M=mn^T$ for some vectors $m$ and $n$. We define the
quadratic expression\footnote{In infinite 
dimensions, this amounts to a renormalization step that is 
conventionally described as ''subtracting infinite constants'' 
arising at a later stage of the development.
Our formulas are renormalized from the outset, and no infinite constants
arise.} 
\[
N(\alpha\beta^T):=
\frac{1}{2}\Big(a(\alpha)a(\beta) - \alpha^TG\beta\Big)\,.
\]
and extend them by linearity (using a basis it is easy to see that the
extension is unique and well-defined). We thus
have $\langle N(f)\rangle=0$ for all quadratic expressions $f$.
We also have
\[
a(\alpha)a(\beta)=2N(\alpha\beta^T) + \tr (G^T\alpha\beta^T)\,.
\]
Remember that in Section \ref{sec-quad-rep} we
considered the symmetric combination
$2E_{ij}=p_iq_j+q_jp_i$. Motivated by this we restrict our attention to
$f=\sum_i\alpha_i\beta^{T}_{i}$ such that $f^T=\pm f$. 
For a single term $f=\alpha\beta^T=\pm f^T$, we find
\beqar
2[N(f),a(\gamma)] &=& a(\alpha)a(\beta)a(\gamma) -
a(\gamma)a(\alpha)a(\beta)\nn\\
&=& \pm a(\alpha)a(\gamma)a(\beta)+a(\alpha)\beta^TJ\gamma \mp
a(\alpha)a(\gamma)a(\beta)-a(\beta)\gamma^T J\alpha\nn\\
&=&a(\alpha\beta^T J\gamma)-a(\beta\alpha^T J^T\gamma)\nn\\
&=& a(fJ\gamma-f^TJ^T\gamma)\,,\nn
\eeqar
so that by linearity,
\[
[N(f),a(\gamma)]=N(f)a(\gamma)- a(\gamma)N(f)=a(fJ\gamma)
\]
for $f=\pm f^T$. Similarly for $g=\gamma\delta^T=\pm g^T$ we find
\beqar
2[N(f),N(g)]&=&[N(f),a(\gamma)]a(\delta)+a(\gamma)[N(f),a(\delta)]\nn\\
&=&a(fJ\gamma)a(\delta)+a(\gamma)a(fJ\delta)\nn\\
&=&2N(fJ\gamma\delta^T)+\tr
(G^TfJ\gamma\delta^T)+2N(\gamma(fJ\delta)^T)+\tr
(G^T\gamma(fJ\delta)^T)\nn\\
&=& 2N(fJg-gJf)+\tr(G^T(fJg-gJf)),\nn
\eeqar
so that again by linearity,
\beqar
[N(f),N(g)]&=&N(fJg-gJf) + \frac{1}{2} \tr(G^T(fJg-gJf))\nn\\ 
&=& N(fJg-gJf)+\frac{1}{2}\tr((G\pm G^T)fJg)\,.\nn
\eeqar
Writing 
\[
(s,\alpha,f)=N(f)+a(\alpha)+s1,
\]
we find for the
bosonic case
\[
[(s,\alpha,f),(s',\alpha',f')]=\left(\shalf\tr((G+
G^T)fJf')+\alpha^TJ\alpha',fJ\alpha'-f'J\alpha,fJf'-f'Jf\right)
\]
and for the fermionic case
\[
[(s,\alpha,f),(s',\alpha',f')]=
\left(
S(f,f',\alpha,\alpha'),
fJ\alpha'-f'J\alpha,fJf'-f'Jf+2(\alpha\alpha'^T-\alpha'\alpha^T)
\right)\,,
\]
where 
\[
S(f,f',\alpha,\alpha')=\shalf\tr\Big((G-G^T)fJf'\Big)+\alpha^TJ\alpha'.
\]

\bfi{Fermionic case.}
To exploit these formulas, we first focus on the fermionic case, and 
assume that $J$ is nondegenerate; without loss of generality, we may
choose $J$ to be the $2n\times 2n$ identity matrix, $J=1$. 

We consider the Lie algebra $\Lz$
defined by the quadratic elements modulo the constant term; that is,
we factor out the center. We write $(\alpha,f)$ for the equivalence 
class of $(s,\alpha,f)$. 
The quadratic
expressions $f$ are antisymmetric, $f^T=-f$, and thus correspond
to $so(2n,\Cz)$. Let us consider the map $u:\Lz\to sl(2n+1,\Cz)$
defined by
\[
u:(\alpha,f)\mapsto \pmatrix{f&2^{-1/2}\alpha\cr
2^{-1/2}\alpha^T &0\cr}\,.
\]
The map $u$ is injective and preserves the Lie product and thus is an
isomorphism onto its image. The image under $u$ of $\Lz$ is the
Lie algebra $so(2n+1,\Cz)$. It can easily be seen that matrices
in the image satisfy
\[
\pmatrix{f&\alpha\cr \alpha^T &0\cr}^T\hat J + \hat
J\pmatrix{f&\alpha\cr \alpha^T &0\cr} =0\,,~~~\hat J =\pmatrix{J&0\cr
  0 &-1\cr}\,,
\]
since, for fermions, $J$ is the $2n\times 2n$ identity
matrix. Restricting this basis to the real numbers we obtain the
real form $so(2n,1)$. Summarizing, we have thus established that the
quadratic elements (with center) form a central extension of
$so(2n+1,\Cz)$. The purely quadratic expressions (no linear and
constant terms) form the Lie algebra $so(2n,\Cz)$. Note that the
group $O(2n,\Cz)$ is the automorphism group of the algebra
defined by the relation
\lbeq{eq.anticcr}
b_k b_{l}^*+b_{l}^*b_{k}=\delta_{kl}\,.
\eeq
Going to the `real' basis $c_k=b_k + b_{k}^{*}$,
$c_{k+n}=i(b_k-b_{k}^{*})$,
we see that the real Lie group $O(2n)$
preserves the relations \gzit{eq.anticcr}. 

For a finite number of generators, the canonical anticommutation 
relations have a unique faithful unitary representation. Therefore as 
in Section \ref{s.bargmann} we can say something interesting about the 
automorphism group of the algebra defined by \gzit{eq.anticcr}. 
Performing a rotation 
$b_k\mapsto b_k':=\sum g_{kl}b_l$ with $g=(g_{kl})$ an element of 
$SO(2n)$ on the generators $b_k$ we get another representation of the 
canonical anticommutation relation, but since this representation is 
unique, there exists a unitary transformation $U(g)$ that relates the 
obtained representation with the original representation: 
$ b_k' = U(g)b_k U(g)^{-1}$, where we simply wrote $b_k$ for the 
representation of $b_k$. Again, $U(g)$ is not unique
for a given $g$, since $-U(g)$ also does the job. In this way we get a
double cover of the group $SO(2n)$, called the {\bf\idx{spin group}}
\idx{$Spin(2n)$}, just as in Section \ref{s.bargmann} we obtained the
metaplectic cover.

\bigskip
\bfi{Bosonic case.}
For the bosonic case we may proceed in an analogous way. 
Again, we assume that $J$ is nondegenerate; this time, the normal
form can be taken without loss of generality as an antisymmetric
$2n\times 2n$-matrix $J$ that squares to $-1$.

We again form the Lie algebra $\Lz$ of inhomogeneous quadratic 
expressions and factor out the center. 
We then apply the map to the equivalence classes $(\alpha,f)$
\[
u:(\alpha,f)\mapsto \pmatrix{Jf&\alpha\cr 0 &0\cr}\,.
\]
It is clear that
\[
(Jf)^TJ+J(Jf)=-f^T+f=0\,,
\]
so that the map $u$ is an isomorphism from $\Lz$ to $isp(2n)$. We thus
see that the inhomogeneous quadratic quantities form a central
extension of the Lie algebra $isp(2n)$.

\chapter{Highest weight representations}\label{c.highest}

This chapter discusses highest weight representations, providing tools 
for classifying many irreducible representations of interest.
We extend the ladder technique used in Section
\ref{section-reps-heis} for determining the unitary representations
of the oscillator algebra to some other small Lie algebras of interest,
and indicate how the ideas generalize further. 

The basic ingredient is a triangular decomposition, which exists for all
finite-dimensional semisimple Lie algebras, but also in other cases
of interest such as the oscillator algebra, the Heisenberg algebra 
with the harmonic oscillator Hamiltonian adjoined.

We look in detail at 4-dimensional Lie algebras with a nontrivial
triangular decomposition (among them the oscillator algebra 
and $so(3)$), which  behave almost like the oscillator algebra. 
As a result, the analysis leading to Fock spaces generalizes without 
problems, and we are able to classify all irreducible unitary 
representations of the rotation group. Various related material
concerning $SO(3)$ and its universal covering group $SU(2)$ is also
included.

\section{Triangular decompositions}

Let $\Lz$ be a Lie $*$-algebra. A {\bfi{triangular decomposition}}
\index{decomposition!triangular}
of $\Lz$ consists of Lie subalgebras $\Lz_-$, $\Lz_0$ and $\Lz_+$
of $\Lz$ satisfying the properties\footnote{
Note that the present concept of a triangular decomposition
is less demanding and hence more general than in the treatment
by \sca{Moody \& Pianzola } \cite{MooP}. Their additional restrictions
allow them to extend much of the finite-dimensional semisimple theory 
outlined below to the infinite-dimensional case.
} 

(T1)~~~$\Lz ~=~ \Lz_- \oplus \Lz_0 \oplus \Lz_+$,

(T2)~~~$\Lz_0 \lp \Lz_\pm ~\subseteq~ \Lz_\pm$,

(T3)~~~$\Lz_0^* = \Lz_0$,~~$\Lz_\pm^* = \Lz_\mp$,

(T4)~~~$\Lz_0$ is abelian and contains the center $Z(\Lz)$.

Triangular decompositions generalize the properties of annihilation
and creation operators in the oscillator algebra $os(1)$ to more
general Lie algebras.
The terminology derives from the following motivating examples.

\begin{expls}
(i)
In the Lie algebra $\Lz = gl(n,\Cz) = \Cz^{n\times n}$, we can define
a triangular decomposition by defining $\Lz_0$ to be the Lie subalgebra
of diagonal matrices, $\Lz_+$ to be the Lie subalgebra of strictly
upper triangular matrices, and $\Lz_-$ to be the Lie subalgebra of
strictly lower triangular matrices. Verification of the axioms is
straightforward.

(ii) The oscillator algebra $os(1)$ has a triangular decomposition,
given by
\[
\Lz_- = \Cz\, a\,, ~~~ \Lz_0 =\Cz\, 1 +\Cz\, n\,, ~~~ 
\Lz_+= \Cz\, a^*\,.
\]
\end{expls}

A {\bfi{triangulated Lie algebra}}\index{Lie algebra!triangulated}
is a Lie $*$-algebra with a distinguished triangular decomposition.
We call the number $\rk\Lz:=\dim\Lz_0/Z(\Lz)$ the {\bfi{rank}},
and $\deg\Lz:=\dim\Lz_\pm$ the {\bfi{degree}} of the triangulated
Lie algebra $\Lz$. The elements of the dual space\footnote{Since in
  the context of Lie $*$-algebras, the notation $V^*$ for the dual of
  $V$ is ambiguous, we use in this section a prime to indicate the 
  dual.}
$\Lz_0'$ are called {\bfi{weights}}.
A {\bfi{highest weight representation}}
\index{representation!highest weight} is a representation $J$ of $\Lz$
on a vector space $\Vz$ with a distinguished element $1$, called the
{\bfi{ground state}}\footnote{In a quantum field theory context,
the ground state is referred to as the {\bfi{vacuum}}.}, such that

(HW1)~~~$J(\alpha) 1 = 0$ for all $\alpha\in\Lz_-$, and

(HW2)~~~$J(\alpha) 1 \in \Cz$ for all $\alpha\in\Lz_0$.

The elements of $\Lz_-$ thus behave like annihilation operators.
The defining properties imply that
\[
w(\alpha):=J(\alpha)1\,, \for \alpha\in\Lz_0\,,
\]
defines a weight $w\in\Lz_0'$, called the {\bfi{highest weight}}
of the representation. A highest weight representation is irreducible 
if and only if the elements
$a_1^*\dots a_k^*1$ with $a_1,\dots, a_k\in\Lz_-$ span a dense
subspace of $\Vz$. In an irreducible highest weight representation
with highest weight $w$, all Casimir elements $C$ of $\Lz$
have a fixed value $C(w)\in\Cz$.

The {\bfi{spectrum}} of $\Lz$ is the set $\Sigma(\Lz)$ of weights 
$w$ for which a unitary group \at{which group?} representation exists,
whose associated infinitesimal
representation is a highest weight representation of $\Lz$ with highest
weight $w$. The spectrum of $\Lz$
determines the possible spectra of each Casimir element $C$ in
arbitrary unitary representations of the universal covering group of
$\Lz$, since the possible eigenvalues are precisely the possible $C(w)$
where $w$ ranges over the spectrum of $\Lz$.

Note that a weight $w$ belongs to the spectrum of $\Lz$ iff there is 
a unitary (cf. Definition \ref{d.lie-rep}) highest weight 
representation of $\Lz$ with highest weight $w$. In
this case, there is a Euclidean inner product on $\Vz$, and without 
loss of generality, the ground state 1 may be assumed to be normalized.

\bigskip
\bfi{The semisimple case.}
There are many examples of triangulated Lie algebras, related to
finite-dimensional \idx{semisimple} Lie algebras (see the outline below)
and to important classes of infinite-dimensional Lie algebras.

We mention without proof (which can be found in many places, e.g., 
\sca{Fuchs \& Schweigert} \cite{fuchsandschweigert}, 
\sca{Fulton \& Harris} \cite{fultonharris}, \sca{Humphreys} 
\cite{humphreys}, \sca{Jacobsen}
\cite{jacobsen}, \sca{Knapp} \cite{knapp}, \sca{Kirillov}
\cite{kirillovLiegroup}) a number of facts about 
finite-dimensional semisimple Lie algebras.

All finite-dimensional semisimple real Lie algebras have a triangular
decomposition, which is unique up to automorphisms. In this case,
$\Lz_0$ is a Cartan subalgebra (a maximal abelian subalgebra generated 
by diagonal matrices in the adjoint representation, for some choice of 
basis), which is unique up to conjugation, and the Lie algebra $\Lz$ 
decomposes as
\[
\Lz = \Lz_0 \oplus \bigoplus_{\alpha\in \Delta} \Lz_{\alpha}\,,
\]
where $\Delta\subset \Lz_{0}'$ is the set of {\bfi{roots}}. 
The roots are nonzero elements of the dual of the Cartan subalgebra 
such that the Cartan subalgebra acts diagonally on $\Lz_\alpha$:
\[
h \lp x = \alpha(h)x\,,~~~h\in\Lz_0,~~x\in \Lz_\alpha\,.
\]
$\Lz_\alpha$ is always 1-dimensional; any nonzero element in 
$\Lz_\alpha$ is called a {\bfi{root generator}}. For each root 
$\alpha\in \Delta$ the negative $-\alpha$ is also a root: 
for all $\alpha\in\Lz^*$, if $\Lz_\alpha \neq 0$, then 
$\Lz_{-\alpha}\neq 0$. Therefore there exists a choice of ordering 
such that $\Delta$ can be written as the union of the set of positive 
roots $\Delta^+$ and the set of negative roots $\Delta^-$ and 
$\Delta^-=-\Delta^+$, in such a way that the cone of nonnegative 
linear combinations from $\Delta^+$ and $\Delta^-$ intersect in 0 only.
One defines
\[
\Lz_\pm = \bigoplus_{\alpha\in \Delta^{\pm}}\Lz_\alpha\,,
\]
and finds (using further properties of the roots) that the semisimple 
Lie algebra is a triangulated Lie algebra.

\begin{expl}
Take $\Lz = sl(n,\Cz)$, the Lie algebra of $n\times n$
matrices with trace zero. Let us we write $E_{ij}$ for the matrix that
is $1$ on the $(i,j)$-entry and zero everywhere else. Then the diagonal
matrices that have trace zero make up the Cartan subalgebra, which is
thus spanned by the matrices $E_{ii}-E_{i+1,i+1}$ for $1\leq i\leq
n-1$ so that the rank is $n-1$. We have for $h={\rm
  diag}(h_1,\ldots,h_n)\in \Lz_0$ in the Cartan subalgebra and for
$E_{ij}$ with $i\neq j$ 
\[
h \lp E_{ij} = (h_i - h_j)E_{ij}\,.
\]
Hence the roots are of the form $\lambda_{i}-\lambda_{j}$ where
$\lambda_i$ reads off the $i$th diagonal entry of an element of the
Cartan subalgebra. We can choose a root $\lambda_i-\lambda_j$ to be
positive if $i< j$. Then $\Lz_+$ are the upper triangular matrices,
and $\Lz_-$ the lower triangular matrices. The positive root
generators are $E_{ij}$ with $i<j$. 
\end{expl} 

Associated with each semisimple Lie algebra is a
{\bfi{weight lattice}}, which is a discrete additive subgroup of
$\Lz_0'$ and whose elements are called {\bfi{integral weights}}.
Additionally, there is a distinguished subset of the weight lattice,
which is closed under 
addition and whose elements are called \bfi{dominant integral weights}.
In terms of these:

(i) For each weight $w$, there is a Lie representation with $w$ as
highest weight.

(ii) A highest weight representation is finite-dimensional if and only
if the highest weight is dominant and integral.

(iii) For compact finite-dimensional Lie algebras, that is
finite-dimensional Lie algebras with a negative definite 
Cartan--Killing form (these are automatically semisimple, see 
Lemma \ref{lem.cartan}), a highest weight representation is
unitary if and only if it is finite-dimensional. The inner
product is then uniquely determined
by the requirement that the ground state 1 is normalized.

(iv) The Lie algebra induces a unitary representation of the
universal covering group $\Gz$ if and only if $w$ is a dominant
integral weight. Thus the spectrum of $\Lz$ consists of all
dominant integral weights of $\Lz$.

In the context of an integrable classical theory associated with $\Lz$,
(iv) is equivalent to the \bfi{Bohr--Sommerfeld quantization
  condition}. (This folklore result is never stated in a precise form,
but see, e.g., \sca{Voros} \cite{Vor}, \sca{Kochetov} \cite{Koc}, and 
\sca{Gadiyar} \cite{Gad}.)

\section{Triangulated Lie algebras of rank and degree one}
\label{s.triang.rd1}

We have seen that the oscillator algebra $os(1)$ has a triangular
decomposition of rank and degree 1. 
A general triangulated Lie $*$-algebra of rank and degree 1 with 
center $\Cz$ must be the direct sum of the algebras
\[
\Lz_-= \Cz a\,, ~~~ \Lz_+=\Cz a^*\,,~~~ \Lz_0=\Cz+\Cz h\,,
\]
where $h$ is a fixed element in $\Lz_0\setminus\Cz$.
The center $\Cz$ commutes with everything, but $h$ in
general does not, which is the case we consider here. Then
we may rescale $h$ to obtain
\[
a\lp h=ia\,.
\]
The operation $*$ then gives
\[
a^*\lp h=-ia^*\,.
\]
For the Lie product of $a$ and $a^*$ we introduce complex numbers $u$
and $v$ and write
\[
a\lp a^* = i(uh +v)\,,
\]
but noting that $(a\lp a^*)^*=-a\lp a^*$ we see that $u,v\in \Rz$.
It is easy to check that for all $u,v\in\Rz$ the Jacobi identities
are fulfilled and hence for all real numbers $u,v$ we have a
Lie $*$-algebra. 

For the two-parameter family of Lie $*$-algebras just defined there are
essentially four different cases;
\begin{enumerate}
\item
$u=v=0$.
This is the Lie $*$-algebra $iso(2)\oplus \Cz$.

\item
$u=0$, $v=\pm1$.
If $u=0$ we can rescale the $a$ and $a^*$ as $a\mapsto \lambda a$ and
$a^*\mapsto \lambda^*a^*$ to get $v=\pm 1$. By complex conjugation of
the algebra we then can choose the sign of $v$ and we find the Lie $*$-algebra  $os(1)$. For the oscillator algebra we have $h\sim a^*a$.

\item
$u=1$ and $v=0$.
This Lie $*$-algebra is $so(2,1)\oplus \Cz$.
If $u$ and $v$ are both nonzero, we can redefine $h$ as $h \to
\alpha h+\beta$ for some $\alpha,\beta\in \Cz$ to obtain this case
or the next one.

\item
$u=-1$ and $v=0$.
This is the Lie $*$-algebra $so(3)\oplus \Cz$.
\end{enumerate}

Note that the elements $a$ and $a^*$ are
abstract
vectors from the point of view of Lie algebras. That means that we
cannot say that $a^*$ is the conjugate of $a$; it is only in
Lie $*$-algebras, in the $*$-Poisson algebras and in their unitary
representations that we can say that $a^*$ is the Hermitian conjugate
of $a$. It is for these reasons that we have treated case 3 and case 4
separately. In a unitary representation we have $J(f\lp
g)=\frac{i}{\hbar }(J(f)J(g)-J(g)J(f))$, so that $J(f)^*=J(f^*)$ makes 
sense. 

As alluded before $so(2,1)$ and $so(3)$ are isomorphic as complex Lie
algebras. If we define in $so(2,1)$ the elements $r=ia$, $s=ia^*$ we
obtain the relations
\[
h\lp r = -ir\,,~~~ h\lp s = is\,,~~~ r\lp s=-a\,,
\]
which defines case 4 of the list above:
$so(3)$. However, the map from $so(2,1)$ to $so(3)$ does not preserve
the $*$-operation, since $r^* = (ia)^* = -ia^* \neq s$. That means
that $so(2,1)$ and $so(3)$ are not isomorphic as Lie
$*$-algebras.

Among the triangulated Lie algebras  of rank and degree 1 listed above,
the most interesting cases for both classical and quantum mechanics are
$os(1)$ and $so(3)\oplus \Cz$.
As we have seen, the oscillator algebra $os(1)$ is related to the
harmonic oscillator. The algebra $so(3)\oplus \Cz$ involves
infinitesimal ordinary rotations and arises when dealing
with the spinning top, as explained in Chapter \ref{c.harmonic}.
The algebra $so(2,1)\oplus \Cz$ is less prominent in classical mechanics
although it arises in the analysis of the celestial 2-body
problem. The algebra $so(2,1)\oplus \Cz$ has important applications
to exactly solvable problems in quantum mechanics, and even appears
in so-called gauged supergravity theories.

\section{Unitary representations of $SU(2)$ and $SO(3)$}\label{s.su2rep}

We now discuss the unitary representations of the Lie groups $SU(2)$
and $SO(3)$. 
The method presented below is often encountered in quantum physics
textbooks. In Section \ref{s.highest} we discuss the highest weight
representations of triangulated Lie algebras of rank and degree 1,
which shows a great similarity with the discussion here.

Since the group $SU(2)$ is compact it has an invariant {\bfi{Haar
measure}} $d\mu(g)$. \at{give the Haar measure!! But do you
really need Haar measure here? Can't you do all the following stuff in
this section by just insisting that the $SU(2)$ generators are
self-adjoint?}
Therefore we can integrate over the group in an
invariant way; invariance of the Haar measure means
$\sint f(hg)d\mu(g)=\sint f(g)d\mu(g)$ . If $SU(2)$ acts on a vector
space $V$  with an inner product $\<,\>_0$ we can integrate
over the group to get an invariant inner product;
\[
\< v,w\rangle = \int_{SU(2)} \< g\cdot v,g \cdot w\rangle_0 d\mu(g)\,,
\]
where we denoted the action of $g\in SU(2)$ on $v\in V$ by $g\cdot v$.
It is a direct consequence of the invariance of the Haar measure that
the inner product $\<,\rangle $ is $SU(2)$-invariant. Hence we have
realized $SU(2)$ by unitary matrices; every representation of $SU(2)$
is equivalent to a unitary representation.

Since the group $SU(2)$ is compact and simply connected 
\at{define this in the Lie group chapter!}
there is a one-to-one correspondence between the representations
of the group and the representations of the Lie algebra 
$su(2)\cong so(3)$.
The Lie algebra consists of antihermitian matrices but multiplying
them by $i$ we obtain Hermitian matrices and we may
use the Pauli matrices to describe $su(2)$. Finding all representations 
of the Lie algebra $su(2)$ therefore gives all representations
of $SU(2)$. 

We put $t_i = \shalf \sigma_i$ for $i=1,2,3$ and define
$L_\pm = (t_1\pm i t_2)$ and obtain a triangulated algebra with
trivial center;
\[
t_3 \lp  L_\pm =\pm L_\pm \,, ~~~ L_+\lp L_- =2t_3  \,.
\]
In a unitary representation we require that $t_3$ is Hermitian and
$L_{\pm}^{*}=L_\mp$. If $v$ is an eigenvector of $t_3$ with eigenvalue
$\alpha$, then $L_-v$ is an eigenvector of $t_3$ with eigenvalue
$\alpha-1$. For a finite-dimensional representation we cannot lower
the eigenvalue forever and hence there exists
a vector $v$ with $L_-v=0$. 
\at{That seems an unnecessarily obscure way to do it. (It depends on,
among other things, the fact that eigenvectors of a self-adjoint
operator corresponding to distinct eigenvalues are orthogonal.)
Instead, if one examines simultaneous eigenvalues of $\J^2$ and $J_z$,
the existence of such a vector can be derived in a few lines as an
elementary consequence of being in a unitary representation.} 
Assume that we have $t_3 v=\alpha v$
for some complex number $\alpha$. Acting on $v$ with $L_+$ we get
vectors with eigenvalues $\alpha,\alpha+1,\alpha+2,\ldots$. Again this
series has to terminate. Thus, there is an eigenvector $w$ with
eigenvalue $\alpha+N$ that is annihilated by $L_+$. Since $t_3$
is Hermitian, vectors with different eigenvalues are orthogonal and
hence linearly independent. Thus the $N+1$ vectors with eigenvalues
$\alpha,\ldots,\alpha+N$ form an irreducible representation. The trace
of $t_3$ is zero, since $\tr t_3=-i\tr (t_1t_2-t_2t_1)$. But
then the sum of the eigenvalues should vanish:
\[
0=\sum_{n=0}^{N}\alpha+n=(N+1)\alpha +\frac{1}{2} N(N+1)
= \frac{1}{2}(N+1)(2\alpha + N)\,.
\]
It follows that $\alpha=-N/2$. Therefore the eigenvalues are the
integers
$-\frac{N}{2},-\frac{N}{2}+1,\ldots,\frac{N}{2}-1,\frac{N}{2}$.
Conversely, for all integers $N$ we find a
representation by giving vectors $e_\alpha$ with $-N/2\leq \alpha \leq
N/2$ and defining the action of $t_3$ and $L_\pm$ by the above
rules. We then recover $t_1$ and $t_2$ by
$t_1=\frac{1}{2}(L_++L_-)$ and
$t_2=\frac{1}{2i}(L_+-L_-)$. We can thus label the
finite-dimensional
representations of $su(2)$ by half-integers $j =0,1/2,1,3/2,2,\ldots$.
We denote them by $D_j$; note that we have $j=N/2$. The
dimension of the representation $D_j$ is $2j+1$ and the eigenvalues of
$t_3$ are
$-j,-j+1,\ldots,j-1,j$. The
Casimir $J^2$ defined by
\[
J^2 = t_1t_1+ t_2t_2 +  t_3t_3 = L_+L_-
- t_3 +t_3 t_3
\]
has the value $j(j+1)$ on the representation $D_j$ since acting on the
state $v$ with eigenvalue $-j$
\[
J^2 v = (L_+L_-
- t_3 +t_3t_3)v= (0+j+ j^2)v= j(j+1)v\,.
\]
The number $j$ is called the \idx{spin} of the representation. Clearly
$D_j$ is irreducible.

The representations that correspond to nonintegral $j$ cannot be 
lifted to
representations of $SO(3)$. Although the Lie algebras $su(2)$ and
$so(3)$ are isomorphic, the groups $SO(3)$ and $SU(2)$ are not!
As mentioned before, $SU(2)$ is the universal covering group of
$SO(3)$. In fact, we have $SO(3)=SU(2)/\Zz_2$. That means that there is
an action of $\Zz_2$ on $SU(2)$, such that $SO(3)$ is the manifold
$SU(2)$ with the points that are related by the $\Zz_2$-action
identified. In Section \ref{s.rotSU2} we gave details on how
$SO(3)$ and $SU(2)$ are related by a 2-1 map $SU(2)\to SO(3)$.
\at{rephrase!}
If a representation of $SU(2)$ is such that $\Zz_2$-related
points have the same image under the representation we have a
well-defined representation for $SO(3)$; this thus precisely
corresponds to the $\Zz_2$-invariant representations. It turns out
that only the representations with integer $l$ correspond to
$\Zz_2$-invariant representations. In physics, particles are
represented by fields that take values in an
$su(2)$-representation. The representations $D_j$ for $j=1/2,
3/2,5/2,\ldots$ correspond to \idx{fermions} and for $j=0,1,2,\ldots$
to \idx{bosons}.

\section{Some unitary highest weight representations}\label{s.highest}

For the quantum theory one considers the unitary highest weight
representations. We investigate the unitary highest weight
representations for the triangulated Lie algebras  of rank and degree 1
listed in Section \ref{s.triang.rd1}. We thus look for a realization
of operators $a$, $a^*$, $h$ and $1$ such that $1$ acts as the
identity, $h$ acts diagonally and is Hermitian, $a^*$ is the
adjoint of $a$ and the 
following relations hold (see \gzit{e.liequant} and Definition
\ref{d.lie-rep}): 
\[
[a,n]=\hbar a\,,~~~[a^*,n]=-\hbar a^*\,, ~~~ [a,a^*]=\hbar (un+v)\,.
\]
Furthermore, we assume there is a vector $|0\>$ with
\[
a|0\>=0\,,~~~ n|0\> = \alpha |0\>\,.
\]
By acting with $a^*$ on $|0\>$ we obtain the other vectors in the
representation. We define
\[
| k \> = \frac{(\hbar a^*)^k}{k!}|0\>\,,
\]
so that 
\[
a^* | k -1\> = \hbar k | k\>\,.
\]
It follows that
\[
n |k\> = \hbar (k + \alpha) |k\>\,.
\]
We have $a|k\> = c_k |k-1\>$ and we want to determine $c_k$. Since
$aa^*=[a,a^*]+a^*a$ we find
\[
\hbar kc_k|k-1\>= aa^* |k-1\> 
=\hbar \Bigl(  (uh+v) + (k-1)c_{k-1}\Bigr)|k-1\>\,,
\]
from which it follows
\[
kc_k - (k-1)c_{k-1} = \hbar (uk + u\alpha +v)\,,
\]
which is solved by 
\beqar
kc_k &=& 0+(1c_1 - 0c_0)+(2c_2-1c_1)+ \ldots + (kc_{k} -
(k-1)c_{k-1})\nn\\
&=&\hbar k \Bigl(\hbar \alpha u + v +\shalf \hbar u(k+1)\Bigr)\,,\nn
\eeqar
so that
\[
a|k\> = \Bigl( \alpha \hbar u + v +\shalf \hbar u(k+1)\Bigr)|k-1\>\,.
\]
The vectors $| k \>$ are orthogonal, as in
the case of the harmonic oscillator. So we suppose
$\< j | k\> = N_k\delta_{jk}$, and calculate $\< j | a^* | k\>$ 
in two ways:
\[
\< j | a^* | k\> = (k+1) \< j|k+1\>,
\]
\[
\< j | a^* | k\> 
=  (v + \bar \alpha \hbar u +  \shalf \hbar u(k+1)) \<j-1| k\>.
\]
Choosing $k=j-1$ we find
\[
j\hbar N_j = 
\left(v+u\hbar \bar \alpha +\shalf(j+1)u\hbar  \right) N_{j-1}\,.
\]
For a representation we require that $N_j\geq 0$ for all $j$. 
We may normalize
$N_0=1$ and it follows that we must have $\alpha\in\Rz$. To have a
faithful representation we need $|1\>\neq 0$ and thus $N_1>0$. 
We distinguish further two cases:

{\sc Case 1:} $u\geq 0$. By assumption
  $\psi_0$ is a nonzero vector and thus has a positive norm. But then
  all $N_j$ are positive. Hence we find nonzero vectors for all $j\in
  \Nz_0$. An example of this case is given by $so(2,1)$, which is a
  noncompact Lie algebra. More generally, noncompact Lie algebras
  (defined by having a
  Cartan--Killing form that is not negative definite) do not admit a
  finite-dimensional unitary representation.

{\sc Case 2:} $u<0$. In this case $N_j$ can
      become negative, unless it becomes zero for some integer
      $j_m$. Thus $ (j_m+1) + 2(\bar \alpha + \frac{v}{\hbar u})$. In
      this case we thus have a finite-dimensional unitary
      representation for every integer $j_m = 0,1,2,\ldots$. The
      dimension of the representation is $j_m+1$. If $j_m=0$ the
      vector $|1\>$ is already zero, and hence $a^*$ operates as $0$
      in this representation. Therefore, if $j_m\geq 1$ the
      representations are faithful.

For the triangulated Lie algebras of rank and degree 1, there is
a Casimir operator of the form $C=\hbar a^*a - q(n)$ for some
quadratic $q(n)$. From Section \ref{s.su2rep} we know that $so(3)$ has
the Casimir $J^2=2aa^*+n^2+in$. And for the algebra $os(1)$ the
element $C=\hbar a^*a-n$ 
is a Casimir. For the harmonic oscillator we then have
$C=0$, since $n$ is precisely $\hbar a^*a$. For
$so(2,1)$ this does not work; there is no analogue of the
number operator with only integer eigenvalues. 
That is, the Lie algebra $so(2,1)$ does not admit a discrete Casimir.
\at{check!}

\chapter{Spectroscopy and spectra}\label{c.spec}

This final chapter applies the Lie theoretic structure to the analysis
of quantum spectra. After a short history of some aspects of 
spectroscopy, we look at the spectrum of bound systems of particles.
We show how to obtain from a measured spectrum the spectrum of the
associated Hamiltonian, and discuss qualitative results on vibrations 
(giving discrete spectra) and chemical reactions (giving continuous 
spectra) that come from the consideration of simple systems and the 
consideration of approximate symmetries. The latter are shown to 
result in a clustering of spectral values.

The structure of the clusters is determined by how the irreducible 
representations of a dynamical Lie algebra split when the algebra is 
reduced to a subalgebra of generating symmetries. 
The clustering can also occur in a hierarchical fashion with fine 
splitting and hyperfine splitting, corresponding to a chain of 
subgroups.
As an example, we discuss the spectrum of the hydrogen atom.

\section{Introduction and historical background}

In this chapter we show some features of spectra and
spectroscopy. In the preceding chapters we discussed properties of
systems. The Hamiltonian of a system has a spectrum consisting of the
eigenvalues, but in practice we don't see this spectrum, but the 
energy differences. One perturbs the system by shining
light on it for example and then observes some response. The responses
give rise to the observed spectrum, the study of which is spectroscopy.

To study the structure of molecules and atoms, we often rely
on destructive methods. The destructive nature of the experiments in
chemistry was taken as a primitive distinction between chemistry and
physics. Nowadays the situation is different. In
high-energy physics, one also shoots particles at each other such
that the original particles are destroyed and energy is converted into
the creation of other particles. On the other side, in chemistry new
laser-techniques are used where molecules are kept intact,
and information about the structure of the molecular bonds
is obtained.

With spectroscopy one can study properties of materials and
mixtures without destructing the sample. There are crudely speaking
two kinds of spectra, relying on
different experimental methods. An emission spectrum is obtained by
putting a system in a state of high energy. The system then falls
back to a state with lower energy, and the energy difference is
emitted in the form of light. Of course, in order to emit light, the
system needs to interact with light. The kind of
interaction then dictates which transitions are possible and hence
which frequencies are emitted. For the absorption spectrum one more or
less does the converse. One puts a system into a beam of (nearly) white
light. The system then absorbs light and re-emits it again, but then
in all directions.

\begin{figure}[htb]
\begin{pspicture}(1,-0.5)(18,4.5)
\psspectrum[element=He](1,2.2)(18,4.2)
\psspectrum[absorption,element=He](1,0)(18,2)
\end{pspicture}
\caption{The emission (upper) and absorption (lower) spectrum of
Helium.}\label{he-spec}
\end{figure}
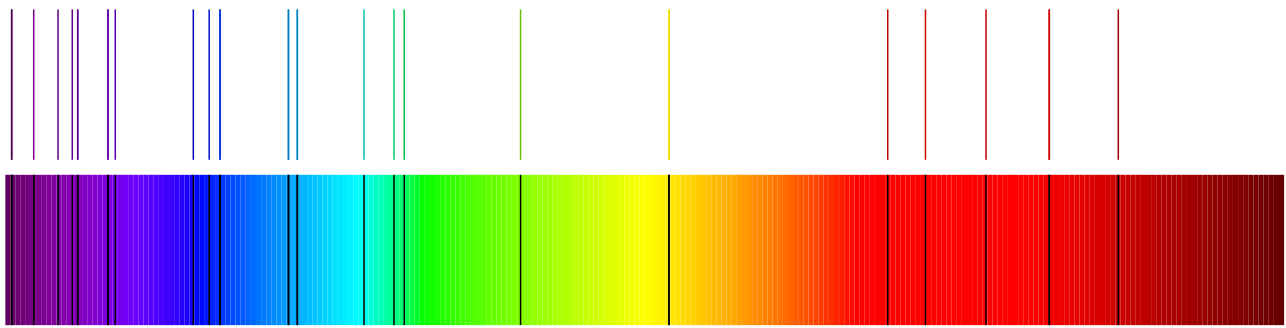

In the 19th century Kirchhoff used an invention of the German chemist 
Robert Bunsen to heat up elements in a flame to study the emitted
light. He passed light through a prism to study 
the intensity of light at different wavelengths. It turned out
that the emitted spectrum of an element had
quite clearly defined lines at certain wavelengths. In 1859
Kirchhoff pointed out that all the elements that he had been studying
had a different emission spectrum. Hence
disentangling the lines of an emission spectrum can help in finding the
components an unknown mixture is made of. Figure \ref{he-spec} gives 
as example an emission and an absorption spectrum of Helium.

Already much earlier, Isaac Newton had used in 1670-1672 a prism to 
study the decomposition of
white light into a spectrum of different colors. In 1814 Joseph von
Fraunhofer invented the spectroscope and identified 574 dark lines in
the light of the sun. In fact, the Fraunhofer experiment can already
be done with primitive equipment. On a sunny, cloudless day one
sits in a dark room with one little hole through which the sun
shines. In the beam of sunlight one places a prism and lets the
light after the prism fall onto a white piece of paper. The observed
spectrum can be seen to display dark lines; in Figure \ref{he-spec}
the lines corresponding to helium are displayed. The Fraunhofer lines
are a manifestation of the absorption 
spectrum. It was Kirchhoff who later explained the origin; light from
the sun has to pass the atmosphere of the sun. In the atmosphere the
elements that are present absorb certain parts of sunlight, at
well-defined frequencies and re-emit it later, but then in all
directions. Therefore the sunlight going in the forward direction --
that is, away from the core of the sun -- has lost intensity at certain
well-defined frequencies. In this way, Kirchhoff showed that the
atmosphere of the sun contained among others hydrogen and sodium.
The reason why the sunlight is almost white before
entering the atmosphere of the sun we will not explain. When the light
of the sun reaches the earth it is already so diluted that the
elements in the earths' atmosphere give almost unobservable absorption
lines. Therefore, the dark lines in the spectrum of the sun are due to
the suns' atmosphere and not the earths' atmosphere.

In 1868, the French astronomer Pierre-Jules-Cesar Janssen observed a
line in the spectrum of the sun that did not match any element known
by then.
The reason he observed it and not Fraunhofer was because Janssen
used the better observing circumstances that a solar eclipse
offers. Normally, the sun is too bright, but when the moon blocks the
solar disc, one sees solely the atmosphere of the sun. The astronomer
Joseph Norman Lockyear concluded that the new line must represent a
new element. They tossed the name helium, from the Greek word
``helios'', which means sun. It was not until 1895 that the physicist
John William Strutt, Lord Rayleigh -- or in short John Rayleigh --
proved that helium is also present on earth; he found it in samples of
the mineral clevite. He exposed the mineral to some
acids that reacted with the material thereby producing gasses. Then
he studied the contents of the gas mixtures, and he found that helium
was present. The reason why he found helium was
explained later. Clevite is a mineral that contains uranium. The element
uranium is radio-active; it can emit $\alpha$-particles, which are the 
nuclei of helium atoms.

\section{Spectra of systems of particles}

We distinguish two kinds of spectra:\\
1. The spectrum in the sense of spectroscopy is the collection of
frequencies emitted or absorbed by the system in its interaction with
light or other electromagnetic (infrared, radio, X-ray) radiation.\\
2. The spectrum of a physical system is the collection of allowed
energy values -- the set of eigenvalues of the associated Hamiltonian.

The relation between the two is as follows. The observed
spectrum (of spectroscopy) consists of the energy differences of the
system: the observed spectra are of the form $\hbar
\omega_{mn}=E_m-E_n$, where the energy levels of the system are
$E_n$. In most systems the spectrum is discrete. Hence also the
observed spectrum is discrete.

For systems that
are made of  constituents that can break apart, the spectrum contains
continuous parts. Consider for
example a molecule of two atoms like $H_2$. At a
certain frequency the molecule can break apart. Then the energy of
the photon can also be put into the kinetic energy of both $H$-atoms,
which is a continuous parameter.

If the Hamiltonian has some imaginary eigenvalues $\lambda$, then $\im
\lambda<0$ and the modes corresponding to $\lambda$ are decaying
modes. In a dissipative environment this results in energy loss, and
the system can move from higher energy to lower energy.

On the other hand, a system can also be excited. It then absorbs
energy from the environment. A typical example of
excitation is an atom interacting with light. The energy levels of the
atom are discrete, and hence only with a fine-tuned frequency the atom
can absorb a photon and attain a state with more energy. The energy
difference between the ground state and the state with the second
lowest energy is called the energy gap. If a photon has the frequency
with the energy corresponding to the energy gap, it can be absorbed by
the atom and the atom can be excited to the state above the ground
state.

An excited atom cannot move down to a state with lower energy due to
energy conservation, unless there is interaction with light.
Incorporating interaction with light into the Hamiltonian makes the 
energies acquire a
small imaginary part, representing the possibility to decay. If an
atom jumps down in energy, it emits a photon with the same
energy. This process is called spontaneous emission. The nice feature
of spontaneous emission is that we can observe it.

The interaction with light is not just any arbitrary interaction. The
interaction term $V_\fns{int}$ in the Hamiltonian
\[
H_\fns{tot}=H_\fns{atom}+H_\fns{env}+V_\fns{int}
\]
needs to respect some symmetries
like Galilean invariance. \at{explain or avoid}
The result is that not all transitions
but only a selected set of transitions is allowed. The rules that
dictate which transitions are allowed are therefore called
{\bfi{selection rules}}.

The interaction is often treated as a perturbation. The justification
is that the interaction term in the Hamiltonian is small compared to
the other terms. One introduces a
dimensionless variable $\lambda$ and re-writes
$V_{int}$ as $V_{int}(\lambda)= \lambda V_{int}$. One recalculates the
spectrum and expands it in $\lambda$ to find
\[
E_k(\lambda)= E_k(0)+\lambda \Delta E_{k}^{1}+\lambda^2 \Delta
E_{k}^{2}+\ldots\,.
\]
Since the interaction is small, the first order correction often
gives the interaction with light accurately enough.
Using the techniques of perturbation theory
one then finds the possible transitions, i.e. the selection rules, and
the probabilities of the transitions. The probabilities gives the
dominance in the observed spectrum; if a transition A is more probable
than a transition B this will result in more spontaneous emission
along transition A. Therefore the peak in the spectrum corresponding
to A is bigger than the peak corresponding to B.

\begin{figure}
\begin{center}
\includegraphics[scale=0.35]{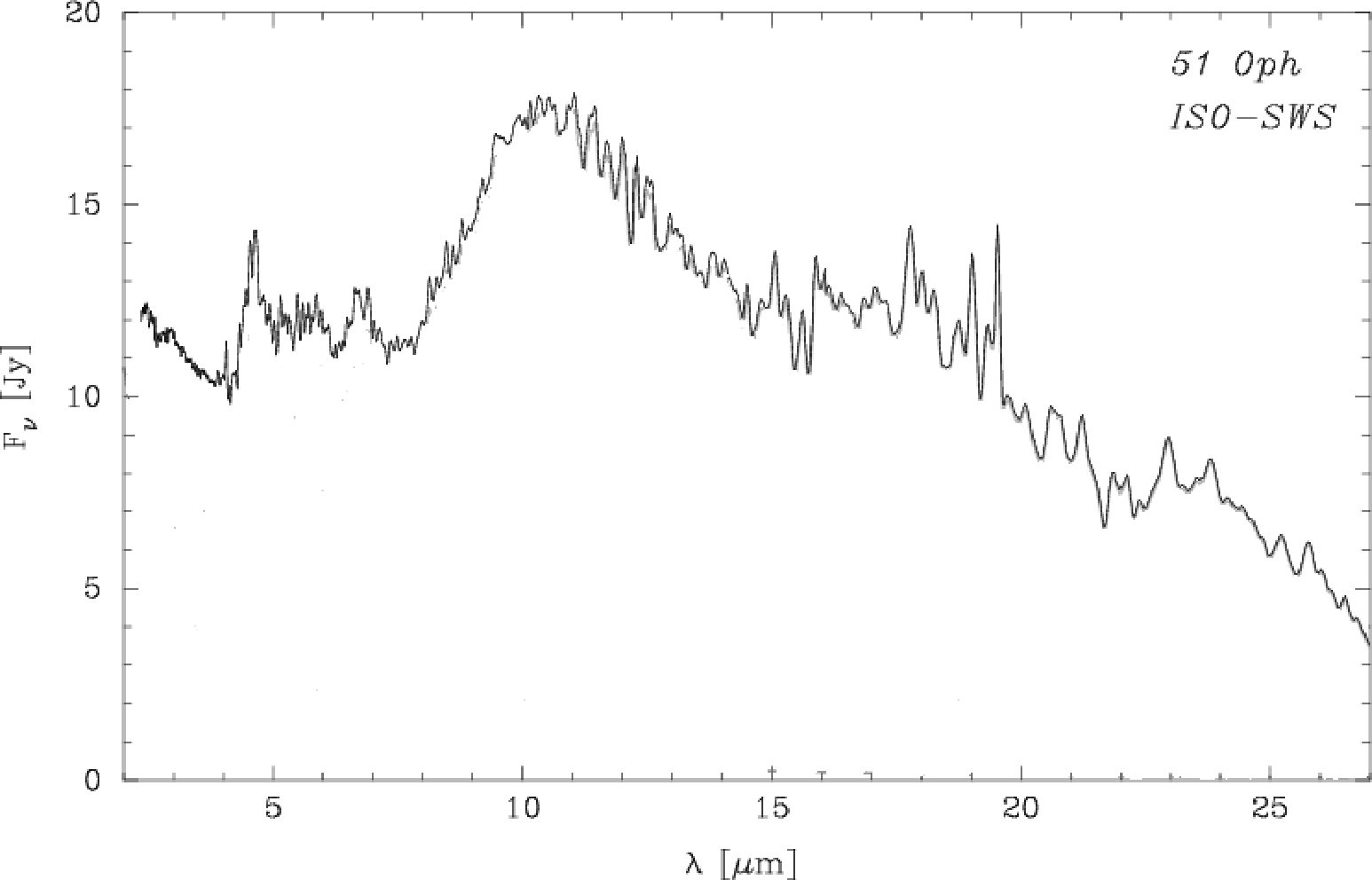}
\caption{An example of a spectrum.}\label{spec1}
\end{center}
\end{figure}

Observed spectra are often displayed by plotting, 
as in Figure \ref{spec1}, \at{poor figure quality}
on the horizontal
axis the frequency and on the vertical axis the observed
intensity. Due to imperfections in measuring methods one never
observes a real peak, but always a smeared out peak, that is, peaks
have a width. However, there can be many reasons why a peak has a
certain width. Imagine for example that one measures the spontaneous
emission of a gas contained in  cylinder. The gas atoms are moving
around in the cylinder, with different velocities with respect to the 
measuring device. For each atom the spectrum is shifted
due to the Doppler effect, known from a similar effect with sound,
which can be observed when an ambulance passes by. Since one measures 
the emission of a whole population of atoms, the measured peak is a 
superposition of peaks that are distributed around a certain frequency. 
That is, the Doppler effect broadens a peak. 

Technical imperfections of the measuring device also broaden peaks.
Making the measuring equipment more and more accurate one can try to 
get a better and better resolved spectrum. Doing this one might see that
broad peaks resolve into a group of smaller peaks. One sees therefore
more structure.

The result of a measurement is a list of data, the
frequencies $\omega_l$. Using the data one wants to obtain information
of the system under study. If one knows the system already quite well,
for example if one knows the parametric form of the Hamiltonian but 
not the precise values of the
parameters, one may fit the measured energies to obtain a set of
parameters that describes the measurements best. One therefore 
has to solve a data analysis problem.
For each label $l$ one has to find energies $E_{k(l)}$ and $E_{j(l)}$
with
\[
\omega_l \approx E_{j(l)} - E_{k(l)},
\]
within the experimental accuracy. Therefore, one solves the
least-squares problem of minimizing the sum
\[
S(E,j,k):=
\sum_l q_l \Big( \frac{E_{j(l)} - E_{k(l)}}{\hbar \omega_l} -1\Big)^2\,,
\]
for some weight factors $q_l $ related to the inverse of the accuracy
of the measurement of $\omega_l$.

In general, both the list $E$ of energy levels $E_i$ and the functions
$j,k$ which determine the {\bfi{assignment}} of spectroscopic 
lines to
transitions are unknown, and must be determined by minimizing
$S(E,j,k)$. Usually, one starts with a preliminary list $E$ of energy
levels, and assigns each line $l$ to a transition which minimizes
the $l$th term, breaking ties arbitrarily. This defines preliminary
assignment functions $j,k$. Fixing these turns the problem of minimizing
$S(E,j,k)$ into a least squares problem for finding the energy levels,
resulting in an improved $E$. Clearly, each cycle decreases the
value of $S(E,j,k)$. The process is stopped when the assignments
no longer change. Then $S(E,j,k)$ has reached a local minimum.
Multiple lists of trial energy levels may be used to increase the
likelihood that the assignment found corresponds to a global minimum.
Frequently, one first assigns a subset of lines to a subset of levels
to find good starting values.

\section{Examples of spectra}

The geometry of the molecule or atom under consideration strongly
influences the spectrum, since the geometry determines the potential.

\begin{figure}
\begin{center}
\includegraphics[scale=0.7]{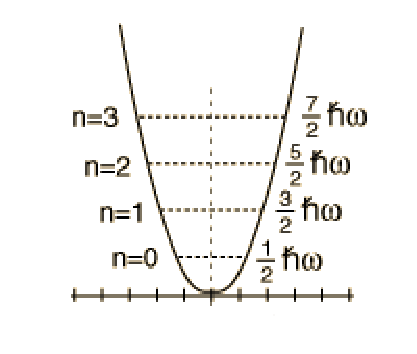}
\caption{Harmonic oscillator potential, with the eigenvalues indicated.}
\label{spec2}
\end{center}
\end{figure}

Consider a molecule of two atoms. We assume that the excitations
inside each atom are of another magnitude than the excitations of the
bond between the atoms. In that case we may consider the molecule as
two balls connected by a spring. The spectrum is as in
Figure \ref{spec2}, and the observed spectrum consists of one peak.
 \at{poor figure quality}

\begin{figure}
\includegraphics[width=0.4\linewidth]{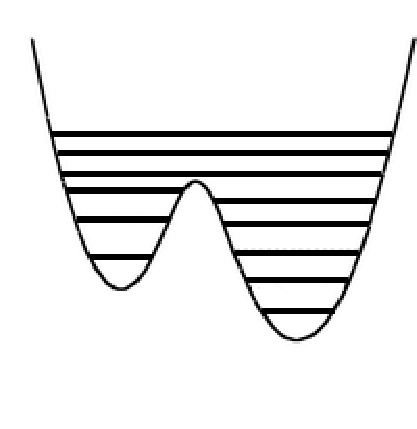}
\hfill
\includegraphics[width=0.50\linewidth]{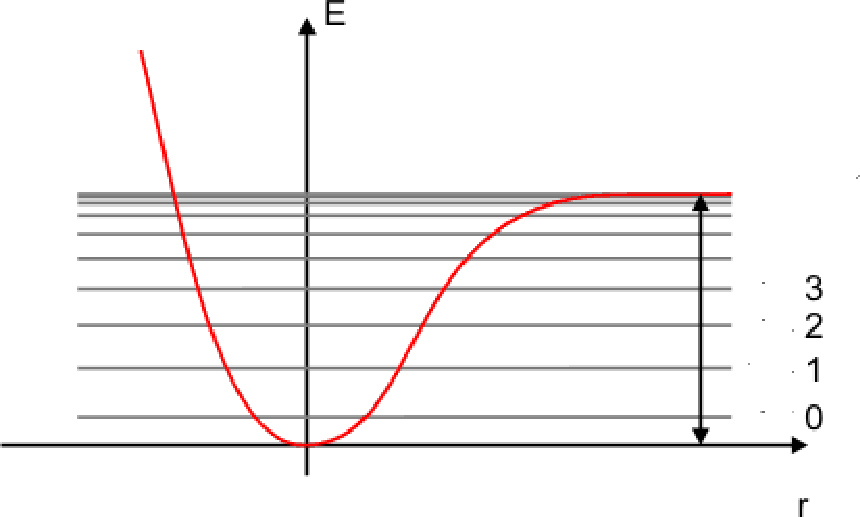}
\caption{On the left a double well potential with the first energy 
levels indicated. On the right the Morse potential; 
the bound states have discrete energy but above the dissociation 
energy the spectrum is continuous.}\label{spec3}
\end{figure}

Consider now a system that has two local minima. An example of this
would be a molecule $C_2H_4$ of which two versions exist, the cis and
trans molecules. The molecular bond between the two $C$-atoms then
behaves around each local minimum as a harmonic oscillator in some
approximation. For higher energies however the two states start to
interact and the molecule can change from cis to trans and vice
versa. A typical spectrum then looks like Figure \ref{spec3}.

When there are asymptotically free states, one says that the system
admits {\bfi{dissociation}}. Free states have continuous kinetic
energy and hence the spectrum contains continuous parts. 
A potential showing
dissociation is the Morse potential given by
\[
V(r) = \alpha(e^{-\beta r} - \gamma)^2 -\alpha\gamma^2\,, r\geq 0
\]
where $r$ is the atomic distance and $\alpha$, $\beta$, and $\gamma$ 
are positive parameters, see Figure \ref{spec3}. 
The potential of the $H_2$ molecule discussed above is another example.
Above the dissociation energy the spectrum is continuous; the bound 
states have discrete energy.

\begin{center}
 \begin{figure}
  \input{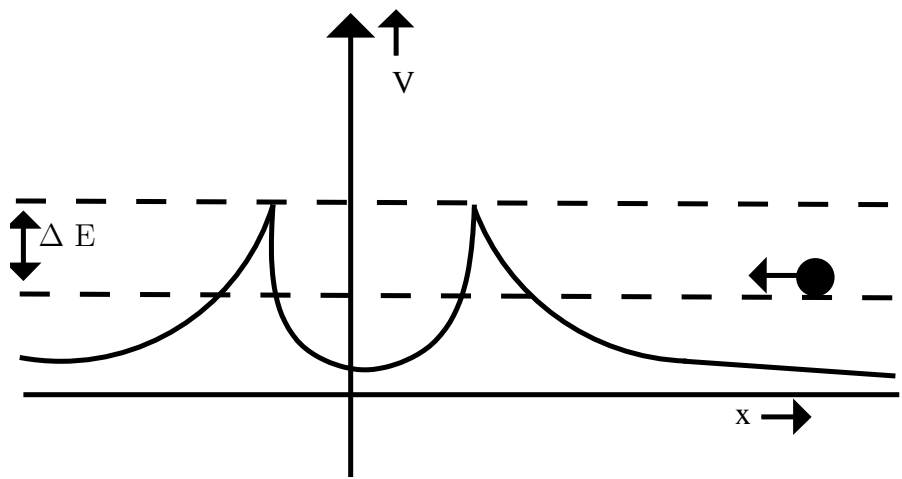}
  \caption{Sketch of the potential a proton experiences in the force
 field of a nucleus.}\label{spec5}
 \end{figure}
\end{center}

Quantum physics has a remarkable feature compared to classical 
mechanics, called {\bfi{tunneling}}. If a particle is in a local 
minimum at energy $E_1$ and another minimum is available with energy 
level $E_0 < E_1$, then (in a semiclassical particle view) there is a 
nonzero probability that the particle ``travels through the barrier'' 
and ends up in the local minimum with lower energy. 
For example, the potential of the $C_2H_4$-molecule 
discussed above admits tunneling since the potential has two local
minima. The probability of tunneling decreases with the height of the
barrier between the two energy levels. Another example where tunneling
occurs is in nuclear physics; the potential of Figure \ref{spec5}
represents the energy a proton feels in the potential field of a 
nucleus. The diameter of the nucleus is roughly the distance between 
the two peaks in Figure \ref{spec5}. The difference to the 
$C_2H_4$-molecule is here that the tunneling takes place between two
states one of which is not integrable. Tunneling can go in two
different directions; one direction is where the proton is shot 
at the nucleus with too little energy to classically penetrate the
nucleus, the other direction is where the proton is inside the nucleus
and classically cannot get out. In the latter case, there is a certain
probability that the proton escapes the nucleus. This explains
qualitatively the stochastic behavior of radio-active decay. 

\begin{figure}
\begin{center}
\includegraphics[scale=0.5]{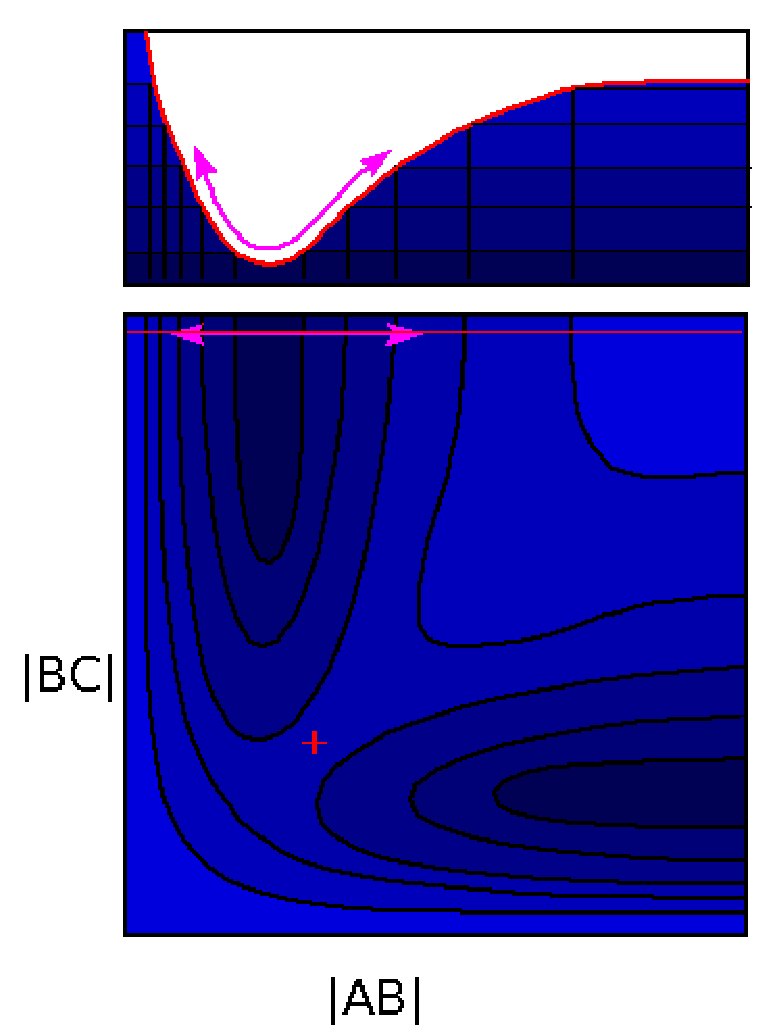}
\caption{ 2D-Plot of the potential experienced in a chemical reaction 
$AB+C\to A+BC$. The potential 
depends on the distances $|AB|$ and $|BC|$. The upper part shows a 
cross-section of the potential. The red marker is the saddle 
point.}\label{spec6}
\end{center}
\end{figure}

\at{the figure produces a spurious space in the text!}
As another example, consider a chemical reaction of the form
$AB+C\to A+BC$, that is, the molecule $AB$ splits off a part $B$ that
then attaches to $C$ to form $BC$. Here there are two important
parameters. The distance $|AB|$ between $A$ and $B$ and the
distance $|BC|$ between $B$ and $C$. A possible potential is plotted
in Figure \ref{spec6}. The plot shows two valleys separated 
by a saddle point, marked by a red cross. 
\at{red not visible in print; also in the caption}
The horizontal valley 
corresponds to $|BC|$ constant, hence to the state $A+BC$. 
The other valley corresponds to $AB+C$, and at the saddle point part 
$B$ is exchanged.

\at{Figure \ref{spec6} is black in print!}

\section{Dynamical symmetries}\label{s.dyn}

As discussed before, when one looks at a poorly resolved spectrum,
one sees some rough features of the system under study. Improving the
resolution allows one to study more structure of the system.

A similar process happens when one studies a hydrogen atom in an
external magnetic field. Upon increasing the magnetic field one sees 
that many lines of the original spectrum split into several close
lines. Thus what first seems to be one state in
fact turns out to be an agglomeration of different states. The states
first had energies that were so close together that they could not be
recognized as belonging to different states -- indeed, they have
exactly the same energy. As we shall see, that these states (seemingly)
agglomerate to one single state is due to symmetry reasons.

The rotational symmetry implies that the energies of different states
related by a rotation have the same energy; more pictorially, whether
an electron circles around the proton with the rotation axis in the
$z$-direction or the $y$-direction gives the same energy.
Turning on the magnetic field results in breaking the symmetry;
then the different states that first agglomerated to form a single
state are disentangled and can be observed separately in the spectrum.

But as with the increasing resolution, taking a closer look at the
hydrogen atom reveals more and more structure. In a first
approximation, the electron in the hydrogen atom can be treated
nonrelativistically. Treating the electron relativistically, one gets
a correction to the spectrum. The first order corrections of special
relativity go under the name of the {\bfi{first radiative 
corrections}}.

We shall look in some detail at the hydrogen atom once we have 
clarified the general principles.

\bigskip
\bfi{Symmetry and broken symmetry.}
The most symmetric physical systems, in particular the standard 2-body
problems (the classical Kepler problem and the quantum hydrogen atom)
are exactly solvable. The Helium atom is already a three-body problem
and is not exactly solvable.

A physical system is called {\bfi{exactly solvable}} 
(or {\bfi{integrable}}, or {\bfi{completely  integrable}}) 
if it has ``enough'' constants of motion. Equivalently, if the 
centralizer $C_\Ez(H)$ of the Hamiltonian $H$ in the algebra $\Ez$ 
of observables is ``large enough''. 
The effect of having enough central elements is that the
system has enough conserved quantities to explicitly solve the
differential equations of the system. 
\at{give references to exact definitions in the classical case}

A {\bfi{dynamical algebra}} of a classical physical system is a Lie
algebra $\Lz$ that one can associate to the system such
that the Hamiltonian $H$ is contained in the Lie--Poisson
algebra $ C^\infty(\Lz^*)$. 
An extensive treatment of the role of Lie algebras in 
infinite-dimensional classical integrable systems (field theories in 
one and two space dimensions) see \sca{Roy Chowdhury} \cite{RoyC}.

In this section, we are however, only interested in the application to
spectroscopy and hence concentrate on the quantum case. For a quantum
mechanical system the requirement defining a {\bfi{dynamical algebra}}
is that $H$ is contained in the closure of the universal enveloping
algebra $\mathcal{U}(\Lz)$ of $\Lz$, equipped with a locally convex
topology such that potentials of the form $e^{-x^2}$ are allowed.

For example, the Heisenberg algebra $h(n)$ is the dynamical
algebra of symplectic classical systems with $n$ position degrees of
freedom, and of traditional Schr\"odinger quantum mechanics.
The hydrogen atom has additional rotational symmetry, and the
special properties of the Coulomb potential imply that one can in
fact find a fairly big dynamical algebra, namely $so(2,4)$, see
e.g. \sca{Wybourne} \cite{wybourne}. 

Now consider any Lie algebra $\Lz$ as a dynamical algebra. Call $\Ez$
the Lie--Poisson algebra associated to $\Lz$ for a classical case or
the universal enveloping algebra of $\Lz$ in the quantum case. The {\bf
\idx{symmetry algebra}} is the centralizer of the Hamiltonian in $\Ez$,
written $C_\Ez(H)$. In the `nicest' case one has $\Ez=C_\Ez(H)$, which
means that $H$ is a Casimir of $\Lz$. Normally, the Lie algebra
$\Lz$ describes the symmetries of the (unperturbed) system and thus
one would expect that the nicest case is the general case.

However, a very symmetric system is rarely studied in isolation, and
realistic systems are at best perturbations of nice systems. In this
case one gets broken symmetries, meaning that the Hamiltonian is only
almost a Casimir. Note that it might happen that the classical theory
has a symmetry, but that in the quantum version of the theory the
symmetry gets broken.
In case of a broken symmetry, one usually first tries to solve the
symmetric problem and then perturb the solutions to get approximate
solutions to the problem with broken symmetry.
We will not go into details about the mathematics of perturbation
theory, since this topic is amply treated in every book on quantum
mechanics. But we will consider some of its qualitative implications.

Suppose we have solved a symmetric problem. Then the solutions are
described as elements of some Hilbert space $\Hz$ on which $\Lz$ acts
unitarily. We can decompose the
Hilbert space into a direct sum of eigenspaces of the Hamiltonian;
$\Hz=\oplus_\lambda \Hz_\lambda$. Let $\psi$ be some eigenstate in
$\Hz_\lambda$ of the Hamiltonian and let $f\in \Lz$, then we see
\[
Hf\psi = [H,f]\psi + fH\psi =  [H,f]\psi +\lambda f\psi\,.
\]
Since $[f,H]=0$ we see that $\Lz$ maps each eigenspace into
itself. Thus all $\Hz_\lambda$ are $\Lz$-modules.

We call the eigenvalue $\lambda$ {\bf
  nondegenerate}\index{nondegenerate eigenvalue} if the dimension 
of $\Hz_\lambda$ is 1, and \bfi{degenerate}\index{degenerate eigenvalue}
if it is bigger than 1. (Dimension zero means that $\lambda$ is not an
eigenvalue.)

If $\lambda$ is degenerate, $\Hz_\lambda$ has
many essentially distinct bases of eigenvectors of $H$. One of
these is usually distinguished by the concrete representation used to
describe the module; $H$ and usually a distinguished
part of $\Lz$ act diagonally. In general the perturbed Hamiltonian no
longer acts 
diagonally on $\Hz_\lambda$, and as a result the level $\lambda$
usually splits into several distinct levels.
The energy level $\lambda$ splits into new levels that
are of the form $\lambda+\epsilon_j$ for some different but small
values $\epsilon_j$, giving rise to a fine structure. In
general, a fine structure implies that either a symmetry is broken
(the system reached a nonsymmetric state) or an external force that
broke the symmetry explicitly has been applied.

The induced representation of $\Lz$ on $\Hz_\lambda$
is unitary. Therefore knowing the irreducible unitary representations
of $\Lz$ can give information about the system under study.

\section{The hydrogen atom}

A hydrogen atom is a bound state of a proton (the nucleus) and an 
electron. It is most easily described by treating the much heavier 
nucleus as fixed (which amounts to neglecting recoil effects) and 
considering the electron as moving in the spherically symmetric 
electrostatic Coulomb field generated by the nucleus.

The electron is a spin $1/2$ particle, a fermion, meaning that it is
described by the spin $1/2$ representation of $so(3)$ on the
Hilbert space $L^2(\Rz^3,\Pz_1) \cong L^2(\Rz^3)\otimes\Pz_1$ 
defined in Section \ref{s.spinor}. 
Below we first discuss the orbital part of the wave
functions, i.e. the $L^2(\Rz^3)$-part. Then we discuss the dynamical
symmetries and how they get broken.

The orbital quantum states are
labeled by integers $n$, $l$ and $m$. The integer $n$ takes the
values $1,2,3,\ldots$ and the number $l$ takes for each fixed
value of $n$ the values $0,1,2,\ldots,n-1$. Finally, the number $m$
takes for each $l$ the values $-l,-l+1,\ldots,l-1,l$. 
Hence the (orbital) state of an electron is described by a state
\begin{equation}
\vert n,m,l\rangle ~~~ {\rm where}~~~ n\geq 1\,, ~~
0\leq l< n\,,~~ -l\leq m\leq l\,.
\end{equation}
The {\bfi{quantum number}} $n$ determines (to a first approximation)
the energy of the state:
\begin{equation}
E_n = -\frac{13.6 eV}{n^2}\,.
\end{equation}
The abbreviation $eV$ means electron Volt and is a unit for
energy. The quantum number $l$ specifies a representation of
$so(3)$. Thus we can make use of the representation theory of
$so(3)$ developed in Section \ref{s.su2rep}.

The electrostatic potential of the
hydrogen atom is $SO(3)$-invariant, hence it is not too surprising
that $SO(3)$-representations plays a role; the orbital part of the
electron wave function can be decomposed in representations of $SO(3)$.
The quantum 
number $l$ corresponds precisely to the irreducible representation
of $so(3)$ of integral spin $l$, that is, precisely to the 
representations of $so(3)$
that lift to $SO(3)$-representations. The quantum number $m$ labels 
the $\sigma_3$-eigenvectors of the representation and
corresponds to the eigenvalue $m$. 
The quantum number $n$ thus determines
which $SO(3)$-representations are allowed, and the $l$ and $m$ then
specify the representation and an eigenvector in this representation.

Now we shortly describe the relation between the quantum numbers and
the orbital wave function of the electron in the hydrogen atom. We can
give the hydrogen atom a coordinate system as follows. We put the
proton in the center and describe the position of the electron by a
radial coordinate $r$ measuring the distance between the proton and the
electron and by two angles $\theta\in [0,\pi ]$ and $\phi\in
[0,2\pi)$. The solutions to the Schr\"odinger equation for the
hydrogen atom are then given by
\[
\psi(r,\theta,\phi) = R_{n,l}(r)Y_{l,m}(\theta,\phi)\,.
\]
The radial part of the wave function $R_{n,l}$ is completely
determined by the quantum numbers $n$ and $l$ and is given by
\[
R_{n,l}=C_{n,l}e^{-\rho/2}\rho^l L_{n-l-1}^{2l+1}(\rho).
\]
Here $C_{n,l}$ is some constant such that $R_{n,l}$ is
normalized to integrate to one, the $L_{q}^{p}(\rho)$ are
generalized Laguerre polynomials (one of the well-known families of 
special functions);
$\rho $ is the normalized radius $\rho = \frac{2r}{na_0}$, and
$a_0$ is a constant called the {\bfi{Bohr radius}}. 
The angular part $Y_{l,m}$ of the wave function is given by
\[
Y_{l,m}(\theta,\phi) = K_{l,m}P_{l}^{m}(\cos \theta)e^{im\phi}\,,
\]
where the $K_{l,m}$ are normalization constants, and the $P_{l}^{m}$ 
are the {\bfi{associated Legendre polynomials}} given by
\[
P_{l}^{m}(x)=\frac{(-1)^m}{2^ll!}(1-x^2)^{m/2}
\left(\frac{d}{dx}\right)^{l+m}(x^2- 1)^l\,.
\]

\bigskip
\bfi{Symmetries and symmetry breaking}.
Nonrelativistically the electron in an electromagnetic field is treated
with the {\bfi{Pauli equation}}. The Pauli equation looks like the
Schr\"odinger equation, but has some extra terms, describing the
coupling of a spin $1/2$ particle to the electromagnetic
field. We now indicate why, in the case where the external
electromagnetic field is switched off, the symmetry group of the
Hamiltonian is $SO(4)\times SO(3)$.

The second factor in the symmetry group, the $SO(3)$, is the
symmetry group that acts on the spin of the electron. That is, it acts
on the $D_{1/2}$ part of $L^2(\Rz^3)\otimes D_{1/2}$.

The first factor in the symmetry group, the $SO(4)$, acts on the
space-part $L^2(\Rz^3)$ of the wave function. The Hamiltonian
of the hydrogen atom is rotationally invariant. 
Infinitesimal rotations are generated by the angular momentum 
$\J = \mathbf{r}\times\mathbf{p}$, where $\mathbf{r}$ is the
radius and $\mathbf{p}$ is the linear momentum and hence the angular 
momentum components describe the Lie algebra $so(3)$. However, there
exists an additional vector whose length is conserved: the length of
the {\bfi{Lenz--Runge vector}}. (Some people call it the
Laplace--Runge--Lenz vector, or even Laplace vector.) This leads to the
bigger group $SO(4)$; see e.g. \sca{Goldstein} \cite{goldstein}. 

\at{explain why dynamical groups
exist that are significant larger than one might naively expect.
The Lenz--Runge vector is one aspect of that story.}

To treat the electron relativistically one uses the Dirac
equation for a spin $1/2$ particle coupled to an electromagnetic
field. The coupling to the electromagnetic field can be done
in a quite easy way. Starting with the Dirac equation
\[
\left( \hbar\gamma\cdot \partial + mc\right)\psi=0
\]
one simply replaces the derivatives with $\partial_\mu - iqA_\mu$
 where the zeroth component of $A$ gives the Coulomb
potential and the spatial $\mathbf{A}$ components contain the magnetic
field via $B = \Nabla\times \mathbf{A}$; the parameter $q$ is
interpreted as the charge. We obtain
\[
\left( \hbar\gamma\cdot \partial -iq\hbar \gamma\cdot A +
  mc\right)\psi=0\,,
\]
where $\gamma\cdot A=\gamma^\mu A_\mu$.

The effect of having the fully relativistic coupling terms is that 
there is a coupling between the spin of the electron and the
orbital angular momentum of the electron.
The additional coupling terms in the
Hamiltonian are called {\bfi{spin-orbit coupling} terms}.
\at{expand:
intrinsic and orbital angular momentum (which remained distinct in the
non-relativistic case) cannot be invariantly distinguished in the
relativistic case.}

Due to the
coupling the separate $SO(3)$ of the spin gets destroyed; without
coupling there is a rotational symmetry group acting separately on the
orbit and on the spin and due to the coupling, the two rotational
symmetries are no longer independent. The angular momentum
$\mathbf{L}$ and the spin $\mathbf{S}$ are no longer separately
conserved in magnitude, but $(\mathbf{L}+\mathbf{S})^2$ is
constant. The symmetry group of the
relativistic hydrogen atom is therefore $SO(4)$. The spectrum that is
observed is called the {\bfi{fine structure spectrum}}.

Going even further and treating the hydrogen atom with quantum field
theory results in a further breakdown of the symmetry to the group
$SO(3)$. The group $SO(4)$ is isomorphic to $SO(3)\times SO(3)$ 
(see Section \ref{s.conn4}) and
corrections from quantum field theory break it down to the diagonal
subgroup $SO(3)$. The observed spectrum is called the {\bf
  \idx{hyperfine structure spectrum}}.

\section{Chains of subalgebras}\label{s.chains}

In more realistic situations,  the Hamiltonian is not
invariant under the total dynamical algebra $\Ez$, the universal
enveloping algebra of $\Lz$. In this case, the
Hamiltonian is not in the center of $\Ez$, but we can consider the
centralizer of $H$ in $\Lz$. The centralizer of $H$ in $\Lz$ is a
subalgebra of $\Lz$, and is therefore a Lie subalgebra of $\Ez$ and we
denote it by $\Lz_1$. We thus have $H\in C_{\Ez}(\Lz_1)$. The Lie
subalgebra $\Lz_1$ generates a subalgebra of $\Ez$, which we denote by
$\Ez_1$.

In simple applications, it often happens that the Hamiltonian $H$ is a
function $H(C_0,C_1)$ where $C_0$ is a Casimir of $\Lz$ (that is, it
is a central element of $\Ez$) and where $C_1$ is a Casimir of $\Lz_1$
(in the center of $\Ez_1$). In more complicated applications we have a
series of approximations to the problem, as explained for the hydrogen
atom before, where relativistic and quantum field theory effects
modify the Hamiltonian. In each step one modifies the
Hamiltonian by adding terms with fewer and fewer symmetries, and the 
symmetry algebra is reduced to correspondingly smaller
subalgebras. We thus have a sequence of subalgebras
\[
\Lz = \Lz_0 \supseteq \Lz_1\supseteq \ldots \supseteq \Lz_n=\hat
\Lz\,.
\]
The final subalgebra $\hat \Lz$ commutes with $H$. 
The generated subalgebra
of $\Ez$, denoted $\hat \Ez$ centralizes $H$ in $\Ez$. If the
Hamiltonian is a function $H=H(C_0,\ldots,C_n)$ where $C_k$ is a
Casimir of $\Lz_k$, the scheme gives
explicitly solvable problems. For example, for the nonrelativistic
hydrogen atom without spin, one finds a series
\[
so(4)\supset so(3) \supset so(2)\supset 1 \,.
\]
Of course, there are many Hamiltonians that cannot be represented
as functions of a chain of Casimirs, but the above scheme covers
many applications, and is a starting point for a perturbative treatment
of many others.

In classical symplectic mechanics one relates the Lie algebra
$\hat \Lz$ to so-called {\bfi{action variables}} and the steps to
$\Lz_0$ are constructed using conjugate {\bfi{angle variables}}.
We will not go into the details defining variables and the related 
techniques. \at{refs}

Consider the situation where $H=H(C_0,C_1)$, that is, the
simple application. We write $H=H_0+H_1$ where $H_0$ is only a
function of $C_0$ and $H_1$ depends on $C_0$ and $C_1$. As before,
we suppose we have realized the
elements of $\Ez$ (and thus of $\Lz$) as operators on some Hilbert
space $\Hz$. We assume that the subspaces $\Hz_\lambda$ on which the
Hamiltonian $H_0=H_0(C_0)$ acts diagonally are finite-dimensional.
This is for example the case for the hydrogen atom. We furthermore
split up $\Hz_\lambda$ in irreducible representations of $\Lz_0$
so that we may assume that $\Hz_\lambda$ is irreducible.
Modifying $H_0$ to $H_0+H_1$ means that the symmetry algebra becomes
smaller; it 
becomes $\Lz_1$. We can restrict the representation of $\Lz_0$ on
$\Hz_\lambda$ to the subalgebra $\Lz_1$ to obtain a representation of
$\Lz_1$. In most cases this representation is reducible and we
write the decomposition of $\Hz_\lambda$
into $\Lz_1$ irreducibles as
\[
\Hz_\lambda  = \bigoplus_{\mu} \Hz_{\mu}^{(1)}\,.
\]
More generally, suppose we have a sequence of subalgebras $\Lz_0\supset
\Lz_1\supset \ldots\supset \Lz_n$ and related Casimirs $C_i\in
Z(U(\Lz_i))$. It follows that the $C_i$ commute among each other;
$C_1$ is in the center of $U(\Lz_1)$, which contains $U(\Lz_k)$ for
$k\geq 2$ and hence $C_1$ commutes with $C_2$, $C_3$, and so on. Thus
on the irreducible representations of $\Lz_1$ the Casimirs act
diagonally. Hence we can assign to each representation of $\Lz_n$
appearing in the decomposition of the original representation $\Hz$
numerical values $\theta_0, \ldots, \theta_n$, corresponding to the
eigenvalues of the $C_i$. Given a physical state $v$ in a
representation of $\Lz_n$ corresponding to the label
$(\theta_0,\ldots,\theta_n)$ we see that the Hamiltonian acts as
\[
H(C_0,\ldots,C_n)v= H(\theta_0,\ldots,\theta_n) v\,,
\]
where on the right-hand side the Hamiltonian is an operator and on the
right hand side $ H(\theta_0,\ldots,\theta_n)$ is a number.

\bigskip
{\bfi{Branching rules}.}
In a lot of favorable cases, for example when the $\Lz_i$ are simple,
the decomposition of an irreducible representation of the large
algebra into irreducible representations of a maximal subalgebra
is known. These decompositions go under the name of \bfi{branching
rules}; splitting up a representation
under a subgroup or subalgebra is called \bfi{branching}.

Let us give an example of a branching rule and look at the
fundamental representation of $su(3)$, that is, the $3$-dimensional
representation of $su(3)$ that defines the Lie algebra $su(3)$. 
The Lie algebra elements are faithfully represented as $3\times
3$-matrices $X$ that are antihermitian; $X^\dagger+X=0$. Now we
consider the Lie subalgebra $su(2)$. There are different ways we can
embed $su(2)$ into $su(3)$, but it turns out that all of them are
equivalent. We can always choose a basis $e_1$, $e_2$, $e_3$ in
$\Cz^3$ such that $su(2)$
only acts nontrivially on the subspace spanned by $e_1$ and $e_2$. We
thus realize $su(2)$ inside $su(3)$ as the following matrices in
$su(3)$
\[
\pmatrix{ a & b & 0 \cr c & d& 0 \cr 0 & 0 & 1  }\,,~~~\mbox{where}~~
\pmatrix{ a & b \cr  c & d  }\in  su(2)\,.
\]
We see that the three-dimensional representation
of $su(3)$ splits into two irreducible representations of $su(2)$, the
trivial one, spanned by $e_3$ and the two-dimensional (fundamental)
representation spanned $e_1$ and $e_2$. One writes this in
shorthand as: $\mathbf{3\to 2+1}$ under $su(2)\subset su(3)$. In the
reference \sca{Slansky} \cite{slansky}, one can find tables of
branching rules. 

{\bfi{Clebsch--Gordan coefficients}.}
In an important special case one can relate the branching rules
to the so-called Clebsch--Gordan coefficients, which are widely used
in physics.

Let us explain 
the Clebsch--Gordan coefficients for $su(2)$. Given two
representation $D_l$ and $D_k$ (see Section \ref{s.su2rep}), 
we can form the tensor 
product $D_k\otimes D_l$. An $su(2)$-element $x$ acts on $v\otimes w$ 
by mapping 
$v\otimes w$ to $x(v)\otimes w + v\otimes x(w)$, where we 
write $x(v)$ for the action of $x$ on $v$. In general, the
representation $D_k\otimes D_l$ is not irreducible, and  we have
\[
D_k\otimes D_l
= D_{k+l}\oplus D_{k+l-2}\oplus \ldots \oplus D_{|k-l|}\,.
\]
The precise decomposition of a vector $v\otimes w$, where $v$ and $w$
are eigenvectors of $\sigma_3$, into the
irreducible components is given by the Clebsch--Gordan
coefficients. For the vector in the $D_k$ representation with
$\sigma_3$ eigenvalue $m$ and norm one we write $|k,m\rangle$. If the 
representation $D_K$ is inside the tensor product of $D_{k_1}$ and
$D_{k_2}$, we can decompose any vector $|K, M\rangle$ as a sum of
vectors of the form $|k_1,m_1\rangle\otimes |k_2,m_2\rangle$ and the
Clebsch--Gordan coefficients are then the coefficients in the
decomposition
\[
|K,M\rangle = \sum_{k_1,k_2m_1,m_2} C_{k_1k_2m_1m_2}^{KM}|k_1,m_1
\rangle \otimes |k_2,m_2\rangle \,.
\]
More generally, the Clebsch--Gordan decompositions say how the tensor
product of two irreducible representations $V_1$ and $V_2$ of a
compact Lie group (or its Lie algebra) decompose into irreducible
representations $W_i$ as $V_1\otimes V_2=\oplus_j W_j$. The
Clebsch--Gordan coefficients are the numerical coefficients in the
projection from one of the summands $W_j$ to $V_1\otimes V_2$.

Now suppose that $\Lz_0=\Lz'\oplus \Lz'$ decomposes into two copies 
of the same Lie algebra $\Lz'$. An important choice of $\Lz_1$ is
the diagonal Lie subalgebra given by elements of the form $(a,a)$ with
$a\in  \Lz'$. Then as a Lie algebra $\Lz_1\cong \Lz'$. 
The irreducible representations of $\Lz_0$ are given by
tensor products of representations of $\Lz_1$. Therefore, decomposing
the irreducible representations of $\Lz_0$ with respect to $\Lz_1$
amounts to giving the Clebsch--Gordan decompositions.

Much more could be said on the topic of symmetries and broken 
symmetries in physics. A nice overview is given in a paper by
\sca{Bijker} \cite{Bij}. It shows how the symmetry concept organizes
not only the world of atoms and molecules that we considered here,
but also that of elementary particles. The isospin symmetry between 
protons and neutrons has a symmetry group $SU(2)$, which extends to the
flavor symmetry group $SU(3)$ for the three light quarks. 
Applications to molecular spectra and the \bfi{interacting boson model}
for modelling atomic nuclei are discussed extensively in the book by
\sca{Frank \& van Isacker} \cite{FraI}. 

Quantum field theory, culminating in the standard model,
is also based on symmetries, namely the space-time symmetries of 
the Poincar\'e group, and a gauge group $U(2)\otimes SU(3)$ which
combines the broken symmetry group $U(2)=U(1)\otimes SU(2)$ of the weak 
interaction (of which only a diagonal subgroup $U(1)$ encoding
the electromagnetic charge is unbroken) with the unbroken color 
symmetry group $U(3)$ of the strong interaction. 

While these topics lie far beyond the scope of this book,  
the interested reader will take the next step and consult deeper 
work of others who studies this in depth. Our journey is finished.


    \backmatter
    \addcontentsline{toc}{chapter}{References}
    \bibliographystyle{plainurl}
    \bibliography{QML}

    \printindex[aut]
    \printindex     

\end{document}